\documentclass[final,1p,times]{elsarticle}

\usepackage{amssymb}
\usepackage{color}

\journal{Physics Reports}

\newcommand{\degree}{{\rm o}}
\newcommand{\im}{\mbox{Im }}
\newcommand{\re}{\mbox{Re }}
\newcommand{\mps}{M_\pi^2}

\newcommand{\gev}{{\mbox{GeV}\,}}
\newcommand{\mev}{{\mbox{MeV}\,}}
\newcommand{\kev}{{\mbox{keV}\,}}
\newcommand{\imag}{{\mbox{Im}\,}}

\newcommand{\dd}{\hbox{d}\,}
\newcommand{\dslash}{D\!\!\!\!\!\!\not\,\,\,}
\begin{document}

\begin{frontmatter}



\title{From controversy to precision on the sigma meson:\\ 
a review on the status of the non-ordinary $f_0(500)$ resonance}


\author{Jos\'e R. Pel\'aez}

\address{Departamento de F\'{\i}sica Te\'orica II. Universidad Complutense. 28040 Madrid. SPAIN}

\begin{abstract}
The existence and properties of  the sigma meson have been 
controversial for almost six decades, 
despite playing a central role in the spontaneous chiral symmetry of QCD or 
in the nucleon-nucleon attraction. This controversy has also been fed by the
strong indications that it is not an ordinary quark-antiquark meson.
Here we review both the recent and old experimental data and the
model independent dispersive formalisms which have provided precise determinations
of its mass and width, finally settling the controversy and leading to its new name: $f_0(500)$. 
We then provide a rather conservative average of the most recent and advanced dispersive determinations of its pole position $\sqrt{s_\sigma}=449^{+22}_{-16}-i(275\pm12)$.
 In addition, after
comprehensive introductions, we will review within the modern perspective of effective theories
and dispersion theory, 
its relation to chiral symmetry, unitarization techniques, its quark mass dependence, popular models,
as well as the recent strong evidence, obtained from the QCD $1/N_c$ expansion
or Regge theory, for its non ordinary nature 
in terms of quarks and gluons. 
\end{abstract}


\end{frontmatter}



\tableofcontents
\section{INTRODUCTION}
\label{sec:intro}

\begin{flushright} {\it ``Nothing clears up a case so much as stating it to another person.''\\  
Sherlock Holmes Quote.\\Silver Blaze. Sir Arthur Conan Doyle, 1892.}
\end{flushright}

For researchers outside the field, it may be surprising that 
despite having established Quantum Chromodynamics (QCD) as
 the fundamental theory of the Strong Interaction 40 years ago, 
the spectrum of lowest mass states, and
particularly that of scalar mesons, 
may be still under debate. 
Actually, light scalar mesons have been a puzzle in our understanding of 
the Strong Interaction for almost six decades.
This may be even more amazing given the fact that
they play a very relevant role within nuclear and hadron physics, as in the nucleon-nucleon attraction and in the spontaneous breaking of chiral symmetry, both of them fundamental features of the Strong Interaction. 
The relatively poor theoretical understanding of hadrons at low energies
causes little surprise since it is textbook knowledge that QCD becomes 
non-perturbative at low energies and does not allow for precise calculations of the light hadron spectrum. 
However, young and not so young \color{black} people outside the field
\color{black}
are often unaware of the fact that even basic empirical properties 
such as the existence of many of the lightest mesons and resonances are still actively discussed, even if they were suggested much before QCD was proposed. 
Moreover, it is often the case that older non-practitioners think that 
no rigorous conclusions or progress can be made about light scalars, particularly about the lightest one, traditionally known as $\sigma$ meson
and nowadays called the $f_0(500)$. This attitude is due to the fact that
the situation on how many states exist, if they exist at all, what are their masses, widths, etc... has remained rather confusing for many decades. Admittedly, the way that, for instance, the lightest meson---the $\sigma$ resonance---has been listed in the Review of Particle physics (RPP) \cite{PDG96,PDG12}, which until 2010 considered it a ``well-established'' state despite quoting its mass in a range between 400 and 1200 MeV, 
did not help in conveying the rigorous efforts pursued both by theoreticians and experimentalists within the light hadron physics community. The continuous efforts of this community have considerably clarified the situation and the modern formalisms have allowed us to slowly  move away from a time of confusion to a time of precision studies. Actually, these efforts have been recently recognized in the last RPP 2012 edition \cite{PDG12}, at least for $\sigma$ particle; perhaps the most controversial light meson for many years, whose mass uncertainties have been reduced by a factor of 5 and, accordingly, has even changed its name to $f_0(500)$.

The purpose of this work is to provide a review of the present status of the $\sigma$ meson and its relevant role within particle and nuclear physics paying particular attention to the developments that triggered this major revision in the RPP. Thus, after a general and historical introduction in this section, the recent developments that have led to this final acceptance of the $\sigma$ parameters will be reviewed in detail in Sec.\ref{sec:parameters}. Next, Sec.\ref{sec:chiralsigma} will be devoted to the role of the $f_0(500)$ in the spontaneous breaking of chiral symmetry by the Strong Interaction and the modern approaches based on Chiral symmetry and effective theories. The last section will be dedicated to the spectroscopic classification and the growing evidence supporting a non-ordinary nature  of the $\sigma$, i.e. its predominantly non quark-antiquark composition.

\subsection{Historical perspective}
\label{subsec:history}

In order to  illustrate the confusing situation of light scalars over the last decades
and to gain perspective on the significance of recent progress, it is instructive 
to review briefly the History of the $\sigma$ meson. 
\color{black}
Unfortunately the description of a confusing situation might result confusing to the reader, since at some given time two conflicting results could coexist or some previous advances
may not have received full recognition. Some relevant results might be forgotten for some time and resurrected later. Thus, the history of the $\sigma$ has advanced forward and backwards,
After this historical perspective, Subsecs.~\ref{subsec:statusparameters} and \ref{subsec:statusspectroscopy} 
will summarize the present situation of the $\sigma$, 
where the controversy and confusion has disappeared in many aspects.
In Sections 2,3 and 4 the historical perspective will be abandoned and 
we will describe in detail the methods that have lead us to the present situation.
\color{black}

\subsubsection{The pre-QCD era: Nucleon attraction, isospin and chiral symmetry}
On the theory side, a relatively light ``neutral scalar meson''
was introduced by Teller and Johnson as early as 1955 \cite{Johnson:1955zz} in order to explain the nucleon-nucleon attraction.
Very soon such a field was incorporated by Schwinger \cite{Schwinger:1957em}
 into a unified description of the known particles in terms of isotopic spin, of which this field was a singlet, and he called it $\sigma$. 
He already pointed out that while the pions, which form a triplet, were known, such 
a $\sigma$ field had not been observed.  Nevertheless Schwinger already remarked that 
  if its mass was above the two-pion threshold  it would be highly unstable and not easily observable. 
This is indeed what happens and, of course, the origin
of the long debate about the existence of the $\sigma$ meson.

\begin{figure}
\centering
\includegraphics[scale=0.49]{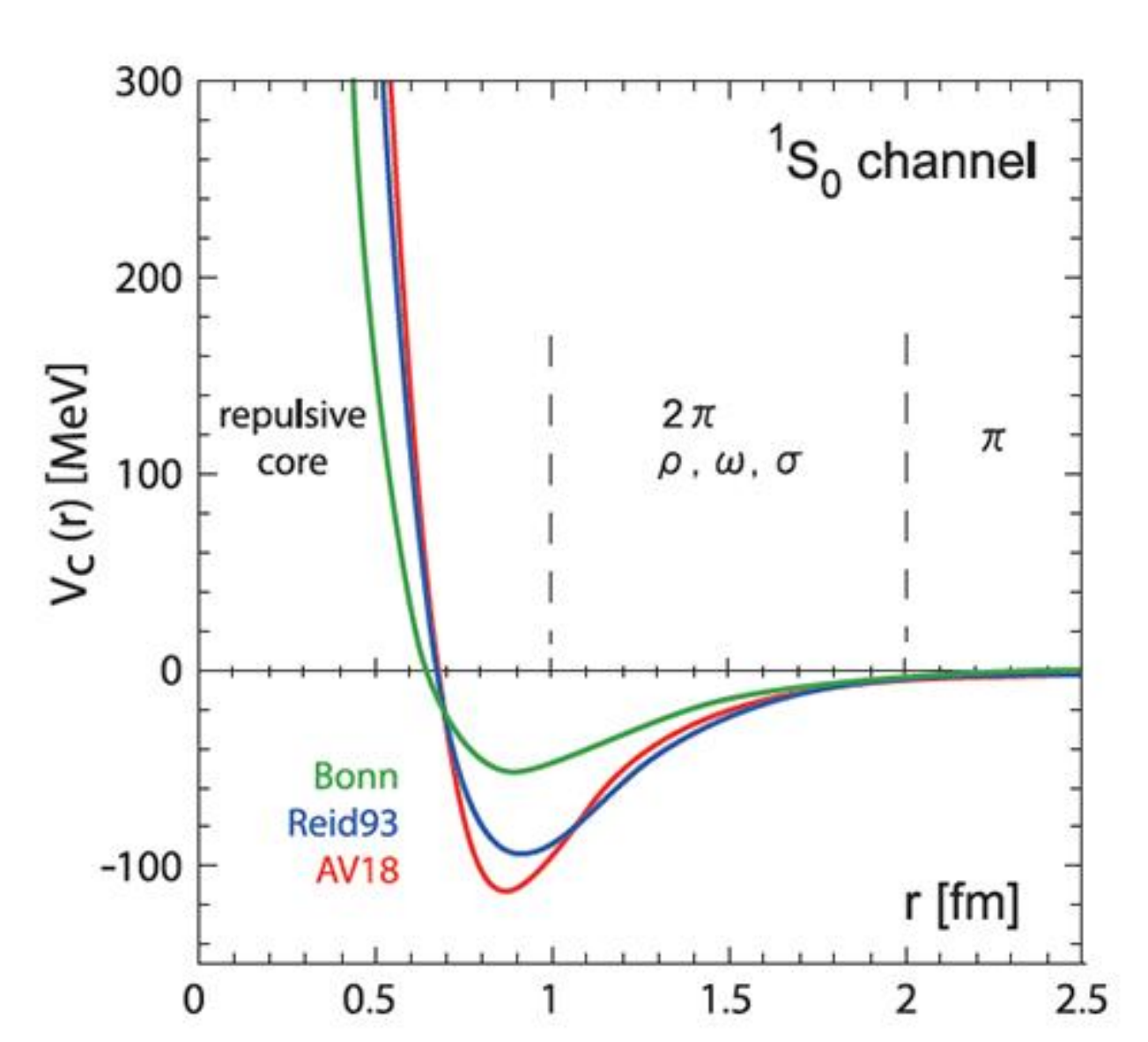}
 \caption{\rm \label{fig:NNPotential} 
A field with the quantum numbers of the $\sigma$ meson was first proposed in order to explain the nucleon-nucleon attraction \cite{Johnson:1955zz}.
The figure, from \cite{Ishii:2006ec}, shows the shape of the NN potential in the $^1S_0$ channel within different modern phenomenological models. Typically, the attractive intermediate part is dominated by the $\sigma$-meson or some sort of correlated
two-pion exchange with the $\sigma$ quantum numbers.
The curves come from \cite{Machleidt:2000ge} (Bonn), \cite{Stoks:1994wp} (Reid93) and \cite{Wiringa:1994wb} (AV18).
Reprinted with permission from N.~Ishii, S.~Aoki and T.~Hatsuda,
  Phys.\ Rev.\ Lett.\  {\bf 99}, 022001 (2007). Copyright 2007 by the American Physical Society. \cite{Ishii:2006ec}}
\end{figure}

In the early sixties an isoscalar-scalar meson with a mass around 540 MeV was already being considered 
within ``one-boson-exchange models'' to explain nuclear forces in relative detail (see for instance the review in \cite{Taketani}).
Even at present, as we can see for example in Fig.\ref{fig:NNPotential}, the most common purely phenomenological models of the nucleon-nucleon interaction \cite{Machleidt:2000ge,Stoks:1994wp,Wiringa:1994wb}, are based on the exchange of bosons and contain a component due to the $\sigma$ meson, which provides the main part of the strong attraction in the 1 to 2 Fermi range. Sometimes the $\sigma$ contribution is referred to as ``correlated two-pion'' exchange \cite{Reuber:1995vc}, ``complicated two-pion exchange'', etc... Note that other mesons 
dominate other parts of the potential, like the $\omega$ producing a short range repulsion, 
or the pion being responsible for the long distance tail, whereas the $\rho$ has a relatively smaller contribution.
The fact that the $\sigma$ plays such an important role in nucleon attraction, and therefore in nuclei formation, makes it particularly relevant for Cosmological and Anthropic considerations that will be addressed in Sec.\ref{sec:chiralsigma}.

Back to History,  the $\sigma$ meson was 
incorporated in the early sixties into simple models of the Strong Interaction, 
like the Linear Sigma Model \cite{GellMann:1960np} (L$\sigma$M), that we will study in Sec.\ref{subsec:lsm}. 
Very briefly, in this model the $\sigma$ belongs to a multiplet of four scalars that 
suffers a spontaneous symmetry breaking such that the other three become massless Nambu-Goldstone bosons (NGB).
In contrast, the $\sigma$ field remains massive. If this broken symmetry is identified 
with the SU(2) chiral symmetry 
that QCD would have if its two lightest quarks $u$ and $d$ were massless, then
the three NGB  could be identified with the pions. In reality the two
lightest quarks have a small mass that can be added to this picture perturbatively and as a consequence  
pions inherit a mass which is small compared to typical hadronic scales. For this reason they are called pseudo-NGB. 
This implementation of chiral symmetry breaking 
is rather simple due to the linear realization of its symmetries and has become very popular, capturing many essential features of pion physics at energies lower than the mass of other resonances not included in the model. 
It is also easily extended to take into account the somewhat heavier strange quark, thus yielding eight pseudo-NGB which, besides the three pions, include the four kaons and the eta meson. In this case there is not just the sigma, 
but a nonet of scalars to which the sigma belongs together with other light scalar mesons 
nowadays called the $f_0(980)$,
the $a_0(980)$  and the $K^*_0(800)$ or $\kappa$ meson. The latter is in many aspects very similar to the $\sigma$, 
including its extremely short lifetime. Since some of these states are controversial too, this spectroscopic classification has also been the subject of a longstanding debate, which seems much clearer today and we will address in the last section of this review. As we will see in Sec.\ref{sec:chiralsigma},
the L$\sigma$M is not the correct effective theory of QCD at low energies, but it has played a very important role in our phenomenological and more intuitive description of low energy hadronic interactions. The L$\sigma$M has its successes and shortcomings, leading to multiple variations (see, for instance 
\cite{Gasiorowicz:1969kn,Basdevant:1970nu,Schechter:1993tc,Chan:1974ra,Achasov:1994iu,Tornqvist:1995kr,Tornqvist:1995ay}
and references therein), some of which we will review in Subsecs.\ref{subsec:other}
and \ref{subsec:models}.

Not only within the L$\sigma$M, but on general grounds, 
the $\sigma$, which has the quantum numbers of the vacuum,  
plays a very important role in the dynamics of the QCD spontaneous chiral symmetry breaking. 
Actually, a relatively light
 scalar-isoscalar meson coupling strongly to $\pi\pi$, and therefore very wide,
is also generated in a Nambu-Jona-Lasinio (NJL) model or its modifications within
hadron physics \cite{NJL}. 
Still, both the L$\sigma$M and NJL are simple models that, despite capturing
many relevant features, neither provide a systematic
description of low energy hadron physics nor a clear connection to QCD, 
usually requiring some ad-hoc modifications.

Therefore, from the theoretical side, from the late 50's to the late 60's it became clear that
the existence and properties of the $\sigma$ were very relevant for our 
understanding of the nucleon-nucleon attraction and 
chiral symmetry in Strong Interactions. These are already strong motivations for the interest on the $\sigma$ meson.
But, in addition, we will see later that with the advent of quarks, gluons and QCD in the 70's, the $\sigma$ and other scalar-isoscalar mesons 
acquired further interest related to the the identification of glueballs and the composition of mesons. 

\subsubsection{To be or not to be: The $\sigma$ History in the Review of Particle Properties.}

Of course, in order to address these issues properly, a reliable and precise determination 
of the $\sigma$ mass, width and couplings is needed. Moreover, 
the experimental situation of the $\sigma$ 
has been so confusing that even its 
very existence has been frequently challenged, 
although it is now firmly established. 

In the next subsections  this controversial status \color{black} through time
will be illustrated in detail. \color{black}
However, before getting into such a detailed explanation,
\color{black} let us \color{black} sketch it briefly following the different developments 
as compiled in the Review of Particle Properties (RPP). This review 
provides every two years what could be considered as a conservative, 
consensual summary of the state of the art. In the next subsections 
further comments and details not always present in the RPP \color{black} will also be included.

Thus, let  us briefly summarize the sigma evolution on the RPP. \color{black}
Very soon after it was suggested theoretically, a narrow $\sigma$ already appeared in the 1964 edition
of the RPP---at that time called ``Data on Elementary particles and Resonant states'' \cite{PDG64}.
Several references were listed with a mass  around 390 MeV and a width between 50 and 150 MeV. 
The possibility that it was a broad resonance
was already being considered in the RPP 1967 edition \cite{PDG67}, following \cite{Lovelace66}.
From the first ``Review of Particle Properties'' edition in 1969 \cite{PDG69} 
until 1973 \cite{PDG73}, a $\sigma$ meson or some other relatively light scalar-isoscalar meson resonance
 under different names 
($\epsilon$, $\eta_{+-}$), appeared in the particle listings, sometimes simultaneously, although 
they were not considered well established states. Actually, in the 1969 edition there were two
$J^P=0^+$ states: a wide $\sigma(410)$ and a narrower  
$\eta_{0+}(720)$, also called $\epsilon$ sometimes. 
From the 1976 edition \cite{PDG76}, such states disappeared from
the tables for 20 years, returning in 1996 \cite{PDG96} under the name of ``$f_0(400-1200)$ or $\sigma$'', with a 
large mass uncertainty ranging from 400 to 1200 MeV and a similarly large range, 
from 500 to 1000 MeV, for the width. Its entry was accompanied with one 
footnote stating that ''the interpretation of this entry as a particle is controversial'',  whereas
another footnote in the ``Non-$q\bar q$ candidates'' section  
stated that it was
``considered as well-established''. Despite keeping these large uncertainties on its parameters, and the same comments about its status, its name was changed to $f_0(600)$ in the 2002 edition. 
Moreover, both the large uncertainties and the name were kept until the 2010 edition. 
As commented before, the $\sigma$ 
has finally suffered a major revision in the 2012 edition, 
changing its name to $f_0(500)$ and reducing its 
mass uncertainties by a factor of 5. In addition, there was a rather large change in 
the central value of the width 
and a very substantial reduction on its uncertainties.

There are several reasons for the very large uncertainties and this very confusing coming in 
and going out of the tables for scalar-isoscalar states below 1 GeV.  Let us recall that it was already Schwinger in his 1957
work \cite{Schwinger:1957em} who pointed out that the sigma could be wide and difficult to observe.
Actually, despite being first suggested within the context of nucleon-nucleon attraction, 
this interaction is not very sensitive to the details of the particles exchanged, 
even less so if they are very wide \cite{Flambaum:2007xj}, as it is the case of the $\sigma$
 meson. Intuitively this can be understood as follows:  phenomenologically
boson exchanges contribute
 to the nucleon-nucleon potential in the $t$-channel. 
This means that the exchanged bosons do not resonate as it happens in other processes 
(like resonant annihilation)
where the direct $s$-channel exchange of a resonance is possible. 

Therefore,  many of the lightest mesons have been traditionally studied in meson-meson scattering, 
where such resonances can be produced in the $s$-channel. For the quantum numbers we are interested in,
this means studying pion-pion scattering, 
or systems were $\pi\pi$ scattering is needed as a part of a larger process. 
Unfortunately,  $\pi\pi$ scattering is also not observed directly and its indirect extraction from data needs some modeling and 
has 
large systematic uncertainties leading to many data sets incompatible among themselves. As we will see throughout this review,
a great amount of work has been needed both on the experimental 
side to measure
scattering phases and inelasticities and on the theoretical side
to sort out what 
data sets are fully consistent with fundamental requirements.
But even when this is settled, the identification and determination of the $\sigma$ parameters is hindered by its large width and unusual shape.

\subsubsection{The 70's and meson-meson scattering phases. No Breit-Wigner resonance found.}

Experimental claims of several narrow $\pi^+\pi^-$
resonances in the 400 to 500 MeV region were made as early as 1962 \cite{samios}. However, pretty soon 
it became clear that such narrow resonances (50-70 MeV widths) were not confirmed.
The first suggestions that there might actually be a behavior characteristic of a wide resonance
in the phase shifts of the scalar-isoscalar sector in the 650 to 800 region 
were given somewhat later in the 60's. Quite a few works were dedicated to this issue ( see
 \cite{sigmainpipipre70} and references therein) following the first attempts to measure
the scattering
phase shift, although the broad $\sigma$ scenario had already been advocated 
in \cite{Lovelace66} from a dispersive study of $\pi N$ elastic scattering. 
In these works,
there was some
evidence for the phase to reach $90^\degree$ around 750 MeV, giving some weak support for a
so called ``$\epsilon$ resonance''. However, there was no 
evidence for a fast phase motion, which therefore implied a broad state, if any, 
perhaps combined with another $\sigma(400)$ broad meson.  This is why, as commented above,
in the first RPP 1969 edition \cite{PDG69}, a $\sigma(410)$ and a narrower  $\eta_{0+}(720)$ (or $\epsilon$)
coexisted. However, in the 1971 edition \cite{PDG71}, the latter was changed to $\eta_{0+}(700-1000)$ (or $\epsilon$),
although at the same time a $\eta_{0+}(1070)$ or $S^*$ was listed too. 
In the 1973 edition \cite{PDG73} the $\sigma$ disappeared from the tables but
an $\epsilon(600)$ resonance was listed instead, with a note saying that the existence of its
 pole was ``not established".  In the same 1973 edition the $\eta_{0+}(1070)$ became the $S^*(1000)$, which was considered ``well established".
 Nowadays this state, which is very narrow, is known as the $f_0(980)$ and is still listed in the RPP.

\begin{figure}
\vspace*{-3.3cm}
  \centering
 \includegraphics[width=0.5\textwidth]{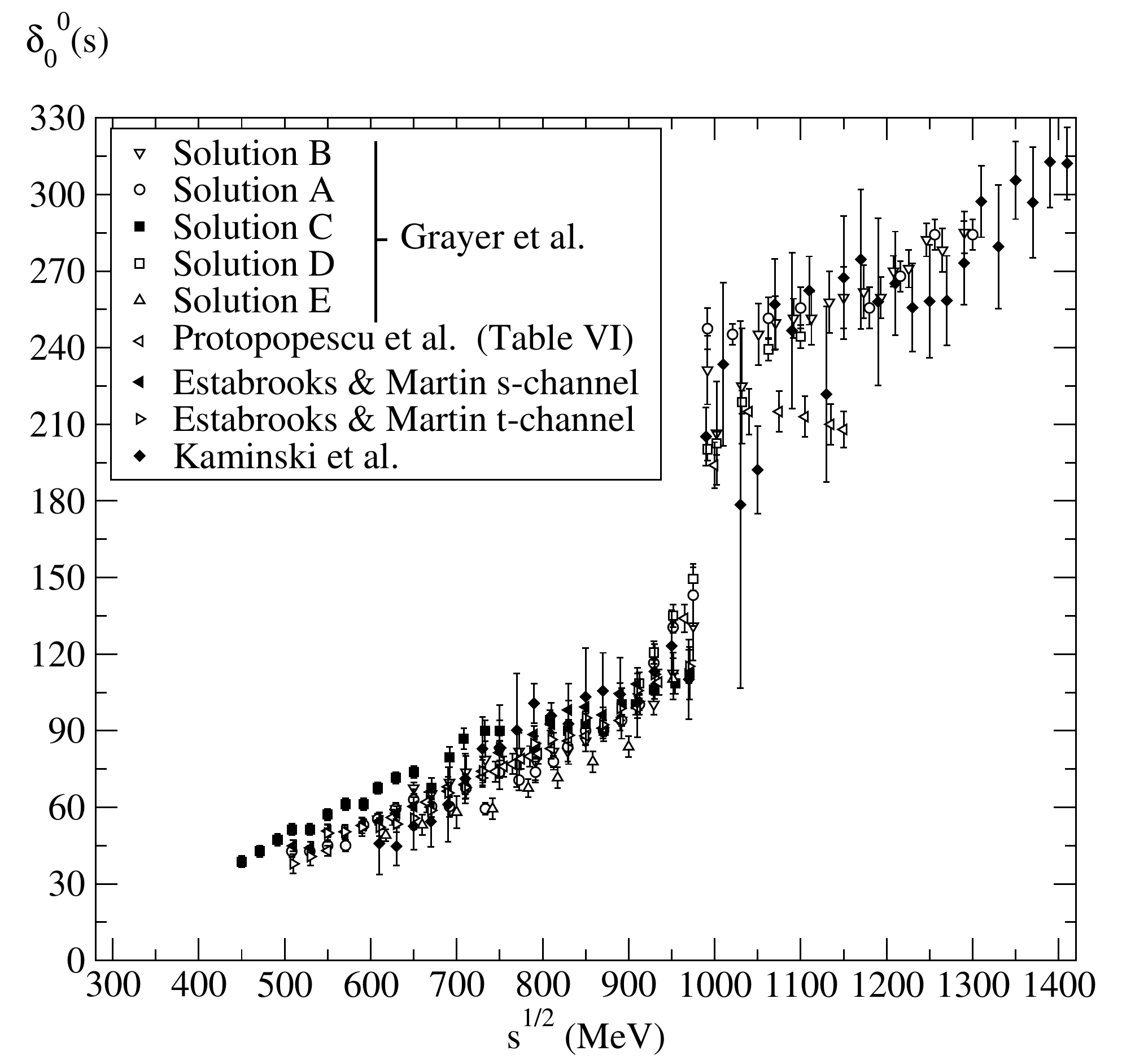}
\hspace*{-.2 cm}
 \includegraphics[width=0.5\textwidth]{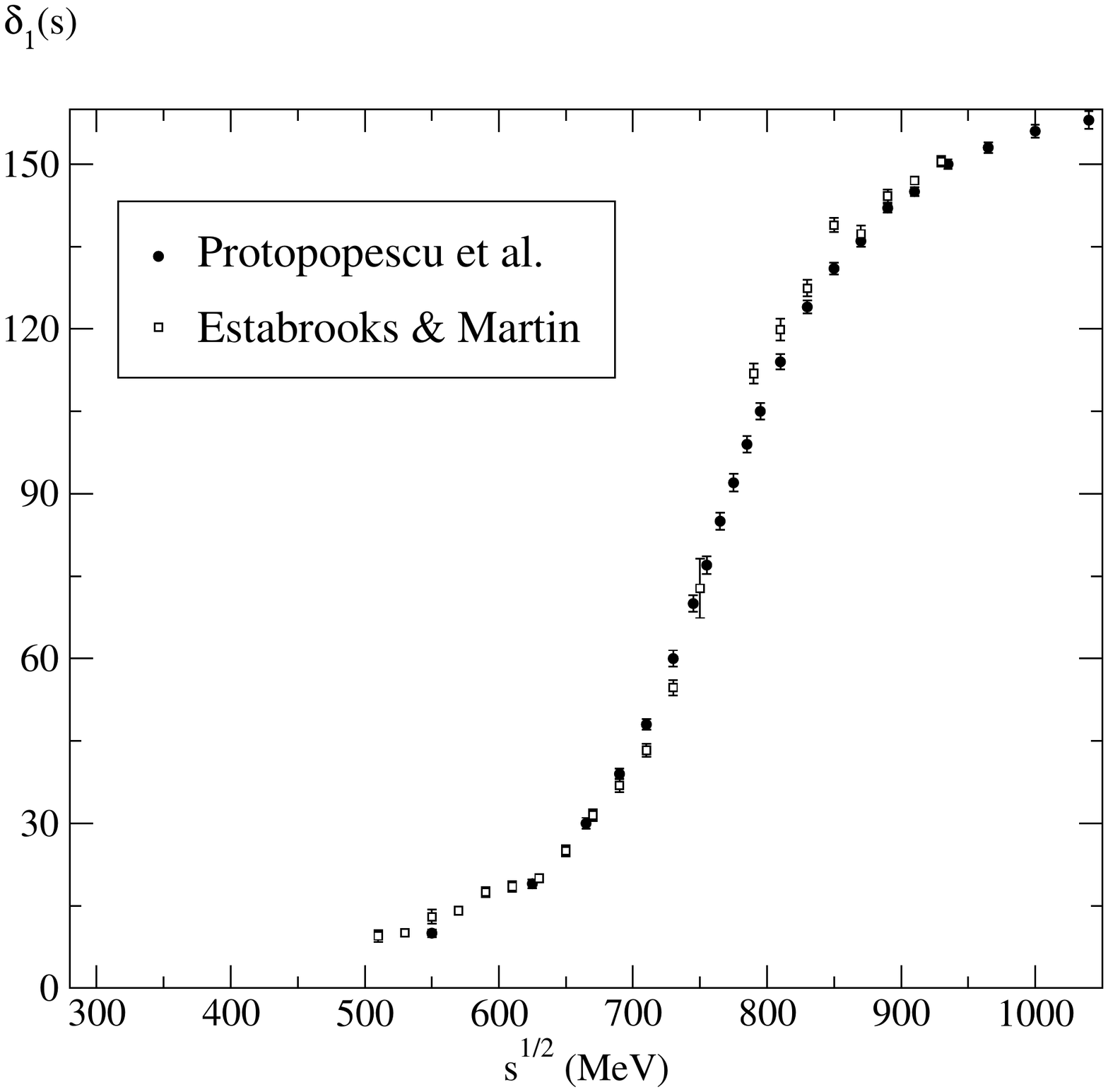}
\vspace*{-.5cm}
  \caption{ Data on $\pi\pi\rightarrow\pi\pi$ scattering phase shifts: Protopopescu et al. from \cite{Pr73}, Grayer et al. from \cite{Cern-Munich} (Solution B also from \cite{Hyams:1973zf}), Estabrooks and Martin from \cite{Estabrooks:1974vu}, Kaminski et al. from \cite{Kaminski:1996da}. 
Left panel: The scalar-isoscalar  phase shift $\delta_0^{(0)}$. Note the 
very large differences due to systematic uncertainties, which exist even within data sets 
from the same experimental collaboration \cite{Cern-Munich} (Something similar happens with \cite{Pr73}, but we only show the most commonly used and consistent data set). Please note that there is no Breit-Wigner-like 
sharp increase of $180^\degree$ on the phase between threshold and 800 MeV. Such sharp phase increase is seen around 980 MeV, corresponding to the $f_0(980)$ meson, although starting  over a background phase of about $100^\degree$ degrees. Right panel: For comparison we also show the vector-isovector $\delta_1$ phase shift, where the $\rho(770)$ resonance can be seen to follow the familiar Breit-Wigner shape \cite{Breit:1936zzb} to a very good degree of approximation.
}\label{fig:00data}
\end{figure}

Experimental analysis of $\pi\pi$ scattering phase shifts from $\pi N\rightarrow\pi\pi N$
became available
in the very early seventies \cite{firstphaseshifts}.
However, most of those were soon superseded by the best known experimental determinations of 
the $\pi\pi$ scattering scalar-isoscalar scattering phases by 
Protopopescu et al. \cite{Pr73} 
and the CERN-Munich  Collaboration, which has three works:
Hyams et al. in 1973 \cite{Hyams:1973zf}, Grayer et al. in 1974 \cite{Cern-Munich}
and Hyams et al. in 1975 \cite{Cern-Munich-high}. 
As we will see, these are the most
extensive phase shift analyses, covering a range of energies from above 500 to well beyond 1.5 GeV
and have been widely used together with some re-analyses 
by other groups who had access to raw data \cite{Estabrooks:1974vu,Kaminski:1996da}.
These references also provided information on other partial waves, particularly the 
vector-isovector, whereas the experimental information on the isospin 2 waves can be found in \cite{pipiscatteringI2}.
In principle
$\pi\pi\rightarrow\pi\pi$ scattering data can be extracted from 
$\pi N\rightarrow \pi\pi N$ because in certain kinematic regions
the latter is dominated by the one-pion exchange process.
Together with the initial pion and the two pions in the final state, this exchanged pion 
forms the $\pi\pi\rightarrow\pi\pi$ subsystem. This amplitude
has to be extracted through a complicated analysis in which some background mechanisms
(like absorption, $A_1$ exchange, etc...) have to be modeled.
This technique has ambiguities that can lead to different solutions for the scattering amplitude. 
\color{black} These are due, for instance, to the fact that in certain energy regions
one is only sensitive to the S-wave from the S-P interference, which
essentially determines the 
$\vert\delta_0^0-\delta_1/2-\pi/4\vert$ phase shift combination.
Thus a two-fold ambiguity appears even if the P-wave is known (this one is called up-down or top-down ambiguity), 
or a four-fold ambiguity for the overall amplitude. 
We will only show here data once these ambiguities have been resolved. 
In Subsec.\ref{subsec:pwbelow} we will comment on
how later reanalyses with polarized targets and the use of dispersive formalisms
settled this issue \cite{Kaminski:1996da,Kaminski:2002pe}.
A detailed account of the method to extract meson-meson amplitudes
and a partial resolution of these ambiguities before the use of dispersive techniques
can be found in the textbook \cite{libropipi}. 
However, even within the same ambiguity solution the results are 
plagued with systematic uncertainties that are due to the use of different models to isolate
the $\pi\pi\rightarrow\pi\pi$ amplitude. Actually, even within the same experimental collaboration, and thus using the same $\pi N\rightarrow \pi\pi N$ data, different and even conflicting
 data sets  of $\pi\pi\rightarrow\pi\pi$ data have been provided.
For instance, in the same publication of Grayer et al. \cite{Cern-Munich}
by the CERN-Munich Collaboration, five different 
sets of $\pi\pi\rightarrow\pi\pi$ data are shown, labeled A to E, 
and some of them are incompatible with the others, as can be seen in Fig.\ref{fig:00data}.
\color{black}
Moreover, with the notable exceptions of \cite{Pr73,Cern-Munich-high}, 
it seems that providing Tables with their results was not fashionable in those days and that
only very vague statements about systematic uncertainties were made. Sometimes, as in \cite{Cern-Munich}, statistical uncertainties were provided for each set of solutions. However, since 
these data sets are incompatible among themselves within statistical uncertainties,
 the differences between sets should be interpreted as an indication of the systematic uncertainty.
As an example, the left panel of Fig.\ref{fig:00data} displays the data on $\pi\pi\rightarrow\pi\pi$ 
scattering phase shifts of the scalar isoscalar wave. 
Note the large differences even within data sets coming from the same experiment \cite{Cern-Munich} (Solution B 
was published first in \cite{Hyams:1973zf})
due to systematic uncertainties. Something similar happens with \cite{Pr73}, but we only show the most commonly used data set,
since it will be seen later that the others are even more inconsistent with fundamental dispersive constraints. 

Another relevant indication of the interest on $\pi\pi$ scattering 
in the early seventies was 
the appearance of $K_{e4}$  experiments
\cite{Zylbersztejn:1972kd,Rosselet:1976pu}. These correspond to the $K\rightarrow \pi\pi e \nu$ decay and provide an indirect
measurement of the $\delta^0_0-\delta_1$
phase combination well below 500 MeV, a region that could not be reached with $\pi N\rightarrow\pi\pi N$ 
experiments.
At that time these low energy data were not very determinant in the $\sigma$ discussion, but 
we will see that recent $K_{e4}$ experiments have actually been decisive to enter the precision era 
for light scalars.

At this point, and in view of Fig.\ref{fig:00data} it is important to emphasize that the $\sigma$ is so 
wide that right from the very beginning it was clear that the familiar Breit-Wigner description \cite{Breit:1936zzb}, 
valid for narrow isolated resonances, is not appropriate to describe the S-wave data.
Indeed, note in Fig.\ref{fig:00data} that there is no isolated Breit-Wigner shape around 500-600 MeV, corresponding to a $\sigma$ or $f_0(500)$ resonance. This means that the $\sigma$ resonance does not appear as a peak in the $\pi\pi\rightarrow\pi\pi$ cross section nor in many other amplitudes which contain in the final state two pions with the quantum numbers of the $f_0(500)$.
Of course, a Breit-Wigner-like shape over a background phase of about 100 degrees is seen around 980 MeV
in Fig.\ref{fig:00data}, corresponding to the $f_0(980)$, but even that shape is somewhat distorted by the nearby $\bar KK$ threshold. As an illustration of how the phase of a Breit-Wigner resonance looks like,  we display in the right panel of Fig.\ref{fig:00data}
the vector-isovector channel, where the shape of the $\rho(770)$ resonance is seen. 
Clearly, nothing like that is seen around 500 MeV in the left panel.

This issue is of importance, because the use of simple Breit-Wigner formulas to fit the modulus of the $\sigma$ 
contribution to an amplitude that contains two pions in the scalar isoscalar channel
is one of the biggest sources of confusion when determining the $\sigma$ parameters. 
As a matter of fact, Watson's final state theorem \cite{Watson:1952ji} implies that the phase of such amplitude should
be the same as that in the left panel of Fig.\ref{fig:00data}, 
but that data is not well reproduced with a 
Breit-Wigner shape.
Trying to determine the $\sigma$ parameters 
by the position of peaks or poles in Breit-Wigner
amplitudes can actually lead to different 
parameters when looking to different processes.
But if such techniques can be misleading,
what is then a rigorous definition of a resonance?

\subsubsection{Poles, resonances and dispersion relations... ignored}

The process independent and mathematically sounded definition of a resonance
is made by means of its associated 
pole in the unphysical (or second) Riemann sheet of the complex energy (squared) plane.
Actually, already in the 1973 RPP edition \cite{PDG73}, it was commented that {\it ``It is clear that 
the behavior of the $\delta^0_0$ is much too
complicated to allow a description in terms of one or
several Breit-Wigner resonances. We therefore list
the positions of the poles of the T matrix''}.
Nevertheless, it is still customary to identify
the pole position $s_R$ with the resonance mass and width 
as follows: $\sqrt{s_R}\simeq M_R-i \Gamma_R/2$, much as it would be 
done within a Breit-Wigner  
notation. 
In what follows, the $\sigma$ parameters will always refer 
to the T-matrix pole-mass and pole-width, unless explicitly stated otherwise.

As a matter of fact, in the very early seventies there were works showing that 
analyticity and unitarity together with crossing constraints,
required the existence of a broad $\sigma$ pole,
despite the poor existing data before the Grayer et al. \cite{Cern-Munich} and the Protopopescu et al. \cite{Pr73} results.
Unfortunately these were not completely consistent since 
$m_\sigma\simeq 410\,$ MeV and $\Gamma\simeq 380\,$MeV were obtained in \cite{LeGuillou71},
whereas in \cite{Basdevant:1972uu} 
an $\epsilon$-resonance with a mass below 650 MeV but a width $\Gamma>650$ MeV was found.

 A very important technique used in the recent theoretical works that triggered the major revision of the RPP in 2012
was developed by S.M. Roy in 1971 \cite{Roy:1971tc}.
He was able to incorporate crossing and analyticity into a set of
exact integral equations for $\pi\pi$ scattering involving only {\it physical region} partial waves.
Unfortunately, these rigorous methods 
were  ignored to a large extent  for more than a decade, with few valuable exceptions \cite{roy70,Pennington:1973xv,Pennington:1973hs,Pennington:1974kp},
between the 70's and the early 90's. Roughly, these years coincide with the time when the $\sigma$ resonance
was absent from the RPP (actually, one can even check the citation gap of \cite{Roy:1971tc,Basdevant:1972uu}
during those years). As we will see in Sec.\ref{subsec:dispersiveapproaches} these equations were resurrected in the late 90's and the early 2000's and
have been crucial to our present understanding and precise determination of the $\sigma$ parameters.
A possible explanation is that around that time QCD appeared and the attention of the
hadronic community was shifted away from other topics. 
In addition, these analytic methods require
powerful mathematical techniques, which many people prefer to avoid.
As we will see in Sec.\ref{sec:chiralsigma}, there are of course simpler and reasonable approaches, 
often related to some approximated dispersion relation,
which still provide very decent results. But 
 dispersive or analytic considerations have often been avoided
in favor of too simple models, which usually work fine for narrow resonances, 
but which may sometimes lead to artifacts or not very rigorous determinations when poles are deep in the complex plane. 
This has been one of the main sources of confusion
on the discussions about the $\sigma$ meson. 

In any case, in 1976 both the $\epsilon$ and $\sigma$ disappeared from the RPP \cite{PDG76}. The 
$\pi\pi$ isoscalar S-wave motion was interpreted as mainly due to the 
narrow $S^*(993)$ (nowadays $f_0(980)$) 
and a broad $\epsilon(1200)$ (nowadays called the $f_0(1300)$ resonance), which replaced all previous scalar mesons below 900 MeV. 
According to the RPP \cite{PDG76} this replacement was motivated  by the need to have the right number of scalar-isoscalar
resonances to form an SU(3) multiplet together with the $\delta(993)$ 
and the $\kappa(1250)$, following \cite{Morgan75}.

\subsubsection{Three quarks for Muster Mark! ... but how many for the $\sigma$?}

With the advent of QCD and the interpretation of quarks and gluons as physical entities,
a new question arose concerning the composition of hadronic resonances. 
Within the quark model mesons were interpreted as quark-antiquark states
that could be grouped into nonets, corresponding to representations 
of the flavor symmetry group. Each member of the multiplet corresponds to different flavor states
of the constituent quark and antiquark. A similar pattern was followed by baryons, made of three quarks, which could be grouped into different flavor multiplets.

However QCD is much more than just the quark model and 
in principle other configurations might be possible, including 
states with more than one valence quark-antiquark pair or even
confined states of gluons, called glueballs,
 which are a striking 
feature of a confining non-abelian gauge theory like QCD.

\begin{figure}[htbp]
  \centering
  \includegraphics[width=\textwidth]{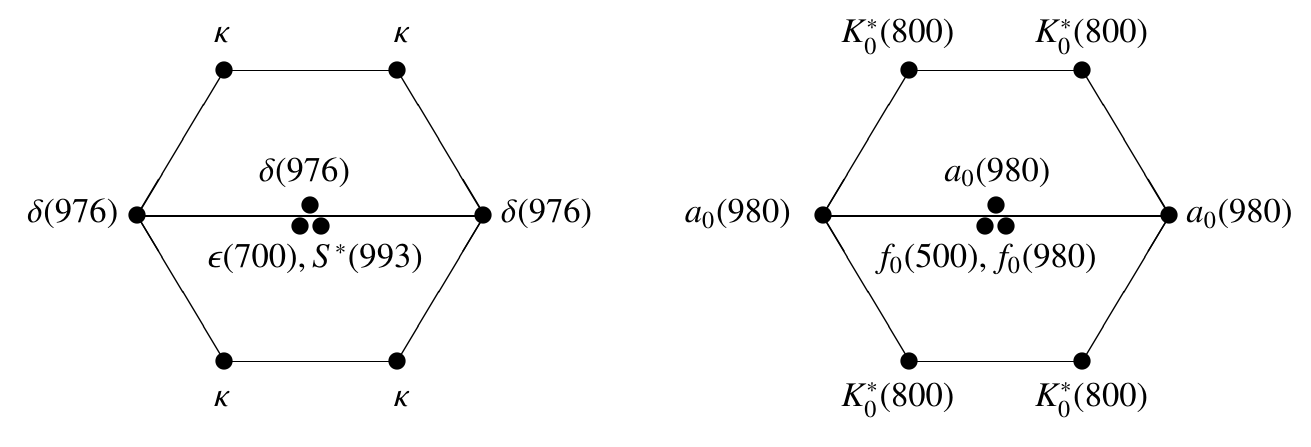}
\vspace*{.1cm}
  \caption{ Left, light scalar nonet as suggested in \cite{Jaffe:1976ig} with the 1976
notation.  Right: the same multiplet in the present notation.
}
  \label{fig:scalarmultiplet}
\end{figure}

As a matter of fact, the interest on light scalars increased because 
they are difficult to fit within the ordinary 
quark-antiquark scheme. Actually, one could form  an SU(3) nonet
of light scalars with the $\epsilon(700)$  (one of the names of the $\sigma$ back then) 
together with the $\delta(976)$ (nowadays $a_0(980)$), the $S^*(993)$ 
and a light $\kappa$ (now called $K_0^*(800)$, still not listed in the RPP summary tables, 
but included in the particle listings). Such a nonet is shown in Fig.~\ref{fig:scalarmultiplet}:
on the left  in the 1976 version and on the right in the modern notation.
However, this nonet identification implies an {\it inverted hierarchy} with respect to the standard quark-antiquark 
composition.  For instance, if these scalars were quark-antiquarks, then the $\kappa$ resonance, which has strangeness, should 
contain a strange quark or antiquark and should be about 200 MeV heavier than the $\delta(976)$, which contains no strangeness.
But, as seen in Fig.~\ref{fig:scalarmultiplet}, the opposite hierarchy is observed.

Nevertheless, since the existence of the $\kappa$ was unclear \color{black} 
it was still possible to assume that it simply did not exist. 
In such scenario the $\delta(976)$ and $S^*(993)$
should belong to a heavier nonet. But since only one strange scalar was seen around 1500 GeV
(now called $K_0(1430)$)
then there were too many scalar-isoscalar states
above 1 GeV for only one nonet.  Therefore this classification
 required to assume that one scalar-isoscalar resonance 
(today called $f_0(1300)$), whose existence 
was also controversial, did not exist either. 
Once the full nonet between 950 MeV and 1500 MeV was identified, 
the  much lighter $\epsilon/\sigma$ was an extra singlet state outside that heavy nonet. 
With these assumptions
the glueball interpretation might seem appealing for the $\epsilon/\sigma$, since
a pure glueball carries no flavor and the lightest
glueball is also expected to have no spin. \color{black}
However, we will see
in Sec.\ref{sec:nature} that this scenario is
not favored by lattice calculations, large $N_c$ arguments,
chiral symmetry and the growing evidence for the existence of a $\kappa$ or $K_0^*(800)$ meson. 
Of course, in order to identify a glueball, 
it is still nowadays essential to understand the $\sigma$ and identify
all light scalar mesons within their multiplets 
and see if there are extra states or not.

An alternative proposal
was advanced in 1976 by Jaffe, within the framework of the MIT bag model,
suggesting that the lightest scalars were not
ordinary quark-antiquark mesons. He found that
 their features, particularly the inverted hierarchy, could be 
better explained within a tetraquark model \cite{Jaffe:1976ig}.  
Within this scheme, the glueball, if it existed, should be heavier than 1 GeV.

Thus, by the end of the 70's the $\sigma$ became even more interesting due to QCD: it was now important for the
identification of the lightest glueball and it was also a strong candidate for a different class of mesons.
Naively speaking, it now became relevant to know how many quarks or antiquarks made up a $\sigma$ meson: none, two or more.
Sec.\ref{sec:nature} will be dedicated to this issue, in which some significant progress has been achieved, but is still 
the subject of intense debate.

\subsubsection{The 80's and early 90's. Chiral Perturbation Theory and $\pi\pi$ scattering.}

Following important developments in the late seventies
on
the Effective Theory approach \cite{Weinberg79},
 the eighties gave rise 
to Chiral Perturbation Theory (ChPT) \cite{chpt1,chpt2}, 
which is the low energy effective theory of QCD.
It provides a systematic low energy  and small mass expansion of a Lagrangian made of pions, kaons and eta mesons, which is consistent with all QCD symmetries and particularly with its spontaneous symmetry breaking pattern. In this formalism the pions are taken into account into a so-called non-linear representation as opposed to the linear representation of the L$\sigma$M, and the $\sigma$ meson is not included explicitly as a degree of freedom.
The leading order is universal in the sense 
that it only depends on the symmetry pattern, 
the particle masses and the scale of spontaneous symmetry breaking,
whereas the next orders contain a set of 
phenomenological parameters known as low energy constants (LECs), which 
take into account the underlying dynamics of QCD.
This expansion has the virtue of providing a model-independent connection with QCD and 
systematically improvable results at low energies, which was a particularly difficult region for hadron physics. 
Soon it was  shown that the phenomenological values of the LECs
can be understood by the exchange of heavier resonances \cite{chpt1,Ecker:1988te}, 
predominantly by the exchange of vectors. There is a small
contribution from  scalars, {\it but whose masses should be $\simeq 1 GeV$  or heavier}. 
No scalar below 1 GeV contributed to these LECs, which at that time 
seemed to play against the 
existence of a light $\sigma$ meson. 
However, we will see later on that this is just evidence about the $\sigma$ not being an ordinary meson, not against its existence.

Despite a relative fading of interest on a light $\sigma$ in the late seventies, in the eighties
there were also attempts to fit quark models to meson-meson results \cite{QMmeson}. 
In addition there were other proposals concerning the nature of light scalars. 
For instance, in \cite{isgur1} they were interpreted as ``weakly bound states of two color singlet mesons'', 
i.e. ``meson molecules'', 
or in \cite{Delbourgo:1982tv} it was shown that
a light sigma, as a member of a light scalar nonet, was obtained within NJL-like models of low-energy QCD, or in \cite{vanBeveren:1986ea}
the pure quark model states were shown to be dramatically modified
when final state interactions of mesons were taken into account through unitarization. 
\color{black} Unitarity is an essential ingredient of 
$\pi\pi$ scattering, whose amplitude saturates the unitarity bound
and becomes resonant already in the elastic region of the scalar and vector partial waves and not too far from threshold.\color{black}

Furthermore, it was even pointed out that there was no need 
to include the $\sigma$ meson
in a Lagrangian (as a tree-level resonance exchange) 
to explain several effects where the $\sigma$ resonance seemed to 
appear, but that all those effects could be mimicked with just
the final state rescattering of two pions \cite{Meissner:1990kz}.
However, as we will see, this necessarily implies the existence of a 
light $\sigma$ pole in the amplitude, although very deep in the complex plane. 

\color{black} Actually, in the late eighties and early nineties, 
ChPT was combined with dispersion relations, giving rise to what has become known as Unitarized Chiral Perturbation Theory (UChPT).\color{black}
The first instance of this formalism was derived
using a dispersion relation for the inverse of the 
amplitude \cite{Truong:1988zp,Dobado:1992zs}, which allows for an 
exact implementation of  elastic unitarity in $\pi\pi$ 
scattering while respecting the ChPT low energy expansion 
\cite{Dobado:1989qm,Dobado:1992ha}.
The interest of this approach, known as the Inverse Amplitude Method (IAM) 
is that, while respecting the systematic low energy expansion of ChPT 
at low energies and therefore the connection with QCD, 
it can simultaneously generate resonant shapes 
{\it without introducing them as explicit degrees of freedom in the Lagrangian}.
In addition, since this formalism is based on a dispersion relation 
for the inverse amplitude, it has all the analytic structures of 
elastic amplitudes in the physical sheet of the complex plane, namely, 
the unitarity or physical cut implemented exactly and the crossing or left 
cut as an approximation. Analyticity and unitarity, which are 
the dominant features of resonance scattering, are thus imposed exactly, but
at the cost of an approximated crossing symmetry. This allowed for the search for poles over the next years,
including that of the $\sigma$.

\color{black} Within the ChPT context, the process of unitarization 
is often reinterpreted as resumming, among others, the diagrams in which the final mesons 
rescatter an arbitrary number of times. 
Formally, these are higher order diagrams, but their 
resummation may be very relevant numerically, particularly close to resonances. 
In the literature it is frequent to abuse the language so that
``resumming higher order diagrams with interactions between the final meson legs''
is also referred to as ``including final state interactions''.
\color{black}

\subsubsection{The mid 90's: The resurrection of a confusing $\sigma(400-1200)$.}

There were still works \cite{Tornqvist:1995kr} trying to build the lightest scalar nonet
as quark-antiquark mesons without the $\sigma$ although 
almost immediately it was found that the poles of that model \cite{Tornqvist:1995ay} contained a  $\sigma$.

According to the ``Note on scalar mesons'' in the 1996 edition
of the RPP \cite{PDG96},
the works that triggered the ``resurrection/confirmation'' of the $\sigma$
were not dispersive analyses nor made any analysis in terms of ChPT,
but  they were re-analyses of $\pi\pi$ S-wave scattering 
within relatively simple 
phenomenological models 
 \cite{Achasov:1994iu,Tornqvist:1995ay,Au:1986vs,Zou:1993az,Zou:1994ea,Janssen:1994wn}
together with newer analyses of $p \bar p\rightarrow 3 \pi$ data \cite{Amsler:1995gf,Amsler:1995bf} 
(which claimed to demonstrate
that the $\sigma$ and the $f_0(1300)$ were two different poles).
Unluckily, as nicely remarked in that ``note on scalar mesons'',
there was a ``large spread in the resonance parameters obtained by these groups''
which was ``due less to differences in the data used than to differences in the models employed''. Hence, the $\sigma$ became the $f_0(400-1200)$. 
Unfortunately, this large 800 MeV uncertainty
in the $\sigma$ mass remained in the RPP until as recently as 2010, despite the strong
evidence for a light $\sigma$ which piled up during almost two decades.

\begin{figure}
  \centering
  \includegraphics[width=0.7\textwidth]{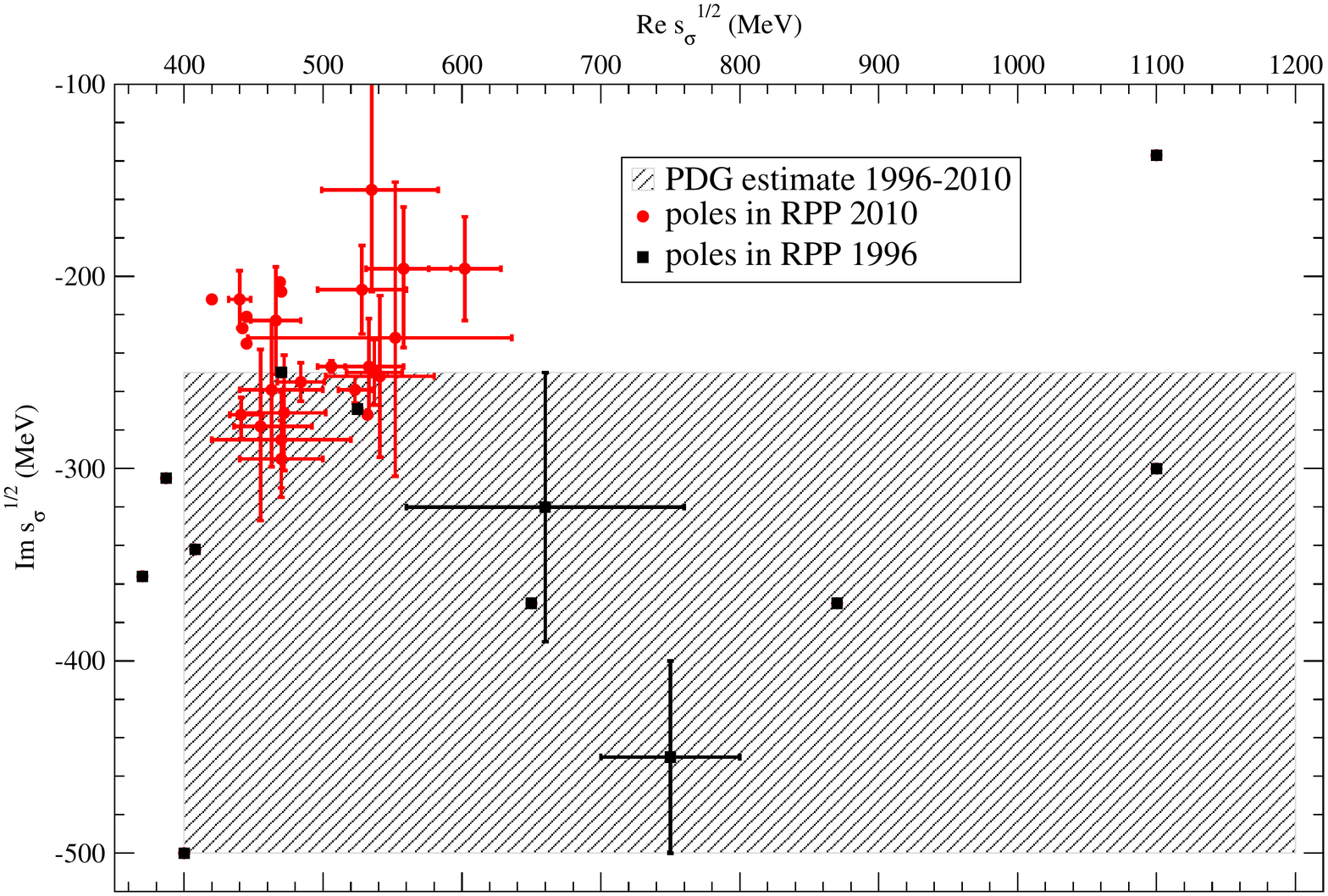}
  \caption{The $\sigma$ or $f_0(400-1400)$ resonance poles listed in the RPP
1996 edition (Black squares) together with those also cited in the 2010 edition 
\cite{PDG10} (Red circles). Note the much better consistency of the latter and the 
general absence of uncertainties in the former.
The large light gray area corresponds to the uncertainty band assigned to the $\sigma$ 
from 1996 to 2010. 
}
  \label{fig:poles}
\end{figure}

In order to illustrate this spread of values, and the confusing situation that prevailed
until 2010 according to the RPP, 
let us recall that the 1996 RPP provided again the so-called ``T-matrix'' pole
(Unfortunately, it also provided, and still does, a 
Breit-Wigner pole, which as was commented 
before might be 
process dependent and confusing for its interpretation). Thus Fig.\ref{fig:poles}
shows the compilation of $\sigma$ poles in 
the complex-energy plane in the 1996 RPP edition (Black squares) \cite{PDG96} together 
with the additional poles included in the next editions until 2010 \cite{PDG10}. 
Note that with few exceptions, the poles in the 1996 edition had no error estimate.
One of the most striking features of this plot is 
the large light gray area that corresponds to the 
uncertainty assigned to the $\sigma$ pole 
in the RPP from 1996 until 2010. 
However, this large uncertainty 
band was driven by a few points which, since the 1996 edition, were scattered
very far from the rest. As emphasized before,  
a very significant part of the apparent disagreement between 
different poles in Fig.\ref{fig:00data} 
is not coming from experimental uncertainties when extracting the data, 
but from the use of models in the interpretation of those data 
together with unreliable extrapolations to the complex plane. 
Actually, different analyses of the same experiment 
could provide dramatically different poles, depending on the parameterization or model 
used to describe the data and its later interpretation in terms of poles and resonances. 
Maybe the most radical example are the three poles from the 
Crystal Barrel collaboration, lying 
at $(1100-i 300)\,\mev$ \cite{Amsler:1995gf}, 
$(400-i 500)\,\mev$ and $(1100-i 137)\,\mev$ \cite{Amsler:1995bf}, 
corresponding to the highest masses and widths in that plot. These poles
were compiled together in the RPP although
they even lie in different Riemann sheets. Moreover we will 
see in Sect.\ref{sec:parameters} that all three lie
outside the region of analyticity of the partial
 wave expansion (Lehmann-Martin ellipse \cite{Lehman-Martin}).

Therefore it should be now clear that in order to extract the
parameters of the $\sigma$ pole, which lies so deep in the complex plane and has no 
evident fast phase-shift motion, it is not enough to have a 
good description of the data. 
As a matter of fact, many functional forms could fit very well the data in a given region, 
but then differ widely from each other when extrapolated outside the fitting region. 
For instance, if all data were consistent (which they are not) one can always find a
good data description using polynomials, or splines, which have no poles at all.
Hence, to look for the $\sigma$ pole, the correct analytic extension to the complex plane, 
or at least a controlled approximation to it, is needed. Unfortunately that 
has not always been the case in 
many analyses, and thus the poles obtained from poor analytic extensions of an 
otherwise nice experimental analysis are at risk of being artifacts or just plain wrong determinations.
This, together with the large uncertainty attached to the $\sigma$ in the RPP,
 is what made many people outside the community to think that no progress was made
in the light scalar sector for many decades.

However, progress was being made and the other remarkable feature 
of Fig.\ref{fig:00data} is that by 2010 
most determinations agreed on a light sigma with a mass 
between 400 and 550 MeV and a half width between 200 and 300 MeV. 
As we will see next, most of these and other  ``light sigma'' poles were obtained 
much before 2010 within many different approaches.
In all these, unitarity played a very important role.

\subsubsection{Poles from Unitarity and  Chiral Perturbation Theory}

For instance, in the early
90's a broad $\sigma$ meson was again found when describing
$\pi\pi$ scattering  within a relativistic treatment 
of coupled channels in a Lippmann-Schwinger formalism applied to a 
separable potential \cite{Kaminski:1993zb}. Only two-body states
were considered, and two-body unitarity was implemented. 
A $\sigma\sigma$ channel was introduced to mimic the $4\pi$
contribution, but it was shown to be relevant only above 
1400 MeV \cite{Kaminski:1997gc,Kaminski:1998ns}. In another example
\cite{Harada:1995dc}, a broad $\sigma$
was included in  a simple model, which was nevertheless 
unitary and crossing symmetric,
containing the lowest-order chiral (contact) term and 
explicit resonances
at tree level, in order to describe $\pi\pi$ scattering. 
Note that in this work \cite{Harada:1995dc} some ad-hoc deformation of
the usual Breit-Wigner shape for the $\sigma$ was carefully implemented.

In the UChPT front beyond the tree level, by 
1996 the poles of the $\rho$ and the $\sigma$ were found
simultaneously using the elastic IAM described 
above \cite{Dobado:1996ps}.
The relevance of this result is that both poles were generated
{\it without a priori assumptions about their existence or nature}.
They result from a data fit of a fully renormalized and unitarized
effective field theory expression, which 
depends only on the four next-to-leading-order (NLO)  
ChPT low energy constants
but contains {\it  no other spurious parameter nor explicit resonances in the Lagrangian}.
Both poles come from an analytic extension to the complex plane
of an amplitude obtained from a dispersion relation 
for the inverse amplitude that has an exact elastic unitarity cut
and approximated left and inelastic cuts.
In addition, this elastic formalism could be 
extended to SU(3) ChPT, by including the strange quark, 
and applied to $K\pi$ scattering \cite{Dobado:1992ha}, generating the $K^*(892)$ vector resonance  pole
\cite{Dobado:1996ps} with the same chiral parameters used for the $\sigma$ and the $\rho$ description.
(The $\kappa$ can also be generated this way but went unnoticed at the time).

Simultaneously, 
it was shown \cite{Oller:1997ti} that using just the leading order (LO) ChPT 
within a coupled channel Lippmann-Schwinger like resummation 
with a natural cutoff, it was possible to generate not only the $\sigma$
but all scalar resonances below 1 GeV 
that are needed to identify a complete SU(3) nonet.
As suggested before, these are the isoscalar $\sigma$, the $f_0(980)$, 
the three isovector $a_0(980)$ and the four
isospin 1/2 strange $\kappa$ resonances. This time
they were generated as a consequence of chiral symmetry, unitarity and analyticity, without any a priori assumption about their existence
or their classification into one multiplet.
This formalism has become known as the ``Chiral Unitary Approach''
and it was soon shown to be closely related to the IAM and UChPT 
\cite{Oller:1997ng}. Indeed, when extended to higher orders it
could also generate the poles associated to 
the vectors $\rho(770)$ and $K^*(892)$ \cite{Oller:1998hw} (as well as the octet component of the $\phi(1020)$, see \cite{Oller:1999ag}). Similar results were 
also obtained for the scalars within the closely related formalism
of Bethe-Salpeter equations combined with ChPT \cite{Nieves:1998hp}.

Nevertheless, as we will see in Sec.\ref{sec:chiralsigma},  for the vectors, which fit well into the 
ordinary $q \bar q$ quark model description,
it is absolutely necessary to include the NLO 
ChPT low energy constants, since a natural 
size cutoff was not enough. Hence, the fact that 
the $\sigma$ can be generated just 
from the LO ChPT contribution, which is universal 
and given just by the spontaneous symmetry breaking scale, together with a 
cutoff of a natural hadronic size, is a strong 
suggestion of its non-ordinary nature.
This result also explains why the $\sigma$ does not contribute 
to the LECs, since the dominant dynamics that produce  it
seem to depend more on physics at the hadronic scale than 
on the underlying quark dynamics
that dominate the LECs. Actually, by 
the explicit introduction of resonances in 
a chiral Lagrangian combined with a 
dispersive approach it was soon shown \cite{Oller:1998zr} that this picture 
is consistent with the coexistence of the light scalar nonet 
described in the previous paragraph with another scalar nonet above 1 GeV. 
The former would be ``dynamically'' 
generated, meaning 
it is generated mostly by the unitarization of the purely mesonic 
dynamics and not by ``preexisting'' resonances that contribute to the ChPT LECs in the Lagrangian 
(like vectors or scalar mesons above 1 GeV). These ``preexisting'' states are also called genuine, and we will see later how 
a definition can be provided within QCD within the $1/N_c$ expansion.

Both UChPT or the chiral unitary approach, as they start from ChPT, 
are sometimes referred as ``non-linear chiral realizations'', 
where light scalar resonances
are not included explicitly. 
But in the late nineties, further support for the existence 
of a broad and light sigma around 500 MeV, 
forming a nonet with the $f_0(980)$, $a_0(980)$ and $\kappa$,
was also obtained from models 
where the scalar resonances are included explicitly 
either as poles in the amplitudes \cite{Ishida:1995xx} or 
in the Lagrangian in a linear representation of a chiral multiplet 
\cite{Delbourgo:1993dk,Black:1998wt,Ishida:1999qk,Tornqvist:1999tn}.
Nevertheless, the confusion \color{black} lingered on, \color{black} 
since there were some analyses that produced a 
narrow sigma around 700 \cite{Svec:1995xr},
although the phase-shifts would differ substantially 
from other analyses like those in Fig.\ref{fig:00data}.

\subsubsection{The new millennium:
 Experimental confirmation from heavy meson decays and general agreement on an $f_0(600)$}

Further experimental evidence for a light $\sigma$ 
came at the turn of the century
 from decays of heavier particles into 
final states with at least two pions in the
scalar-isoscalar channel. Given the
fact that systematic uncertainties in these 
experiments were radically different from those found in $\pi\pi$ scattering,
these heavy particle decays provided a very 
strong support for the existence of a light $\sigma$ 
between 450 and  550 MeV. Actually, when the 
results from \cite{asner00,Aitala:2000xu} 
were included in the RPP 2002 edition,
the name of the $\sigma(400-1200)$ was changed to $f_0(600)$.

\begin{figure}
  \centering
  \includegraphics[width=0.75\textwidth]{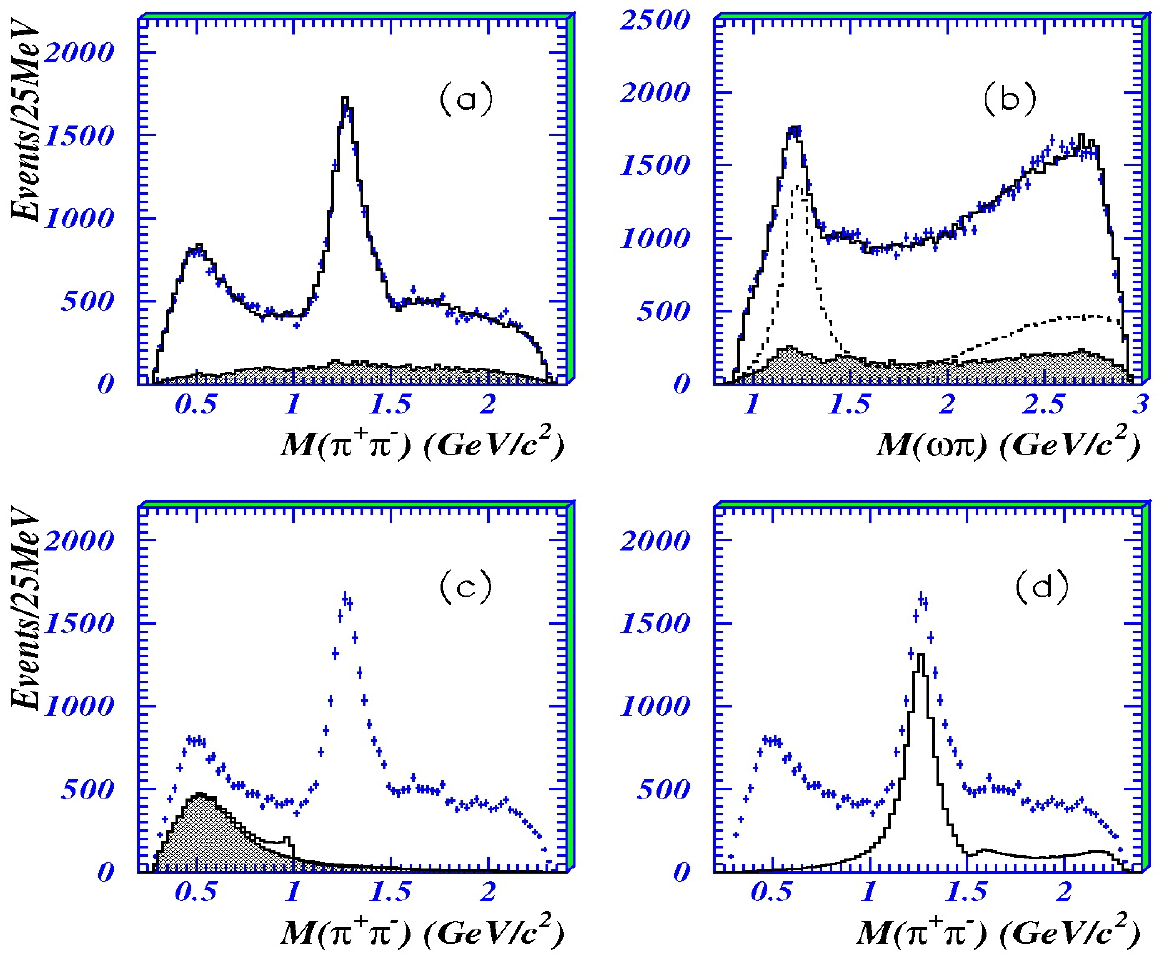}
  \caption{Figure taken from \cite{Ablikim:2004qn}. Panels  (a), (b)  show the $\pi^+\pi^-$ and $\omega\pi$ invariant mass projections of data from the BES 
analysis of $J/\Psi\rightarrow \omega \pi^+\pi^-$. Panels 
(c) and (d) show the $J^{PC}=0^{++}$ and $2^{++}$ projections, respectively. The shaded area in (c) corresponds to the $\sigma$ contribution. The contribution of the $f_0(980)$ can be seen as a small peak 
on top of the shaded area right below 1 GeV. Compare the height of the peak and the
asymmetric shape of the $\sigma$ in (c)
to the height and shape of the $f_2(1275)$ resonance in (d). 
Reprinted from  Phys.\ Lett.\ B {\bf 598}, 149, 2004, M.~Ablikim {\it et al.}  [BES Collaboration], ``The sigma pole in $J/\psi \to \omega \pi^+ \pi^-$,''. Copyright 2004, with permission from Elsevier.
}
  \label{fig:BESdata}
\end{figure}

The decays that contributed to this change were
$\tau\rightarrow\pi^-\pi^0\pi^0\nu_\tau$ \cite{asner00}, $D\rightarrow 3\pi$ \cite{Aitala:2000xu} (these two do not give T-matrix poles and are thus not included in Fig.\ref{fig:00data}), 
$J/\psi\rightarrow\omega\pi^+\pi^-$ \cite{Ablikim:2004qn}, $D\rightarrow \pi^-\pi^+\pi^+$ \cite{bomvicini07} or $\psi(2S) \rightarrow \pi^+\pi^- J/\psi$ \cite{ablikim07}.
Even though part of these results were subject to some criticism
over the parametrizations used to describe 
the $\sigma$ pole, which is so deep in the complex plane,
 it was really clear 
that a significant resonant contribution was needed in the $\sigma$ channel. 
There is actually something like a ``peak'' seen in many of these experiments
and that definitely made a much more convincing case for the $\sigma$ meson.
As explained for instance in \cite{Bugg:2003kj},
a peak or bump is seen in these decay experiments and not in 
meson-meson scattering partly because the latter is distorted
by an ``Adler zero'' \cite{Adler:1964um} 
below threshold, due to chiral symmetry. 
Such a peak can be seen  
in panel (c) of Fig.\ref{fig:BESdata}, which shows the $\sigma$
 contribution needed to explain the $\pi\pi$ invariant mass 
distribution of events in \cite{Ablikim:2004qn}.
Note its asymmetric shape compared 
to that of the $f_2(1275)$ resonance in panel (d) in the same figure.

In general these decay analyses tend to 
yield a pole mass between 500 and 550 MeV, which is
somewhat higher than the results of  
dispersive approaches that we will discuss below. 
This might be due to the fact that
these decays are usually analyzed with
simple models and not dispersion theory. 
These approaches are less rigorous and sometimes 
even inconsistent with some fundamental requirements
of unitarity, analyticity, etc. as it may happen
with certain isobar models with superimposed 
Breit-Wigner terms, which violate 
unitarity, or with K matrices, which should also 
incorporate information on meson-meson 
scattering in one way or another. 
Moreover, these experiments usually refer to the ``Breit-Wigner''
shape, which as we have seen might even be process dependent,
and may explain in part this 50 to 100 MeV 
mass shift from the dispersive determinations.
Anyway, this information from decays improved considerably the 
attitude towards the $\sigma$  
and closed the case in favor of the existence of a 
light and broad $f_0(600)$ meson. 
But for precise pole determinations 
and resonance parameters,
these analyses are still somewhat model-dependent.
Rigorous, precise and model independent pole determinations appeared 
some years later and triggered the major revision that has motivated this report.

In addition, by the end of the nineties and throughout the first 
decade of the new millennium, 
virtually all approaches that fitted either heavy 
meson decays and/or $\pi\pi$ scattering phases containing
at the very least some minimal implementation of chiral symmetry, 
unitarity and analyticity obtained a pole
around 430 to 550 MeV corresponding to a broad resonance 
whose width lied between 400 to 600 MeV
(for references on these years see \cite{Kaminski:1993zb,Kaminski:1997gc,Kaminski:1998ns,Bugg:2003kj,Ishida:2001pt,CGL,Black:2000qq,Gallegos:2003gq,Pelaez:2004xp,Napsuciale:2004au,Zhou:2004ms}). This can be checked in Fig.\ref{fig:poles}.

\subsubsection{The QCD connection via the $N_c$ behavior}

Around that time there was a relevant 
development concerning the nature 
and spectroscopic classification of the $\sigma$ meson
and its scalar partners. 
In particular, when using the IAM only the NLO ChPT parameters were needed to generate all light scalars and vectors.
Recall that those parameters encode the 
underlying QCD dynamics but their precise values are not calculable from perturbation QCD. Nevertheless, by means
of the $1/N_c$ expansion,
their dependence on the QCD number of colors $N_c$ \cite{chpt2} had been previously calculated in a model independent way. 
The  $1/N_c$ expansion also predicted the scaling of the mass and width of
ordinary $q\bar{q}$ mesons \cite{'tHooft:1973jz,Witten:1980sp}. 
Thus, by re-scaling the
ChPT low energy parameters with $N_c$ 
it was possible to calculate the $N_c$ behavior
of the generated resonances
in the vicinity of $N_c=3$.  
Actually, 
it was shown \cite{Pelaez:2003dy} that the $\rho(770)$ 
followed nicely the expected
behavior of ordinary mesons but the $\sigma$ (and its scalar partners) did not.
Since the IAM  {\it does not make any a priori assumption about 
the existence or nature of 
resonances}, this result was a very 
strong piece of evidence against the 
ordinary nature of the $\sigma$ meson
and provided a relatively direct link to QCD.

In addition, these results were consistent with a previous publication \cite{Sannino:1995ik}, 
in which $\pi\pi$ scattering was studied 
within a model which included ``all (large $N_c$ leading) resonances in the range of interest'' . This work
suggested that the $\sigma$ should  be included too, although presumably it was
``not of the simple qq type", and hence its
exchange ``should be of subleading order in the large $N_c$
limit''. 

\subsubsection{Precision in Theory and experiment}

Therefore, by the mid 2000's the collective experimental and theoretical effort had led to 
a relatively well accepted 
picture of a light sigma around 450 to 550 MeV, 
which most likely was part of a light nonet of non-ordinary mesons.
However, this still had to make it to the particle 
and nuclear physics community at large.
In order to  settle the 
issue of the $\sigma$ existence
and improve the precision on its parameters, a model independent approach
is needed. For this task, two 
parts are necessary: 
first, a consistent set of data to be used as input, and
second,
a rigorous analytic extension to the complex plane, 
to avoid artifacts, 
in which the pole is derived as a consequence of data and 
not included a priori in the amplitude with a specific 
functional form. 
Both criteria are met within 
the 
dispersion relation formalism for $\pi\pi$ scattering. 

In particular, the 
dispersive formalism is based on \color{black} the first principles of causality and crossing, which imply strong analytic constraints on the amplitudes.
These provide the correct 
analytic extension to the complex plane. The dispersive integrals 
can be used to, on the one hand, discard data sets which are inconsistent with 
basic requirements as causality and crossing ---together with 
isospin and chiral symmetries as well as unitarity---, 
but, on the other hand,
they can also be used \color{black} to obtain data fits or solutions that are constrained 
by these requirements.
Dispersion relations are simpler
and much more powerful for $\pi\pi\rightarrow\pi\pi$ scattering than for other processes like
heavy meson decays, due to the fact that
one is dealing with two-body states
and also because  all particles involved
have the same mass in the isospin limit. 
As we have seen, dispersive approaches had already been used in the 70's, but
one of the main problems was the poor quality of data, particularly at very low energies.
This was partially alleviated by new $K_{l4}$ experiments in 2001 by the BNL-E865 Collaboration \cite{Pislak:2001bf} or by the use of Chiral Perturbation Theory to fix the low energy part of the amplitude 
\cite{Dobado:1992ha,Dobado:1996ps,Oller:1998zr,CGL,Pelaez:2004xp,Zhou:2004ms,Guerrero:1998ei,GomezNicola:2001as} 
(parameterized by a set of ``subtraction constants'' in the integral formalism, as we will see in the next Section).
All these works provided very consistent $\sigma$ poles around 500 MeV and
implemented unitarity, chiral symmetry as well as analyticity,
 although crossing symmetry was an approximation, which was a common criticism. 
As we will see in the next sections, this crossing symmetry is responsible for an analytic 
structure (called the ``left cut'') which is much closer to the $\sigma$ meson pole than 
to any other resonance. From the left cut approximations implemented
in the previous references it seemed that its contribution was small and
it would not affect much the existence or position of the $\sigma$ pole.
However, for a precise and rigorous determination of the pole parameters, 
an accurate determination of the left cut contribution was needed.

As commented above, the formalism to implement crossing symmetry into 
partial wave dispersion relations had already been derived in 1971 by S.M. Roy \cite{Roy:1971tc}.
These equations were revisited in the late 90's and early 2000's. In particular solutions to these 
equations below 800 MeV  were worked out for the S and P-waves in \cite{ACGL} 
(the data above that energy and on the other waves are input) 
and later refined with ChPT constraints at low energy \cite{CGL}.
These solutions were then used in 2006 \cite{Caprini:2005zr} to extend 
the partial waves to the complex plane, As a result it was shown that the $\sigma$ pole could be found with remarkable precision at $M=441^{+16}_{-8}- i 272^{+9}_{-12.5}$ MeV,
having good control over the left cut contribution.
Moreover, it was demonstrated that such a $\sigma$ pole
lied within the region of convergence 
of the partial wave expansion (the so-called Lehman-Martin ellipse \cite{Lehman-Martin}). The latter is an important result, since, as commented above, 
some of the poles calculated with other methods and shown in 
Fig.\ref{fig:poles} do not fall within this region.

\begin{figure}[htbp]
  \centering
  \includegraphics[width=0.49\textwidth]{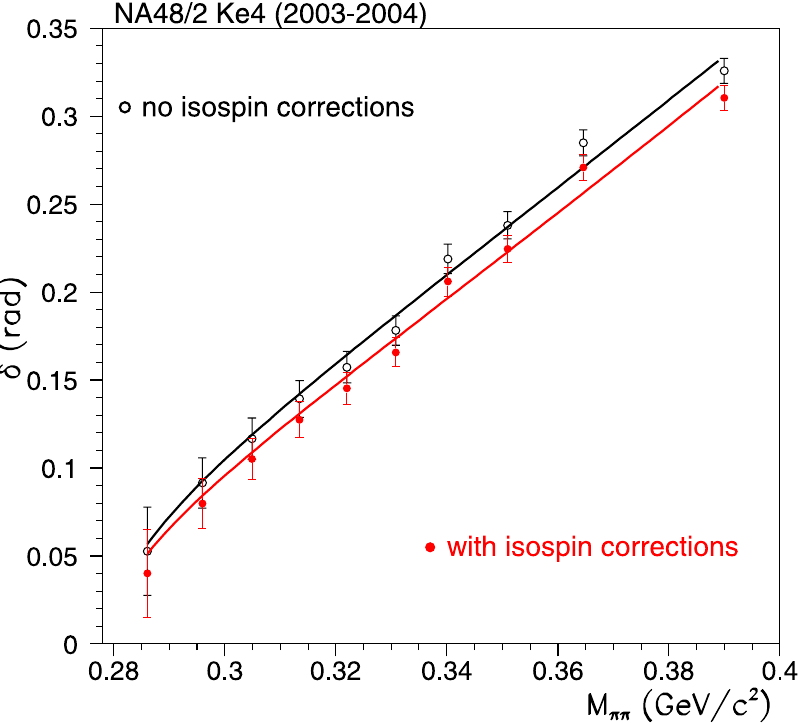}
\vspace*{.1cm}
  \caption{NA48/2 results from $K_{e4}$ decays. On a first approximation
they measure the 
difference $\delta\simeq\delta_0^0-\delta_1^1$ between $\pi\pi$ scattering phase shifts, although the precise relation will be discussed in Sect.\ref{subsec:data} and it requires some isospin corrections shown in the plot. Figure taken from the NA48/2 Collaboration reference \cite{Batley:2010zza}.
}
  \label{fig:NA48data}
\end{figure}

On the experimental front, the NA48/2 Collaboration at CERN \cite{Batley:2010zza} performed a new measurement of $K_{l4}$ decays 
which was much more accurate than the previous ones. Already with the preliminary results
it was possible to show \cite{Yndurain:2007qm,Caprini:2008fc} 
that with very simple parametrizations respecting unitarity and analyticity
in the elastic region, a light $\sigma$ pole was necessary
and fairly consistent with the one just discussed from ChPT+Roy equations. 
Actually, with the release of the final NA48/2 data, which we show in Fig.\ref{fig:NA48data},  it was even possible \cite{GarciaMartin:2011cn} to make a data analysis by fitting, instead of solving, 
once-subtracted Roy equations without ChPT input, called GKPY equations. The resulting $\sigma$ pole \cite{GarciaMartin:2011jx}
at $457^{+14}_{-13}-i279^{+11}_{-7}$ MeV  was, once again, very consistent with previous determinations.
Further confirmation came from the additional Roy-equation analysis of \cite{Moussallam:2011zg} extended
to the two-kaon threshold, but also from other 
approaches using the so-called analytic K-matrix formalism 
\cite{Mennessier:2008kk,Mennessier:2010xg}.

Moreover a similar pole was also found in \cite{Albaladejo:2008qa} within a model of a coupled channel unitarized chiral Lagrangian, this time with an explicit $\sigma\sigma$ channel 
to take into account the $4\pi$ contribution that should be present in $\pi\pi$ scattering, 
which is neglected below $K\bar K$ threshold 
in almost all other approaches, including those based on Roy Eqs.,
since it is very 
small and has not been measured. 
As it happened in \cite{Kaminski:1997gc,Kaminski:1998ns},
it was found that such a $4\pi$ contribution does not alter significantly 
the position of the $\sigma$ pole, which was found at $(456\pm6-i241\pm7)$ MeV. 
Actually, it had also been shown before \cite{Bugg:1996ki} that the amplitude $\pi\pi\rightarrow4\pi$
is very small below 1.4 GeV and can be modelled by contributions from the $f_0(1370)$ and the $f_0(1500)$ resonances. 
Of course, that channel is of relevance at higher energies.
The work \cite{Albaladejo:2008qa} also excluded the glueball interpretation of the $f_0(500)$ in favor of the $f_0(1710)$
and an important contribution to the $f_0(1500)$.

\subsection{Present Status of the $f_0(500)$ parameters: A major RPP revision}
\label{subsec:statusparameters}

\begin{figure}
  \centering
  \includegraphics[width=0.5\textwidth]{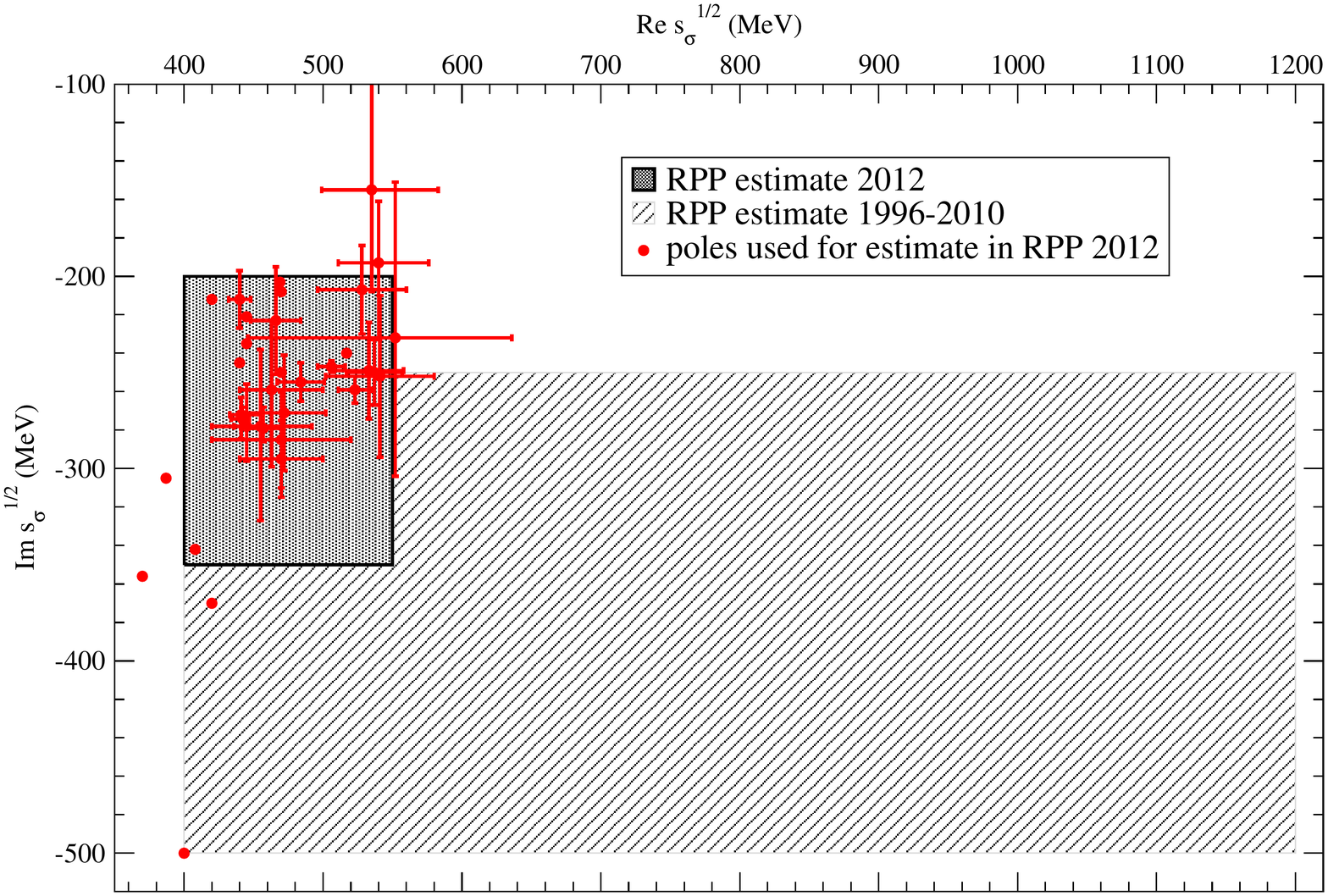}
\hspace*{-.3 cm}
  \includegraphics[width=0.5\textwidth]{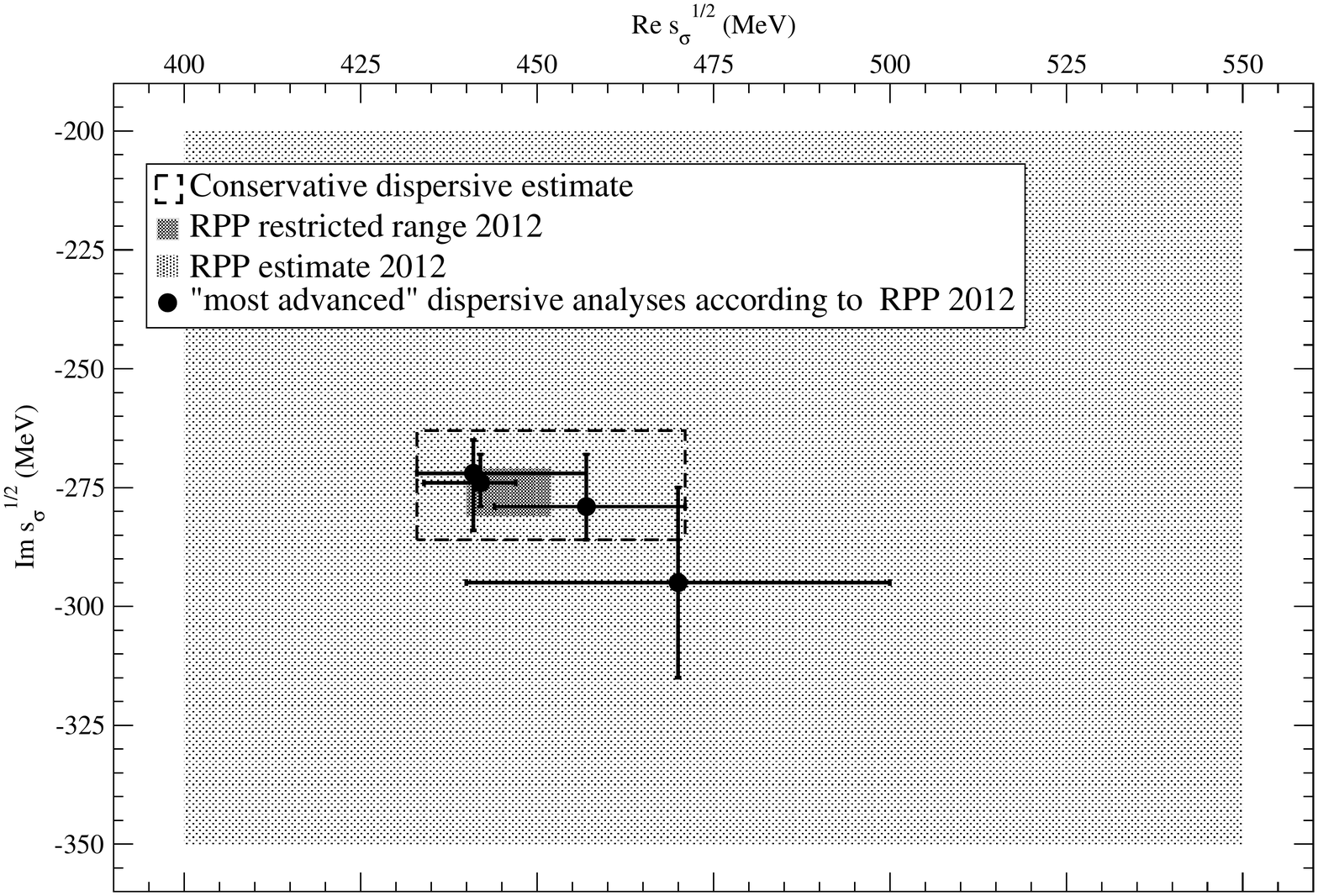}
\vspace*{.1cm}
  \caption{Left panel: Compilation of $f_0(500)$ or $\sigma$ resonance poles in the 2012 RPP edition. The light gray area stands for the
uncertainty assigned to the $\sigma$ poles from 1996 until 2010. The darker gray 
rectangle is the new uncertainty
estimated in the summary tables, Eq.(\ref{rpp2012sigmapole}). 
Right panel: Detail of the new uncertainty band. We only show the four ``most advanced dispersive analyses'' according to the 2012 RPP and as a darker area the ``more radical'' and
``restricted  range of $f_0(500)$ parameters'', Eq.(\ref{rppradicalpole}), 
if one averages those four analyses. The dashed rectangle corresponds to 
the ``conservative dispersive estimate'' advocated here: $\sqrt{s_\sigma}=449^{+22}_{-16}-i(275\pm12)\, {\rm MeV}$ which, as explained in the text, takes
 into account that the differences
between these four dispersive approaches are of systematic nature.
}
  \label{fig:dispersivepoles}
\end{figure}

As a consequence of these rigorous and precise analyses together 
with results from heavy meson decays and all previous 
analyses which consistently found a light $\sigma$ over the previous decades that we have described above, the 
Particle Data Group decided to make a major revision of the $f_0(600)$ parameters in 
the RPP 2012 edition \cite{PDG12}.
In particular, the large 400-1200 MeV uncertainty in the mass was reduced by almost a factor of 5 to to 400-550 MeV, whereas the very large uncertainty of 600-1000 MeV for the width was both shifted downwards
and reduced to become 400-700 MeV. Hence, the RPP T-matrix pole was estimated to be
\begin{equation}
\sqrt{s_\sigma}=(400-550)-i(200-350)\,{\rm MeV}\qquad {\rm (RPP2012 \ estimate)}.
\label{rpp2012sigmapole}
\end{equation}
In the left panel of Fig.\ref{fig:dispersivepoles} we show
this dramatic improvement: the light gray area corresponds to the uncertainty in the RPP since 1996 until 2010, whereas the smaller and darker rectangle represents the new uncertainty listed in the summary tables of the RPP. For consistency with the new uncertainties, even the name of the resonance has changed from $f_0(600)$ to $f_0(500)$.
By comparing with Fig.\ref{fig:poles} 
it can be noticed that some of the poles previously
listed in the RPP have been removed, since now only analyses consistent 
with the NA48/2 final data are taken into account.

However, the darker region can still be considered rather conservative. 
As explained above,
many of the poles shown in that figure correspond to simple 
models without the full analytic properties, 
which did not aim to a precise determination of 
the $\sigma$ parameters, some of them 
correspond to fits with simple functional forms whose analytic continuation is therefore not reliable and some others
are using data sets which have been shown to be inconsistent with dispersion relations.
Only those works which aimed at an accurate determination of the $f_0(500)$ pole based on data descriptions 
consistent with dispersive constraints, and whose poles are obtained from sound analytic extrapolations to the complex planes, should be taken into account. Of course, this was well known to the RPP authors 
and this is why in the 2012 RPP ``Note on scalar mesons below 2 GeV'', they suggested that {\it ``One might also take the more radical point of view and just average the most advanced dispersive analyses, \cite{CGL,Caprini:2005zr,GarciaMartin:2011jx,Moussallam:2011zg}...''}, which we show in the right panel of Fig.\ref{fig:dispersivepoles}, ``{\it which provide a determination with minimal bias}''. By averaging the values obtained in those four references a more restricted range of parameters is estimated at the 2012 RPP:
\begin{equation}
\sqrt{s_\sigma}=(446\pm6)-i(276\pm5)\, {\rm MeV}\qquad {\rm (RPP2012 \ restricted\ range)}
\label{rppradicalpole}
\end{equation}
In the left panel of Fig.\ref{fig:dispersivepoles}, the area covered by this ``restricted'' uncertainty would be almost imperceptible within the darker rectangle, and hence we show in the right panel an expanded view of 
the darker rectangle and just the ``most advanced dispersive analyses'' according to the 2012 RPP. Thus
the ``restricted range of parameters'' corresponds to the smallest
 and even darker rectangle in the middle of the plot. 

\color{black} Now, as we will see in Sect.\ref{sec:parameters}
these uncertainties may now be considered too small \color{black}, since the differences between those
four determinations are more of a systematic than statistical nature. 
Thus, weighting them as if the uncertainties and differences 
were statistical 
to obtain an even smaller uncertainty is somewhat optimistic. 
Moreover, an uncertainty of 
less than 3\% is hard to achieve due to isospin breaking effects, which are not incorporated into these formalisms (except maybe in the experimental uncertainties), and to the absence of $4 \pi$ channels, 
although we have seen above that this effect is very small. A suggestion  would be to 
take as a {\it conservative dispersive estimate} 
the band that covers \cite{Caprini:2005zr} and 
\cite{GarciaMartin:2011jx}, since the pole 
in \cite{CGL} was not really calculated with the analytic extension 
of Roy equations but from the phenomenological representation and 
can be considered superseded by the results of \cite{Caprini:2005zr}. In addition
the result of \cite{Moussallam:2011zg} 
lies within this estimate (it uses results from the \cite{CGL,Caprini:2005zr} group as input). That is:
\begin{equation}
\sqrt{s_\sigma}=449^{+22}_{-16}-i(275\pm12)\, {\rm MeV}\qquad {\rm (Conservative\ dispersive\ estimate)}
\label{myestimate}
\end{equation}
In Fig.\ref{fig:dispersivepoles} this corresponds to the rectangle enclosed within the dashed line.

Thus, as we have just seen, even by the conservative and cautious standards of the RPP
the $\sigma$ was already well established by 1996 settling the issue about its existence.
In 2012, also within the conservative approach of the RPP, the $\sigma$ parameters are 
known well enough to settle the issue of its mass and with: It is definitely light $\simeq 500$ MeV and almost as wide as massive.
But there is still an ongoing debate over other properties that we address next.

\subsection{Present Status of the $\sigma$ spectroscopic classification and nature.}
\label{subsec:statusspectroscopy}

Still under some discussion are the classification of the $\sigma$ into multiplets and its composition in terms of quarks and gluons.
The situation about these two features has also been clarified considerably over the last years and some firm statements can already be made. 

\subsubsection{The nonet of light scalar mesons}
\label{subsec:introscalarnonet}
The problem with spectroscopy 
is to which U(3) multiplet the sigma should be assigned. {\it There is 
a very strong evidence
for the sigma resonance to belong to a nonet with the 
 $f_0(980)$, $a_0(980)$ and $K^*_0(800)$ } \cite{Jaffe:1976ig,Delbourgo:1982tv,vanBeveren:1986ea,Kaminski:1997gc,Kaminski:1998ns,Oller:1997ti,Oller:1997ng,Oller:1998zr,Black:1998wt,Pelaez:2004xp,Close:2002zu}. 

At present, the most controversial among these resonances is the $K_0^*(800)$
or $\kappa$ meson, 
which seems to be following
 a recognition process in the RPP very similar to that of the $\sigma$, although somewhat delayed.
 At present the $K_0^*(800)$ is not yet in the RPP summary tables, but it is present in the 
 particle listings. However, according to the RPP, as of today the $K_0^*(800)$  still ``Needs confirmation''.
Nevertheless, over the last years, the evidence for it has been accumulating 
and is now compelling.
Actually some recent developments are very similar
to those described in previous sections for the
sigma and it seems likely that its situation in the RPP may change relatively soon\footnote{
It is planned to have a revision of the $K_0^*(800)$ or $\kappa$ 
and possibly move it to the summary tables in one of the next RPP
editions \cite{privateHanhart}.}.
In particular: i) it was obtained within UChPT 
in its different variants \cite{Oller:1997ti,Oller:1998hw,Pelaez:2004xp} 
and therefore without a priori assumption about its existence or nature;
ii) it was shown that it should be below 900 MeV \cite{Cherry:2000ut};
iii) its pole was obtained within a very rigorous dispersive approach \cite{DescotesGenon:2006uk}
\color{black} using Roy-Steiner equations, which
are a modification of Roy equations taking into account different masses.  \color{black} The result is
$M_\kappa=658\pm13\,$MeV  and $\Gamma_\kappa=557\pm12\,$MeV.
\color{black} 
A similar pole has also been recently found in conformal parametrizations constrained with Forward Dispersion Relations \cite{Pelaez:2016tgi}.
iv) As it happened with the sigma, 
there is also a strong support for a $\kappa$ from heavy meson decays.
For example, recent results on $J/\Psi\rightarrow K^0_S K^0_S\pi^+\pi^-$ 
decays from the BES2 Collaboration \cite{Ablikim:2010ab} 
require a $K_0^*(800)$ contribution
whose $S$-matrix pole would have $M_\kappa=764\pm164^{+71}_{-54}\,$MeV  
and $\Gamma_\kappa=612\pm298^{+286}_{-170}\,$MeV. Other experiments
also seem to require such a wide resonance, although they usually quote 
model dependent Breit-Wigner parameters and then the mass appears around 800 MeV,
which is why the $\kappa$ is nowadays called $K_0^*(800)$ (see the RPP \cite{PDG12}
for a compilation).
\color{black}
Here it is important to remark that the work in \cite{Cherry:2000ut}
has been often misrepresented \cite{Napsuciale:2004au,Close:2002zu} as refuting the existence of the $\kappa$,
when what it actually does is just to rule out a $\kappa$ at 900 MeV,
but, as emphasized in the very abstract, not a lighter one below 825 MeV. 
Therefore, the existence of a  light $\kappa$
 is not disputed by the authors in that work
\footnote{M. Pennington, private communication.}.

The other members of the possible light scalar multiplet, the $a_0(980)$ and $f_0(980)$ 
are not questioned. 

Note also that the classification of states into multiplets is affected by mixing.
This can be of two kinds: on the one hand, 
the assignment of the scalar-isoscalar positions in the nonet center
is affected by octet-singlet mixing \cite{Oller:2003vf} 
and, on the other hand, all members of the multiplet
 could be affected from mixing with  
members of more massive multiplets \cite{Black:1998wt,Fariborz:2005gm,Achasov:2005hm,Giacosa:2005zt,Giacosa:2006tf}.

These issues will be discussed in detail in Sect. \ref{sec:nature}

\subsubsection{What the $f_0(500)$ is not.}

More controversial is the nature of this resonance, i.e., its composition in terms of quarks and gluons. Of course, one should interpret with care what is meant by ``composition", since the intuitive Fock space decomposition is not always well defined. 
Unfortunately this makes statements about the 
$\sigma$ nature somewhat model-dependent.
Nevertheless, authors usually refer to some kind of ``constituent", 
``dressed" or ``valence" quarks and gluons much the same as in quark models 
one says that  $uud$ is the composition of the proton. 
With this caveat in mind, what seems to be quite 
well established is that {\it the $\sigma$ is not  an ordinary meson 
in the sense that it cannot be interpreted as predominantly 
made of a quark and an antiquark}.
This is also most likely true of its multiplet partners.

This statement is not only due to the long and well recognized 
fact that it cannot be described with a usual Breit-Wigner formalism,
a fact that was revisited also in relatively recent years \cite{Giacosa:2007bn}.
We have already commented that it was long ago 
suggested by Jaffe \cite{Jaffe:1976ig} that the ordinary identification of
mesons as $q \bar q$ states did not match well the light scalars due to their inverted hierarchy as well as their large widths.  But there are other pieces of evidence.

For instance, the large $N_c$ limit is a regime in which the
composition of a meson can be well defined within QCD, 
which also dictates the leading behavior of masses, 
widths and other observables in the $1/N_c$ 
expansion. This is why a strong support
for the non-ordinary nature of the $\sigma$ meson came from 
 the $1/N_c$ behavior of poles generated from unitarized ChPT 
\cite{Pelaez:2004xp,Pelaez:2003dy,Pelaez:2006nj}.
In particular, the $\rho$ pole was shown to behave as expected from QCD
for an ordinary $q\bar q$ meson,
whereas the $\sigma$ pole dependence was a 
rather different one. Let us once again emphasize that these 
poles are generated without introducing them explicitly nor assuming they have any 
specific nature or belong to a particular multiplet. Actually, by 2007 
it was even considered by Jaffe that the different leading $1/N_c$
 behaviors of light scalars found in \cite{Pelaez:2003dy}  were {
\it ``... the only reliable identifications of observed effects that may be
examples of a different class of hadrons"}. 
Subsequently, he considered these different behaviors to illustrate his sound
definition of ordinary and extraordinary hadrons in terms of their 
$N_c\rightarrow\infty$ limit 
\cite{Jaffe:2007id}.
It is also important to remark the difference \cite{Pelaez:2010er} 
between the leading behavior of the $1/N_c$ expansion around $N_c=3$, 
which gives us information on the physical effects of the
$\sigma$ that we observe, and the large $N_c$ limit, of formal interest,
where some results are also available for the $f_0(500)$ \cite{Nieves:2009kh,Nieves:2009ez,Sun:2004de}. Of course, the most
relevant results for the physical resonance are obtained from the former.
The $1/N_c$ non-ordinary behavior was also observed 
in $\pi\pi$ scattering by explicit introduction of resonant states \cite{Nieves:2011gb}. 
Moreover, it has been recently shown that
the same conclusion can be obtained in a model independent way \cite{Nebreda:2011cp},
 without  using any particular model, nor UChPT, but just from data, the pole positions and the general behavior of $q\bar q$ states in QCD.
 We will dedicate Sects.\ref{subsec:nc}, \ref{subsec:ncmodelindep} 
and \ref{subsec:ncUChPT} to 
explain in detail this $1/N_c$ behavior for the $\sigma$.

An additional piece of evidence for a predominantly
 non $q\bar{q}$ nature comes from Regge theory. It is well known that ordinary $q\bar{q}$ mesons
can be classified into straight Regge trajectories which relate linearly their spin $J$ and mass squared, with an almost universal slope $\sim 1 \,\gev^{-2}$. However,
it was long ago pointed out that the ``enigmatic $\sigma$ meson" did not fit into the Regge systematics of $q\bar{q}$ states 
and was supposedly ``alien to this classification" \cite{Anisovich:2000kxa}. Some 
authors avoided this issue by considering its very large pre-RPP2012 width as an uncertainty, 
and with such a large width it can almost fit anywhere.
But Regge trajectories are complex and the width of a 
resonance is not an uncertainty, but should be taken properly into account. 
Actually, it has been possible to {\it calculate} instead of fitting, 
the complex Regge trajectories
of resonances poles that appear in the elastic 
scattering of two mesons \cite{Londergan:2013dza}.  
When calculated for widely accepted $q\bar{q}$ states, 
the resulting trajectories are almost real, linear
and have the universal slope, whereas the $\sigma$
trajectory  is not almost real, but even its real part is non-linear and with a
slope 
much smaller than $\sim 1 \gev^{-2}$. This 
strongly supports its predominantly non-$q \bar{q}$ nature and even hints at a 
hadronic energy scale for the dynamics that generate the resonance.
This will be detailed in Sect.\ref{subsec:reggesigma}

 Note that we have repeatedly 
emphasized that the {\it dominant} behavior is not that of a $q\bar{q}$ state. But that does not exclude the possibility that  a $q\bar{q}$ state could 
appear in its composition as a subdominant component. Actually, 
there is even some evidence that this might be the case and the $\sigma$ might contain some 
small admixture of $q\bar q$ states, which usually are much heavier 
than the physical $f_0(500)$,
as suggested by linear sigma models \cite{Black:1998wt,Close:2002zu},
unitarized quark model calculations \cite{vanBeveren:2006ua},
UChPT and the leading $1/N_c$ behavior \cite{Pelaez:2006nj,Nieves:2011gb},
instanton induced mixings in tetraquark models \cite{Hooft:2008we}
 or semi-local duality arguments \cite{RuizdeElvira:2010cs}.

\color{black}
Note that so far it has not been possible to quantify in a reliable and model independent way how dominant is one component over the others. This is still a matter of debate.
Actually, in the next subsection 
we will see that it is not completely settled what the non-ordinary component may be. In Sec.\ref{sec:nature} we will review the most significant models in detail as well as how to quantify the deviation of the $\sigma$ from the ordinary $N_c$ behavior.
\color{black}

In conclusion, although there might be some subdominant $q\bar{q}$ component in it,
it seems well established that {\it the dominant $\sigma$ 
component is not a $q\bar{q}$ state. 
We thus know what it is not. But the debate is now focused on what it is}.

\subsubsection{What the $f_0(500)$ might be. }

First of all, the glueball interpretation is still advocated by two groups, one in Marseille
\cite{Narison:2000dh,Mennessier:2010xg} and another 
one in Bern-Munich \cite{Minkowski:1998mf,Ochs:2013gi}, who collaborated in $\gamma\gamma$ decays of the sigma \cite{Mennessier:2008kk}. However, this interpretation 
is very hard to support since the very large $\sigma$ width
cannot be accommodated naturally within the expected dependence on the QCD number of colors $N_c$, 
as its width to two mesons should be
 suppressed by $1/N_c^2$, i.e., one more power of $N_c$ than $q \bar q$ mesons, 
which are already narrower than the $\sigma$. 
It has been recently shown \cite{Nebreda:2011cp}, 
within a model independent formalism, that 
very unnatural cancellations, of several orders of magnitude,
 should occur between different orders in the $1/N_c$ expansion
for the $\pi\pi$ elastic phase to be explained 
with glueball-like $N_c$ behavior for the $f_0(500)$ \cite{Nebreda:2011cp}.
The glueball $N_c$ behavior is also at odds with the $N_c$ behavior found in UChPT,
particularly the width.
Concerning the mass, lattice calculations (either in pure Yang-Mills or full QCD calculations)
find the lightest scalar glueball around 1.5-1.8 GeV \cite{latticeglueball}, 
very far from the 500 MeV region where the $\sigma$ is found.
In addition, in many scalar meson models the glueball
can be easily identified as one of the 
main components of the $f_0(1500)$ or $f_0(1710)$
\cite{Albaladejo:2008qa,Bugg:1994mg,Amsler:1995td,Chanowitz:2005du,Giacosa:2005qr,Fariborz:2006xq}.
Moreover, in the $\sigma$ as a glueball scenarios
one has to get rid of the $K_0^*(800)$ or $\kappa$ meson \cite{Ochs:2013gi}
(and of the $f_0(1370)$ as well), which is again in conflict with 
model independent and rigorous dispersive approaches \cite{DescotesGenon:2006uk}
or UChPT \cite{Oller:1997ti,Oller:1998hw,Oller:1998zr,Pelaez:2004xp}. 
The reason is that the $\kappa$ is very similar to the $\sigma$ but with strangeness. However, a glueball 
cannot have strangeness and it would be extremely unnatural 
not to fit the $\sigma$ and $\kappa$ in the same nonet, 
or to have one mechanism explaining the formation of the 
$\sigma$ and a different one for the $\kappa$. Once again, of course,
the above arguments do not exclude the existence of some 
glueball component in the $f_0(500)$ 
but only as long as it is subdominant.

The most extended interpretation is in terms 
of so-called tetraquarks in the sense that
there are two valence quarks and two antiquarks 
almost bounded to form a color neutral resonance.
As we have seen this was first advocated by Jaffe \cite{Jaffe:1976ig})
who showed that such an arrangement can give rise to a nonet of 
light  scalar-isoscalar mesons.
Many authors have followed this suggestion with different variants, 
particularly including the description of meson-meson 
scattering \cite{Black:1998wt,Achasov:2005hm}
or decays \cite{Giacosa:2006rg}. There are also different mechanisms
explaining the quark-level dynamics responsible for the formation of
scalar mesons, like the original ``bag model'' plus one gluon exchange
\cite{Jaffe:1976ig}, diquark-antidiquark configurations \cite{Maiani:2004uc},
even including instanton effects \cite{Hooft:2008we}, which can be used to 
explain the mixing between light tetraquark states and some
$\bar{q} q $ states \cite{Hooft:2008we}.
Recently, a drawback has been found in the pure tetraquark picture (or pure
mixture of tetraquark and quark-antiquark) since 
Weinberg has shown that in the $N_c\rightarrow$ limit
the standard tetraquark is as narrow 
as $q\bar{q}$ ordinary mesons. The very large $\sigma$ width seems to be 
related to some other kind of composition. This means that even 
in these scenarios
some interplay with \color{black} two meson-states, 
whose diagrams \color{black} are subdominant at large $N_c$, is needed.

Nevertheless that is not quite a big problem,
because in the literature
there are several meanings of the
word ``tetraquark''. 
Namely, 
in some contexts the word 
tetraquark only refers to the configuration in which there are no 
color neutral substructures, 
whereas the term ``meson-molecule" or ``composite-mesons''
is used to refer to the case when
each quark-antiquark pair forms a color neutral meson 
and then the resulting two mesons are also quasi-bound.
From the point of view of quantum numbers or valence quarks,
these configurations are indistinguishable, and some general features,
like the inverted hierarchy are common to both pictures. 
\color{black} The use of these words is very informal and that is why I will
frequently use them within quotation marks. \color{black}

Thus, the ``molecule'' name is better suited when the resonance is 
close to a two-meson scattering 
threshold, as it happens with the $f_0(980)$ and the $K\bar K$ threshold \cite{isgur1},
but is a somewhat less natural name for the $\sigma$.
For this reason such a  component is sometimes called the ``pion cloud"
or just ``final state meson-meson interactions''.
Once again, these configurations can mix.
Actually it has even been proposed that the $\sigma$ 
may have a layer structure with some
inner tetraquark structure (with some small admixture 
of $q\bar q$ in a P-wave)
 which rearranges in the outer layers into a
pion-pion state \cite{Close:2002zu}. Some tetraquark models
include this pion-pion state phenomenologically, for instance
with an additional parameter
that describes the'' tunneling'' 
between a diquark structure and the $\pi\pi$ state \cite{Maiani:2004uc}
and is responsible for the widths of scalars.
Predictions about couplings and decays can be made 
but these are often very model dependent and 
the uncertainties are very hard to estimate. 
 Therefore it  seems more adequate not to interpret 
the sigma as a pure elementary or composite object.
The very large width and the correct mass must be reproduced
taking also into account
the hadronic level interaction of the two pions in which the sigma decays. 
Even Jaffe recently argued that ``there 
is no clear distinction between a meson-meson molecule and a $\bar q \bar qqq$ state''
\cite{Jaffe:2007id}.

In addition, apart from the  mixed ``tetraquark/molecule/pion cloud'' 
dominant component, there is also  evidence about some subdominant $q\bar{q}$
component. However, it also seems that this one by itself may be more massive than the 
physical $f_0(500)$.
For instance this is the case of the unitarized quark models 
in which
a quark-antiquark state above 1 GeV 
can be deformed into the physical $\sigma$ meson around 500 MeV by
making a $\pi\pi-q\bar{q}$ interaction sufficiently 
strong \cite{vanBeveren:2006ua}.
There has also been a considerable amount of work studying the mixing of light
tetraquark-like and 
heavier $q \bar{q}$ states within constituent quark-models \cite{Vijande:2005jd} or within the framework of chiral effective Lagrangians
\cite{Black:1998wt,Fariborz:2005gm,Giacosa:2006tf}, leading to very successful phenomenology. 
Studies within the Schwinger-Dyson/Bethe Salpeter approach also seem to suggest a tetraquark/molecule interpretation, with the ligtest $q\bar q$ configuration much heavier than 500 MeV. This will be discussed 
in detail in Sec.\ref{subsec:SDBS},

Moreover, as commented before, within the framework of UChPT it was found that by 
increasing $N_c$ not far from $N_c=3$
the $\sigma$ width did not decrease as for an ordinary meson, but increased instead.
But this is not the whole story, since if one keeps increasing $N_c$, 
it is also possible for the width to decrease again and to
behave as for ordinary $q\bar q$ mesons, although 
at a mass around or above 1 GeV \cite{Pelaez:2006nj}. 
As we will see in Sect.\ref{subsec:semilocal}
this behavior seems to be favored by semi-local 
duality arguments \cite{RuizdeElvira:2010cs} 
and could also be indicative of a sub-dominant  $q\bar q$
component, although once again at a much higher mass than the physical $f_0(500)$.

And last, but definitely not least, there are lattice QCD calculations of scalar mesons. The possibility for tetraquarks to bound into scalar mesons
was shown in \cite{Alford:2000mm}, although for large quark masses.
There are several works that find a $\sigma$ meson \cite{Kunihiro:2003yj,Mathur:2006bs,Prelovsek:2010kg,Wakayama:2014gpa} with a large tetraquark/molecular component. Of course, these are still calculations at high pion masses so that the $\sigma$ cannot decay and have a width, but the results are very encouraging.
Moreover, some lattice works also find a $\kappa$ light meson
with a predominant tetraquark/molecular interpretation \cite{Prelovsek:2010kg,Dudek:2014qha} .

In conclusion, the nature of the $f_0(500)$ is the less robust part of our present knowledge.
With all the caveats about the poor definition of Fock space
within a relativistic context for light quarks, 
 and with a rather informal 
use of language, it can be summarized as follows:
in the most widely accepted scenario the $f_0(500)$ 
is {\it predominantly} made of some $qq\bar{q}\bar{q}$ arrangement.
This could be in the form of conventional
tetraquarks or as a $\pi\pi$-molecule/cloud state or, more likely, a mixture of them. 
In addition, there is some evidence for a subdominant, but non-negligible, admixture 
of a $q\bar{q}$ state whose mass would be above 
1 GeV if its interaction with the tetraquark/molecule/cloud
 component was turned off. 
This subdominant $q\bar q$ component is most likely related to the existence
of another nonet of scalars above 1 GeV, which might be 
predominantly of a $q\bar {q}$ nature.
Sect.\ref{sec:nature} will include a detailed account 
on this whole issue.

Once the historical perspective and 
the present status of the $\sigma$ meson have been provided,
a detailed explanation of the different 
topics and most significant results will be given, starting from
the recent precise dispersive determination of the $\sigma$ parameters.

\section{PRECISE $f_0(500)$ PARAMETERS FROM DISPERSION THEORY}
\label{sec:parameters}

As we have reiterated in the previous section, 
the most reliable extractions of the $\sigma$ parameters---those which have triggered the major revision of the $\sigma$ mass and width in the RPP--- 
have been obtained with dispersive analyses of 
$\pi\pi\rightarrow\pi\pi$ amplitudes. 
After introducing the appropriate 
notation in Sec.\ref{subsec:notation}, data on $\pi\pi$ scattering phase shifts, 
inelasticities and total cross sections will reviewed in
Sec.\ref{subsec:data}. 
The theoretical tools 
will be introduced later on in Sec.\ref{subsec:dispersiveapproaches}. 
In particular the connection between poles and resonances 
and the two most relevant dispersive approaches will be presented in detail.
The application of these techniques to describe the data
will be discussed in Sec.\ref{subsec:cfd} and Sec.\ref{subsec:precisepoles}
will be devoted to the extraction of the $f_0(500)$ pole parameters.

\subsection{Notation}
\label{subsec:notation}
Excellent  pedagogical introductions
to the topics reviewed in this section, including Regge theory,
can be found in the books \cite{MartinSpearman} and \cite{libropipi},
the latter being dedicated solely to $\pi\pi$ scattering,
as well as the relatively recent 
review in \cite{Yndurain:2002ud}.
Here just the very basics 
of different topics of relevance for  $f_0(500)$
pole determinations will be provided, since  
the detailed derivations can be found in the original references. 

\begin{figure}
  \centering
  \includegraphics[width=0.35\textwidth]{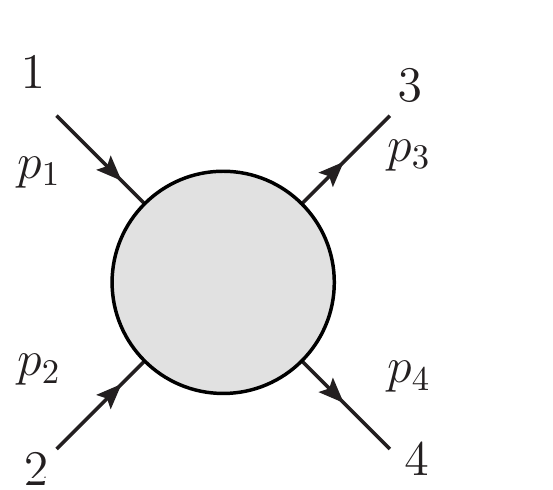}
\vspace*{.1cm}
  \caption{ $\pi_1(p_1)\pi_2(p_2)\rightarrow\pi_3(p_3)\pi_4(p_4)$ scattering.
}
  \label{fig:12-34}
\end{figure}

For $\pi_1(p_1)\pi_2(p_2)\rightarrow\pi_3(p_3)\pi_4(p_4)$ scattering, 
where
$\pi_k=\pi^0,\pi^\pm$ and $p_i$ are the four-momenta of the particles in the process, as illustrated in Fig.\ref{fig:12-34}, 
we will define the usual 
Mandelstam variables as $s=(p_1+p_2)^2$, $t=(p_1-p_3)^2$ and $u=(p_1-p_4)^2$. 
Only two of these variables are independent,
since $s+t+u=4M_\pi^2$, and frequently 
the $u$ dependence will be omitted if not needed, i.e. 
the scattering amplitude may be denoted as $T(s,t,u)$ or $T(s,t)$. 
Note also that the isospin limit 
$M_{\pi^\pm}=M_{\pi^0}\equiv M_\pi$ will be considered
unless explicitly stated otherwise. 
Thus, in practice, it is most convenient to work with
amplitudes of definite isospin $I$ in the $s$-channel:
\begin{eqnarray}
    T^{I=0}(s,t,u)&=&3A(s,t,u)+A(t,s,u)+A(u,t,s), \label{ec:isospinbasis}\\
    T^{I=I}(s,t,u)&=&A(t,s,u)-A(u,t,s),\nonumber\\
    T^{I=2}(s,t,u)&=&A(t,s,u)+A(u,t,s),\nonumber
\end{eqnarray}
where
$A(s,t,u)=T(\pi^+\pi^-\rightarrow \pi^0\pi^0)$.
In addition, it is also very useful to work with partial waves of
definite angular momentum. Unfortunately, in the literature
there are various normalizations and notations when writing the amplitudes and
the partial wave expansion, which 
we summarize next:
\begin{eqnarray}
&&T^{(I)}(s,t)=32 \pi \sum_\ell (2\ell+1)P_\ell(\cos\theta) t^{(I)}_\ell(s),\quad F^{(I)}(s,t)=4 \pi^2 T^{(I)}(s,t),\label{ec:pwdef}\\
&&t^{(I)}_\ell(s)=\frac{\sqrt{s}}{2 k}\hat f^{(I)}_\ell(s)=\frac{ \hat f^{(I)}_\ell(s)}{\sigma(s)}, \quad
\hat f^{(I)}_\ell(s)=\frac{\eta_\ell^{(I)}(s)e^{2i\delta_\ell^{(I)}(s)}-1}{2i},
\nonumber 
\end{eqnarray}
where $\delta_\ell^{(I)}(s)$ and $0\leq\eta_\ell^{(I)}(s)\leq1$ are the phase shift 
and inelasticity of the $I$, $\ell$ partial wave, $\ell$ is the angular momentum, 
$k^2=s/4-M_\pi^2$ is the center of mass momentum and we have defined for convenience 
$\sigma(s)\equiv 2k/\sqrt{s}$. Note that we have introduced 
an overall factor of 2 in the first equation due to pions being identical in the isospin limit.

Let us now remark that partial waves will be 
extended to the complex energy squared plane in search
of resonance poles. The region of convergence of the partial wave series 
is known as the Lehman-Martin ellipse in the $t$ plane
\cite{Lehman-Martin}, which is
derived from axiomatic field theory plus the 
positivity condition $\im t^{(I)}_\ell(s)\geq0$, 
implied by unitarity.  
In the case of $\pi\pi$ scattering the ellipse has
foci at $t=0$ and $t=\mps-s$, and the right extremity
is at $t=r(s)$, where $r(s)=16s/(s-4\mps)$ for 
$4\mps\leq s\leq 20 M_\pi^2$ and 
$r(s)=4s/(s-16\mps)$ for $s\geq 20 M_\pi^2$ \cite{Caprini:2005zr}. 
For textbooks on this issue see \cite{libropipi,libromartin}.

In the elastic case, $\eta=1$ and 
\begin{equation}
\hat f^{(I)}_\ell(s)=\sin\delta_\ell^{(I)}(s)\,e^{i\delta_\ell^{(I)}(s)},
\end{equation}
so that 
\begin{equation}
\im \hat f^{(I)}_\ell(s)= \vert \hat f^{(I)}_\ell(s)\vert^2,
\end{equation}
which is nothing but the elastic unitarity condition.

Note that $I=0,1,2$ and since all pions are 
considered identical bosons in the isospin limit,
 whenever $I$ is even (odd) 
then $\ell$ is even (odd). Thus the 
isospin index for odd waves will be omitted, i.e.,
$\delta_1^{(1)}$ will be denoted by $\delta_1$.
In addition, we may refer to partial waves either by their $I$, $\ell$ quantum numbers or by the
usual spectroscopic notation S0, S2, P, D0, D2, F, G0, G2, etc...
In particular, the scalar-isoscalar partial wave, where the $f_0(500)$ appears, is called the S0-wave.
In addition, since we are dealing with scattering of spin zero mesons, 
the angular momentum $\ell=J$, where $J$ is the total angular momentum,
and thus we will often write $t_J^{(I)}$ instead of $t_\ell^{(I)}$, since both notations are frequent in the literature.

Later on we will be interested in threshold parameters defined from the following expansion
\begin{eqnarray}
  \frac{s^{1/2}}{2 M_\pi k^{2\ell+1}}{\rm Re}\, \hat{f}_\ell^{(I)}(s)\simeq
  a_\ell^{(I)}+b_\ell^{(I)} k^2+O(k^4).
  \label{eq:defthreshold}
\end{eqnarray}
Note that $a_\ell^{(I)}$ and $b_\ell^{(I)}$ are the usual scattering lengths and slope parameters.
Customarily these are given in
$M_\pi$ units.

\subsection{The $\pi\pi$ scattering data}
\label{subsec:data}

\color{black}
We separate the discussion of experimental information on $\pi\pi$ scattering
in three energy regions. Partial wave analyses providing experimental information on $s$ and $t$ dependence exist up to 1.8 GeV. The general features in the whole energy range will be discussed first in Sec.\ref{subsec:generalpw}, 
whereas the specific features of the elastic and inelastic regions
will be discussed in Secs.\ref{subsec:pwbelow}
and \ref{subsec:pwabove}, respectively. 
Finally, in section \ref{subsec:highenergydata} 
we discuss the experimental information on total cross sections at higher energies
and how it is described in terms of Regge theory, from which the $t$
behavior is inferred.

\subsubsection{General features of $\pi\pi$ partial wave determinations.}
\label{subsec:generalpw}

\color{black}

The best and most widely used data on partial-wave $\pi\pi$ scattering phases and inelasticities 
are obtained from two kind of processes, $\pi N\rightarrow\pi\pi N'$ 
scattering and $K\rightarrow \pi\pi e \nu$, also known as the $K_{e4}$ decay.

For the S0-wave, the $\pi N\rightarrow\pi\pi N'$ data comes from: i) the $\pi^+ p\rightarrow\pi^+\pi^-\Delta^{++}$ reaction ( and $\pi^+ p\rightarrow K^+K^+\Delta^{++}$ for the inelasticity) by Protopopescu et al. \cite{Pr73},
which provides several solutions; ii) the high statistics study of the $\pi^- p\rightarrow\pi^-\pi^+ n$ reaction at 17.2 GeV
performed by Grayer et al. \cite{Cern-Munich}, which  provides 5 different solutions;
and iii) the Krakow group reanalysis \cite{Kaminski:1996da} 
of the CERN-Krakow-Munich Collaboration on $\pi^- p\!\uparrow \rightarrow\pi^-\pi^+ n$
using a polarized target \cite{cern-munich-crakow}.
\color{black}
The extraction of the $\pi\pi$ amplitude is a complicated procedure.
For a detailed description we refer the reader to the excellent book \cite{libropipi} and the review lectures in
\cite{Petersenlectures}, besides the original experimental references.
In this section we will only provide a sketch of the method and its caveats, together with a briefly commented list of relevant references.

The key observation is that the above reactions 
are dominated by the exchange of one pion, 
which together with the pion in the
initial state and the two in the final state
 form  the $\pi\pi \rightarrow\pi\pi$ subprocess. 
The main assumptions
 are that in this one-pion exchange 
mechanism the $\pi\pi$ scattering amplitude factorizes in the
total amplitude and that partial waves can be extracted 
by measuring angular distributions of the final di-pion state.
However, there are several complications in this picture.

The first complication is the existence of phase-shift ambiguities that 
affect differently the elastic and inelastic regions, and
yield different mathematically acceptable solutions for the S0 wave.
These ambiguities will only be discussed very briefly in the following 
subsections because they were solved a long time ago and dedicated
books or reports
\cite{libropipi,Petersenlectures} already exist on this issue.
For our discussion, the relevant observation is that
there is one widely accepted solution of these ambiguities
in the whole energy range of interest, which is the one already shown 
in Fig.\ref{fig:00data}. 

However, the second complication 
is that, even after sorting out these ambiguities,
the resulting solution 
is still plagued with further systematic uncertainties
which for long prevented a precise determination of the amplitude. 
These were due to:
\begin{itemize}
\item the fact that the exchanged pion is not on-shell and an on-shell extrapolation is needed.
\item exchanges of more pions, known as absorption (or Reggeized $\pi$-exchange). 
\item other resonance exchanges with ``natural'' parity (i.e. the same as for the pion),
like the $a_1(1260)$, generically called $A_1$ exchanges.
\item contributions from other resonances (which could also be Reggeized)
with `unnatural'' parity, like the $a_2(1320)$, also called $A_2$ exchanges.
\end{itemize}
These non-trivial contributions, which may spoil factorization,
 were implemented by means of models. As a consequence, 
even within the same experiment 
and using the same solution of the ambiguities, 
systematic uncertainties dominate the result, as it was already seen in 
Fig.\ref{fig:00data},
 and make a purely statistical analysis
 meaningless. One of the aims of this report is to review the recent 
dispersive techniques used to obtain precise results despite the existence of these large systematic uncertainties.

Let us now comment on the particular features of each energy region.

\subsubsection{Partial waves below $K\bar{K}$ threshold}
\label{subsec:pwbelow}

In this region
phase-shift ambiguities appear when there are more helicity amplitudes than observables
available from the $\pi N\rightarrow\pi\pi N'$ reaction.
This happened in the first studies  
when only a few observables were used, either due to low  statistics in some component of the angular distributions, the inability to
separate the one-pion exchange from other contributions,
or the absence of polarization measurements.

For instance, before the advent of full amplitude analyses, 
only one angular distribution was used, because all others were polluted 
by contributions other than one-pion exchange. But this observable was
sensitive to $\sin(\delta_P-2\delta_S)$ and only
$\vert \delta_S-\delta_P-\pi/4\vert$ could be determined. Even though the S2 and P waves
were relatively well known, this led to the so-called ``up-down'' ambiguity in the 
S0 phase shift. Actually, there was an ambiguity on each side of the energy 
where the above combination vanishes, which occurs near the $\rho(770)$, 
so that for a time there was a four-fold ambiguity up-up, up-down, down-up and down-down. The ambiguity
below the $\rho(770)$ was easily resolved by 
refining the amplitude analyses with more angular distribution observables or for instance with $K_{\ell4}$ data. 

Once again, a detailed description of the different ambiguities 
on the S0 phase-shift arising
from different analysis methods and how they are resolved is well beyond the scope of this review. Thus, we just recommend the pedagogical and extensive references 
\cite{libropipi} and \cite{Petersenlectures} and
provide in the next paragraph a very brief account of 
some later discussions and references.
\color{black}

The ``up" solution above the $\rho(770)$ was actually behind some claims of a narrow $\epsilon$ resonance around 750 MeV.
Until the 90's it was a general belief that the issue had been resolved in the 70's in favor of the ``down" solution by the observation of a very rapid S-wave amplitude decrease  between 950 and 980 MeV 
\cite{Hyams:1973zf,Swaveanomaly} 
\footnote{which roughly corresponds to the evidence for the $S^*$, now $f_0(980)$, being on top of a background so that the cross section changes from
unitarity saturation to almost vanishing within a small energy interval.}, as well as the observation of a wide enhancement in the $\pi^0\pi^0$ system around 800 MeV, hard to reconcile with the ``up" solution \cite{pi0pi0noup}.
In addition, Forward Dispersion Relations \cite{Morgan:1969ca} and Roy equations \cite{Pennington:1973hs} seemed to imply that only the ``down" solution was consistent with dispersion theory.
However,  the ``up-down" controversy was briefly
resurrected and killed again in the 90's \cite{Svec:1995xr}    
when a narrow $\sigma$ was suggested around 750 MeV 
in a reanalysis of the CERN-Krakow-Munich polarized data.
The key point was the observation that the $a_1$ resonance exchange, ignored in most previous analyses,  
was important to extract $\pi\pi$ scattering from $\pi N\rightarrow\pi\pi N'$.
The subsequent Krakow group reanalysis of \cite{Kaminski:1996da} 
confirmed that the $a_1$ 
contribution was sizable and that
 such a solution with a narrow $\epsilon$ was 
indeed possible. 
 Actually, it was found that each ``up" and ``down" solution could be either be ``steep" or ``flat" and that the previous Roy 
 equations analysis had only excluded the ``steep-flat" ambiguity, but it was still possible to have either ``up-flat"
 or ``down-flat" solutions.
The very same Krakow group found later on \cite{Kaminski:2002pe} 
 that only the ``down-flat" solution was reasonably consistent with Roy equations. Therefore, since this ``up/down-flat/steep" issue is
settled, Fig.\ref{fig:00data} only shows
data sets consistent with the ``down-flat'' solution, which from now on
will be the only one considered here.

Even though the sigma appears in the S0-wave,  P, D and F-waves are also relevant for dispersive analyses. 
In general, their structure is much simpler. The P-wave is  dominated by the $\rho(770)$, as seen in 
right panel of Fig.\ref{fig:00data}, and the D0-wave is dominated by the $f_2(1270)$.
The I=2 waves, which are repulsive,
were also extracted from $\pi N\rightarrow\pi\pi N'$
but with two charged pions in the final state \cite{pipiscatteringI2}. The F wave is very tiny and little is known about it.

\begin{figure}
  \centering
  \includegraphics[width=0.45\textwidth]{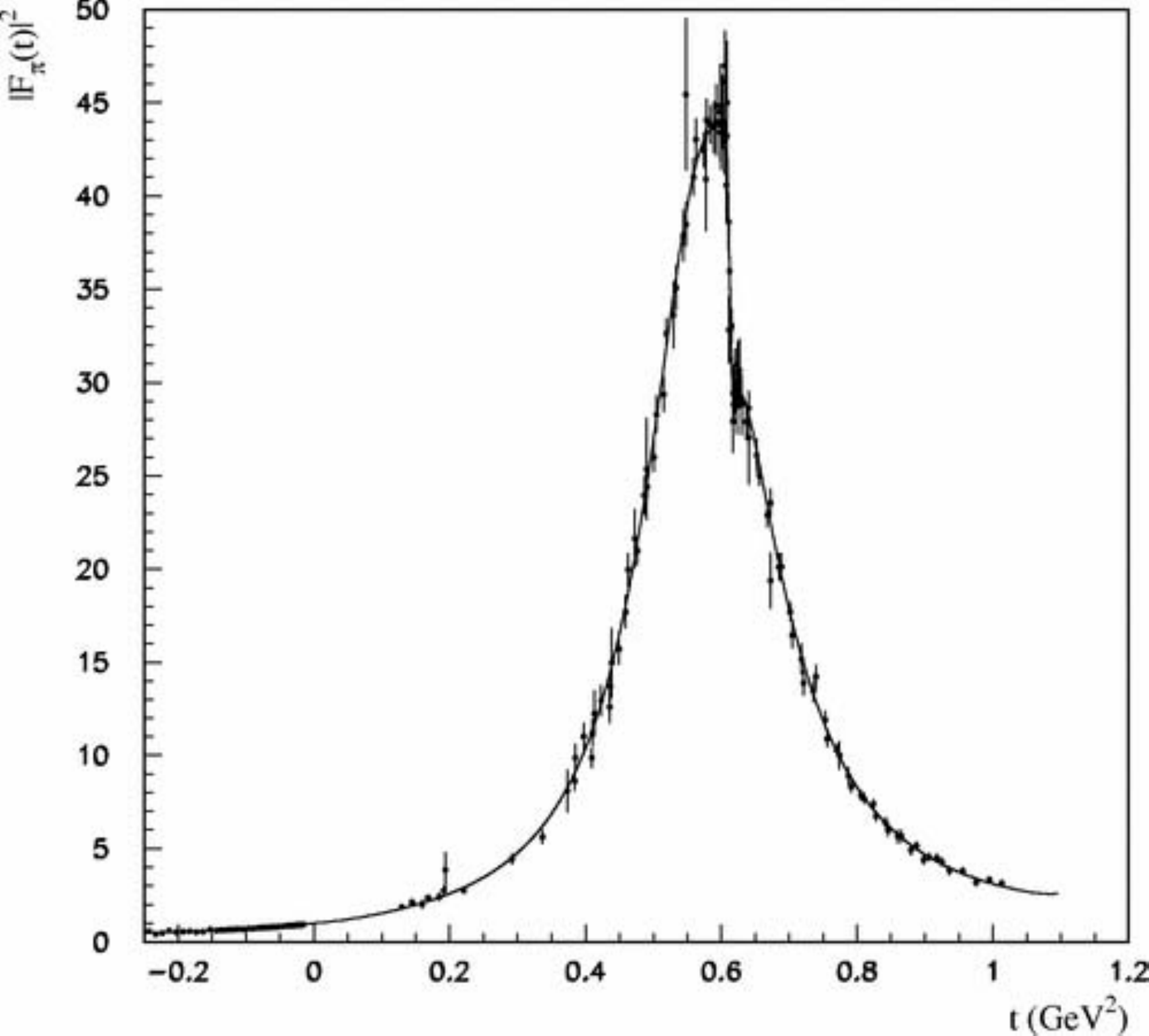}
  \includegraphics[width=0.54\textwidth]{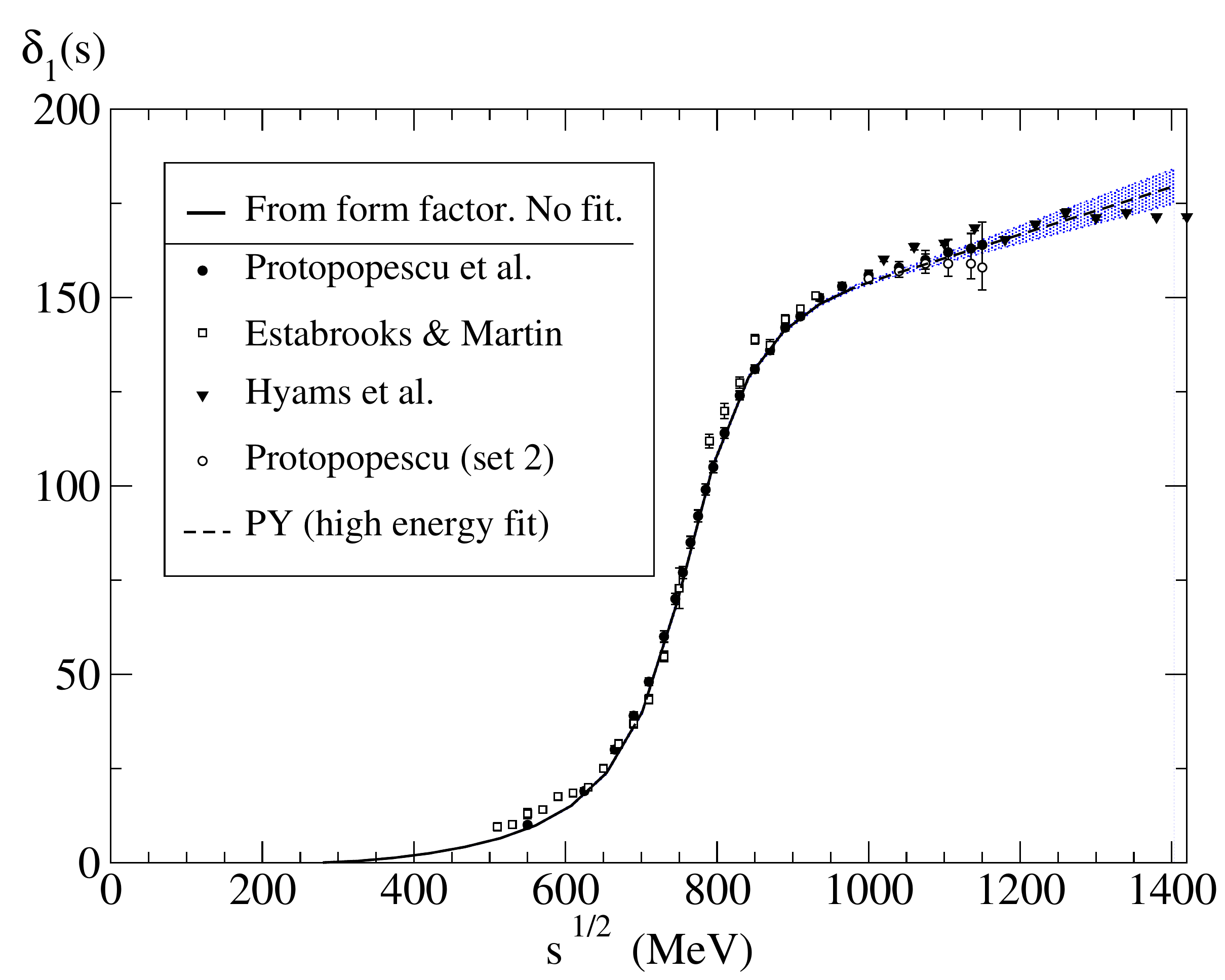}
\vspace*{.1cm}
  \caption{Left panel:  Precise pion
form factor description. 
Right panel:  Phase shift data sets from $\pi N\rightarrow\pi\pi N'$ \cite{Pr73,Hyams:1973zf,Estabrooks:1974vu}. The curve \cite{Pelaez:2004vs}
 below $K\bar K$ threshold is {\it not} fitted to those data, 
but predicted from the form factor 
shown in the left panel. In contrast, above $K\bar K$ threshold the dashed curve
is a fit to scattering data. 
Left figure from \cite{de Troconiz:2004tr}, reprinted with permission from J.~F.~De Troconiz and F.~J.~Yndurain,
  Phys.\ Rev.\ D {\bf 65}, 093001 (2002). Copyright 2002 by the American Physical Society.
Right figure taken from \cite{Pelaez:2004vs}.
}
  \label{fig:FF}
\end{figure}

The P-wave deserves particular attention since it will
 be needed when dealing with $K_{e4}$ data.
This is the best known partial wave, although not from $\pi N\rightarrow\pi\pi N'$, but from the pion form factor measured 
in $e^+e^-$ scattering or, more recently, in $\tau$-decays (not contaminated by $\omega$ production). 
By Watson's final state theorem \cite{Watson:1952ji},
in the elastic region the phase of the form factor is exactly that of $\pi\pi$ scattering in the vector-isovector channel.
Electromagnetic corrections can be accounted for and a very rigorous treatment
can be performed \cite{De Troconiz:2001wt,de Troconiz:2004tr}. 
Actually, on the left 
panel of Fig.\ref{fig:FF} we show the description 
of the form factor provided in \cite{de Troconiz:2004tr},
which below the $K\bar K$ threshold leads to the {\it prediction} 
for the scattering phase shown in the right panel versus  the data 
obtained from $\pi N\rightarrow\pi\pi N'$. 
As we can see, the prediction is 
extremely good and much more reliable than a fit to those 
$\pi N\rightarrow\pi\pi N'$ data sets, which indeed 
are incompatible among 
each other if taking into account just their statistical uncertainties.
Therefore, at least in the elastic region, there is a very precise and 
reliable description of the P-wave data. Above $K\bar K$ threshold
one has to rely again on 
$\pi N\rightarrow\pi\pi N'$ data and therefore 
the uncertainties become much larger \cite{Pelaez:2004vs}.

\begin{figure}
  \centering
  \includegraphics[width=0.5\textwidth]{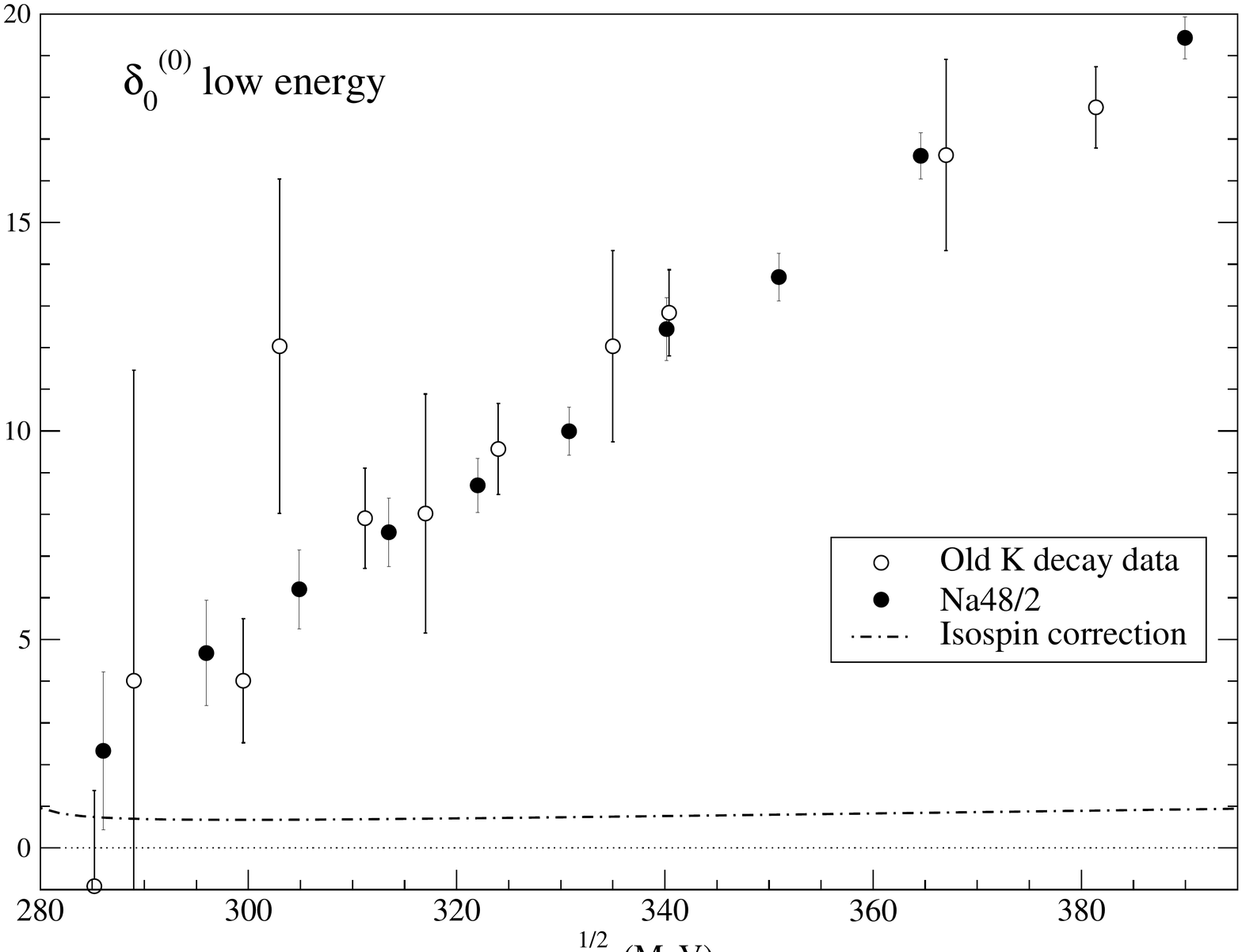}
\vspace*{.1cm}
  \caption{ Data on S-wave $\pi\pi$ scattering from $K_{e4}$ decays.
The NA48/2 data of 2010 comes from \cite{Batley:2010zza} and is much more precise
than previous measurements \cite{Rosselet:1976pu,Pislak:2001bf}.
Note that all these data have been plotted after subtracting the isospin correction 
derived in \cite{Colangelo:2008sm} 
and adding the $\delta_1$ phase obtained in \cite{de Troconiz:2004tr} from the pion form factor.
}
  \label{fig:Kl4data}
\end{figure}

 The other relevant source of data for $\pi\pi$ scattering comes from $K_{e4}$ decays. 
As we saw in the historical introduction, 
the first experiments were already carried out in the 70's \cite{Zylbersztejn:1972kd,Rosselet:1976pu}
but they have been repeated later in the new millennium with much larger statistics: 
in 2001 by the BNL-E865 Collaboration \cite{Pislak:2001bf} and in 2010 by the CERN NA48/2 Collaboration \cite{Batley:2010zza}. They provide the most reliable and accurate
information about S-wave $\pi\pi$ scattering below 400 MeV
and it confirms the accepted solution of the ``up-down'' ambiguity at low energies.
Once again Watson's final state theorem implies that,
in the isospin limit,
 the phase of this weak decay is given by the phase 
of the $\pi\pi$ rescattering in the final state. 
 These pions can be produced with total isospin 0 or 1, 
and therefore what is measured is just the interference
 of both isospin amplitudes, i.e., $\delta_0^{(0)} -\delta_1$, as seen in Fig.\ref{fig:NA48data}. Of course, the real world is not exactly isospin symmetric and the masses of the charged 
and neutral pions differ by roughly 3\%. This is the typical size of isospin violations in pion systems,
but it is not necessarily a good estimate 
close to the $2\pi$ threshold as in the $K_{e4}$ case. 
The reason is that in the 
real world there are actually two thresholds: one for $\pi^0\pi^0$ 
about 10 MeV below that for $\pi^+\pi^-$. 
In between these two thresholds only real $\pi^0$
pairs can exist, and this makes isospin violation effects much larger than naively expected in other energy ranges. 
In 2008 \cite{Colangelo:2008sm} it was realized that this was 
an important correction to $K_{e4}$ and 
a relatively simple equation  was provided in time
to correct for isospin and extract the isospin symmetric  $\delta_0^{(0)} -\delta_1$
combination from the NA48/2 experiment. The difference between taking into account this correction
or not can be seen in Fig.\ref{fig:NA48data}, where the red data points correspond to 
the isospin corrected results. \color{black} The net effect is to decrease 
the value of $\delta_0^{(0)} -\delta_1$ by a small correction  between $0.65^{\rm o}$
and $0.88^{\rm o}$, depending on the point. (Detailed values for each energy
are given by the NA48/2 
Collaboration in Table 12 of their work \cite{Batley:2010zza}). \color{black}
Of course, for previous $K_{e4}$ experiments 
the correction given in \cite{Colangelo:2008sm}
 should also be applied.
Thus, in Fig.\ref{fig:Kl4data} we show the data on $K_{e4}$ 
for the S-wave,
that is, including the isospin correction and adding the $\delta_1$ phase shift obtained from the form factor \footnote{even including an overall
 3\% uncertainty for isospin violation, the uncertainty
in the P-wave is smaller than the uncertainty in the NA48/2 results}. 
The size of the 
isospin correction that has been subtracted to the data is also shown in the figure.  The latest 
and very precise NA48/2 data \cite{Batley:2010zza} 
have been highlighted with respect to the older experimental results, because they have turned out to be 
very decisive in the RPP major revision of the $\sigma$ pole. 
Previous experiments are certainly less precise.

Before reviewing the data above 1 GeV, it is worth mentioning that there are other experimental results
concerning the S-wave at threshold. These come from
pionic atoms \cite{Adeva:2007zz,Adeva:2005pg}, yielding:
$  a_0^{(0)}-a_0^{(2)}=0.280\pm0.013{\rm (St.)}\pm0.008{\rm (Syst.)}\,M_\pi^{-1}$ and
$ a_0^{(0)}-a_0^{(2)}=0.264^{+0.033}_{-0.020}\,M_\pi^{-1}$, respectively; or from
or  $K_{3\pi}$ decays, also studied by the NA48/2 Collaboration \cite{Cabibbo:2005ez}, which yield:
$  a_0^{(0)}-a_0^{(2)}=
0.268\pm0.010{\rm (St.)}\pm0.041{\rm (Syst.)}\pm 0.013{\rm (Ext.)}\,M_\pi^{-1}$.
In general, any approach describing the NA48/2 $K_{e4}$ data is also compatible with these results.

\subsubsection{Partial waves above $K\bar K$ threshold}
\label{subsec:pwabove}

Although in the 500 MeV energy region $\pi\pi$ 
scattering is almost purely
elastic, the sigma pole is deep in the complex 
plane and in order to determine it
accurately and  reliably, information 
from the inelastic region is also needed.
Up to 1.8 GeV data 
on partial waves have
 been obtained in terms of 
the phase shift and inelasticity. It is important to remark
that this inelasticity takes into account within its uncertainties 
{\it any possible channel} to which $\pi\pi$ can couple. 
Therefore, using this phase-inelasticity
parameterization {\it all channels are included}, 
as long as they provide a contribution to the inelasticity,
contrary to what can happen with specific models
which start from a dynamical description (either from a Lagrangian or explicit inclusion of resonances) and include just a few states in order to calculate the inelastic amplitude.

The inelasticity to $4\pi$, $6\pi$... to which $\pi\pi$ can couple, 
seems to be very small below the $K\bar K$ threshold 
and has not even been observed. Therefore, for an analysis 
based on experimental data, 
we have to consider that it has been 
included within the systematic uncertainties, 
which, as we have commented above, are rather large for other reasons.
Nevertheless, S-wave inelasticity has been measured and
becomes relevant around the $K\bar K$ threshold,
close to the $f_0(980)$ resonance, whose decay to $K\bar{K}$ is seen and the corresponding branching
ratio is of the order of a few tens of a percent \cite{PDG12}.

\color{black}
Once inelasticity sets in, further ambiguities arise when extracting partial waves from data.
Note that the phase-shift and the modulus of the partial wave 
become independent
and also that in practice the partial wave series in Eq.\ref{ec:pwdef} is truncated at some $\ell_{max}$.
Thus, the amplitude becomes a polynomial 
in $z=\cos \theta$, which 
can always be written as $T(s,z)=c \exp(i\phi) \Pi^{\ell_{max}}_{i=1} (z-z_i)$, where $c$ is a real constant and $z_i$ are its zeros. However, from experiment one determines $\vert T \vert^2=c^2 \Pi^{\ell_{max}}_{i=1} (z-z_i)(z-z_i^*)$ and does not know whether 
the amplitude has a zero at $z_i$ or $z_i^*$. 
These are the so-called Barrelet ambiguities \cite{Barrelet72} 
on the sign of each $\im z_i$.
As a consequence, even the same experimental group can provide several solutions in the inelastic regime.

As already commented, a full analysis of these ambiguities is well beyond the scope of this report, 
which focuses on the sigma and the precise analyses once 
these ambiguities have been sorted out (see for instance \cite{PDG76,Froggatt:1975me,PDG84} as well as the
book  \cite{libropipi} and the report \cite{Petersenlectures}
and references therein). 
Here we will only comment briefly on
the experimental situation
 and why one particular solution, called $(- - -)$, became widely accepted as the correct one. 
We will also see in Secs.\ref{subsec:cfd} and \ref{subsec:precisepoles} that, 
for the $f_0(500)$ precise determination, the most severe
situation with ambiguities, which occurs above 1.4 GeV, has been circumvented
by using Regge  instead of partial waves.
The results for the sigma are very compatible using either Regge theory or the $(- - -)$ partial-wave solution. 

Thus, the most elaborated partial wave studies, 
set $\ell_{max}=3$ and 
the analyses involve three zeros, ordered by increasing $\re z_i$, and consequently three possible sign ambiguities. 
However, it can be shown that $\im z_3<0$
below 1.8 GeV. 
\color{black}
Concerning the data above $K\bar K$ threshold, 
those from Protopopescu et al. \cite{Pr73} only reach 1150 MeV, whereas the CERN-Munich 
Collaboration in its latest work (Hyams et al. \cite{Cern-Munich-high}) made a full study dedicated to $\pi\pi$ from 1 to 1.8 GeV.
Their analysis allowed for two different solutions below 1.4 GeV and four different solutions above that energy. 
\color{black}
These are labeled by the signs of $\im z_i$
as $(+ - -),(+ + -),(- + -)$ and $(- - -)$.
Several methods were combined to select one solution over the others, namely:
i) an slightly better  $\chi^2$, ii) consistency
with $\pi^0\pi^0\rightarrow\pi^0\pi^0$ data, iii) absence of anomalous behaviors or unphysical artifacts and,
once again, iv) fixed-$t$ dispersion relations and Roy equations.
For example:
\color{black}
Hyams et al. \cite{Cern-Munich-high}, showed that two of their solutions were ``somewhat less favored by $\chi^2$'' and some may contain unphysical artifacts 
(like the $(+ - -)$ solution ). 
They also indicate that one of the most favored solutions,
 the so-called ``$(- - -)$'' solution 
is almost identical to the one provided in a previous work 
of the collaboration \cite{Hyams:1973zf} below 1.4 GeV.
In addition, in \cite{Shimada:1974jv} it was shown 
that the only solution that ``does not induce anomalous behavior
of the amplitude zeros in other charge configurations'' 
is precisely the $(- - -)$. Moreover, this solution is also 
compatible with those of \cite{Kaminski:1996da}, where polarized data is also analyzed. Furthermore, it has been recently shown
that, with $\pm 5$ degrees added as a systematic uncertainty,
 the $(- - -)$ solution is
compatible with Roy equations up to 1.1 GeV 
and Forward dispersion Relations up to 1.4 GeV \cite{GarciaMartin:2011cn}, 
although only when accompanied by the ``dip'' solution for the inelasticity, that we comment next. 
Therefore, below 1.4 GeV there is a general consensus 
on considering the $(- - -)$ solution as the correct one.

\begin{figure}
  \centering
  \includegraphics[width=0.5\textwidth]{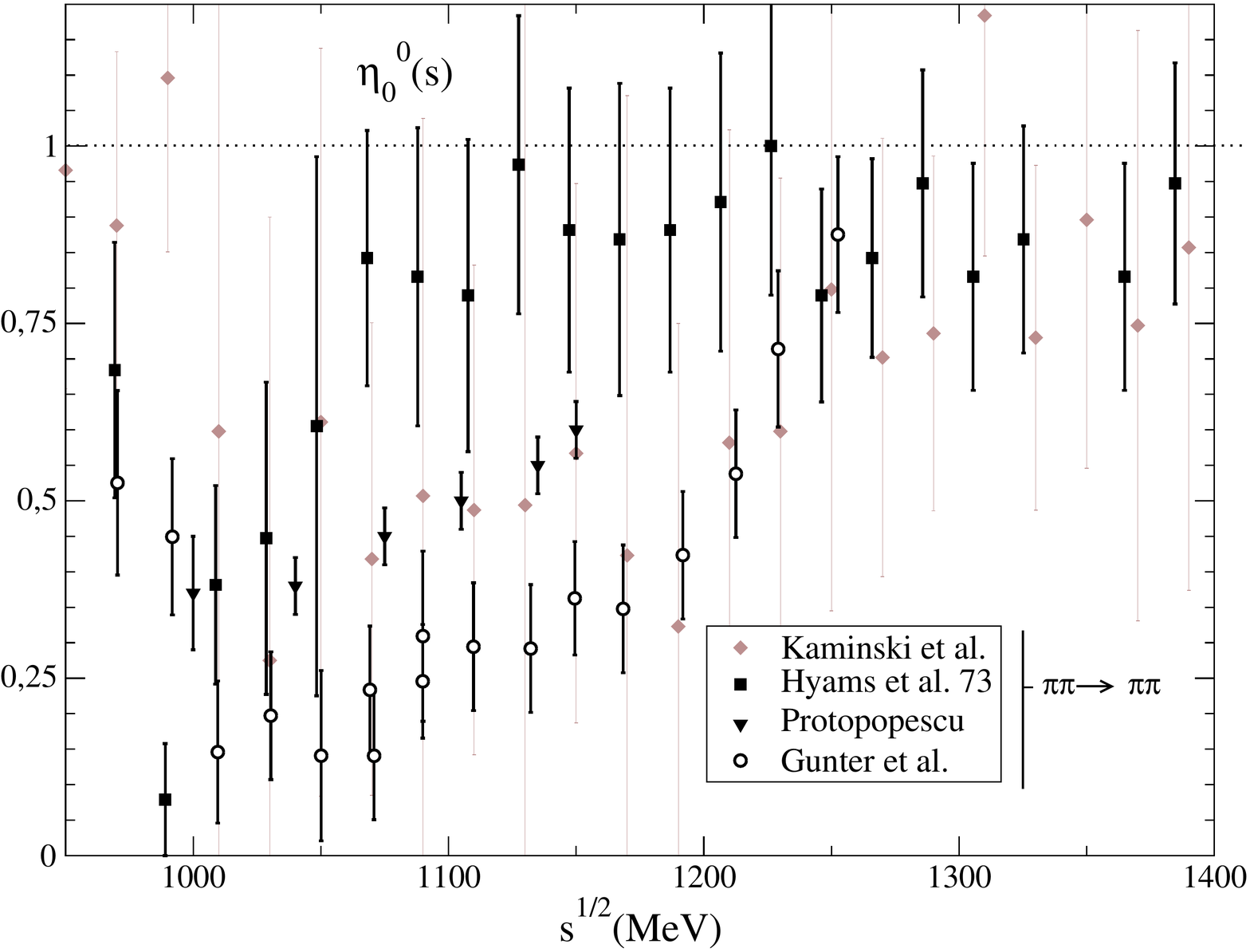}
\hspace*{-.3 cm}
  \includegraphics[width=0.5\textwidth]{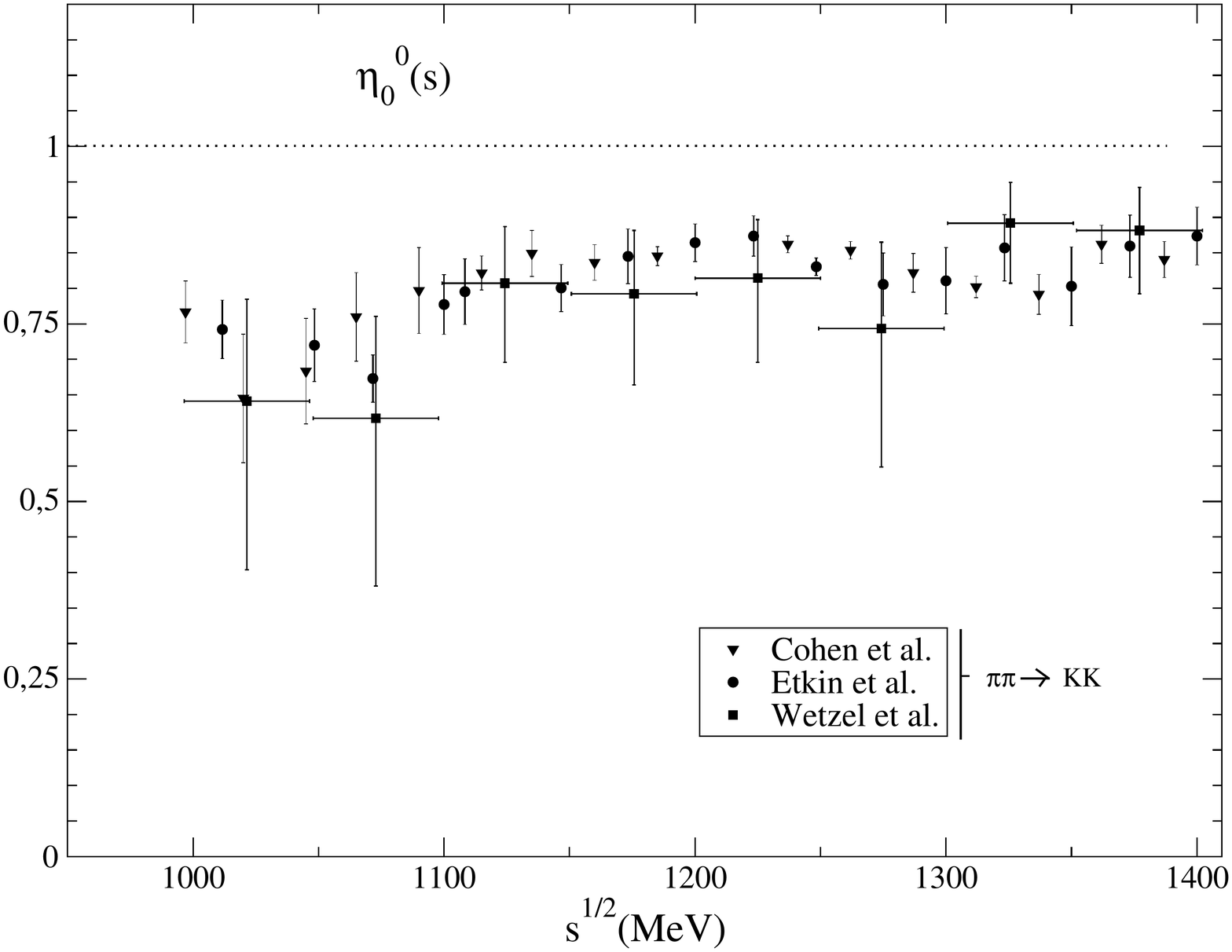}
  \caption{$\pi\pi$ inelasticity above the 1 GeV region. In principle it
should be $0\leq\eta_0^{(0)}\leq1$. Left panel:  Measured 
from the subprocess $\pi\pi\rightarrow\pi\pi$ \cite{Pr73,Hyams:1973zf,Gunter:1996ij}. Note that the value of $\eta_0^{(0)}$ 
suffers a  drastic drop between 1 and 1.1 GeV. We have plotted lighter the results of Kaminski et al.
\cite{Kaminski:1996da} because 
they have so large uncertainty bands that would obscure the existence of a dip.
Right panel: Measures from the subprocess $\pi\pi\rightarrow K\bar K$ \cite{inelpipiKK}.
Note the data does not show the dip structure of the left panel.
}
  \label{fig:ineldata}
\end{figure}

Actually, there has also been a 
longstanding controversy about the size of the inelasticity 
right above the 1 GeV region 
\cite{Au:1986vs,Zou:1993az,Morgan:1993td,Bugg:2006sr}. This controversy is illustrated
in Fig.\ref{fig:ineldata} where we show on the left panel the inelasticity measured
in $\pi\pi$ final states \cite{Pr73,Hyams:1973zf,Gunter:1996ij} 
\footnote{\cite{Gunter:1996ij} corresponds to some proceedings of the E852 Collaboration, for phase shifts their uncertainties are not competitive with those published in other references and that is why they have not been shown in Fig.\ref{fig:00data}}, where a ``dip'' structure is seen between 1 and 1.1 GeV,
whereas on the right panel we show the data obtained from $\pi\pi\rightarrow K\bar K$ \cite{inelpipiKK},
where such a pronounced dip is not observed. We will see that dispersion relations have 
recently settled this issue in favor of the ``dip-solution''. 

Above 1.4 GeV, apart from having four solutions, 
the experimental results are much less reliable.
First of all, because of the convergence of the partial wave expansion, since it can be shown that around 1.7 GeV the F-wave is already as large as the P-wave, the D0 as large as the S0 and the D2 as large as the S2. 
In addition, the Hyams et al. results \cite{Cern-Munich-high} seem to imply that above 1.5 GeV the amplitude is almost elastic, or at least the elastic part is bigger than the inelastic one, contrary to all other processes like $\pi N$, $NN$. Even more, the inelastic $\pi^+\pi^-$ cross section seems to decrease between 1 and 1.7 GeV. Similar concerns can be stated for other waves, and for a thorough discussion of the caveats on 
the $\pi\pi$ phase shifts above 1.4 GeV we refer the reader to \cite{Pelaez:2004vs,Pelaez:2003eh}. \color{black} It is also worth remarking that W. Ochs, a member of the CERN-Munich collaboration, has presented a recent reanalysis in \cite{Ochs:2013gi} supporting the $(-+-)$ solution above 1.4 GeV.
Fortunately, this is identical to solution $(---)$ below 1.4 GeV
which is the most relevant region for our purposes. \footnote{ I thank W. Ochs for these observations.} \color{black}

Thus, Fig.\ref{fig:00data} shows the
data on the $\pi\pi$ phase shift 
up to 1.4 GeV, obtained from $\pi N\rightarrow \pi\pi N'$. 
Recall that below 1.4 GeV 
the Hyams et al. data from \cite{Hyams:1973zf} 1973 (called ``Hyams 73'' in the plot) is
almost identical to the  $(- - -)$ solution by the same group in Hyams et al.
\cite{Cern-Munich-high} obtained in 1975,
which is the only one plotted up to that energy.

\color{black}
We will see that due to the above caveats, some dispersive analyses 
\cite{Pelaez:2004vs,GarciaMartin:2011cn,GarciaMartin:2011jx}
do not make use of partial waves above 1.4 GeV, but use Regge theory instead.
\color{black}
Let us nevertheless remark that beyond 1.4 GeV some authors  
\cite{CGL,ACGL,Moussallam:2011zg} still make use of the Hyams et al. data
\cite{Cern-Munich-high}
 (although the impact of this energy region in their results is rather small). 
\color{black}
We will also see that both approaches give compatible results for the $f_0(500)$ parameters.
\color{black}
In 
any case at some given
high energy the partial wave formalism has to be abandoned and one has to 
rely on full amplitudes and a different description of high energy data.
This is done by means of Regge theory, which we comment next.

\subsubsection{High energy data}
\label{subsec:highenergydata}

In principle, dispersive integrals extend to infinite energies, but
since our interest is on the $f_0(500)$ region we will suppress 
the higher energy region by means of subtractions, which
will be explained in detail below. Nevertheless,
in order to claim precision, the high energy contribution should be taken into account and for this we will now review the existing data
above the 1.4 GeV region. In
practice the data description up to a few GeV is more than enough.

As we will see later, the most relevant dispersion relations for our purposes
are of two types: those with
fixed $t=0$, called Forward Dispersion Relations, and those for partial waves, where $t$ is integrated out.
For the former type, we only need the $\pi\pi$ forward 
amplitude that, by the optical theorem, is proportional 
to the total $\pi\pi$ cross section, for which there is data at high energies. In contrast,
partial-wave dispersion relations also need 
the $t$ dependence, for which data is not available at high energies but can be obtained from other processes assuming factorization.

\begin{figure}
  \centering
  \includegraphics[width=0.48\textwidth]{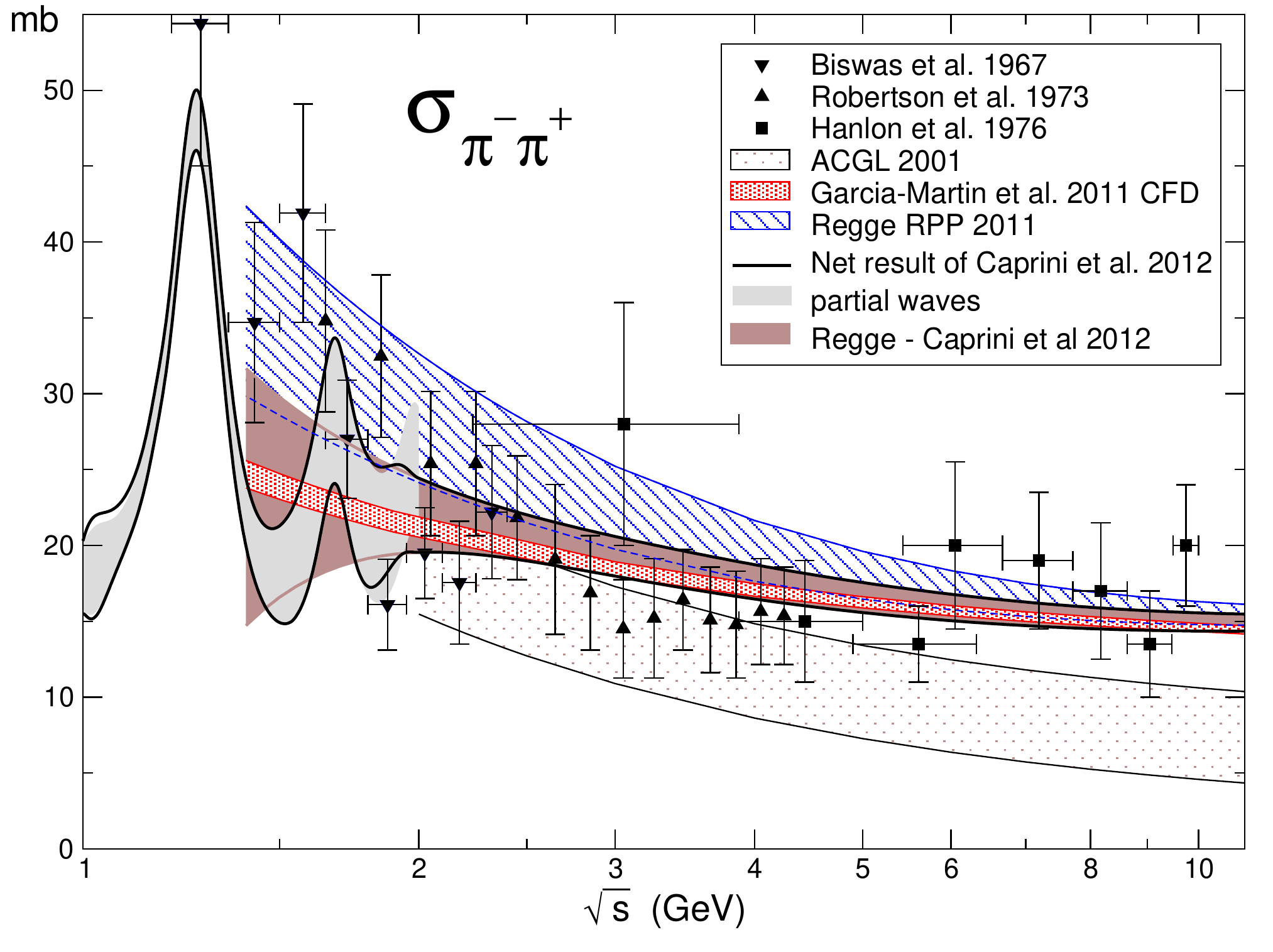}
  \includegraphics[width=0.48\textwidth]{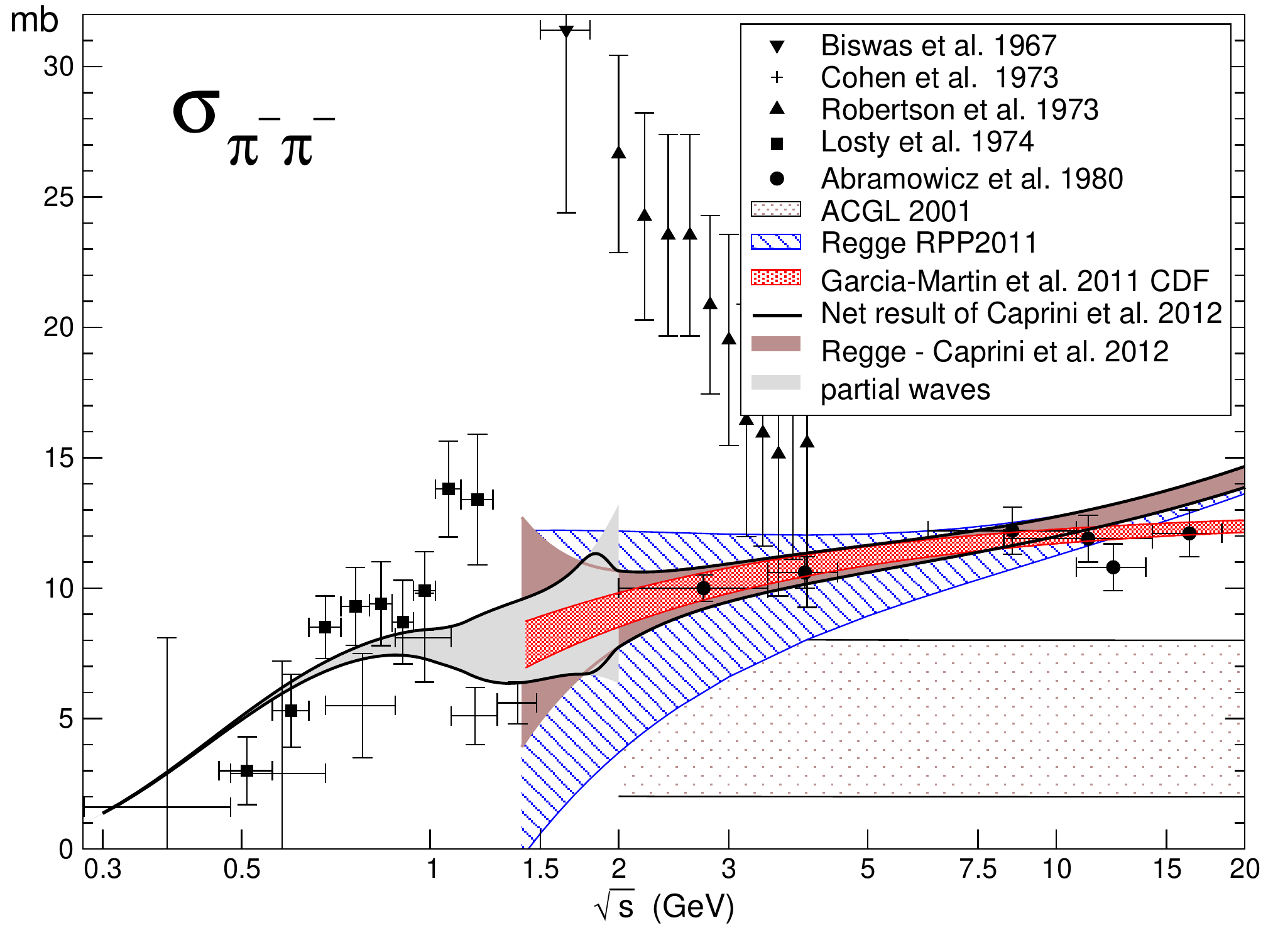}
  \caption{ Compilation of data on $\pi\pi$ total cross sections versus some Regge parameterizations: 
ACGL \cite{ACGL}, the 2011 update of the 2010 edition of the RPP \cite{PDG10},
Garc\'{\i}a-Mart\'{\i}n et al. \cite{GarciaMartin:2011cn} (update from 
\cite{Pelaez:2003eh,Pelaez:2003ky}), the Regge result from \cite{Caprini:2011ky}.
The grey band is the cross section as reconstructed from the partial wave analyses in \cite{Caprini:2011ky} and the Losty et al. data come from \cite{Lostycrosssection}.
Left: $\pi^-\pi^+$ data from \cite{Biswasregge,Robertsonregge,hanlonregge}.
Right: $\pi^-\pi^-$ data from \cite{Biswasregge,Cohenregge,Robertsonregge,Abramowitzregge}.
The author thanks I.~Caprini, G.~Colangelo and H.~Leutwyler for kindly providing their 
tools for making these plots.
}
  \label{fig:reggedata}
\end{figure}

\begin{figure}
  \centering
  \includegraphics[width=0.49\textwidth]{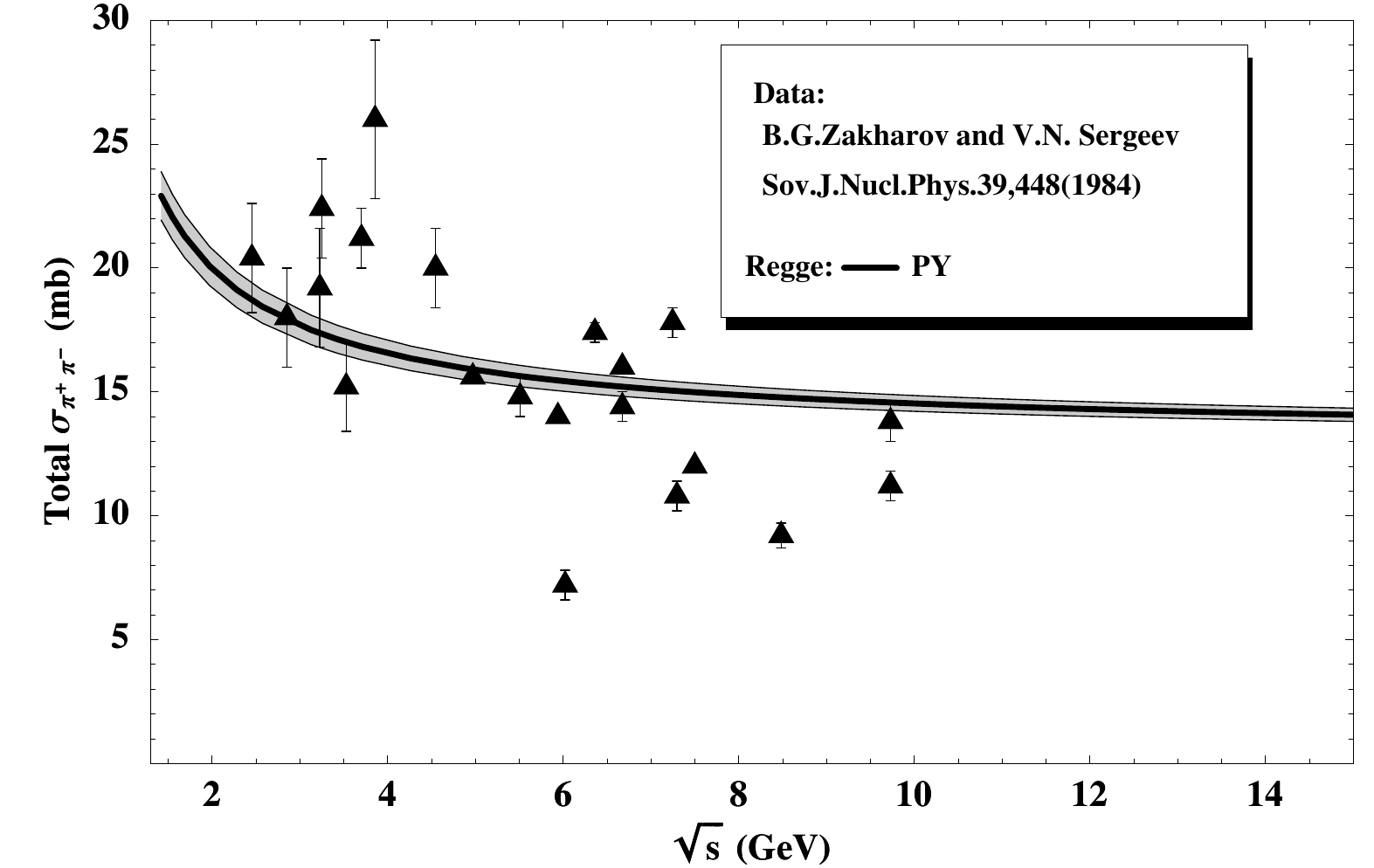}
  \includegraphics[width=0.49\textwidth]{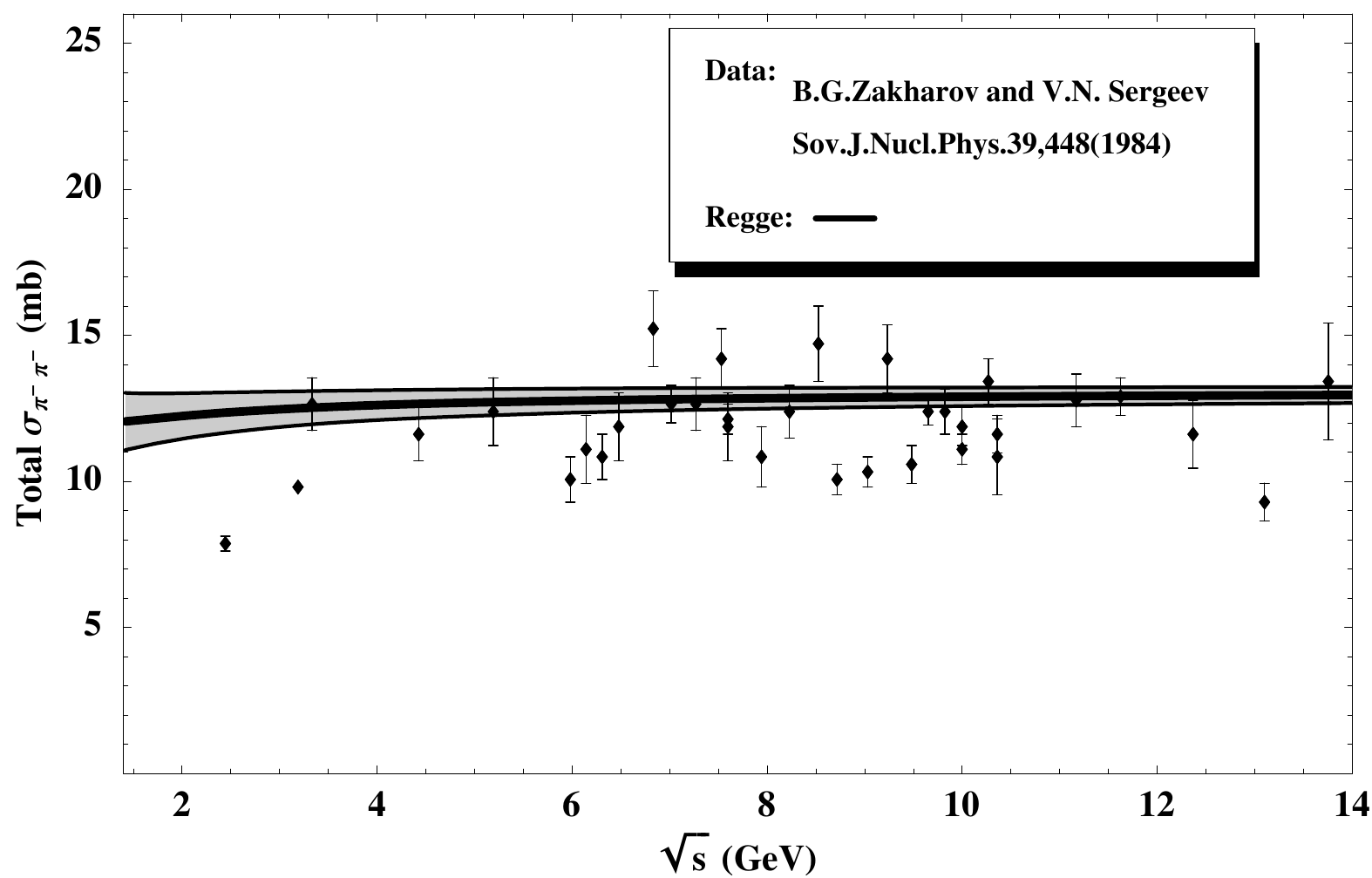}
\vspace*{.1cm}
  \caption{ $\pi\pi$ total cross sections from \cite{Zakharov}.
Although not shown, a 7-10\% systematic uncertainty should be added to
these data according to the authors of \cite{Zakharov}.
The curves are not fits, but correspond to the Madrid-Krakow parameterization of \cite{Pelaez:2003eh,Pelaez:2003ky}. Figures taken from 
 ``Regge description of high energy pion pion total cross sections,'' J.~R.~Pelaez,
  Int.\ J.\ Mod.\ Phys.\ A {\bf 20}, 628 (2005). Copyright 2005 World Scientific Publishing Company.
}
  \label{fig:reggerussians}
\end{figure}

Let us first discuss the total cross sections, which have been measured 
for the following initial states: $\pi^+\pi^-$ \cite{Biswasregge,Robertsonregge,hanlonregge,Zakharov}, $\pi^-\pi^-$ \cite{Biswasregge,Robertsonregge,Cohenregge,Abramowitzregge,Zakharov} and $\pi^0\pi^-$ \cite{Biswasregge}. The data for the latter are very scarce and provide little information. 
These data were obtained from the following reactions:
$\pi^- p\rightarrow\pi^+ \pi^- n$ \cite{Biswasregge,Robertsonregge,Cohenregge},
$\pi^\pm p\rightarrow \Delta^{++} X$, 
as well as $\pi^\pm n\rightarrow p X$ \cite{hanlonregge,Abramowitzregge,Zakharov}.
We show these data in Figs.\ref{fig:reggedata} and  \ref{fig:reggerussians}. Note that, particularly
below the 2 GeV, these data sets are not very consistent with each other.

The description of high energy $\pi\pi$ scattering cannot be done in terms of 
the partial wave expansion because, as we have seen, somewhere between roughly 1.5 and 2 GeV
that expansion breaks down.
High energy hadron-hadron scattering is well described in terms of Regge theory
and other dispersive techniques like sum rules, 
which were developed in the 60's and 70's when they were also applied to
$\pi\pi$ scattering 
\cite{Basdevant:1972uu,roy70,Pennington:1973xv,Pennington:1974kp,Froggatt:1975me,lsv,Olsson,Wanders}.
Introductory textbooks can be found in \cite{Donnachie:2002en,CollinsRegge}.
As we explained here in the introduction, for some time later these techniques were practically abandoned for $\pi\pi$ scattering
(see \cite{Caprini:2011ky} for a nice account on the developments
 of Regge phenomenology for this reaction).

However, the interest in $\pi\pi$ scattering was revived in the 2000's due to the new uses of
Roy equations. The data had been somewhat forgotten, but
in \cite{Pelaez:2003eh,Pelaez:2003ky} it was found that these total cross sections
could be described fairly well {\it on the average} with 
Regge theory, by including the following four Regge trajectories: the Pomeron, the P' (or $f_2$), 
the $\rho$ and some isospin 2
double $\rho$ exchange. \color{black}
Let us recall that there is a well-known
equivalence, called ``average duality'' or ``semi-local duality'', 
between the direct channel resonances and the crossed channel Regge poles. This allows for a dual description of amplitudes in terms of Regge exchanges that extends from the asymptotic region 
down to the resonance region, although in the latter the equivalence is not local in energy but ``on the average'' over an energy region.
For an introduction to duality in Regge theory 
we refer once again to \cite{Donnachie:2002en,CollinsRegge}.
Note that this  
{\it on the average description} over a certain energy region, 
is enough \color{black} if one wants to use those expressions 
inside integrals, as it is the case of dispersion theory.
Most of the parameters of these trajectories can be obtained from a fit to 
data on total cross sections for $\pi N$ and $NN$, and then used for $\pi\pi$ using factorization.
This approach is basically an update of the Regge parameterizations suggested by Rarita {\it et al.} \cite{Rarita}.
The rest of the parameters were determined from $\pi\pi$ scattering at low energies by means of 
some sum rules, as the one derived by Olsson \cite{Olsson} or those derived
in \cite{ACGL} from crossing symmetry,
which were forced upon the fit. To avoid the conflicting data sets, $\pi\pi$ scattering was only fitted above 2 GeV
whereas below that energy  an extrapolation of the Regge formulas was used. Note that the data of \cite{Zakharov} was not included in the fits, since it was rediscovered a posteriori. In addition, to reach the TeV range one should use other parameterizations, which include some logarithmic growth, provided in \cite{Pelaez:2003ky}. For our purposes the simpler parameterizations in \cite{Pelaez:2003ky} can be used, which extend to 10-20 GeV and tend to 12-13 mb 
at very high energies.
Actually, these simple parameterizations and fits
have been gradually improved in subsequent works by the same Madrid-Krakow group \cite{Yndurain:2007qm,Pelaez:2004vs,Kaminski:2006qe}
whose latest version is \cite{GarciaMartin:2011cn}. In Fig.\ref{fig:reggedata} these parameterizations correspond to the blue dashed area. It has also been shown in \cite{Pelaez:2004ab} that these parameterizations describe the data from \cite{Zakharov}, as shown in Fig.\ref{fig:reggerussians} (once a 7-10\% systematic uncertainty is added to those data 
following the estimate of the authors of \cite{Zakharov}).

The description of these data was revisited 
later by the Bern-Bucharest group in \cite{Caprini:2011ky}, going beyond the simple average description and including non-leading Regge contributions to ensure a smooth transition 
between the energy regions where partial wave or Regge formalisms are used.
This led to the continuous light blue area in Fig.\ref{fig:reggedata}, which certainly
matches the partial wave representation at 2 GeV. Below that energy, 
it should still be considered an average description.
It can be noticed that 
 results from \cite{Caprini:2011ky} are very compatible with those from the
 Madrid-Krakow group \cite{Pelaez:2003eh,Pelaez:2003ky}, although the inclusion of subleading terms
produces a larger uncertainty and they can be considered more conservative.
Recall that none of the parameterizations have been obtained by fitting 
the \cite{Zakharov} data, although as seen in Fig.\ref{fig:reggerussians} these data are 
reasonably well described too. If these data were to be included in the fits  of 
the Bern-Bucharest group, their uncertainties would decrease \cite{reggeinprep} and become somewhat closer to those
used by the Madrid-Krakow group in \cite{GarciaMartin:2011cn}.

This whole picture was also revisited by Halzen and Ishida in \cite{Halzen:2011xc}, 
with slightly different
functional forms. Their result is once again 
compatible with \cite{Pelaez:2003eh,Pelaez:2003ky}. They actually note that even though extending Regge theory down to 1.4 GeV might not be guaranteed to work a priori,
 it still provides a ''fairly good'' description
at those low energies.

In conclusion, there is a fairly good  agreement between the data and different models on total 
$\pi\pi$ cross sections above 1.4 GeV, although some approaches are 
somewhat more conservative on the uncertainty estimates. As remarked above, this input is needed both for 
forward or partial wave dispersion relations.

However, for partial wave dispersion relations, the $t$ dependence of amplitudes at
high energies is also needed.  Unfortunately there is no data here
and one has to rely on pure Regge theory and several 
sum rules that constraint the possible $t$
behavior of the residues of the different Regge trajectories. 

In particular the behavior of the rho reggeon exchange \color{black} at $t=4 M_\pi^2$
\color{black}
is constrained by the
following sum rule \cite{Pelaez:2004vs}
\begin{eqnarray}
I\equiv \int_{4M^2_\pi}^\infty\!\!\dd\! s\,
 \frac{\imag F^{(I_t=1)}(s,4M^2_\pi)-\imag F^{(I_t=1)}(s,0)}{s^2}-
  \int_{4M^2_\pi}^\infty\dd \!s\,\frac{8M^2_\pi[s-2M^2_\pi] \,\imag F^{(I_s=1)}(s,0)}{s^2(s-4M^2_\pi)^2}=0.\quad
\label{eq:Isumrule}
\end{eqnarray}
It can be checked that the contributions of the S-waves cancel and only the P and D  waves 
contribute (we also include F and G waves, but they are negligible). 

\color{black} Another sum rule constraining the $t$ dependence of the Regge contribution
to the amplitude \color{black}
was given in \cite{ACGL} and requires the vanishing of
\begin{eqnarray}
 J\equiv\int_{4M^2_\pi}^\infty\dd s\,\Bigg\{
 \frac{4\imag F'^{(0)}(s,0)-10\imag F'^{(2)}(s,0)}{s^2(s-4M^2_\pi)^2}
 -6(3s-4m^2_\pi)\,\frac{\imag F'^{(1)}(s,0)-\imag F^{(1)}(s,0)}{s^2(s-4M^2_\pi)^3}
 \Bigg\}=0.\label{eq:Jsumrule} 
\end{eqnarray}
\color{black} Note that the $t$ dependence has been recast in terms of   
$F'^{(I)}(s,t)\equiv\partial F^{(I)}(s,t)/\partial\cos\theta$. \color{black}
At high energy, the integral is dominated by isospin zero Regge trajectories, thus
this sum rule is more appropriate to constrain the Pomeron and the $P'$ reggeon.

\begin{figure}
  \centering
  \includegraphics[width=0.49\textwidth]{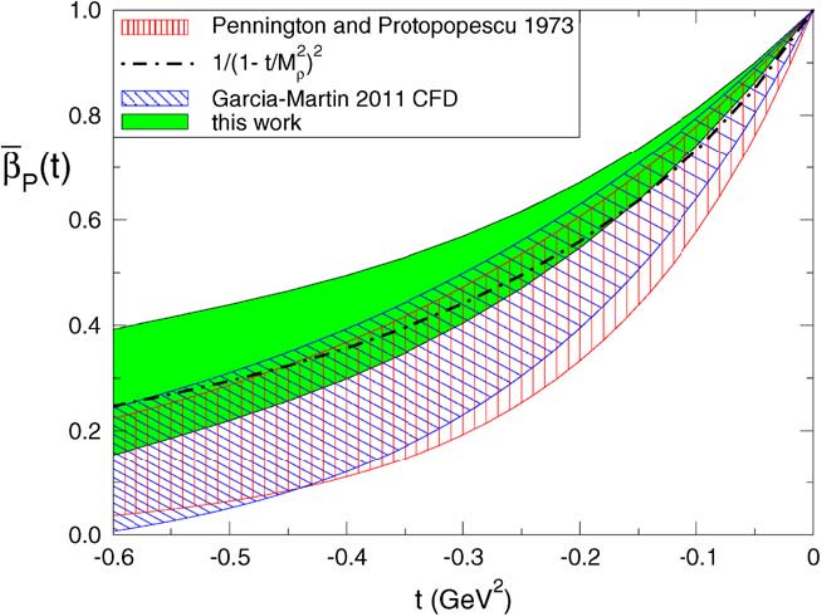}
  \includegraphics[width=0.49\textwidth]{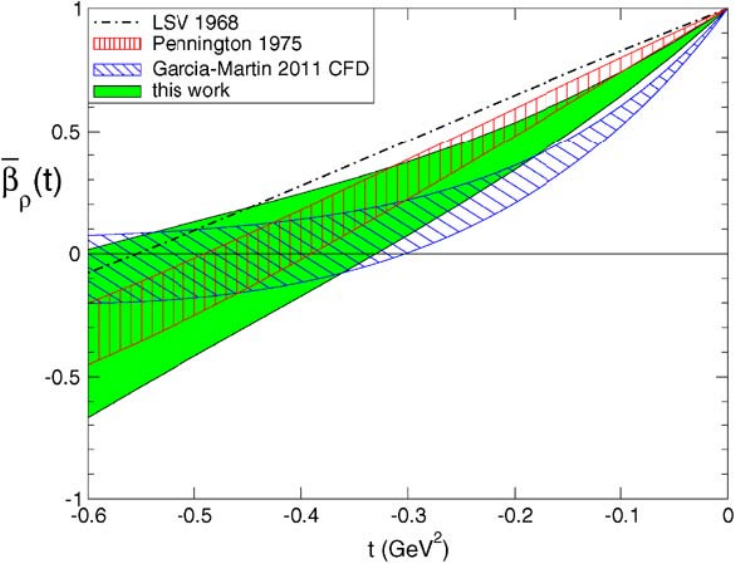}

  \includegraphics[width=0.49\textwidth]{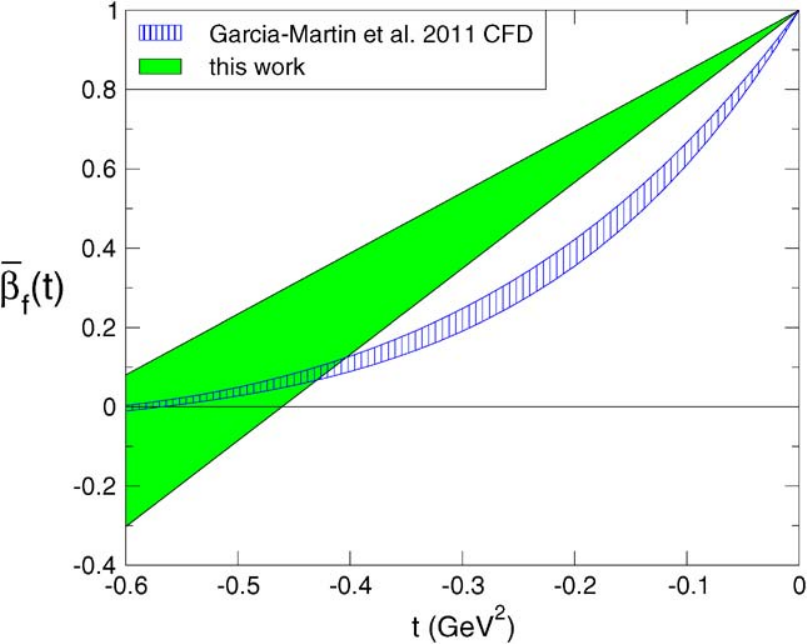}
  \includegraphics[width=0.49\textwidth]{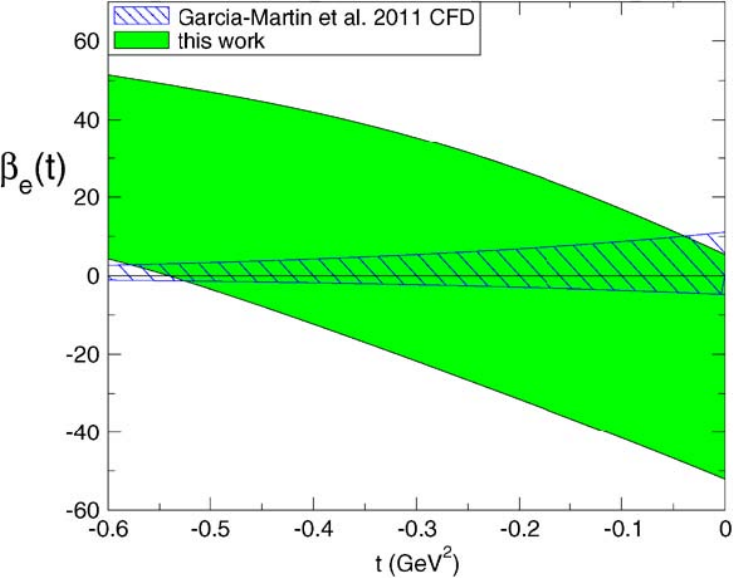}
\vspace*{-.1cm}
  \caption{  
Regge residues responsible for the $t$ dependence of $\pi\pi$ scattering at high energies.
The figures represent the profiles of these residues, i.e., once the $t=0$ 
value has been factored out: $\beta(t)=\beta(0)\bar \beta (t)$.
The Regge parameterizations shown correspond to 
the Lovelace-Shapiro-Veneziano (LSV) model \cite{lsv,ACGL}, 
Pennington and Protopopescu \cite{Pennington:1973xv}, Garc\'{\i}a-Mart\'{\i}n et al. \cite{GarciaMartin:2011cn} (update from \cite{Pelaez:2003eh,Pelaez:2003ky}),
and as green bands those of \cite{Caprini:2011ky}.
The agreement for the leading Pomeron trajectory is fairly reasonable, as well as for the 
$\rho$ residue. Apart from the larger uncertainties of \cite{Caprini:2011ky},
the only significant disagreement with the parameterizations in
 \cite{Pelaez:2003eh,Pelaez:2003ky,GarciaMartin:2011cn} is in the residue of the
subleading $f$ trajectory.
Figures taken from \cite{Caprini:2011ky}.
}
  \label{fig:reggeresidues}
\end{figure}

The $t$ dependence of the Regge trajectories, \color{black} including the previous constraints,
\color{black}
 was provided in 
\cite{Pelaez:2003eh,Pelaez:2003ky} (once more the latest update is in \cite{GarciaMartin:2011cn}) 
and later revisited in \cite{Caprini:2011ky}.
We show the respective residues in Fig.\ref{fig:reggeresidues} in the region of interest 
for later dispersive approaches. The general agreement of different models 
is fair except for the subleading $f$ trajectory, which always appears together with the Pomeron trajectory.
As before, the uncertainties of \cite{Caprini:2011ky} are much larger than for 
\cite{Pelaez:2003eh,Pelaez:2003ky}. 
Nevertheless, since these differences between models appear in the high energy part of the partial-wave dispersive integrals, their effect is not larger than a few MeV in the $\sigma$ pole
determination \cite{reggeinprep}, which is well covered by the uncertainties in the conservative 
dispersive estimate given in Eq.\ref{myestimate}. As we will see below, the $\sigma$ pole determinations from the Bern-Bucharest \cite{Caprini:2011ky}
and Madrid-Krakow groups  \cite{GarciaMartin:2011jx} are very compatible.

\subsection{Analyticity: cuts and poles}
\label{subsec:dispersiveapproaches}

\subsubsection{Analyticity from causality in non-relativistic scattering}
\label{subsubsec:nonrel}
Dispersion relations are the mathematical consequence of the analytic
properties of the amplitude $T(s,t,u)$ in the complex plane, 
which in turn are a consequence of causality. \color{black} Rigorous proof of this connection
between causality and analyticity only exists within non-relativistic scattering and we 
only sketch it briefly here. In addition non-relativistic scattering provides
an intuitive interpretation of poles in the amplitude.
We recommend \cite{Nussenzweig} for a rigorous and detailed but pedagogical
account. For relativistic scattering there is no general proof beyond
axiomatic field theory or perturbation theory and is therefore an hypothesis (Mandelstamm hypothesis). 
\color{black}

Let us first define the non-relativistic quantum scattering problem, projected in partial waves:
On the one hand, the system should be described by a solution of the 
radial Schr\"odinger equation:
\begin{equation}
\frac{d^2 u_l(k^2,r)}{d r^2}+\Big[k^2-2V(r)-\frac{\ell(\ell+1)}{r^2}\Big]u_l(k^2,r)=0,
\label{ec:Sch}
\end{equation}
where we have set $m=\hbar=1$ units and $V(r)$ is real spherically symmetric potential.
Note that  the center of mass momentum $k^2\equiv 2E$ always appears squared.
On the other hand, to be solutions of a scattering problem 
for partial waves, asymptotically they must be a superposition 
of incoming and outgoing spherical waves, i.e., for $r\rightarrow \infty$:
\begin{equation}
u_\ell(k^2,r)\rightarrow [\Phi_\ell^-(k^2) e^{ikr}+\Phi_\ell^+(k^2)e^{-ikr}]\sim \frac{A_\ell(k^2)}{2ik}[S_\ell(k^2)e^{ikr}-(-1^\ell)e^{-ikr}],
\end{equation}
where  $A_\ell$ is the normalization factor of the incoming wave and $S_\ell(k^2)=(-1)^{\ell+1}\Phi^-_\ell(k^2)/\Phi^+_\ell(k^2)$ is called the $S$-matrix partial wave. If there was no interaction $S_\ell(k^2)=1$.
Note that, although the $u_\ell(k^2,r)$ are functions of $k^2$, we have now introduced some explicit
dependence on $k$, defined as $k=\sqrt{k^2}=\sqrt{2E}$, which is a double valued function of $k^2$ or $E$.
This implies that we need two $E$ sheets to map $k$ on the $k^2$ or $E$ plane. Thus, if we define $k=\kappa^{1/2}({\rm cos} \alpha/2 + i\sin \alpha/2)$
with $\kappa$ real and positive we have:
\begin{itemize}
\item[-]Sheet I, called physical or first Riemann sheet, with $0\leq \alpha\leq 2\pi,\ \im k>0$.
\item[-]Sheet II, called unphysical or second Riemann sheet, with $2\pi\leq \alpha\leq 4\pi,\ \im k<0$.
\end{itemize}
The names are due to the convention that the observable or physical $S$-matrix should be recovered
 as $S(\re k + i \im k))\rightarrow S_{physical}(k)$ for  $\im k\rightarrow 0^+$.
Note that the information on both sheets is redundant, because $\Phi^+_\ell(k)=\Phi^-_\ell(-k)$, which implies that 
$S_\ell^I(k^2)=1/S^{II}_\ell(k^2)$.
\color{black}
In practice, there will be an incoming packet $\Phi_{in}(r,t)\equiv-\int_0^\infty dE\, A(E)e^{-ikr-iEt}$
and a similar outgoing packet. The scattering wave is defined as
the difference of the outgoing packet when there is interaction minus the packet if there was no interaction, which can be written as:
\begin{equation}
\Phi_{sc}(r,t)=\int_0^\infty dE\, A(E)[S(E)-1]e^{ikr-iEt}=2\pi\int_0^\infty dE\, A(E)e^{-ikr-iEt}G(r,E)
\end{equation}
where $G(r,E)=[S(E)-1]\exp(2ikr)/2\pi$. Then, by considering its Fourier transform  
$g(r,\tau)=\int_\infty^\infty G(r,E)\exp(-i E\tau)dE$, we can rewrite:
\begin{equation}
\Phi_{sc}(r,t)= \int_{-\infty}^{\infty} dt' g(r,t-t')\Phi_{in}(r,t'),
\end{equation}
so that the scattering wave at time $t$ has been written 
as a functional (called causal transform) of the incident wave at a different time $t'$.
Now, causality demands that the scattering wave at time $t$ cannot be influenced by 
the incident wave at time $t'>t$, that is $g(r,\tau)$=0 if $\tau=t-t'<0$. But this means that
\begin{equation}
G(r,E)=\frac{1}{2\pi}\int_0^\infty d\tau g(r,\tau) e^{iE\tau}.
\end{equation}
Note that due to the integral starting at $t=0$, if $E=E_R+i E_I$, with $E_I>0$, then
there is a factor $\exp(-E_I t)$, which for sufficiently well-behaved $g(r,\tau)$
ensures the convergence of the integral.
Therefore, it follows that $G(r,E)$ has a regular analytic continuation in the upper half plane of $E$,
and so does $S(E)$.

\begin{figure}
  \centering
  \includegraphics[width=0.8\textwidth]{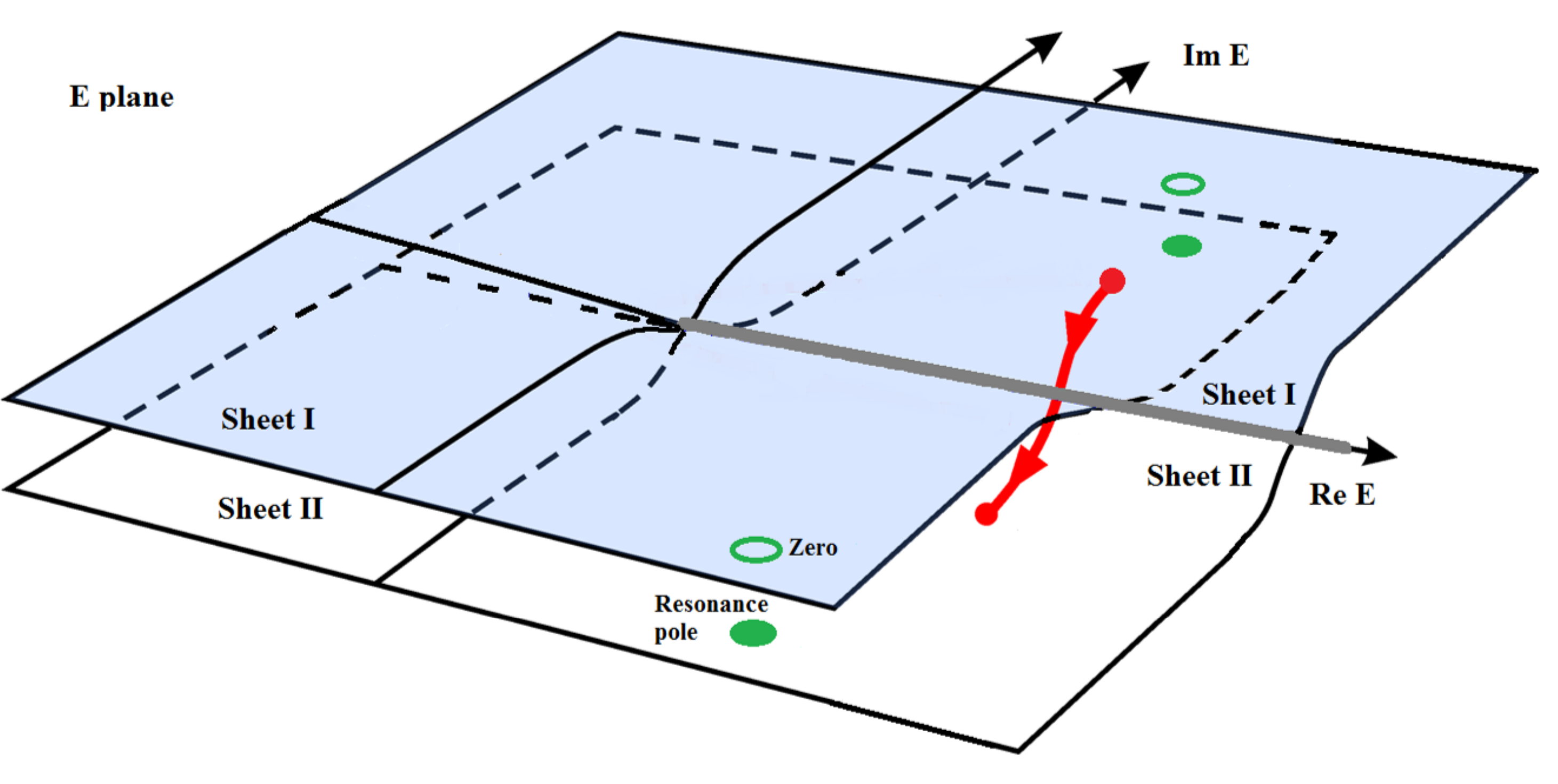}
  \caption{ Analytic structure of the non-relativistic scattering $S$-matrix.
It is analytic in the first Riemann sheet with a cut along the real axis that connects continuously with the second sheet 
(as illustrated with the red curve). Resonance poles correspond to conjugated pairs of poles in the second Riemann sheet 
or zeros on the first. Poles in the negative real axis of the first sheet, corresponding to bound states, can in principle occur, but they 
are absent in $\pi\pi$ scattering under the isospin-conserving strong interaction.
}
  \label{fig:twosheets}
\end{figure}

Now, since the coefficients of Eq.\ref{ec:Sch} are real,
$\Phi^+_\ell(k^{2*})=[\Phi^+_\ell(k^2)]^*$ and $\Phi^-_\ell(k^{2*})=[\Phi^-_\ell(k^{2})]^*$,
so that $S(E^*)=S(E)^*$ on each sheet. This property is sometimes referred to 
as a Schwartz reflection. 
Therefore there is also an analytic continuation to the second sheet. \color{black}
This restricts all singularities to lie on the real axis, where there is actually a cut from zero energy to infinity
which connects the two Riemann sheets and, possibly, poles at negative energies corresponding to bound states.
Note also 
that when $S_\ell^{II}$ has a pole at $k_0^2<0$, then  
$u_\ell(k_0^2,r)\rightarrow \Phi^-_\ell(k_0^2) e^{ir\, {\rm Re} k_0} e^{-r\, {\rm Im} k_0}$ with $\im k_0>0$. That is, the wave function 
decreases exponentially with $r$  and is therefore normalizable and localized in space, representing a bound state of mass $\re k_0$
and binding energy $k_0^2/2$. Since we are working in the isospin limit neglecting electromagnetism there are no bound states
in $\pi\pi$ scattering and we will not find such poles in this report. We will nevertheless find poles in the second Riemann sheet,
which correspond to non-normalizable quasi-bound states or resonances. Note that due to the Schwartz reflection, these poles  will appear as conjugated pairs in the second sheet or as zeros in the first sheet. We have illustrated this analytic structure in Fig.\ref{fig:twosheets}.
Note that the amplitude in the real axis is continuous with the lower
half of the second sheet and therefore it can feel the influence of poles in that half plane. If a pole is isolated and far from the threshold,  its effect is seen as a bump or peak in the 
physical amplitude in the region closest to the pole. 

Finally, it is important to remark that 
this two-sheet analytic structure is inherited by the scattering amplitude or T-matrix,
where poles can only occur in the second Riemann sheet or in the negative real axis on the first sheet.
Of course, the T-matrix in the second sheet is no longer the inverse of the T-matrix on the first, but 
they are still related. For partial waves the relation will be provided in Eq.\ref{ec:firsttosecondsheet} when discussing 
resonances in 
Sect.\ref{subsec:poles}. 

Moreover, although we have been discussing the partial wave formalism,
the analytic structure is also similar for the case of a fixed scattering angle.

\subsubsection{Analyticity and crossing in relativistic $\pi\pi$ scattering}
\label{subsubsec:ancross}

Within relativistic scattering, 
 the energy and the scattering angle roles are played
by the Lorentz invariant Mandelstam variables $s$ and $t$, respectively. 
The analytic properties of the relativistic $S$ matrix  
can be now derived perturbatively in terms of diagrams and in some cases from axiomatic field theory
(see also the textbooks \cite{Edenbook} and \cite{MartinSpearman}). However, 
in general the analytic properties are derived 
from the so-called Mandelstam hypothesis \cite{Mandelstam:1958xc}, 
based on crossing symmetry, which relates $\pi\pi$ scattering in 
the so-called $s$, $t$ and $u$ channels. 

Let us look back at the scattering process in Fig.\ref{fig:12-34}.
We have been considering that the initial state is on the left, and the final one on the right,
so that it describes the process $\pi_1(\vec p_1)\pi_2(\vec p_2)\rightarrow\pi_3(\vec p_3)\pi_4(\vec p_4)$. Here $\pi_k=\pi^\pm, \pi^0$ and particles are on-shell.
Following the definition of Mandelstam variables in Sec.\ref{subsec:notation}, we say
this is the {\it $s$-channel}, because the energy of the process
is $s=4 E^2$, and we write the amplitude as: $T_{12\rightarrow34}(s,t,u)$.
For this process to occur physically, we need $s\geq 4 M_\pi^2$, whereas $t,u\leq 0$.

\begin{figure}
  \centering
  \includegraphics[width=0.49\textwidth]{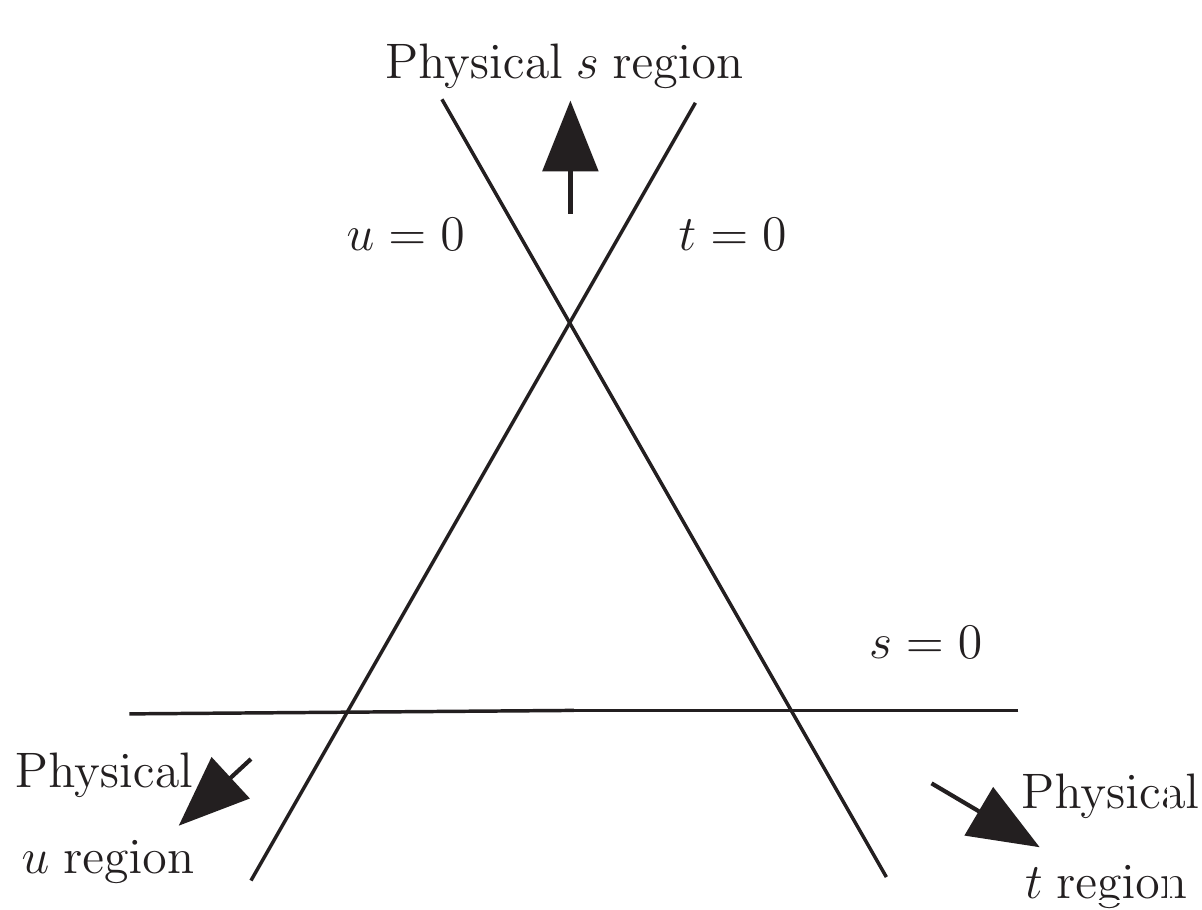}
  \includegraphics[width=0.49\textwidth]{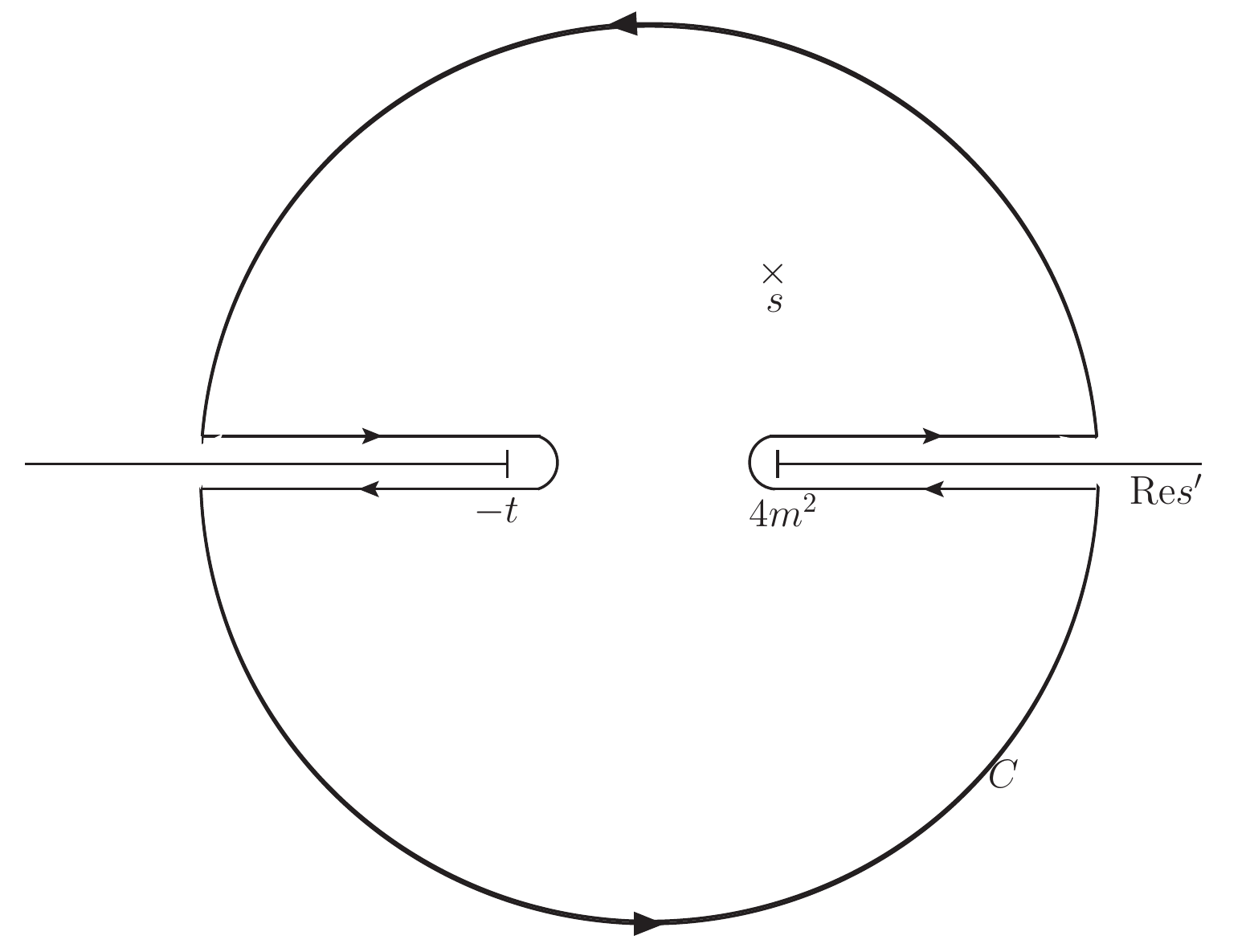}
\vspace*{.1cm}
  \caption{ Left: The Mandelstam plane and the physical regions for $\pi\pi$ scattering.
Right: Analytic structure of fixed $t$ amplitudes on the $s$ plane, with a right cut from $4M_\pi$ to $+\infty$ 
and a left cur from $-\infty$ to $-t$.  Note that there are no poles in the first Riemann sheet for 
$\pi\pi$ scattering. We also show the integration contour for fixed-$t$ dispersion relations. }
  \label{fig:stu-plane}
\end{figure}

Now, the {\it $t$-channel} corresponds to looking at the picture 
from top to botton instead of left to righ. Of course, now $\pi_3$ is coming out, although it is in the initial state, but we can interpret it as the antiparticle $\bar \pi_3$ coming in with $-\vec p_3$.
We can do similarly with $\pi_2$ and then this is just the process
$\pi_1(\vec p_1)\bar \pi_3(-\vec p_3)\rightarrow\bar\pi_2(-\vec p_2)\pi_4(\vec p_4)$.
 We would then write the amplitude as $T_{1\bar3\rightarrow\bar24}(t,s,u)$. 
However, now the process only occurs if $t\geq 4 M_\pi^2$, whereas $s,u\leq 0$.
The $u$-channel is similarly defined but now exchanging the second particle with the fourth instead of the third one.
In the left panel of Fig.\ref{fig:stu-plane} we have plotted the $s-t$ plane showing the physical 
regions where the $s$, $t$ and $u$-channels can occur.

The Mandelstam hypothesis states that the $T_{1\bar3\rightarrow\bar24}(t,s,u)$ amplitude
is the analytic continuation of $T_{12\rightarrow34}(s,t,u)$, and similarly 
for 
the $u$-channel. In particular, it states that there is a unique analytic function $T(s,t,u)$
that satisfies
\begin{equation}
  T(s,t,u)=\left\{
\begin{array}{l}
T_{12\rightarrow 34}(s,t,u),\qquad 
  s \geq 4m^2,\quad  t\le 0,\quad u\le 0,
  \nonumber\\
  T_{1\bar 3\rightarrow \bar 24}(t,s,u),\qquad 
  t \geq 4m^2,\quad s\leq 0,\quad u\le 0,\\
  T_{1\bar 4\rightarrow 3\bar 2}(u,t,s),\qquad 
  u \geq 4m^2,\quad s\le 0,\quad t\le 0.\nonumber
\end{array}\right.
\end{equation}
That this is the case can be shown diagrammatically 
in perturbation theory 
 and so far no counterexample is known, 
but a general proof beyond perturbation theory is lacking.

Together with the Mandelstam hypothesis one also assumes that singularities in the $T(s,t)$ amplitude are just those
demanded by some physical cause. Part of the analytic structure is similar to the non-relativistic
case, although now formulated in terms of the $s$ variable, while keeping $t$ fixed.
Hence, there is a ``physical'' or ``right'' cut on the real axis from threshold $s=4M_\pi^2$
to $+\infty$. 
In addition, \color{black} process where bound states exist
 present a corresponding pole below threshold, \color{black}
but they do not exist in the case of $\pi\pi$ scattering.
Once again, there can be poles associated to resonances, but since 
the amplitude also satisfies a Schwartz reflection:
\color{black}
\begin{equation}
T(s^*,t)=T^*(s,t),
\label{ec:Schwartzreflection}
\end{equation}
for fixed real $t$ and since resonance pole positions have imaginary parts, 
then these resonance poles must come in conjugated pairs that lie on the second or unphysical Riemann sheet. 
Once more, and for fixed $t$, the second sheet is defined for values of $s$ over the cut as $S^{II}(s-i\epsilon,t)=S(s+i\epsilon,t)$. But
in the elastic case unitarity reads $S(s,t)^{-1}=S(s,t)^*$ and by the Schwartz reflection we find
$S^{II}(s+i\epsilon,t)=S^{-1}(s+i\epsilon,t)$. 
\color{black}
Since both sides of this equation are analytic, this equality can be extended to the whole cut complex plane.
However, a  fundamental difference with the non-relativistic case is pair creation, so that the physical cut 
is now a superposition of cuts which open up  at the threshold 
of any state that can be created once sufficient energy is available in the process. Consequently the $S$ matrices have now
more Riemann sheets accessible when crossing continuously the additional thresholds.

In addition, due to crossing symmetry, relativistic amplitudes have other cuts in the real axis.
This is due to the Mandelstam hypothesis, since now
the same $T(s,t,u)$ amplitude also 
represents the $u$-channel process that has a physical threshold for $u\geq4 M_\pi^2$,
which for fixed $t$ means $s\leq -t$. In other words, for fixed $t$ there is a cut from $s=-\infty$ to $s=-t$.
The analytic structure of $\pi\pi$ fixed-$t$ amplitudes is shown in the right panel of Fig.\ref{fig:stu-plane}.

As a final remark on crossing, let us point out that 
if we gather the amplitudes in the isospin basis of Eqs.\ref{ec:isospinbasis} 
into a three isovector $(T^{(0)},T^{(1)},T^{(2)})$, 
crossing relations can be rewritten in terms of the simple matrices:
\begin{equation}
  \label{eq:param:crossing-matrices}
  C_{st} =\left(
  \begin{array}{rrr}
    1/3 & 1 & 5/3 \\
    1/3 & 1/2 & -5/6 \\
   1/3 & -1/2 & 1/6
  \end{array}\right),
  \:\:
  C_{su} =\left(
  \begin{array}{rrr}
    1/3 & -1 & 5/3 \\
    -1/3 & 1/2 & 5/6 \\
    1/3 & 1/2 & 1/6
  \end{array}\right),
  \:\:
  C_{tu} =\left(
  \begin{array}{rrr}
    1 & 0 & 0 \\
    0 & -1 & 0 \\
    0 & 0 & 1
  \end{array}\right),
\end{equation}
in the sense that, for instance $T^{(I)}(t,s,u)=\sum_{I'} C_{st}^{II'}T^{(I')}(s,t,u)$,
and similarly for the rest. Note that $C_{st}^2=C_{su}^2=C_{tu}^2=1$. 

\begin{figure}
  \centering
  \includegraphics[width=\textwidth]{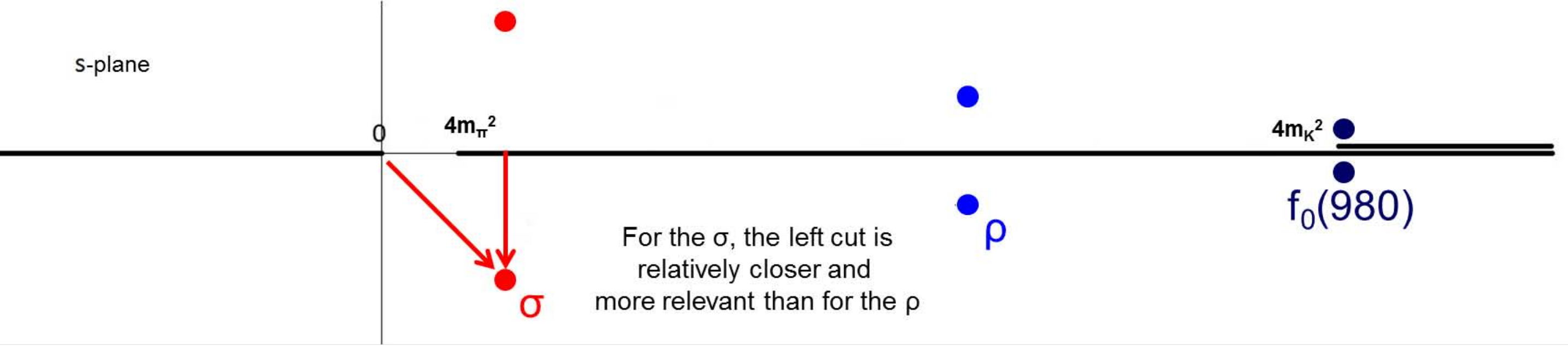}
  \caption{ Most relevant analytic structures of $\pi\pi$ partial waves below 1.5 GeV
the  right cut from $4M_\pi^2$ to $+\infty$, which is actually a superposition of cuts that open at each new threshold, of which the most relevant one is that of $K \bar K$ at $4M_K^2$, which is shown explicitly.
We also show the approximate position of the conjugated pairs of poles of the lightest resonances (that would be in the second Riemann sheet). Note that the distances
of the $\sigma$ pole to the left cut and to the physical cut are relatively similar.
}
  \label{fig:pwcutsandpoles}
\end{figure}

So far, we have been discussing the analytic structure of the whole amplitude $T(s,t,u)$,
but we have seen that the usual way of dealing with $\pi\pi$ scattering, and particularly so when looking for resonances, is by means of
partial waves $t_J^{(I)}(s)$ of definite isospin and angular momentum. 
The analytic structure of these partial waves follows from that of $T(s,t,u)$ after integration 
over the scattering angle, which is nothing but an integration over $t$.
The general case can be rather complicated, see \cite{MartinSpearman}, but since 
we are dealing with $\pi\pi$, where there are no bound states, and in the isospin limit, 
where all particles involved have the same mass, then
the resulting analytic structure of $\pi\pi$ partial waves is quite simple.
They obviously inherit the right-hand or physical cuts at each new physical threshold, 
whereas upon integration over $t$, the left cut
falls again into the negative axis, but covering it from $s=0$ to $-\infty$ \cite{MartinSpearman}. 

This is depicted in Fig.\ref{fig:pwcutsandpoles}, where we show the relevant 
cuts in the complex plane for $\pi\pi$ partial waves in the region of interest for 
the lightest resonances, which are the $\pi\pi$ cut, the $K \bar K$ cut and the left cut. 
In principle, there are other right cuts opening at $4 M_\pi^2, 16 M_\pi^2,...$ but 
we will see that the inelasticity has only been measured  close or above the $K\bar K$ threshold.
For illustration we show the positions where the conjugated pairs of poles of the $\sigma$ and $\rho$
and $f_0(980)$ 
resonances, to be discussed below, lie in the second Riemann sheet.
 
\subsubsection{Poles and resonances}
\label{subsec:poles}

Before embarking on the details of dispersive calculations
 we should review the meaning  of the word ``resonance" in connection with the $f_0(500)$. 
Three definitions are customarily used for the mass and width
of an elastic, background free and {\it narrow} resonance:  
\begin{itemize}
\item[i)] In a region where the phase shift raises from almost zero to $\simeq180^{\rm o}$,  the mass is identified with 
the energy $s^{1/2}$ at which the phase shift crosses $90^{\rm o}$.
 The width corresponds to the distance from the mass to the energy where $\delta_\ell=45^\degree$, neglecting $\Gamma^2$ terms.
These definitions are easily identified in the usual  Breit-Wigner parameterization and therefore they are 
sometimes called Breit-Wigner mass and width. Note that the modulus of the amplitude 
reaches its maximum possible value precisely at this resonance mass. Hence this definition
corresponds to the intuition that a resonant behavior 
is where the interaction becomes as strong as possible since 
at that point the cross section reaches a maximum, 
which usually corresponds to a ``peak'' in the  cross section.
 
\item[ii)] Once again, in a region where there is sharp increase 
of the phase shift by $\simeq 180^{\rm o}$,
the mass $M$ is identified with the energy where the derivative of the phase shift $\delta(s)$ is a maximum 
and the width with the inverse of that derivative in appropriate units: $\Gamma=(M d\delta/ds)^{-1}_M$.
This definition identifies a resonance as a metastable state, 
whose lifetime is the inverse of
the width (Wigner's time delay theory). 

\item[iii)] The mass and width of the resonance are identified from the associated pole position $\sqrt{s_p}$ of the partial wave amplitude in
the  {\it unphysical} (or second) Riemann sheet, as $\sqrt{s_p}=M-\,i\, \Gamma/2$.
This a well defined mathematical statement, but maybe less intuitive physically.

\end{itemize}
In the case of {\sl narrow and isolated} resonances all three
definitions  coincide to first order in $\Gamma/2M$.
However,  the $\sigma$ is very broad and  
the two first definitions are inappropriate. Let us see why.

Concerning the first definition, we have already shown in Fig.\ref{fig:00data}
the  S0-wave phase shift for $\pi\pi$ scattering $\delta_0^{(0)}(s)$
and it does indeed cross 
$\pi/2$, but it does so at an energy of $s^{1/2}\simeq800\,$MeV, where there might be a 
maximum of the modulus, but nothing that resembles a peak.
 Moreover, as we have already commented,
the position of such a  peak can change from one process to another, 
it is not a universal feature.   For instance, the presence of a background phase or a zero in the vicinity
of the resonance, as it actually happens with the $f_0(500)$ as seen from $\pi\pi$ scattering, can distort its shape. 
When the $\sigma$ is produced in heavy-particle decays this zero is not present
and some vague peak is seen, but around 550 MeV.

 The second definition can also be distorted by a background,
but it has the advantage over the first one
that it may still provide a good approximation
if this background is created by other non-resonant dynamics and varies slowly.
This is for instance the $f_0(980)$ case, as we can see in Fig.\ref{fig:00data}
which has fast increase of $\simeq 180^{\rm o}$ around 980 MeV, but starting from $\simeq90^{\rm o}$
around 850 or 900 MeV (in addition, the shape is also somewhat distorted by the nearby $K\bar K$ threshold).
However, there is nothing like a sharp increase by $\pi$
from threshold to 900 MeV in the S-wave (as compared to that 
of the $\rho(770)$ meson seen in the right panel of  Fig.\ref{fig:00data}). 
Moreover, the energy {\sl derivative} of  
$\delta_0^{(0)}(s)$ is definitely not maximum near 500 nor 800 MeV.
Actually, the phase seems to raise at a rather constant rate.

All these considerations are nothing but a re-statement that,
as we have repeatedly commented, the partial wave amplitude does not resemble a
Breit-Wigner shape with a clear peak and a simultaneous steep rise in the phase. 
Nevertheless, and as already advanced in the introduction, there exists a pole 
in the second Riemann sheet, but
at very low mass and with very large width. 
Thus, we are left with the third definition, which has the advantage that
a divergence cannot be removed or displaced by adding any background or other
non-resonant contribution.  Under very general assumptions \cite{Edenbook}
the same pole should be present when 
observing the S0-wave in any process.

In order to look for the pole on the second Riemann sheet associated to
a resonance $R$ which appears in
the $\pi\pi$ elastic partial wave $t_J^{(I)}$ let us recall 
that, for elastic scattering, the $S$-matrix in the second Riemann sheet is the inverse of the 
$S$-matrix on the first. Hence, a pole located  
at $\sqrt{s_R}=M_R-\,i\, \Gamma_R/2$ in the second sheet,
 corresponds exactly with the position of a {\sl zero} 
of the S-matrix partial wave $S_J^{(I)}(s)$,
\begin{equation}
S_J^{(I)}(s)=1+2\,i\, \hat{f}_J^{(I)}(s)=e^{2\,i\,\delta_J^{(I)}},
\label{ec:resonancecondition}
\end{equation}
in the first (or {\sl physical}) Riemann sheet.
 Thus, we can find the location of the
resonance by looking for the solutions of 
$S_J^{(I)}({s}_R)=0$ in the complex plane.
This zero condition may be written in a simpler manner as
\begin{equation}
\cot\delta_J^{(I)}(s_R)=-\,i\,,
\end{equation}
where, of course, $\cot\delta_J^{(I)}$ now corresponds to a function in the complex plane
which has all the singularities
of the amplitude, except for the cut along the real axis above threshold
where it coincides with the physical $\cot\delta_J^{(I)}$.

In practice, since the $t$-matrix partial 
waves are related to the $S$-matrix partial waves by 
$S_J^{(I)}(s)=1-2 i \sigma(s) t_J^{I}(s)$, we can write 
the amplitude in the second Riemann sheet $t^{II}(s)$
in terms of the one in the first Riemann sheet $t^{I}(s)$, as follows:
\begin{equation}
t^{II}(s)=\frac{t^{I}(s)}{1+2 i \sigma(s)t^{I}(s) },
\label{ec:firsttosecondsheet}
\end{equation}
where the determination of $\sigma(s)$ is chosen such that  $\sigma(s^*)=-\sigma(s)^*$ to ensure the Schwartz reflection
symmetry of the amplitude.  In other words, on the upper half $s$ plane we can take $\sigma=+\sqrt{1-4m^2/s}$ as usual,
 whereas on the lower half $s$ plane we must then take $\sigma(s)=-\sigma(s^*)^*$.

In addition, the coupling of a resonance to two pions
can be defined from its pole residue as:
\begin{equation}
g^2=-16\pi \!\!\lim_{s\rightarrow s_{pole}}\!\!(s-s_{pole})\,t_{J}(s)\,(2J+1)/(2k)^{2 J}.
\label{ec:defresidue}
\end{equation}
This residue is of interest for models of the
spectroscopic nature of these resonances.

Back to Fig.\ref{fig:pwcutsandpoles}, we want to emphasize that, in contrast with 
the $\rho$ resonance pole, the distance of the
$\sigma$ pole to the left cut is comparable to the distance to the right cut and smaller than to the inelastic $K\bar K$ cut. This is why,
in order to claim a precise and rigorous determination of the $\sigma$ parameters, 
the left cut must be under control, whereas its influence is negligible for other resonances.
At the end, the contribution of that cut is not too large, but its size has to be established firmly if one wants to claim precision.

\subsubsection{Dispersion Relations}
\label{sec:disprel}

Now that we know the analytic structure of $\pi\pi$ scattering amplitudes on the first Riemann
sheet, we can use Cauchy's theorem to relate the value  of the amplitude at one given 
energy to an integral of the amplitude over a contour. However, Cauchy's theorem applies to one complex variable whereas in scattering there are two
independent variables: the energy and the scattering angle, or $s$ and $t$. 
We are interested then in single variable dispersion relations and this is achieved by 
fixing one of them, or by integrating $t$ to obtain partial waves that depend on $s$ alone.

We have already seen the singularity structure of amplitudes for fixed $t$ and thus
the amplitude at any
given complex $s$ in the analyticity domain can be obtained as 
an integral over the contour $C$ depicted in the right panel of Fig.\ref{fig:stu-plane},
as follows: 
\begin{equation}
T(s,t,u)=\frac{1}{2\pi i}\oint{ds^\prime \frac{T(s^\prime,t,u')}{s^\prime-s}}.
\label{Tcauchy}
\end{equation}

The contour C is made of two parts: 
a circle of radius $R$ centered at $s^\prime =0$,
and straight lines parallel 
to the left and right cuts but infinitesimally above or below them by some $\epsilon'$.
Now, if $T(s,t,u)$ 
goes to zero faster than $1/s$ as $\vert s\vert\rightarrow \infty$
and we let $R\rightarrow \infty$
then the contribution from the circular part of the contour vanishes.
Thus, we are  left with the contributions 
along the cuts. Let us now recall that $u'=4m^2-t-s'$. Thus, remembering also that Eq.\ref{ec:Schwartzreflection}, \color{black} implies 
$T(s'+i\epsilon',t,u'-i\epsilon')=T^*(s'-i\epsilon',t,u'+i\epsilon')$ \color{black} and observing that 
the lines above and below each cut 
run in opposite directions, we find
\begin{eqnarray}\label{Tcauchy-2}
T(s,t,u)
&=&\frac{1}{\pi}\int_{4m^2}^{\infty}{ds^\prime 
\frac{\mathrm{Im}T(s^\prime,t,u')}{s^\prime-s}}+\frac{1}{\pi}\int_{-\infty}^{-t}{ds^\prime \frac{\mathrm{Im}T(s^\prime,t,u')}{s^\prime-s}},
\label{ec:predisp0}
\end{eqnarray}
\color{black} where, as usual, we have taken the $\epsilon'\rightarrow0$ limit \color{black}
and on the real axis we define $T(s^\prime,t,u)\equiv T(s^\prime+i0^+,t,u+i0^-)$. 

The above equation is valid everywhere in the complex plane
except on the singularities. If we want the dispersion relation for the real axis, 
where we have the cut singularity, we must actually consider the amplitude at $s+i\epsilon$ 
with $s$ real, 
and use the relation:
\begin{equation}
\frac{1}{s^\prime-s-i\epsilon}=PV\frac{1}{s^\prime-s}+i\pi\delta(s^\prime-s),
\end{equation}
where $PV$ denotes the principal value. Note that the effect of $i\pi\delta(s^\prime-s)$ on Eq.\ref{ec:predisp0}
is to extract $i \im T (s,t,u)$ out of the first integral, which cancels exactly with the imaginary part on the left side. Hence on the real axis we find:
\begin{equation}\label{Tcauchy-real}
\mathrm{Re}T(s,t,u)=\frac{1}{\pi}PV\int_{4m^2}^{\infty}{ds^\prime \frac{\mathrm{Im}T(s^\prime,t,u^\prime)}{s^\prime-s}}+\frac{1}{\pi}\int_{-\infty}^{-t}{ds^\prime \frac{\mathrm{Im}T(s^\prime,t,u^\prime)}{s^\prime-s}}.
\end{equation}
Therefore, for real values of $s$  {\it dispersion relations provide the real part of the amplitude from its imaginary part}.

If the amplitude $T(s,t,u)$ does not tend to zero 
fast enough at $\infty$, 
the circular contribution of the contour $C$
will not vanish.  However, if we subtract the amplitude 
evaluated at another point $s_0$, we can write
\begin{equation}\label{Tcauchy-1s}
T(s,t,u)-T(s_0,t,u)=\frac{1}{2\pi i}(s-s_0)\oint{ds^\prime \frac{T(s^\prime,t,u^\prime)}{(s^\prime-s)(s^\prime-s_0)}},
\end{equation}
and now for the circular part to vanish it is enough to require $T(s,t,u)/s$
to tend to zero at $\infty$ faster than $1/s$. The so-called ``once subtracted''
dispersion relation now reads:
\begin{equation}\label{Tcauchy-onesubstracted}
T(s,t,u)=T(s_0,t,u)+\frac{s-s_0}{\pi}\int_{4m^2}^{\infty}{ds^\prime \frac{T(s^\prime,t,u^\prime)}{(s^\prime-s)(s^\prime-s_0)}}+\frac{s-s_0}{\pi}\int_{-\infty}^{-t}{ds^\prime \frac{T(s^\prime,t,u^\prime)}{(s^\prime-s)(s^\prime-s_0)}}.
\end{equation}
The price to pay is that one should know the amplitude at the subtraction point $s_0$. If that is not enough for convergence, one can make another subtraction, typically at the same point, to find
\begin{eqnarray}\label{Tcauchy-two-substracted}
T(s,t,u)&=&T(s_0,t,u)+(s-s_0)\frac{\partial}{\partial s_0}T(s_1,t,u)+\frac{1}{2\pi i}(s-s_0)^2\oint{ds^\prime \frac{T(s^\prime,t,u^\prime)}{(s^\prime-s)(s^\prime-s_0)^2}},\nonumber
\end{eqnarray}
which is called  a ``twice subtracted'' dispersion relation, etc. 
In principle, due to the Froissart bound \cite{Froissart61},  $\sigma_{tot}(s)<c (\log s)^2$, two subtractions 
are enough to ensure convergence, although sometimes more subtractions can be convenient for particular purposes.

At this point it is important to remark that the contribution from the ``left cut'' is usually
the most difficult one to calculate. However, 
for $\pi\pi$ scattering amplitudes,
 which are highly symmetric since all particles involved are pions,
there are certain cases in which the left cut contribution
can be recast in terms of amplitudes in the physical region, 
namely for Forward Dispersion Relations, Roy and GKPY equations, that we will discuss below.

For textbook level references on $\pi\pi$ scattering and dispersive theory 
we refer again to \cite{MartinSpearman,libropipi} 
together with the relatively more recent review on \cite{Yndurain:2002ud}.
What follows next is a brief summary of the topics which 
are discussed at length in those references.

\subsubsection{Forward Dispersion Relations for $\pi\pi$ scattering}

A straightforward application of the previous equations 
occurs for forward ($t=0$) $\pi\pi$ scattering. This is relevant 
not only because the equations are much simpler, but also because forward scattering amplitudes
are proportional to total cross sections which are relatively easier to measure compared to other observables. Actually, we have already seen in Sec.\ref{subsec:highenergydata} that these are the only available data for
$\pi\pi$ scattering above 2 GeV.

A simplification occurs due to the fact that we can write any $\pi\pi$ scattering amplitude
in terms of three amplitudes which are symmetric or antisymmetric under 
$s\leftrightarrow u$ crossing. The symmetric ones are: $\pi^0\pi^0\rightarrow\pi^0\pi^0$ and
$\pi^0\pi^+\rightarrow\pi^0\pi^+$, 
whereas the antisymmetric one is the amplitude
where isospin one is exchanged in the $t$-channel $T^{I_t=1}$.
In terms of the usual $s$-channel isospin states, these amplitudes are written as:
\begin{equation}
  \label{fdr:basis}
  T^{00}=\frac{1}{3}(T^{(0)}+2T^{(2)}),\quad
  T^{0+}=\frac{1}{2}(T^{(1)}+T^{(2)}),\quad
  T^{I_t=1}=\frac{1}{3}T^{(0)}+\frac{1}{2}T^{(1)}-\frac{5}{6}T^{(2)}.
\end{equation}

\color{black}
An interesting remark when writing a forward dispersion relation is that, 
since $t=0$, we can change variables in the left cut integral from $s'$ to $u'=4M_\pi^2-s'$.
In particular, for $T^{00}$ we can write:
\begin{eqnarray}
\mathrm{Re}T^{00}(s,0,u)-T^{00}(4M_\pi^2,0,0)&=&\frac{(s-4M_\pi^2)}{\pi}\left[PV\int_{4M_\pi^2}^\infty{ds^\prime
  \frac{\mathrm{Im}T^{00}(s^\prime,0,u^\prime)}{(s^\prime-s)(s^\prime -4M_\pi^2)}}\right.\nonumber\\
&&\left.+\int_{4M_\pi^2}^\infty{du^\prime
\frac{\mathrm{Im}T^{00}(s^\prime,0,u^\prime)}{(u^\prime-u)u^\prime }}\right],
\end{eqnarray}
where the principal value is taken so that we can use the above Eq.\ref{Tcauchy-onesubstracted}
for $s$ in the real axis.
Note that at high energies this amplitude is dominated 
by Pomeron exchange so that at very high $s'$ or $u'$ 
it grows as a logarithm (although as we have seen in Sec.\ref{subsec:highenergydata}
it can be approximated to a constant up to $\sim 20\,$GeV).
Thus, in order to make each integral convergent
we have used a once-subtracted dispersion relation
with the subtraction point chosen at threshold $s_0=4M_\pi^2$. 

Now, since $u'=4M_\pi^2-s'$ and $u=4M_\pi^2-s$, it is customary to
omit the $u$ and $u'$ variables unless they are both needed to
illustrate some symmetry explicitly, and so we will do in what follows
to ease the notation.
\color{black}

Therefore, since this amplitude is $s\leftrightarrow u$ symmetric then $T(s,0,u)=T(u,0,s)$
and we can rewrite the integral over the left hand cut as an integral over the right cut, i.e.:
\begin{equation}\label{t00}
\mathrm{Re}T^{00}(s,0)=T^{00}(4M_\pi^2)+\frac{(s-4M_\pi^2)}{\pi}PV\int_{4M_\pi^2}^\infty{ds^\prime
  \frac{(2s^\prime-4M_\pi^2)\mathrm{Im}T^{00}(s^\prime,0)}{s^\prime(s^\prime-s)(s^\prime -4M_\pi^2)(s^\prime+s-4M_\pi^2)}}.
  \label{ec:FDR00}
\end{equation}
Note that this equation is written just in terms of the $s$-channel $T^{00}$ amplitude
over its physical region $s'\geq 4 M_\pi^2$, $t=0$.
\color{black}
It will also be interesting to evaluate this dispersion relation 
at $s=2 M_\pi^2$, where the amplitude is real and the principal value is not needed. The result is the following sum rule:
\begin{equation}
K\equiv T^{00}(2M_\pi^2,0)-T^{00}(4M_\pi^2,0)+\frac{8M_\pi^4}{\pi}\int_{4M_\pi^2}^\infty{ds^\prime
  \frac{\mathrm{Im}T^{00}(s^\prime,0)}{s^\prime(s^\prime -2M_\pi^2)(s^\prime-4M_\pi^2)}}=0.
\label{ec:srK}
\end{equation}
\color{black}

In the same way we obtain a Forward Dispersion Relation for $T^{0+}$, since it
is also $s\leftrightarrow u$ symmetric \color{black}
and dominated at high energies by Pomeron exchange\color{black}. It reads:
\begin{equation}\label{t0+}
\mathrm{Re}T^{0+}(s,0)=T^{0+}(4M_\pi^2)+\frac{(s-4M_\pi^2)}{\pi}PV\int_{4M_\pi^2}^\infty{ds^\prime
  \frac{(2s^\prime-4M_\pi^2)\mathrm{Im}T^{0+}(s^\prime,0)}{s^\prime(s^\prime-s)(s^\prime -4M_\pi^2)(s^\prime+s-4M_\pi^2)}}.
\end{equation}
\color{black}
Once again at $s=2 M_\pi^2$ we obtain a sum rule:
\begin{equation}
L\equiv T^{+-}(2M_\pi^2,0)-T^{+-}(4M_\pi^2,0)+\frac{8M_\pi^4}{\pi}\int_{4M_\pi^2}^\infty{ds^\prime
  \frac{\mathrm{Im}T^{+-}(s^\prime,0)}{s^\prime(s^\prime -2M_\pi^2)(s^\prime-4M_\pi^2)}}=0.
\label{ec:srL}
\end{equation}
\color{black}

It is important to remark that, by looking at Eq.\ref{fdr:basis} we see that the integrand of each of the two Forward Dispersion Relations 
in Eqs.\ref{t00} and \ref{t0+}
depends on only two $s$-channel isospin amplitudes. Now, since the imaginary parts of these $s$ channel isospin amplitudes
are positive above threshold, they cannot cancel against each other anywhere inside the integral. This
property is known as ``positivity" and will be of
relevance to keep uncertainties relatively small. 

The Forward Dispersion Relation for the $T^{I_t=1}$ amplitude is obtained
following the same steps, except for the fact that it is antisymmetric, $T(s,0,u)=-T(u,0,s)$,
and therefore one arrives to 
\begin{equation}
  \label{eq:fdr:fdrIt1}
  \mathrm{Re}T^{I_t=1}(s,0) = \frac{2s-4M_\pi^2}{\pi}\,PV
  \int_{4M_\pi^2}^\infty ds'\,
  \frac{\mathrm{Im} T^{I_t=1}(s',0)}{(s'-s)(s'+s-4M_\pi^2)}.
\end{equation}
Note that in this case \color{black} no subtractions are needed,  since the leading
Reggeon for the $I=1$ exchange is the $\rho$ which implies that $\mathrm{Im}\, T^{I_t=1}(s',0)$ 
falls off as  $\sim 1/s'$ \color{black}. Now there is no positivity in the integrand,
which depends on the three isospin amplitudes with different signs, as seen in Eq.\ref{fdr:basis}.
Once again, the whole relation is written in terms of the $T^{I_t=1}$ amplitude
in its $s$-channel physical region.

\subsubsection{Roy and GKPY Equations}
\label{subsubsec:royeqs}

Roy equations are  an infinite set of coupled integral equations for partial waves, 
obtained from the integration of fixed $t$ dispersion 
relations plus $s\leftrightarrow u$ crossing symmetry, written in such a way that all 
amplitudes appearing inside the dispersive integrals are calculated over the physical region,
including those initially coming from left cut contributions.
There is an  extensive literature on Roy equations, starting with the original derivation by Roy in 1971 \cite{Roy:1971tc}, including phenomenological applications \cite{Basdevant:1972uu,roy70,Pennington:1973xv,Pennington:1973hs,Pennington:1974kp,Froggatt:1975me} as well as more formal works on the structure of the equations, properties of the integral kernels, uniqueness of solutions, extension to higher energies, 
etc \cite{Mahoux:1974ej,Auberson:1974in,Pomponiu:1975bi}. These were mostly developed in the 70's and nice introductions can be found as a collection of lectures in \cite{Petersenlectures} or in textbooks of that time \cite{libropipi} as well as the review by the very S.M. Roy \cite{Roy:1990hw}. However, the interest in Roy equations faded away until the late 90's and early 00's. Our aim here is just to make a very brief introduction 
to Roy equations and then focus on the works used to obtain the $f_0(500)$ pole. Hence, for a detailed explanation of different developments and  a complete list of references before 2001 we recommend the Roy-equation review in \cite{ACGL}.
For us it suffices to say that the renaissance of Roy equations
 was mainly motivated by a renewed interest in
threshold parameters  
\cite{ACGL,CGL,Ananthanarayan:1998hj,Gasser:1999hz,Wanders:2000mn,DescotesGenon:2001tn}. This was in part due to a discussion about the size of these parameters in connection with 
the pion mass counting in Chiral Perturbation Theory (ChPT) and the size of the chiral condensate \cite{Fuchs:1989jw}, but also fostered 
by the proposals to obtain a better experimental determination 
from pionic atom experiments (which were finally carried out successfully and gave the results already commented at the end of Sec.\ref{subsec:pwbelow} \cite{Adeva:2007zz,Adeva:2005pg}).  Hence, in the 00's, the whole $\pi\pi$ scattering was revisited with Roy equations \cite{ACGL} 
including sometimes ChPT input \cite{CGL} and solving
the old ``up-down" ambiguity \cite{Kaminski:2002pe}. The use of ChPT is interesting because it is the
low energy effective theory of QCD and provides strong constraints on the subtraction terms
in Roy equations \cite{CGL}. Based on this approach a precise and rigorous dispersive determination of the $\sigma$ was published in 2006
\cite{Caprini:2005zr}.  

\color{black} More recently, a new set of equations 
has been derived \cite{GarciaMartin:2011cn,Kaminski:2008fu,Kaminski:2008rh}, 
which are similar to those of Roy but with  
one subtraction instead of two. These are known as GKPY equations and
have allowed \color{black} for a precise
determination of the $f_0(500)$ pole just from experimental data, without using ChPT but relying at low energies on the recent and precise NA48/2 data 
already reviewed in Sec.\ref{subsec:pwbelow}. 
Another pole determination using standard twice-subtracted Roy equations
and some input from  \cite{ACGL} has been obtained in \cite{Moussallam:2011zg}.
 As we have repeatedly emphasized, these approaches are in fairly good agreement (also with the non-dispersive determination in \cite{CGL})
and have triggered the radical $f_0(500)$ revision
in the RPP 2012 edition \cite{PDG12}.

The idea behind Roy \color{black} and GKPY \color{black} equations is to project into partial waves the fixed $t$ dispersion relations.
In this way one obtains a dispersion relation for each given partial wave, which
in the case of $\pi\pi$ scattering have the usual right cut, plus a left cut starting from 0 and extending to $-\infty$.
This is relatively straightforward and actually it is 
the basis for the unitarization techniques that we will review in  
Sec.\ref{subsec:uchpt}. However, if one only does that, there is still a left cut where the partial wave at unphysical values, $s\leq 0$,
are needed.
Roy \color{black} and  GKPY \color{black} equations make use of $s\leftrightarrow u$ crossing to 
recast the whole left cut of a given partial wave as a sum over all partial waves in the physical channel. At the end, all input in the dispersive integrals is over 
the physical region, but the price to pay is that 
all partial wave equations are now coupled. This is an infinite set of coupled integral equations, but if one does not consider very high energies one can just keep the lowest partial waves in the equation.

Let us then sketch the derivation of \color{black} both Roy and GKPY equations.
As explained above, the only real difference is that the former have one subtraction and the latter have two. Since the standard twice-subtracted Roy equation have been much more extensively treated in the literature, we will sketch here the once-subtracted GKPY case, but the derivation of Roy eqs. is parallel. Let us then start from a fixed $t$ dispersion relation, once-subtracted as in Eq.\ref{Tcauchy-onesubstracted}, \color{black} but choosing the subtraction point at $s_0=0$, and recasting the left cut contribution in terms of
the variable $u'=4M_\pi^2-s'-t$, i.e.,
\begin{equation}
   \label{eq:gkpy:1s-dr}
  T^{(I)}(s,t) = T^{(I)}(0,t) 
  + \frac{s}{\pi} \int_{4M_\pi^2}^\infty ds'
  \left[
    \frac{\im T^{(I)}(s',t)}{s'(s'-s)}
    - \frac{\im T^{(I)}(u',t)}{u'(u'-s)}
  \right].
\end{equation}
Each of these two integrands by itself would yield a divergent integral due to Pomeron contributions to $\im T$,
which grow like $\sim s'$, however when taken together such contributions cancel against each other \cite{GarciaMartin:2011cn}. In the twice subtracted case, \color{black} i.e. the Roy equations as originally derived by Roy \cite{Roy:1971tc}, \color{black} one does not even have to worry about this issue.

Neither the subtraction constant, nor the second piece of the integral are evaluated at physical values of $s\geq 4M_\pi^2$. But this can be fixed by using the crossing matrices defined
in Eq.\ref{eq:param:crossing-matrices}, i.e.
\begin{equation}
T^{(I)}(u',t,s')=\sum_{I'} C_{su}^{II'}T^{(I')}(s',t,u'),\quad 
T^{(I)}(0,t)=\sum_{I''} C_{st}^{II''}T^{(I'')}(t,0).
\end{equation}
The subtraction term is still reexpressed in terms of
$T^{I''}(t,0)$ which depends on $t$, 
but then we can write another dispersion relation for $T^{I''}(t,0)$
  \begin{eqnarray}
    \label{eq:gkpy:t-dr}
    T^{(I'')}(t,0) = T^{(I'')}(t_0,0) +
    \frac{t-t_0}{\pi} \int_{t_0}^\infty ds'
    \left[
      \frac{\im T^{(I'')}(s',0)}{(s'-t)(s'-t_0)}\right.\left. -
      \frac{\sum_{I'''}C_{su}^{I''I'''}\im T^{(I''')}(s',0)}
      {(4\mps-t-s')(4\mps-s'-t_0)}
    \right],
  \end{eqnarray}
which for convenience is subtracted at $t_0=4M_\pi^2$. 
By defining $a_0^{(1)}=0$ we can write
$T^{I}(\mps,0)=a_0^{(I)}$ and gather all these results as follows:
 \begin{eqnarray}
 \label{eq:gkpy:final-dr}
     T^{(I)}(s,t) &=&
     \sum_{I'} C_{st}^{II'} a_0^{(I')}\\ \nonumber
     &+& \frac{s}{\pi} \,\int_{4\mps}^\infty ds' \left[ \frac{\im
         T^{(I)}(s',t)}{s'(s'-s)} - \frac{\sum_{I'} C_{su}^{II'} \im
         T^{(I')}(s',t)} {(s'+t-4\mps)(s'+s+t-4\mps)} \right] 
\\ \nonumber
     &+& \frac{t-4\mps}{\pi} \,\int_{4\mps}^\infty ds' \sum_{I''}
     C_{st}^{II''} \left[ \frac{\im T^{(I'')}(s',0)}
{(s'-t)(s'-4\mps)} -
       \frac{\sum_{I'''}C_{su}^{I''I'''}\im
         T^{(I''')}(s',0)} {s'(s'+t-4\mps)} \right].\label{eq:preroy}
   \end{eqnarray}
Thus the whole input into the dispersive integrals 
has been rewritten in terms of the $s$-channel amplitude 
in its  physical region $s'\geq 4\mps$. 
This is an equation that the $\pi\pi$ amplitude must satisfy exactly.
Note that $s\leftrightarrow u$ 
crossing for fixed $t$ has been imposed in the 
amplitude and is therefore guaranteed. However, symmetry 
under $s\leftrightarrow t$ is not ensured, since it has only been used to 
recast the subtraction constant. Thus, it is frequent to implement 
\color{black} Roy and GKPY equations \color{black} together with some supplementary conditions like $s\leftrightarrow t$ crossing sum rules, or only on a subset of amplitude parameterizations that satisfy unitarity, since that property is also not ensured by \color{black} Roy or GKPY equations \color{black}  alone.

In principle, everything can now be recast 
in terms of partial waves by using the
partial wave expansion in Eq.\ref{ec:pwdef} for each one of the 
$T^{(I)}$, $T^{(I')}$ and $T^{(I'')}$. The equation for a particular $\ell$ partial wave 
can be obtained by using the orthogonality of Legendre polynomials to extract the desired partial wave on the left hand side by multiplying by the appropriate $P_\ell(t)$ and integrating over $t$ between 0 and 1.
However, this creates a rather lengthy expression in the right hand side of the equation, whose derivation we will not follow here, but just show the final \color{black} expression  which is valid for 
both Roy and GKPY equations \color{black}
 as it is customarily presented \cite{Basdevant:1972uu, roy70,ACGL,GarciaMartin:2011cn}:
  \begin{eqnarray}
  \label{Royeqs}
    t_\ell^{(I)}(s) =\overline{ST}_\ell^{I}(s)
    +\sum_{I'=0}^2 \sum_{\ell'=0}^{\ell_{max}}
    \int_{4\mps}^{s_{max}} ds'  \overline{K}_{\ell\ell'}^{I I'}(s,s') \im t_{\ell'}^{I'}(s') + \overline{DT}^{I}_\ell(s) ,
  \end{eqnarray}
 where $\overline{ST}_\ell^{I}$ stand for the subtraction terms, which as we have seen only depend on the scattering lengths of the S0 and S2-waves. 
For the \color{black} GKPY once-subtracted case, which we have used to illustrate the derivation here, 
 they are just constants, whereas for Roy equations they are first order polynomials in $s$. \color{black}
The second term contains the dispersive integral of 
the partial waves in the physical region multiplied by
some integration kernels $\overline{K}_{\ell\ell'}^{I I'}(s,s')$, which for
 the \color{black} once-subtracted GKPY case
can be found in \cite{GarciaMartin:2011cn} and for the twice subtracted Roy equations 
in \cite{Basdevant:1972uu, roy70}.
 It is worth remarking
that at large energies these kernels behave as $\sim 1/s^2$ for GKPY equations 
and as $\sim 1/s^3$ for Roy equations. Thus the weight of the high energy region is somewhat larger for the former.
\color{black}
However, note that only the  $\ell'\leq\ell_{max}$ appear in the kernel term, which is only
integrated up to a maximum energy $s_{max}$,  \color{black} typically chosen between 800 and 1100 MeV. 
In some works this $s_{max}$ is also called the ``matching energy $s_0$''. \color{black}
The rest of waves and the high energy contribution above $s_{max}$
are collected into the so-called driving terms $\overline{DT}^{I}_\ell(s)$.

The reason for separating the lowest waves $\ell\leq\ell_{max}$ from the rest is that 
\color{black} Roy and GKPY \color{black} equations are an infinite set of coupled integral equations for all $\ell$,
but in practice they are only used on the lowest waves, 
whereas the other waves are considered as input. At low energies higher partial waves are 
 subdominant  and one expects that describing them with simple data fits should be accurate enough. 
For the works dealing with the determination of the sigma $\ell_{max}=1$ and
thus only S and P-waves are treated with Roy equations 
\cite{CGL,ACGL,Caprini:2005zr,GarciaMartin:2011cn,GarciaMartin:2011jx,Moussallam:2011zg},
 whereas D waves and higher are included in the driving terms. 
Recently \cite{Kaminski:2011vj}
the D  and F waves have also been analyzed with Roy equations 
and when the resulting waves are used 
inside the $S$ and $P$ wave driving terms, the changes are 
within the uncertainties of previous calculations \cite{GarciaMartin:2011cn}.

In addition, the energy integral is split into two parts and the region above $s_{max}$ 
is also included
into the driving terms, due to several reasons. In particular, as already discussed in Sec.\ref{subsec:highenergydata}, there are no data on partial waves above 2 GeV 
and even above 1.5 GeV the available
 data are not very reliable. Moreover the partial wave truncation may simply be a bad approximation above those energies. 
Furthermore, 
Roy equations do not have a unique solution in the whole $s$ range, although in certain regions
one can guarantee the uniqueness of the solution by providing, for instance, the value of the S-wave scattering lengths and the phase \color{black} at \color{black} 
$s_{max}$ (see \cite{ACGL,Caprini:2005zr}). Thus, a possible approach is to consider everything above $s_{max}$ as input while solving  Roy equations below $s_{max}$, obtaining a single solution. 
In practice this has been done in \cite{CGL,ACGL,Moussallam:2011zg} for $S$ and $P$ waves and twice subtracted Roy equations, considering all other waves as input.

For the twice-subtracted case, which is the most widely used and the one derived in 
Roy's original work \cite{Roy:1971tc}, 
the form of the equations is the same as in Eq.\ref{Royeqs},
although now the subtraction terms are first order polynomials in $s$, 
whose coefficients, once again, only depend on the scattering lengths of the S0 and S2-waves. 
The corresponding kernels can be found in \cite{ACGL} and, as commented above, at large energies behave as $\sim 1/s^3$.
On the one hand, this means that for the twice subtracted equations the high energy input 
has a much lesser weight on the results. On the other hand, since the subtraction terms are polynomials, the uncertainty in the threshold parameters propagates to the results but
grows with the energy, until they become very large. For this reason 
the most relevant and precise results using twice subtracted 
dispersion relation make use of additional ChPT information at low energies \cite{CGL}.
A comparison of the size of different terms using only experimental input 
and one or two subtractions can be found in \cite{GarciaMartin:2011cn}.

\begin{figure}
  \centering
  \includegraphics[width=0.54\textwidth]{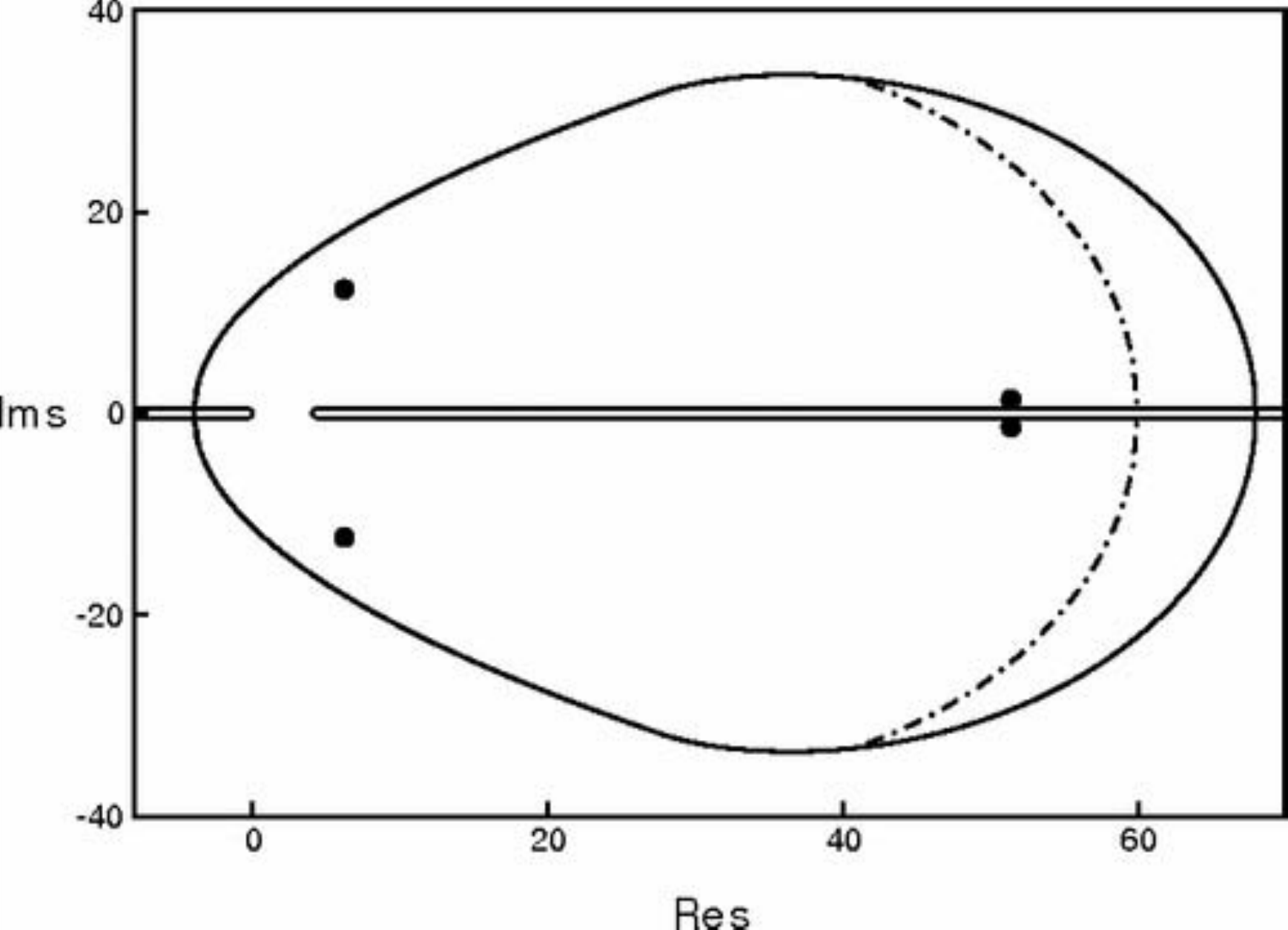}
  \caption{ Domain of applicability of Roy (and GKPY) equations in the complex $s$ plane (from \cite{Caprini:2005zr}), in $M_\pi^2$ units .
The continuous contour corresponds to using $s$ and $t$ inside the
Martin-Lehmann ellipse, which ensures convergence of the partial wave expansion. The domain defined by the discontinuous line, established using just axiomatic 
field theory, is only slightly smaller (\cite{Mahoux:1974ej,libromartin} and see \cite{Caprini:2005zr} for details).
The dots correspond to the positions of the conjugated poles of the $f_0(500)$ and the $f_0(980)$, all them well inside the analyticity domain. 
Reprinted figure with permission from  I.~Caprini, G.~Colangelo and H.~Leutwyler,
  Phys.\ Rev.\ Lett.\  {\bf 96}, 132001 (2006). Copyright 2006 by the American Physical Society.
}
  \label{fig:reddragon}
\end{figure}

Concerning the validity of Roy  equations, they 
require the use of the partial-wave series for $\im T$,
which converges 
inside the Lehman-Martin ellipse. 
However, by assuming Mandelstam analyticity the Roy equations
are valid in  a somewhat larger region (see \cite{Roy:1990hw} for a review).
The resulting domain of validity of Roy equations in the $s$ complex plane
has been calculated in \cite{Caprini:2005zr} and is shown in Fig.\ref{fig:reddragon}.
In this figure the positions of the $f_0(500)$ and $f_0(980)$ conjugated poles are shown, 
and they clearly lie inside this domain, showing the full consistency of the approach
when applied to determine the pole positions, to be reviewed in Sec.\ref{subsec:precisepoles}.
\color{black} Comparing Fig.\ref{fig:reddragon} with Fig.\ref{fig:poles} it is also evident that some of the poles obtained from some old naive models lie well beyond the Lehman-Martin ellipse. \color{black}

Note also that, if one assumes Mandelstam analyticity, in the real axis the applicability region 
\cite{Roy:1971tc} is $-4\mps\leq s\leq60\mps$ i.e. $\sqrt{s}\leq1150$ MeV (see \cite{Roy:1990hw} for a review).
In the seventies \cite{Mahoux:1974ej,Auberson:1974in,Auberson:1977re} it was shown that this $\sim1.1$ GeV applicability bound could be overcome by choosing different trajectories in the complex plane of kinematic variables, although at the price of a considerable increase in complexity
so that they have been rarely used. As one of the very few examples, a first extension up to $\sim1560$ MeV \cite{Mahoux:1974ej}, using manifestly crossing symmetric dispersion
relations in the variables $x\sim st + tu + us$ and $y\sim stu$, 
has been used in \cite{Ananthanarayan:1998hj} in connection with the low energy $\pi\pi$ expansion and $\ell\leq2$ resonances.  
 There are even more complicated dispersion relations over hyperbolae in the $(s,t)$ plane that extend the validity to almost 1800 MeV \cite{Auberson:1974in} and the existence of solutions up to infinity has also been studied in \cite{Auberson:1977re}. However,
 due to the complexity of these formalisms, all applications relevant to extract the $\sigma$ pole use the 
 original formalism, similar to the one presented here, and therefore 
 everything above 1.1 GeV is input and part of the driving terms.·

Let us now concentrate on the use of dispersion relations to obtain reliable descriptions of data that will later be used 
to determine accurately the $\sigma$ pole parameters.

\subsection{$\pi\pi$ amplitude from dispersion relations}
\label{subsec:cfd}

There are several ways in which dispersion relations can be used: 

\begin{enumerate}
\item[1)] as consistency checks in order to 
discard inconsistent data sets, 
\item[2)] as constraints when fitting data, yielding what 
has been traditionally known as ``energy dependent data fits"
or ``constrained fits to data", 
\item[3)] solving them in a certain physical region, to {\it predict}
the amplitudes there, using as input data from another region or from other channels,
\item[4)] to extend the amplitude from the physical region into 
the complex plane to look for resonance poles (or sometimes zeros of dynamical interest like Adler zeros).
\end{enumerate}

Focusing on the the most relevant developments for the major revision
of the $\sigma$ parameters in the RPP 2012 edition,
this subsection will review the first three uses, which concern the amplitude for physical values of $s$,
and the next subsection will be dedicated to 
pole determinations for unphysical values of $s$.

\subsubsection{Dispersion relations as consistency checks of data}

We have already commented that Forward Dispersion Relations \cite{Morgan:1969ca} and Roy equations \cite{Pennington:1973hs} 
had been used to show the inconsistency of ``up" solutions. After some controversy and the resurrection of the
problem in the late 90's, the ambiguity was definitely
solved in \cite{Kaminski:2002pe} in favor of the
``flat-down" scenario, which was the only one fully consistent with Roy equations.
That is why we only plotted such kind of solutions in Fig.\ref{fig:00data} and have discarded other possibilities.

However, even within that class of solutions there are different data sets, which as shown in Fig.\ref{fig:00data} are still
inconsistent with each other within their statistical uncertainties. The consistency of these different data
sets below 900 MeV was checked in \cite{Pelaez:2004vs} against
 the  $T^{I_t=1}$ and $T^{00}$ Forward Dispersion Relations in Eqs.\ref{t00} and \ref{eq:fdr:fdrIt1}. 
 For this purpose, very flexible conformal expansions were fitted to the existent $K_{e4}$ data and 
 different sets of the $S0$-wave phase shifts in the elastic region. The parameterization ensured
 elastic unitarity, the correct threshold behavior and maximal analyticity domain in the $s$ complex plane
 with a right and a left cut. Those conformal expansions converged very rapidly and only required few parameters
 to attain an acceptable $\chi^2/d.o.f.$
\color{black}
Similar fits were performed for every other $\pi\pi$ scattering
partial wave and used inside the dispersion relations. These are called {\it Unconstrained Fits to
Data}, or UFD for short. These fits are uncorrelated from wave to wave,
therefore they can be very easily changed if new and
more precise data ever become available in a particular wave. 
The functional form of these UFD fits is relatively simple and is detailed in 
\cite{Pelaez:2004vs}. 
In addition a Regge 
description of high energy $\pi\pi$ scattering similar to 
that explained in Sec.\ref{subsec:highenergydata} 
was also considered as input in the dispersive integrals.
\color{black}

Next, in order to quantify how well the dispersion relations
are satisfied, one has to compare the input curve versus  the
dispersive solution over a given energy range. 
For this purpose $\Delta_i(s)$ is defined as the difference
between the left and right sides of each dispersion relation $i$ that one wants to check,
whose uncertainties are called $\delta\Delta_i(s)$. 
Next, the average discrepancies are defined as follows:
\begin{equation}
\bar{d}_i^2\equiv\frac{1}{\hbox{number of points}}
\sum_n\left(\frac{\Delta_i(s_n)}{\delta\Delta_i(s_n)}\right)^2 ,
\label{avdiscrep}
\end{equation}
where in \cite{Pelaez:2004vs} the values of $\sqrt{s_n}$ were taken at intervals of 25 MeV \color{black} in the 
$\sqrt{2}M_\pi\leq\sqrt{s}\leq 0.925\,$GeV range. That is, 
they were calculated even below threshold. \color{black}
Note the similarity of $\bar{d}_i^2$ with an averaged $\chi^2/( d.o.f.)$, so that
$\bar{d}_i^2\leq 1$ means that the two curves are close and therefore there is
a good fulfillment of the corresponding
dispersion relation within uncertainties. 
\color{black} However, dispersion relations have not been imposed in the experimental analyses and, as we will see next, large
deviations will occur. \color{black}

The results in terms of $\bar{d}_i^2$ are summarized in Table~\ref{tab:FDRPY05}, 
where we observe that many of
the most commonly used data sets were very inconsistent
with Forward Dispersion Relations and $K_{e4}$ data, which 
in principle was the most reliable set 
(Note that the very precise NA48/2 data was not available yet, so that the uncertainties in $K_{e4}$ were still rather large). 
Actually, the data sets or parameterizations 
in the last six rows of the Table can be considered inconsistent 
even from a conservative point of view, since the average of the two $\bar{d}_i^2$ 
for the two dispersion relations was larger than 3 and in particular
the $\bar{d}^2$ for the $T^{00}$ dispersion relation
was larger than 4. Moreover, the Grayer et al. set E, could also be discarded because its good consistency is only apparent, due to having a much larger uncertainty than other sets and it only satisfies the dispersion relation 
as long as it overlaps with other sets within a couple of standard deviations.
This came as a relatively big surprise since some of these data sets 
were among the ones most widely used in the literature and therefore 
the conclusions of any work relying on them should be taken very cautiously. 

\begin{table} 
   \centering 
 \begin{tabular}{|l|cc|} \hline 
Data sets& $\bar{d}^2$ for $T^{I_t=1}$ & $\bar{d}^2$ for $ T^{00}$\\ \hline
Global fit from \cite{Pelaez:2004vs}&0.3&3.5\\
$K_{e4}$+ {\rm Grayer et al. B} & 1.0 &2.7 \\
$K_{e4}$+ {\rm Grayer et al. C} & 0.4 &1.0 \\
$K_{e4}$+ {\rm Grayer et al. E} & 2.1 &0.5 \\
$K_{e4}$+ {\rm Kaminski et al.} & 0.3 &5.0 \\ \hline
$K_{e4}$+ {\rm Grayer et al. A} & 2.0 &7.9 \\
$K_{e4}$+ {\rm EM, $s$ channel} & 1.0 &9.1 \\
$K_{e4}$+ {\rm EM, $t$ channel} & 1.2 &10.1 \\
$K_{e4}$+ {\rm Protopopescu et al. VI} & 1.2 &5.8 \\
$K_{e4}$+ {\rm Protopopescu et al. XII} & 1.2 &6.3 \\
$K_{e4}$+ {\rm Protopopescu et al. VIII} & 1.8 &4.2 \\
\hline
 \end{tabular} 
\caption{ Fulfillment of the $T^{I_t=1}$ and $T^{00}$ Forward Dispersion Relations, Eqs.\ref{eq:fdr:fdrIt1} and \ref{t00}, by fits to the $K_{e4}$ data in \cite{Rosselet:1976pu,Pislak:2001bf}
when combined with one phase shift analysis below 900 MeV: Note there are different solutions of the CERN-Munich experiment by 
Grayer et al. 
\cite{Cern-Munich} (solution D starts above that energy),
Kaminski et al. corresponds to \cite{Kaminski:1996da}, EM to the solutions of Estabrooks and Martin in \cite{Estabrooks:1974vu} and Protopopescu et al. to the phase shift sets given in different Tables of \cite{Pr73}. The data corresponding to the eight sets that satisfy best these constraints are actually those plotted in Fig.\ref{fig:00data}. The ``Global fit" parameterization in \cite{Pelaez:2004vs} was obtained from 
a fit to those $K_{e4}$ data and some average of different sets in the 870 to 970 MeV region, where they seem to be most compatible among themselves, when considering systematic uncertainties of the order of 10$^\degree$.
 \label{tab:FDRPY05}
} 
 \end{table}


Finally, in \cite{Pelaez:2004vs} a so-called global fit was obtained by fitting the $K_{e4}$ data and a set of averaged
data points in the 870 to 970 MeV region, where they seem to be most compatible among themselves, to which a systematic uncertainty of the order of 10$^\degree$ was added to cover the different sets. 
As it can be seen in Table~\ref{tab:FDRPY05}, the average of the two FDR's was also less than 3 and this is why it has been included in the best five fits. 
It is worth noticing that it was very close to solution C 
and not far from B. Since it contained information from several fits
it was considered 
to be more realistic concerning uncertainties.


\subsubsection{Dispersion relations to constrain data fits}

This is the most common use of dispersion relations in $\pi\pi$ scattering.
Among others, single channel dispersion relations 
for partial waves, in which the left cut and or the subtraction constants
are treated within some approximation
have been widely used in the literature to fit multiple data sets. 
Of particular interest are those works  that implement 
two body unitarity exactly (which is {\it not directly} implemented in Roy or Forward Dispersion Relations)
and include Chiral Perturbation Theory 
to approximate parts of the amplitude like the left cut or the subtraction constants. 
Most of these are generically
known as Unitarized Chiral Perturbation Theory.
As we will see at the end of this section, this approach yields $\sigma$ poles which are very consistent with those obtained from Roy and GKPY equations, but their left cut and  high energy parts are just approximations, so that they do not aim at precise determinations of $\sigma$ parameters.
Actually, although they existed for long  and were certainly good enough to determine
the $\sigma$ pole  existence in the 400 to 500 MeV region, they did not
 trigger the recent radical change in the RPP 2012 $\sigma$ parameters. Nevertheless
they are some of the most relevant methods in order to understand 
the $f_0(500)$ spectroscopic classification and nature.  For these reasons we will 
dedicate to them the whole Sec.\ref{subsec:uchpt}.

Thus, let us focus on the 
four works \cite{CGL,Caprini:2005zr,GarciaMartin:2011jx,Moussallam:2011zg} considered ``the most advanced dispersive approaches" by the 2012 RPP edition, which were used to obtain the ``restricted range"
of sigma parameters in Eq.\ref{rppradicalpole}. 
Of those, only the pole determination of the Madrid-Krakow group \cite{GarciaMartin:2011jx}
was obtained from a dispersion relation using as input a constrained data fit, whereas
in \cite{CGL,ACGL,Moussallam:2011zg}
the pole was extracted from {\it solutions} 
of Roy equations below 800 MeV, together with some theoretical constrains on threshold parameters.
Thus, we will review now the constrained fits of \cite{GarciaMartin:2011cn} and in the next subsection we will comment on the solutions of \cite{CGL,ACGL,Moussallam:2011zg}.

In the previous section we have seen how even some of the most widely used phase shift
data sets are in strong conflict with dispersion relations. 
This should not come as a surprise since these data have been extracted using models that contain large systematic uncertainties. 
Therefore one can try to obtain consistent fits by
using the dispersion relations not just as checks but as constraints on the fit parameters. 
Since one is interested in both consistency and a relatively good description of data,
these constrained data fits are to be obtained from data sets 
which by themselves are not too inconsistent with dispersion 
relations and require just a small improvement. 

\color{black} Thus, on a first work \cite{Pelaez:2004vs} 
the three Forward Dispersion Relations
in Eqs.\ref{t00}, \ref{t0+} and \ref{eq:fdr:fdrIt1}
together with the two sum rules in Eqs.\ref{ec:srK} and \ref{ec:srL}
were imposed upon the five fits in the first rows of Table~\ref{tab:FDRPY05}.
 This was achieved by minimizing:
\begin{equation}
 \sum_{i=T^{+-},T^{00},T^{I_t=1}} \!\!\!\bar{d}_i^2+\bar{d}_K^2+\bar{d}_L^2+\sum_k\left(\frac{p_k-p_k^{\rm exp}}{\delta p_k}\right)^2,
\label{ec:minimizeFDRs}
 \end{equation}
where $\bar d_{K,L}^2$ are the discrepancies of the two sum rules in Eqs.\ref{ec:srK} and \ref{ec:srL}.
As before in the dispersive integrals 
the UFD description was used for all other waves.
The $p_k$ are all the parameters of the UFD sets, which are therefore allowed to 
vary, but not very far from their unconstrained values to ensure that the data is still reasonably well described. 

The results of the best fits in Table~\ref{tab:FDRPY05} after this procedure
are summarized in Table~\ref{tab:FDRPY05Improved}, where one can see that the new constrained fits can be made to satisfy rather well the two Forward Dispersion relations that involve $I=0$.
We also show the  $K$ sum rule in Eq.\ref{ec:srK}, which also
has contributions from $I=0$ but it can be noticed that only the 
constrained Global fit and the one from Solution C satisfy it well. For this reason these two were considered the 
two best fits. Moreover, since it was shown in \cite{Pelaez:2004vs}
that the central value from the constrained fit to Solution C lied
inside the uncertainty band of the Global fit, but the uncertainties of the latter were considered more realistic, from that moment the 
Global fit was considered as the benchmark for a precise description of $\pi\pi$ data consistent with Forward Dispersion Relations.

\begin{table} 
   \centering 
 \begin{tabular}{|l|ccc|} \hline 
Data Fits Constrained with FDR& $\bar{d}^2$ for $T^{I_t=1}$ & $\bar{d}^2$ for $ T^{00}$& $K$\\ \hline
Global fit from \cite{Pelaez:2004vs}&0.4&0.66&1.6$\sigma$\\
$K_{e4}$+ {\rm Grayer et al. C} & 0.37 &0.32 &1.5$\sigma$\\ \hline
$K_{e4}$+ {\rm Grayer et al. B} & 0.37 &0.83 &4.0$\sigma$\\
$K_{e4}$+ {\rm Grayer et al. E} & 0.6 &0.09 &6.0$\sigma$\\
$K_{e4}$+ {\rm Kaminski et al.} & 0.43 &1.08 &4.5$\sigma$\\ 
\hline
 \end{tabular} 
\caption{ Fulfillment of the $T^{I_t=1}$ and $T^{00}$ Forward Dispersion Relations, Eqs.\ref{t00} and \ref{eq:fdr:fdrIt1}, and the sum rule $K$ in Eq.\ref{ec:srK}, by 
the best fits in Table~\ref{tab:FDRPY05} after they have been constrained to
satisfy Forward Dispersion Relations and sum rules. The fulfillment 
of the sum rule is expressed in standard deviations.
Note that the only fits that once improved 
satisfy the Forward Dispersion Relations
and simultaneously the sum rule are the Global fit and that of solution C. 
 \label{tab:FDRPY05Improved}
} 
 \end{table} 

\color{black}

Actually, the Global fit just described 
was the starting point 
of the Madrid-Krakow group for further constraining 
the data fits to $\pi\pi$ data with Roy and GKPY equations,
and extending the analysis to higher energies,
although the functional form was slightly 
improved throughout a series of works 
\cite{Yndurain:2007qm,GarciaMartin:2011cn,Pelaez:2004vs,Kaminski:2006qe,Kaminski:2006yv}. The improvements were made in all waves and included more flexible parameterizations, differentiable matching conditions, sum rules, etc.
as well as an update on new data, either at higher energies, or
the latest NA48/2 data. 
In Table~\ref{tab:UFDCFDdiscrepancies} we show how well the
updated and improved
UFD, which include the improved 
S0 ``global-fit'', satisfy the three
Forward Dispersion Relations, three Roy Equations and three GKPY equations. 
Note that the UFD fit does not always 
satisfy very well these dispersion relations. In particular
the GKPY equations for the S0-wave in the region above 932 MeV are
satisfied poorly. There is clear room for improvement.

On a second stage, nine dispersion relations were imposed on the
previous fits as additional constraints. This is achieved by minimizing:
\begin{equation}
 \sum_i W^2_i \bar{d}_i^2
 +\bar{d}^2_I+\bar{d}^2_J+\sum_k\left(\frac{p_k-p_k^{\rm exp}}{\delta p_k}\right)^2,
\label{tominimize}
 \end{equation}
\color{black} where now the $\bar d_i^2$ include not only the
averaged discrepancies of the three FDRs up to a given energy, but also those of the three Roy equations and the three GKPY equations. 
The discrepancies of Roy and GKPY equations are defined in a similar way as we did for the FDRs in Eq.\ref{avdiscrep}, but starting from threshold and up to 992 MeV or 1100 MeV. The discrepancies in the FDRs are calculated this time up to 932 GeV or 1420 GeV.
This is done to check the fulfillment in different regions.
\color{black}
In addition,
$\bar d_{I,J}^2$ are the discrepancies of the two sum rules in Eqs.\ref{eq:Isumrule} and \ref{eq:Jsumrule}.  There is no need to impose the $K$
and $L$ sum rules, because they are well satisfied.
The $W_i$ are numbers, typically between 3 and 7, to weight the dispersive
constraints against the data description. They are varied in some energy intervals to ensure
that the dispersion relations are well satisfied in all energy regions while not spoiling data.
The price to pay
is that now all the waves are correlated. 
However, these new {\it Constrained
Fits to Data} (CFD for short) are much more
 reliable than the UFD set, being consistent
with analyticity, unitarity, crossing, etc. This can be checked in 
Table~\ref{tab:UFDCFDdiscrepancies} where the average discrepancies of 
this CFD set for the nine dispersion relations are given up to different energy regions. Note that 
Forward Dispersion Relations were calculated up to 1420 MeV (we saw that beyond that energy phase-shift analyses became much less reliable), whereas Roy and GKPY equations are limited 
to $\sqrt{s}<1100 \mev$ as discussed in the previous section.

\begin{table}
\centering
\begin{tabular}{|l|c|c|c|c|}
\hline
$\qquad\bar d_i^2$ & UFD & CFD & UFD & CFD\\
\hline
\hline
FDRs&\multicolumn{2}{c|}{\small $s^{1/2}\leq 932\,$MeV}&\multicolumn{2}{c|}{\small $s^{1/2}\leq 1420\,$MeV}\\
\hline
$\pi^0\pi^0$& {0.31} & 0.32 &  {2.13}& 0.53 \\
$\pi^+\pi^0$& {1.03} & 0.33 &  {1.11}& 0.43\\
$I_{t=1}$& {1.62}  & 0.06 &  {2.69}& 0.22\\
\hline
\hline
Roy Eqs.&\multicolumn{2}{c|}{\small $s^{1/2}\leq 992\,$MeV}&\multicolumn{2}{c|}{\small $s^{1/2}\leq 1100\,$MeV}\\\hline
S0&  {0.64}& 0.02  &{0.56}& 0.04 \\
S2&  {1.35}& 0.21  &{1.37}& 0.26 \\
P&   {0.79}& 0.04  &{0.69}& 0.12 \\
\hline
\hline
GKPY Eqs.&\multicolumn{2}{c|}{\small $s^{1/2}\leq 992\,$MeV}&\multicolumn{2}{c|}{\small $s^{1/2}\leq 1100\,$MeV}\\
\hline
S0& {1.78} & 0.23 &{2.42} & 0.24 \\
S2&  {1.19} & 0.12 &{1.14} & 0.11\\
P & {2.44} & 0.68 &{2.13} & 0.60 \\
\hline
\hline
Average& 1.24 & 0.22 & 1.58 &0.28\\
\hline
\end{tabular}
\caption{ Average discrepancies $\bar d_i^2$ of the 
unconstrained data fits (UFD set) and the constrained ones (CFD) for each 
dispersion relation up to two different energies \cite{GarciaMartin:2011cn}.
Note the clear improvement in consistency for the CFD set, as all $\bar d_i^2<1$
in both energy regions.
\label{tab:UFDCFDdiscrepancies}}
\end{table}

\begin{figure}
  \centering
 \includegraphics[width=\textwidth]{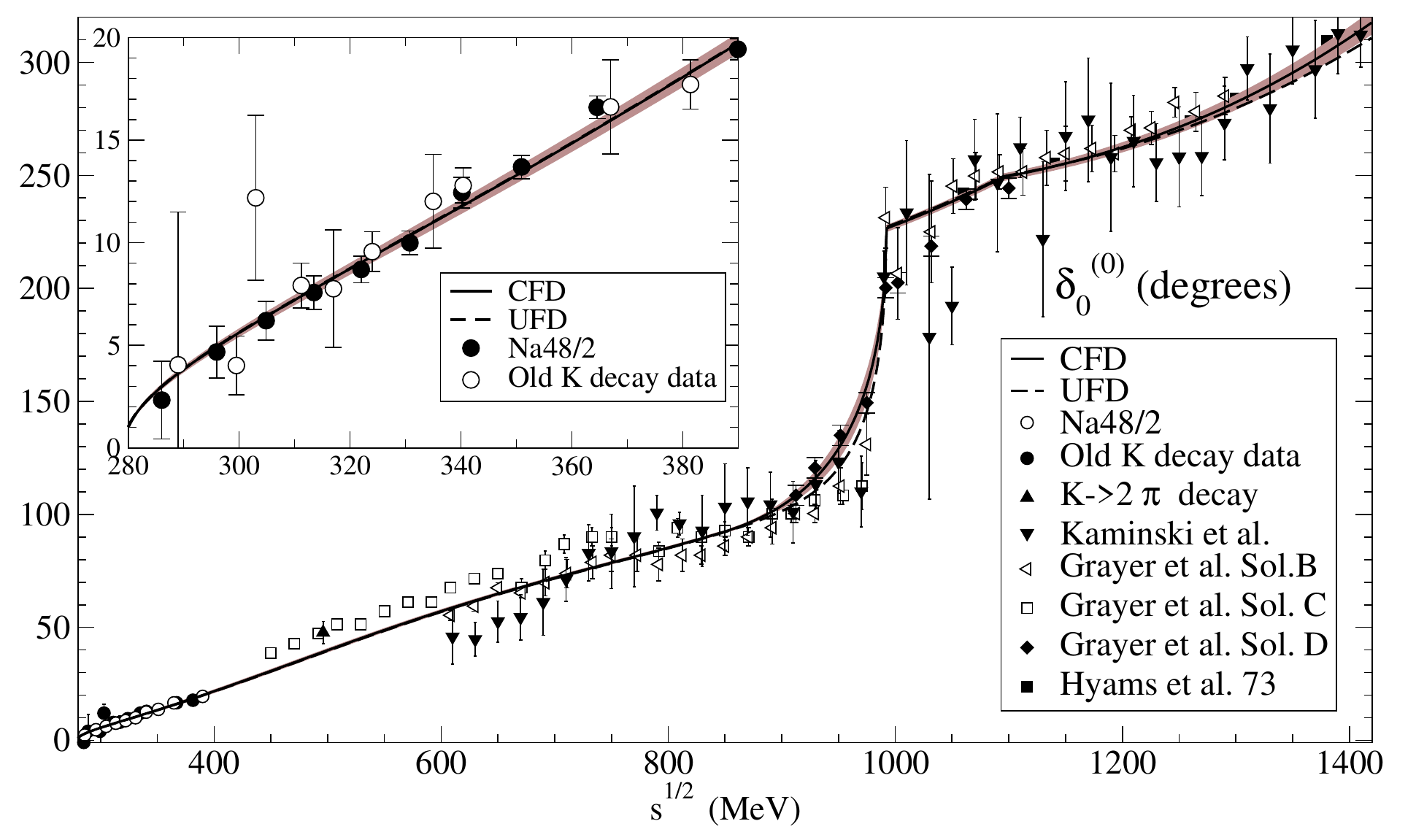}
  \caption{Fits to $\pi\pi$ scattering data from \cite{GarciaMartin:2011cn},
both unconstrained (UFD) and constrained by dispersion relations (CFD) as explained in the text.
Note that the CFD parameterization is just a relatively small variation of the UFD, consistent within uncertainties, except in the 1 GeV region. Compared to Fig.\ref{fig:00data}, 
data sets that do not satisfy relatively well Forward Dispersion Relations and sum rules
are not shown \cite{Pelaez:2004vs}. The small uncertainty band is mainly driven by the precise
$K_{e4}$ data of NA48/2 \cite{Batley:2010zza}. The ``Old K-decay data'' correspond
to older $K_{e_4}$ experiments\cite{Rosselet:1976pu,Pislak:2001bf}. 
As in Fig.\ref{fig:Kl4data}, all $K_{e4}$ experiments have been corrected for isospin and subtracted the $\delta_1$ phase as explained in Sec.\ref{subsec:pwbelow}. Figure taken from \cite{GarciaMartin:2011jx}. }
\label{fig:S0waveCFD}
\end{figure}

\begin{figure}
  \centering
 \includegraphics[width=0.7\textwidth]{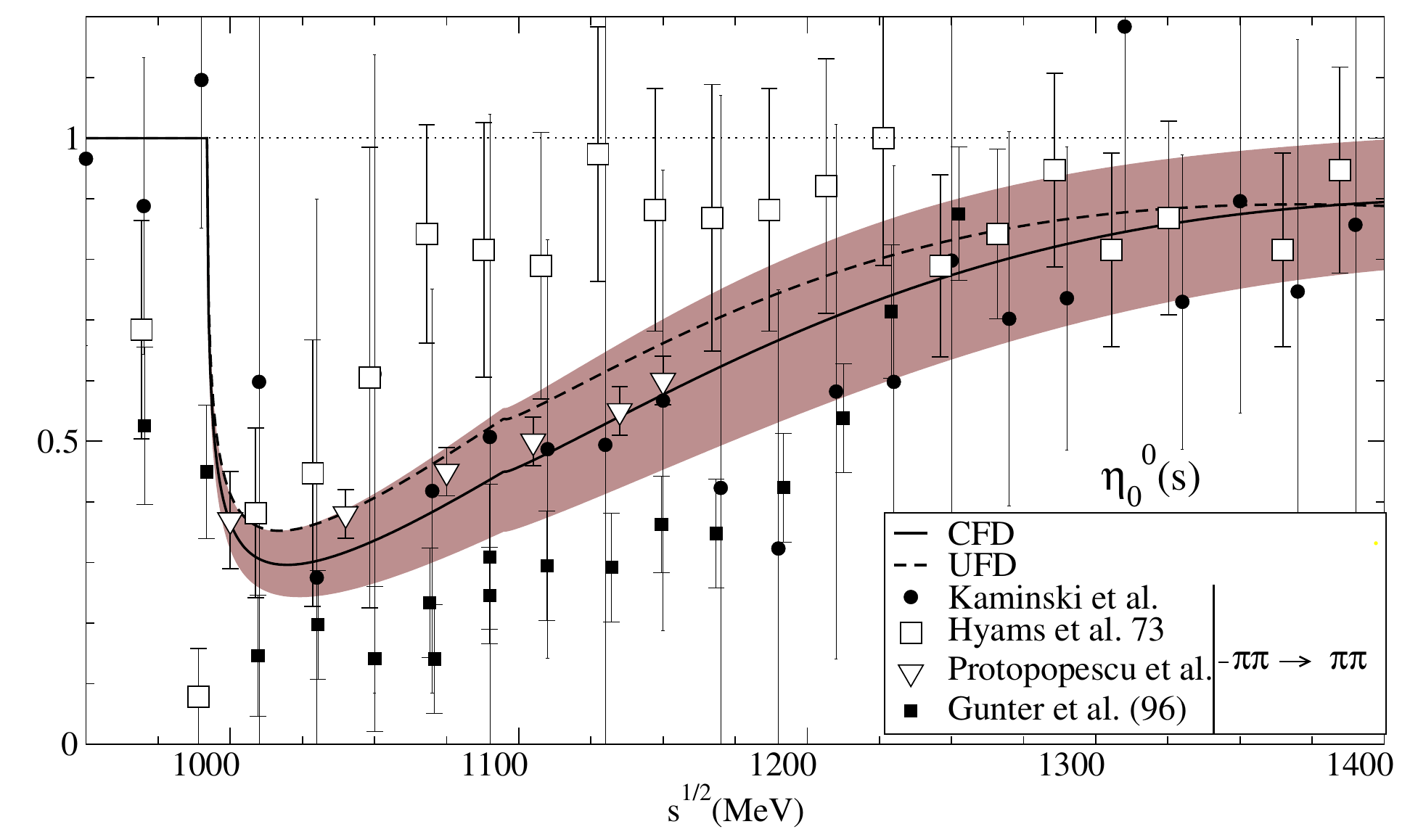}
  \caption{Constrained (CFD) and unconstrained (UFD) fits to the S0 inelasticity from \cite{GarciaMartin:2011cn}. 
By comparing to Fig.\ref{fig:ineldata}, it is clear that the ``dip-scenario'' is the favored one.
The data are as in Fig.\ref{fig:ineldata}. Figure taken from \cite{GarciaMartin:2011jx}.}
\label{fig:S0inelCFD}
\end{figure}

The resulting constrained S0-wave and inelasticity are shown in Fig.\ref{fig:S0waveCFD}
and Fig.\ref{fig:S0inelCFD}, together with the unconstrained results. It can be noticed that the change in the S0-wave phase shift 
from unconstrained to constrained in the S0-wave is very small, 
only sizable in the $f_0(980)$ region or above. When comparing with
Fig.\ref{fig:00data} note that below 900 MeV only the 
data which are roughly  
consistent with dispersion relations and low energy $K_{e4}$ decays are shown.
The very small uncertainty band is driven by the very precise low energy data of NA48/2.
Note also that the smallest the uncertainty band the more difficult to satisfy 
the dispersion relation, so the results of Table~\ref{tab:UFDCFDdiscrepancies} are even more impressive. The latest 
CFD S0-wave is closer to the Grayer et al. solution B, although this solution was slightly disfavored compared to solution C when only Forward Dispersion Relations and the ``Old $K_{e4}$ decay data`` were considered. This is largely
due to the recent NA48/2 data and the isospin correction \cite{Colangelo:2008sm}
discussed in Sec.\ref{subsec:pwbelow}, see Fig.\ref{fig:NA48data}.

Concerning the inelasticity in Fig.\ref{fig:S0inelCFD}, the dispersive constraints clearly support
the ``dip-scenario'' since the inelasticity decreases to $\eta_0^{(0)}\simeq 0.3$  
slightly above 1 GeV. In \cite{GarciaMartin:2011cn} it was actually shown that forcing
the non-dip scenario into the dispersive constraints requires a 
phase that does not describe the data. The Gunter et al. data set, which stands for some preliminary results of the E852 Collaboration \cite{Gunter:1996ij} has not been included in the fit, but also favors the dip solution.
That the ``non-dip'' scenario is disfavored by Roy equations was also confirmed in \cite{Moussallam:2011zg}. As we will see below, settling the issue of the inelasticity above $\bar KK$ threshold is relevant
for a precise determination of the $f_0(980)$ pole, and indirectly for the $\sigma$.

\begin{figure}
  \centering
 \includegraphics[width=0.325\textwidth]{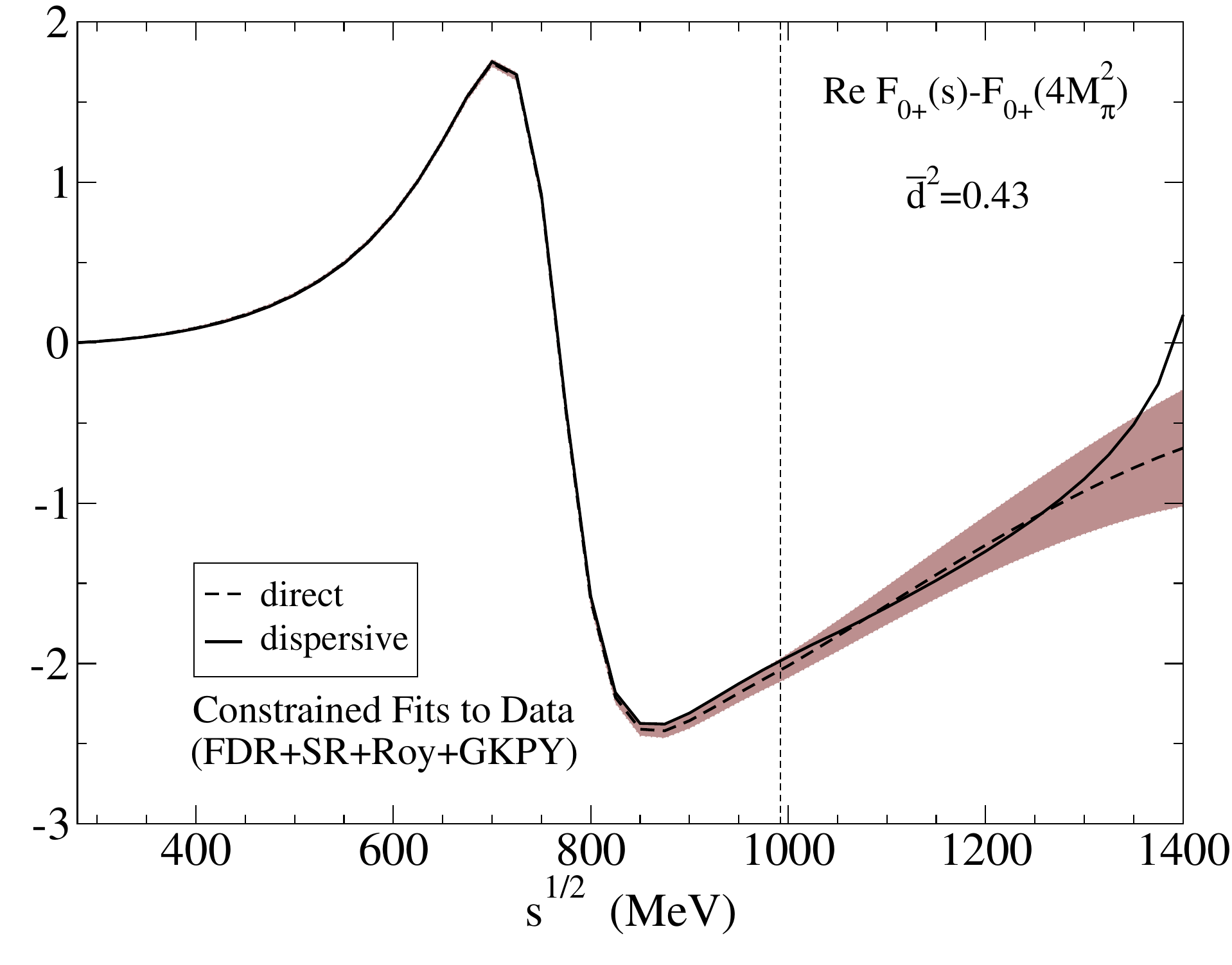}
 \includegraphics[width=0.32\textwidth]{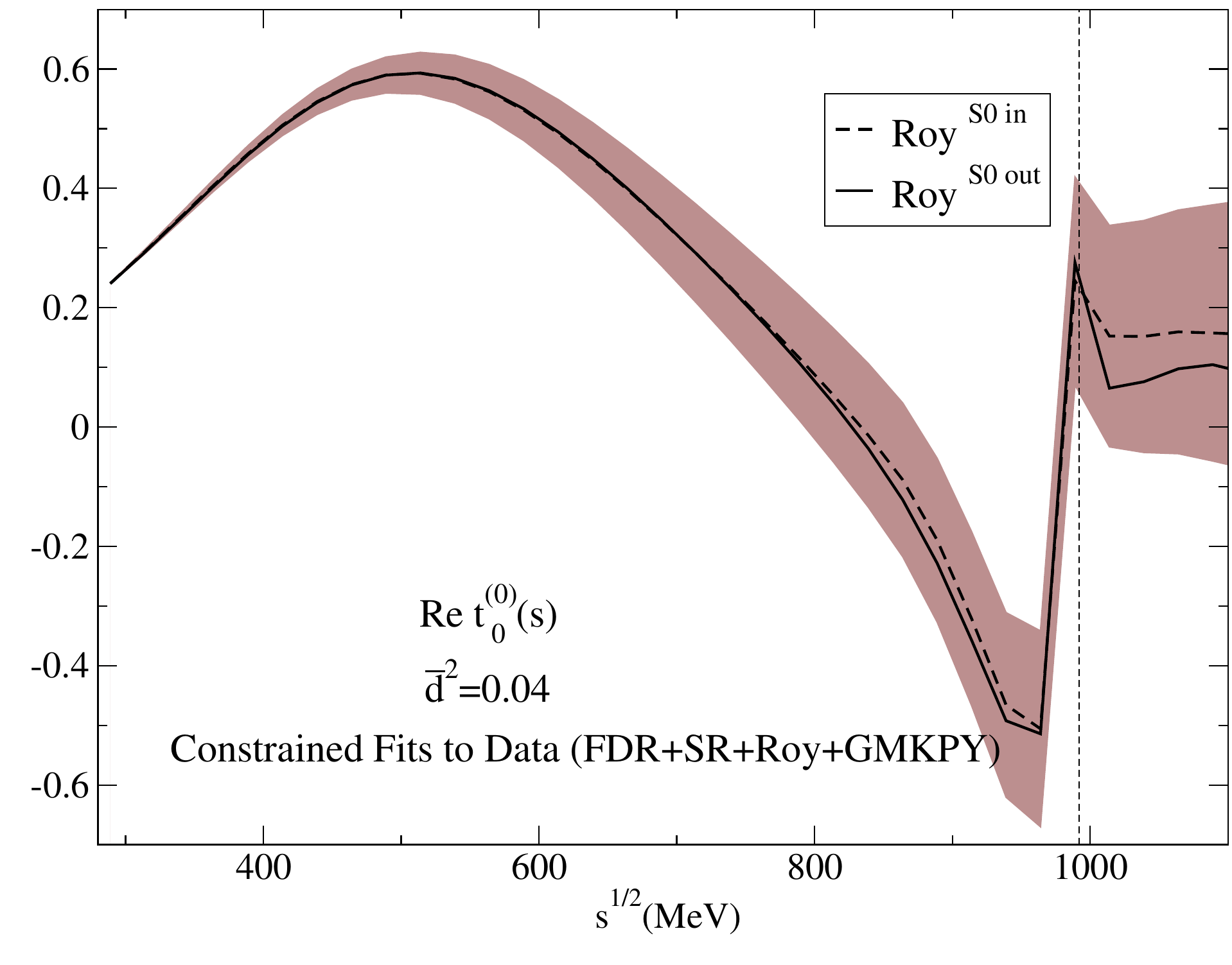}
 \includegraphics[width=0.32\textwidth]{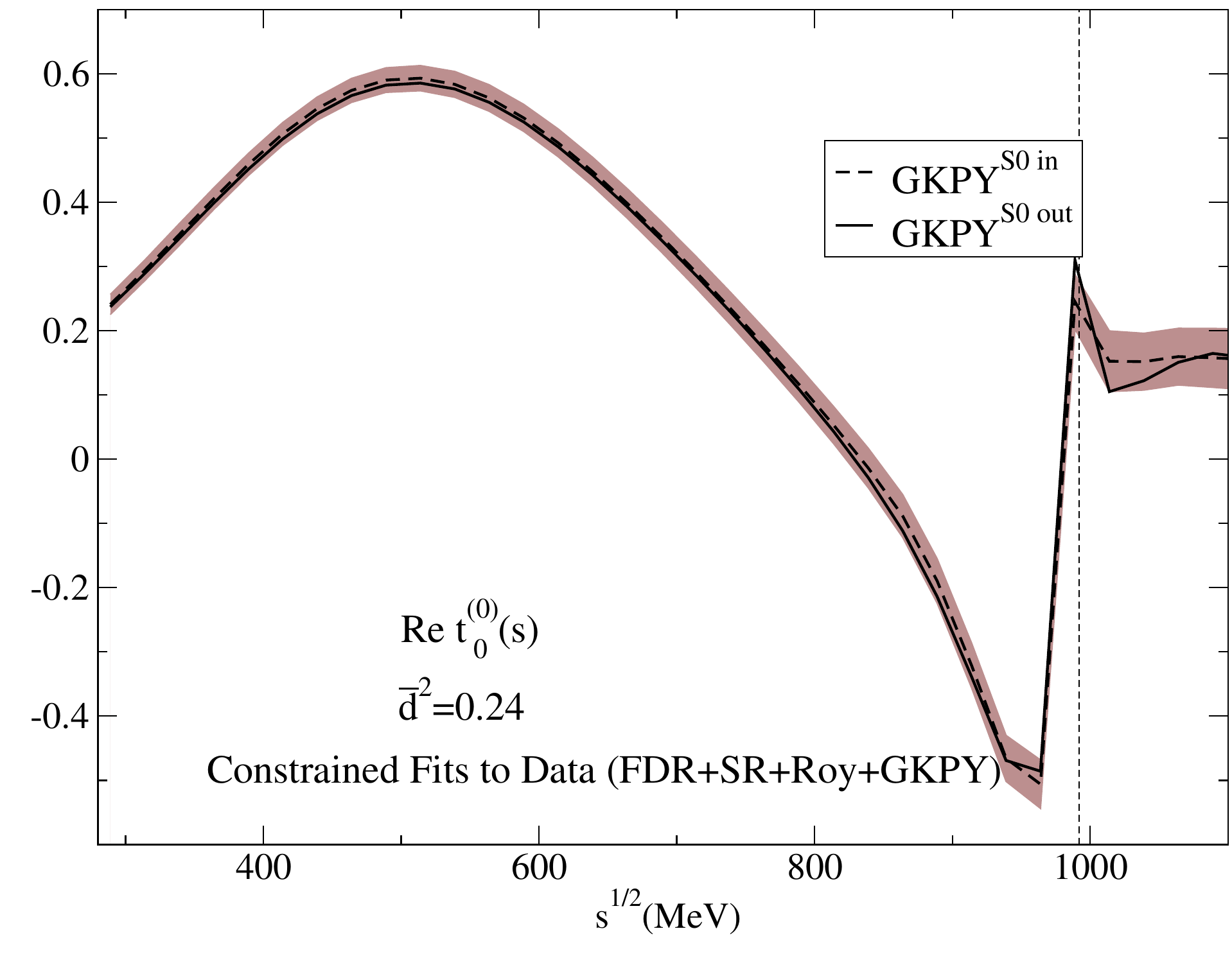}
 \includegraphics[width=0.325\textwidth]{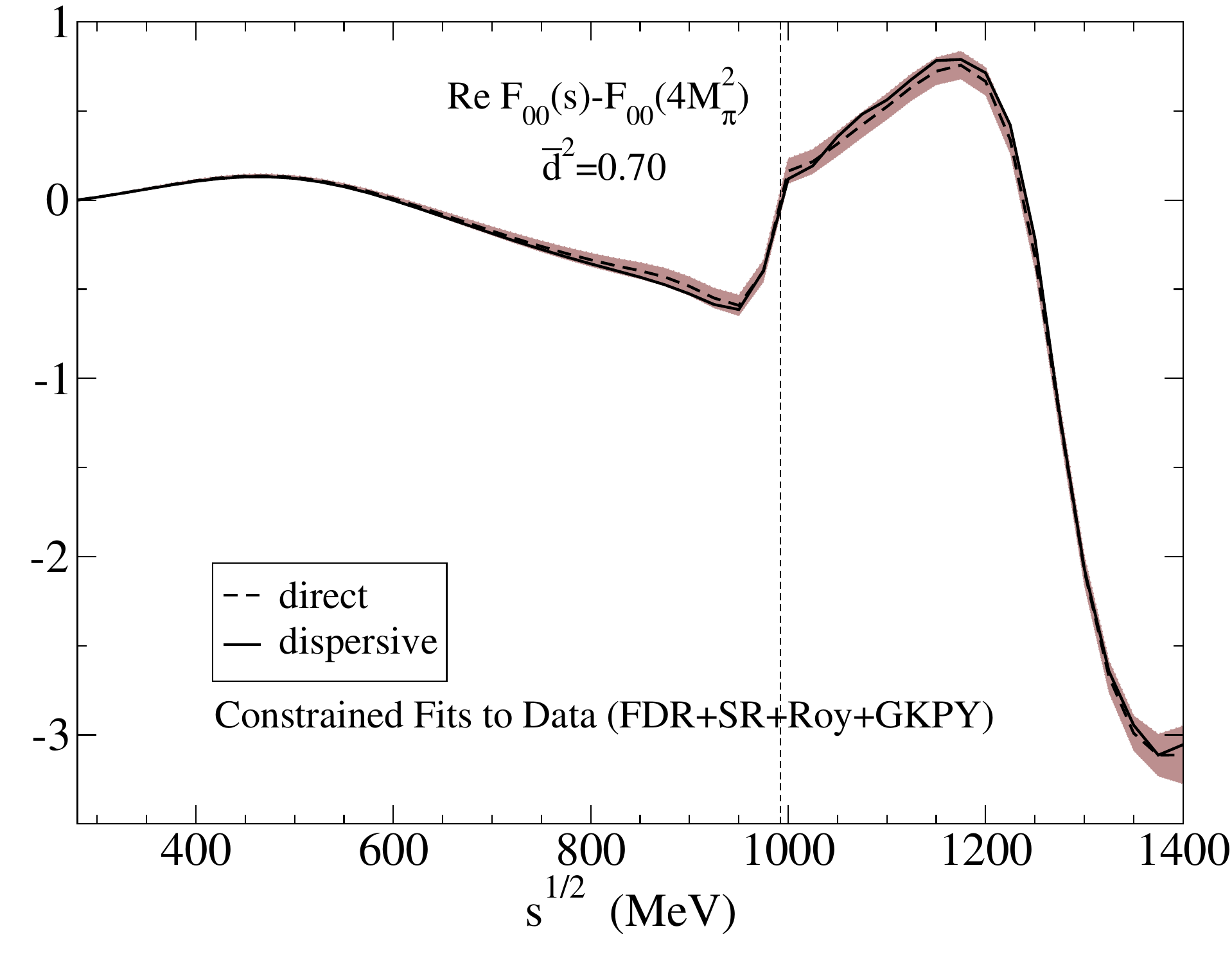}
 \includegraphics[width=0.32\textwidth]{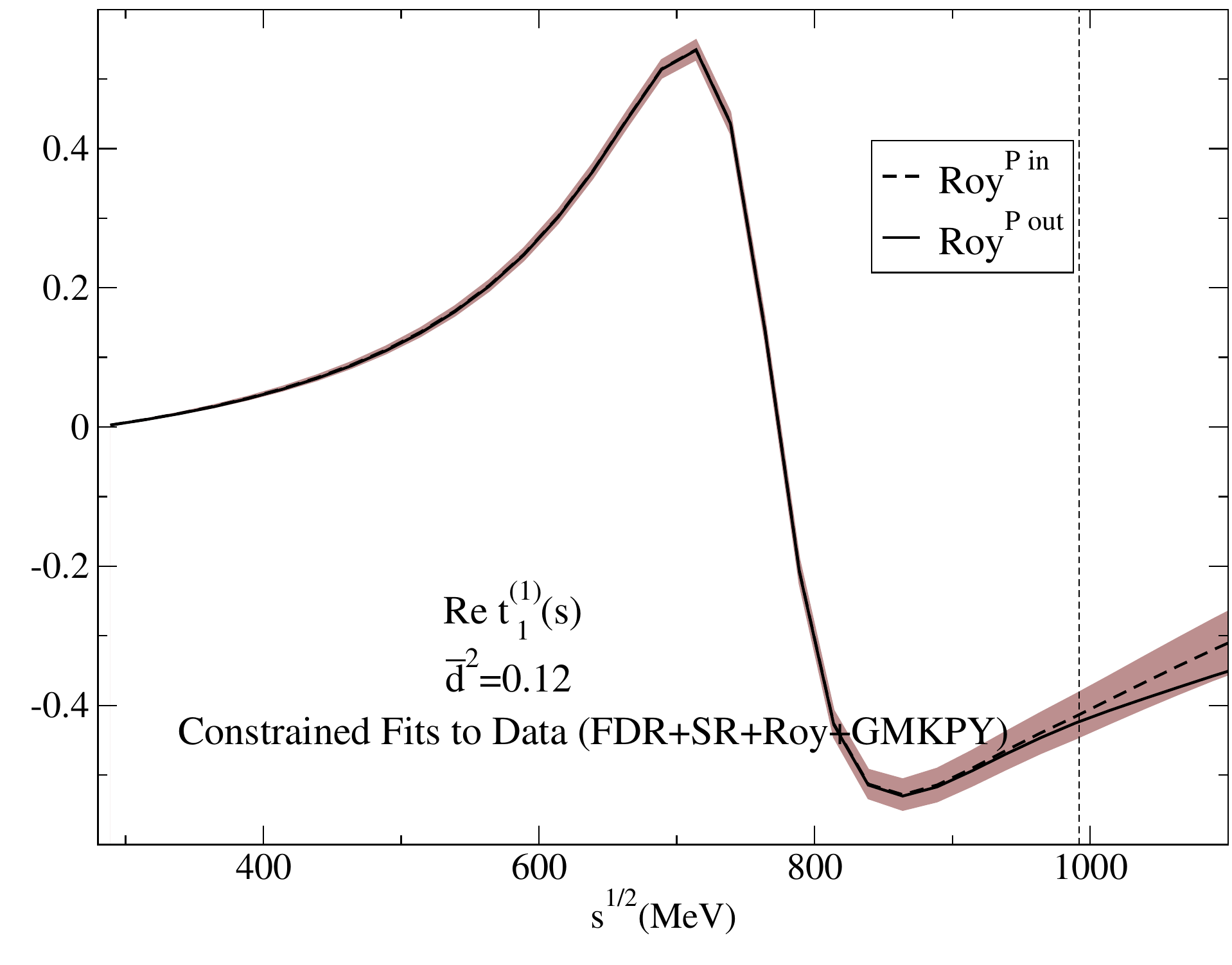}
 \includegraphics[width=0.32\textwidth]{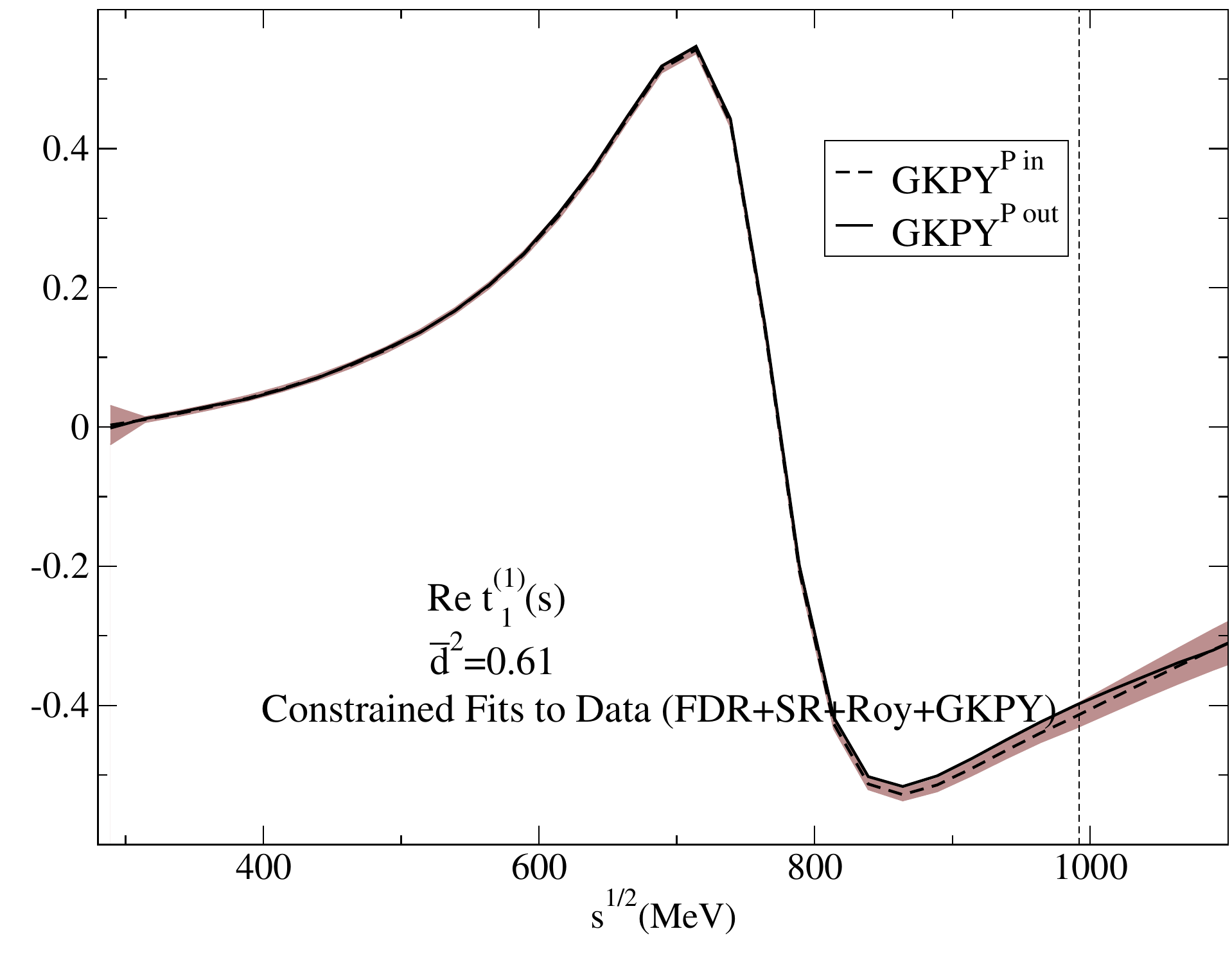}
 \includegraphics[width=0.325\textwidth]{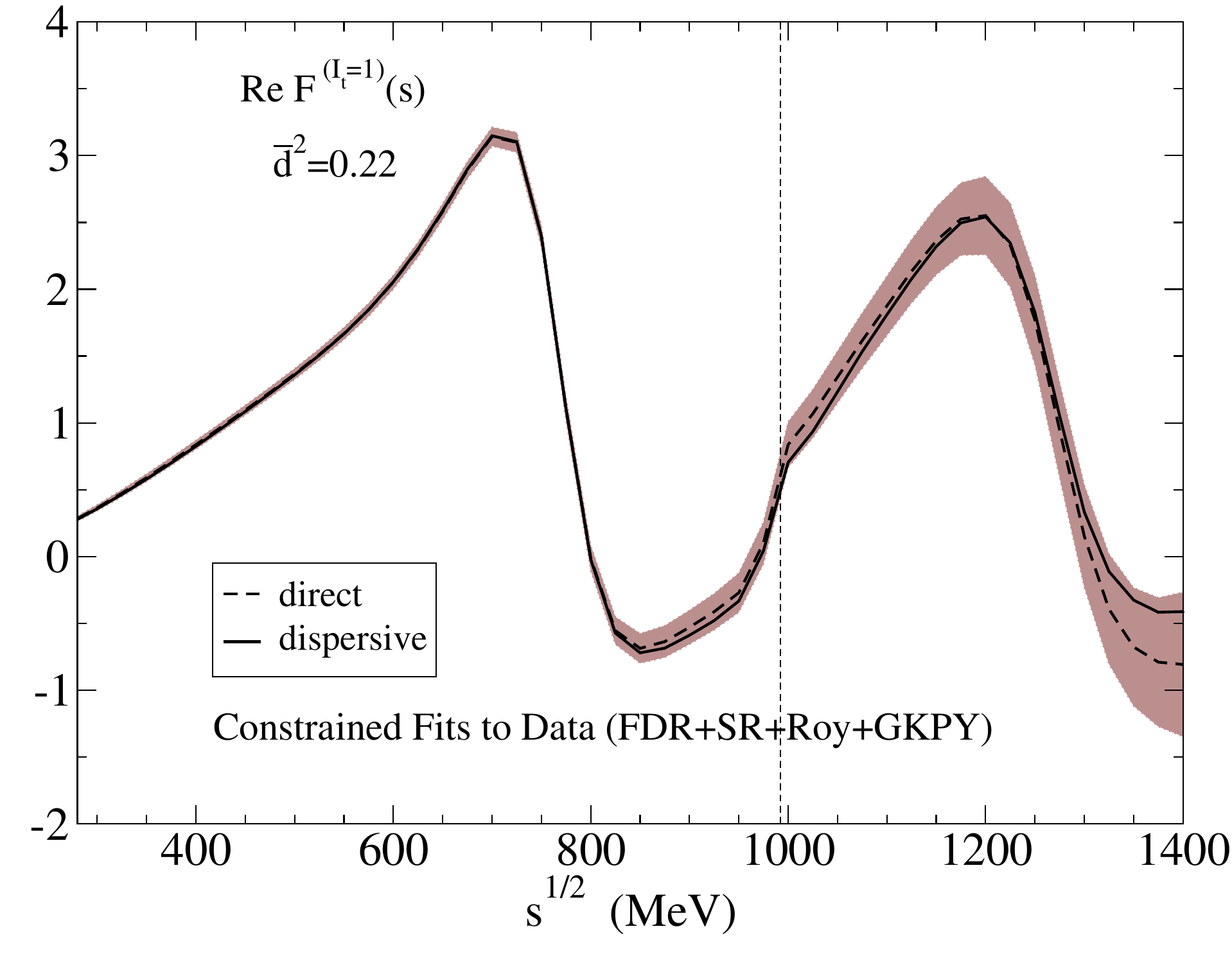}
 \includegraphics[width=0.32\textwidth]{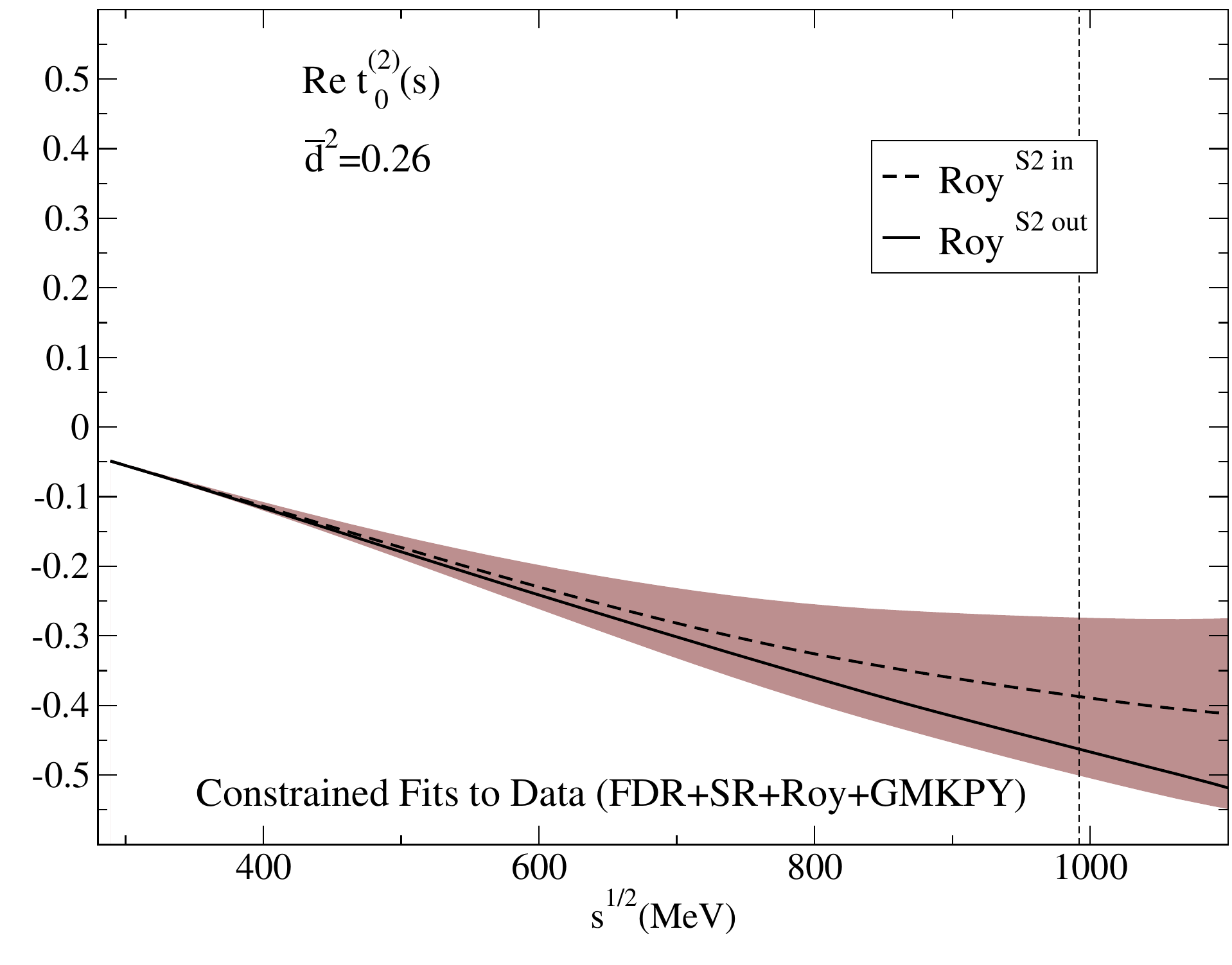}
 \includegraphics[width=0.32\textwidth]{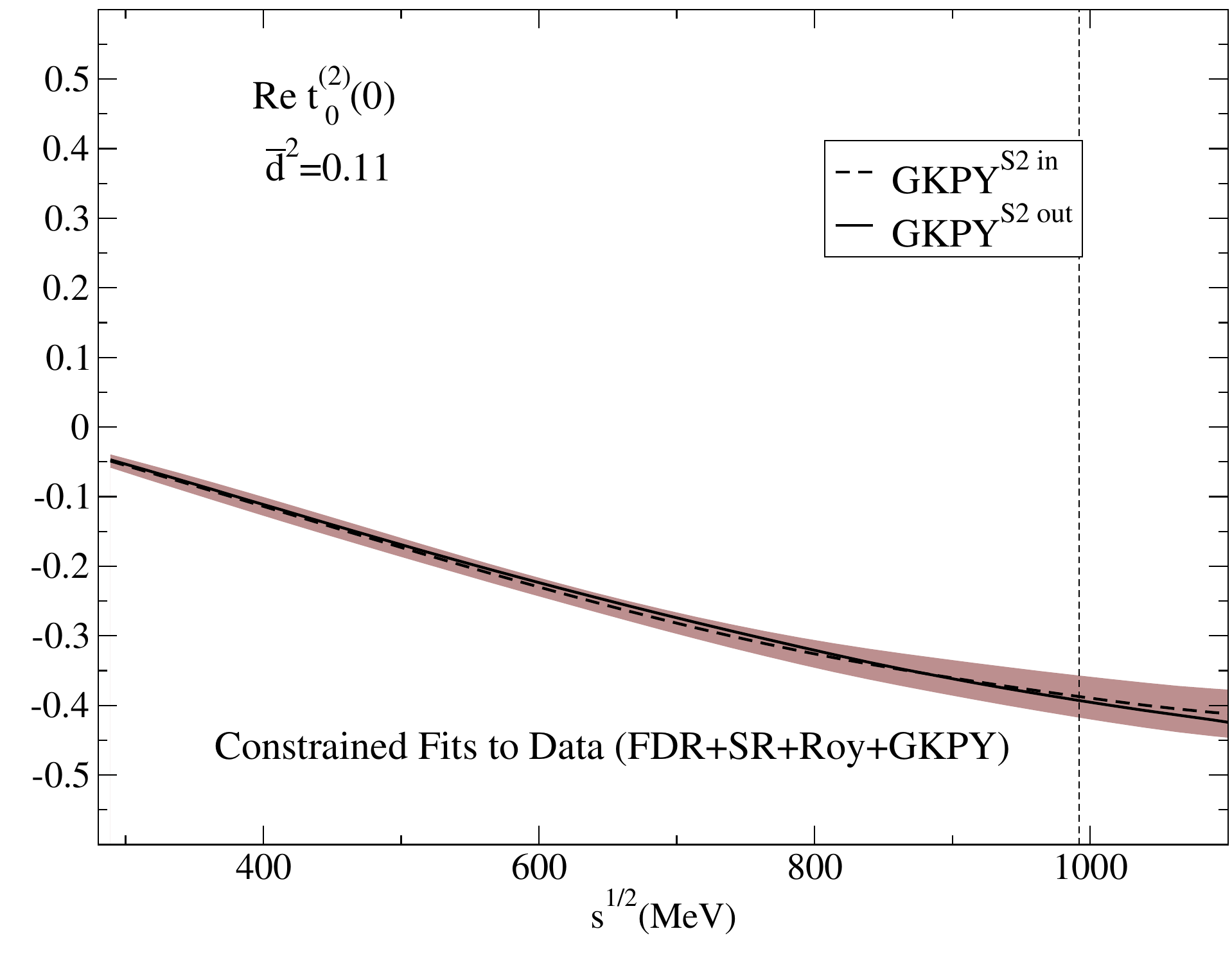}
  \caption{Consistency of the Constrained Data Fit in \cite{GarciaMartin:2011cn}
with respect to dispersion relations. Left column: 
Forward Dispersion Relations, direct means the input amplitude
and dispersive stands for the result of the dispersive integral. 
Central column: Roy equations for the S0, P and S2-waves
Roy Equations. Right column:  GKPY equations for the S0, P and S2-waves.
In Roy and GKPY equations, ``in'' means that the curve is calculated with the CFD parameterizations whereas ``out'' stands for the dispersive result using CFD as input.
The dotted vertical line marks the opening of the $K\bar K$ threshold.
The dark bands represent the uncertainty. Figures taken from \cite{GarciaMartin:2011cn}.
}
\label{fig:fulfillment}
\end{figure}

In Fig.\ref{fig:fulfillment} we show the comparison between the CFD input and its corresponding
dispersive output from Forward Dispersion Relations (left column),
Roy equations (central column) and GKPY equations (right column).
As usual, Forward Dispersion Relations extend up to 1400 MeV, whereas Roy and GKPY equations are valid only below 1100 MeV. It can also be noticed that {\it given this same input}
Roy equations are a much more stringent constraint than GKPY at low energies, whereas GKPY equations are more stringent above 400 or 450 MeV. This is mainly due to the linear propagation in $s$ (quadratic in energy) of the uncertainties of the 
scalar threshold parameters and the main motivation for the derivation of GKPY equations \cite{GarciaMartin:2011cn,Kaminski:2008fu,Kaminski:2008rh}, with just one subtraction instead of two. Note also that the CFD set is consistent 
within uncertainties throughout the whole energy region.
Moreover, in case one could doubt the data selection, a recent statistical analysis \cite{Perez:2015pea} has shown that the UFD set, despite its assumptions on the systematic uncertainties
of the data, only violates marginally the  normality requirements of the residual distributions, which would be satisfied with
 rather tiny modifications of the data selection, leading to almost identical results perfectly consistent with statistical requirements, thus reinforcing the results for the UFD set,
 which was the starting point of this whole approach. 
As we will see below, this very consistent and precise data description will give rise to a reliable and accurate
$f_0(500)$ pole determination. But first let us comment on the other use of the dispersive approach.

\subsubsection{Solutions of Dispersion relations}

Dispersion relations can also be solved to predict the amplitude
in a certain physical region from data or other constraints
in other regions. With different variations, this is the approach 
followed in the other three ``most advanced'' dispersive analysis 
\cite{CGL,Caprini:2005zr,Moussallam:2011zg}
quoted in the 2012 RPP edition. The renewed interest in Roy equations in the 00's was actually fostered
by a detailed work along these lines in \cite{ACGL}, which was then used as a starting point or basic reference for subsequent works.

The first issue to address within this approach is whether a  solution exists and is unique
once the driving terms and scattering lengths are known.
The answer is positive for the S and P-wave Roy equations in a region
between threshold and a ``matching'' energy $s_0$ 
chosen between $m_\rho\simeq770\,\mev$ 
and the energy where $\delta_0^{(0)}=\pi/2$.
Since the latter occurred around 860 MeV
for the energy-independent analysis of Estabrooks and Martin \cite{Estabrooks:1974vu}
and the CERN-Munich solution B of Grayer et al. \cite{Cern-Munich}
(which is the only one in the previous Hyams et al. \cite{Hyams:1973zf} work by the same collaboration),
 in \cite{ACGL} the matching energy was chosen at $\sqrt{s_0}=800$ MeV 
\footnote{As seen in Fig.\ref{fig:00data}
other experimental data sets, like solution C, may reach $\pi/2$ before 800 MeV, 
but the choice is also consistent with the later findings for the UFD and CFD fits in \cite{GarciaMartin:2011cn}.}.
As explained in Sec.\ref{subsubsec:royeqs}, the amplitude above that energy as well as higher partial waves are considered input and included in the driving terms.
Since at threshold the phase vanishes, the values of the S and P phase shifts at 
the matching point define a boundary problem together with the Roy equations.
Note also that since Roy equations are solved exactly one is setting the $4\pi$
contribution, which should open up at 560 MeV, exactly to zero. As commented in previous sections,  such inelasticity 
has not been observed below 1 GeV, 
and some estimates of its size in some models \cite{Kaminski:1997gc,Kaminski:1998ns,Albaladejo:2008qa,Bugg:1996ki}
show that it is indeed very small and does not alter significantly the position of the $\sigma$ pole.

Then, in \cite{ACGL} this problem was solved numerically by minimizing the difference between the right and left hand of the Roy equations at 22 points. This was made by varying the parameters of the simple parameterizations
first suggested in \cite{Schenk:1991xe}:
\begin{equation}
\tan \delta_{J}^{(I)}=\sigma(s)k^{2J}[a_J^{(I)}+B_J^I k^2+C_J^I k^4+D_J^I k^6]\Big(\frac{4\mps-s_J^I}{s-s_J^I}\Big)\,,
\label{eq:Schenk}
\end{equation}
where one of the parameters is fixed from the value of the phase at the matching point.
The scattering length $a_0^{(2)}$ was tuned to avoid cusps at the matching point.
Note that for a given value of $a_0^{(0)}$ 
(which in the above equation is given in $M_\pi$ units) and a given input,
Roy equations admit just one solution without cusps for one value of  $a_0^{(2)}$.
As shown by Morgan and Shaw \cite{Morgan:1969ca}, this produces a ``universal curve'' which is almost a straight line in the $(a_0^{(0)},a_0^{(2)})$ plane. However, due to the uncertainties in the input this universal curve becomes a ``universal band". Interpolation
parameterizations of the $B_J^I,C_J^I,D_J^I$ and $s_J^I$ in terms of $(a_0^{(0)},a_0^{(2)})$
can be found in \cite{ACGL}. Note that by parameterizing the elastic phase shift, elastic unitarity is guaranteed.

\begin{figure}
  \centering
 \includegraphics[width=0.8\textwidth]{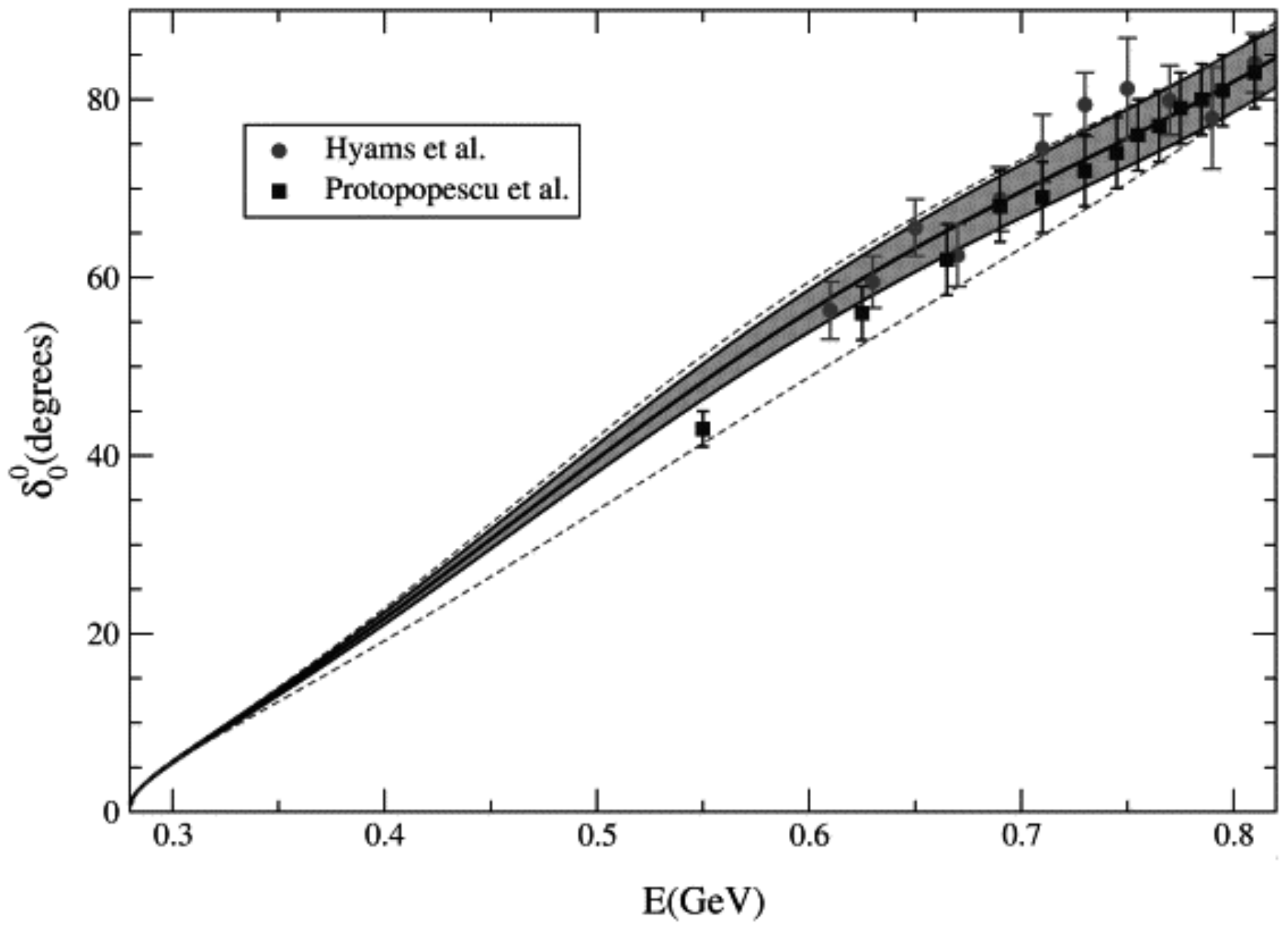}
  \caption{$\pi\pi$ scattering S0-wave phase shift.  
Solutions of Roy equations below 800 MeV
with or without 
imposing chiral symmetry constraints (shaded region \cite{CGL} or 
region between dashed curves \cite{ACGL}, respectively). 
Reprinted from Nucl.\ Phys.\ B {\bf 603}, 125 (2001).  G.~Colangelo, J.~Gasser and H.~Leutwyler,
  ``$\pi \pi$ scattering''. Copyright 2001, with permission from Elsevier. }
\label{fig:CGL}
\end{figure}

The functional space spanned by these functions was large enough to reach differences 
between the left and right sides of the Roy equations of the order of $10^{-3}$, much smaller than 
the experimental uncertainties there.  Note that this is a very different approach to that in \cite{GarciaMartin:2011cn},
which we described in the previous subsection, since there dispersion relations were only imposed
within experimental uncertainties in a fit to data, not solved. Thus, traditionally the UFD and CFD  of 
\cite{GarciaMartin:2011cn} would have been
called  ``energy dependent data analyses''. In contrast, 
in the case of \cite{CGL,ACGL,Moussallam:2011zg}
{\it no data on the S or P phases shifts are used as input below 800 MeV} 
and should be considered as predictions.
We show in Fig.\ref{fig:CGL} the resulting S0-wave phase shifts from \cite{ACGL}
as the region between the dashed curves.

The predictions of Roy equations can be made even more accurate by imposing further
theoretical constraints from chiral symmetry, which implies 
relations between amplitudes. In particular, at low energies chiral symmetry implies that 
in a low energy and pion mass expansion of the amplitudes, only certain terms are allowed, which depend on a few phenomenological constants at each order.
This expansion is known as Chiral Perturbation Theory (ChPT) \cite{chpt1,chpt2} and will be explained in detail
in Sec.\ref{subsec:chpt}. 
What is relevant now is that this chiral expansion can be matched to the 
phenomenological expansion of the amplitudes provided by the Roy equations,
thus providing further constraints on the solutions of the Roy equations.
This was done in \cite{CGL} using the phenomenological representation obtained from Roy equations in \cite{ACGL}. The matching procedure, which was carried out up to order $k^6$ (or $M_\pi^6$, which both correspond to a two-loop calculation in ChPT)  is rather technical and the reader is referred to \cite{CGL} for details as well as for the implications on the low energy parameters of ChPT.
\color{black}
As a matter of fact, one of the main results obtained by combining Roy  equations and ChPT concerns the s-wave $\pi\pi$ scattering lengths. Actually, the sharp predictions obtained within the framework of Roy equations
triggered an intense experimental program. This effort culminated in the NA48/2 precise measurement of 
the $\pi\pi$ scattering
phase at low energies, which not only provided precise values of threshold parameters, but also turned out to be very relevant 
for the dispersive determination of the $\sigma$ parameters, as we have seen in the previous section, and
for the RPP selection of $\sigma$ results. The verification of these ChPT+Roy equations predictions ---at a time when lattice methods were not yet providing significant results--- was crucial for the acceptance of the theoretical picture that underlies the analysis 
of ChPT. It showed, for instance, that the theoretical alternatives developed in \cite{Fuchs:1989jw}, although logically coherent, are not viable as they are not consistent with what is observed.
\color{black}

Nevertheless, for this $\sigma$ review the most relevant 
result is the  S0-wave phase shift from \cite{CGL}, which is also shown 
in Fig.\ref{fig:CGL} as a continuous line.  
Note that the uncertainties 
of the Roy equation solution with chiral 
symmetry constraints (gray band in Fig.\ref{fig:CGL} are smaller than without chiral constraints, although, of course, 
 they lie within the uncertainty of the solutions of Roy equations alone (area between dashed lines). 

\begin{figure}
  \centering
 \includegraphics[width=0.8\textwidth]{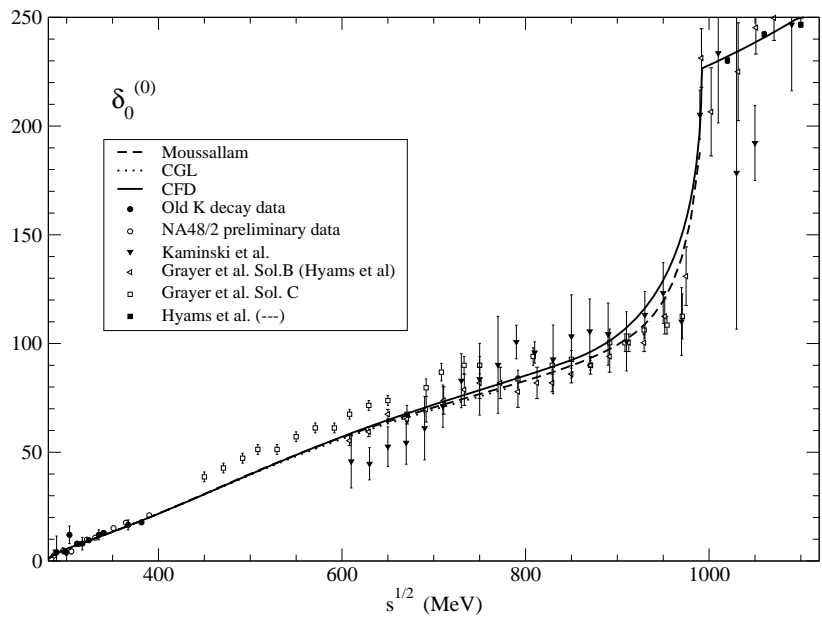}
  \caption{$\pi\pi$ scattering S0-wave phase shift. 
We show the central value of the Roy equation solution 
with chiral symmetry constraints in \cite{CGL} (CGL) and \cite{Moussallam:2011zg} (Moussallam)
versus the central values of the Constrained Fits to Data in 
\cite{GarciaMartin:2011cn} (CFD). Note
the fairly good agreement of these approaches.}
\label{fig:S0wavecomparison}
\end{figure}

These solutions were revisited first in \cite{DescotesGenon:2001tn} 
in view of the almost simultaneous  $K_{e4}$ decay results of the E865 collaboration 
and a general agreement with the solutions in \cite{CGL}
was found, with discrepancies never much beyond the one standard deviation level. Note that those $K_{e4}$ results were practically superseded by the recent NA48/2
data and by the need to include a sizable isospin correction as explained in Sec.\ref{subsec:pwbelow}. \color{black}
After some debate \cite{Pelaez:2003eh,Caprini:2003ta}, the Roy equation solutions 
in \cite{CGL} were also shown to be compatible
with the CFD fits of the Madrid-Krakow group \cite{GarciaMartin:2011cn}
described in the previous section, 
although only after the final NA48/2 data, corrected from isospin, 
were included in the latter analysis
and the value of $\delta_0^{(0)}(m_K)-\delta_0^{(2)}(m_K)$ 
 was removed from the CFD fits (see \cite{Pelaez:2003eh,Caprini:2003ta,Aloisio:2002bs,Colangelo:2007df,Cirigliano:2009rr} for details).
\color{black}
This fairly reasonable agreement between the CFD parameterization and the \cite{CGL,ACGL} 
predictions can be seen in  Fig.\ref{fig:S0wavecomparison}.


In addition, in \cite{Moussallam:2011zg} the S0-wave 
solution of Roy equations in \cite{ACGL} has been extended up
to the $K \bar K$ threshold, where the new matching point is located.
The P and S2-waves are taken as input from the results in \cite{ACGL}
but including the latest S-wave scattering lengths provided by NA48/2.
Note that setting the matching point  at the $K\bar K$ threshold\footnote{as usual, isospin breaking is neglected and $m_K$ is the average of the charged and neutral masses.} does not imply a unique solution. Thus,  for the same input  as in \cite{ACGL}, 
including the phase at 800 MeV,
another condition has to be imposed on the derivative of the phase shift at the two-kaon threshold. Roy equations are solved below 800 MeV, but between this energy and 
$4m_K^2$, data on the phase shift is fitted. 
For this, a small modification of the S0-wave Schenk parameterization in Eq.\ref{eq:Schenk} is used to ensure the derivative condition at $K \bar K$ threshold. Below 800 MeV the results 
are very similar to those in \cite{ACGL} and \cite{GarciaMartin:2011cn}, as can also be seen in Fig.\ref{fig:S0wavecomparison}.
From 800 MeV up to the two-kaon threshold, the difference between the UFD or CFD parameterizations and this Roy equation solution is somewhat larger, as can be seen in Fig.\ref{fig:CGL}, but not dramatic. As we will see in the next subsection, this will be in part the cause of the relatively small different values of the pole positions obtained when using the constrained fits or the Roy equation solutions.

\subsection{Precise determination of the $f_0(500)$ pole  from dispersion relations}
\label{subsec:precisepoles}

Once we have seen that there are accurate descriptions of 
$\pi\pi$ scattering amplitudes which are consistent with data and 
dispersion theory, we are in a position to review the 
dispersive extraction of the $\sigma$ pole.
The very same dispersion relations that have been imposed or solved 
to ensure the consistency of the data description in the real axis 
can now be used to calculate the value of the amplitude at any point on 
the {\it first Riemann sheet} of the complex $s$ plane.

However, since we are interested in resonances and their associated poles 
we need the value of the amplitude on the second Riemann sheet. 
Fortunately, 
we have already seen in Sec.\ref{subsec:poles} that {\it in the elastic regime}
the $S$ matrix partial wave 
on the second Riemann sheet is the inverse of the $S$ matrix partial wave
on the first, so that we are actually looking for zeros of the $S$ matrix on the first Riemann sheet, namely $S_J^{(I)}(s)=1-2 i \sigma(s) t_J^{I}(s)=0$.
Note that this relation does not hold for the forward $S$ matrix, which in
order to be inverted would require knowledge of other values of $t\neq0$.
Hence, one can continue it to the first Riemann sheet, but its values on the second Riemann sheet require more information that just $t=0$ amplitudes. 
\color{black} Moreover, even if we were able
to continue the Forward Dispersion Relations to the second sheet and find poles,
we would not be able to determine the spin of the associated resonance (at most we could determine its isospin 
depending on which FDRs the pole appears, and tell if the spin is even or odd).
Therefore, when looking for poles and resonances one uses
partial wave dispersion relations, in which we have seen that the partial wave unitarity condition allows for a
straightforward continuation to the second sheet and the spin of the resonance is completely determined.
\color{black}
 Of course, once these
equations had been implemented in the physical region, 
the analytic extension is straightforward, or even easier, since the principal values
which were needed on the real axis are no longer present. 
In addition, the residue of the pole in the second Riemann sheet can be easily calculated and
 related, via Eq.\ref{ec:defresidue}, to the coupling of the resonance to two pions $g_{\sigma\pi\pi}$.

At this point it is important to emphasize that making the 
analytic extension 
to the complex plane by means of a partial wave dispersion relation 
is {\it model independent} because in that way a  \color{black} {\it continuation to the complex plane by means of a particular functional form or model} is avoided. 
The only relevant issue is to have a data description 
which is consistent with dispersion relations in the physical region and then the analytic continuation 
to complex values of $s$ is performed with the dispersive integral, whose only input lies on the real axis. \color{black} 
Usually one uses 
physically motivated functional forms,
or simple polynomials at different energies which are carefully 
matched onto each other at different physical regions, but a spline or a 
curve drawn by hand would equally do as long as it satisfied dispersion 
relations and described the data. Note that 
these functional forms may not have an analytic continuation to the complex 
plane if they are made by matching pieces, or, if made from models,
these may have different analytic structures depending on what resonances or 
poles one starts from. Therefore, when using models or particular functional forms instead of dispersion relations, the results for the poles
can be very model dependent. Of course, if a resonance is relatively narrow and well isolated, simple parameterizations can provide a good approximation, but the $\sigma$
is extremely wide and thus the analytic continuation has to be made very carefully. In particular, as discussed in Sec.\ref{subsubsec:ancross}, it is important to have the left cut under control to
claim precision.
Certainly there are many models that make very reasonable 
analytic extensions and approximations, many of which we will review in the following sections, but unfortunately other models do not and get plain wrong 
poles and artifacts.
It is therefore advisable not to rely blindly on any model 
or specific parameterization to perform the analytic continuation in search for the $\sigma$ pole.
One has to check the analytic properties first.
Doing otherwise has been one of the main sources
of confusion when dealing with the existence and the parameters 
of the $\sigma$ meson
and that is why the appearance of dispersive model-independent approaches has triggered the major revision of the $\sigma$ pole in the 2012 RPP edition.

Hence, we have collected in Table~\ref{tab:dispersivepoles}
the poles and couplings that have been obtained by using Roy or GKPY equations to 
perform the analytic extension to the complex plane.
On the one hand, the poles in the first two rows are obtained 
from solutions \cite{Caprini:2005zr,Moussallam:2011zg} 
of Roy equations and they basically share the same input, 
except that the solutions of  \cite{Caprini:2005zr} made use of theoretical 
ChPT constraints 
at low energy, whereas the most recent one \cite{Moussallam:2011zg} 
includes the NA48/2 data and fits data from 800 MeV to the two-kaon threshold instead.
Also, in \cite{Caprini:2005zr}, the input $\delta_0^{(0)}=(82.3^{+10}_{-4})^{\degree}$ at 800 MeV was taken conservatively
in order to include more data sets, whereas in \cite{Moussallam:2011zg}
those additional sets were ignored and $\delta_0^{(0)}=82.3^{\degree}\pm3.4^{\degree}$ was assumed as input, as done in \cite{ACGL}. This explains the smaller and much less asymmetric  uncertainties in
\cite{Moussallam:2011zg} compared to those in 
\cite{Caprini:2005zr}. 
On the other hand, the last two rows in Table~\ref{tab:dispersivepoles}
are not obtained from solutions but from data fits constrained 
to Roy and  GKPY equations as well as Forward Dispersion Relations.
 Then either Roy or GKPY equations are used for the analytic continuation. Note that these two pole extractions are very consistent, 
although the one from GKPY equations is more accurate, since the 
subtraction term uncertainties do not grow with energy as in Roy equations.
Despite differing in the approach and the input above $K \bar K$ threshold and other waves,
the results in all four rows are quite consistent.
The same fair agreement is found for the coupling to two pions.

\begin{table}
\centering
\begin{tabular}{|c|c|c|c|}\hline
&Eqs.&$\sqrt{s_\sigma}$ (MeV)&$\vert g_{\sigma\pi\pi} \vert$\\\hline
\rule[-2mm]{0mm}{6mm}Caprini, Colangelo, Leutwyler \cite{Caprini:2005zr,Leutwyler:2008xd} & Roy& $441^{+16}_{-8}-i (272^{+9}_{-12.5})$&$
3.31\color{black}^{+0.17}_{-0.08}\color{black}$\\ \hline
\rule[-2mm]{0mm}{6mm}Moussallam. \cite{Moussallam:2011zg} & Roy& $442^{+5}_{-8}-i (274^{+6}_{-5})$&-\\ \hline
\rule[-2mm]{0mm}{6mm} Garc\'{\i}a-Mart\'{\i}n, Kaminski,  & Roy& $(445\pm25)-i(278^{+22}_{-18})$&$3.4\pm0.5$ GeV\\
\rule[-2mm]{0mm}{6mm} Pel\'aez, Ruiz de Elvira \cite{GarciaMartin:2011jx} & GKPY& $(457^{+14}_{-13})-i(279^{+11}_{-7})$& $3.59^{+0.11}_{-0.13}$ GeV\\
\hline
\end{tabular}\caption{Poles and residues of the $f_0(500)$ resonance obtained from the analytic continuation
to the complex plane using dispersive methods. The input in the integrals only 
comes from  the physical region in the real axis, where it is  consistent with data and dispersion relations.
The two first rows use as input solutions of Roy equations: either below 800 MeV \cite{Caprini:2005zr} or solutions below 800
and constrained fits up to $2M_K$ \cite{Moussallam:2011zg}. Since they basically share all other input, they are almost identical.
The last two rows are obtained \cite{GarciaMartin:2011jx} from the analytic continuation using Roy or GKPY equations of an input which
is not a solution but a fit to data constrained by Roy and GKPY equations up to 1.1 GeV and Forward Dispersion Relations up to 1.42 GeV.
}
\label{tab:dispersivepoles}
\vspace*{-.3cm}
\end{table}

Therefore, in order to be  conservative and include the differences between the 
results in Table~\ref{tab:dispersivepoles} as systematic uncertainty, 
the following band was suggested in Sec.\ref{sec:intro}
\begin{equation}
\sqrt{s_\sigma}=449^{+22}_{-16}-i(275\pm12)\, {\rm MeV},
\qquad \vert g_{\sigma\pi\pi} \vert=\color{black} 3.45^{+0.25}_{-0.22}\,\color{black} {\rm GeV}
\label{myfullestimate}
\end{equation}
as a ``conservative dispersive estimate''. This has been obtained by combining the first and fourth rows of Table~\ref{tab:dispersivepoles}.
This estimate covers completely the result in the second row. 
The third row is less relevant since it is just a less accurate check 
of the fourth row, performed by the same authors \cite{GarciaMartin:2011jx}. 
 Following a similar procedure to combine uncertainties, a conservative dispersive estimate for the coupling is also provided in Eq.\ref{myfullestimate}.
The RPP 2012 edition includes the result of \cite{CGL} as one of the ``most advanced dispersive results", 
but as the very authors explain in \cite{CGL} their calculation  
is obtained from the specific parameterization, whose functional form was 
given here in Eq.\ref{eq:Schenk}, not from the dispersive formalism.
Some of these authors with another collaborator performed the corresponding dispersive pole extraction
in \cite{Caprini:2005zr}, thus superseding that of \cite{CGL},
which is nevertheless listed in the third row of Table \ref{tab:otherpoles}.

At this point it is important to remark that the effect of removing completely the contribution of the left-hand cut was also studied in \cite{Caprini:2005zr} leading to a pole  at 
$\sqrt{s_\sigma}=500-i 260\,$MeV. So the whole left hand cut contribution 
decreases the $f_0(500)$ mass by $\delta M\sim 60\,\mev$ but increases the width by $\delta\Gamma=24\,\mev$.

Of course, good approximations to these poles were already obtained
before the full treatment with Roy or GKPY equations became available. 
As a matter of fact, as long as the model or parameterization does not have artifacts, is fitted to a reasonable set of data 
and has some minimal analyticity requirements, a $\sigma$ pole is always found 
at a fairly reasonable place around 400 to 550 MeV in mass 
with a large width. Some models even have an approximation to the left cut and then the resulting pole
is very close to the dispersive result.

\begin{table}
  \centering
  \begin{tabular}{|c|c|c|}
\hline
 & $\sqrt{s_\sigma}$ (MeV) & $\vert g_{\sigma\pi\pi}\vert$ (GeV) \\ \hline 
\rule[-1.5mm]{0mm}{5mm}Dobado, Pel\'aez \cite{Dobado:1996ps} & $440-i\,245$ & - \\ \hline 
\rule[-1.5mm]{0mm}{5mm}Oller, Oset \cite{Oller:1998zr} & $445-i\,221$ & - \\ \hline
\rule[-1.5mm]{0mm}{5mm}Colangelo, Gasser, Leutwyler \cite{CGL}& $470\pm30-i (295\pm20)$& - \\ \hline
\rule[-1.5mm]{0mm}{5mm}Oller \cite{Oller:2003vf}& $(443\pm2)-i(216\pm4)$ & $2.97\pm0.04 $\\\hline
\rule[-1.5mm]{0mm}{5mm}Zhou {\it et al.}\cite{Zhou:2004ms}& $470\pm50-i (285\pm25)$& -\\ \hline
\rule[-1.5mm]{0mm}{5mm}Garc\'{\i}a-Mart\'{\i}n, Yndur\'ain, Pel\'aez \cite{Yndurain:2007qm}& $474\pm6-i (254\pm4)$& $3.58\pm0.03$\\ \hline
\rule[-1.5mm]{0mm}{5mm}
Caprini \cite{Caprini:2008fc}& $463\pm6^{+31}_{-17}-i (259\pm6^{+33}_{-34})$& -\\ \hline
\rule[-1.5mm]{0mm}{5mm}Mennessier, Narison, Wang \cite{Mennessier:2010xg}& $452\pm13-i (259\pm16)$& $2.64\pm0.10$\\ \hline 
\rule[-1.5mm]{0mm}{5mm}Pel\'aez, R\'{\i}os \cite{Pelaez:2010fj} (fit D) & $453-i\,271$ & 3.5 \\
\hline 
\rule[-1.5mm]{0mm}{5mm}\color{black} Caprini {\it et al.}\color{black} 
 \cite{Masjuan:2014psa} 
&\color{black} $457\pm28-i\,(292\pm29$)\color{black} & $3.8\pm0.4$ \\
\hline
  \end{tabular}
  \caption{Other relatively recent determinations of the $\sigma$ pole 
using some form of analytic properties or data constrained by Roy equations and chiral symmetry \cite{CGL}. The different methods are briefly explained in the text.
\label{tab:otherpoles}}
\vspace*{-.3cm}
\end{table}

For instance, in Table \ref{tab:otherpoles} 
we have collected a sample of poles obtained
from several parameterizations that meet some of the previous requirements.
From top to bottom they are ordered from older to more recent.
One should not be misguided by the tiny uncertainties attached to some of these results,
since   systematic uncertainties associated to the model are rarely evaluated and
 these may come from the approximations on 
the left cut or high energy contributions, the data selection, the analytic continuation, etc... 
Hence, when uncertainties in that table are comparable or even smaller than the uncertainties of the
dispersive determinations, they should be considered just statistical uncertainties mainly
 coming from the fit to the data that has been fitted. These errors are expected to be much amplified by the analytic
extrapolation to a pole situated deep in the complex, since analytic
continuation is an ill-posed (or unstable) problem in the Hadamard
sense \cite{Masjuan:2014psa,Hadamard}.
 
With this caveat about uncertainties in mind, let us comment on the different entries of Table \ref{tab:otherpoles}.
The result in the first row was obtained from a dispersion relation for the 
inverse amplitude in the elastic approximation \cite{Dobado:1996ps}, 
which ensures elastic unitarity and
where the left cut and the subtraction constants are approximated within NLO ChPT.
In the second row the $N/D$ method was used. This method consists of one dispersion relation for 
the numerator of the amplitude, containing an approximated left cut,
and another dispersion relation for the denominator, containing the
 physical cut where unitarity is imposed exactly \cite{Oller:1998zr}.
The result in the third row is obtained, as just commented above,
 from the phenomenological parameterization of the solutions of Roy equations in \cite{CGL}.
The fourth row is a reanalysis \cite{Oller:2003vf} with the methods of \cite{Oller:1998zr}
where the coupling is also provided. The fifth row corresponds to the analysis using  a partial-wave dispersion relation plus chiral and crossing constraints, where the left cut is approximated by ChPT to two loops and a cut-off as an additional parameter.
Both the sixth \cite{Yndurain:2007qm} and seventh \cite{Caprini:2008fc} rows use conformal mappings, which
maximize the analyticity domain of the phenomenological parameterization, to fit the recent $K_{\ell4}$ and other data.
The eighth row \cite{Mennessier:2010xg} corresponds to an analytic K-matrix model
with a form factor shape.
The ninth row is again the Inverse Amplitude Method but to two loops \cite{Pelaez:2010fj}
with additional constraints on the pion mass dependence from lattice.
In the last row we show the result of
 an analytic method to extract the pole without using the integral representation,
which uses Pad\'e approximants and the CFD parameterizations of \cite{GarciaMartin:2011cn} as input.
As seen in Table~\ref{tab:otherpoles} 
all these approaches and models provide a fairly consistent 
description of the $f_0(500)$.

The interest of many of these models goes beyond the simple determination of the $f_0(500)$ parameters and
may provide an understanding of the dynamics that generates the $\sigma$ meson, 
how it relates to other resonances and QCD, and even how an $f_0(500)$
description in other processes can be related to its properties in
$\pi\pi$  scattering. In particular, 
the poles obtained using ChPT constraints will be treated in more detail in the next section.

\section{CHIRAL SYMMETRY AND THE $f_0(500)$}
\label{sec:chiralsigma}

As we have shown in previous sections, the  existence of a light and very broad $\sigma$ meson has been 
firmly established for almost two decades and its parameters have been determined with relatively good precision by means of dispersive analyses and data within the last decade. In this section the connection of this resonance with chiral symmetry and QCD will be reviewed. In particular me will make extensive use of the modern techniques of effective Lagrangians.
Recommended textbooks on this modern perspective are 
\cite{Donoghuebook,Weinbergbook,ourbook,Schererbook}.

\subsection{Chiral Symmetry}
\label{subsec:chiralsymmetry}

Since the physics we are interested in occurs below 1 or 1.5 GeV at most, 
we can  restrict the QCD Lagrangian to the lightest $N_f=3$ quark flavors,
 $q_j=u,d$ and $s$. For brevity we have omitted the quark color indices by gathering the $N_c$ quark fields with different colors into a vector $q_j$. 
Then the QCD Lagrangian can be written as 
\begin{eqnarray}\label{QCDLagrangian}
\mathcal{L}_{QCD}&=&\sum_{j=1}^{N_f}\bar q_j(x)\left(i\dslash-m_j\right)q_j(x)-\frac{1}{4}\sum_{a=1}^{N_c^2-1}G^a_{\mu\nu}(x)G_a^{\mu\nu}(x)\\
G_{\mu\nu}^a(x)&=&\partial_\mu A_\nu^a-\partial_\nu A^a_\mu+gf^{a}_{bc}A^b_\mu A^c_\nu,\qquad
\dslash_\mu=(\partial_\mu-ig t_a A^a_\mu/2)\gamma^\mu,\nonumber
\end{eqnarray}
where $A^a_\mu$, with $a=1...(N_c^2-1)$ are the gluon fields, $\gamma^\mu$ the Dirac matrices, $g$ is the strong coupling constant, $f^a_{bc}$ are the structure constants of the $SU(N_c)$ group and $T^a$ are the $SU(N_c)$ 
generators. We have kept $N_c$ arbitrary, because the study of the $N_c$ dependence will be very relevant in Sec.\ref{sec:nature}. However, in real life
$N_c$=3 and  $T^a=\lambda_a$, i.e. the $SU(3)$ Gell-Mann matrices. 

In practice, the so-called ``current-quark masses'' \cite{PDG12} $m_u=2.3^{+0.7}_{-0.5}\, \mev$, 
$m_d=4.8^{+0.7}_{-0.3}\, \mev$, $m_s=95\pm5\, \mev$  that appear in this Lagrangian\footnote{Actually these are renormalized masses within the
$\overline{MS}$ renormalization scheme at a scale $\mu= 2\,\gev$. }
are much smaller than typical hadronic scales of order $\sim 1\,\gev$.
Thus it makes sense to consider first the ``chiral limit'' of vanishing quark masses for the $N_f=2$ or $N_f=3$ lightest quarks,
in which the Lagrangian becomes invariant under the  $SU(N_f)_L\times SU(N_f)_R$ group of
transformations $L$ and $R$ given by:
\begin{equation}\label{chiral-transformations}
q_{L,R}\longrightarrow U_{L,R}P_{L,R} q=\exp{\left( -i\theta^{L,R}_{j} \frac{T^j}{2}\right)}q_{L,R},
\end{equation}
where now $T^j$ are the generators of the $SU(N_f)$ group, 
namely $T^j=\lambda^j$ for $N_f=3$
and $T^j=\tau^j$, the Pauli matrices, for $N_f=2$. In addition, $q_{L,R}$ stand for the quark field projections:
\begin{equation}
q_{L,R}=P_{L,R} \,q=\left(\frac{1\mp\gamma_5}{2}\right)q. 
\end{equation}
While parity transforms these projections into each other, this cannot be achieved with a rotation, thus
it seems as if they had some ``handedness''. That is the reason why they are known as the 
left and right components of the quark field
and the symmetry is called a ``chiral symmetry'' \footnote{From the Greek ``$\chi\epsilon\iota\rho$" (kheir), for ``hand''.}.
Classically, this would imply the conservation of the charges associated to the following vector and axial currents:
\begin{equation}
V_\mu^j=\bar q \gamma_\mu T^j q, \quad A^j_\mu=\bar q \gamma_\mu \gamma_5 T^j q.
\label{chiralcurrents}
\end{equation}
Note that the set of ``vector" transformations corresponding
to $\theta_a\equiv\theta_a^L=\theta_a^R$ 
is a subgroup called $SU(3)_{L+R}$ or
 $SU(3)_{V}$. Axial transformations
correspond to $\theta_a=\theta_a^L=-\theta_a^R$ and despite not forming a subgroup, the notation is frequently abused and their set is  
called $SU(3)_A$.

Thus, if $N_f$ light quarks were massless, hadrons would 
appear in multiplets of $SU(N_f)_L\times SU(N_f)_R$. Of course, 
quarks are not exactly massless and small perturbations within this classification
are expected;
of the order of a few MeV for $u$ and $d$ quarks and of the order of a hundred for the strange one. 
 However, only multiplets of $SU(3)_V$ are actually observed. For instance, 
there is a clear $J^{P}=1^-$ nonet of vector mesons
 formed by the three $\rho(770)$, the four $K^*(892)$, the $\omega(770)$ and
the $\phi(1020)$. Their mass difference can be easily explained by the strange quark
in the $K^*(892)$ or the two strange quarks in the $\phi(1020)$. But the
closest $J^{P}=1^+$ axial-vector to the $\rho(770)$ is the $a_1(1260)$, almost 500 MeV heavier. 
These two particles do not have strangeness and their mass difference cannot be explained by the tiny masses of the $u$ and $d$ quarks. 
Given the fact that there are no other parameters that break 
the $SU(3)_L\times SU(3)_R$ chiral symmetry in the QCD Lagrangian, 
this and other similar examples imply that it 
 must be spontaneously broken down to $SU(3)_V$. 
This means that axial charges do not annihilate the vacuum. That is, there are states $\vert \phi^k(p_\mu)\rangle$
such that:
\begin{equation}
\langle0\vert A_\mu^j\vert \phi^k(p_\mu)\rangle=i f_j p_\mu \delta^{jk}\neq0, 
\label{prePCAC}
\end{equation}
Therefore, the vacuum is not invariant under chiral symmetry transformations 
and the main measure \cite{Colangelo:2001sp,CGL} of this spontaneously broken
invariance is the non-vanishing chiral 
condensate $\langle 0 | \bar qq | 0\rangle$.
In the chiral limit, Goldstone's theorem then implies the appearance of an
octet of massless Nambu-Goldstone bosons (NGB)---one per broken symmetry generator. 
Due to the small quark masses these NGB acquire a small mass
and are called pseudo-NGB, but the relevant fact is that they
are still separated from the other mesons by a large gap. 
As a matter of fact, for $N_f=2$ these pseudo-NGB are nothing but the 
three pions, whereas for $N_f=3$ they also comprise the four kaons and the eta, which form an octet several hundreds of MeV
lighter than any other octet of mesons. Thus, Eq.\ref{prePCAC} implies that the $f_j$ are to be identified with the 
pion, kaon, and eta decay constants.
In addition, since the pseudo-NGB are massive, Eq.\ref{prePCAC} above implies:
\begin{eqnarray}
\langle0\vert \partial^\mu A_\mu^j\vert \phi^k(p_\mu)\rangle= f_j m_{\phi^j}^2 \delta^{jk},\label{PCAC}
\end{eqnarray}
which is known as the Partially Conserved Axial Current relation (PCAC), since this current 
is conserved except
for the the explicit chiral symmetry breaking induced by the small quark masses.

Note also that if one restricts the Lagrangian to the two lightest quarks, the same pattern is reproduced, 
this time with $SU(2)$  instead of $SU(3)$
groups. The $SU(2)_V$ subgroup is nothing but isospin symmetry.

The existence of NGBs and a mass gap
has important consequences on the dynamics of mesons at low energies,
particularly for the $f_0(500)$, that will be reviewed throughout this section.
The rigorous and systematic way to implement these chiral constraints is through the low energy effective theory of QCD, which is known as
 Chiral Perturbation Theory (ChPT). Unfortunately, ChPT is an energy expansion and cannot describe resonances and their associated poles, but just their low energy effects. However ChPT can be combined with dispersion theory, giving rise to an approach generically known as unitarization, which is able to describe resonances while incorporating chiral constraints at low energy. All these issues will be the subject of the next subsections. However, we will start describing a simple and widely used
 model that captures many, {\it but not all} of the main low energy features of chiral symmetry. 
It will be used to introduce some basic concepts and notation, but the 
issues in which it deviates from QCD and the experimental observations will be pointed out too.

\subsection{The Linear Sigma Model}
\label{subsec:lsm}

A very simple and popular model to deal with the $\sigma$ and chiral symmetry breaking is the Linear Sigma Model (L$\sigma$M), proposed in 1960 by Gell-Mann and Levy \cite{GellMann:1960np}. 
 Nucleons will be omitted since they are much heavier than 
pions and the $\sigma$ meson and therefore not relevant for our purposes. We will also neglect quark masses 
and their explicit quiral symmetry breaking 
for the moment and introduce them later. Thus, if we group the pions and the sigma meson into 
$\phi^A=(\sigma,\pi^a)$ with $A=0,1,2,3$ and $a=1,2,3$, then the Lagrangian reads:
\begin{eqnarray}
{\cal L}_{L\sigma M}&=&\frac{1}{2} \partial_\mu \phi^A \partial^\mu \phi^A- V(\phi)=\frac{1}{2} \partial_\mu \phi^A \partial^\mu \phi^A+\frac{\mu^2}{2} \phi^A\phi^A-\frac{\lambda}{4}(\phi^A\phi^A)^2\\
&=&\frac{1}{2} \partial_\mu \sigma \partial^\mu \sigma+\frac{1}{2} \partial_\mu \pi^a\partial^\mu \pi^a
+\frac{\mu^2}{2} (\sigma^2+\pi^a\pi^a)-\frac{\lambda}{4}(\sigma^2+\pi^a\pi^a)^2,
\end{eqnarray}
where $\lambda>0$ and summation over repeated indices is assumed. One of the most interesting features of this model is that it is renormalizable.
The $\phi^A$ notation is useful to make explicit
the $O(4)$ invariance of this Lagrangian, i.e., rotations of the $\phi^A$ field. The fact that $O(4)$ rotations 
are {\it linear} transformations among $\pi^a$ and $\sigma$ 
fields, is why the L$\sigma$M is called ``linear''.
In addition, the $\phi^A$ notation makes it easy to  identify
the  $\mu^2<0$ case with a $\lambda \phi^4$ theory for $\phi^A$ fields
of mass $\mu^2$. Note that in this case the potential $V(\phi)$ has a unique minimum
at $\sigma=\pi^a=0$. 

However, if $\mu^2>0$ the potential has a degenerate set of minima at $\sigma^2+\pi^a\pi^a=\mu^2/\lambda$. Note that this is the equation of a 4-sphere that has $O(3)$ degeneracy.
If these minima are identified with the vacuum, this implies that one combination of the $\sigma$ and $\pi^a$ fields has a non-vanishing vacuum expectation value.
Without loss of generality we can choose 
$\langle \sigma \rangle=\sqrt{\mu^2/\lambda}\equiv v$, 
since other combinations describe the same physics and can be 
brought to this case by renaming the fields. Then $\langle \pi^a\rangle=0$.
By defining the shifted field $\tilde\sigma=\sigma-v$, the Lagrangian can be rewritten as:
\begin{equation}
  \label{eq:lsmshifted}
  {\cal L}=\frac{1}{2}\partial_\mu \tilde\sigma \partial^\mu \tilde\sigma +\frac{1}{2} \partial_\mu \pi^a\partial^\mu \pi^a-\frac{1}{2}(2\mu^2)\tilde\sigma^2-\lambda v \tilde\sigma(\tilde\sigma^2+\pi^a\pi^a)-\frac{\lambda}{4}(\tilde\sigma^2+\pi^a\pi^a)^2+\frac{\lambda v^4 }{4}.
\end{equation}
Note that the $\tilde \sigma$ field 
has acquired a mass $M^2=2\mu^2=2 \lambda v^2$ whereas 
the three $\pi^a$ remain massless.
Let us also remark that in order to increase the $\tilde\sigma$ mass
while keeping the vacuum expectation $\nu$ constant, the $\lambda$ coupling should be
made larger. 
This new Lagrangian no longer exhibits explicitly 
the $O(4)$ symmetry, but just an $O(3)$ symmetry, i.e., only $\pi^a$ field rotations.
That is, there is an $O(4)\rightarrow O(3)$ spontaneous symmetry breaking. 

The above symmetry breaking 
pattern is isomorphic to $SU(2)_L\times SU(2)_R\rightarrow SU(2)_{L+R}$. 
To see this it is enough to recast our fields into a $2\times2$ matrix
$\Sigma=\sigma \mathbb{I}+i \tau^a \pi^a$, where $\tau^a$ are the Pauli matrices.
Then, the massless L$\sigma$M Lagrangian becomes
\begin{equation}
  \label{eq:LSMwithSigma}
  {\cal L}_{L\sigma M}=\frac{1}{4}Tr(\partial_\mu \Sigma^\dagger \partial^\mu \Sigma)
+\frac{\mu^2}{4}Tr(\Sigma^\dagger \Sigma)-\frac{\lambda}{16}[Tr(\Sigma^\dagger \Sigma)]^2,
\end{equation}
where in the first line it is now apparent that the Lagrangian is invariant under 
$SU(2)_L\times SU(2)_R$
transformations $\Sigma\rightarrow L\Sigma R^\dagger$, with $L$ and $R$ unitary
$SU(2)_L$ or $SU(2)_R$ matrices, respectively. 
Note that these are still linear transformations.
Once again the minimum of the
potential is not unique. It requires $Tr(\Sigma\Sigma^\dagger)=2 v^2$ 
and, as before, the $\Sigma$ does not vanish in vacuum, i.e.
 there is a spontaneous symmetry breaking.
Now, defining $\tilde \Sigma\equiv \Sigma -v \mathbb{I}=\tilde\sigma\mathbb{I}+i\tau^a\pi^a$, 
the vacuum condition reads $Tr(\tilde \Sigma\tilde \Sigma^ \dagger)=0$.
But note that if $L=R$ then $\tilde\Sigma\rightarrow L\tilde \Sigma L^\dagger$, 
so that the vacuum is invariant under $L=R$ transformations, 
which form the subgroup $SU(2)_{L+R}$. 
In other words, as already advanced, the $O(4)\rightarrow O(3)$ spontaneous symmetry breaking pattern
can be recast as $SU(2)_L\times SU(2)_R\rightarrow SU(2)_{L+R}$.

In addition, since $\Sigma\Sigma^\dagger=(\sigma^2+\pi^a\pi^a)\mathbb{I}$,
we can write $\Sigma(x)= S(x) U(x)$, where $U(x)$ is a unitary matrix
and $S(x)$ a real field such that $S^2(x)=v^2$ in vacuum.
As before, we can now redefine the field 
separating its vacuum expectation value as $S(x)=v+\hat\sigma(x)$. Then
we can rewrite the L$\sigma$M in terms of $\hat\sigma(x)$ and $U(x)$ as follows:
\begin{equation}
  \label{eq:LSMwithSandU}
  {\cal L}_{L\sigma M}=
\frac{1}{2}\partial_\mu \hat\sigma \partial ^\mu \hat\sigma-\frac{1}{2}(2\mu^2)\hat\sigma^2-\lambda v \hat\sigma^3
-\frac{\lambda}{4} \hat\sigma^4
+\frac{(v+\hat\sigma)^2}{4}Tr(\partial_\mu U^\dagger \partial^\mu U),
\end{equation}
where some irrelevant constant terms have been dropped. Once more the scalar field $\hat\sigma$ has a mass squared of $2\mu^2$.
Being unitary, the $U(x)$ matrix can be parameterized as 
$U(x)=\exp(i \tau^a \hat\pi^a/v)$. By reexpanding $U(x)$ in powers of $1/v$ one can check that the $\hat\pi^a$ remain massless and thus
correspond to the 3 NGB associated to the 3 spontaneously broken symmetry generators.
Note that the relations between the $\hat\sigma, \hat\pi^a$ and 
$\tilde \sigma, \pi^a$ fields
are not linear.  Moreover, $L$ and $R$ acting on the $\hat\pi^a$ fields are no longer linear 
transformations, which is why the $\hat \pi^a$ fields are called the non-linear realization of chiral symmetry. 

Nevertheless, the relation between $\hat\sigma, \hat\pi^a$ and 
$\tilde \sigma, \pi^a$ fields, when expanded 
in powers of $1/v$, reads $\hat\sigma=\tilde \sigma+...$ and 
$\hat\pi^a=\pi^a+...$. Under these conditions there are theorems  ensuring that
the same observables result from both Lagrangians \cite{Haag}.
For instance, we have already seen 
that $M_{\hat\sigma}=M_{\tilde\sigma}$ and this 
holds to all orders for the physical mass of the scalar-isoscalar field in the model. Thus, for simplicity
 we will just use the generic name $M_\sigma$.
For other parameterizations also used in the literature we refer to \cite{Donoghuebook,ourbook}.

As we commented in the previous section, in QCD these NGB are identified with pions  and then 
$v=f_0$, the pion decay constant in the chiral limit. In real life $f_0\simeq f_\pi=92.3$~MeV.
Of course, the L$\sigma$M would be at most an effective theory of QCD since not all 
hadrons are present. 
In particular, within $SU(2)$ there are only four explicit degrees of freedom: a scalar isoscalar singlet $\tilde \sigma$ and
a pseudoscalar-isotriplet, which are the pions. 
The full $SU(2)_L\times SU(2)_R$ chiral symmetry of QCD would require the existence of degenerate multiplets with opposite parity. However, since this full symmetry is spontaneously broken
these multiplets are no longer degenerate but occur with larger masses.
For the $SU(3)$ case there is a scalar singlet and a pseudoscalar octet of NGB. Once more
the companion multiplets with opposite parity would be much heavier due to the spontaneous symmetry breaking. 
Therefore, the L$\sigma$M could be considered at most as an effective theory 
valid for energies well
below the mass of these heavier opposite-parity multiplets, or any other resonance like the $\rho(770)$. Furthermore, we will see that with the present precision
the L$\sigma$M does not reproduce the observed low energy behavior beyond the leading order
in a momentum expansion. 

An interesting feature of the L$\sigma$M Lagrangian written as in Eq.\ref{eq:LSMwithSandU}
is that the $\lambda$ coupling has disappeared from the term containing pions.
We already saw that the large $\hat\sigma$ mass limit implied 
a large coupling $\lambda$ and strong interactions for the $\sigma$ field.
However, the pion self couplings,
which only see the $v$ constant, do not become strong in this heavy $\sigma$ limit. Moreover, since 
$\hat\pi^a$ always appear in a term with derivatives, their interactions at low energies 
are always weak, indeed vanishing at zero momentum (for the massless case).
Furthermore, the $M_\sigma\rightarrow\infty$ limit implies that  $\hat\sigma\rightarrow 0$, and the only term that survives in Eq.\ref{eq:LSMwithSandU}
is 
\begin{equation}
  \label{eq:LSMheavylimit}
  {\cal L}_{M_\sigma\rightarrow 0}=\frac{v^2}{4}Tr(\partial_\mu U^\dagger \partial^\mu U).
\end{equation}
This means that in the $M_\sigma\rightarrow\infty$ limit the $\hat\sigma$ decouples completely from the pions
and we are left with a pure theory of NGB.
Note that once the $\hat \sigma$ field has been removed, 
there is no way to recast the $\hat \pi^a$ fields into a linear representation of the chiral group.
Therefore the Lagrangian in this limit and its extensions with other
chirally symmetric Lagrangians written in terms of just
the $U(x)$ field are generically called
{\it non-linear sigma models}.

In practice, if the $\hat\sigma$ field has a finite mass $M_\sigma=2\mu^2$,  one can ``integrate it out'' and
obtain an effective low energy theory of pions in terms of their quadrimomenta or derivatives
 over $M_\sigma$. For a textbook introduction to ``integrating out'' heavy fields within the path integral formalism 
 the reader is referred to  \cite{Donoghuebook,ourbook} whereas detailed calculations with heavy resonances can be found in
 \cite{chpt1,Ecker:1988te}.
Here it suffices to explain that intuitively this ``integrating out" the $\sigma$ amounts to 
expanding in powers of momenta over $M_\sigma$ all the Feynman diagrams where the $\sigma$ is exchanged.
In the $SU(2)$ case, the resulting Lagrangian, which we write
 in the exponential notation \cite{Donoghuebook}, is:
 \begin{equation}
 \label{eq:LSMLO}
 {\cal L}_{L\sigma M}\simeq  \frac{f_0^2}{4}Tr(\partial_\mu U^\dagger \partial^\mu U)+ \frac{f_0^2}{8 M_\sigma^2}[Tr(\partial_\mu U^\dagger \partial^\mu U)]^2+...,
\end{equation}
where the dots stand for higher orders in the derivatives.
The leading order term, corresponding to $M_\sigma\rightarrow\infty$ is common 
to all theories with the same symmetry breaking pattern, but the 
second one is specific of the L$\sigma$M. We will see below that the size of that term for $M_\sigma\simeq 500\,$MeV
and the absence of other terms at that order is incompatible with observations. 
This has been well known for long \cite{chpt1} in the literatur, and well described at the textbook level \cite{Donoghuebook}.
Therefore, the L$\sigma$M provides only a first approximation to low energy QCD, which is not
sufficient given the present level of precision. Nevertheless it is still interesting
because it can provide some qualitative understanding, correct to leading order,
 of hadron physics at very low energies.
Of course, with additional extensions the agreement can be improved
and many modifications exist in the literature, which will be reviewed in Subsec.\ref{subsec:other}. 

The matrix parameterizations are interesting for our purposes because 
it is straightforward to generalize the L$\sigma$M to $SU(3)_L\times SU(3)_R\rightarrow SU(3)_{L+R}$
by simply using unitary $3\times3$ matrices. In this case
$U (x)=\exp (i \sqrt{2} \Phi(x))$ where
\begin{equation}
\Phi (x) \equiv
\left(
\begin{array}{ccc}
\frac{1}{\sqrt{2}} \pi^0 + \frac{1}{\sqrt{6}} \eta & \pi^+ & K^+ \\
\pi^- & - \frac{1}{\sqrt{2}} \pi^0 + \frac{1}{\sqrt{6}} \eta & K^0 \\
K^- & \bar{K}^0 &  - \frac{2}{\sqrt{6}} \eta
\end{array}
\right).
\end{equation}
In the $SU(3)$ case there are eight NGB which remain massless corresponding to the 
eight broken generators, which are now identified with the pions, the kaons and the eta.
In principle, one can still keep just one scalar field $S$. However, there are interesting
generalizations by promoting the $S$ field to a unitary matrix $\hat S(x)$, that now represents a full 
scalar nonet. This is a ``generalized'' $SU(3)$ L$\sigma$M \cite{Schechter:1993tc} and will be discussed in Subsec.\ref{subsec:other}. Note that by integrating out all scalar fields in the $\hat S(s)$ matrix it leads again to the same low energy expansion as in Eq.\ref{eq:LSMLO},
 where $M_S$ would be now the mass of the scalar nonet in the $SU(3)$ limit of equal quark masses $\hat m=m_s$.
  Of course, once one starts including other fields, more chirally symmetric terms are possible. However, we will see in 
the next section that the generic terms 
produced in the low energy expansion to NLO by a heavier
scalar nonet respecting the QCD spontaneous symmetry breaking pattern
have been studied in \cite{Ecker:1988te} and they are again inconsistent with getting a 
contribution from a light scalar around 500 MeV. The correct description of the $\sigma$ should 
therefore avoid this kind of NLO unobserved contributions to the low energy expansion. 
Any scalar field contributing to the low energy effective Lagrangian at next to leading order must be much heavier than 500 MeV.
Since as we have seen in previous sections a $\sigma$ around 500 MeV exists,  it must be introduced in a way that 
does not contribute to the low energy expansion to that order. In the following sections 
we will see how this consistent description can be achieved.

So far we have ignored quark masses, which are the only source of explicit chiral symmetry breaking in the QCD Lagrangian. 
As commented in the previous section, 
these are sufficiently small to be treated as a perturbation. This means that the 
$\sigma$ direction
is no longer an arbitrary choice but corresponds to the direction slightly favored by the small
explicit breaking. Conversely, the mass term is in the direction of $\sigma$ and it is then natural to write:
\begin{equation}
{\cal L}_{mass}=c\sigma=
\frac{c}{4}Tr(\Sigma^\dagger+\Sigma)
=\frac{c(v+\tilde \sigma)}{4}Tr(U^\dagger+U).
\end{equation}
By expanding the $U(x)$ matrix, one sees that the pions acquire a common  mass $M_\pi^2=c/f_\pi$, since this is still the isospin-conserving formalism.
In contrast, within the $SU(3)$ formalism one has different masses for the non-strange and strange quarks, which are introduced as follows:
\begin{equation}
{\cal L}_{mass}=
\frac{1}{4}Tr(M_0(\Sigma^\dagger+\Sigma)),
\end{equation}
where $M_0=2 B_0\, diag(\hat m,\hat m, m_s)$. 

Note that, in principle, other choices of symmetry breaking terms are possible and indeed will be present in the general chiral low energy effective theory to be reviewed in the next subsection.
For instance one could include non-linear terms on the quark mass but, since this is a perturbation to the Lagrangian in the chiral limit,
a linear term is supposed to provide a good first approximation.
Moreover, this one
already ensures the fulfillment
of the Gell-Mann-Okubo relation \cite{GMO}: $4 M_{0\,K}^2-M_{0\,\pi}^2-3 M_{0 \eta}^2=0$, which 
is rather well satisfied by the physical masses. 

Up to here, we have been discussing the Lagrangian or tree-level
mass of the $\sigma$ meson. However, its value is corrected at higher orders in 
the coupling constant expansion. In particular, when taking into account interactions the $\sigma$ decays to two pions and acquires a width. This means that the one-loop $\sigma$ propagator has a complex pole. Being a renormalizable model, the position of this pole can be calculated perturbatively to NLO \cite{chpt1,Masjuan:2008cp}. In the chiral limit one finds:
\begin{equation}
  \label{eq:LSMsigmaNLO}
  s_\sigma=
(2\lambda f_\pi^2)\left[ 1+ \frac{3\lambda}{16 \pi^2}\left(-\frac{13}{3}+\pi\sqrt{3}-i\pi \right)+O(\lambda^2)\right],
\end{equation}
where the constant $-13/3+\pi \sqrt{3}$ appears due to the choice of renormalization scheme and therefore $f_\pi$ is the decay constant renormalized to NLO within the same scheme.

If, as usual, we rewrite the pole position as $s_\sigma=(M_\sigma-i\Gamma_\sigma/2)^2$,
this means that up to NLO within the L$\sigma$M the pole position is at:
\begin{equation}
  \label{eq:masswidthLSMNLO}
  M_{\sigma\, NLO}^2=(2\lambda f_\pi^2)\left(1+\frac{3\lambda}{16 \pi^2}(-\frac{13}{3}+\pi\sqrt{3})\right),\quad
\Gamma_{\sigma\, NLO}=\frac{(2\lambda f_\pi^2)}{M_{\sigma\, NLO}}\frac{3 \lambda}{16\pi}.
\end{equation}
Let us do some simple numerics. The observed 
$M_\sigma\simeq 450\, \mev$ in Eq.\ref{myestimate} is obtained by setting $\lambda\simeq 10$, which then yields
 $\Gamma_\sigma\simeq 225$, i.e., about a factor of 2 too small with respect to 
the conservative dispersive estimate in Eq.\ref{myestimate} or the RPP2012 estimate in Eq.\ref{rpp2012sigmapole}.
If on the contrary one imposes $\Gamma_\sigma\simeq 550\, \mev$, then $\lambda\simeq 19$ and $M_\sigma\simeq 670 \,\mev$, 
which is a much higher pole mass than obtained by the precise dispersive determinations or the RPP2012 Eq.\ref{rpp2012sigmapole}. For $\lambda\simeq 15$ one finds $M_\sigma \simeq 400\,\mev$ and $\Gamma_\sigma\simeq 570\, \mev$, which would fall within
 the RPP2012 estimate but outside the conservative dispersive estimate.
One should nevertheless take into account 
that the interaction is becoming somewhat strong, as the NLO correction to
$M_\sigma$ amounts to $\simeq 15\%$ of the total. 
Out of the chiral limit one expects the width to be even narrower 
since there is considerably less phase space to decay into two pions
and therefore one would need an even larger coupling constant to account for the large width, although this would now lead to a
heavier $\sigma$, when one would actually prefer otherwise.
A large $\lambda$ may lead to concerns about the convergence of the theory.
For this reason the $\sigma$ pole has also been
calculated with resummations of L$\sigma$M diagrams and Pad\'e unitarization techniques \cite{Basdevant:1970nu}. This will be commented in Subsec.\ref{subsec:other} below.  

In summary, the NLO description of the pole is not consistent 
with the most precise dispersive determination of the $\sigma$ pole,
although it still  can  be considered a relatively fair
qualitative approximation. 

Before concluding this section, a couple of comments are in order. First  ,
the L$\sigma$M is very relevant in electroweak physics 
because when the subgroup $SU(2)_L\times U(1)_Y$ is gauged
 (the $Y$ subscript referring to the third generator
of $SU(2)_R$) 
 it provides a representation of the electroweak spontaneous symmetry breaking sector 
of the Standard Model. In that case the spontaneous symmetry breaking scale is $v\simeq 250$~GeV
and the NGB are not seen because they become the longitudinal components of the massive $W^\pm, Z^0$ 
electroweak bosons through the Higgs mechanism. The massive $\tilde \sigma$ field is actually the Higgs boson.
For this reason the $\sigma$ meson is sometimes referred to as ``the Higgs boson of QCD''.
This is very misleading, not only because the L$\sigma$M {\it is not} the correct effective theory of QCD at low energies,
but because there is no Higgs mechanism in QCD and the NGB are actually seen in the spectrum as pions. 
Moreover it is easily checked that by rescaling the $\sigma$ mass by $v/f_\pi$ the Higgs  would come out
with a mass higher than 1 TeV and with a large width of a similar order. 
This is at odds with the observation of a $\sim$125 GeV candidate.

Second, an observation in relation to Nambu-Jona-Lasinio or extended Nambu-Jona-Lasinio-like models. 
In principle these are inspired in QCD by integrating out the gluonic degrees of freedom, which
leads to quartic interactions between quarks that generate a mass gap (see the reviews in \cite{NJLreviews}).
This gives quarks a constituent mass of a few hundred MeV and spontaneous symmetry 
breaking of QCD occurs. Therefore NJL 
contain Nambu-Goldstone modes which can be identified with pions. 
However, to make contact with the hadronic world, an effective interaction theory with diagrams 
similar to those in the L$\sigma$M is extracted from resummations, or there is some other kind of
bosonization leading generically to 
a L$\sigma$M. If it is the simplest L$\sigma$M, all the above considerations apply again.
Of course, it can also lead to extended versions of the L$\sigma$M,
 or chiral Lagrangians with just mesonic degrees of freedom, for instance including vector mesons.
 Thus we are led again to effective theories of mesons that will be commented below.

In particular, the L$\sigma$M and any other Lagrangian in terms of mesons which respects chiral symmetry can be recast (see for instance 
\cite{Ecker:1988te} and \cite{NJLreviews}) 
at low energies 
within a general expansion in terms of NGB known as Chiral Perturbation Theory.
This is a model independent and systematic framework to deal with chiral symmetry at low energies, which provides a well defined connection with QCD and that we will briefly review next, since it plays a key role in the present understanding of the $f_0(500)$ meson.

\subsection{Chiral Perturbation Theory}
\label{subsec:chpt}
The existence of  a mass gap between the pseudo-NGB and the rest of hadrons means that, at low energies, pions, kaons and the eta are the 
only relevant degrees of freedom. Therefore the low energy effective Lagrangian of QCD
 can be written in terms of only these fields, as a chiral expansion in their derivatives (i.e., momenta) and masses.  
Due to the symmetries of QCD, including its spontaneous chiral symmetry breaking, only 
 a finite number of independent terms exist at each chiral order.
This systematic approach expansion is known as Chiral Perturbation Theory (ChPT) \cite{Weinberg79,chpt1,chpt2}. 
Apart from the original references, for pedagogical introductions the following books \cite{Donoghuebook,Weinbergbook,ourbook,Schererbook} 
and reviews  \cite{Meissner:1993ah,Pich:1995bw}
are recommended.

The leading order (LO) $SU(3)$ ChPT Lagrangian in the absence of other sources or fields (like photons, weak bosons, etc, which can also be introduced in the Lagrangian but are not relevant for our present discussion) only depends on 
the meson masses and the symmetry breaking scale given by the meson decay constants.  
 Since to this order
all decay constants are the same,  $f_0=f_\pi=f_K=f_\eta$, the LO Lagrangian reads: 
\begin{equation}
{\cal L}_2 = \frac{f_0^2}{4} Tr( \partial_{\mu} U^{\dagger}
\partial^{\mu} U +
M_0 (U + U^{\dagger})),\quad M_0=2 B_0 \,{\rm diag}(\hat m,\hat m, m_s),
\label{ec:Lag2}
\end{equation}
where as usual $f_0$ is the pion decay constant in the $SU(3)$ chiral limit
and $M_0$ is the tree level mass matrix, which 
 is considered a perturbation due to the small quark masses.
As discussed in Subsec.\ref{subsec:lsm}, this mass term implies that the Gell-Mann--Okubo relation
\cite{GMO} is obtained at leading order.
Thus, the chiral expansion is performed in terms of momenta and 
NGB masses, with the following counting: $O(p^2)=O(M_0^2)$.
\color{black}
With this counting for the mass terms, sharp predictions were
obtained in the framework of Roy equations combined with ChPT \cite{CGL}, which were confirmed by the NA48/2 collaboration \cite{Batley:2010zza}, disfavoring  an alternative counting scheme for mass terms developed in \cite{Fuchs:1989jw}.
This was indeed one of the main motivations for the renewed interest on Roy equations in the early 2000's.
\color{black}

Note that the Lagrangian above is universal in the sense that, since it only depends on $f_0$ and 
meson masses, the leading order of the low energy expansion
of any model respecting the QCD symmetries, having the same spontaneous 
chiral breaking scale and yielding the GMO relation at leading order,
will necessarily lead to this very same Lagrangian. 
Actually, we have already seen in Subsec.\ref{subsec:lsm} that the above Lagrangian can be obtained from the L$\sigma$M
in the heavy $\sigma$ limit.

Being universal, the scattering partial waves from this Lagrangian are frequently called Weinberg's low energy theorems
and read:
\begin{equation}
t_0^{(0)}= \frac{2s-M_\pi^2}{32\pi f_\pi^2},\quad
t_1^{(1)}= \frac{s-4 M_\pi^2}{96\pi f_\pi^2},\quad
t_0^{(2)}= \frac{2M_\pi^2-s}{32\pi f_\pi^2}.\quad
\label{ec:LET}
\end{equation}
At this point we may recall that already in the early sixties it was shown that
one consequence of chiral symmetry was the existence of Adler zeros \cite{Adler:1964um} at energies of $O(M_\pi^2)$ in the scalar waves. It is easy to check that 
at this order they actually occur at $s=M_\pi^2/2$ for
 $I=0$  and at $s=2 M_\pi^2$ for $I=2$, both them certainly below the two-pion threshold $4M_\pi^2$. 

In contrast, higher order terms are not universal. Actually,
each higher order Lagrangian term is multiplied by a Low Energy Constant (LEC), whose
value is not fixed by $f_0$ and the masses alone, but by the specific underlying dynamics.
These LECs can be interpreted as interaction vertices or couplings. 
For instance, within SU(3) the NLO Lagrangian, in the absence of external sources, reads:
\begin{eqnarray}\label{L4}
\mathcal{L}_4&=&\!\! L_1Tr \left( \partial^\mu U^\dagger \partial_\mu U\right)^2+L_2Tr  \left( \partial^\mu U^\dagger \partial^\nu U\right)Tr\left( \partial_\mu U^\dagger \partial_\nu U\right)+L_3Tr\left( \partial^\mu U^\dagger \partial_\mu U\partial^\nu U^\dagger \partial_\nu U\right)\\
&+&\!\! L_4Tr\left(\partial^\mu U^\dagger \partial_\mu U\right)Tr\left(M_0 U+ M_0 U^\dagger\right) 
+L_5Tr\left( \partial^\mu U^\dagger \partial_\mu U(M_0 U+U^\dagger M_0)\right) \nonumber\\
&+&\!\! L_6\left[Tr\left(M_0 U +M_0
  U^\dagger\right)\right]^2+L_7\left[Tr\left(M_0 U - M_0
  U^\dagger\right)\right]^2+L_8Tr\left( M_0 U M_0 U + M_0 U^\dagger
  M_0 U^\dagger\right), \nonumber
\end{eqnarray} 
where the equations of motion from the LO Lagrangian have been used to eliminate some other possible terms in favor of these eight alone.
Note that for the SU(3) case the relevant eight NLO LECs are called $L_i$. There would be more LECs if one considered the couplings to other fields,
but these are irrelevant for our purposes in this review.  
The observed values of these LECs are specific of QCD since, in principle, other models would yield different values.

It is important to remark that $L_1,L_2$ and $L_3$ appear in terms with derivatives but no meson masses and therefore survive in the chiral limit, whereas the other constants appear in terms containing meson masses. Actually, the latter are responsible for the renormalization of the masses and decay constants. Eight parameters may seem a lot, but not all LECs are equally important numerically for meson-meson scattering, which generically is driven by the three first and is not very sensitive to the values of $L_4$ and $L_6$. Moreover, the role of $L_7$ is very minor, since, as we will see below is related to 
the exchange of the very heavy $\eta'(960)$ meson. 

There is an $SU(2)$ version of ChPT \cite{chpt1} which has less terms
because the mass matrix is proportional to the identity
and also because some of the above operators are not independent.
For instance, in SU(2) one finds $2Tr(\partial_\mu U\partial^\mu U^\dagger \partial_\nu U\partial^\nu U^\dagger)
=[Tr(\partial_\mu U\partial^\mu U^\dagger)]^2$. Thus the $L_1$ and $L_3$ terms 
in $SU(3)$ combine into a single $2L_1+L_3$ term in $SU(2)$.
In this way, the eight independent terms in the $SU(3)$ NLO ChPT Lagrangian above
become  just four independent terms in $SU(2)$. The associated  LECs, now called $l_i$, 
are related to the $SU(3)$ ones by: 
\begin{eqnarray}
l_1=4L_1+2L_3-\nu_K/24, \qquad l_2=4L_2-\nu_K/12,\label{eq:Ltol}\qquad\qquad\\
l_3=-8L_4-4L_5+16L_6+8L_8-\nu_\eta/18, \qquad l_4=8L_4+4L_5-\nu_K/2.\nonumber 
\end{eqnarray}
The $\nu_P=(\log M_P^2/\mu^2+1)/(32\pi^2)$ terms are due to integrating out the kaons and etas, with $P=K,\eta$ respectively. In this review we will use the $SU(3)$ notation.

Within ChPT, observables can be calculated as in any Quantum Field Theory, using for instance Feynman diagrams.
The relevant observation is that there is a well defined {\it power counting} for each diagram, which gives its energy power as $D=2+\sum_n N_n(n-2)+2N_L$, where $N_n$ is the number of vertices from the Lagrangian terms with $n$ derivatives (or masses) and $N_L$ is the number of loops.
For each observable, infinities appearing in loops from diagrams of a given chiral order are renormalized by combinations of LECs
coming from the ChPT Lagrangian at that order. In this way, although ChPT is not a renormalizable theory, it provides
finite results for all observables order by order in the energy expansion.
It is easy to check that each loop suppresses a diagram by a factor of $p^2/(4 \pi f_0)^2$, or $M_i^2/(4 \pi f_0)^2$, with $M_i=M_\pi,M_K, M_\eta$. 

\begin{figure}
  \centering
 \includegraphics[width=\textwidth]{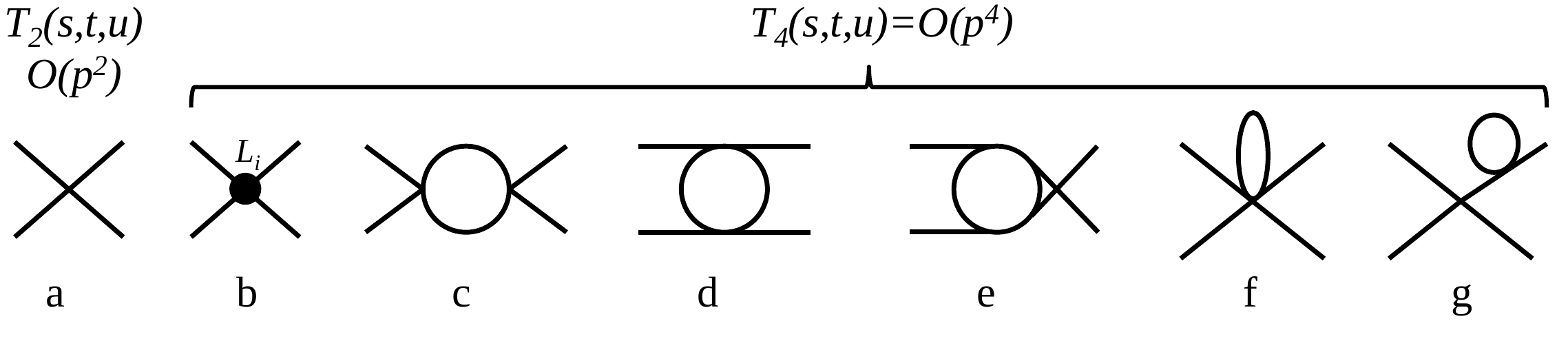}
  \caption{Classes of diagrams that appear in the
  NLO ChPT calculation of meson-meson scattering. }
\label{fig:ChPTdiagrams}
\end{figure}

In this review we are mainly interested in $\pi\pi$ scattering but,
 later on, when studying the spectroscopic classification  and nature of the $\sigma$,
 other meson-meson NLO amplitudes will be of interest.
Thus, for later use it is convenient now to
illustrate the ChPT approach with the NLO meson-meson scattering amplitude calculation. 
Within ChPT,
scattering amplitudes are obtained as an expansion $T(s,t,u)=T_2(s,t,u)+T_4(s,t,u)+T_6(s,t,u)...$,
where $T_{2k}$ stand for the $O(p^{2k})$ contributions, corresponding to $k-1$ loops. 
The generic diagrams contributing up to NLO are shown in Fig.\ref{fig:ChPTdiagrams}.
Diagrams like ``a" provide the $O(p^2)$ leading order, are constructed with vertices from Eq.\ref{ec:Lag2}, 
and are completely determined from meson masses and $f_0$.
All other diagrams contribute at NLO, i.e. $O(p^4)$. 
The loop diagrams ``c", ``d" and ``e" are responsible for the $s$, $t$ and $u$ cuts. In particular, in the s-channel and for partial waves, diagrams ``c" are responsible for the physical cut and unitarity, whereas diagrams like ``d" and ``e" provide the left cut.
``Tadpole" diagrams like ``g" are due to mass and field renormalizations.
The divergences in the loop diagrams cancel against the
combination of $L_i$ LECs that appear in tree level diagrams like ``b" with ${\cal L}_4$ vertices.
As usual in perturbative Quantum Field Theory calculations, divergent loop integrals are regularized by introducing a dependence on some arbitrary regularization scale. To preserve chiral symmetry explicitly throughout the calculation,
dimensional regularization to dimension $D$ is used.
Then, the divergent parts are renormalized by the LECs
\begin{equation}
L_i=L_i^r(\mu)+\Gamma_i \frac{\mu^{D-4}}{32\pi^2}\Big[\frac{2}{D-4}+\gamma_E\-\log(4\pi)-1 \Big].
\end{equation}
This is the so-called $\overline{MS}-1$ scheme. In the above equation $2\Gamma_1=2\Gamma_2=3\Gamma_4=\Gamma_5=3/8$, $\Gamma_6=11/144$ and $\Gamma_8=5/48$. Since $\Gamma_3=\Gamma_7=0$, $L_3$ and $L_7$ are finite and need no renormalization, but all other
LECs become scale dependent:
\begin{equation}
L_i^r(\mu)=L_i^r(\mu_0)+\frac{\Gamma_i}{16\pi^2}\log\Big(\frac{\mu_0}{\mu}\Big).
\label{ec:renormLECs}
\end{equation}
In particular the relations between $SU(3)$ and $SU(2)$ LECs above refer to their renormalized value at the same scale $\mu$.

For all observables, the $\mu$ dependence
cancels order by order against the regularization scale dependence of loop diagrams.
Note that to NLO the masses and the decay constants must be renormalized and thus depend on the LECs, since they are also an expansion in powers of the quark masses (but not the momenta). In particular, the 
decay constants $f_\pi$, $f_K$ and $f_\eta$ are no longer equal beyond leading order.

Similarly, the NNLO $T_6(s,t,u)$ contribution contains two loop diagrams with LO vertices, one-loop diagrams with one LO and one NLO vertex,
as well as tree diagrams with NNLO vertices. The number of NNLO LECs in ChPT needed for all processes is very large, but within $SU(2)$ only six combinations are needed in $\pi\pi$  scattering calculations.

The NLO ChPT calculation of $\pi\pi$ scattering was given within the SU(2) formalism in \cite{chpt1}.
Within the SU(3) formalism, $\pi\pi$, $\pi K$ and $\pi \eta$ elastic scattering to NLO
were first given in \cite{Bernard:1990kx} and the two independent
$KK$ scattering amplitudes in \cite{Guerrero:1998ei}. The remaining three independent NLO amplitudes $\eta\eta$, $K\eta$ and $K\eta\rightarrow K\pi$
where given in \cite{GomezNicola:2001as}, purging and unifying the truncation conventions of the previous five 
which were also recalculated. 
As we will see in the following subsections, all these calculations have been widely
used within unitarization formalisms to generate resonances, and in particular the $\sigma$.
To NNLO, which corresponds to two-loop calculations, only the SU(2) $\pi\pi$ scattering amplitude, which  was given in \cite{Knecht:1995tr,Bijnens:1995yn},
has been unitarized to study the $\sigma$, although
SU(3) NNLO calculations exist for $\pi\pi$ and $\pi K$ scattering \cite{Bijnens:2004eu}.

Let us now comment on the LECs. 
Unfortunately, these cannot be calculated using perturbative QCD, 
but have to be determined phenomenologically either from experiment or 
models (see for instance \cite{LlanesEstrada:2003ha}), or from theory using lattice calculations. The latter are periodically reviewed by the FLAG working group \cite{Aoki:2013ldr}
and there is a fair agreement between the phenomenological and theoretical calculations. 
Since here we are mostly interested in phenomenological applications, 
in Table~\ref{tab:LECs} we provide several sets of phenomenological determinations of the eight NLO LECs. 
Note that all LECs are of the order of $10^{-3}$ or smaller. Recall that, due to renormalization, all LECs 
except $L_3$ and $L_7$ acquire a 
renormalization scale and scheme dependence, so that the values in the Table correspond to  $\mu=M_\rho\sim\, 770\,\mev$ within the $\overline{MS}-1$ scheme, which is the usual one in ChPT \cite{chpt1,chpt2}. Dimensional regularization is used to preserve chiral symmetry explicitly throughout the calculations.

\begin{table}
\footnotesize
\centering
\begin{tabular}{|l|c|c|c|c|c|c|c|c|c|c|c|}
\hline\hline
$10^{3}$&
GL& 
NNLO& 
NLO &
IAMIII & 
FitI  & 
FitII & 
RS&
$V$&
$S$&
$S_1$&
\\ 
&
\cite{chpt2}& 
\cite{Bijnens:2014lea}& 
\cite{Bijnens:2014lea} &
\cite{Pelaez:2004xp}& 
\cite{Nebreda:2010wv} & 
\cite{Nebreda:2010wv} & 
\cite{Ecker:1988te}&
&
&
&
\\ 
\hline
\hline 
$L_1^r$ &
0.7(3)&
0.53(06) &
1.0(1) &
0.60(9)&
1.10&
0.74&
0.6&
0.6&
-0.2&
0.2&
$O(N_c)$
\\
\hline 
$L_2^r$ &
1.3(7)&
0.81(04)&
1.6(2)&
1.22(8)&
1.11&
1.04&
1.2&
1.2&
0&
0&
$O(N_c)$
\\
\hline 
$L_3$ &
-4.4(2.5)&
-3.07(20)&
-3.8(3)&
-3.02(6)&
-4.03&
-3.12&
-3.0&
-3.6&
0.6&
0&
$O(N_c)$
\\
\hline 
$L_4^r$ &
-0.3(5)&
$\equiv 0.3$&
0.0(3)&
$\equiv0$&
-0.06&
0.00&
0.0&
0&
-0.5&
0.5&
$O(1)$
\\
\hline 
$L_5^r$ &
1.4(5)&
1.01(06)&
1.2(1)&
1.90(3)&
1.34&
1.26&
1.4&
0&
1.4$^{(a)}$&
0&
$O(N_c)$
\\
\hline 
$L_6^r$ &
-0.2(0.15)&
0.14(05)&
0.0(4)&
-0.07(20)&
0.15&
-0.01&
0.0&
0&
-0.3&
0.3&
$O(1)$
\\
\hline 
$L_7$ &
-0.4(2)&
-0.34(09)&
-0.3(2)&
-0.25(18)&
-0.43&
-0.49&
-0.3$^{(b)}$&
0&
0&
0&
$O(1)$
\\
\hline 
$L_8^r$ &
0.9(3)&
0.47(10)&
0.5(2)&
0.84(23)&
0.94&
1.06&
0.9&
0&
0.9$^{(a)}$&
0&
$O(N_c)$
\\
\hline \hline 
\end{tabular}
\caption{Values of the NLO LECs multiplied by $10^{3}$.
Columns two to seven provide renormalized values in the $\overline{MS}-1$ renormalization 
scheme at the scale $\mu=M_\rho$. The second column corresponds to \cite{chpt2}, which became  the reference values for
many years. Columns three and four come from the recent review \cite{Bijnens:2014lea}, one using NNLO formulas and the other just the NLO expressions.
IAMIII comes from a fit to phase shifts and inelasticities
\cite{Pelaez:2004xp} with the coupled channel IAM (only statistical uncertainties shown), whereas Fit II is an elastic IAM fit which also includes lattice information on mass dependences \cite{Nebreda:2010wv}. 
The ``RS" column corresponds to the Resonance Saturation estimates obtained in \cite{Ecker:1988te}. In the next columns we have made explicit 
the vector ``$V$", scalar octet ``$S$" and scalar singlet ``$S_1$" contributions to the total RS estimate.
The values (a) are input and (b) is saturated from a heavier pseudoscalar singlet resonance, basically the $\eta'(960)$.
In the last column we show the
leading behavior in the $1/N_c$ expansion as calculated in \cite{chpt2,chptlargen}.
Note that in some of the phenomenological determinations $L_4^r$ is set to 0, or to $0.0\pm0.3$,
 since it is very small from resonance saturation, is suppressed at large $N_c$ and it is hard to determine precisely from data. }
\label{tab:LECs}
\end{table}

Since we are concerned with the $\sigma$ meson, 
one of the most interesting ways to understand the values of the LECs is the fact 
that they are saturated by the effect of the lowest lying resonances.
As a matter of fact, ChPT can be understood as the effective mesonic Lagrangian obtained 
by integrating out all particles and resonances heavier than pions, kaons and etas.
Intuitively, this is nothing but expanding the heavier particle propagators $\sim 1/(s-M_R^2)\simeq -1/M_R^2...$ at low energies $s\ll M_R^2$. Thus, all possible hadronic interactions between NGB where one of these heavier particles is exchanged
are then approximated by contact terms between NGB. But if these respect all QCD symmetries it is then possible to recast these contact
terms order by order in terms of the LECs. Conversely, the values of the LECs can be understood as the
sum of all contributions from heavier resonances expanded at low energies. This is the Resonance Saturation hypothesis.
Moreover, since heavier resonances are more suppressed
due to their larger masses, one expects the dominant contribution to come 
from just the lowest multiplets of each kind. This is sometimes called the Single Resonance Approximation (SRA).
In particular, in \cite{Ecker:1988te} it was found that 
apart from $L_7$, which is saturated by the $\eta'(960)$ meson, and is not very relevant for our purposes here, all other $L_i$
NLO LECs were fairly well understood as 
$L_i=L_i^V+L_i^S+L_i^{S_1}$, where $L_i^V$ is the contribution from the lightest octet of vector resonances,
$L_i^S$ is the contribution from the lightest octet of scalar resonances  and
 $L_i^{S_1} $ is that from the lightest scalar singlet. These contributions read \cite{Ecker:1988te}
\begin{eqnarray}
&&L_1^V=\frac{G_V^2}{8 M_V^2}; \quad L_2^V=2 L_1^V;\quad L_3^V=-6L_1^V,\nonumber \\
&&L_1^S=-\frac{c_d^2}{6M_S^2},\quad L_3^S=-3L_1^S;\quad L_4^S=-\frac{c_dc_m}{3M_S^2},
\quad L_5^S=-3L_4^S,\quad L_6^S=-\frac{c_m^2}{6 M_S^2},\quad L_8^S=-3L_6^S,\nonumber \\
&&L_1^{S_1}=\frac{\tilde c_d^2}{2 M_{S_1}^2},\quad L_4^{S_1}=\frac{\tilde c_d\tilde c_m }{M_{S_1}^2},\quad L_6^{S_1}=\frac{\tilde c_m^2}{2 M_{S_1}^2}
\end{eqnarray}
where $G_V$, $c_d$, $c_m$, $\tilde c_d$ and $\tilde c_m$ are the coupling constants of the vector and scalar resonances to NGB fields
allowed by the QCD symmetries (see \cite{Ecker:1988te} for definitions), which can be determined phenomenologically.
By setting  $M_V=770\,\mev$ and $M_S=M_{S_1}=983\,\mev$ the size of the different contributions can be read in Table~\ref{tab:LECs}
under the ``$V$", ``$S$" and ``$S_1$" columns. 
The sum of these contributions can be read under the ``RS" column and it can be seen that 
it provides a fairly good approximation to the phenomenological parameterizations. 
 Of course, one has to keep in mind that these contributions are obtained from a tree level interpretation of a resonance model and carry no renormalization scale dependence,
 but intuitively they are expected to
provide a good approximation to the phenomenological evaluated at a scale in the range 0.5 to 1 GeV \cite{Ecker:1988te}, as it happens indeed.
This observation will be very relevant in Subsec.\ref{subsec:ncUChPT}, when dealing with the $N_c$ behavior of amplitudes.
In Table \ref{tab:LECs} we can also observe that, in general, the largest contributions come from the vector exchange and that the scalar  singlet barely contributes or its contributions are largely canceled by the scalar octet. This is nothing but a manifestation of the classic Vector Meson Dominance
approach proposed by Sakurai \cite{Sa69} before the advent of QCD.
Concerning the scalars, the ``$S$" and ``$S_1$" contributions were obtained with $M_S\simeq 1 \,\gev$.  Note that trying to identify $M_{S_1}=M_\sigma=450\, \mev$
would lead to very different values from those observed since, apart from yielding contributions 4 times larger than those provided in the table, the scalar octet 
contributions will not cancel against the singlet ones in $L_1$, $L_4$ and $L_6$. 

Therefore, we are now in position to compare with the expectations from integrating out the $\sigma$ in the L$\sigma$M.
This is a textbook exercise (see \cite{Donoghuebook}), 
corresponding to $c_m=\tilde c_m=c_d/2=\tilde c_d/2$ and $c_d^2=3 f_\pi^2/14$ above.
We already  provided the expansion of 
the massless L$\sigma$M up to four derivatives in Eq.\ref{eq:LSMLO}.
This gives  the L$\sigma$M prediction for the 
three LECs that survive in the chiral limit, $L_1,L_2,L_3$.
By looking at Eq.\ref{eq:LSMLO}, we see that $L_2=0$ whereas $2L_1+L_3=f_\pi^2/4 M_\sigma^2>0$.
This is already at odds with the non-vanishing value of $L_2$ 
and the negative sign of $2L_1+L_3$ in Table~\ref{tab:LECs}.
Actually, the prediction for all LECs when integrating out the $\sigma$ in the L$\sigma$M is:
\begin{equation}
2L_1+L_3=2L_4+L_5+8L_6+4L_8=\frac{f_\pi^2}{4 M_\sigma^2},\qquad L_2,L_7=0.
\end{equation}
It is easily checked that this does {\it not} correspond to the observed values 
of the LECs, not even qualitatively, due to the different signs and different hierarchy
pattern of the observed LECs.
Therefore, {\it  the L$\sigma$M is not the correct low energy effective theory of QCD} since it already differs from it at NLO,
although it is true that it reproduces the LO.
In particular, {\it the $\sigma$ from the L$\sigma$M does  not correspond to the $f_0(500)$ observed in nature},
since integrating it out would give the wrong LECs. Of course, it can be observed that most, but not all, of the problem comes from the absence of vector mesons, which as we have already seen dominate the NLO contributions, which therefore cannot be mimicked with an scalar only.
In addition the $2L_4+L_5+8L_6+4L_8$ combination, where vectors do not contribute, comes too large, by at least a factor of 2 or 3
when using $M_\sigma\simeq 500\,$MeV. 

The last column in Table \ref{tab:LECs} provides the 
leading behavior of the NLO LECs according to the $1/N_c$ expansion, obtained from \cite{chpt2} and from \cite{chptlargen} for $L_7$.
The interest of the $1/N_c$ expansion \cite{'tHooft:1973jz,Witten:1980sp}
is that it is the only perturbative 
approach to QCD that is valid in the whole energy regime. 
This approach will be briefly reviewed in Subsec.\ref{subsec:nc}.
Here it suffices to say that the QCD coupling $g$ scales as $1/\sqrt{N_c}$, the number of gluons as
$N_c^2$ and the number of quarks per flavor as $N_c$.  
At leading order in this expansion only planar diagrams contribute 
and only those without quark loops. Since quarks are the only ones carrying flavor and can only be
external lines in meson diagrams  as those of ChPT, the $N_c$ behavior of any operator 
in the ChPT Lagrangian can be calculated by counting the flavor traces.
This allows for a model independent extraction of the 
$N_c$ dependence of each operator in ${\cal L}_4$ and correspondingly the $L_i$ behavior
\cite{chpt2,chptlargen}.

Before finishing this subsection it is worth mentioning that there is a recent 
extension of the usual SU(2) ChPT formalism  to allow for a $\sigma$ degree of freedom \cite{Soto:2011ap}.  Note that this theory has a well-defined power counting in powers of momenta
as well as pion and $\sigma$ masses, over a chiral symmetry breaking scale $\Lambda_\chi$.
The LO Lagrangian compared to that of the L$\sigma$M differs in
two aspects: the self-interactions of the scalar field are set to zero (in principle these are reconstructed perturbatively at higher orders) and it has four low energy constants 
describing the most general
 interactions between the $\sigma$ and the pions to that order.
The LO description is, of course, rather crude and needs input from lattice-QCD  to make predictions for the $S$-wave parameters, which still do not come out very close to the experimental determinations. A considerable improvement is expected at NLO, although so far this 
formalism has only been explored to NLO to compare with lattice results
on the quark mass dependence of  $f_\pi$ and $M_\pi$.

Let us then summarize this subsection: we have seen that ChPT provides {\it the} 
QCD low energy effective theory that contains all possible Lagrangian terms 
containing pions, kaons and the eta, which are compatible with QCD symmetries. 
Unfortunately this is just an expansion and by itself cannot reproduce 
resonances although ChPT can provide useful information about heavier resonances  from their contribution to the values of the LECs. Surprisingly, despite being the lightest resonance, the $\sigma$ does not contribute to the LECs, which already suggests that it has a different nature from other resonances.

It would nevertheless be desirable to implement the chiral symmetry constraints
in the description of resonances, but before doing that it is important to recall the relation between 
resonances and unitarity.

\subsection{Unitarity}

The impossibility of describing resonances with 
the ChPT expansion is closely related to the fact that since the energy behavior of ChPT 
amplitudes at large $s$ is dominated by
polynomials, their modulus can grow indefinitely and eventually they will violate unitarity.
The elastic unitarity condition is particularly simple for elastic partial waves. 
In Sec.\ref{subsec:notation} we wrote it for $\hat f^{(I)}_\ell(s)$
and we recast it here for the $t(s)=\hat f(s)/\sigma(s)$ normalization, which is more usual in the ChPT context,
\begin{equation}
\im t(s)= \sigma(s) \vert t(s)\vert^2,
\label{ec:unit}
\end{equation}
where we recall that  $\sigma(s)=2k/\sqrt{s}$ 
and that 
$\sigma(s)\simeq1$ as soon as the energy is sufficiently above threshold. Therefore, the following bounds hold
\begin{equation}
\vert t(s)\vert\leq 1/\sigma(s),\quad \re t(s)\leq 1/2\sigma(s).
\label{ec:unitbounds}
\end{equation}
Note that for simplicity we have suppressed the isospin and angular momentum indices.
A definition of a strong theory is precisely one that saturates the unitarity bounds.
Actually, as we have already seen, one of the intuitive characterizations of resonances is that 
they saturate the unitarity bounds above (this is definition ``i'' in Sec.\ref{subsec:poles}),
although that only applies to well isolated, elastic and narrow resonances.

Still, within ChPT partial waves are obtained as an expansion $t(s)=t_2(s)+t_4(s)+t_6(s)...$
where $t_{2k}\sim O(p^{2k})$. However, once this series is truncated it cannot
satisfy elastic unitarity exactly
 since the highest powers on the left and right sides of Eq.\ref{ec:unit} cannot match.
Of course, ChPT satisfies elastic unitarity perturbatively:
\begin{eqnarray}
\im t_2(s)=0, \quad \im t_4(s)=\sigma(s) t_2(s)^2,\quad \im t_6(s)=2 \sigma(s)t_2(s) \re t_4(s), ...
\label{ec:pertuni}
\end{eqnarray}

The extension of the unitarity condition to the case when two  two-body channels $\vert 1\rangle$ and  $\vert 2\rangle$,
are coupled and open (i.e. accessible at a given energy) is straightforward. 
For this purpose we define:
\begin{equation}
T(s)=\left(
\begin{array}{cc}
t_{11}(s) & t_{12}(s) \cr
t_{12}(s) & t_{22}(s)
\end{array}\right),\quad 
\Sigma(s)=\left(
\begin{array}{cc}
\sigma_1(s) & 0 \cr
0 & \sigma_2(s)
\end{array}\right),
\label{ec:pwmatrices}
\end{equation}
where $t_{ij}$ is the partial-wave amplitude between states $\vert 1\rangle$, $\vert 2\rangle$
and $\sigma_i=2 k_i/\sqrt{s}$, with $k_i$ the CM momentum of the particles in the corresponding  $\vert i\rangle$ state.
Then the unitarity relation for energies where only those two channels are open 
reads:
\begin{equation}
\im T(s) = T(s) \,\Sigma(s) \,T^*(s), 
\label{ec:matrixunit}
\end{equation}
which is nothing but the matrix version of Eq.\ref{ec:unit}.
Let us remark that this form of the unitarity relation is also valid
for an arbitrary number $k$ of open two-body states, in the energy region where they 
are the only accessible states. It is enough to define the corresponding $T$  and $\Sigma$ matrices as $k\times k$ matrices.

Once more, the ChPT partial waves cannot satisfy coupled channel unitarity exactly.
Nevertheless using the $T$-matrix notation, for the case of $k$ open two-body  channels, 
they once again satisfy
\begin{equation}
\im T_2(s)=0, \quad \im T_4(s)=T_2(s) \Sigma(s)  T_2(s),...
\label{ec:matrixpertuni}
\end{equation}
for energies where only those $k$ channels are open.

In the next subsections we will 
review the formalisms  that can describe resonances by implementing unitarity while
simultaneously incorporating the ChPT expansion at low energies up to a given order.
These are generically known as unitarization methods.

\subsection{Unitarization: Elastic $\pi\pi$ scattering}
\label{subsec:uchpt}

We have just seen that unitarity plays a very important role as soon as 
the interaction becomes strong and therefore in the resonance region. For
elastic meson-meson scattering it is most convenient to recast the unitarity condition.
Eq.\ref{ec:unit},
in terms of the inverse amplitude, since then
\begin{equation}
\im \frac{1}{t(s)}=\im \frac{t(s)^*}{\vert t(s)\vert ^2}=-  \frac{\im t(s)}{\vert t(s)\vert^2}=-\sigma(s),
\label{ec:inverseunit}
\end{equation}
where in the last step we have used Eq.\ref{ec:unit}. This means that {\it the
imaginary part of the inverse of an elastic partial wave is known exactly}. Therefore, a unitary
elastic partial wave above threshold simply reads:
\begin{equation}
t(s)=\frac{1}{\re t(s)^{-1}-i \sigma(s)}
\label{ec:generalunit}
\end{equation}

The inelastic case of several coupled two-body states can be treated similarly starting from the matrix unitarity relation 
 Eq.\ref{ec:matrixunit}, which can be recast as $\im T(s)^{-1}=-\Sigma(s)$. In other words, the
imaginary part of the inverse of the partial wave matrix is known exactly.
Therefore a unitary matrix of partial waves in the energy region where only $k$ 
two-body states are open
reads:
\begin{equation}
T(s) = [\re T(s)^{-1}-i \,\Sigma(s)]^{-1}, 
\label{ec:unitarizedmatrix}
\end{equation}
which is just the matrix form of Eq.\ref{ec:generalunit}.

A very popular and relatively simple approach is the so-called $K$-matrix method, 
\begin{equation}
t(s)=\frac{K(s)}{1-i\sigma(s)K(s)},
\label{ec:K-matirx}
\end{equation} 
where $\re t(s)$ has been approximated by $K(s)$, which is a real function for real $s$, 
typically a polynomial or rational function in $s$.
It is especially popular in the coupled channel case, where $K(s)$ becomes a real matrix.
This method may work fine in the real axis, but one has to be careful on its extensions to the complex plane.
For instance, since $\sigma(s)$ has a pole at $s=0$ then Eq.\ref{ec:K-matirx} 
forces all partial waves to vanish at $s=0$, which is unphysical. In addition, it generically lacks a left cut. Nevertheless,
 in the physical axis above threshold, or even close to  it, the $K$-matrix may provide more reasonable results.
 Note, however, that as we already saw in in Fig.\ref{fig:pwcutsandpoles}, the $\sigma$ pole, which is the one we are interested in, is not very close to the physical axis and is relatively close to the left cut and the $s=0$ region.

As we will see, {\it in the physical $s$ axis,} all elastic unitarization methods can be recast into Eq.\ref{ec:generalunit}
and all the coupled channel ones (with only two-body accessible states)
into Eq.\ref{ec:unitarizedmatrix}.
The problem, of course is to calculate $\re T(s)^{-1}$.
In the literature this is sometimes presented as an arbitrariness in the unitarization procedure, but the issue is 
just to obtain the best possible approximation to $\re t(s)^{-1}$ and to determine  in what range of $s$ it is valid. In principle,
the more constraints are incorporated
in the calculation of $\re t(s)^{-1}$, the better. These constraints are symmetries, more terms from ChPT at low energies, analyticity, etc. 
We mention here analyticity, because in order to study resonances it is not only important that the approximation
describes the data,  but it is also essential that it has a sound analytic extension of $t(s)$
into the complex  plane. Recall that Eqs.\ref{ec:generalunit} and  \ref{ec:unitarizedmatrix} are only valid in the physical part of the real axis.
For instance, it is better that the $\sigma(s)$ factor in Eq.\ref{ec:generalunit} comes from an analytic function
which does not have the spurious analytic structure that is present in simple approaches like the $K$-matrix.

\subsubsection{The simplest unitarized model with chiral symmetry}
\label{subsec:simplest}

As an illustration,
let us work out the very simple approach 
where we approximate $\re t(s)^{-1}\simeq 1/t_2(s)$,
i.e., by just the LO ChPT result, to find
\begin{equation}
  t(s)\simeq\frac{1}{1/t_2(s)-i\sigma(s)}=\frac{t_2(s)}{1-i\sigma(s)t_2(s)}.
\label{ec:simplestunit}
\end{equation}
Of course, by re-expanding the above equation for small $s$ 
the LO ChPT result is recovered.
Note also that not only the algebraic unitarity condition is implemented in Eq.\ref{ec:simplestunit},
but also a right cut is inherited from $\sigma(s)$,
although only elastic unitarity is fulfilled. This implies the existence of a second Riemann sheet, where we can look for poles associated to resonances.
Using Eq.\ref{ec:firsttosecondsheet}, the pole condition
in the second Riemann sheet
is easily obtained:
\begin{equation}
t^{II}(s)\simeq\frac{t(s)}{1+2i\sigma(s) t(s)}\rightarrow\infty,\quad \Rightarrow
\quad 1+i\sigma(s)t_2(s)=0,
\end{equation}
where, as discussed in Sect.\ref{subsec:poles}, on the upper half $s$ plane $\sigma=+\sqrt{1-4M_\pi^2/s}$ as usual,
 whereas on the lower half $s$ plane we must then take $\sigma(s)=-\sigma(s^*)^*$.

The solution to the pole condition depends on $t_2(s)$, which is different for each isospin and angular momentum. 
Let us recall that $t_2(s)$ are the low energy theorems
that we already provided in Eq.\ref{ec:LET}.
Thus, within this approximation for the scalar isoscalar wave we find a pole at $s_\sigma$ 
as the solution of 
\begin{equation}
\sigma(s_\sigma)(2 s_\sigma-M_\pi^2)=i 32\pi f_\pi^2.
\end{equation} 
With $f_\pi=92.3\, \mev$ and $M_\pi=139.57\, \mev$ 
the pair of conjugated poles is found at  $\sqrt{s_\sigma}\simeq (493\pm i441)\,\mev$.
For later purposes it is very relevant to notice that in the chiral limit $M_\pi=0$, this simple model yields \cite{Donoghuebook}:
\begin{equation}
\sqrt{s_\sigma}=(1-i)\sqrt{8\pi} f_\pi,
\label{ec:simplestpolechirallimit}
\end{equation}
which numerically translates into $\sqrt{s_\sigma}\simeq (467\pm i467)\,\mev$.

Remarkably, the mass of these estimates is already within the RPP12 estimate in Eq.\ref{rpp2012sigmapole} and not far from the more stringent ``Conservative dispersive estimate'' in Eq.\ref{myestimate}.
The width, however, comes out too large by somewhat less than a factor of 2.
This result is fairly good, taking into account that this is just a 
purely theoretical prediction of 
the simplest  unitarized model which implements some chiral constraints.
Recall that no data has been fitted, there are no NLO terms, no left cuts, inelasticities, etc.

In contrast, if we follow similar steps
for the vector channel, the pole is now a solution of  $\sigma(s_\rho)(s_\rho-4M_\pi^2)=i96\pi f_\pi^2$, namely $s_\rho\simeq (1170\pm i1119)\,\mev$. This is a very bad approximation of the 
$\rho(770)$, whose mass is 400 MeV lighter and its half width is about 15 times smaller.
The $\rho(770)$ is not well approximated within this simple approach.

Straightforward caveats to this method are that:
i)   
the approximation $\re t(s)=t_2(s)$ has been used 
well beyond the applicability range of ChPT,
ii) crossing symmetry is violated because there is no left cut, iii) there are no inelastic effects, iv) spurious poles can also appear in the first Riemann sheet.
In the next subsections, we will review more 
elaborated unitarization techniques which improve
on these caveats. In addition, we will consider higher orders of ChPT.

\subsubsection{The elastic Inverse Amplitude Method}
\label{subsubsec:IAM}

\begin{figure}
\includegraphics[scale=0.475]{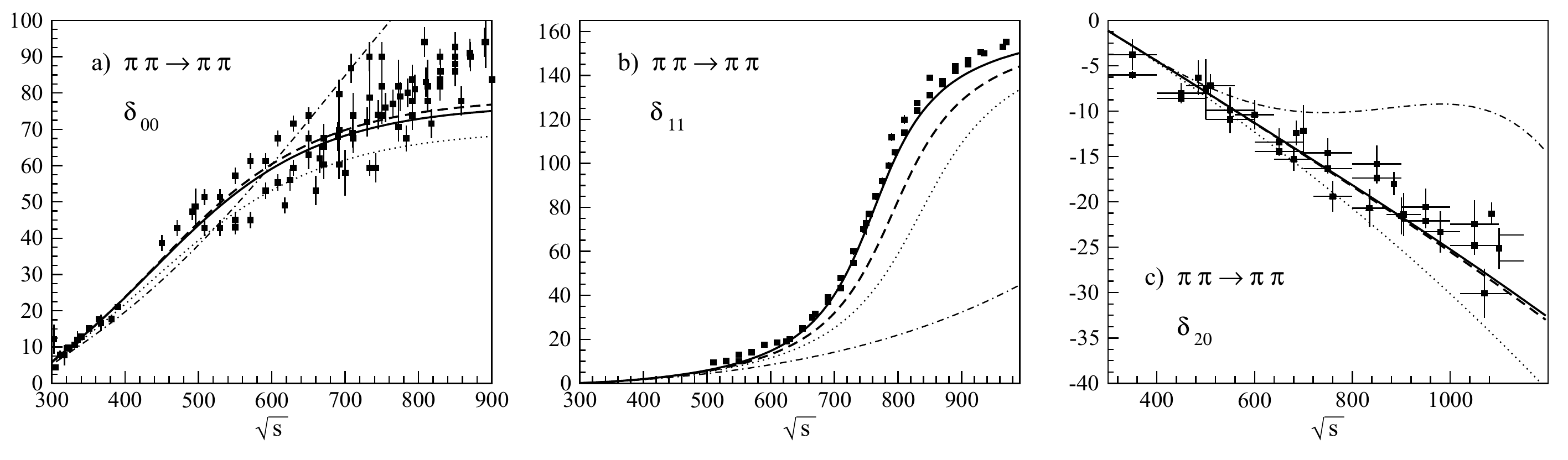}
\vspace{-1cm} 
\caption{\rm \label{fig:IAMelastic} 
Results from IAM fits to elastic $\pi\pi$, $\pi K$
scattering data and lattice results on the quark mass dependence of the masses and decay constants \cite{Nebreda:2010wv}. The continuous and dashed lines
 correspond to the IAM with
the FitI and FitII LECs in Table~\ref{tab:LECs}, respectively.
For comparison, the  dotted lines stand for the results of the IAM with the ChPT
LECs obtained from the two-loop analysis of $K_{e4}$ decays of \cite{Amoros:2001cp}.
The results of standard NLO
ChPT with the same set of LECs are shown as dot-dashed lines. 
Figures taken from \cite{Nebreda:2010wv}}
\end{figure}

Since we are particularly worried about having sound analytic properties, 
the best way is to implement unitarity
from Eq.\ref{ec:inverseunit}, by writing a dispersion
relation for the {\it inverse amplitude} instead of the amplitude itself \cite{Truong:1988zp,Dobado:1992zs,Dobado:1989qm,Dobado:1996ps}.
For convenience, and since
$t_2(s)$ is real, instead of $1/t(s)$ we define 
$G(s)=t_2(s)^2/t(s)$, that also has a right cut and
a left cut ($LC$), which has the same form for the integrand as the right cut. 
Since we have already seen that scalar waves  
have dynamical Adler zeros \cite{Adler:1964um} in the low energy region below threshold,
we also allow for a pole contribution $PC(s)$ in $G(s)$.
Therefore, a dispersion relation for $G(s)$ reads:
\begin{equation}
  \label{disp1/t}
  G(s)=G(0)+G'(0)s+\frac{1}{2}G''(0)s^2+
  \frac{s^3}{\pi}\int_{RC}ds'\frac{\im G(s')}{s'^3(s'-s)}+
  LC(G)+PC(s).
\end{equation}
In the elastic approximation, unitarity in Eq.\ref{ec:inverseunit} together with
Eq.\ref{ec:pertuni} 
allow us to evaluate \emph{exactly} 
$\im G=-\sigma t_2^2=-\im t_4$ on the right cut. In addition,
since in the low energy region $G=t_2^2/(t_2+t_4+...)\simeq t_2-t_4+...$
we can also approximate at low energies $\im G(s)\simeq -\im t_4(s)$
on the left cut, and write
$LC(G)\simeq -LC(t_4)+...$.  Note that there are three 
subtractions, i.e. the factor of $1/s'^3$, because $t_4$ grows as $s^2$.
These subtractions suppress the 
high energy part and in particular the inelastic 
contributions. Hence, the integrals are 
dominated by the  low energy region where
it is  justified to use ChPT. 
As explained in Sec.\ref{sec:disprel}, the price to pay
for the three subtractions is that the dispersive integrals only determine
the amplitude up to a second order polynomial
$G(0)+G'(0)s+\frac{1}{2}G''(0)s^2$. However, its coefficients
are nothing but the values of the inverse amplitude or its derivatives at $s=0$,
for which the chiral expansion can be safely applied. In particular, to one-loop,
$G(0)\simeq t_2(0)-t_4(0)$, $G'(0)\simeq t_2'(0)-t_4'(0)$ and $G''(0)=-t_4''(0)$, since $t_2''(0)$ vanishes.
Let us neglect for the moment the pole contribution $PC(s)$, 
which is of higher order and
only numerically relevant below threshold. 
Then one finds that all contributions can be recast as:
\begin{eqnarray}
  G(s)&\simeq& t_2(0)+ t_2'(0)s-t_4(0)-t_4'(0)-t_4(0)\frac{s^2}{2}-
  \frac{s^3}{\pi}\int_{RC}ds'\frac{\im t_4(s)}{s'^3(s'-s)}
  -LC(t_4)\nonumber\\
&=&t_2(s)-t_4(s),
\end{eqnarray}
where in the final step we have used that the first two terms are 
just $t_2(s)$ whereas the rest of the terms are nothing but an
exact dispersion relation for $-t_4(s)$.
Recalling that we defined $G(s)=t_2(s)^2/t(s)$, we arrive at 
the elastic formula for the Inverse Amplitude Method (IAM)
 \cite{Truong:1988zp,Dobado:1992zs,Dobado:1989qm,Dobado:1996ps}:
\begin{equation}
  \label{ec:IAM}
  t(s)\simeq \frac{t_2^2(s)}{t_2(s)-t_4(s)}\,.
\end{equation}
Remarkably, this simple equation ensures elastic unitarity, inherits the analytic properties of the dispersive integral, and 
at low energies matches ChPT to NLO.
In addition, as seen in Fig.~\ref{fig:IAMelastic} it
describes fairly well data up to somewhat 
less than 1 GeV. The IAM curves in that plot
were obtained in a recently updated fit to elastic $\pi\pi$, $\pi K$
scattering data, plus lattice results on the quark mass dependence of the masses and decay constants \cite{Nebreda:2010wv}.
Note that the continuous and dashed lines  correspond to the FitI or FitII sets of LECs
in Table~\ref{tab:LECs}, which are fairly compatible
with the pure ChPT determinations. 
Some small differences on the LECs can be expected
since the IAM fits data up to higher energies than one would use for plain ChPT
and also because the IAM contains
the part of the higher order ChPT contributions that are needed for exact unitarity.
For these reasons the IAM LECs can be expected to lie somewhere in between the NLO and NNLO determinations of the LECs with pure ChPT.
In the figure we also show the result of a naive extrapolation of ChPT, which seems to work in all waves up to roughly 500 MeV, although it does much better in the scalar isoscalar channel.  One of the striking features of that plot is that the Breit-Wigner like resonance
shape of the $\rho(770)$, which is not seen in standard NLO ChPT,
 is naturally generated within the NLO IAM, since even with the standard values of the parameters the resonant shape is clearly visible, although somewhat displaced unless
the LECs are refitted. 

Incidentally, recalling that the perturbative unitarity of ChPT
implies $\im t_4=\sigma t_2^2$, i.e., Eq.\ref{ec:pertuni}, we can recast the NLO IAM
in Eq.\ref{ec:IAM} as
\begin{equation}
  \label{ec:IAMNLOtoLO}
  t(s)=\frac{t_2^2(s)}{t_2(s)-\re t_4(s)- i \im t_4(s)}=
\frac{1}{1/t_2(s)-\re t_4(s)/t_2^2- i \sigma(s)}.
\end{equation}
which is of the general form in Eq.\ref{ec:generalunit}.
Thus it is usual to present naively the NLO IAM as nothing but including
the NLO expansion of $\re 1/t(s)= 1/t_2(s)-\re t_4(s)/t_2^2...$ 
into the general elastic unitary form in Eq.\ref{ec:generalunit}. 
Note however that Eq.\ref{ec:generalunit} is only valid on the real axis, whereas 
from the previous derivation it is clear that the IAM is valid in the whole complex plane,
although, of course, it is a better approximation near the unitarity cut, and becomes worse near the left cut or the
very small region where $PC(s)$ is sizable.
This naive way of presenting the IAM does not make explicit its analytic properties,
since taking the real part of an analytic function is not generally 
an analytic function. In addition it seems that the 
NLO expansion of the amplitude is being used for all 
values of $s$ when in fact it is only needed at $s=0$ for 
the subtraction constants and at low energies for the left cut,
since the right cut is exact, at least in the elastic regime.
In particular, the IAM differs from the usual $K$-matrix approach
in that, apart from having a left cut,
 it only has the form of Eq.\ref{ec:generalunit} 
on the physical cut, but is rather different outside. For instance, it does not have the generic zero at $s=0$ of the $K$-matrix unitarization.

Moreover since the NLO IAM can also be written as
$t_2/(1-t_4/t_2)$
it is usually called a Pad\'e approximant. Let us recall that the 
[m,n] Pad\'e approximant of a function $f(z)$ is a rational function
$R(s)=(a_0+a_1 z+...a_m z^m)/(1+b_1z+...b_n z^n)$ whose first (n+m) derivatives at $z=0$ are equal to those of $f(z)$. Indeed, if $f(z)$ can be expanded as 
$f(z)=f(0)+ f'(0) z+...$ the [0,1] Pad\'e approximant reads $f(0)/(1+f'(0)z/f(0))$, similarly to 
the NLO IAM. However, contrary to $f(0)$ and $f'(0)$, which are numbers,
$t_2(s)$ and $t_4(s)$ are functions of $s$ which contain not only polynomial 
but logarithmic functions with cuts. Therefore the IAM cannot be interpreted as a Pad\'e approximant in the $s$ variable. If anything, it would have the same formal expression of a Pad\'e approximant in powers of $1/f_0^2$, the inverse of the decay constant.

\begin{figure}
  \centering
  \includegraphics[width=0.49\textwidth]{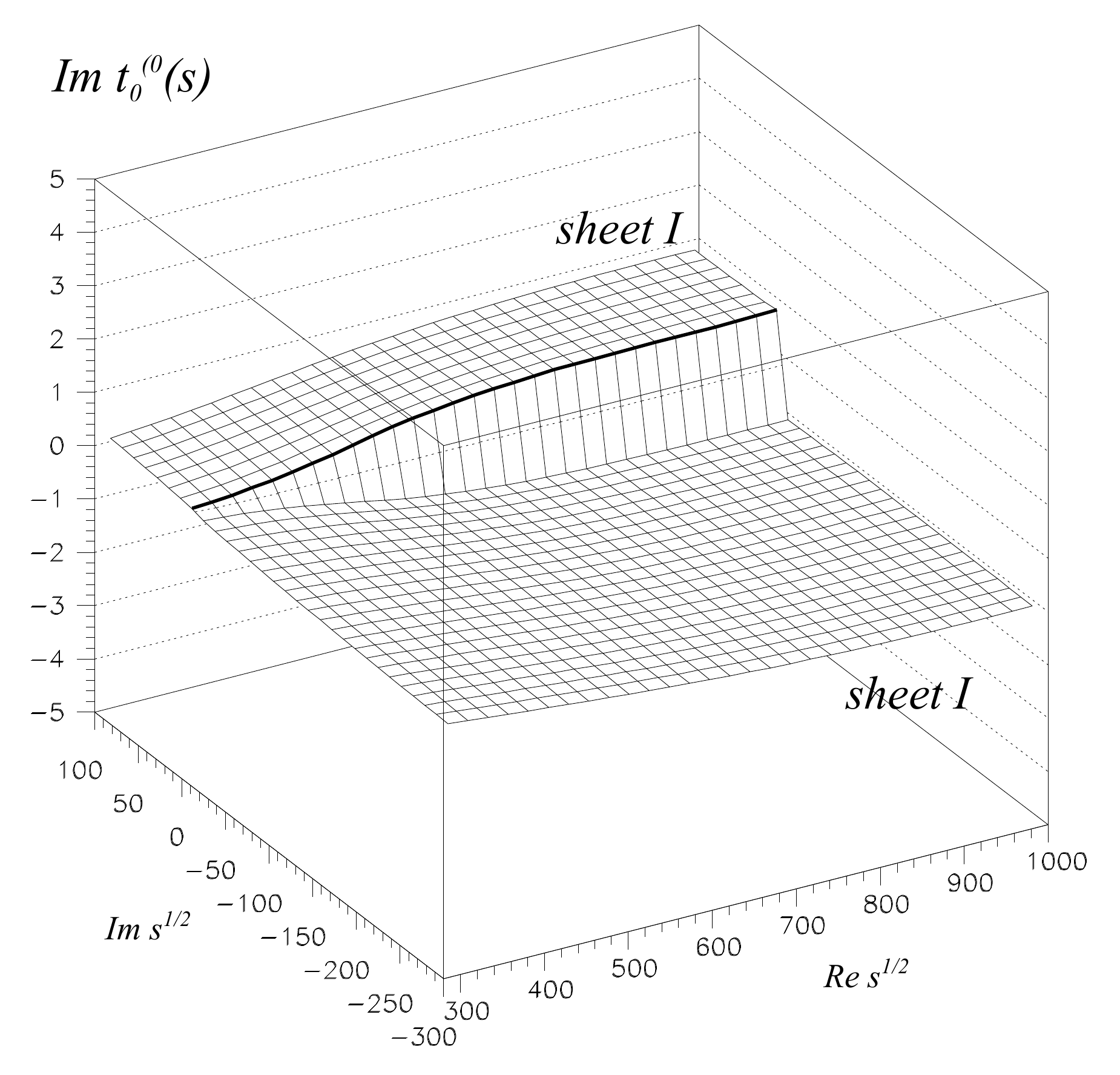}
  \includegraphics[width=0.49\textwidth]{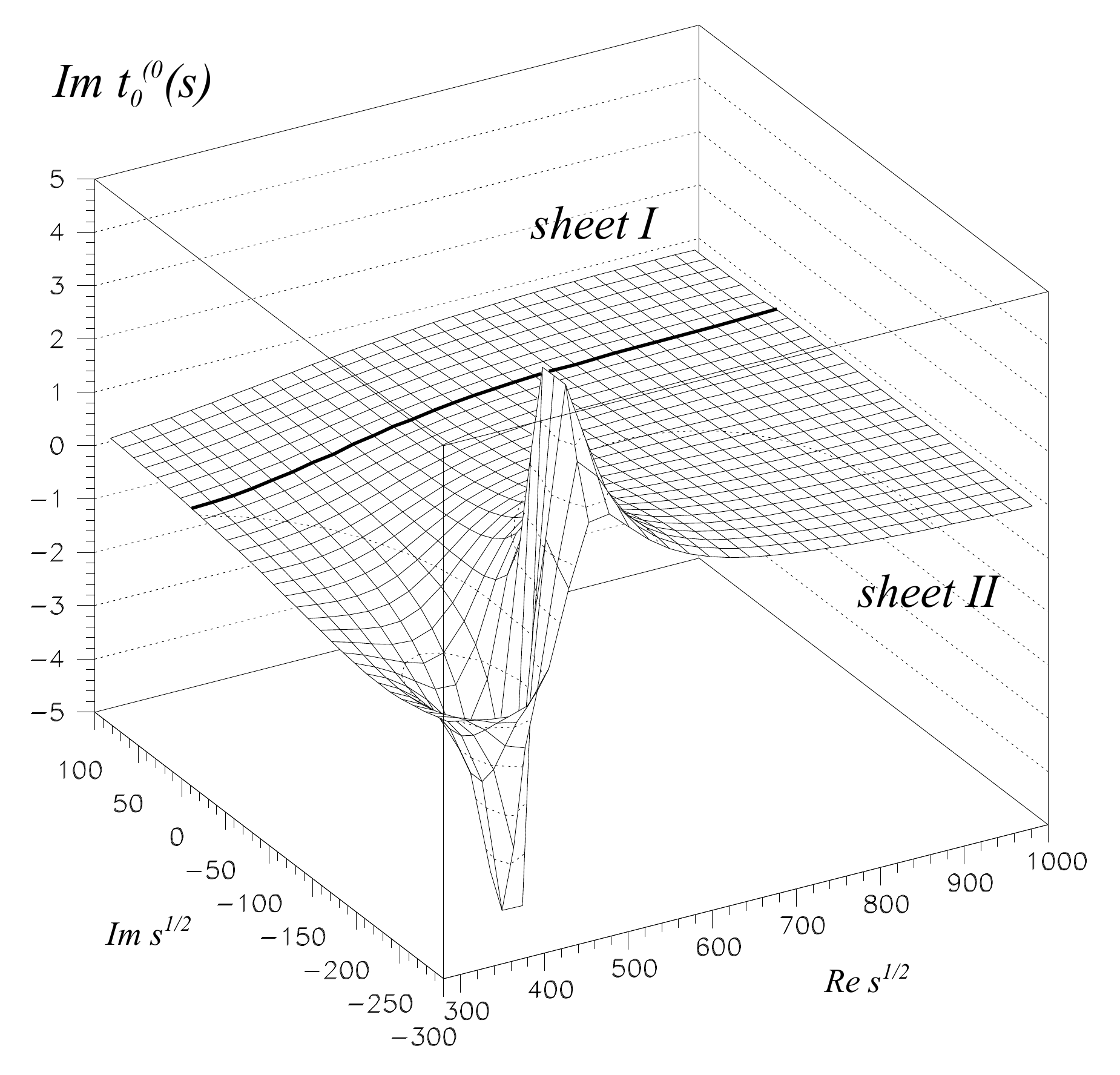}
\includegraphics[width=0.49\textwidth]{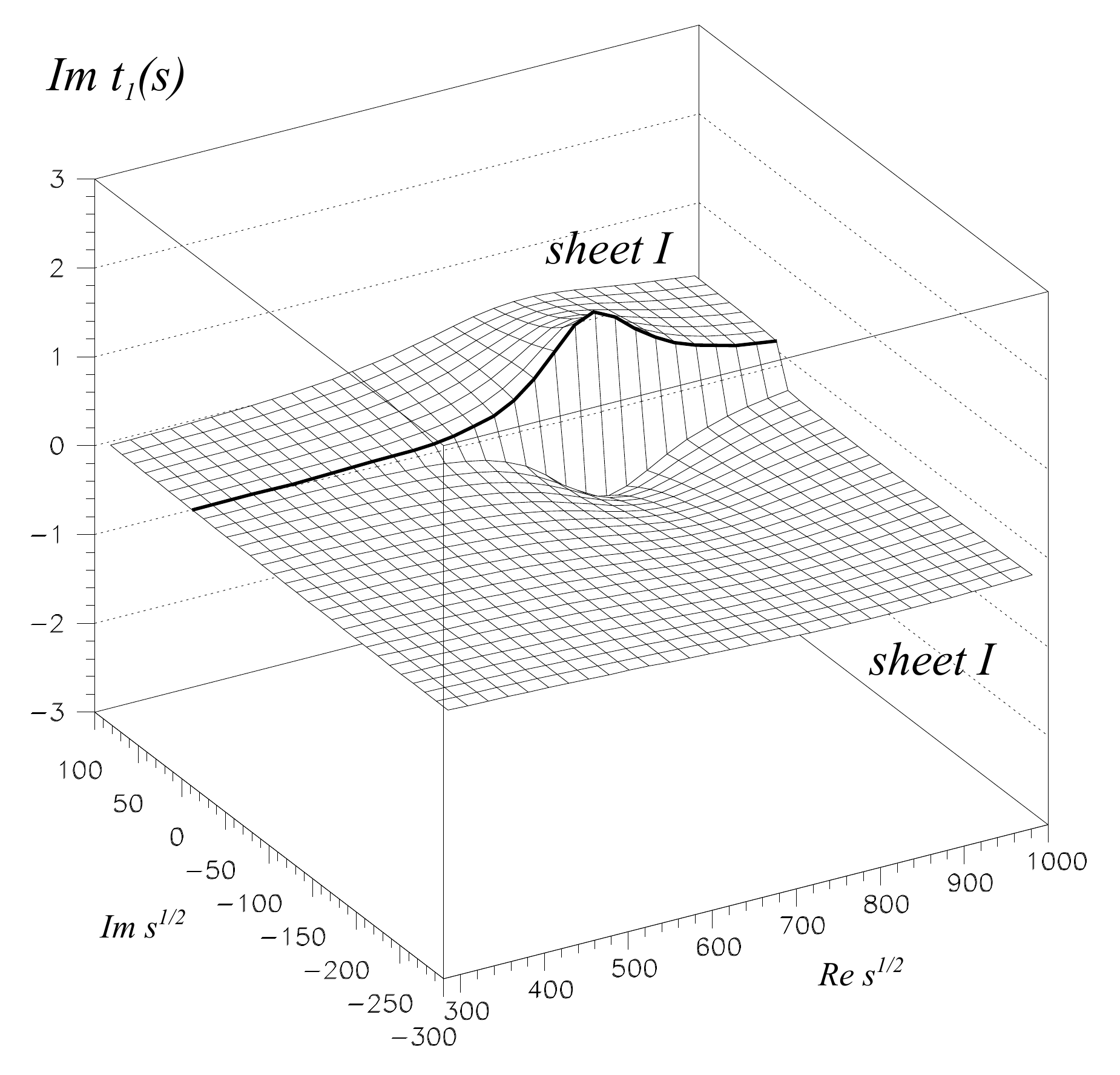}
  \includegraphics[width=0.49\textwidth]{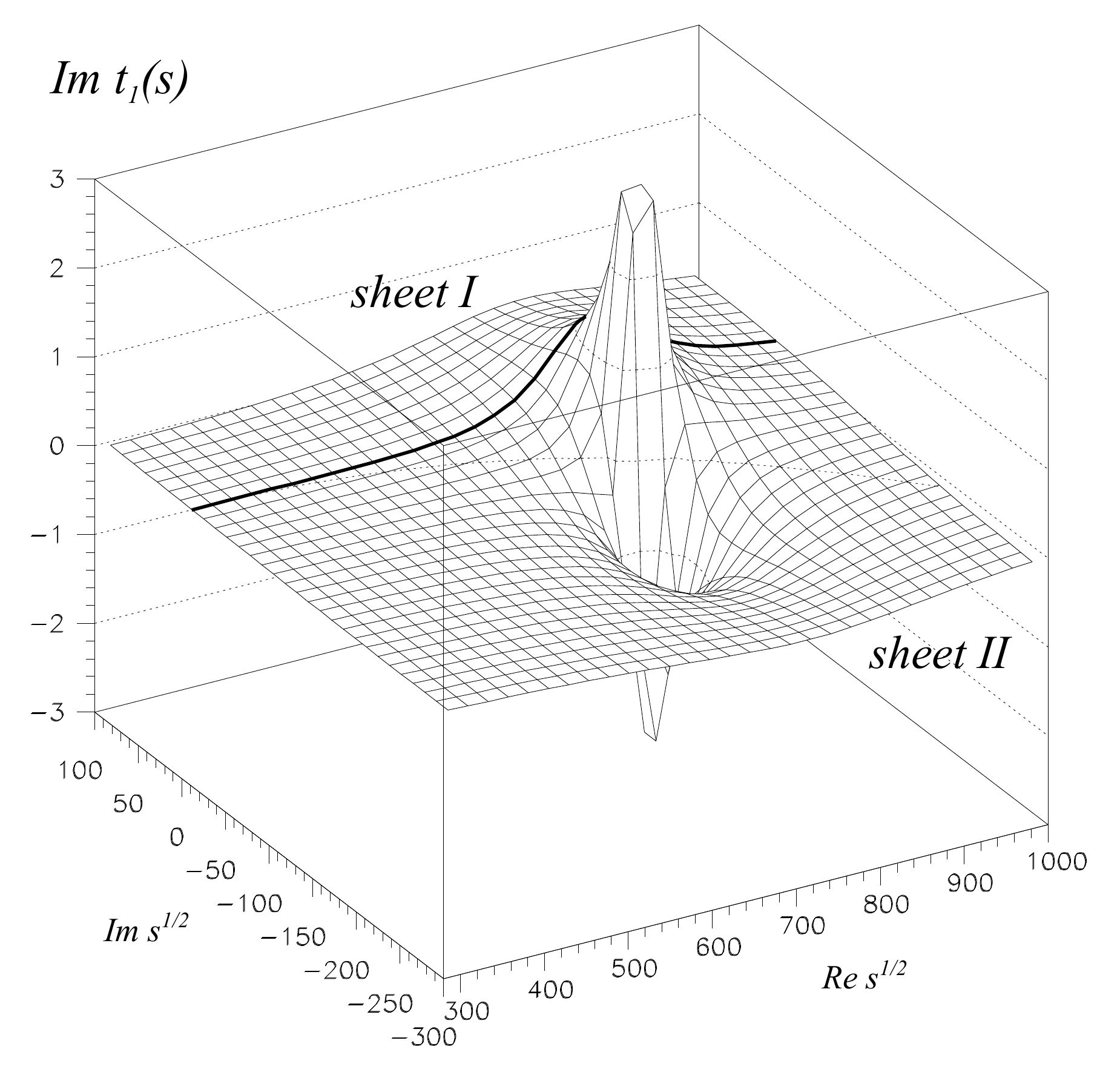}
\caption{ $\im t_0^{(0)}(s)$ and resonance poles calculated with the elastic NLO IAM \cite{Dobado:1996ps}.
 On the left panels we show the first Riemann sheet, where the cut in the real axis can be observed between the upper and lower halves of the complex $s$ plane (note we give units of $\sqrt{s}$ for a better visualization of the 
 $\sqrt{s_{pole}}=M_R-i\Gamma/2$ relation). 
 If from the first sheet on the upper half the cut is crossed continuously,
one arrives to the second Riemann sheet, where resonance poles can be found.
The upper panels correspond to the scalar-isoscalar wave, where the $\sigma$ pole 
is located deep in the complex plane. For comparison, the vector wave is shown
in the lower panels, where the $\rho(770)$ pole can be seen to lie much closer to the real axis than that of the $\sigma$. 
Figures taken from \cite{Dobado:1996ps}.
}
  \label{fig:nicepoles}
\end{figure}

Now, the IAM is an analytic function with the same cut structure of $t(s)$,
namely, a left cut approximated to NLO ChPT as well as
a right or unitarity cut, which is calculated exactly within the elastic approximation.
This right cut can be observed in the left panels of Fig.\ref{fig:nicepoles},
where we plot $\im t(s)$ approximated with the NLO IAM.
Thus, if we cross continuously the cut
 from the upper half plane of the first sheet, we end up
in the lower half of the second sheet, where we can look for poles.
By so doing, two poles are found, 
one corresponding to the $\sigma$ resonance in the 
scalar-isoscalar wave and another one for the $\rho(770)$ in the vector wave.
Already in 1996 \cite{Dobado:1996ps}, before the inclusion of the $f_0(600)$ in the RPP,  it was found
 $\sqrt{s_\rho}=(760-i 75)\,\mev$ whereas $\sqrt{s_\sigma}=(440-i245)\,$MeV,
in remarkable agreement with the most recent determinations.
A more recent update \cite{Pelaez:2010fj}, 
including modern $K_{e4}$ data, 
yields $M_\sigma=(453-i271)\,\mev$ 
in good agreement with the 
Conservative Dispersive Estimate given in Eq.\ref{myestimate}.
Since these poles are obtained from dispersion relations, they were already 
listed in Table~\ref{tab:otherpoles}.
Note that no uncertainties are given because the systematic uncertainties of higher orders, or of approximating the left cut and keeping the elastic approximation, are not easy to quantify. But looking at Fig.\ref{fig:IAMelastic} and the curves obtained without fitting
the LECs, these can be estimated to be of the order of 10 or 15\%.

These poles can be found in the right panels of Fig.\ref{fig:nicepoles}. We do not see the conjugated pole in the upper half plane because it lies in the second sheet, whereas in those plots the upper half plane is still in the first sheet when it is 
continuously connected to the lower half in the second, so that the cut is not visible and the amplitude looks continuous across the real axis.

The $\rho$ and the $\sigma$ poles can be generated either with 
the SU(2) or the SU(3)
elastic NLO IAM. However, when using the SU(3) formalism, it is also possible
to describe the elastic $K \pi$ scattering  
\cite{Dobado:1989qm,Dobado:1992zs,Dobado:1996ps} in which the $K^*(892)$ 
and $\kappa$ (or $K_0^*(800)$) associated poles are also found.
Moreover, the IAM has also been applied to
elastic $\pi N$ scattering, within NLO and NNLO Heavy Baryon ChPT, 
where it is able to describe the $\Delta (1232)$ resonance \cite{GomezNicola:1999pu}.

The IAM can be easily and systematically
extended to higher orders of ChPT. 
The dispersive derivation is analogous and, for instance, to NNLO one arrives at
\cite{Dobado:1996ps}:
\begin{equation}
t(s)\simeq\frac{t_2(s)^2}{t_2(s)-t_4(s)+t_4(s)^2/t_2(s)-t_6(s)},
\end{equation}
which satisfies all the properties already discussed for the NLO IAM,
but when re-expanded reproduces the ChPT series up to NNLO.
The two-loop ChPT was thoroughly studied in \cite{Pelaez:2010fj,Nieves:2001de}.
As a matter of fact, the updated IAM pole  
commented above and listed in Table~\ref{tab:otherpoles} corresponds to a 
NNLO IAM fit \cite{Pelaez:2010fj}. Note that since $D$ and higher waves have $t_2(s)=0$,
for their unitarization with the IAM, the NNNLO, i.e., $O(p^8)$ would be necessary. No such calculation exists, but by considering
only the $O(p^8)$ leading term in the chiral limit, which just adds one more free parameter, it has been shown that
the $f_2(1270)$ resonance could also be generated with relatively reasonable values of the LECs \cite{Dobado:2001rv}.

For completeness, and even though it will be 
negligible except for high quark masses
when studying the quark mass dependence of  
resonances in Sec.\ref{subsec:qm} below, 
let us now include the pole contribution $PC$
ignored so far. Its contribution can be calculated explicitly from
its residue  \cite{GomezNicola:2007qj} and, to one-loop,  we
find a modified IAM (mIAM) formula:
\begin{eqnarray}
  \label{ec:mIAM}
  t^{mIAM}&=& \frac{t^2_2}{ t_2-t_4 +A^{mIAM}}\\
  A^{mIAM}&=&t_4(s_2){-}\frac{(s_2{-}s_A)(s{-}s_2)
    \left[t'_2(s_2){-}t'_4(s_2)\right]}{s{-}s_A},\nonumber
\end{eqnarray}
where $s_A$ is the position of the Adler zero in the $s$-plane, and
$s_2$ its LO approximation.
The standard IAM
is recovered for $A^{mIAM}=0$, which holds exactly for
all partial waves except for the scalar ones. In the usual IAM
derivation  \cite{Dobado:1996ps} that we have followed above,
$A^{mIAM}$ is neglected,
since it formally yields a NNLO contribution and  is numerically
very small, except near the Adler zero, where it diverges. However,
if $A^{mIAM}$ is neglected, the IAM Adler zero occurs at $s_2$,
correct only to LO, it is a double zero instead of a simple one and
a spurious pole of the amplitude
appears close to the Adler zero. All of these caveats are removed
with the mIAM. The differences in the 
physical and resonance region between the IAM and the mIAM are less
than 1\%. Thus, the only real application of the mIAM
will be in Sec.\ref{subsec:qm} for large $M_\pi$, since as we will see
 the $\sigma$ 
poles ``split'' into two virtual poles below threshold, 
one of them approaching the unphysical Adler zero region, where the standard IAM fails. 

In summary, we have introduced the elastic Inverse Amplitude Method,
which  combines ChPT and dispersion relations,
within the elastic approximation. Within this formalism
the elastic cut and unitarity are implemented exactly and 
the ChPT expansion is recovered up to the desired order.
In its derivation ChPT is only used 
for the subtraction constants or the left cut, 
which is justified, since the former correspond to 
values of the amplitude at very low energies and the second is subtracted
to suppress the high energy contribution. 
Also, crossing symmetry is not exact, but holds up to the order to which the left cut
contribution is being calculated within ChPT.

We have dedicated a long section to this approach,
not only because it has been very successful in describing
elastic meson-meson and pion-nucleon scattering, but 
because it is able to generate the poles of the 
resonances that appear in these elastic processes {\it without any a priori assumption about their existence or nature}. 
Moreover, note that 
{\it no additional parameters are needed beyond the LECs of ChPT},
which encode the underlying dynamics QCD.
In particular the result is still fully renormalized and there is no residual dependence on the regularization scale. 
This will allow us to investigate the dependence of resonance properties
on QCD parameters such as quark masses and the number of colors
in Secs.\ref{subsec:qm}, \ref{subsec:nc}, \ref{subsec:ncmodelindep} 
and \ref{subsec:ncUChPT}. 

\subsubsection{Chiral unitarization method with crossing symmetry constraints}

As we have seen, as soon as one imposes elastic unitarity exactly on the physical cut, since one is dealing differently
with the left cut, crossing symmetry is violated. However, in \cite{Zhou:2004ms} a unitarization method was combined with crossing symmetry constraints in order to determine the $\sigma$ parameters. Unitarity was guaranteed  by writing
the S matrix as a product $S=\Pi_i S^{p_i} S^{cut}$, where $S^{p_i}$ contained simple unitary expressions for isolated poles.
The number of poles to use in this part of the S-matrix has to be decided from the outset,
whereas $S^{cut}=\exp(2i\sigma(s)f(s)$. The $f(s)$ function represents the non-resonant part and has a left and a right cut, so that
it obeys a dispersion relation similar to that in Eq.\ref{Tcauchy-real}, 
although for just one variable. Being  $f(s)$ basically  the logarithm of the $S$ matrix, its dispersion relation only needs one subtraction.
The left hand cut is then calculated using ChPT. Much as it happened with the IAM, this approximation 
is only valid at low energies. In the IAM the use of the inverse function meant that the left cut converges with the three subtractions, but in this case, with only one subtraction, the authors imposed a cutoff of the order of 2 GeV$^2$ in both the left and right-cut integrals.

Of course, this does not ensure crossing, but the authors of \cite{Zhou:2004ms} then performed fits to data and
to the the so-called Balachandran-Nuyts-Roskies
crossing symmetry relations \cite{BNR} until they were satisfied at the level of 1\%.
 This turns out into a simpler dispersive approach than Roy or GKPY equations, while improving 
the treatment of crossing symmetry from most unitarized models, although nevertheless crossing symmetry is not exact.

The analytic properties of the model allow for a sound continuation to the complex plane and 
the determination of  a $\sigma$ pole at
$M_\sigma=470\pm50\,$MeV and $\Gamma_\sigma=570\pm50\,$MeV, which we included in the fifth line of Table \ref{tab:otherpoles}.
This result has larger uncertainties than other entries in that table, but it is actually more reliable than many others due to 
a careful analysis of the systematic errors induced by various theoretical uncertainties, like different cutoffs, the treatment
of the scattering data in the vector partial wave, how to weight the constraints from crossing symmetry, the inclusion of two or three poles in $S^{p_i}$, etc.

Interestingly, it was also shown that without the constraints from crossing sum rules, this unitarization method would lead to 
$M_\sigma=542\pm50\,$MeV and $\Gamma_\sigma=546\pm50\,$MeV, showing once more the relevance of crossing symmetry for a precise determination of the $\sigma$ pole, which was achieved later by means of Roy and GKPY equations, as explained in the previous section.

\subsection{Unitarization: Meson-meson coupled channels}
\label{subsec:coupleduchpt}

So far we have only studied the elastic channel.  In practice  this is
quite enough to have a maybe not very precise but 
fair representation of the $f_0(500)$ meson,  since
$\pi\pi$ scattering is elastic 
up to the $K\bar K$ threshold, i.e. $\sim$ 1 GeV.

However in order to understand other properties of this meson,
like its classification in SU(3) multiplets, it is also relevant
to take into account other channels
and even some other meson-meson scattering 
amplitudes, either because they couple to $\pi\pi$ scattering
or because it is in those other channels where the multiplet partners of the $\sigma$ appear.
For all means and purposes, it will be enough to
consider coupled scattering of two-meson states.
As we have already commented $4\pi$, $6\pi$, ... states, 
are almost negligible in practice below 1.4 GeV.

We already presented in Eq.\ref{ec:unitarizedmatrix} the general form
of the $T$ matrix on the real axis that satisfies coupled channel unitarity when several
two-body states are energetically accessible. We only need 
an approximation to $\re T^{-1}$. 
In what follows we will present different coupled channel methods, 
which  on the real axis can be understood as different approximations to  the
ChPT series  $\re T^{-1}=T_2^{-1}(1+T_4 T_2^{-1} ...) $. Note that these are  called ``approximations to the series" because 
for simplicity and practicality some of these methods neglect parts of 
the contributions to higher terms in these series. Typically, 
these  methods differ in their treatment 
of left cuts and/or the 
inclusion of additional parameters besides those of ChPT, being in this respect
generally more relaxed that in the elastic case. 

\color{black}
Before describing in the following subsections
the particulars of each coupled channel unitarization technique,
let us emphasize that
all them provide very similar results for meson-meson scattering. In particular, they all
find a light scalar nonet made of the $\sigma$, $\kappa$, $f_0(980)$ and $a_0(980)$ whose
generation is dominated by the meson-meson loop dynamics, which is driven by the leading order
interactions of the ChPT Lagrangian.  In contrast, vector meson poles require information on the NLO LECs of ChPT 
or, alternatively, have to be introduced by hand from a Lagrangian.
Those unitarization methods reaching beyond 1.2 GeV
also find that the data can be described with just one ``preexisting'' (surviving in the large $N_c$ limit) scalar nonet around 1 GeV. 
These results are very robust and the effects of the neglected or 
approximated left cuts and the many-meson intermediate channels 
do not change this basic scenario.
\color{black}

\subsubsection{The $N/D$ Method}

The classic $N/D$ method was derived in 1960 by Mandelstam and Chew \cite{Chew:1960iv}, precisely to explain the $\pi\pi$ low energy interaction. 
Although the derivation starts 
from the double-dispersive representation it finally yields relatively simple integral equations. These lead to unitary amplitudes on the real axis. As usual, the left cuts are approximated and more often neglected. Therefore although based on dispersion relations, in practice it is not as well suited for precision studies
as Roy-like or Forward Dispersion Relations. However it is very useful for understanding the origin and nature
of resonances.  Let us briefly sketch the derivation here.
\color{black} In principle, all other methods can be recast into this one, 
although this might involve further integral equations and approximations,
and is more practical to use directly the formulas provided by the methods themselves.
\color{black}

The $N/D$ method provides a solution to the unitarity condition in
Eqs.\ref{ec:unit} or \ref{ec:matrixunit}, depending on whether one is in an elastic or inelastic regime, respectively. 
For simplicity let us concentrate first
on the elastic case. In order to cancel explicitly the vanishing
threshold behavior of the partial wave, let us define the ``reduced'' amplitude 
$\hat t_{IJ}(s)=t(s)/k^{2J}$.
For the sake of brevity, let us also drop the $IJ$ subindices. 
The reduced amplitude can then be written as $\hat t(s)=\hat N(s)/\hat D(s)$,
where $\hat D(s)$ only contains the right hand cut
and $\hat N(s)$ only the left hand cut of $t(s)$.
Of course we are free to multiply $\hat D(s)$ and $\hat N(s)$ by the same factor
without affecting $\hat t(s)$. 
Thus, any pole 
can be removed from either $\hat D(s)$ or $\hat N(s)$ by introducing in the other function the corresponding zeros
\color{black} and customarily $\hat N(s)$ is chosen free of poles, which can nevertheless appear in $\hat D(s)$. \color{black}

Now, due to their analytic structure in the complex $s$ plane, in the real axis below threshold
$\hat D(s)$ is real 
whereas $\hat N(s)$ is real outside the left cut.
In addition, above threshold $\im \hat D(s)=\hat N(s) \im \hat t(s)^{-1} =-\sigma(s)\hat
N(s) k^{2J}$  due to unitarity. In contrast $\im \hat N(s)=\hat D(s) \im \hat t(s)$ over the left cut.
This implies that the following dispersion relations can be written:
\begin{eqnarray}
\label{N-disp}
\hat N(s)&=&\sum_{m=0}^{n-J-1}{a^\prime_ms^m}+\frac{(s-s_0)^{n-J}}{\pi}\int^{s_{L}}_{-\infty}{ds^\prime\frac{\hat
    D(s^\prime) \im\hat t(s^\prime)}{k^{2J}(s^\prime)(s^\prime-s)(s^\prime -s_0)^{n-J}}},\\
\label{D-disp} \hat D(s)&=&\sum_{m=0}^{n-1}{\hat a_ms^m}-\frac{(s-s_0)^n}{\pi}\int_{s_{th}}^{\infty}{ds^\prime\frac{k^{2J}(s^\prime)\sigma(s^\prime)\hat
    N(s^\prime)}{(s^\prime-s)(s^\prime -s_0)^n}}+\sum_i \frac{\gamma_i}{(s-s_i)}.
\end{eqnarray}
where $s_L$ is the branching point of the left cut (0 for $\pi\pi$ scattering but could be different for other processes), $n$ is the number of subtractions required to have
$\hat N(s)/s^{n-J}\rightarrow 0$ when $s\rightarrow \infty$. 
Apart from $\mathrm{Im}\, \hat t(s)$ on the
left hand cut, the subtraction constants of  $\hat D(s)$
and $\hat N(s)$ make all the input needed. \color{black}
The poles in $\hat D(s)$ are known as 
Castillejo-Dalitz-Dyson (CDD) poles \cite{Castillejo:1955ed} and lead to zeros in the partial wave.
Of course, it is always possible to multiply $N(s)$ and $D(s)$ by a polynomial to convert these poles into subtraction constants
\cite{libropipi,MartinSpearman}. \color{black}

Frequently, Eqs.~\ref{N-disp} and \ref{D-disp} are solved by 
 setting $\im \hat t=0$ on the left hand cut, which is therefore neglected. 
 In such case it is possible to take $\hat N(s)=1$,
by including all the possible zeros of the polynomial
$\sum_{m=0}^{n-J-1}{a^\prime_ms^m}$ as CDD poles in $\hat D(s)$, so that:
\begin{equation}\label{ND-no-LC}
\hat t(s)=\frac{1}{\hat D(s)},\quad 
\hat D(s)=\sum_{m=0}^{J}{a_ms^m}+\sum_i^{M_J}{\frac{R_i}{s-s_i}}-\frac{(s-s_0)^{J+1}}{\pi}\int_{s_{th}}^{\infty}{ds^\prime\frac{k^{2J}(s^\prime)\sigma(s^\prime)}{(s^\prime-s)(s^\prime -s_0)^{J+1}}},
\end{equation}
where we have taken into account that $n=J+1$, 
and the subtraction constants $a_m$ and the CDD pole residues have been redefined and renamed. With these conventions,
Eq.~\ref{ND-no-LC} is the most general structure of an elastic partial wave
of angular momentum $J$ when the left hand cut is neglected \cite{Oller:1998zr}. 

Once again, in the real axis above threshold  Eq.\ref{ND-no-LC} can be recast into the 
general form of the elastic unitary partial wave provided in Eq.\ref{ec:generalunit}.
It is enough to note first that the imaginary part of the integral in Eq.\ref{ND-no-LC} 
is such that $\im \hat D(s)/k^{2J}=-\sigma(s)$ above threshold.
Second, we identify  $\re t^{-1}(s)=\re \hat D(s)/k^{2J}$.
Of course, the advantage of the $N/D$ method is that we have a dispersive
representation where the right cut of the inverse amplitude is treated exactly within the elastic approximation.
It is also free of the $s=0$ singularities of the K-matrix, since the right-cut structure is that of the
dispersive integral, whose imaginary part only coincides with $\sigma(s)$ on the real axis above threshold.
If the left cut had not been neglected, it should have been introduced perturbatively through $\hat N(s)$, which is rather cumbersome.

At this point is where QCD dynamics comes into play to give meaning to the
subtraction constants in  Eq.~\ref{ND-no-LC}. 
In particular, in \cite{Oller:1998zr} 
these were split into $a_m=a_m^L+a_m^{SL}$,
where  the term $a_m^L$ is $O (N_c)$ and $a_m^{SL}$ 
is $O(1)$ in the QCD $1/N_c$ expansion.  
As we will see in Subsec~\ref{subsec:ncmodelindep}, 
this is due to the fact that in the $1/N_c$ expansion the meson-meson amplitude is generically $O(1/N_c)$. 
In addition, the integral of Eq.~\ref{ND-no-LC} is $O(1)$. 
Therefore when $N_c\rightarrow \infty$, Eq.~\ref{ND-no-LC}
becomes:
\begin{equation}\label{Dinf}
 \hat D^\infty(s)=\sum_{m=0}^{J}{a_m^Ls^m}+\sum_i^{M_J^\infty}{\frac{R^\infty_i}{s-s_i}}\equiv k^{2J}/t^\infty(s),
\end{equation}
where $a_m^L$ and $R^\infty$ are  the $N_c$ leading parts of $a_m$ and $R_i$, respectively, and $M_J^\infty$ the
number of leading CDD poles. Now, defining 
\begin{equation}\label{gND}
g_J(s)k^{2J}\equiv\sum_{m=0}^{J}{a^{SL}_ms^m}-\frac{(s-s_0)^{J+1}}{\pi}\int_{s_{th}}^\infty{ds^\prime\frac{k^{2J}\sigma{s^\prime}}{(s^\prime-s)(s^\prime
    -s_0)^{L+1}}},
\end{equation}
we arrive at
\begin{equation}\label{ND-t}
t(s)=\left[1/t^\infty(s)+g_J(s)\right]^{-1}.
\end{equation}
Note that the loop contributions are all inside $g(s)$
and therefore all the terms in $t^\infty$ should correspond 
to tree level structures, which in addition are leading in the $1/N_c$ expansion. Crudely speaking, these structures {\it exist before} unitarization and thus the poles in $t^\infty(s)$ are informally called
``preexisting resonances'', 
which survive in the $N_c\rightarrow\infty$ limit.
The unitarization is accomplished through the function $g_J(s)$,
since $\im 1/t(s)=\im g_J(s)=-\sigma(s)$, which on the real axis leads to the general form of a unitary amplitude in Eq.\ref{ec:generalunit}. Note that this $g(s)$ function is divergent and therefore depends on a renormalization scale. Dimensional regularization, instead of a cutoff, is customarily used, but the scale dependence remains in $g_J(s)$.

This formalism can be easily generalized to
scattering processes with multiple coupled channels by
employing the usual matrix notation \cite{Oller:1998zr}. 
Then $t$,  $t^\infty$
become $T$, $T^\infty$ and $g_J$ is a diagonal matrix
whose elements are given by Eq.~\ref{gND},
evaluated with the appropriate masses for each channel.
Once again $\im g_J(s)=-\Sigma(s)$ and the $N/D$ method neglecting the left cut can be easily recast into the general matrix form for a unitary $T$-matrix in Eq.\ref{ec:unitarizedmatrix}. 

\begin{figure}
  \centering
  \includegraphics[width=0.45\textwidth]{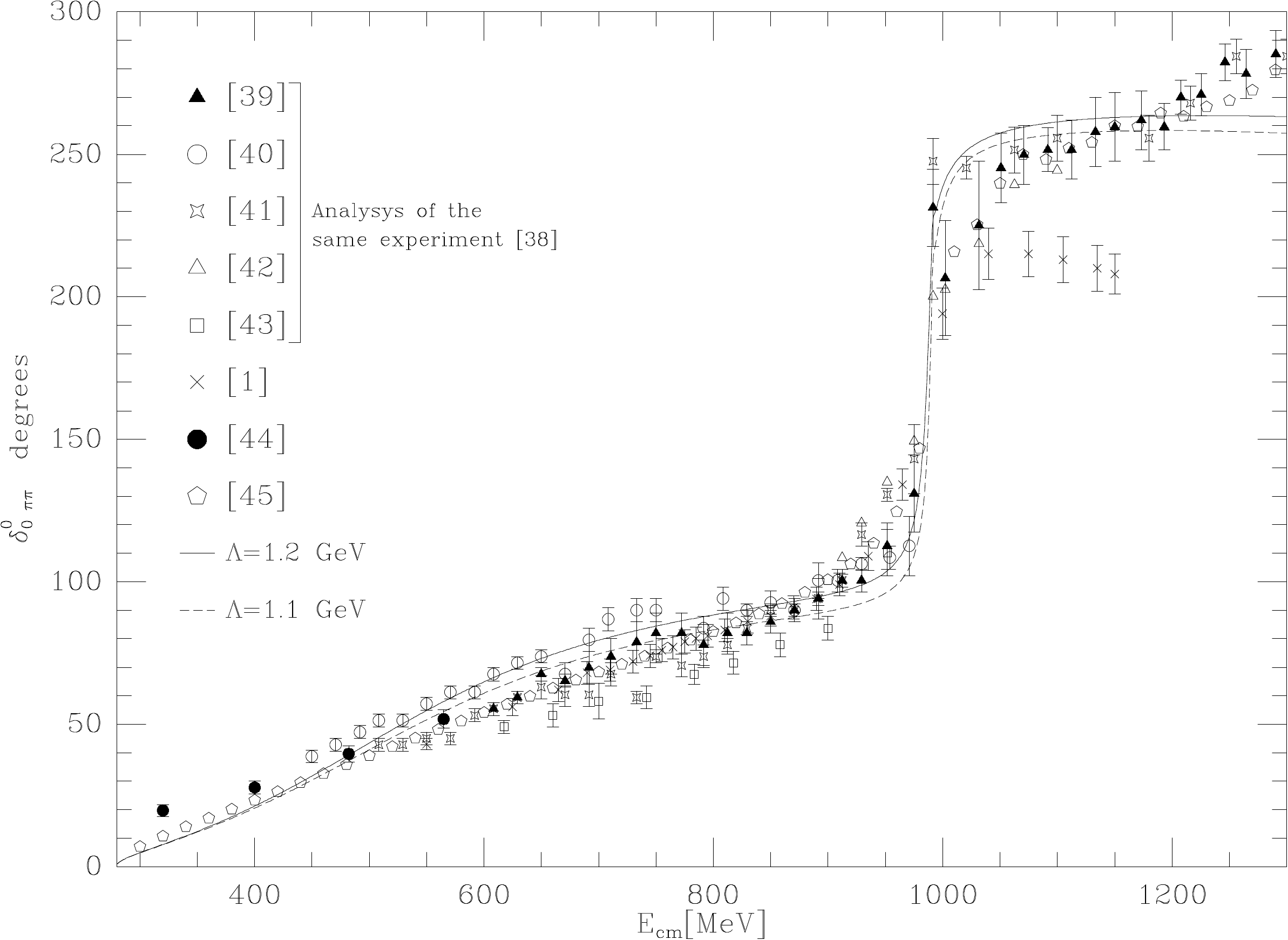}
  \includegraphics[width=0.45\textwidth]{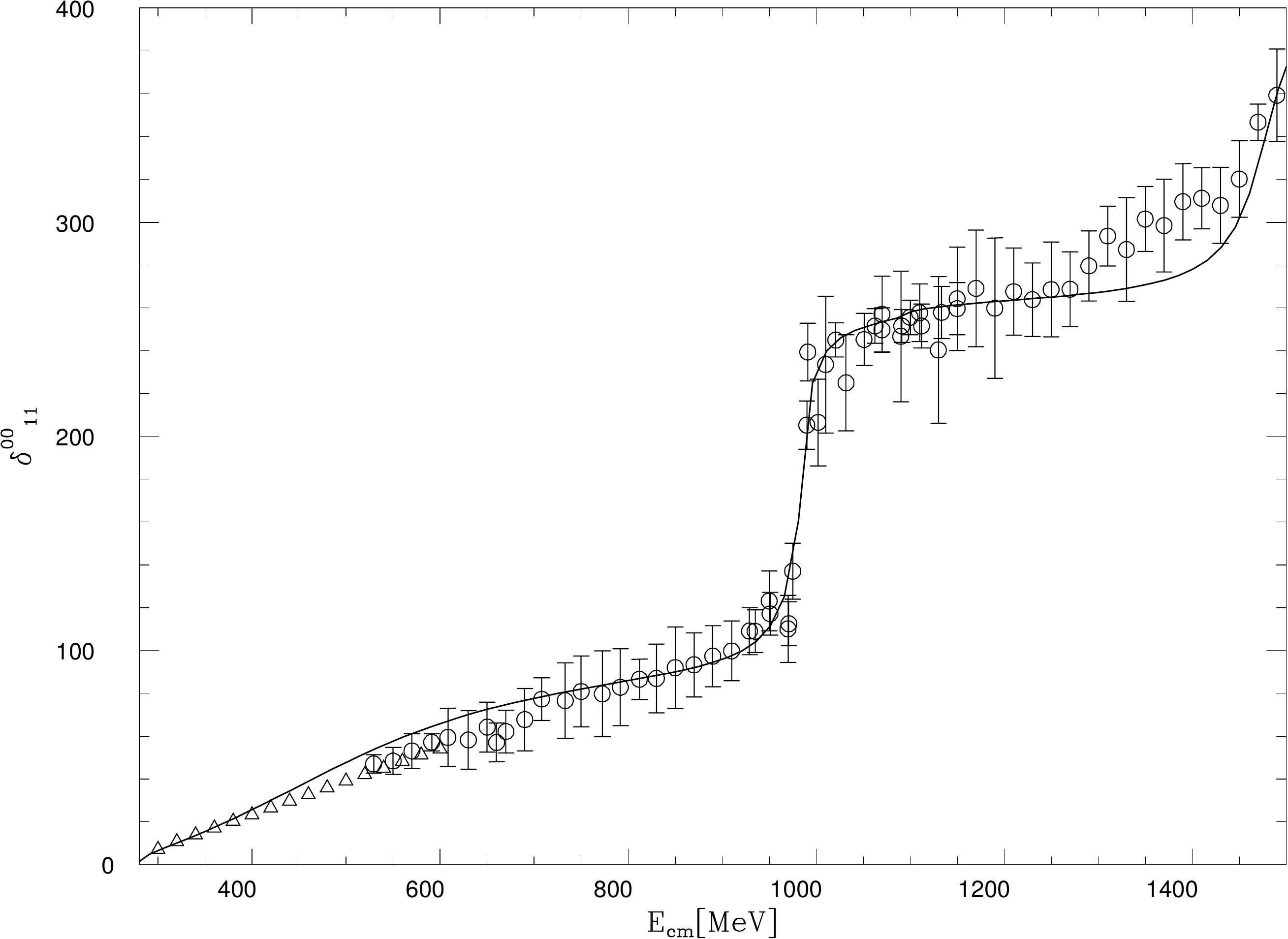}
\caption{ $S0$-wave $\pi\pi$ scattering. Left (from \cite{Oller:1997ti}): described by unitarizing LO ChPT
with the Chiral Unitary Approach, Eq.\ref{LS}. Note that  the cutoff is  the only 
free parameter and comes out of order $4\pi f_\pi\simeq 1.2\gev$. The citations
in the plot correspond to the original figure in \cite{Oller:1997ti}.
Right (from \cite{Oller:1998zr}): the same wave unitarized with the $N/D$ method \cite{Oller:1998zr}, which includes LO ChPT
as well as an explicit multiplet of ``preexisting'' scalar resonances above 1 GeV, which allows 
for a fair description of data up to 1500 MeV.
Left figure reprinted from  Nucl.\ Phys.\ A {\bf 620}, 438 (1997), J.~A.~Oller and E.~Oset,
  ``Chiral symmetry amplitudes in the S-wave isoscalar and isovector  channels
 and the $\sigma$, $f_0(980)$, $a_0(980)$ scalar mesons". Copyright 1997, with permission from Elsevier.
 Right: reprinted figure with permission from    J.~A.~Oller and E.~Oset,
  Phys.\ Rev.\ D {\bf 60} (1999) 074023. Copyright 1999 by the American Physical Society.
}
  \label{fig:ChUAND}
\end{figure}

In \cite{Oller:1998zr},  $t^\infty$ was obtained 
from the lowest order ChPT Lagrangian
and the poles 
from the exchange of resonances at tree level in the $s$-channel.
Namely $T^\infty=T^{(2)}(s)+T^{\mathrm{RES}}$.
The tree level exchange of resonances was  described
with  the same Lagrangian used to study the Resonance 
Saturation Hypothesis \cite{Ecker:1988te}
that we discussed in Subsec.\ref{subsec:chpt}, although their couplings and masses were fitted to data
after unitarization. 
The exchange of resonances
in crossed channels was not included, consistently with 
neglecting the left cut, and the residual 
dimensional regularization scale was set to 
the natural value $\mu=M_\rho$.
The very nice result for the S0-wave, which is the one of interest for this review, is 
shown on the right panel of Fig.\ref{fig:ChUAND}. There we can see that the 
inclusion of explicit resonances in the unitarization process has allowed to extend the description 
of $\pi\pi$ scattering up to 1.5 GeV. The price to pay, of course, is the presence of explicit
resonances and their couplings in the Lagrangian, in contrast to the simple Chiral Unitary Approach
that we will review next, which only has one free parameter (the cutoff).

Now, apart from the relatively nice description of the data,
the relevance of this approach \cite{Oller:1998zr} is that, apart from 
 the ``preexisting'' poles, 
which have acquired a width due to unitarization, new poles appear.
These new poles are sometimes called ``dynamically generated'', which
can be confusing for people outside the field. Of course the only dynamics is that of QCD. However, the name ``dynamically generated'' 
is often used in the community meaning that the dominant dynamics that generates these
poles is mainly due to unitarization, namely, to loops with LO ChPT vertices,
 which intuitively means meson-meson physics rather than quark and gluon physics. 
 The resonances  generated predominantly
by this kind of effects do not survive in the large-$N_c$ limit 
(at least with parameters similar to those in $N_c=3$, since
they could contain a small mixture of ``preexisting'' states, that would survive that limit, although at a rather different position). \color{black} In hindsight \footnote{I thank J.A. Oller and E. Oset for this clarification.}, within this approach, calling a state ``preexisting'' actually means that it cannot be ``dynamically generated'' {\it from the scattering of two light NGB-pseudoscalars in the scalar channel}. But this does not exclude the possibility that it might be dynamically generated from other channels. Of course, the whole point of this discussion is that the $\sigma$ and the other light scalar resonances {\it can be naturally described as``dynamically generated'' from the scattering of two light NGB-pseudoscalars in the scalar channel}. Whether the preexisting states are genuinely preexisting or dynamically generated from other channels is interesting, but accessory to the main topic of this report. The relevant point is that in scalar meson-meson scattering they have to be included explicitly. \color{black}

The results from \cite{Oller:1998zr} showed that including two preexistent nonets was not demanded by data, since one of them would almost decouple, would have an almost undetermined mass and would yield narrow unobserved resonances. 
Thus, only one ``preexisting'' nonet was needed, and its mass was slightly above 1 GeV, clearly separated from
the physical mass of the $\sigma$ meson around 500 MeV. In addition the $N/D$ method also yielded some
poles dynamically generated from the loops. 
The  clearly ``dynamically generated'' poles were the $\sigma$, the $a_0(980)$ and the $\kappa$, whereas the $f_0(980)$, being so close to the ``preexisting'' nonet, was generated from a mixture of dynamics of the two kinds. 
The effect of the exchange of resonances in crossed channels was also 
estimated in \cite{Oller:1998zr} and it did not seem to alter significantly the results. The method was also extended to the vector sector, where both the $\rho(770)$ and $K^*(892)$ were identified  as ``preexisting'' resonances.

Therefore, the spectroscopic picture of two light scalar nonets 
generated with different dynamics
received a strong support. The lightest one is predominantly 
due to rescattering effects, or unitarization, whereas the 
heavier one, slightly above 1 GeV is predominantly 
made of  ``preexisting'' states that will survive 
the large $N_c$ limit, although becoming very narrow.

This two-nonet approach has been revisited 
and/or refined in the literature, for instance 
from the BSE perspective in \cite{Nieves:1998hp}, as already commented in Subsec.\ref{subsec:ChUABSE}, even including perturbative left cuts, or by adding $\sigma\sigma$ and $\rho\rho$ states
in \cite{Albaladejo:2008qa}. Moreover, there are extensions to $U(3)$ 
that include the $\eta'$ and then study the $\sigma$ and other light s-wave resonances within this unitarization approach
 \cite{Beisert:2003zd,Guo:2011pa}. In some cases the cutoff dependence is 
recast in the form of subtraction constants.
In any case, the basic picture of two nonets
is always found, although the specifics of the more massive states
can change slightly. This has led, for instance, to an identification of a possible glueball candidate around 1.7 GeV with these methods \cite{Albaladejo:2008qa}. 

Of course, one of the main ingredients of this approach is the explicit inclusion of the ``preexisting resonances'' 
and the assumption of their specific $N_c$ dependence, as well as the minimal couplings in \cite{Ecker:1988te}. In 
the subsections that follow we will
review other unitarization methods, which do not introduce 
explicit resonances but just the LECs of ChPT without any spurious parameters. 
Of course this will limit the applicability to 1.2 GeV at most, 
but will be useful also for other purposes.

\color{black}
As a final remark, let us note that, in principle, 
the techniques to be described next can be recast into an $N/D$ form.
This is straightforward in the elastic case, since these methods provide a 
relatively simple parameterization of each partial wave $t(s)$ and its phase shift
$\delta(s)$. In such case, a $D(s)$ function, with the correct right-hand cut structure, is given by
the Mushkelishvili-Omn\'es function \cite{Omnes}\footnote{In Sec.\cite{Omnes} we 
show will present the Mushkelishvili-Omn\'es method in more detail
to describe form factors for $\sigma\rightarrow\gamma\gamma$ decays.}:
\begin{equation}
D(s)=\exp\left(-\frac{s}{\pi}\int_{s_{th}}^\infty ds'\frac{\delta(s')}{s'(s'-s)}\right),
\end{equation}
where the integral can be subtracted or CDD pole terms can be added if needed.
Hence we define $N(s)=t(s) D(s)$, thus carrying the left-hand cut singularities if the 
method has not neglected them. One can deal with the coupled channel case similarly, but in matrix form.

In practice, of course, one uses the simpler expressions given by each one of the particular 
approaches that are described next.
\color{black}

\subsubsection{The Chiral Unitary Approach and Bethe-Salpeter Equations}
\label{subsec:ChUABSE}

We already saw in Subsec.\ref{subsec:simplest} that a 
qualitatively correct $\sigma$ pole could be generated 
with the simplest unitarized model, Eq.\ref{ec:simplestunit},
which uses just the LO ChPT result. 
The mass came surprisingly close to its actual value 
although the width came a factor of two too large. 
In contrast, both the mass and the width of the vector meson obtained with the same method were too different from the physical values for the $\rho(770)$. This suggests that the NLO ChPT contribution in the IAM 
plays a relatively small role for the scalar waves, 
but an important one for the vector ones.

Thus, if one is only interested in the scalar waves, one could 
make the radical approximation that the only interaction 
vertices that are needed are those of LO ChPT \cite{Oller:1997ti}. 
Let us first study this approximation and then how to extend the treatment to include higher orders.

If only $T_2$ amplitudes are to be considered, since they come from tree level diagrams that do not have logarithms
nor cut singularities, they can be identified
with the potential $V$ in the 
Bethe-Salpeter equations  (BSE, similar to the
Lippmann-Schwinger equations but in a relativistic framework), which read 
$T=T_2+\overline{T_2GT}$. Here,
 $T_2(p_1,p_2;q)$ is the matrix of $O(p^2)$ ChPT partial waves
for two incoming mesons
with four-momenta $p_i$ and two outgoing mesons, one of them with four-momentum $q$. The second term in the equation is defined as:
\begin{equation}
(\overline{T_2GT})_{il}=i \int \frac{d^4q}{(2 \pi)^4} 
\frac{T_{2\;ij}(p_1,p_2;q)}{q^2-m^2_{1j}+i\epsilon} 
\frac{T_{jl}(q;p'_1,p'_2)}{(P-q)^2-m^2_{2j}+i\epsilon}.
\label{VGT}
\end{equation}
The subindex $i$ corresponds to the $\vert i\rangle$ state, which consists of two
mesons with total four-momentum $P$, one with mass $m_{1i}$ and initial four-momentum $p_1$ whereas 
 the other one has mass $m_{2i}$ and initial four-momentum $p_2$.

A relevant remark \cite{Oller:1997ti} is that if $T_2$ 
is separated in an on-shell part plus an off-shell term, the latter, 
when used inside Eq.\ref{VGT}, does not have to be calculated,
since to a very good approximation 
it can be reabsorbed numerically into the definition of masses and decay constants. 
This is called the ``on-shell 
factorization'' approximation of $T_2$ and $T$ from Eq.\ref{VGT} 
reducing the BSE to pure algebraic relations, i.e. $T=T_2+T_2 G T$,
where
$G$ is a diagonal matrix given by
\begin{equation}
G_{ii}=i \int \frac{d^4 q}{(2 \pi)^4} \frac{1}{q^2-m^2_{1i}+i\epsilon}
\frac{1}{(P-q)^2-m^2_{2i}+i\epsilon}. 
\label{Gii}
\end{equation}
Diagrammatically this function corresponds to the ``bubble'' created by the two circulating mesons
in diagrams of type ``c'' in Fig.\ref{fig:ChPTdiagrams}. The on-shell
factorization has simply factorized the external legs out from this bubble.
\color{black} A different derivation of this ``on-shell factorization'' 
starting from unitarity and using a dispersion relation for $T^{-1}$
can be found in \cite{Oller:1998zr,Oller:2000fj}. \color{black} 
Note that the above integral diverges and has to be regularized. 
 A cutoff was used in the seminal works \cite{Oller:1997ti,Oller:1997ng}, but it can be easily translated into a dimensional regularization scale \cite{Oller:1998hw}. For simplicity we will often refer to this scale as ``cutoff''. 
The solution of these algebraic equations is:
\begin{equation} 
T=[1-T_2 G ]^{-1} T_2,
\label{LS}
\end{equation}
which is known as the Chiral Unitary Approach \cite{Oller:1997ti}.
\color{black}
Having zero range interactions, this is actually equivalent to the 
so-called the Chew-Mandelstam method \cite{Chew:1960iv,libropipi} rather than 
to what is usually understood by BSE. 
\color{black}
Formally, it can be reinterpreted as the geometric series 
$T=T_2+T_2 GT_2+T_2GT_2GT_2+T_2GT_2GT_2GT_2....$, so that this method
can be diagrammatically
interpreted as the {\it resummation} of all diagrams with one, two, three ... bubbles
in the s-channel, with only $O(p^2)$ vertices to connect them.
Since $\im G=\Sigma=-\im T^{-1}$ this method is very similar 
to the simplest unitarization method of Eq.\ref{ec:simplestunit} in matrix form. 
Unitarity in coupled channels is therefore ensured. 
In addition, and in contrast to the K-matrix approach, it is free from $s=0$ singularities
thanks to using the $G(s)_{ii}$ functions.
Actually, Eq.\ref{LS} above can be recast into the
generic coupled channel form,  Eq.\ref{ec:unitarizedmatrix}.
First one has to neglect all diagrams in the chiral series
depicted in Fig.\ref{fig:ChPTdiagrams}, except
the ``c'' or ``bubble'' diagrams, i.e., $T_4\simeq T_4^c$. In a second step 
one uses the ``on-shell'' factorization to obtain
$T_4\simeq T_4^c\simeq T_2 G T_2$ with everything written
in terms of physical masses and coupling constants.
Thus the Chiral Unitary approach amounts to the following approximation: 
$\re T^{-1}= T_2^{-1}- \re G ....$.

\begin{figure}
  \centering
  \includegraphics[width=0.4\textwidth]{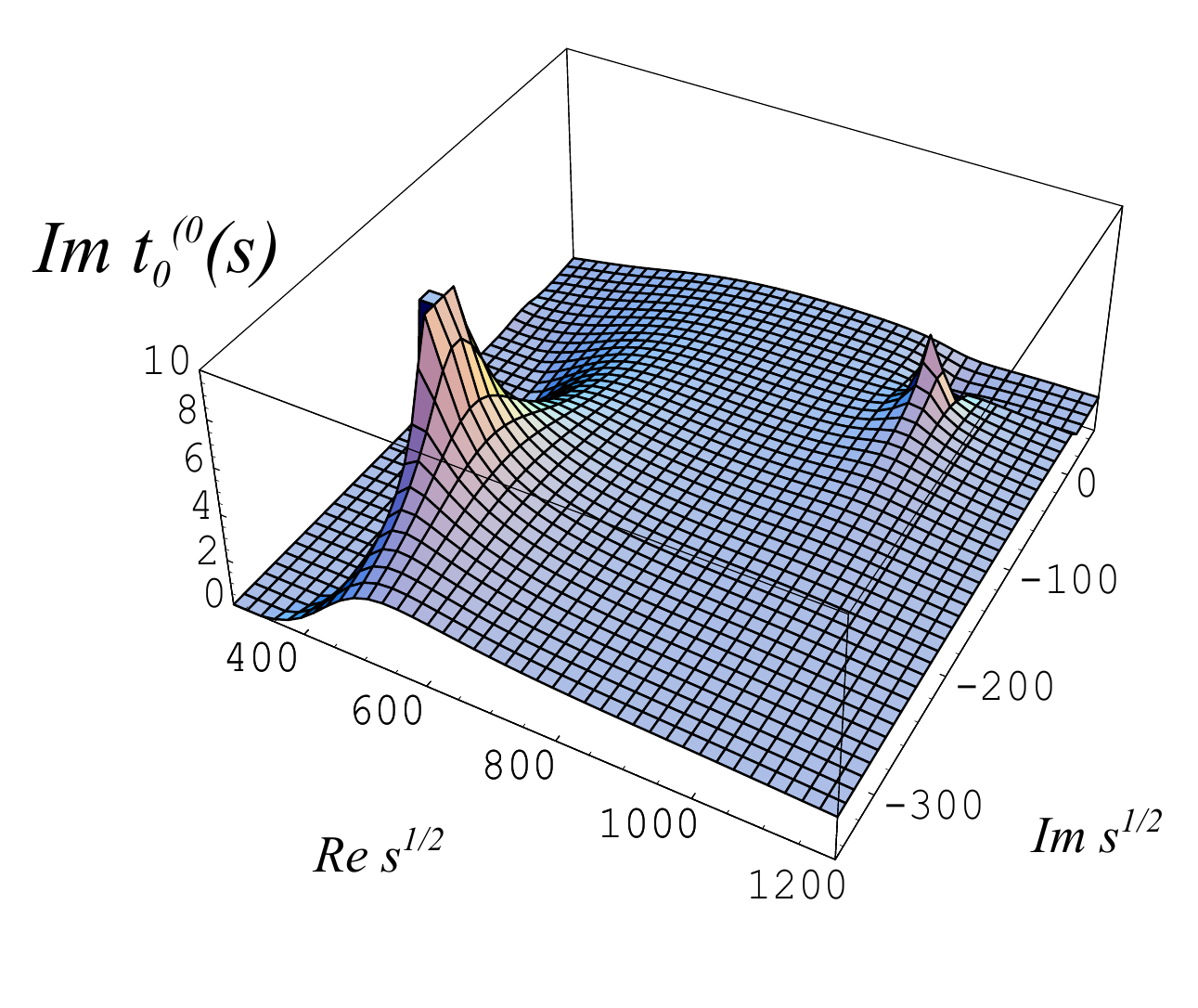}
  \includegraphics[width=0.4\textwidth]{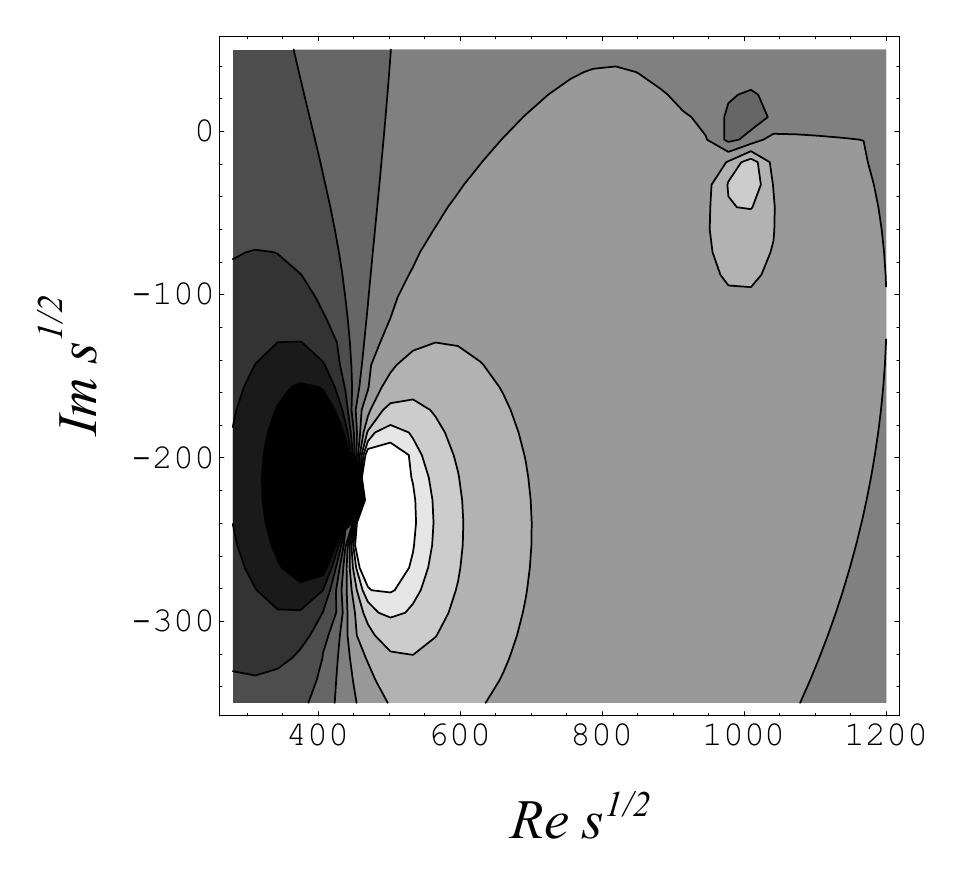}
\vspace*{-.5cm}
\caption{ Surface and contour plots showing the imaginary part of the $\pi\pi$ scattering scalar-isoscalar partial wave. Note
the poles of the $\sigma$ and $f_0(980)$ in the lower half plane of the second Riemann sheet and how far from the real axis the former is compared to the latter. This result is generic of 
coupled channel formalisms for unitarizing ChPT.
Figure taken from \cite{Oller:1998hw}.  
}
  \label{fig:OOP-Fig9}
\end{figure}

Note that 
$\re G(s)$ provides an additional higher order  contribution
beyond the simple LO that appeared in the simplest unitarization scheme
and that this new contribution depends on a cutoff. 
Now, recall that we have already seen that the simplest LO unitarization
provided a fair qualitative approximation to the scalar channels, so that the
 NLO and higher order contributions should be relatively small. 
However, this small new contribution made it possible  in \cite{Oller:1997ti}  to describe 
remarkably well the existing data 
on the isoscalar and isovector 
scalar waves of $\pi\pi\rightarrow\pi\pi$ and $\pi\pi\rightarrow K\bar K$
scattering by means of Eq.\ref{LS} up to 1250 MeV
with a unique cutoff of order 1 GeV for all partial waves. 
 The result for the $\pi\pi$ $S0$-wave can be seen on the left panel of 
 Fig.\ref{fig:ChUAND}.
\color{black} In \cite{Kaiser:1998fi}, It was noted that in the region below 1 GeV, the resulting phase seems to prefer the highest data 
points, which is in some tension with Roy Equation analyses. However, the result is remarkable taking into account that it only depends on one parameter. \color{black}
In addition, the poles of the $\sigma$ and $f_0(980)$  mesons,  shown in Fig.\ref{fig:OOP-Fig9},
 were generated simultaneously in the scalar-isoscalar wave whereas that of the $a_0(980)$ appeared in the scalar-isovector wave.
\color{black} In a follow up of this approach \cite{Locher:1997gr}, 
the same two pairs of conjugated poles were found and a''molecular'' nature of the $f_0(980)$ was advocated from the position of additional poles in other Riemann sheets. \color{black}

 In contrast, it was not possible to generate the vector mesons with such a natural cutoff only, since they required taking into account the ``polynomial contribution'' from the ``b'' diagrams in Fig.\ref{fig:ChPTdiagrams}, which contains the $L_i$. In the literature these two different 
generation mechanisms have been 
called ``dynamical generation'' 
versus ``dynamical reconstruction'' \cite{Giacosa:2009bj}.
This emphasizes that 
in the former case, i.e. for the $\sigma$, 
it is  mainly the loop dynamics that generate
the resonance, with very little or at least not explicit
information about it in the Lagrangian. In contrast,
in the latter case, some  
information on the resonance is needed before unitarization,
which can  be either contained in the LECs, as we have just seen 
with the IAM and the $\rho(770)$,
or, as we will see later, 
by including explicitly in the Lagrangian 
some 'bare' approximation to the
state, for instance its large-$N_c$ limit. 
For a general discussion on these two mechanisms, see \cite{Giacosa:2009bj}.

Actually it was soon shown \cite{Oller:1997ng} that by considering also the tree level NLO ``b" diagrams in Fig.\ref{fig:ChPTdiagrams}
it was possible to describe not only the isoscalar and isovector S-waves, but the P waves and isotensor waves simultaneously
and up to 1.2 GeV. In terms of the graphs in 
Fig.\ref{fig:ChPTdiagrams} this amounts to considering the approximation $\re T\simeq T_2+T_4^b+T_2 GT_2...$ to the full NLO result. 
Interestingly, the $\rho(770)$ and $K^*(892)$ vector meson poles where found
simultaneously with those of the full light scalar nonet,
comprising the $\sigma$, $f_0(980)$, $a_0(980)$ and $\kappa$ resonances.  This was achieved with a cutoff of order 1 GeV,
which guaranteed the description of the scalars,
and fairly reasonable $L_i$ constants, which ensured the description of the vectors. 
The fact that all these resonances could be obtained with this crude approximation to the full NLO ChPT amplitudes
triggered the interest in completing the full NLO ChPT calculation, which is needed for the coupled channel IAM that will be reviewed below.

The interest of these works \cite{Oller:1997ti,Oller:1997ng}, apart from generating simultaneously the members 
of the light  scalar nonet from simple implementations of chiral symmetry, 
unitarity and analyticity, is that they showed that the dynamics responsible for the scalars is very different from that 
responsible for the vector mesons. The former are dominated by meson loops with LO ChPT interactions, which 
only contain information on the size of the spontaneous symmetry breaking scale, i.e a mesonic scale, whereas vectors are dominated
by the effect of the NLO $L_i$, which contain the details of the underlying QCD dynamics, i.e., quark and gluon degrees of freedom.

Due to its simplicity and effectiveness, the method, in its many variations, 
has become extremely popular and successful in other instances, like meson-nucleon interactions \cite{Oset:1997it},
where one wants to study resonances whose generation 
is dominated by two-particle loops that can be described with chiral Lagrangians
(see the review in \cite{Oller:2000ma}).
Still, being so simple, the Chiral Unitary approach has some straightforward caveats, 
which are the complete absence of left cuts, the fact that it is 
not completely renormalized and therefore has a residual cutoff dependence 
and the use of the on-shell factorization approximation.
For  meson-meson scattering S-waves these contributions
seem to be rather small  and the approach provides very satisfactory results. 
However, these caveats triggered the investigation of more 
elaborated unitarization techniques.

In particular, the effect of the 
on-shell factorization approximation was studied in \cite{Nieves:1998hp}.
In these works the Bethe-Salpeter equations  were actually solved in the ladder approximation without using the on-shell factorization. In addition, this approach allowed
for a systematic expansion of the potential $V$ in which not only the LO, but also the NLO
contributions to the potential could be considered. Moreover, in principle the left cut could also be included
perturbatively. Of course, the price to pay is that the solution is not algebraic anymore.
The results for $\pi\pi$ scattering, which are the ones we are interested in here,
are very similar to those achieved with the Chiral Unitary approach with just the LO for the scalar waves or including the $L_i$ for the P-waves. This supports again that
the dynamics dominating the generation of scalars is different from that generating vector mesons. Finally, it was also shown that, by neglecting the  crossed channel contributions,
the BSE could be solved algebraically. The general solution gave rise to an amplitude of the $N/D$ type, with striking similarities to a Pad\'e approximant, although not exactly of the IAM form. 
This approach to the BSE with chiral Lagrangian constraints  has been recently revisited
for meson-meson scattering in \cite{Nieves:2011gb,Ledwig:2014cla} and it has also been quite successful 
in the generation of resonances in meson-baryon scattering \cite{otherBSE}. \color{black} Moreover, the
algebraic BSE Eq.\ref{LS} is also able to generate \cite{Molina:2008jw}
the poles associated to the $f_0(1370)$ and $f_2(1270)$ resonances in $\rho\rho$ scattering,
 from the contact terms and tree level $\rho$ meson exchange obtained within the ``hidden gauge formalism'',
which is a particular realization of chiral symmetry including heavier resonances.
Extensions of this unitarized formalism also describe successfully decays \cite{Nagahiro:2008um} or 
production of the $f_2(1270)$ in heavier meson decays \cite{MartinezTorres:2009uk}.
\color{black}

\subsubsection{ The coupled channel Inverse Amplitude Method}
\label{subsubsec:coupledIAM}

The generalization of the elastic NLO IAM in Eq.\ref{ec:IAM}
to coupled channels is straightforward. It is enough
 to introduce the 
NLO ChPT expansion $\re T^{-1}=T_2^{-1}(1+T_4 T_2^{-1} ...) $ 
into Eq.\ref{ec:unitarizedmatrix}, recalling that, 
as we already saw in Eq.\ref{ec:matrixpertuni}, $\im T_4= T_2\Sigma T_2$. This leads to:
\begin{equation}
T(s)=T_2(s) \,[T_2(s)-T_4(s) ]^{-1}\,T_2(s).
\label{ec:IAMmatrix}
\end{equation}
This equation is called the coupled channel IAM since it is nothing but the matrix form of Eq.\ref{ec:IAM} \cite{Oller:1997ng,Oller:1998hw,Guerrero:1998ei,GomezNicola:2001as}. 
Formally, it reproduces the low energy 
expansion of ChPT $T\simeq T_2+T_4+...$. On the physical cut, this is similar to the K-matrix approach if one approximates
$K=\re T^{-1}$ by the NLO ChPT expansion. However  in this case $\re T^{-1}$ is not a polynomial matrix, 
but contains logarithmic functions and their corresponding cuts, as required by the ChPT expansion. 

Note that for this approach one needs the fully renormalized NLO calculation of all the elements of the $T_4 $ matrix.
This means once again that no spurious parameters are needed, but just those of NLO ChPT Lagrangian.
However, when this method was first applied \cite{Oller:1997ng,Oller:1998hw}
only  a few of the $T_4$ elements were available:
($\pi\pi$ \cite{chpt1}, $K \pi$ and $K\eta$ scattering \cite{Bernard:1990kx}).
This led to a partial approximation to the NLO IAM that we have called the NLO Chiral Unitary Approach
and has been discussed in Subsec.\ref{subsec:ChUABSE}.
Nevertheless, triggered by the successful 
results of this partial approximation,  
the full one-loop $K \bar K\rightarrow K\bar K$  and $K \bar K\rightarrow K\bar K$
 calculations were soon completed  \cite{Guerrero:1998ei}.

\begin{figure}
  \centering
  \includegraphics[width=0.9\textwidth]{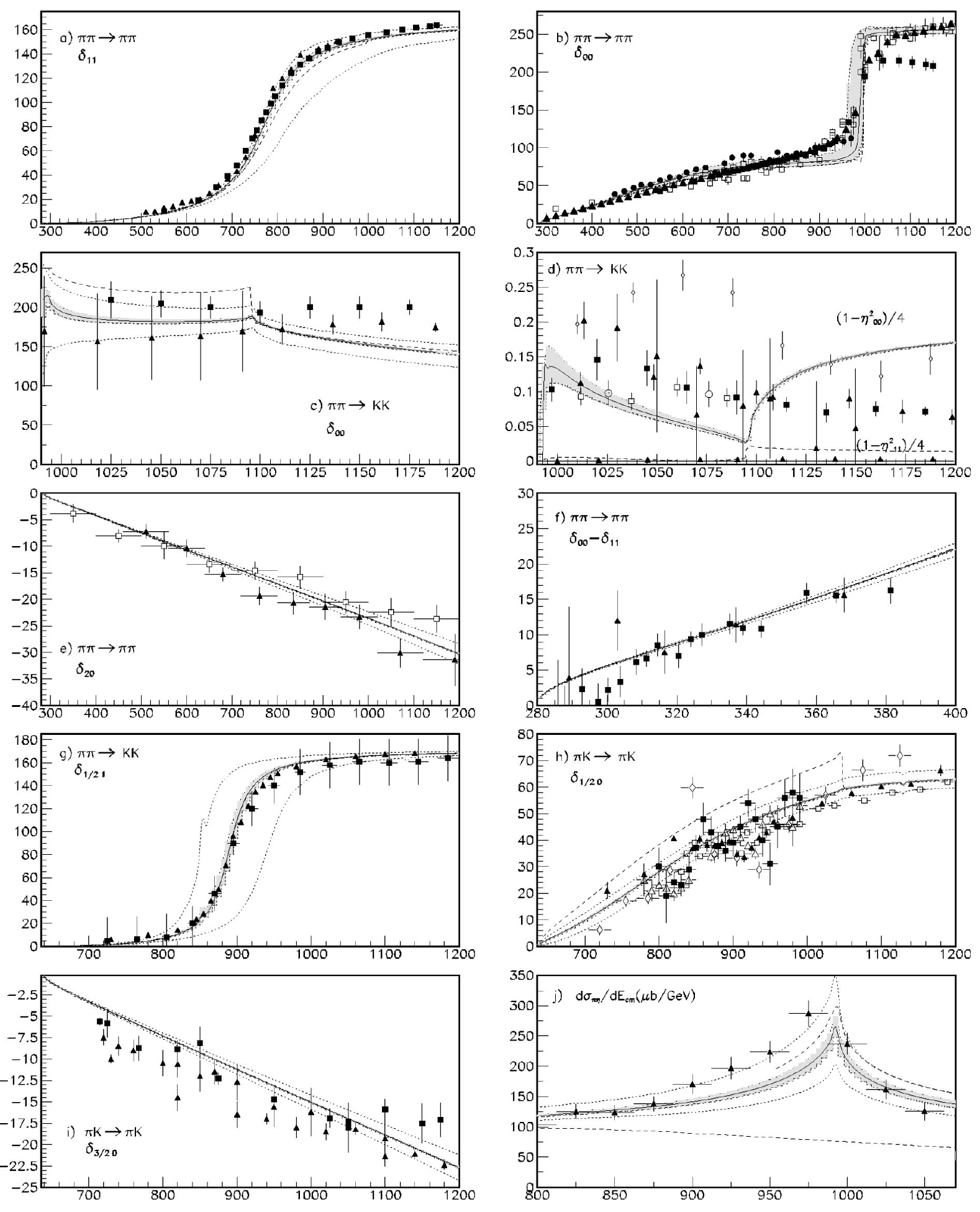}
\caption{Description of meson-meson scattering data with the coupled channel NLO IAM. Figure taken from \cite{GomezNicola:2001as}
}
  \label{fig:IAMcoupled}
\end{figure}

Thus when Eq.\ref{ec:IAMmatrix} was applied to the coupled $\pi\pi$, $K\bar K$ states \cite{Guerrero:1998ei},
the $S$ and $P$ waves up to 1.2 GeV were nicely described
by fitting data with fairly reasonable $L_i$ parameters \cite{Guerrero:1998ei}. 
Since amplitudes were fully renormalized this means that 
the  $\sigma$ and $f_0(980)$ poles were not an artifact of previous approximations, like on-shell factorization, 
a spurious cutoff, etc, but were actually consistent
with the NLO ChPT expansion.
In \cite{GomezNicola:2001as} the NLO ChPT amplitudes between two-NGB states, i.e. $\pi\pi$, $K\bar K$
$K \pi$, $K \eta$, $\eta \pi$ and  $\eta\eta$, were 
calculated with common simplifications and normalizations. When applying to them the coupled channel IAM in Eq.\ref{ec:IAMmatrix}
a fairly good description of all existing data on meson-meson scattering below 1.2 GeV
was achieved, as can be seen in Fig.\ref{fig:IAMcoupled}. 

Once again poles were found \cite{Pelaez:2004xp} for the complete scalar nonet,
i.e. the $\sigma$, $\kappa$, $f_0(980)$ and $a_0(980)$ , as well as those of the $\rho(770)$ and the $K^*(892)$ vectors.
Due to the existence of conflicting data sets, which required the addition of some estimated systematic uncertainties, as well as to the use of  different ChPT conventions equivalent to NLO but differing at higher orders, several sets of $L_i$ could be fitted. Generically the $L_i$ come out somewhere in between the NLO and NNLO determinations of the LECs using standard ChPT. This can be seen in Table~\ref{tab:LECs}, by comparing the IAMIII column with those of NLO and NNLO $L_i$. This is fairly reasonable since the unitarization procedure is somehow taking into account some higher order diagrams.

Thus, coupled channel unitarized ChPT is able to describe well the existing data and to generate the poles of the 
light scalar nonet, with fairly reasonable values of the NLO low energy constants.
However, there are several caveats. As usual, the unitarity cuts are treated exactly within the two-body approximation, whereas left cuts are just approximations. Actually, some spurious left cuts are generated \cite{Guerrero:1998ei,GomezNicola:2001as} when they are treated non-perturbatively. This is a generic and very old problem of unitarization 
formalisms and K-matrix approaches and it also occurs in other coupled channel approaches, not only in the IAM.
The reason is that unitarity cuts are common to all elements of the $T$-matrix, but left cuts can be different for 
diagonal and non-diagonal elements. Although left cuts come out correctly in the ChPT calculation of each $T_4$ element, when obtaining the inverse in the coupled channel IAM, Eq.~\ref{ec:IAMmatrix}, the determinant mixes all left cuts for all elements. This also happens if matching the ChPT series with any approximation to $\re T^{-1}$ in other unitarization techniques. So, rigorously, spurious left cuts appear in the partial waves. Moreover, having the wrong left cuts means that there cannot be a dispersive derivation for the coupled channel IAM
as it exists for the elastic one.
Fortunately though, their numerical effect is very small, of the order of a few per cent \cite{Guerrero:1998ei,GomezNicola:2001as}, and that is why the coupled channel IAM still gives a rather good  description of data and resonances. 

Note that these caveats do not apply to the elastic IAM, which is obtained from a  dispersion relation and in this sense it is much better founded. Fortunately, when dealing with the $\sigma$ meson the elastic formalism is more than enough for many purposes.
But for this reason, we have to emphasize once more that the IAM is not intended for precision calculations.

As a summary to all  meson-meson scattering unitarization sections, we have 
reviewed how the $\sigma$ meson can be generated by unitarization of Chiral Perturbation Theory.
These techniques not only describe the meson-meson scattering data but allow for reasonable analytic continuations
and generate the poles associated to the lightest scalar mesons, which are largely dominated by two-meson-loop dynamics.
The pole of the $\sigma$ comes out very close to that obtained by the dispersive data analyses used for precision determinations of the $\sigma$ properties seen in Sec.\ref{sec:disprel}.
Therefore, the main properties of the $\sigma$ are well described once data, chiral symmetry and unitarity 
are implemented within  a scheme with decent analytic properties.
Of course, when unitarizing certain approximations are made:  on the left cut, 
or neglecting many-body inelastic channels, etc, but these have been investigated and are small.
Thus, although these unitarization methods cannot compete 
in precision with the more rigorous dispersive approaches, they provide
a connection to Chiral Perturbation Theory (ChPT) and therefore to the QCD parameters. 
In particular, in  Subsec.\ref{subsec:qm} below we will review how the $\sigma$ depends on quark masses
 and in Section \ref{sec:nature} we will study how the $\sigma$ depends on the number of QCD colors $N_c$.
But to end the unitarization subsection, let us comment on how unitarized
meson-meson scattering amplitudes can be used to describe the effects of the $\sigma$ meson in
other processes, by providing the final state interactions of decay products. 

\color{black}
\subsection{Unitarization: The $\sigma$ in other processes}
\label{subsec:unitotherprocesses}

Once meson-meson amplitudes are properly unitarized they can be used to describe the final state interactions of two mesons that appear in another process, typically the decay of heavier particles
 like the $\phi$, $J/\Psi$, $D$ or $B$ mesons. In principle, different
 light resonances can be produced in the decay of these mesons, but here we will concentrate in the production of the $f_0(500)$ or $\sigma$.

The very basic idea can be illustrated with the process $A\rightarrow\pi\pi$ in the 
isoscalar-scalar channel and in the elastic regime, which is described by an amplitude, 
or form factor $F(s)$. The initial state $A$ is anything other than two pseudoscalar mesons, so that we are not dealing with meson-meson scattering.
Such processes may be triggered by whatever mechanism, like electroweak interactions,
 but once the pions are produced, they re-scatter strongly and produce the $\sigma$. 
Neglecting electroweak corrections, elastic unitarity demands 
that ${\rm Im} F(s)= \sigma(s)  F(s) t_0^{(0)}(s)^*$, where as usual $\sigma(s)\equiv2k/\sqrt{s}$, and $t_0^{(0)}$ is the scalar-isoscalar partial wave. 
Note that $\vert F(s)\vert$ is not constrained by the previous equation, which can be multiplied by any real function. However, the phase of $F$ must be equal to that of $t_0^{(0)}(s)$ (Watson's theorem). 
Thus, we know the phase of $F(s)$ and that the sigma pole must be present in $F(s)$.
But there is already a function with those properties, which is the unitarized $t_u(s)$, 
so that we can define $F_u(s)\equiv R(s)t_u(s)$, 
with $R(s)$ a real function in the real axis and no poles.
Customarily $F_u(s)$ is also called a unitarized amplitude.
In practice, $R(s)$ is obtained by matching the above expression to the ChPT calculation of $F(s)$.

For instance, when using the ChUA to LO in ChPT and the expansion of $F(s)=F_0(s)+...$, then
$F_u(s)=F_0(s) t_u(s)/V(s)$, where $V(s)$ is the LO calculation of $t(s)$, namely, 
the low energy theorems in Eq.\ref{ec:LET}.  This approach is easily generalizable to coupled channels by considering $F(s)$, $t(s)$ and $R(s)$ as matrices. As we will see next, in the last years
it has been applied to describe a great deal of experimental data in which the $\sigma$ resonance is produced.
The relevance of this approach is that it makes the production process consistent with 
the scattering data, for which we have seen in previous sections that there are
very strong constraints about unitarity, analyticity, as well as for the pole position itself.  
Hence, this 
a much more constrained and complete treatment that just considering sums of 
Breit-Wigners as it is usually done in many experiments.

This approach is also applicable if there are other particles in the final state, namely $A\rightarrow B \pi\pi$,
as long as the  interaction of $B$ with 
the two pions is negligible. This is 
of course the case if $B$ only has electroweak
interactions with the pions, but in practice 
it is also the case of many other processes even if $B$ is made of other hadrons.

For example, a relatively narrow 
peak in the $\sigma$ region is seen in the $J/\psi\rightarrow p\bar p \pi^+\pi^-$ decay 
\cite{Eaton:1983kb} with a shoulder in the $\rho(770)$ region. 
A good description of this process within the coupled channel ChUA was obtained in \cite{Li:2003zi}
including an apparently narrower $\sigma$.

This apparently narrower $\sigma$ is very common 
in production processes compared to the very wide, almost non-resonant shape observed 
in scattering. Compare, for instance the low peak around 450 MeV in the $J/\psi\rightarrow\omega\pi^+\pi^-$ decay observed by the BES Collaboration and shown here 
in Fig.\ref{fig:BESdata}, with the absence of any structure around 450 MeV in the $\pi\pi$ scattering
phase in  the left panel of Fig.\ref{fig:00data}. This narrowness is easily explained 
within the unitarized formalism. To fix ideas, let us use the very simple LO ChUA, following \cite{Roca:2004uc}. 
Here the production vertex $V_P$ gives the LO form factor that we generically called $F_0(s)$ and
the Bethe-Salpeter equations lead directly to the generic amplitude above, that we write 
now as $V_P t_u(s)/ V(s)$. Since in the $\sigma$ region, $2s>>m_\pi^2$, the production amplitude
behaves as $t_u(s)/s$.  It is this $s$ in the denominator that makes the broad structure of the $\sigma$ in 
$t_u(s)$
look narrower. 
Actually, the decays $J/\psi\rightarrow\omega\pi^+\pi^-$ and $J/\psi\rightarrow\phi\pi^+\pi^-$  have been
nicely described within a ChUA matching up to NLO with ChPT
in \cite{Meissner:2000bc} or to LO plus the addition of the exchange of vector and axial-vector mesons \cite{Roca:2004uc}. 

Furthermore, the BES Collaboration has also measured the production of the $\sigma$ in $\psi(2S)\rightarrow J/\psi \pi^+\pi^-$ \cite{ablikim07}. In their analyses they first used different Breit-Wigner forms, 
which are strongly model 
dependent, as we have already discussed, and do not have the correct 
final state phase for the $\pi\pi$ system nor a $\sigma$ pole consistent with dispersion relations.
However, they have also explicitly shown next that their data can be described with a ChUA with a pole fixed at $469-i 203\,$MeV.
The formalism for this decay is basically the same we have just described 
for decays of the $J/\psi$ into a vector and two pseudoscalars and was treated in detail in \cite{Guo:2004dt}.

Note that the structure of the vertex that produces 
the vector-pseudoscalar-pseudoscalar state is responsible 
either for the smallness or even lack of
 $\sigma$ signal in the $\phi\pi^+\pi^-$ final state compared to that of the $f_0(980)$
or to the large $\sigma$ signal in the $\omega\pi^+\pi^-$ and $J/\psi\pi^+\pi^-$ final states.  The sigma or the $f_0(980)$ are generated by rescattering once this ``filtering'' is determined at the production vertex.

The filtering of the initial production vertex is 
also responsible for the tiny contribution of the $\sigma$ meson to $\phi\rightarrow \gamma PP$, where $P$ are pseudoscalars. In this process there is a clear $f_0(980)$ signal,
nicely described by treating the final state interactions with the ChUA \cite{Oller:1998ia,Palomar:2003rb}, 
but a very small one from the $\sigma$ which is also swamped by other effects \cite{Palomar:2003rb} which are
more model dependent. It is very illustrative that
within a L$\sigma$M supplemented with ChPT constraints \cite{Bramon:2002iw}, 
this filtering is due to 
the production vertex being proportional to $m_\sigma^2-m_K^2$, which is very small.
Thus, although a priori this $\phi$ radiative 
decay was considered promising to study scalars \cite{Close:1992ay},
this filtering makes it only suitable for the $f_0(980)$
but not very useful for drawing robust conclusions about the $\sigma$. 
Similarly, this filtering effect has also been recently described \cite{Sekihara:2015iha} in semileptonic decays of $D$ mesons, 
where in $D_s$ decays the $f_0(980)$ is nicely observed but not the $\sigma$
and the opposite occurs in $D^+$ decays.

Finally, this filtering is also relevant to interpret the recent LHCb measurements: the $\sigma$ was clearly seen in $\overline B^0\rightarrow J/\psi \pi^+\pi^-$ data \cite{Aaij:2013zpt}, with little strength for the $f_0(980)$,
 but the opposite situation
was previously measured in $B^0_s\rightarrow J/\psi \pi^+\pi^-$ data \cite{Aaij:2011fx}. 
Once again these effects can be easily accommodated within the ChUA 
approach to model the final state interactions \cite{Liang:2014tia}.

This technique has also been applied to $D^0$ decays 
into $K_s^0$ and two pseudoscalar mesons, 
which were measured in \cite{Muramatsu:2002jp,Rubin:2004cq},
allowing for  a unified study of the $\sigma$ together with the $f_0(980)$ as well as the $a_0(980)$
from the same decay. 

\color{black} 
Of course, the approach just discussed does not directly apply 
without further assumptions
when three particles exist in the final state as in the $D\rightarrow K_s^0 MM$ decays. In particular, in \cite{Xie:2014tma} it is assumed
that the process has several steps: First $D\rightarrow K_s^0 R$, where $R$ is $f_0(500)$ or $f_0(980)$ for $MM=\pi\pi$ and 
$R=a_0(980)$ for $MM=\pi\eta$. Next the $R\rightarrow MM$ decay occurs. The final step
is the rescattering of the $MM$. Note that in the last two steps the $K_s^0$ is assumed to be an spectator.
The process is modelled initially from the quark level 
to obtain the appropriate symmetry factors
to produce the $K_s^0 R$ state, including some hadronization constants and production vertices.
The unitarized scattering amplitude, in this case using the ChUA described in previous sections, is only used in the last step, to describe the $MM$ distribution. The results
reproduce well the basic experimental features. \color{black}
\color{black}

\subsection{Quark mass dependence}
\label{subsec:qm}

The quark mass dependence of the $f_0(500)$ is relevant for our understanding
of the $\sigma$ nature and spectroscopic classification
but also because it allows for a connection with lattice QCD.

ChPT provides a rigorous expansion of NGB  masses 
in terms of quark masses $m_q$, which appear through the lowest order  
NGB meson-mass matrix $M_0^2\sim m_q$, as we saw in Eq.\ref{ec:Lag2}. 
Thus, the dependence of resonance parameters on quark masses can be studied simply
by changing the NGB meson masses in the unitarized amplitudes.

\begin{figure}
  \centering
  \includegraphics[width=0.6\textwidth]{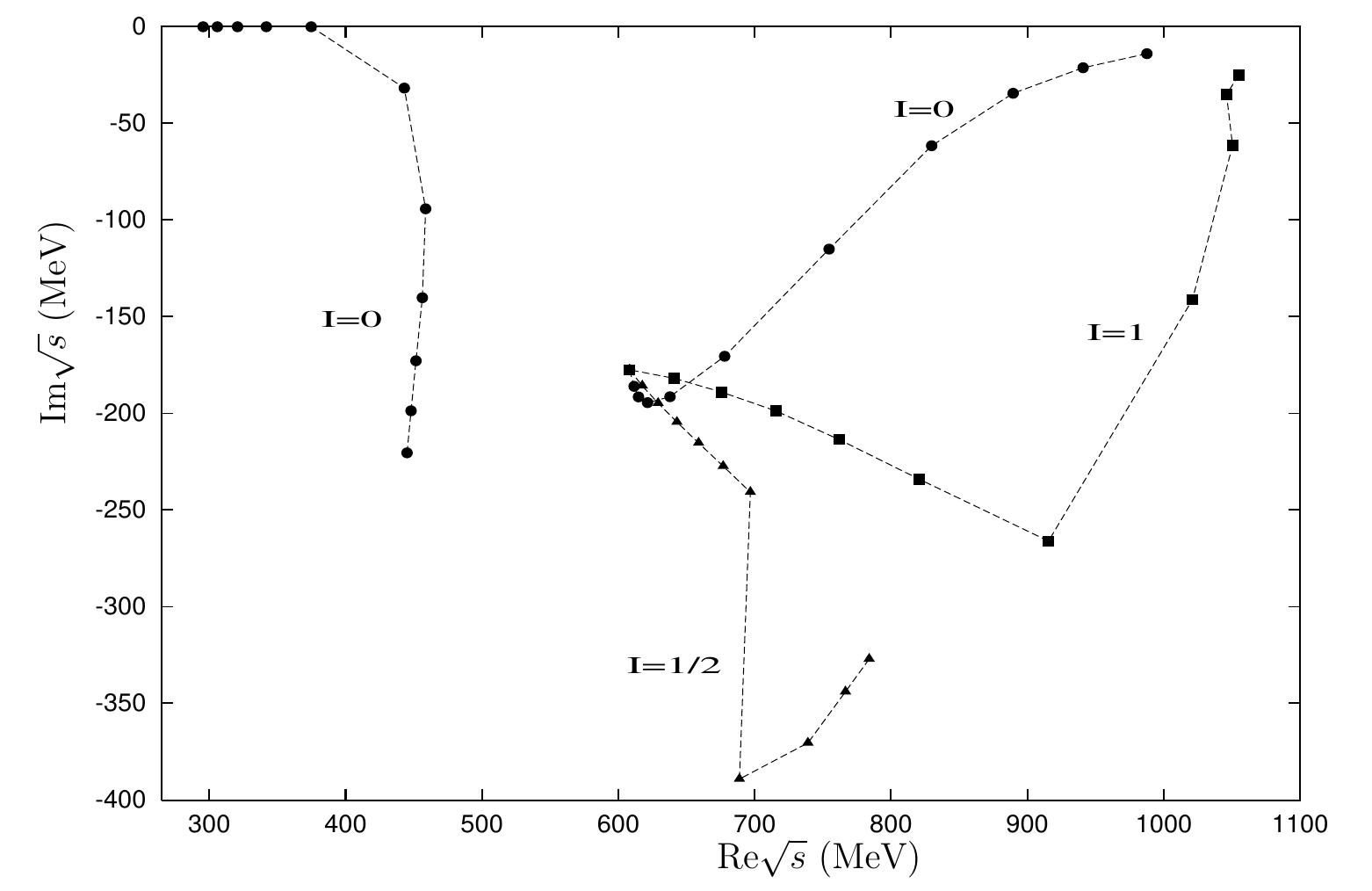}
\vspace*{-.5cm}
\caption{Trajectories of the poles that appear in coupled channel unitarized amplitudes of different isospin
as the pion, kaon and eta masses are varied from their physical values to a common value of 350 MeV \cite{Oller:2003vf}. This shows that the 
lightest scalars actually belong to a nonet in the SU(3) limit.
The two trajectories with $I=0$ correspond to the singlet and octet states, not directly to the poles of the
$\sigma$ or $f_0(980)$ resonances, which are a mixture of these two.
 Figure taken from \cite{Oller:2003vf}. Reprinted from  J.~A.~Oller,
  ``The mixing angle of the lightest scalar nonet,''
  Nucl.\ Phys.\  A {\bf 727}, 353 (2003). Copyright 2003, with permission from Elsevier.
\label{fig:Oller-equalmasslimit}
}
\end{figure}

To LO ChPT   it was shown in \cite{Oller:2003vf} that, by restoring exact SU(3) symmetry 
in an $N/D$ chiral unitarized approach, the  poles of the lightest scalar mesons became degenerated 
into an octet and a singlet.
The poles of the octet were 
those of the $\kappa$, the $a_0(980)$ and one particular combination of the $\sigma$ and $f_0(980)$.
This was achieved by
setting the pion, kaon and eta masses to the same value, not only in the scattering kinematics 
but also in the leading order ChPT vertices. In addition all values of the subtraction constants 
(or the cutoff, since they can be translated into one another) were also made equal.
The pole trajectories in the complex plane 
from their physical positions to those where 
SU(3) is restored
are shown in Fig.\ref{fig:Oller-equalmasslimit}. 
This result was very relevant because it showed explicitly that all those scalar resonances do form a nonet, without
any a priori assumption about their existence or nature, but just chiral symmetry, unitarity, analyticity and a data fit.

Higher order unitarized calculations of the  quark-mass dependence of light mesons 
have only been carried out in the elastic regime,
but as we have already seen, if one is not looking for precision,
  elastic unitarization techniques are quite  enough to deal with the sigma.
The IAM has been used in these studies since it deals 
  approximately with the left cut and because its only parameters are those of ChPT at each order, without the need to make assumptions on 
  subtraction constants or cutoffs where some spurious mass dependence could hide.
In particular, the 
$\rho(770)$ and $\sigma$ parameter dependence on the averaged
non-strange light quark-mass $\hat m$ has been studied within the unitarized 
SU(2) elastic IAM formalism to NLO in \cite{chiralexIAM} and to NNLO in \cite{Pelaez:2010fj}.
In addition, the $\pi\pi$ elastic scattering phases dependence was also studied in \cite{Nebreda:2011di} with the same methods. 
Moreover the $\rho(700)$, $\sigma$, 
 $\kappa(800)$ and $K^*(892)$ dependence on both the non-strange quark mass $\hat m$ and the strange one $m_s$ was
  studied in \cite{Nebreda:2010wv} within the SU(3) NLO IAM formalism. 

If the elastic approximation is to be used, the range of values of $M_\pi$ that 
can be considered should fall within the ChPT
range of applicability and allow for some elastic $\pi\pi$ and $\pi K$
regime below $K\bar K$ or $K\eta$ thresholds, respectively. Being very optimistic,
this means $M_\pi\leq 440$ MeV \cite{Nebreda:2010wv}, but we will see it is even less at NNLO. 
Higher order corrections are expected to become larger as $M_\pi$ is increased, but
this uncertainty can be studied by unitarizing the NNLO.

\begin{figure}
\vspace{-3cm}
  \includegraphics[width=0.41\textwidth]{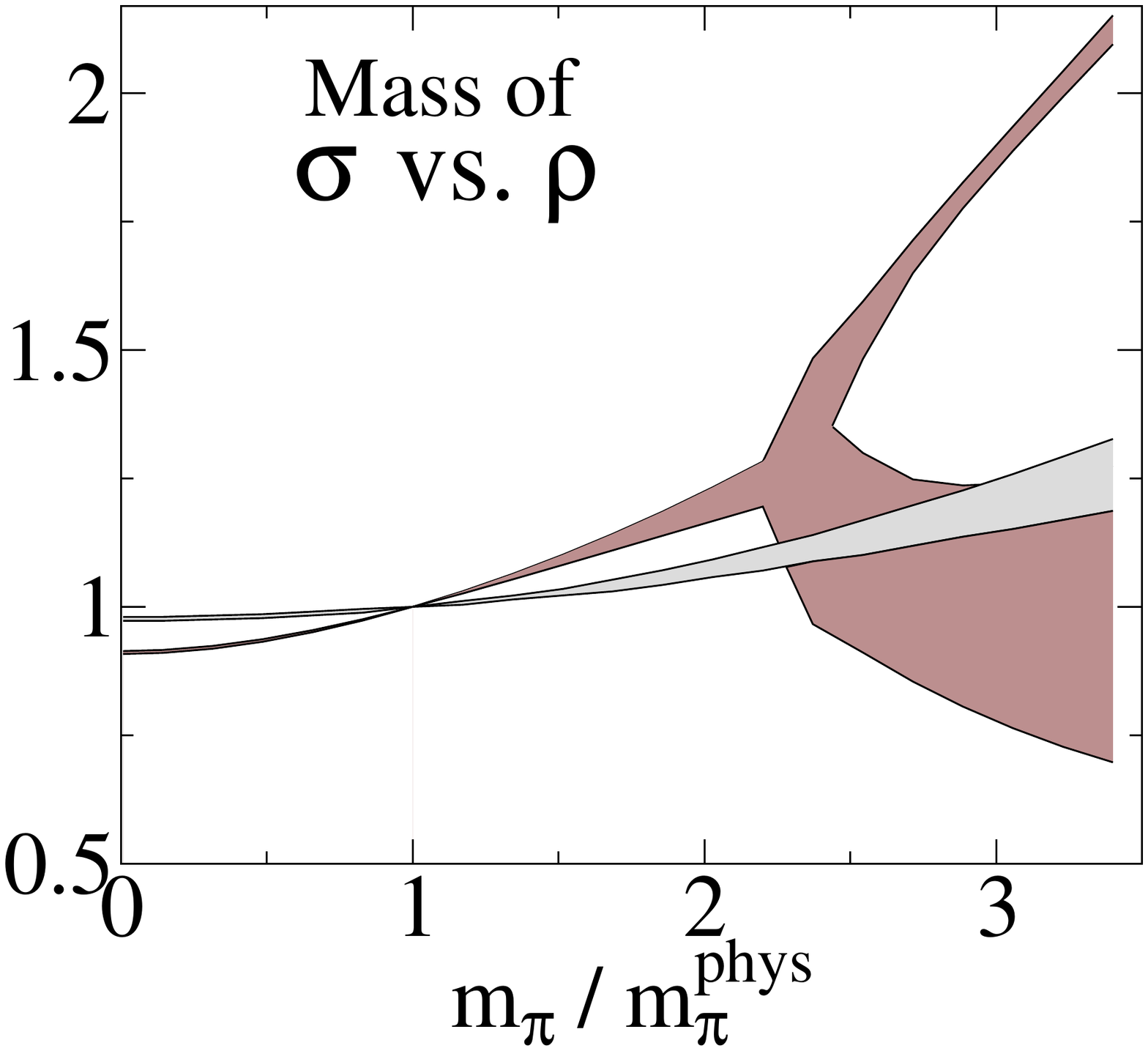}
\hspace{-.4cm}
  \includegraphics[width=0.58\textwidth]{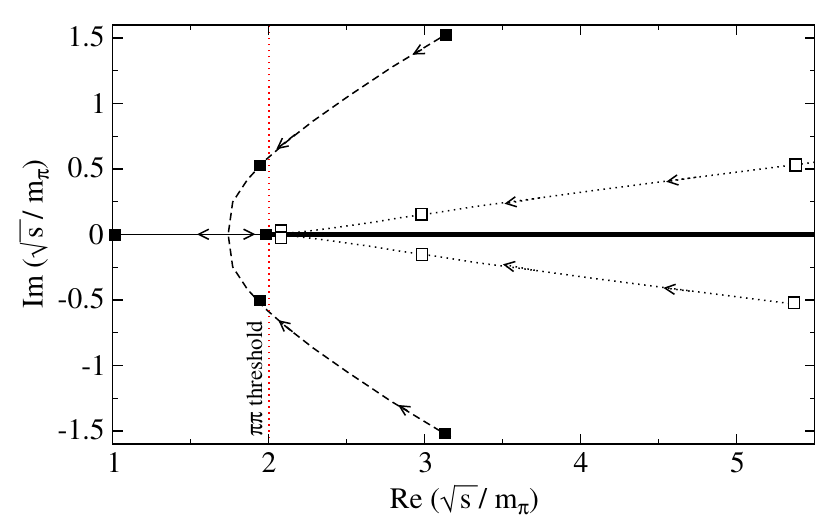}
\vspace{-.3cm}
\caption{Left: $M_\pi$ dependence of the $\sigma$ (dark band) and $\rho(770)$ (light band) masses
       in units of their physical values \cite{chiralexIAM}. 
       The bands cover only the uncertainties in the SU(2) LECs. Right: in $M_\pi$ units
the $\sigma$ (dashed line) and $\rho(770)$ (dotted line) pole trajectories on the second Riemann sheet of the complex plane
as $M_\pi$ is increased from their physical value (first point in their trajectories). 
Both $M_\sigma$ and $M_\rho$ grow with $M_\pi$, but the 2-pion threshold grows faster, 
so that in $M_\pi$ units both resonances seem to approach threshold. 
Note that in the original works \cite{chiralexIAM} the notation $m_\pi$ was used instead of $M_\pi$.
Figures taken from \cite{Pelaez:2010er} (left) and \cite{Pelaez:2010fj} (right).
\label{fig:movimientoplano}
}
\end{figure}

Thus, the left panel of Fig.~\ref{fig:movimientoplano} shows the NLO IAM
results \cite{chiralexIAM} for the $M_\pi$ dependence
of $M_\rho$ and $M_\sigma$ (defined from the pole position
$\sqrt{s_{pole}}=M-i\Gamma /2$), normalized to their physical values.
The bands cover the SU(2) LECs uncertainties only. 
Let us denote in this subsection the physical mass of the pion by $M_\pi^{\rm phys}$.
From $M_\pi=0$ up to $2.4 \, M_\pi^{\rm phys} $
both the sigma and $\rho(770)$ masses are described by
rather smooth curves, monotonously increasing with $M_\pi$. 
Note that $M_\sigma$ grows somewhat faster than $M_\rho$ in that region. 
This effect has also been observed on the lattice \cite{Kunihiro:2008dc}.
However, around $2.4 \, M_\pi^{\rm phys} $ 
there is a striking
 splitting in the curve describing the evolution of the $\sigma$ mass,
which does not occur anywhere for the $\rho(770)$.

This splitting can be understood by looking now at the right panel of the 
same Fig.~\ref{fig:movimientoplano}, which
shows the trajectories followed by  the $\sigma$ and $\rho(770)$
pole positions in the second Riemann sheet of the complex plane
 as $M_\pi$ is varied.
Note the use of  $M_\pi$ units in order to see the pole
movements {\it relative to the two-pion threshold}, which is
fixed at 2 in the figure. Then, since 
the two-pion threshold at $2M_\pi$ grows
faster with $M_\pi$ than both $M_\sigma$ or $M_\rho$, both resonance masses seem to decrease in $M_\pi$ units.
Let us now look at the conjugated pair of $\rho(770)$ poles, which
reach the real axis at the same time they cross threshold. This
is a generic feature of resonances with $J\geq 1$ \cite{Hanhart:2014ssa}.
Then, one of them jumps into the first sheet and stays below
threshold in the real axis as a bound state, while its conjugate
partner remains on the second sheet practically at the very same
position as that in the first. 
In contrast, as seen in the right panel of Fig.~\ref{fig:movimientoplano}, the 
conjugated $\sigma$
poles meet in the real axis {\it below threshold}, becoming virtual states.
Contrary to some common belief, and crudely speaking, scalars
can have a pole-width even if there is no phase-space available 
because their pole-mass is below the two-pion threshold.
This is also a generic feature of poles in $S$-wave amplitudes \cite{Hanhart:2014ssa}. Similar movements were found
in potential models \cite{Kaminski:1998ns,Cannata:1989ub},
unitarized quark models \cite{vanBeveren:2002gy},finite density analysis 
\cite{FernandezFraile:2007fv}, or in general studies of chiral symmetry restoration \cite{Hyodo:2010jp}. 
Moreover, the non-analyticity of the hadron mass when the conjugate poles reach the real axis
has been recently studied in detail within the general formalism of Jost functions in \cite{Hyodo:2014bda,Hyodo:2013iga}. 
The conclusion for the $\sigma$ is a similar warning 
to that raised in Ref.~\cite{chiralexIAM} about  
naive mass extrapolations for states which appear near thresholds
on the lattice, although within a more general framework.

Another example of the pole moving below threshold 
but keeping a finite width may also occur \cite{Hyodo:2010jp} when increasing the
$\pi\pi$ attraction in the $\sigma$ channel 
 relatively to other scales, for instance in 
the process of chiral restoration with temperature or in medium.

Once the pair of poles has reached the real axis, they do not have to be conjugated anymore and, as $M_\pi$ increases, they 
lie on the real axis with two different pole masses. 
What is found for the $\sigma$ \cite{Hanhart:2014ssa} is that one pole
moves towards threshold and for $M_\pi\simeq 3.3 \, M_\pi^{\rm phys}$ it
jumps through the branch point to the
first sheet, staying in the real axis below threshold, very
close to it as $M_\pi$ keeps growing. The other $\sigma$ pole moves
down in energies away from threshold and remains 
on the second sheet.
This is why there is a ''splitting'' on $\sigma$-mass curve on the left panel of 
Fig.~\ref{fig:movimientoplano}, 
each new branch corresponding now to one of the two poles in the real axis below threshold.
Since this lower pole may get close to the Adler zero region, 
 it is important to remark that all calculations were performed with the modified 
IAM of Eq.\ref{ec:mIAM}, which differs very little form the IAM everywhere except near the Adler zero region, where it does not have spurious singularities and reproduces better the Adler zero.
Let us also remark that these  very asymmetric poles could
signal a prominent molecular component \cite{Morgan:1993td,Morgan:1992ge,Baru:2003qq},
for large pion masses, which might be studied on the lattice.

\begin{figure}
\vspace{-3cm}
  \centering
  \hspace{-.6cm}  \includegraphics[width=0.41\textwidth]{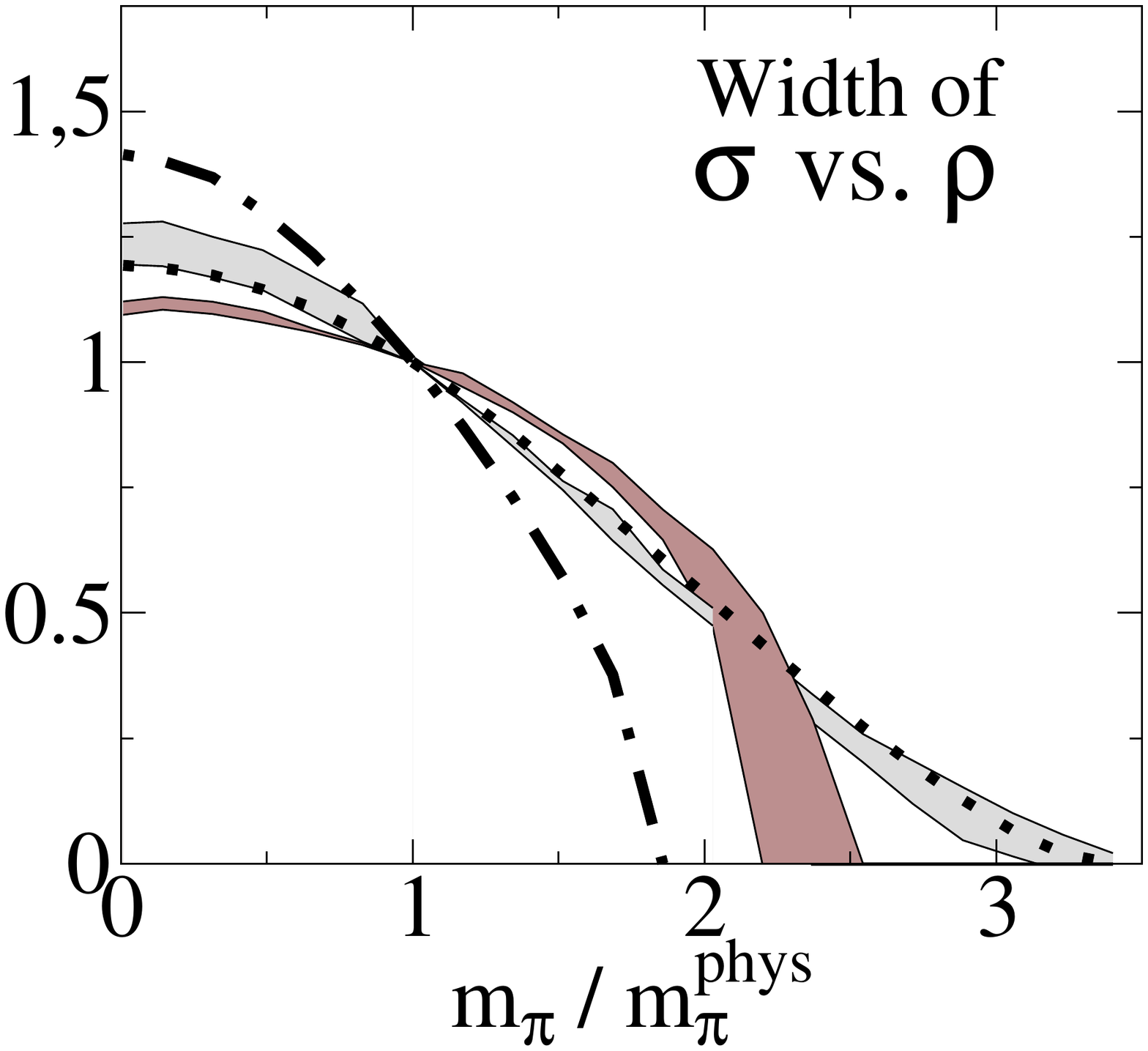}
  \hspace{-.4cm}  \includegraphics[width=0.52\textwidth]{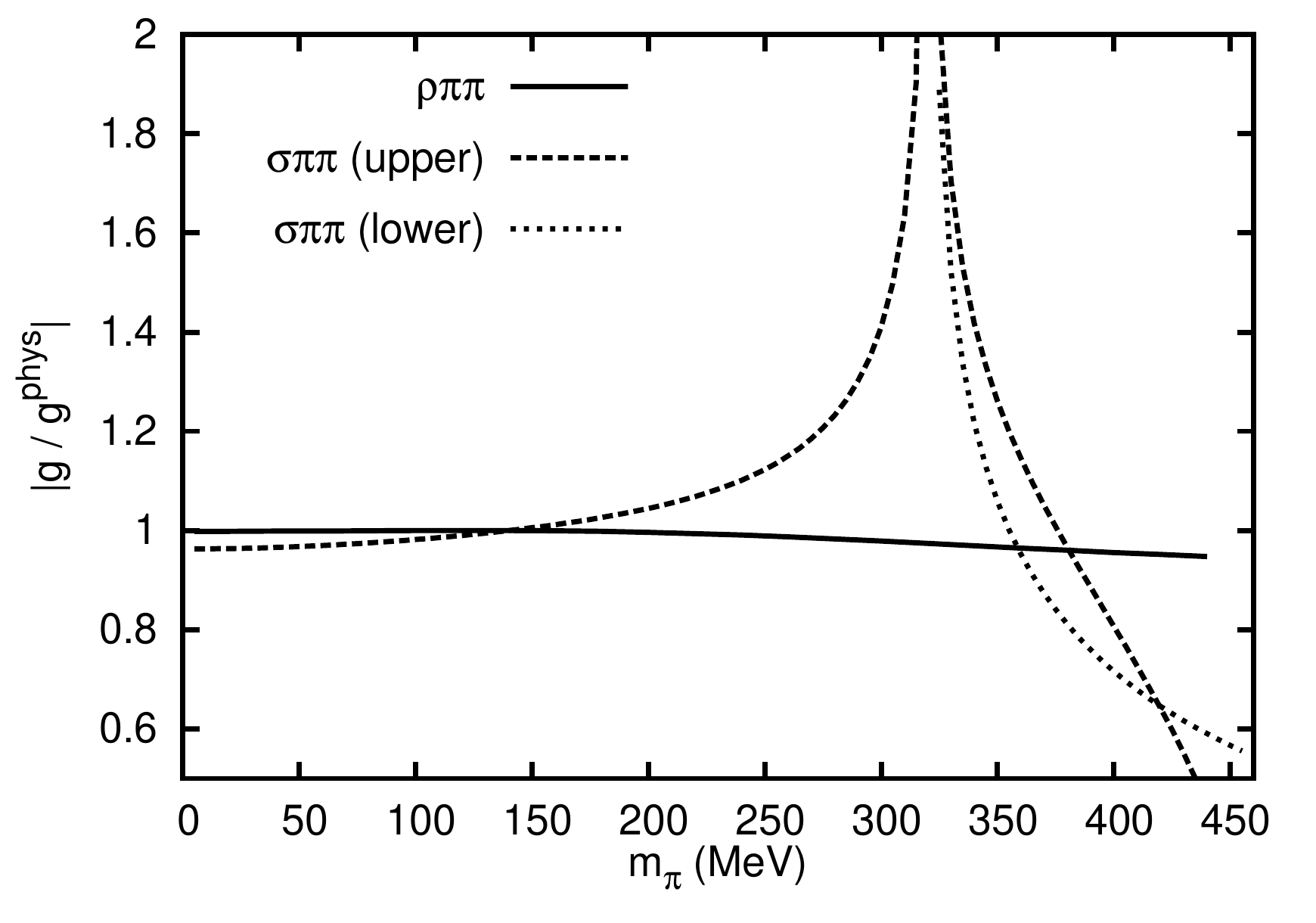} 
  \caption{ 
       Left: $m_\pi$ dependence of the $\sigma$ (dark band) and $\rho(770)$ (light band) widths
       in units of their physical values \cite{chiralexIAM}. 
       The dotted line, almost overlapping with the $\rho(770)$ results,
       shows how the 
       $\rho(770)$ width would change from the phase space reduction only, assuming
       a constant coupling of the resonance to two mesons. In contrast, the behavior of the $\sigma$ width does not follow the variation of phase space only, represented by the dot-dashed line. Right: the $\rho\pi\pi$ and $\sigma\pi\pi$ couplings as obtained from the modulus of their pole residues in $\pi\pi$ scattering calculated within NLO IAM \cite{Nebreda:2010wv,chiralexIAM}. Note the $\rho\pi\pi$ almost constant behavior 
versus the strong $m_\pi$ dependence of $\sigma\pi\pi$ around the splitting point into two branches.
Figures taken from \cite{chiralexIAM} (left) and \cite{Rios:2011vr} (Right).
Right figure reprinted with permission from 
  G.~Rios, C.~Hanhart and J.~R.~Pelaez,
  AIP Conf.\ Proc.\  {\bf 1322}, 452 (2010). Copyright 2010, AIP Publishing LLC.
}
\label{fig:widthandcoupling}
\end{figure}

Next, in the  left panel of Fig.~\ref{fig:widthandcoupling} the $M_\pi$
dependence of $\Gamma_\rho$ and $\Gamma_\sigma$ normalized to their
physical values are compared. Note first that both widths become smaller as $M_\pi$ increases. 
One could then wonder if this decrease is just due to phase space, since the two-pion threshold
is increasing faster than both resonance masses.
Hence this decrease is compared in Fig.~\ref{fig:widthandcoupling} with the expected phase space reduction 
 as resonances approach the $\pi\pi$ threshold.
Note that $\Gamma_\rho$ follows very well
this expected behavior, which
implies that the $\rho\pi\pi$ coupling is almost $M_\pi$ independent,
as shown in the right panel of the same figure (continuous line)
\cite{chiralexIAM,Rios:2011vr}.
The coupling is defined as the modulus of the square root of the pole residue.
This result is very relevant because the very small $M_\pi$ dependence
of the $\rho\pi\pi$ coupling was assumed in a lattice study \cite{Aoki:2007rd} 
but was later found in the lattice calculation \cite{Dudek:2012xn}\footnote{Check the numbers in the Erratum.}, which did not assume it from the start. Moreover
it also seems consistent with other recent lattice calculations \cite{Fahy:2014jxa}
and \cite{Chen:2012rp}. 
Still, this comparison between lattice and ChPT couplings must be taken cautiously, 
because lattice computations are performed at rather 
large pion masses from the unitarized ChPT point of view and their 
couplings are obtained from a phenomenological model 
fitted to the phases obtained on the lattice, not from the residue of the pole. 
However, the fair consistency with the lattice findings for the $\rho(770)$
 gives relative confidence on the
unitarization method and its results for the $\sigma$, which have not been tested on the lattice yet.

Back to $\Gamma_\sigma$, it can also be seen  on the left panel of Fig.~\ref{fig:widthandcoupling}
that it deviates from the
pure phase space reduction expectation. 
Once again the coupling can be extracted from the pole residue and
is also shown on the right panel of Fig.~\ref{fig:widthandcoupling}.
The $M_\pi$ dependence of $\Gamma_\sigma$ is rather mild near the physical region,
but it becomes very strong around the splitting point. Note that this time there are
``upper'' and ``lower'' branches above the splitting point, 
both varying strongly with $M_\pi$. 
This behavior might be a hint of some relevant or even dominant ``molecular'' 
component \cite{Baru:2003qq,weinbergdeuterium,mol} within the $\sigma$ at large $M_\pi$ masses.

\begin{figure}
\vspace{-3cm}
  \centering
  \includegraphics[width=0.52\textwidth]{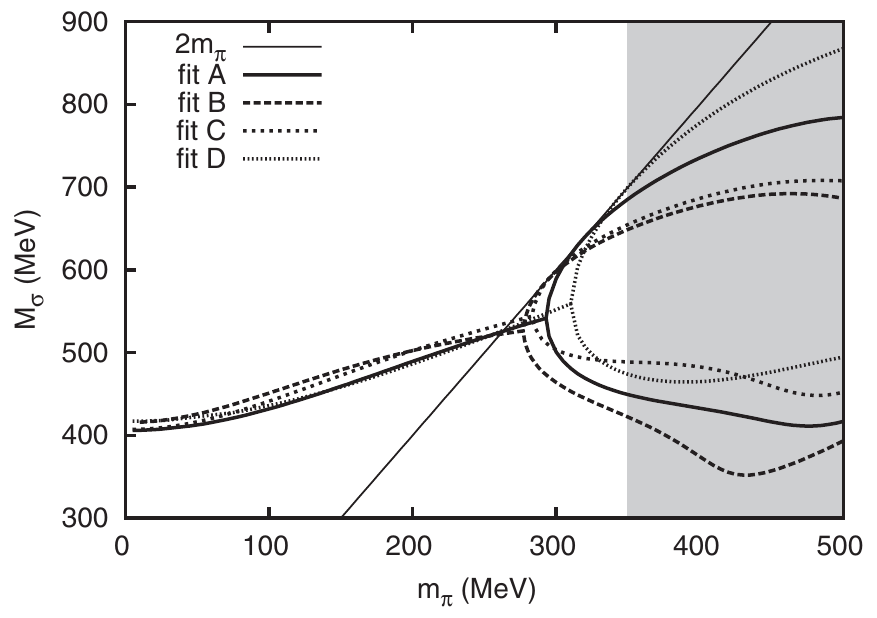}
  \hspace{-.5cm}  \includegraphics[width=0.41\textwidth]{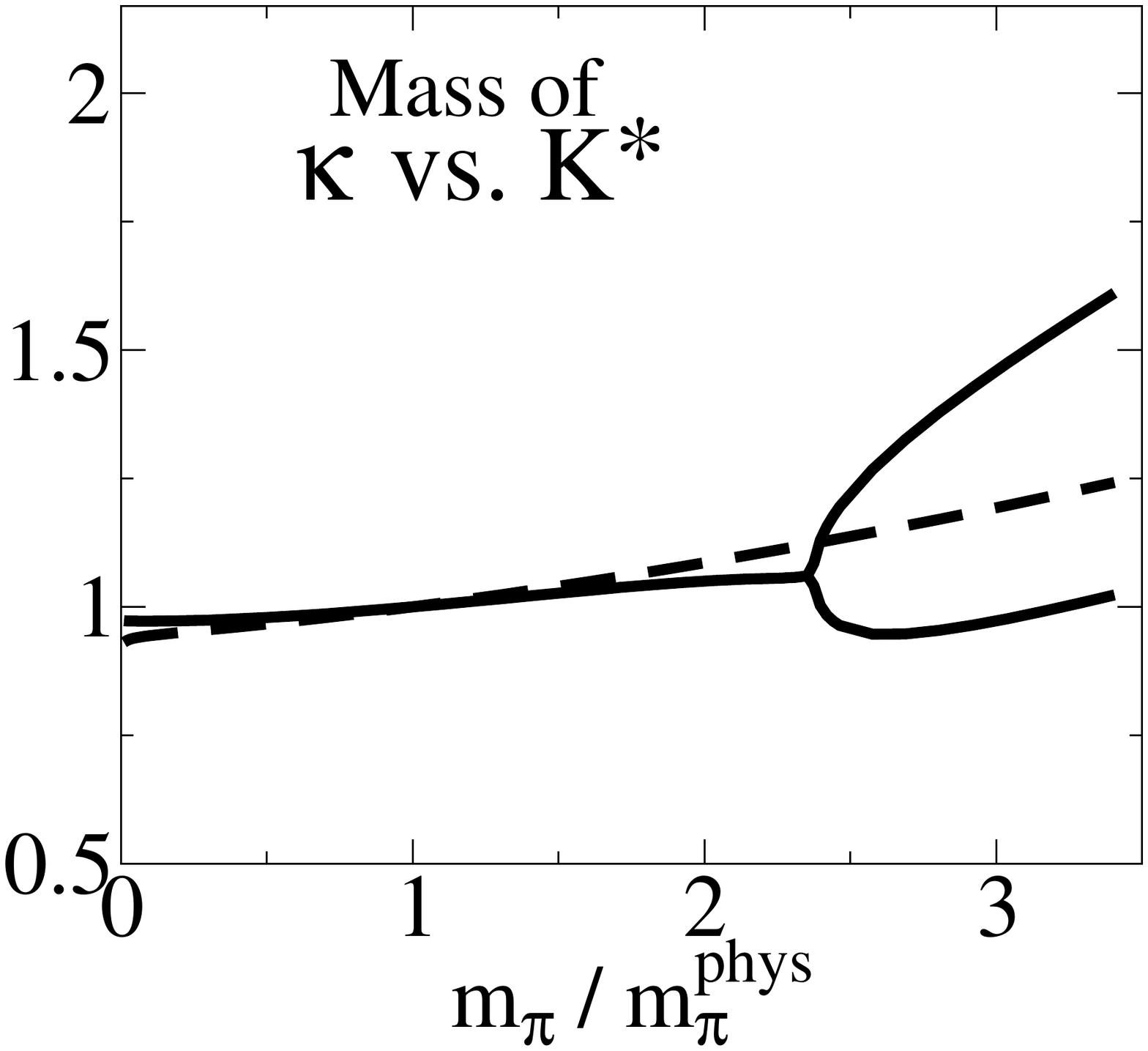}
\caption{
Left: Dependence of the sigma mass $M_\sigma$ on the pion mass,
from the NNLO (two-loops) IAM \cite{Pelaez:2006nj}. Different curves represent different 
fits on \cite{Pelaez:2006nj}. The thin continuous line shows the $2m_\pi$ threshold.
Right: $m_\pi$ dependence of the 
 $\kappa(800)$ (solid line) and $K^*(892)$ (dashed line) masses \cite{Nebreda:2010wv}.
All masses and widths are defined from the pole positions
as obtained from NLO IAM fits. 
Figures taken from \cite{Pelaez:2010fj} (left) and \cite{Pelaez:2010je} (right).
\label{fig:NNLOykappa}
}
\end{figure}

Of course, one might wonder how robust the NLO IAM results are as $M_\pi$ becomes very large.
Hence, the left panel of Fig.~\ref{fig:NNLOykappa} shows the sigma mass $M_\pi$ dependence
calculated with the NNLO IAM \cite{Pelaez:2006nj}, i.e. to two-loops,
 which can be compared with the NLO IAM results already discussed in
Fig.\ref{fig:movimientoplano}. The calculation of uncertainties is more cumbersome at NNLO and 
thus there are four different fits to data and lattice results on the $M_\pi$ dependence of 
$f_\pi$ and the NGB masses \cite{Pelaez:2006nj}. It can be noticed that the NLO and NNLO results look qualitatively
 similar, although the NNLO calculation places the splitting point slightly below $M_\pi=300\,$MeV, instead of the
240 MeV found at NLO. The $\rho(770)$ coupling is still found to be quite independent of $M_\pi$
and the $\sigma$ width and coupling have the same NNLO qualitative behavior than to NLO, although
taking into consideration a lower splitting point.

This is in  qualitative agreement with the lattice
findings  in \cite{Prelovsek:2010kg}, where a bound state
seems to exist
for $M_\pi\simeq325\,$MeV. 
Let us nevertheless recall the caveats
raised from the very authors of \cite{Prelovsek:2010kg}, 
since they cannot calculate
accurately the width, and some possibly relevant
contributions from  ``disconnected contractions''
have not been included in the calculation.
\color{black} While finishing this report the first lattice calculation
of $\pi\pi$ scattering in the scalar isoscalar channel
which takes into account such contribution has appeared \cite{Briceno:2016mjc}.
The amplitude is calculated for $M_\pi=236$ and 391 MeV.
For the higher pion mass a bound state at $758\pm4$ is found,
which is qualitatively consistent with the predictions of the NNLO IAM
we have just described. As seen in the left panel of Fig.~\ref{fig:NNLOykappa},
the upper branch of fit D predicts a bound-state pole
in remarkable quantitative 
agreement with this lattice findings. For $M_\pi=236\,$MeV
the lattice phase shift is described with several parameterizations that 
fulfill unitarity (although without a left cut), like a K-matrix, and all 
them display a pole deep in the second Riemann sheet of the energy-squared
complex plane,
with a very large imaginary part, corresponding to a very wide resonance. 
They do not observe any significant change in the sigma coupling to two pions, but
that is also consistent with the IAM because, as seen in the right panel of Fig.\ref{fig:widthandcoupling}
the coupling at $M_\pi=236$ and 391 MeV is also relatively similar,
since the expected singularity is placed between those two masses but
far enough to be seen in any of them.
\color{black}

In addition, there is another relevant piece 
of support for this whole picture coming from the lattice.
The NLO IAM analysis can also be extended to SU(3) \cite{Nebreda:2010wv} in order to include strangeness.
This allows for a study of the strange quark mass dependence of all the
resonances generated with IAM,
and in particular of the sigma, which is completely negligible. 
But the SU(3) NLO IAM also generates the $\kappa$ scalar meson and the $K^*(892)$ vector meson.
The dependences of the mass of both resonances
on the pion mass are shown as curves on the right panel of 
Fig.~\ref{fig:NNLOykappa}. 
Note, on the one hand, the very similar smooth behavior
of the $K^*(892)$ vector compared to that of the $\rho(770)$ vector in 
Fig.\ref{fig:movimientoplano}. On the other hand, note the
striking similarities between the $\kappa$ in Fig.~\ref{fig:NNLOykappa} and the $\sigma$ in Fig.\ref{fig:movimientoplano}.
 This similar mass dependence could be expected, since we have already argued that the $\kappa$ is very similar to the $\sigma$ but in $\pi K$ instead of $\pi\pi$ elastic scattering. Thus, as for the $\sigma$, 
we see a splitting point, above which 
the $\kappa$ becomes a virtual state. The relevant point is that this virtual $\kappa$ state
has been recently claimed in a lattice calculation of $\pi K$ scattering at high pion masses \cite{Dudek:2014qha}, which once more gives  support for the unitarized ChPT approach that we have reviewed in this subsection.

The dependence of the $\sigma$ properties upon  quark mass variations has also been 
the subject of interest in relation to Cosmological and Anthropic considerations.
 As commented in the introduction, the $\sigma$ exchange is responsible for the most part  
of the nucleon-nucleon attraction and thus a small variation in its parameters 
could lead to different nuclear binding energies, transition rates, etc... In particular the cosmological production
of light nuclei, or the stellar production of carbon (very fine-tuned through the triple-$\alpha$ process)
could be severely affected by changes in the quark masses, leading to a Universe very different to the one we live in.
 On the one  hand
it may also lead to Cosmological bounds on the variability of quark masses. 
On the other hand, this has Anthropic implications which lie far beyond the scope of this review.
Incidentally, some of
the most recent and advanced analyses of these $\sigma$ effects have been estimated by means of some form of unitarized ChPT \cite{Cosmology} and then implemented in different ways inside the NN potential and Cosmological models, which are also beyond the scope of this review.

\color{black}
Finally, while preparing this review a study of the dependence of the mass and width of the $\sigma$ and
$\rho$ resonances on the QCD $\theta$ angle has appeared in \cite{Acharya:2015pya}, following a very similar approach to the one just reviewed here to study the quark mass dependence. The $\rho(770)$ mass only decreases very slightly as $\vert \theta\vert$ grows, whereas the $\sigma$ mass suffers a bigger decrease but only when $\vert \theta\vert > \pi/2$.
The widths of both resonances increase following the larger phase space available due to their respective mass decrease. 
\color{black}

\color{black}
To summarize this section, the quark  mass dependence of scalars calculated within the Chiral Unitary Approach, 
strongly supports their identification within  a single nonet
below 1 GeV.  In addition, one should also keep
in mind that their quark mass dependence is 
a prediction of QCD that could be tested on the lattice. 
For the particular case of the sigma we have seen
that this mass dependence has been calculated at 
higher orders within unitarized ChPT and provides a connection 
with lattice QCD. The approach seems reliable, at least semi-quantitatively, because it is shown to give good results for the $\rho(770)$ 
for which lattice calculations  exist. Moreover, for high quark masses the sigma becomes a bound state, with one pole in the first sheet and another one in the second at somewhat different positions. 
Such a bound-state pole has been very recently found in lattice calculations \cite{Briceno:2016mjc}.
Thus, at those quark mssses, the UChPT sigma
can be interpreted as a meson-meson composite state. 
Similar results are predicted for the $\kappa$ and have been recently corroborated on the lattice.
It is also worth noting that the sigma quark mass dependence is of relevance for Cosmological and Anthropic considerations.
\color{black}

\subsection{Other approaches}
\label{subsec:other}

In the previous subsections we have addressed, by means
of unitarized Chiral Perturbation Theory, the description
of meson-meson scattering, which is the process with strongest 
constraints to determine the sigma parameters and behavior. ChPT provides 
the most general
description up to a given order in terms of pions, kaons and etas.
Then, without a priori assumptions about the existence  and properties 
of light scalars, these are generated from unitarization.

However, there are other approaches in which the scalar mesons,
and sometimes even other fields,
 are introduced from the start. This can be done
by using {\it ad hoc} functional forms that respect unitarity and resemble 
a resonance exchange, or by considering  
some specific form of Lagrangian interactions for these heavier fields
among themselves and  
with pions, kaons and etas. Lacking
a well defined power counting, as in ChPT, the choice of interactions
relies on a judicious choice of terms, based mostly 
on symmetries, lowest dimensionality and simplicity.
Sometimes the choice is also guided by some quark-level model.
In most cases, interactions 
are treated at tree level, by considering that
the parameters of the Lagrangian are some kind of effective couplings,
which absorb loop effects. 
This is usually enough to get a rough description
of masses and decays. Note, however, that very often
the mass and widths given for the $\sigma$ and other scalars do not correspond
to poles in the complex plane but depend on the parameterization used to describe each resonance. When this parameterizations are judiciously chosen 
these parameters come out relatively similar to those obtained from the poles in 
Subsec.\ref{subsec:precisepoles}.

Often these models can be considered extensions of the L$\sigma$M,
which at least includes the leading order ChPT interaction between pions, kaons and etas. However, as soon as meson-meson scattering is included in the picture, some form of unitarization is frequently incorporated.

With the advent of ChPT and dispersive methods, these
simple approaches are mostly considered ``toy-models''\footnote{See for instance the abstract in \cite{Black:2000qq} or the conclusions in \cite{Close:2002zu}. } or qualitative approximations, semi-quantitative at best. However,
these models,
without really aiming at precision, 
have been historically  very important to establish
the existence of the $\sigma$, its differences with ordinary mesons and
its spectroscopic classification.
One advantage of these models is the ability to calculated in a simple
way many other processes beyond scattering, in which scalars are involved. In addition, they frequently set the stage for further discussions about mixing, composition, spectroscopic classification, etc, particularly for heavier scalars.

Unfortunately, the problem with these models is that different choices of 
coupling or mass terms, of the choice of resonances that 
are included in the calculation, the observables fitted, etc,
can sometimes lead to different conclusions about the nature,
mixing, etc... of a state. The existence of precise dispersive determinations of the $\sigma$, $\kappa$ and $f_0(980)$
should help reducing this apparent arbitrariness in the future.

In this subsection the most popular or influential models
based on mesonic degrees of freedom will be briefly reviewed.
Most of them lead to some form of ``effective chiral Lagrangian''.
The only exception is the Krakow-Paris model which uses 
separable potentials in which the QCD chiral symmetry breaking pattern is not incorporated.
In the next Section, a specific subsection will be devoted
to some of ``quark-level'' models.

\subsubsection{Extended and Unitarized $L\sigma$M}
\label{subsec:ulsm}

When 
discussing  the L$\sigma$M
in Subsec.\ref{subsec:lsm} above, we already noted that a Pad\'e type unitarization 
had already been applied back in 1970 within the $SU(2)$ case \cite{Basdevant:1970nu}, in order to sum the strong-coupling
perturbation series and satisfy unitarity. 
Pad\'e approximants were actually applied to the $\pi\pi$ scattering amplitude
computed from the truncated L$\sigma$M to second order in the coupling expansion.
Thus, if a partial wave was calculated perturbatively as $t(s)=\lambda t_1(s)+\lambda^2 t_2(s)+...$,
where we have suppressed the $I,J$ indices, 
that partial wave was then approximated by $t^{[1,1]}=\lambda t_1^2/(t_1-\lambda t_2)$.
This may look similar to the unitarization of ChPT reviewed in the previous subsections, 
but it is not, since there the terms in the series 
correspond to higher orders in the energy expansion, whereas here they correspond to higher orders in the coupling constant of a particular model.
The definition of $M_\sigma^2$ was the value of $s$ (in the real axis)
at which the real part of the Pad\'e denominator vanishes. 
The only two free parameters were $f_\pi$ and the coupling $\lambda$.
When $f_\pi$  was taken at its physical value it was found that $M_\sigma\simeq 425\,$MeV
and its width was $220\,$ MeV, far too narrow for our present knowledge.
In addition a $\rho$-like vector appeared at 600 MeV as well as a second scalar
resonance around 870 MeV, not too far from the present $f_0(980)$. 
Note that these two resonances were not included in the original Lagrangian.
The best fit required a too large $f_\pi=125\,$MeV, yielding  $M_\sigma=530\,$MeV, but
still a too narrow $\Gamma_\sigma=310\,$MeV and a far too narrow $\Gamma_\rho=35\,$ MeV.
Although the qualitative features were there and some crossing symmetry checks 
were shown to be satisfied to good accuracy,  for the  knowledge already available in the 70's 
the unitarized L$\sigma$M was clearly insufficient. 

Nevertheless, the unitarized $SU(2)$  L$\sigma$M has been revisited often \cite{Chan:1974ra,Achasov:1994iu,Achasov:2005hm,Achasov:2007fz}, using different approximations,
and always yielding 
a ``semi-quantitative'' description of $\pi\pi$ scattering, requiring the presence of a broad scalar
around 400-650 MeV. 
It was even shown \cite{Delbourgo:1993dk} that the mesonic $SU(2)$ L$\sigma$M could be generated dynamically
from the one-loop treatment of a quark-level L$\sigma$M Lagrangian, 
although in this case the $\sigma$ mass came around 650 MeV.
It is interesting also to note that if only the 
tree level L$\sigma$M is unitarized, the $\sigma$ resonance mass width comes now somewhat too large,
in the 650-750 MeV range, in contrast to what happened in the Pad\'e approach up to second order.
Also, the Lagrangian or ``bare" $\sigma$ mass comes out 
around 1 GeV, very different from the resonance mass.

The L$\sigma$M extension to three flavors and $SU(3)$ 
was carried out in \cite{Volkov:1984kq,Schechter:1993tc}. 
In \cite{Tornqvist:1999tn} a broken $U(3)\times U(3)$ L$\sigma$M was compared with experimental
data on masses and decays of the lightest scalar and pseudoscalar mesons. The comparison was performed at tree level. The model has six parameters, which included three 
symmetry breaking terms. By 
fixing  $M_\pi$, $M_K$ and $M_{\eta'}^2+M_\eta^2$, $f_\pi$ and $f_K$
 to their physical values, there is just one parameter left, which only changes the $\sigma$ and $f_0(980)$ masses and couplings
to two-meson states.  Only a very crude description was achieved, since
the $a_0(980)$ came at $\sim 1030\,$ MeV, the $\sigma$ at $\sim620-650$ MeV,
the $f_0(980)$ at $\sim 1200\,$MeV and a $K\pi$-resonance at $1123\,$MeV.
At that time the existence of a $\kappa$ meson below 1 GeV was unclear 
and this last pole was identified with the $K^*(1430)$ instead of a light $\kappa$. Such 
a large mass difference between the $a_0(980)$ and the $K^*(1430)$ was complicated to understand if they belonged to the same multiplet \cite{Close:2002zu}. 
Nevertheless,  in a later review \cite{Close:2002zu} the author claimed growing evidence
for a light nonet which comprised a $\kappa$ with an undetermined mass. 
We will comment again about this model in Subsec.\ref{subsec:uqm} below. 

Almost simultaneously an SU(3) L$\sigma$M was also being used as a template for a possible classification of light scalars by S. Ishida and collaborators. By analyzing $\pi\pi$ \cite{Ishida:2001pt} and $K \pi$ \cite{Ishida:1997wn} scattering data they found that $\sigma$ and $\kappa$ poles where necessary. 
In these works an {\it ad hoc} deformation of the Breit-Wigner formalism was used. In a later model \cite{Ishida:1999qk} they considered ``plausible to 
regard the $\sigma(600)$, $\kappa(900)$, $a_0(980)$ and $f_0(980)$, forming with the members of the $\pi$ nonet a linear representation of $SU(3)$ chiral symmetry". As usual, with a tree level analysis, the L$\sigma$M can accommodate qualitatively the masses of these states.
Only with that information it was suggested to assign these states to $\bar qq$ configurations. However, the values of the mass and widths of the $\sigma$ and $\kappa$  do not describe well our present knowledge.

Going beyond just tree level masses and decays, meson-meson scattering 
was also studied within a unitarized SU(3) L$\sigma$M in \cite{Black:2000qq},
which was considered as a ``toy-model'' by the authors. 
The method consisted in unitarizing with a K-matrix the tree level approximation.
The results for the $\sigma$  did not change much with respect to
the $SU(2)$ case. The  sigma mass quoted
is 457 MeV and the width 632 MeV. Once again the ``bare'' or Lagrangian-mass came out
higher, around 850 MeV.
The model was able to describe the $f_0(980)$ shape correctly and 
a fairly good description of data was found, although
it was also shown that giving up the renormalizability  
constraint could be helpful for the fit above 1 GeV.
In the second part of this paper the possibility of including a second nonet
at higher energies within a chiral framework was sketched. We will discuss this issue
in a separate subsection below.

Another recent extension of the L$\sigma$M are the works 
by the Frankfurt group in \cite{Parganlija:2010fz,Parganlija:2012fy}.
On a first instance the two flavor case was considered \cite{Parganlija:2010fz}.
Here vector and axial-vectors were included, although in a less restrictive 
framework that usual, in which these fields ensure global instead of local
$U(2)\times U(2)$ invariance,  guided by the fact that the symmetry is global in QCD as well.  In addition, terms of dimension four were
added in order to calculate 
all two-meson decay widths as well as $\pi\pi$ scattering lengths.
The parameters of the model are then fitted to data on decays.
By choosing an appropriate potential the scalar-isoscalar and scalar-isotensor masses scale
 as $O(1)$ and its width as $O(1/N_c)$. 
The $N_c$ scaling was a consequence of the assumptions that the $I=2$ vector corresponds to the $\rho(770)$, the $I=1$ pseudoscalar corresponds to the  pion and that both are quarkonia.
No further assumptions about the fields were made and that leads to the above $N_c$ behavior,
which, as we will see in Subsec.\ref{subsec:nc} 
is the expected behavior for $\bar qq$
resonances. However, if these two fields were identified with the $\sigma$ and $a_0(980)$, respectively, the width of the $\sigma$ came out too small. The alternative assignment $f_0(1370)$ and $a_0(1450)$ yielded the correct widths but an incorrect scattering length. This was
interpreted as a need for more scalar degrees of freedom and thus the model was extended to three flavors in  \cite{Parganlija:2012fy}. A ``surprisingly good''
fit to meson masses, decay widths and decay amplitudes was obtained, in which
once again the preferred assignment was for $\bar qq$ scalar mesons to be above 1 GeV. 
Note that this time the $\bar qq$ assignment of a strange scalar resonance 
could also be studied, and consistently it also preferred to be identified with the $K^*(1430)$.
According to the authors this indicated the need for an additional nonet of scalars below 1 GeV
that should not match a $\bar qq$ assignment.
In addition, the role of the glueball mixing with quarkonia has been studied within this extended L$\sigma$M and the conclusion is that the $f_0(500)$ glueball component, if any, is strongly subdominant \cite{Janowski:2011gt}, with the $f_0(1710)$ as preferred scalar glueball candidate \cite{Janowski:2014ppa}.
 All in all, the extended version of the L$\sigma$M in \cite{Parganlija:2010fz,Parganlija:2012fy} is a much more complete version including not only scalars and pseudoscalars but  also vectors and axial-vectors within a single theoretical framework. This could be interpreted as an effective approach offering a rather realistic treatment of the meson spectrum, although  given the large number of physical resonances it is of course not complete (but there is an ongoing effort to extend it further\footnote{F. Giacosa and D. Parganlija, private communication.}). Nonetheless, the agreement with data is quantitatively alright in all channels, arguably with the exception of the $I=0$ scalars. The reasons for this are the experimental uncertainties and the various possibilities for mixing scenarios in the isoscalar sector.

In summary, the SU(3) L$\sigma$M has the basic ingredients of the lightest scalar nonet and when applied at tree level 
gives a qualitative, but not accurate, description of the masses. 
Since it is consistent with the LO ChPT amplitudes, when unitarized it can provide a semi-quantitative
description of scattering data. If supplemented with additional resonances with other quantum numbers, or Lagrangian terms, the agreement can even be improved. When sufficient experimental data is included in the picture, and as long as one aims just for a semiquantitative  description, it preferably points out at a non $\bar qq$  nature of the lightest nonet.
However, as carefully warned by many of the practitioners, it should just be considered a toy model.

\subsubsection{Syracuse model}
\label{subsec:Syracuse}

One of the most elaborated approaches for scalars has been developed
over the years by Schechter and collaborators at Syracuse University. They have explored both the linear and non-linear representations and have included explicitly different sets of resonances. 
They contributed very significantly to the $\sigma$ and $\kappa$ comeback in the mid 90's and the identification of the 
members of the lightest scalar nonet as the $\sigma$, $\kappa$,
$f_0(980)$ and $a_0(980)$. They have described not only mass 
relations and decay calculations but meson-meson scattering as well.
A predominantly tetraquark-like configuration for the lightest scalars
is favored in their models, with strong rescattering effects due to ``regularization'', i.e. some form of unitarization.

In particular, in \cite{Sannino:1995ik} the study of violations
of the
partial-wave unitarity bound, Eq.\ref {ec:unitbounds}, was shown
to favor the existence of a sigma  resonance. The approach relied on the
observation that at leading $1/N_c$ $\pi\pi$ scattering reduces
to a tree-level exchange of an infinite number of width-less
 resonances (see Subsec.\ref{subsec:nc} below).
Thus, a model was built 
by including the non-linear LO ChPT Lagrangian in Eq.\ref{ec:Lag2},
plus the tree level  chiral invariant exchange of the $\rho(770)$.
This delayed drastically the onset of unitarity violations but
even the  addition of other leading $N_c$ resonances like the $f_0(1300)$, $f_2(1270)$ and $\rho(1450)$  was not sufficient to restore unitarity. 
However the inclusion of 
a broad and light scalar-isoscalar resonance was enough to 
satisfy the unitarity bound up to 1.3 GeV. 
This broad state was ``presumably not of the 
simple $\bar qq$ type'' and hence its exchange would be subleading in the $N_c$ counting. 
In a later work of the group \cite{Harada:1995dc}, the $f_0(980)$ and the $\bar KK$ threshold were added to the analysis
and  it was possible to obtain a good description of $\pi\pi$ scattering data up to $\sim 1.2\,$GeV. 

In connection
with our previous discussions  it is worth noting that
the $\sigma$ was parameterized by a ``regularized'' description  proportional
to $M_\sigma G/(M_\sigma^2-s-iM_\sigma G')$, where $G\neq G'$. Thus,
this shape {\it was not} identical to a Breit-Wigner. From the data fit, it was found that $M_\sigma\simeq 525-560\,$MeV and $G'\simeq370-470\,$MeV.
A strong support for a $\kappa$ meson below 1 GeV 
was also obtained in \cite{Black:1998zc}
following a relatively similar approach, but applied to $K \pi$ scattering.

Then, with these evidences for both the $\sigma$ and $\kappa$,
an effective Lagrangian was built \cite{Black:1998wt}
incorporating explicitly a scalar nonet into a L$\sigma$M.
In \cite{Black:1998wt}, an analysis of mass terms was carried out.

We will follow here the formalism in \cite{Black:1998wt} because it is a basic template
 for many other models. In particular the following scalar matrix is used:
\begin{equation}
S \equiv
\left(
\begin{array}{ccc}
S_1^1 & a_0^+ & \kappa^+ \\
a_0^- & S_2^2& \kappa^0 \\
\kappa^- & \bar\kappa^0 &  S_3^3
\end{array}
\right); \quad a_0=\frac{S_1^1-S_2^2}{\sqrt{2}}, \quad S_1=\frac{S_1^1+S_2^2+S_3^3}{\sqrt{3}}, \quad S_8=\frac{S_1^1+S_2^2-2S_3^3}{\sqrt{6}},
\label{ec:Sfields}
\end{equation}
where the last two combinations are isosinglets but  only the last one belongs to the octet. Note that
$S_1$ and $S_8$ are expected to mix due to $SU(3)$ breaking.
Following the ``ideal mixing'' idea of Okubo \cite{Okubo:1963fa}, which works well for vector mesons,
but now for $0^+$ instead of $1^-$ fields, the ``ideal mixing model'' allows for the following mass terms:
\begin{equation}
{\cal L}_{m_{ideal}}=-a Tr(SS)-b Tr(SS {\cal M}),\quad {\cal M}= diag(1,1,m_s/\hat m).
\end{equation}
Diagonalizing the fields with this mass Lagrangian
leads to the ``ideally mixed'' mass eigenstates 
$S_3=S_3^3$ and $S_{12}=(S_1^1+S_2^2)/\sqrt{2}$. 
Moreover, 
with only these terms the masses of the fields in Eq.\ref{ec:Sfields}
satisfy the constraints $M^2_a=M^2_{12}=2M^2_\kappa-M^2_3$,
which have two kinds of solutions:
\begin{eqnarray}
M_{3}>M_\kappa>M_a=M_{12},
\label{ec:mhierarchy}\\
M_{12}=M_a>M_\kappa>M_3. \label{ec:minverted}
\end{eqnarray}
Intuitively, the first solution is easily interpreted in terms of $\bar qq$ mesons,
by identifying $S_a^b\sim q_a \bar q^b$, with $q_1,q_2,q_3=u,d,s$. In this way $S_3=s \bar s$
whereas $S_{12}$ has no strangeness, so that
the heavier mass of the strange quark will roughly explain the 
mass hierarchy in Eq.\ref{ec:mhierarchy}. 
However, the observed mass hierarchy of light scalars, $M_f\simeq M_a\simeq 980\,$MeV, $M_\kappa\sim 700-800\,$ MeV, $M_\sigma\sim 500\,$ MeV does not fit into
this pattern, but rather into that of Eq.\ref{ec:minverted}.  

Thus, in the physical case it is Jaffe's tetraquark structure \cite{Jaffe:1976ig}
the one that allows for a more natural quark-level interpretation. In particular,
by defining $T_a=\epsilon_{abc}\bar q^b \bar q^c$ and 
$\bar T^a=\epsilon^{abc}q_bq_c$ then, schematically, one can identify:
\begin{equation}
(S_a^b) \sim (T_a\bar T^b)=
\left(
\begin{array}{ccc}
\bar s \bar d ds  & \bar s\bar dus & \bar s \bar d ud \\
\bar s \bar u ds & \bar s \bar u us & \bar s \bar u ud \\
\bar u \bar d ds & \bar u\bar d us&  \bar u \bar d ud
\end{array}
\right), \quad f_0(980)=S_{12} , \quad \sigma=S_3^3 .
\label{ec:4qmatrix}
\end{equation}
This isoscalar meson identification is referred to as ``dual ideal mixing''.
The observed mass hierarchy is roughly recovered because 
each  additional strange quark increases the mass of the meson by some hundred MeV.
But let us remark that  within the effective Lagrangian
 approach of \cite{Black:1998wt} it is not necessary to 
assume any internal quark structure for the
mesons.

In \cite{Black:1998wt} it was shown that the previous rough agreement
could be improved by adding the following
two terms to the mass Lagrangian: $-c Tr(S)^2-Tr(S)Tr(S{\cal M})$.
Terms quadratic in the quark masses were neglected assuming that quarks masses can be treated as perturbations.
The parameters could now be fixed from the $f_0(980)$, $a_0(980)$
and $\sigma$ masses (for the latter 550 MeV was taken). This resulted
in a prediction $685<M_\kappa<980\,$MeV, in fair agreement with observation.
The price to pay is that now the isoscalar fields do not follow the ideal nor
the dual ideal mixing.
The new mixing scheme can be defined by the angle $\theta_s$, where
\begin{equation}
\left(
\begin{array}{c}
\sigma\\
 f_0 \\
\end{array}
\right)
=
\left(
\begin{array}{cc}
\cos \theta_s & -\sin \theta_s \\
\sin \theta_s & \cos \theta_s\\
\end{array}
\right)
\left(
\begin{array}{c}
S_3\\
S_{12}  \\
\end{array}
\right).
\end{equation}
Ideal mixing corresponds to $\theta=90^\degree$ and dual ideal mixing to $\theta=0^\degree$. By studying the masses only,
two solutions were found: $\theta_s\simeq -21 ^\degree$ and $\theta_s\simeq -89^\degree$. More information was needed
to decide which was the preferred solution.

Thus, in order to take resonance decay data into account, more Lagrangian terms were considered, 
coupling the $S$ matrix to a pair
of pseudoscalar matrix fields containing pions, kaons and etas. These terms contained two 
derivatives to respect chiral symmetry and the NGB nature of pseudoscalars. 
Terms with higher derivatives were neglected for simplicity.
When considering only decays to pions and kaons, the analysis 
involved only two additional parameters, allowing for a description
of  $\kappa\rightarrow K\pi$,
$\sigma\rightarrow \pi\pi$, $\sigma\rightarrow\bar KK$,$f_0\rightarrow \pi\pi$ and  $f_0\rightarrow\bar KK$ decays.
For the width calculation a $M G/(M^2-s-iM G')$
form was again assumed for the $\sigma$ and $\kappa$. This adds another two parameters. The outcome of the analysis was a
preferred mixing angle $\theta_s\simeq -17^\degree\pm4^\degree$. 
This result was closer to the 
``dual ideal mixing''scenario and therefore supporting the predominant
tetraquark interpretation over the regular $\bar qq$ nature.
Nevertheless, both the $\sigma$ and $\kappa$ were 
treated differently from the $f_0(980)$ and $a_0(980)$
due to the  regularized functional form with the $G$ and $G'$ parameters.
This was some kind of unitarization, therefore suggesting that meson loops also play a very relevant role in the nature of the $\sigma$ and $\kappa$. It should be remarked that no dynamical quark-level dynamics were assumed so that 
the tetraquark/molecule discussion was not really addressed, just the mixing pattern.

This angle can be compared with the results obtained in \cite{Oller:2003vf} within the unitary chiral approach already discussed in 
Subsecs.\ref{subsec:ChUABSE} and \ref{subsec:qm}.
Using just the leading order ChPT partial waves and a relatively simple coupled-channel unitarization scheme, the poles associated to 
all the members of the multiplet could be calculated. By changing 
to the singlet/octet basis instead of the charge basis, 
one can calculate the couplings, i.e. the residues of the poles, to these states and infer the mixing angle. In \cite{Oller:2003vf} a mixing angle $\theta=+19^\degree\pm 5^\degree$ 
with respect to the $S_1$, $S_8$ states is found, which translates into $\theta_s\simeq +36^\degree\pm5^\degree$, which 
also disfavors the $\bar qq$ interpretation. 

No description of meson-meson scattering was done within \cite{Black:1998wt,Black:1999yz},
although, as commented in the previous subsection,
  the group has addressed this issue in \cite{Black:2000qq}
by unitarizing the $SU(3)$ L$\sigma$M. The resulting masses of both approaches are fairly consistent. 

A relevant observation emphasized in \cite{Black:2000qq} is that 
$\bar qq$ states, ``tetraquarks'' or ``molecules'' made of four quarks 
but rearranged as two pseudoscalar mesons transform the same under $SU(3)\times SU(3)$
\footnote{$U(1)_A$ transformations differ between these configurations, but their 
effect could be absorbed by some parameters of the model.}.
Thus, from these transformations alone one cannot differentiate the inner structure of scalar mesons. Within these models 
it all reduces once again to the mass hierarchy and its interpretation 
in terms a constituent quarks and their mixings, as it was done in \cite{Black:1998wt}. 

The group has also studied the effect of introducing
a heavier scalar nonet into this picture and has refined the model 
in several subsequent works \cite{Fariborz:2005gm}. This will be commented
in the next subsection together with other models that also 
consider two scalar nonets within a chiral framework. 

\subsubsection{Two-nonet chiral models}
\label{subsec:twononets}

Not only the lightest scalar mesons have difficulties 
fitting into a $\bar qq$ scheme. If one wants to accommodate the
$f_0(1370)$, $f_0(1500)$, $f_0(1700)$, $a_0(1450)$ and $K_0^*(1430)$ into 
such an scheme, the first observation is that there is one more isoscalar than needed to form another nonet of heavier scalars.
As already commented, this could be due to the presence of an additional  glueball
state, which, as we saw in the introduction, is predicted by lattice calculations \cite{latticeglueball} 
to be in the 1.5 to 1.8 GeV region.
But even for those states that cannot mix with a glueball there is some trouble, because the 
$a_0(1450)$ and the $K^*(1430)$ are nearly degenerate, whereas in a $\bar qq$ scheme
the later should be some few hundred MeV heavier, as it would contain one strange quark.
Also, the $f_0(1500)$ is almost degenerate with the $a_0(1450)$. 
Given the fact that many of these states have widths of some hundred MeV and that the
$\sigma$ and $\kappa$ are also very wide, it also seems very likely that they could mix.
A simultaneous treatment of these two multiplets seems then appropriate.

Thus, the Syracuse model was extended \cite{Black:1999yz} 
within the framework of non-linear chiral Lagrangians by incorporating 
another $S'$ matrix containing a conventional $\bar qq$ nonet.  
In this model,  mesons in the the heavier scalar nonet interact with the previous $S$-matrix through the following term
\begin{equation}
{\cal L}_{mix}=-\gamma Tr(S S'). 
\label{ec:instantonmixing}
\end{equation}
This interaction form can be justified within QCD from instanton dynamics \cite{Fariborz:2008bd} and as we will see
in Subsec.\ref{subsec:models} it has also been  used in other approaches.
For simplicity, the analysis is carried out in the ``unmixed'' approximation
for the lightest nonet, which for $\gamma=0$ would then exhibit a ``dual ideal mixing''. Still, 
the masses and two-meson decays 
of all these resonances
can be fitted rather nicely and ``economically''.
Within this scheme the $\sigma$ comes out as mostly tetraquark but with some significant mixing of a 
heavier scalar above 1 GeV. Once again  no dynamical quark-level dynamics were assumed
and different dynamical tetraquark models  (conventional tetraquark, diquark-antidiquark, molecules...) 
could fit within this formalism.

The two-nonet model was also considered in \cite{Close:2002zu,Tornqvist:2002bx} with the same coupling
between the two nonets as in Eq.\ref{ec:instantonmixing}. However a ``Higgs-like'' mechanism
at the hadron level was at work, where the axial-vector mesons  ``ate up''
a nonet of pseudoscalars so that at the end there was only one nonet of pseudoscalars and two nonets of scalars.  Since these models predict a large mixing
between ``tetraquark'' and $\bar qq$ components, but are also dominated by large
meson-meson rescattering, in \cite{Close:2002zu} it was conjectured that
these components may have a different spacial distribution inside the resonance.
In particular it was suggested that ``near the center'' they were dominated
by a diquark-antidiquark structure in an S-wave, mixed with some $\bar qq$ component,
but that further away these four particles would rearrange into a meson-meson state. For softer energies one would then be seeing the outer meson-meson structure, whereas for more energetic probes one might be seeing a tetraquark configuration. The model in \cite{Tornqvist:2002bx} is used as a realization of the conjecture that
the composition of the scalars might be seen different depending on the energy 
used to probe them. However, taking into account that at present we are still discussing
the composition, there is little evidence of its spacial distribution, if there is any. In addition, the spacial distribution could be highly model dependent.

There is another model \cite{Napsuciale:2004au} which, despite having also two chiral nonets,
 finds the opposite 
mixing pattern, namely, that the scalars with a predominantly ``tetraquark'' structure
are the heavy ones. The mixing is, however, very strong, so that the ''tetraquark`` component inside the 
sigma would still be rather large, although not dominant. In this case, apart from the instanton-induced mixing, 
 additional mass terms are included
to provide an inverted hierarchy for the
lightest scalars. As in previous examples, this model yields just a rough approximation to
masses and decays, but, as pointed out by the very author, in this case it is not able to describe properly the decays in the 
isoscalar sector.  Therefore it could actually be interpreted as evidence against the possibility to construct a model 
where the light scalar nonet is mainly a $\bar qq$ and the heavier one mainly a tetraquark.
In addition, as it will be discussed in Subsecs.\ref{subsec:ncmodelindep} and \ref{subsec:ncUChPT}, a predominant $\bar qq$ nature is disfavored 
by the large-$N_c$ behavior of the $\sigma$ found within unitarized NLO and NNLO ChPT.
Such a component can only be subdominant and at large-$N_c$ if its mass is above 1 GeV. 
We will also see in Subsec.\ref{subsec:reggesigma} that a recent calculation of 
the $\sigma$ pole Regge trajectory using dispersion theory also shows that it does not behave as a 
meson predominantly made of a $\bar qq$ state.  

Finally, there is yet another model by the Frankfurt group \cite{Giacosa:2006tf}
in which the mixing between a diquark-antidiquark nonet
and $\bar qq$ nonet has been studied by means of chiral Lagrangians. 
A similar analyses on decay constants, masses, etc, is performed and
yet again the usual conclusion is reached. Quoting the authors: ``scalar states below  1 GeV are mainly four-quark states and the scalars between 1 and 2 GeV quark-antiquark states, probably mixed with the scalar glueball in the isoscalar sector".

\subsubsection{The Krakow-Paris multichannel model}

This is the only model to be reviewed here 
in which chiral symmetry is not implemented at the Lagrangian level \footnote{From 1997, the scalar-isoscalar scattering length from ChPT was imposed 
as a further constraint on the fit, but that did not 
change the results in any significant way. R. Kaminski, private communication.}.
\color{black} We nevertheless include it here because it has received quite some attention and, for our purposes regarding the sigma chiral symmetry and dispersion theory, it yields 
an important estimate on the size of the effect of the 
four pion state on the sigma pole. \color{black}
The formalism makes use of relativistic propagators within a system 
of coupled Lippmann-Schwinger equations to analyze $\pi\pi$ and $\bar KK $ scattering  data.
Thus, it bares some vague resemblance with the BSE method above. 
As far as the $\sigma$ parameters are concerned
the two-channel model \cite{Kaminski:1993zb} is already a good enough approximation, since they
are barely modified by the three-channel formalism \cite{Kaminski:1997gc,Kaminski:1998ns}.
The two-channel model can be 
schematically recast in momentum space as:
\begin{equation}
  \label{eq:LSRobert}
  \langle \vec p\vert T\vert \vec q\rangle= \langle \vec p\vert V\vert \vec q\rangle
+\int  \frac{d^3 k}{(2\pi)^3} \langle \vec p\vert V\vert \vec k\rangle
\langle \vec k\vert G\vert \vec k\rangle
\langle \vec k\vert V\vert \vec q\rangle,
\end{equation}
where $V,T, G$ are symmetric $2\times 2$ matrices  ($1=K\bar K, 2 =\pi\pi$) 
and $\langle \vec k\vert G_{ij}\vert \vec k\rangle=\delta_{ij}/(E-2E_i(s)+i\epsilon)$,
$E$ is the total energy and $E_i(k)=\sqrt{k^2+M_i^2}$. 
The interaction is then parameterized by a separable (non-local) potential:
\begin{equation}
  \label{eq:separable}
  \langle \vec p\vert V_{11}\vert \vec q\rangle=\lambda_{11} g_1(p)g_1(q),\quad
\langle \vec p\vert V_{n 2}\vert \vec q\rangle=\lambda_{n2} g_n(p)g_2(q)+\lambda_{m3} g_n(p)g_3(q), \quad n=1,2,
\end{equation}
where a third index has been included to simplify the notation of the five
coupling constants $\lambda_{ij}$ and the following form factors are used:
\begin{equation}
  \label{eq:gRobert}
  g_i(p)=\sqrt{\frac{4\pi}{M_i(p^2+\beta_i^2)}},\quad M_1=M_K, \quad M_2=M_3=M_\pi,
\end{equation}
which introduce three parameters more, called ``range parameters'', $\beta_i$.
This separable formalism implies that the $T$ matrix is also separable 
and Eq.\ref{eq:LSRobert}
becomes algebraic, which makes the model very manageable. With these 8 parameters a 
remarkable fit to data is achieved up to 1.4 GeV \cite{Kaminski:1993zb}. In addition,
the solutions are analytic and can be continued to the complex plane in search for poles.
The model finds a $\sigma$ pole that would not be present if the relativistic effects had not been included.
In addition, two more poles are found, one around 1400 MeV
and another one corresponding to the  $f_0(980)$ that becomes a 
compact $\bar KK$  molecule  ($\sim 0.7\,$fm) if the coupling to $\pi\pi$ is switched off.

In order to improve the description of the 1300-1600 MeV region, 
an effective $4\pi$ channel, called $\sigma\sigma$, 
was added to the model \cite{Kaminski:1997gc,Kaminski:1998ns}. 
The resulting sigma pole of the latest model update has $M_\sigma=523\pm12\,$MeV 
and $\Gamma_\sigma=518\pm14\,$MeV. These values are well within the RPP 2012 estimate 
but outside the conservative dispersive estimate suggested in Eq.\ref{myestimate} of
the present report. Of course, one has to take into account that this is a very simple model,
with no left cuts, and a very specific choice of potential and form factors, which necessarily only provide an approximation to the actual amplitude in the complex plane.

More interestingly, this model allows for an estimation of the effect
of the $4\pi$ intermediate channel: it only shifts 
 the $\sigma$ mass by 6 MeV and the width by 8 MeV.
This is one of the two instances where this effect has been estimated.
Despite having all the approximations commented above, we consider this rough estimate relevant, because the $4\pi$ state is not included
in the $\sigma$ pole determinations from Roy and GKPY  equations that we have 
reviewed in Sec.\ref{sec:parameters}. When the $\sigma$ parameters are obtained using exact solutions of Roy equations
\cite{Caprini:2005zr,Moussallam:2011zg}, the $4\pi$ is simply neglected below the matching energy, whereas when Roy and GKPY equations are used just as fit constraints,
one merely assumes that the effect might be included in the estimated systematic uncertainties.
Judging by the size of these estimates, the RPP2012 ``radical" or restricted range Eq.\ref{rppradicalpole} seems
to have a too small uncertainty attached.  In contrast, the 
conservative dispersive estimate
proposed here in  Eq.\ref{myestimate} has an uncertainty that covers well enough this source of error. 

\color{black}
\subsubsection{Conformal symmetry in chiral models: A dilaton model and gravity duals}

To end this section, we will briefly summarize two kind of models whose 
driving motivation or main feature is
not chiral symmetry but conformal symmetry, although they are formulated 
consistently wit the requirements of spontaneous chiral symmetry breaking.
First, we will introduce one in which an spontaneous breaking of conformal symmetry is assumed to occur in QCD. Second we will review other models in which a dual correspondence holds between an (almost)
 conformal field theory in our space  
and a gravitational theory in another space.

Thus, let us first comment on how
the recent and precise $\sigma$ determinations also lead to revisit \cite{Crewther:2013vea} the old ideas \cite{Ellis:1970yd} that the vacuum $\langle q\bar q\rangle$ condensate, in the chiral limit, may also break spontaneously scale invariance. 
This spontaneous scale breaking might occur if the (non-perturbative) running of the QCD coupling with three light flavors has 
a fixed point in the infrared limit. In such case an extra NGB appears, which is massless in the chiral limit and has been identified with the sigma. 
An effective chiral invariant Lagrangian, alternative to SU(3) Chiral Perturbation Theory, can be build with an explicit sigma field, which couples to the vacuum via the energy momentum tensor and a counting scheme in which $M_\sigma^2=O(M_K^2)$ has been proposed in \cite{Crewther:2013vea}.
Such couplings for the sigma are like those of a dilaton in the context of nonlinear realizations of conformal symmetry,
and  this is why the sigma is also called a dilaton within this context, but should not be confused with the dilaton in gravitational or similar contexts. It should be noted that the sigma in this formalism becomes a quark-antiquark state, whose width scales as $\Gamma=O(1/N_c)$. The arguments that disfavor the linear sigma model when integrating out the sigma do not apply directly here because the interaction terms are different.
An interesting phenomenological feature of this model is a natural explanation for the $\Delta I=1/2$ rule in kaon decays with the $K_S\sigma$ coupling fixed from $\gamma\gamma\rightarrow\pi\pi$ and $K_S\rightarrow\gamma\gamma$.
The authors of \cite{Crewther:2013vea} claim that the ``lowest order appears to be a good approximation'',
but that this is part of a ``wider program to obtain numerically convergent'' expansions in the scalar-isoscalar sector.
They also point out that more  stringent tests would require explicit calculations of $\sigma\sigma\sigma$, $\sigma\sigma\pi\pi$,... couplings and that some unitarization procedure may check whether it produces small corrections to lowest order results. To my view, it would also be interesting to get a precise description of $\pi\pi$ scattering phase shifts as well as threshold parameters, because we have already seen many examples in which getting the mass and width of the sigma is not enough. Moreover, the proposed $q\bar q$-like  large $N_c$ behavior of the $\sigma$ is rather unnatural as we will discuss in Sec.\ref{subsec:ncmodelindep} 
when discussing model independent observables without using standard ChPT (the unitarized ChPT results of Sec.\ref{subsec:ncUChPT} cannot be used in this context because this scenario follows different assumptions). Also, the $q\bar q$ nature is not naturally reconciled with the Regge behavior to be discussed in Sec.\ref{subsec:reggesigma} obtained without any a priori assumption on the sigma nature. Finally, there is the issue of classifying the rest of scalars in multiplets, and the evident similarities of the $\sigma$ with the $\kappa$ or $K^*_0(800)$ resonance, which could not be a dilaton.

In a second kind of approaches, the fact that in the chiral limit QCD is relatively close to a conformal theory
has raised much interest on the conjectured ``holographic'' correspondence \cite{Maldacena:1997re}, or duality. This occurs between the strongly coupling
limit of a (super-)conformal field theory (CFT) defined on the Anti-de-Sitter (AdS) asymptotic boundary, and the propagation of weakly coupled strings in a higher dimensional AdS space.
This ``AdS/CFT'' holographic description of large-$N_c$ ($N=4$ supersymmetric) gauge theories 
has triggered the study of their dual gravitational picture,
in which the spectrum of the gauge theory strong regime 
might be computed semiclassically in the dual theory weak 
regime.
In practice this conjecture 
has been implemented within a five-dimensional theory holographically dual to strongly
coupled QCD (AdS/QCD) \cite{Erlich:2005qh}. 

Some caveats must be be kept in mind, though. QCD is not supersymmetric nor
really a conformal theory, so that a mass scale related to the typical hadronic scale 
must be incorporated in the dual theory. This can be done with a hard ``wall'', i.e. a cutoff, 
or a soft ``wall''
that effectively cuts off the AdS space in the infrared region .
The ``soft-wall ``scenario is implemented with a dilaton field in the dual space (not to be confused with the dilaton model discussed right above).
Moreover the results of this duality would not correspond 
 to QCD but to its large-$N_c$ limit.
Still, these models may be theoretically attractive because they may provide 
some understanding
of some non-perturbative QCD features, since they could be calculated perturbatively in the dual space. 

In principle, these models can implement the spontaneous chiral symmetry breaking 
pattern of QCD and the existence of eight NGB and could even be reformulated into a Chiral Lagrangian for mesons. This is why we have included them in this Section.
Actually, some of them have been recast into Chiral Perturbation Theory, even giving specific predictions for
some of the Low Energy Constants \cite{DaRold:2005vr}.  

A thorough description of all models based on this formalism is far beyond the scope of this report.
The main reason is that the performance of holographic models is particularly disappointing
for the sigma resonance. Most 
holographic models of meson spectra just {\it avoid} dealing with the sigma and 
concentrate in mesons on linear Regge trajectories 
or heavier scalars, if any scalar at all. In addition, 
when dealing with heavier scalars, very often these are not isoscalar states.
Finally, even when discussing the sigma, only 
the mass is estimated, but not the width, which is its most salient feature. And even the mass is usually
far from the values that have been well-known for long and very far from those established with precision from the dispersive analyses we have reviewed in this report.
Actually, the historical confusing situation around the $\sigma$ meson that 
we have illustrated in Sec.\ref{subsec:history} and the fact 
that the $\sigma$ mass was listed
in the RPP with a 400 to 1200 MeV uncertainty, has lead some authors to consider models
with sigma masses of 100 or 200 MeV \cite{crazyadsmass,Li:2013oda}, well below the two-pion threshold.
Those models do not describe the lightest scalar resonance, even qualitatively.

A first reason for the difficulty of dual models to describe the sigma is that meson masses in these models are usually 
obtained by fitting parameters to Regge linear trajectories.
However, as we will see in Sec.\ref{subsec:reggesigma},
it has been well known for long from phenomenology that the $\sigma$ meson does {\it not} fit well into these linear trajectories \cite{Anisovich:2000kxa} and it has been shown recently that dispersion theory and the precise $\sigma$ pole determinations imply that its trajectory must be non-linear \cite{Londergan:2013dza}. If it was linear, it would give a very bad description of $\pi\pi$ scattering.

Actually, when the sigma mass with a large uncertainty is considered
together with other mesons in dual model fits, either 
 the resulting sigma mass or the fit itself do not come out particularly well \cite{Li:2013oda,Gherghetta:2009ac,Sui:2010ay}.
Some authors are well aware of this problem and have checked that removing the $f_0(500)$ or the $f_0(980)$ from their
fits results ``in a much better fit to the model'' \cite{Gherghetta:2009ac} and that this could be due that these scalar states ``mix with scalar excitations, scalar glueballs,
and possible four quark states'' or that ``it may be that either
the lowest or first radially excited state has been misidentified'' 
(actually these authors do find their lightest scalar state around 800 MeV). 

A second reason is that, in principle, results of 
dual models belong to the large-$N_c$ limit where in general, and
as will be seen in Sec.\ref{subsec:nc}, meson widths tend to zero. 
As a consequence there are no reliable dual estimates of the most salient feature of the sigma: its huge width.  Actually, most dual models that deal with the sigma simply ignore its width.
Moreover, we will also see in Sec.\ref{sec:nature} that 
it is very likely that the lightest scalar $\bar qq$ state might appear 
around 1 GeV in the large-$N_c$ limit. Due to mixing, this 1 GeV state might be a subdominant component of the $\sigma$ at $N_c=3$, but the sigma, as such, would no longer exist around 500 MeV in the large $N_c$ limit.
It is interesting to find that some of the most popular dual models follow this pattern and
find the lowest scalar-isoscalar state around 1 GeV \cite{Ghoroku:2005vt,Colangelo:2008us,dePaula:2009za,Erdmenger:2014fxa}. 
In most cases this state is directly identified with the $f_0(980)$, 
although it was also identified with the sigma \cite{Ghoroku:2005vt,dePaula:2009za} when its mass was still listed as 400-1200 MeV in the PDG, which is not possible any longer. 
However, as pointed out by some of the very authors of these models \cite{Colangelo:2008us,Erdmenger:2014fxa},
for the $f_0(500)$, is that at large-$N_c$ no state is found around 500 MeV.
Therefore they do not describe the $\sigma$.
But it is very plausible that such a state around  1 GeV might correspond to
a subdominant component of the sigma that is seen at large-$N_c$,
and which, being of a $q \bar q$ nature, might have a linear Regge trajectory.

Given the relatively poor results in the description of the $\sigma$ meson from dual models when considering it a $q\bar q$ state, the tetraquark description has also been explored. Interestingly, it has been found \cite{Forkel:2010gu} that
it is possible to find an attractive interaction ``strong enough to render the tetraquark the lightest scalar meson, about 20\% lighter than the $q\bar q$ ground state''. This definitely goes in the right direction, although maybe not enough to describe a $\sim$450 MeV sigma. 
Actually, the author points out that ``To meet quantitative phenomenological expectations would probably require some additional contribution''.

In view of the situation, a description 
of the sigma remains a challenge for holographic models.

\color{black}

\subsection{The $\sigma$ meson and Chiral Symmetry restoration}
\label{subsec:chiralrestoration}

We cannot finish this section about Chiral Symmetry and the $\sigma$ without
mentioning the role it plays in the  Chiral Symmetry Restoration (CSR) 
\cite{Pisarski:1983ms}. This restoration has become a 
basic ingredient of our present understanding of the QCD phase diagram and of hadron physics under extreme conditions of temperature $T$ and baryon density. 
Such conditions are  achieved nowadays in Heavy-Ion and Nuclear Matter experimental facilities  such as RHIC, CERN (ALICE) and soon at FAIR.
This is an extense topic by itself that lies far away from our main scope
and therefore we will provide little more than
a few comments on some basic or recent references.

First of all, lattice simulations  support that deconfinement and CSR take place very close to one another in the phase diagram. Different lattice groups have explored until very recently  the phase diagram and other thermodynamic properties \cite{Borsanyi:2010bp,Cheng:2010fe,Bazavov:2011nk}. For the case of vanishing baryon chemical potential, the QCD transition becomes a crossover for the physical case of 2+1 flavors, at a critical temperature $T_c\sim$ 145-155 MeV. For finite baryon chemical potential, new phases and features arise such as a possible critical line and critical point \cite{de Forcrand:2002ci} and color superconductivity \cite{Alford:2007xm}, which are in principle  realizable in physical systems such as neutron stars, although the  present lattice knowledge in that case is not so well established due to the fermion determinant sign problem. 

An important conclusion of lattice analysis is also that the transition is compatible with the $O(4)$ universality class for two light flavors in the chiral limit  \cite{Ejiri:2009ac}. Actually, one of the simplest model realization of CSR had been historically  the L$\sigma$M within the $O(4)\rightarrow O(3)$ breaking pattern \cite{Hatsuda:1985eb, Bochkarev:1995gi} where the $\sigma$-component of the $O(4)$ field acquires a thermal vacuum expectation value and mass both vanishing at the transition in the chiral limit and  $\pi-\sigma$ mesons degenerate as chiral partners. A relevant role for the $\sigma/f_0(500)$ meson is naturally expected for its sharing of the quantum vacuum numbers.
Over recent years, an understanding of the $\sigma$ role is being achieved also within the modern approach of 
unitarized ChPT, in which one can eschew  the
L$\sigma$M caveats commented in previous sections.  In particular, unitarization techniques for $\pi\pi$ scattering within ChPT at finite temperature allow to define a $I=J=0$ thermal pole at $\sqrt{s_p}= M_p(T)-i \Gamma_p(T)/2$  whose trajectory in the complex plane as $T$ varies shows some interesting features 
\cite{Dobado:2002xf,Cabrera:2008tja}. Thus, the sudden drop of $M_p (T)$ towards the two-pion threshold can be interpreted in terms of CSR, as opposed for instance to the  $I=J=1$ $\rho(770)$-channel where the mass drop is much softer. Actually, it has been recently shown \cite{Nicola:2013vma} that the scalar susceptibility  saturated with this $\sigma$, with squared mass $M_S^2=M_p^2-\Gamma_p^2/4$, develops a maximum near $T_c$ compatible with lattice data, unlike the pure ChPT prediction which is monotonically increasing. On the other hand, chiral partners in the scalar-pseudoscalar sector are understood through degeneration of correlators and susceptibilities \cite{Nicola:2013vma} which is also compatible with the behavior of the lattice thermal masses in the corresponding channels \cite{Cheng:2010fe}.   In addition, introducing nuclear density effects combined with temperature reveals new in-medium decay channels for the $\sigma$ state, which in particular accelerate the migration of the pole to the two-pion threshold \cite{Cabrera:2008tja}. 
The latter analysis extends previous work on 
$\pi\pi$ scattering at finite density in this channel \cite{Cabrera:2005wz}. Moreover, relatively 
similar pole trajectories
to those described in Subsec.\ref{subsec:qm}, where the 
real part of the sigma pole position moved below threshold,
have also been observed when studying a general framework for chiral 
symmetry restoration. These pole movements could in principle
be helpful to discriminate 
different inner structures proposed for the $\sigma$  through the so-called ``$\sigma$-softening'', i.e. its becoming a sharper resonance,
 in a medium with high temperature and/or density \cite{Hyodo:2010jp}.

Furthermore, the role of the $\sigma$ state for CSR could become more complicated if its possible tetraquark component is also considered at finite temperature  \cite{Heinz:2008cv}.

As a final comment, we should recall that after its recent revision in the RPP 2012 edition, 
the sigma has the same status as other meson resonances, with an uncertainty of plus or minus a few tens of MeV.
A priori, it is no longer justified to ignore its presence or to use it with inappropriate parameterizations (Breit-Wigner forms, etc..) when trying to describe the hadron medium.
So far, this has been the case of most analysis and in particular those based on the so-called Hadron resonance Gas  \cite{Andronic:2008gu,Huovinen:2009yb}. 
Thus, it would be desirable to revisit
the effect of neglecting the $\sigma$ or of using
inadequate parameterizations to describe it.
 Some recent studies indicate that its effect is
 crucial in some particular observables regarding CSR \cite{Nicola:2013vma},
whereas the effect of the $\sigma$ has also been shown to compensate largely with 
the isospin 2 channel for other observables like scalar susceptibilities \cite{GomezNicola:2012uc}
. Actually this cancellation between the isoscalar and isotensor channels in hadron resonance gases
was already found in \cite{Venugopalan:1992hy} and in view of the present parameters of the $f_0(500)$ meson
it has been suggested that the sigma meson should not be included in Hadron Resonance Gas models (unless the isotensor wave is included too), as long as one is interested in isospin-averaged observables \cite{Broniowski:2015oha}. 
Furthermore, if the sigma is implemented and the low energy part is important for a given observable,
we have shown in previous sections that the simplest L$\sigma$M, which is the model most widely used for this purposes,
 is insufficient to describe our present knowledge.

\section{THE SPECTROSCOPY AND NATURE OF THE $f_0(500)$}
\label{sec:nature}
\begin{flushright} 
\begin{minipage}{10cm}
{\it ``-'Is there any point to which you would wish to draw my attention?'\\
-'To the curious incident of the dog in the night-time.'\\
-'The dog did nothing in the night-time.'\\
-'That was the curious incident,' remarked Sherlock Holmes.''}
\end{minipage}
 
{\it Silver Blaze. Sir Arthur Conan Doyle, 1892.}
\end{flushright}

This final section will be devoted to the spectroscopic classification of the $f_0(500)$ and to the discussion of its nature. As it happened with the curious incident of the 
dog in the night-time, much of what we know about
the curious features of the $\sigma$ is not about what it does or what it is, but about how it does not behave, what it does not have  or what it is not: it does not behave as an ordinary Breit-Wigner resonance, it does not have a peak as other resonances, it does not have a steep rise of the phase, it does not cross $\pi/2$ at its nominal mass, it does not have the expected $N_c$ behavior, it does not saturate 
the low energy constants of ChPT, there is nothing like ``scalar meson dominance'' and it cannot be fitted into linear Regge trajectories.

Its spectroscopic classification, namely, to what $SU(3)$ multiplet it belongs, has been a matter of debate for almost as long as 
flavor symmetry was proposed. However, it is completely settled by now, once the $\kappa$ or $K_0^*(800)$ pole has been firmly
established, as we already discussed in Subsec.\ref{subsec:introscalarnonet} in the Introduction. 
The evidence for this particle has been mounting over the years, but after the recent determination of its parameters 
 within the rigorous and model-independent Roy-Steiner dispersive formalism \cite{DescotesGenon:2006uk}, questioning the existence of this pole means questioning causality, since the 
 $K\pi$ scattering data in the elastic region is not very controversial. Moreover, we have actually seen how a similar pole is generated
 in unitarized ChPT and is shown to be a requirement to describe scattering data in other chiral approaches and a similar pole has also been recently found in conformal parametrizations constrained with Forward Dispersion Relations \cite{Pelaez:2016tgi}.
The existence of the $\kappa$ forces the lightest scalar nonet to be made
of the $\sigma$, $\kappa$, $a_0(980)$ and $f_0(980)$. When studying the quark mass dependence of unitarized chiral amplitudes in 
 Sec.\ref{subsec:qm} we have actually seen how their combinations  degenerate into an octet and a singlet.

However, the $\sigma$ composition in terms of quarks and gluons is still a matter of debate. 
First of all, as already stressed in the Introduction, the intuitive Fock-space decomposition is not  well defined 
within the gauge field theory relativistic formalism.
Thus,  the $\sigma$ constituents  should be understood in the same sense as when we say within the quark model that the nucleon is made of three quarks. This may refer to  ``constituent", 
``dressed" or ``valence" quarks and gluons. 
In quark models  the $LS$ coupling scheme is customarily used to classify eigenstates of the two-particle Schr\"odinger equation 
describing bound $\bar qq$ pairs. The total spin operator is $\vec S= \vec S_q+\vec S_{\bar q}$, which added to the orbital angular momentum
defines the total momentum $\vec J=\vec L+\vec S$. Parity is also defined as $P=(-)^{L+1}$, so that 
states are labeled by $J^P$. Note that although only 
electrically charged particles are charge conjugation $C=(-)^{L+S}$ eigenstates, $C$ is often used together with $J$ and $P$
to label as $J^{PC}$ the whole multiplet to which the neutral particle belongs.  
In terms of spectroscopic notation where $L=0,1,2...$ is denoted by $S,P, D...$, scalar mesons are $0^+$ states 
so that 
 they correspond to 
$^{2S+1}\!L_J=^3\!\!\!P_0,^3\!\!P_1,^3\!\!P_2...$ states. The lightest $0^+$ mesons are then expected to have the lowest possible momentum,
so that they should be $^3\!P_0$. Note that in order to form scalar $\bar qq$ mesons we need at least angular momentum $L=1$,
which means that scalars are expected to be heavier than $^3\!S_1$ states, which are $J^P=1^-$. 
But the latter are nothing but the $\rho(770)$ and its multiplet partners, 
which are definitely heavier than the $\sigma$.
The lightest scalars are therefore expected to be heavier tan the $\rho(770)$, typically above 1 GeV.
Actually, the lightest tensor $2^+$ state  found is the $f_2(1270)$, which should correspond to a $^3\!P_2$, thus, having the same orbital angular momentum as the scalars.  
Note that the same scheme between the $\sigma$ and $\rho$  
is applicable to the the $\kappa$ and $K^*(892)$.

It is true that the $f_0(980)$ and $a_0(980)$ are heavier than the $\rho(770)$, so one may try to consider them as the light scalars 
of a multiplet with heavier scalar mesons like the $K^*(1430)$ and one of the several $f_0$ states above 1 GeV (see for instance \cite{Tornqvist:1995kr}).
Moreover, if the $\kappa$ or $K_0^*(800)$ was not well established it would make sense to investigate the possibility that the sigma may  be a glueball \cite{Minkowski:1998mf,Mennessier:2008kk,Ochs:2013gi}. However, 
once the $\kappa$ existence is well confirmed in a model-independent analysis \cite{DescotesGenon:2006uk} it needs
some lighter partners without strangeness to form a multiplet and the only candidate is the $\sigma$. Furthermore, there is not a single lattice calculation of the glueball that yield its mass as low as 500 MeV \cite{latticeglueball}, all them actually place the glueball well above 1 GeV, in the 1.5 to 1.8 GeV region.

But then, once the lowest nonet members have been identified as the $f_0(500)$, $K_0^*(800)$, $a_0(980)$ and $f_0(980)$,
the fact that they are lighter than the lightest vector nonet $\rho(770)$, $K^*(892)$, $\omega(770)$ and $\phi(1020)$,
strongly suggests that they do not fit well in a $\bar qq$ scheme. In addition, the expected $\bar qq$ mass hierarchy 
is not observed since the $K_0^*(800)$  (whose pole mass is actually below 800 MeV in several determinations listed in the RPP 2012)
would have one more strange quark and should therefore be heavier than the $a_0(980)$ and at least one of the isoscalars.
At his point is when different models of non-$\bar qq$ configurations come into play. We will review the most popular ones below.

Note however that there is also a molecular interpretation that considers 
 light scalars as primarily made of two bound mesons, in a similar
way as we consider the deuteron made of two nucleons rather than six quarks. This subtlety should be kept in mind 
when when dealing with quark-level models, to be reviewed at the end of this section.

So far we have been commenting the hints of a light-scalar non-ordinary nature basing our reasoning in models at the quark 
or constituent level. However, in this section  we will review first other approaches that also address the nature of the $f_0(500)$, but avoid referring to its constituents. 
In such case one looks for ways in which it deviates from an ordinary feature or behavior followed by most other mesons.
This is the case of the $N_c$ dependence and its classification into Regge trajectories, where we will see that the $\sigma$  deviates from the ordinary behavior. The first one is particularly relevant because it provides a link to QCD parameters.

\subsection{Mesons and the $1/N_c$ expansion}
\label{subsec:nc}

The $1/N_c$ expansion \cite{'tHooft:1973jz,Witten:1980sp}, where $N_c$ is the number of colors in the QCD Lagrangian, Eq.\ref{QCDLagrangian},
 is of interest because in principle it is the only QCD perturbative expansion 
that is also valid in the low energy regime. 
For this report, the most relevant application is that
it provides a clear identification of the 
leading order $1/N_c$ dependence of
different kinds of states.
For our purposes it is enough to
review some very basic results within this formalism.
For pedagogical introductions 
we refer to \cite{Donoghuebook,Coleman,Manohar:1998xv,Pich:2002xy}.

Note that the number of quark fields 
of a given flavor is now $N_c$, whereas the number of
gluons is $N_c^2-1\sim O(N_c^2)$, so that at large $N_c$ gluon exchanges dominate over quark exchanges.
Moreover, in order to have a smooth large-$N_c$ limit for the running coupling constant,
the QCD  coupling $g$ must scale as $1/\sqrt{N_c}$. Under these conditions it is possible to show that the dominant diagrams in any process are
planar. Actually, as long as the diagram is planar, exchanging
any amount of gluons does not change the $N_c$ counting. However, non-planar
exchanges are suppressed by $1/N_c^2$ factors. In addition internal quark
loops are suppressed by factors of $1/N_c$. Therefore, quarks only appear
as external lines, which may define the initial and final states, for instance providing flavor quantum numbers.

In order to study the behavior of mesons, it is useful to define local quark bilinears 
$B=\bar q \Gamma q$ where $\Gamma$ contains some desired Lorenz and flavor structure. 
Then,  a generic n-point function $\langle B_1 ...B_n\rangle$ is $O(N_c)$. 
As it is customary, and unless necessary,  indices, position dependence, hermitean conjugation, time ordering, etc, 
will be suppressed for simplicity, but can be traced back in the original references.
Assuming confinement in color singlets and since intermediate quark loops are $1/N_c$ suppressed, it can be shown that
the  only singularities in $\langle B(k)B(-k)\rangle$ can be one-$\bar q q$-meson poles \cite{Witten:1980sp}. In other words
\begin{equation}
\langle B(k)B(-k)\rangle =\sum_n\frac{f_n^2}{k^2-M_n^2}.
\end{equation}
This has a series of consequences from QCD. First, being a sum of planar diagrams, the left hand side has a smooth large $N_c$  limit, so that 
$\bar q q$-meson masses must also have a smooth limit, independent of $N_c$. Therefore
$\bar qq$ mesons are stable $M_n \sim O(1)$. Second, from asymptotic 
freedom it is known that the left hand side behaves logarithmically for large $k^2$. Third, the matrix elements of $B$ to create a meson are $f_n =\langle 0 \vert B\vert n\rangle \sim O(\sqrt{N_c})$. 
In particular, the pion decay constant $f_\pi$ defined in Eq.\ref{prePCAC}, behaves as $f_\pi\sim O(\sqrt{N_c})$, and the same happens for the decay constants of the kaons and eta, as well as with the
pion decay constant in the chiral limit $f_0$, since mass corrections do not change the $1/N_c$ counting.

Now, the width of a $\bar qq$ meson created from bilinear $B$ currents is obtained 
from the three-point function:
once again $\langle BBB\rangle\sim O(N_c)$. But the creation of each meson 
requires one  $\langle 0 \vert B\vert n\rangle$ term, which all together contribute with a $N_c^{3/2}$ factor. 
Therefore the amplitude that connects these three mesons and describes one meson decaying into the other two 
must be  $\sim1/\sqrt{N_c}$. The decay width, being proportional to the modulus squared of the amplitude
is therefore $\Gamma_n\sim O(1/N_c)$.

From now on we will often call this $M\sim O(1)$, $\Gamma=O(1/N_c)$, 
{\it the $N_c$ ordinary-meson behavior}, because this is 
how the familiar $\bar qq$ mesons behave. As will be commented below
other kinds of mesons, like 
$\bar qqg$ hybrids \cite{Cohen:2014vta}, or the most intuitive tetraquark configurations, 
are indistinguishable from $\bar qq$-mesons just from their $N_c$ behavior.

By analyzing the $N_c$ behavior of the four-point function $\langle BBBB \rangle\sim N_c$
one can study meson-meson scattering amplitudes. Now, four $\langle 0 \vert B\vert n\rangle$
factors are needed to create the four mesons, thus contributing with  $N_c^2$ so that the scattering
amplitude behaves as $1/N_c$. Note that this is consistent with the LO ChPT amplitudes
being $t \sim 1/f_\pi^2$, as we saw in Eq.\ref{ec:LET}.

Concerning glueballs, a similar analysis can be carried out by considering 
gluonic currents $J_G=Tr(G^a_{\mu\nu}G^{\mu\nu}_a)$. Glueball masses are also found to be 
 $M\sim O(1)$. However, glueball decay widths turn out to be $\Gamma\sim O(1/N_c^2)$. In other words, glueballs are 
expected to be even narrower than ordinary $\bar qq$ mesons. As we will see right below,
in Subsec.\ref{subsec:ncmodelindep},
this will play very strongly against the glueball interpretation of the $f_0(500)$.

The analysis can be extended to tetraquark states \cite{Jaffe:1976ig}. 
But here 
one should separate several possible configurations. The first one is just two mesons propagating freely. 
To fix ideas, and since we are interested in the $\sigma$  case, this two-meson state can be denoted by 
$\vert \pi\pi \rangle$. Trivially, it has $M\sim O(1)$ and since the amplitude to itself is $O(1)$, its 
``width to two mesons'' is $\Gamma\sim O(1)$. 
The second configuration is what is most frequently understood by a tetraquark,
in which independently of $N_c$ there are always two valence quarks and two antiquarks present, i.e.
a $q\bar q q\bar q $ configuration. This we will call $T$. 
But for $N_c>3$ there is yet a  third possible generalization
of a $N_c=3$ tetraquark \cite{Jaffe:2007id,Jaffe}, namely a state made of $N_c-1$ quarks and $N_c-1$ antiquarks (with all 
quarks antisymmetrized in their color indices and the same for all antiquarks). This we will call $(N_c-1)\bar qq$
or ``polyquark''. 

The calculation of the usual tetraquark configuration is obtained by considering quark quadrilinears,
but evaluated at the same point, i.e. $Q=\sum_{ij} C_{ij} B_i(x)B_j(x)$, where $C_{ij}$ are coefficients to obtain the desired quantum numbers. This form can always be reached by Fierz transformations \cite{Coleman}.
 In his Erice lectures \cite{Coleman}, S. Coleman maintained that tetraquarks did not exist  
(presumably implying that they were broad) in the large $N_c$ limit, because
the two-point function of the $\bar qq\bar qq$ currents are dominated
by the creation and annihilation of two-meson states. This became common knowledge until very recently,
when  Weinberg pointed out \cite{Weinberg:2013cfa} that
such an argument only applies to leading order disconnected diagrams, whereas 
a possible tetraquark pole should appear in the connected part. Let us then write: 
\begin{equation}
\langle QQ \rangle= \sum_{ijkl} \Big[C_{ij}C_{kl} \langle B_iB_j\rangle\langle B_kB_l\rangle+
\langle B_iB_jB_kB_l\rangle_{connected}\Big].
\end{equation}
The first term counts $O(N_c^2)$ and leads to the original reasoning in \cite{Coleman}.
However, since the tetraquark pole can only exist in the second term, which is $O(N_c)$, then 
the same behavior we found for $B$ also follows for $T$, namely, the tetraquark mass is $M_T\sim O(1)$ and $\Gamma_T\sim O(1/N_c)$ \cite{Weinberg:2013cfa}.
Thus, {\it the most intuitive tetraquark configurations have an ordinary $N_c$ behavior indistinguishable from $\bar qq$ mesons}.
Of course, in certain cases a further suppression can occur and some particular tetraquarks, if they 
exist, should be even narrower \cite{Knecht:2013yqa}.

Finally, there is the polyquark proposed by R. L. Jaffe in \cite{Jaffe:2007id,Jaffe}. Its analysis
is  more complicated because the number of constituents grows with $N_c$ and  diagrams that
by themselves would be suppressed can contribute to the leading order because their number also grows with $N_c$.
This feature is similar to what happens with baryons in the large-$N_c$ limit and the correct scaling properties are 
obtained through a mean-field approximation \cite{Witten:1980sp,Cohen:2011cw}. In such cases, the mass of the state
also grows with the number of constituents, i.e. $M_P\sim O(N_c)$. The polyquark to $\pi\pi$ coupling decreases exponentially \cite{Cohen:2014vta} with $N_c$ because as $N_c$ grows it is harder and harder to annihilate all constituents 
except the two quark-antiquark pairs that make the final pions. However, the decay of the $(N_c-1)\bar qq$ 
into one pion and an $(N_c-2)\bar qq$ is $O(1)$ \cite{Witten:1980sp,Cohen:2011cw}. The couplings among all these different states have been calculated explicitly in \cite{Cohen:2011cw}.

The mass and width behavior of all these states has been collected in Table~\ref{tab:Ncstates}.
In the sections below these results will be used to show that a dominant 
 ordinary composition is very disfavored for the $\sigma$, and a glueball dominant composition even more so.
However, there are hints of a possible mixing with a subdominant ordinary component, although heavier than
the physical mass of the $\sigma$.

 \begin{table} 
   \centering 
 \begin{tabular}{|c|cccccc|} \hline 
 $ $ & $q\bar q$ & $gg$ & $q\bar q g$ & $\pi\pi$ &  $T(q\bar q q\bar q)$ & $(N_c-1)q\overline{q}$\\ 
 \hline 
 $M$ &$O(1)$ & $O(1)$ & $O(1)$  & $O(1)$ & $O(1)$   & $O(N_c)$        \\  
 $\Gamma_{\mathrm{Tot}}$ & $O(1/N_c)$ & $O(1/N_c^2)$ & $O(1/N_c)$& $O(1)$
& $O(1/N_c)$& $O(1)$ 
        \\ 
 \hline 
 \end{tabular} 
\caption{
Leading behavior in the $1/N_c$ expansion of the mass and width 
 for various configurations in QCD. The first three are intrinsic, non-fissible configurations (conventional meson, glueball, hybrid). The last three are states that may break apart into two or more mesons without the need for creating any additional quarks (two mesons, tetraquark, polyquark). Table taken from \cite{Cohen:2014vta}.
 \label{tab:Ncstates}
} 
 \end{table}

So far we have analyzed the $N_c$ behavior of different kinds of mesons.
Let us now turn to the ChPT Lagrangian that we briefly reviewed in Subsec.\ref{subsec:chpt}.
It is written in terms of the $U(x)$ matrices,
which can be expanded in powers of $\pi^a/f_0$ meson fields, which can be considered 
quark bilinears to LO in the $1/N_c$ counting. 
Therefore the LO Lagrangian operators behave as $O(N_c)$ in the $1/N_c$ expansion.
This is the case of terms with a single flavor trace like those 
appearing in the LO ChPT Lagrangian of Eq.\ref{ec:Lag2}, 
consistently with the $f_0\sim O(\sqrt{N_c})$
and $M_ 0\sim O(1)$. The flavor trace corresponds to the only quark line that connects all 
meson operators in leading $1/N_c$ diagrams. 

A similar reasoning applies for the the terms multiplied by  $L_3, L_5, L_8$, which therefore scale as $O(N_c)$. However, the NLO ChPT Lagrangian contains terms with two flavor traces,
like those accompanying the $L_4, L_6$ and $L_8$ low energy constants (LECs).
The additional trace corresponds to an additional quark loop, which as we have already commented 
brings a $1/N_c$ suppression factor. Hence $L_4$ and $L_6$ scale as $O(1)$ in the $1/N_c$ expansion.
Naively, one would think that $L_1$ and $L_2$ should also be $O(1)$, since they appear 
in Lagrangian terms with just one flavor trace. However the 
SU(3) matrix relations were used to eliminate the structure 
$\langle \partial_\mu U^\dagger \partial_\nu U \partial^\mu U^\dagger\partial^\nu U^\dagger\rangle$
in favor of the $L_1$ and $L_2$ terms. Since this term is $O(N_c)$ its behavior is inherited by 
$L_1$ and $L_2$. Still, the combination $2L_1-L_2$ is $O(1)$. Finally, $L_7\sim O(1)$ but 
obtaining this behavior is rather subtle \cite{chptlargen}. The derivation is beyond
our scope, but it is due to the apparent conflict between
integrating out the heavy $\eta'(960)$ meson to recover the $L_7$ term in QCD, while taking the large $N_c$
where the $\eta'(960)$ becomes very light. The behavior of all NLO LECs is provided in the last column of Table~\ref{tab:LECs}.

There is an additional subtlety to the $N_c$ scaling of the LECs, due to renormalization. 
Looking back at Eq.\ref{ec:renormLECs}, it can be noticed that
the logarithmic scale dependence of the LECs is $O(1/N_c)$. Therefore, when stating that some 
$L_i$ behaves as $O(N_c)$ or $O(1)$, the scale $\mu$ at which this behavior is being imposed must be chosen.
Different choices of $\mu$ introduce a subleading uncertainty. 
Typically the $N_c$ scaling is applied for $\mu=0.5$ to 1 GeV. Actually the expected $N_c$ hierarchy 
between LECs is roughly observed in the phenomenological values provided in Table~\ref{tab:LECs}, 
which correspond to $\mu=M_\rho\simeq 770\,$MeV.

In the next subsections we will make use
 these $N_c$ scaling properties to obtain information about the nature of the $f_0(500)$ resonance.

\subsection{Model independent observables with enhanced $1/N_c$ suppression.}
\label{subsec:ncmodelindep}
This subsection is dedicated to an approach that
only makes use of the $1/N_c$ expansion and the properties of 
the $\pi\pi$ scattering partial waves 
obtained from the  dispersive representation of 
the data that we described in Sec.\ref{sec:parameters}.
This approach is therefore model independent and is useful to enhance some non-ordinary meson features.
This is done by studying 
observables whose deviations from unity
are strongly suppressed for ordinary mesons, thus allowing the use
of  the $1/N_c$ expansion at the physical value $N_c=3$. 
As we will see this method
strongly disfavors the identification of the $f_0(500)$ as an ordinary $q\bar q$ resonance
or a Weinberg tetraquark, and very strongly as a glueball. 

Let us recall the three
criteria used to identify resonances, 
reviewed in Subsec.\ref{subsec:poles}: by the phase reaching $\pi/2$ at its mass, by the maximum of the phase shift derivative and by the
pole position. 
Let us then suppose that a resonance has a pole at $s_R=M_R^2-iM_R \Gamma_R$.
\footnote{Note we have slightly changed our notation here, from $s_R=(M_R-i \Gamma_R/2)^2$, to follow the usual notation for these studies in the literature. The $1/N_c$ counting of the mass and width is still the same.
Also, $t(s)$ here has an opposite sign compared to that in \cite{Nieves:2009kh,Nieves:2009ez}.
} 
Now, since an ordinary resonance
behaving as $M_R\sim O(1)$ and $\Gamma_R\sim O(1/N_c)$ becomes narrow in the large $N_c$ {\it limit},
it was shown in \cite{Nieves:2009kh} that
the scattering phase shift of the elastic partial wave where 
the resonance appears satisfies:
 \begin{eqnarray}
 \delta(M_R^2)=\frac{\pi}{2}-\underbrace{\frac{\re t^{-1}}{\sigma}
\Big\vert_{M_R^2}}_{O(N_c^{-1})}+O(N_c^{-3}),
 \label{deltaexpansion}\qquad
\delta'(M_R^2)=-\underbrace{\frac{(\re t^{-1})'}{\sigma}
\Big\vert_{M_R^2}}_{O(N_c)}+O(N_c^{-1}). 
 \end{eqnarray}
The first equation is simply telling us that, according to the first criterion, 
the pole mass $M_R$ tends to the energy where the phase is $\pi/2$ 
in the infinitely narrow resonance limit. But it also tells us the $N_c$ order of the deviations.
The second equation is less intuitive
but tells us that the derivative of the phase at $M_R$ grows with $N_c$
since the resonance becomes narrower and narrower. This criterion 
was studied in 
\cite{Nieves:2009kh} for the $f_0(500)$ with a relatively 
inconclusive result about its assumed 
$\bar qq$ behavior.

However, the previous equations are very useful to build the following
adimensional observables \cite{Nebreda:2011cp}:
\begin{equation}
\frac{\frac{\pi}{2}- \re t^{-1}/\sigma}{\delta}\Big\vert_{M_R^2}
\equiv\Delta_1=1+\frac{a}{N_c^3}..., \qquad\quad
-\frac{[\re t^{-1}]'}{\delta'\sigma}\Big\vert_{M_R^2}\equiv\Delta_2=1+\frac{b}{N_c^2}....
\label{abdef}
\end{equation}
which are 1 up to corrections suppressed by $1/N_c^2$ or $1/N_c^3$, respectively. 
Hence, for an ordinary resonance in the physical world $N_c=3$, 
one would naturally expect 
these observables to be 1 up to 
 $O(4\%)$ 
and $O(10\%)$ corrections, respectively.

\begin{table}
  \centering
  \begin{tabular}{crrrr}
    & $\rho(770)$ & $K^*(892)$ & $\sigma/f_0(500)$ & $\kappa/K^*_0(800)$ \\\hline
\rule[-1mm]{0mm}{5mm}$a$    &    $-0.06\pm0.01$ &  0.02  &  $-252^{+119}_{-156}$ & -2527 \\
\rule[-1mm]{0mm}{5mm}$b$    &  $0.37 ^{+0.04}_{-0.05}$    & 0.16 &$77^{+28}_{-24}$ & 162 \\
\hline
  \end{tabular}
  \caption{Coefficients of the strongly suppressed 
{\it corrections} from unity in the  Eq.\ref{abdef} observables assuming that the resonance has an ordinary nature.
These were obtained in \cite{Nebreda:2011cp} from the dispersive analyses of \cite{GarciaMartin:2011cn,GarciaMartin:2011jx,DescotesGenon:2006uk}.
For $\bar qq$ resonances, $a$ and $b$ are expected to be of order one or less.
This is nicely satisfied by the $\rho(770)$ and the $K^*(892)$, 
but both the $\sigma$ and $\kappa$ would require $a$ and $b$ unnatural values by several orders of magnitude.  
For the glueball interpretation of the $\sigma$ the situation is even more disfavored, see the main text. Table taken from \cite{Nebreda:2011cp}.}
  \label{tab:Nc3results}
\end{table}

For simplicity it is better to express
the deviations in terms of $a,b$, which are naturally expected to be of order one {\it or less} (cancellations with higher order terms can substantially decrease their effective value, but not increase it). 
The calculation of these observables in \cite{Nebreda:2011cp}, 
which we list in Table~\ref{tab:Nc3results},
made use of
the dispersive analysis of 
scattering phase shifts in \cite{GarciaMartin:2011cn}, together with the corresponding 
$\sigma$ and $\rho(770)$ pole positions 
given in \cite{GarciaMartin:2011jx}. Note that all numbers in the table are 
obtained from physical scattering amplitudes at $N_c$=3.
The coefficients of the light vector mesons $\rho(770)$ and $K^*(892)$
come out, as expected for ordinary $\bar qq$ mesons, of order one or less.
However, for the $\sigma$ and the $\kappa$
the parameters are two 
or more orders of magnitude larger than expected for an ordinary $M_R\sim O(1)$ and $\Gamma_R\sim O(1/N_c)$ behavior.

In the same work \cite{Nebreda:2011cp}, a similar study was also carried out for a glueball composition, 
whose width scales as $O(1/N_c^2)$. In such case then:
\begin{equation}
\Delta_1=1+\frac{a'}{N_c^6},\qquad
\Delta_2=1+\frac{b'}{N_c^4}.\label{abpdef}
\end{equation}
Following the same procedure as before 
$a'=-6800^{+3200}_{-4200}$ and $b'=2080^{+760}_{-650}$
were found for the $f_0(500)$.
In other words, a very dominant or pure glueball nature for the 
$f_0(500)$ is strongly disfavored by the $1/N_c$ expansion of QCD,
even more than the $\bar qq$ interpretation. This is because
it would require even more unnatural coefficients, this time too large by
three to four orders of magnitude.

Let us recall that nothing but the $QCD$ $1/N_c$ expansion at $N_c=3$ has been used. This is not a large-$N_c$ result
nor the large-$N_c$ limit. In addition, the input comes from the model independent dispersive analyses of the amplitudes.
{\it This is a model-independent result which strongly
disfavors the existence of a predominant $\bar qq$ or non-glueball component for the $f_0(500)$}. 
A dominant component of the $f_0(500)$ in the form of 
the standard 4-quark generalization of tetraquarks to arbitrary $N_c$
is as disfavored as 
much as the ordinary $\bar qq$ meson. Of course, this does not exclude the possibility
of such subdominant components inside the $f_0(500)$, it only disfavors them being dominant.

Finally, let us remark that this method does not provide the $N_c$ dependence of the $\sigma$ meson. 
As we review next, this can be obtained from unitarized Chiral Perturbation Theory.

\subsection{$N_c$ dependence from unitarized ChPT}
\label{subsec:ncUChPT}

We have seen in Subsections \ref{subsec:uchpt} and \ref{subsec:coupleduchpt}
that the $\sigma$ meson can be nicely described by unitarization of the ChPT amplitudes.  
In particular, some methods do not require further 
parameters than the low-energy constants of ChPT, or LECs, up to a given order.
Now recall that in 
the last column of Table \ref{tab:LECs} the leading $1/N_c$ scaling of the LECs  was provided,
which is model independent. Moreover, 
the leading order $1/N_c$ behavior of the NGB masses is $O(1)$
and the pion decay constant scales as $f_\pi\simeq f_0\sim O(\sqrt{Nc})$.
Hence it is straightforward to obtain the leading $1/N_c$ behavior of the 
resonances generated with unitarization methods 
by simply scaling all ChPT parameters according to their $1/N_c$ leading behavior.

Let us first examine the ``simplest unitarized model with chiral symmetry ''
presented in  Sect.\ref{subsec:simplest}, which only makes use of the 
LO ChPT Lagrangian. Recall that we saw 
in Eq.\ref{ec:simplestpolechirallimit} that in the chiral limit this model yields
$\sqrt{s_\sigma}=(1-i)\sqrt{8\pi} f_\pi\sim O(\sqrt{N_c})$. That is,
the sigma mass and width both grow as $\sqrt{N_c}$. Since NGB masses
are $O(1)$, mass  corrections do not modify this leading $N_c$ behavior.
Of course, this very simple model is only able to generate a crude approximation to a
light sigma, since it comes about a factor of 2 too wide in practice.
But this is already a suggestion that 
the $f_0(500)$ pole may not behave in the expected way for an ordinary meson.
In contrast, recall that this method is not able to yield even a crude description of the $\rho(770)$,
so that we do not get any hint of its behavior in this way.

Actually, in \cite{Sannino:1995ik} it was shown that 
a reasonable description of $\pi\pi$ scattering could be achieved by 
including, besides the LO chiral Lagrangian,  
all tree-level 
exchanges of large-$N_c$ leading resonances, 
except at low energies, where it was shown the need for 
a sigma pole, which had to be included explicitly with an {\it ad hoc} 
unitary functional form.
In that work, it was already claimed that this  $\sigma$ was presumably
``not of the simple $\bar qq$ type" 
and its exchange should be of subleading order in the large $N_c$.

Now, as seen in Subsecs.\ref{subsec:uchpt} and \ref{subsec:coupleduchpt},
a very good description of the $f_0(500)$ and the other light resonances
is achieved when 
the NLO ChPT amplitudes are unitarized. Since we want to avoid
any spurious $N_c$ dependence, but just to keep the one dictated by the LECs and $f_0$,
the most straightforward implementation of the $N_c$ dependence
is through the Inverse Amplitude Method (IAM), reviewed in Subsecs.\ref{subsubsec:IAM} and \ref{subsubsec:coupledIAM}. Recall that this method directly uses the fully renormalized $T_2$ and $T_4$
amplitudes.

However, some subtleties have to be kept in mind. First, {\it a priori}, 
one should be careful {\it not to take $N_c$ too large and in particular
to avoid the $N_c\to\infty$ limit}. \color{black} The reason is that this is a limit in which 
meson loops are more suppressed than direct resonance exchange, 
whereas unitarization methods are devised for strongly interacting
theories where meson loops are very important.
\color{black}
As shown in Sec.\ref{subsubsec:IAM}, 
the derivation of the IAM from a dispersion relation
relies on the fact that the exact elastic 
contribution on the right or unitarity cut dominates the
dispersion relation and that NLO ChPT provides a good approximation to the subtraction constants. 
Since the IAM describes the data and the resonances, 
within, say 10 to 20\% errors, this means that at $N_c=3$ 
other contributions, like those on the left cut, are not
approximated badly.  But meson loops, 
responsible for the unitarity cut, scale as $3/N_c$ whereas the inaccuracies due to the approximations scale partly as $O(1)$.
Thus, we can estimate that those 10 to 20\% errors at $N_c=3$ may become 100\% errors around $N_c\sim30$ or $N_c\sim15$, respectively.
Hence, results \cite{Pelaez:2003dy,Pelaez:2006nj} 
beyond $N_c=30$, and even beyond $N_c\sim15$ should be interpreted 
with care. Another reason to keep $N_c$ not too
far from 3 is that  the SU(3) ChPT formalism does not include
the $\eta'(980)$,
whose mass is related to the $U_A(1)$ anomaly and scales as 
$\sqrt{3/N_c}$.
Nevertheless, by keeping $N_c<30$, its mass
would be $>310\,$MeV and still the pions would be the only relevant degrees
of freedom in the $\sigma$ region.
Of course, there could be special cases in which the IAM 
could still work for very large $N_c$, 
as it is has been shown for the vector channel for QCD \cite{Nieves:2009kh,Nieves:2009ez}. 
But that is not the case for the scalar channel, which, 
if used for too large $N_c$ may lead to poles in the third quadrant
\cite{Nieves:2009kh,Nieves:2009ez}
for some values of the LECs, which lack a clear physical interpretation. 

The second subtlety to keep in mind is that  
the LECs  are renormalized quantities and have a renormalization scale dependence, which of course cancels in physical observables. 
However, one has to decide at what renormalization
scale the $1/N_c$ scaling applies and choosing one 
renormalization scale or another
amounts to shifting part of the observed value of the LECs from 
the leading to the subleading contribution. This causes an uncertainty
that is usually estimated by allowing the renormalization scale $\mu$ 
to vary between 0.5 and 1 GeV, which is the same range 
used to compare the measured LECs to their Resonance Saturation estimates.

\begin{figure}
\centering
  \includegraphics[width=0.33\textwidth]{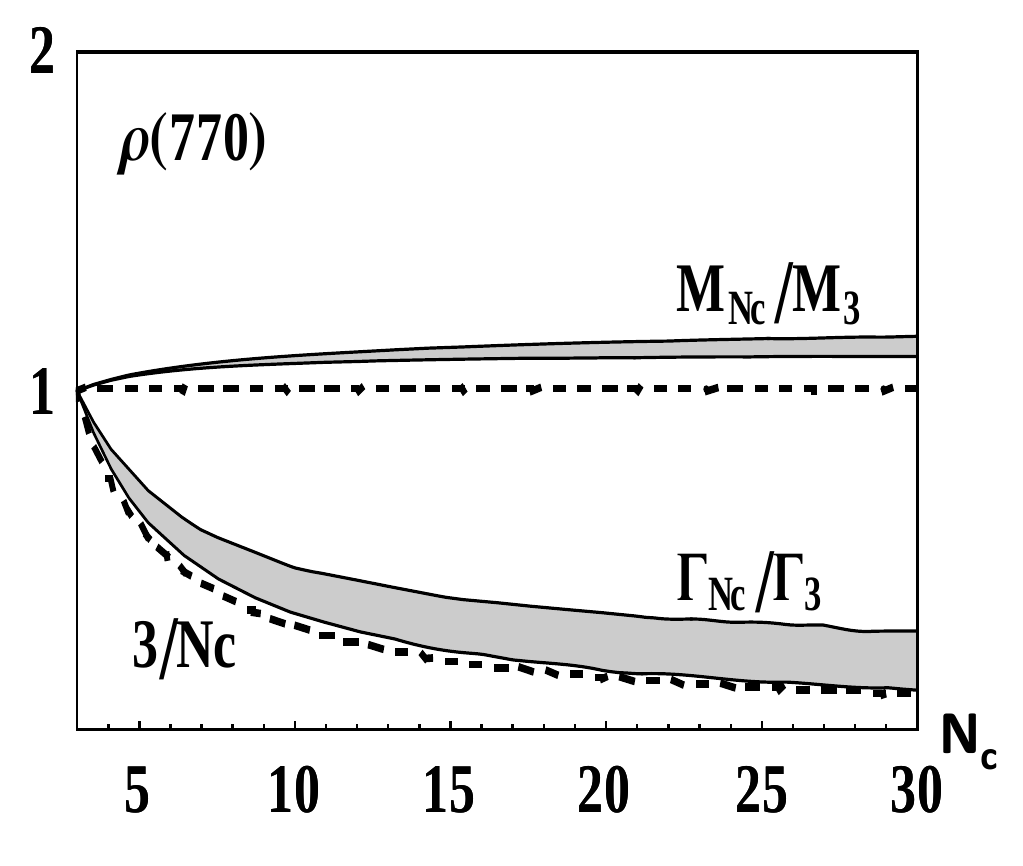}
  \includegraphics[width=0.32\textwidth]{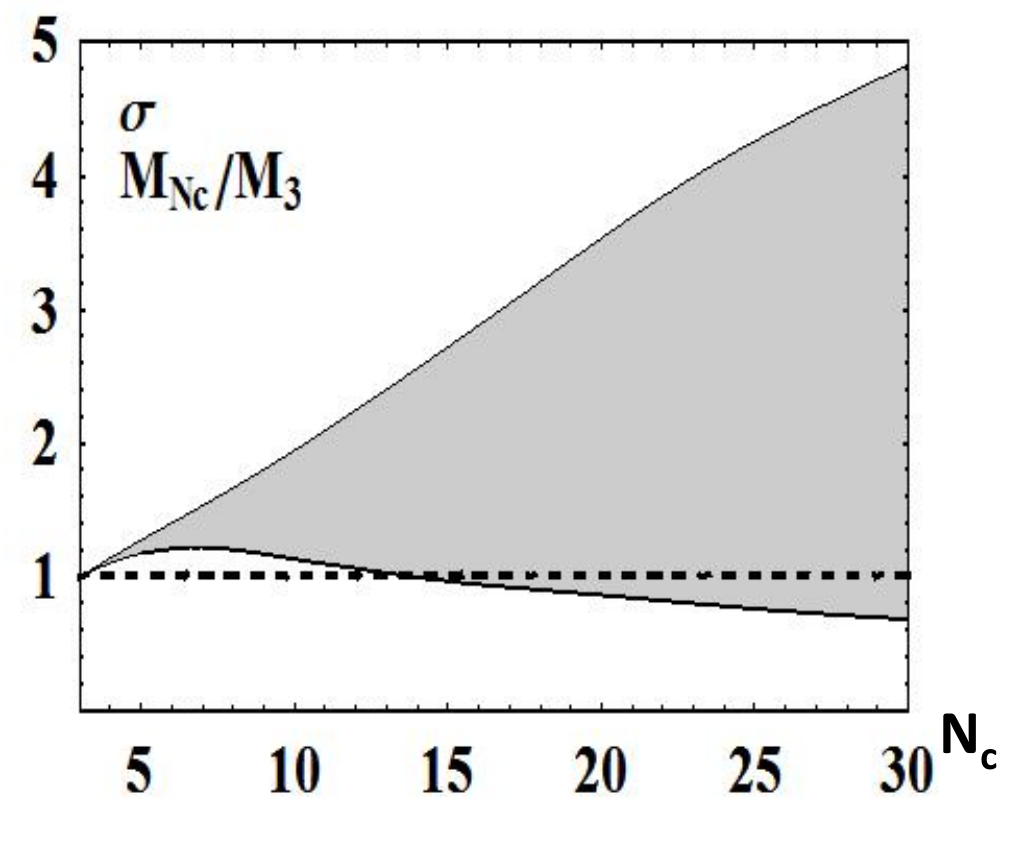}
  \includegraphics[width=0.32\textwidth]{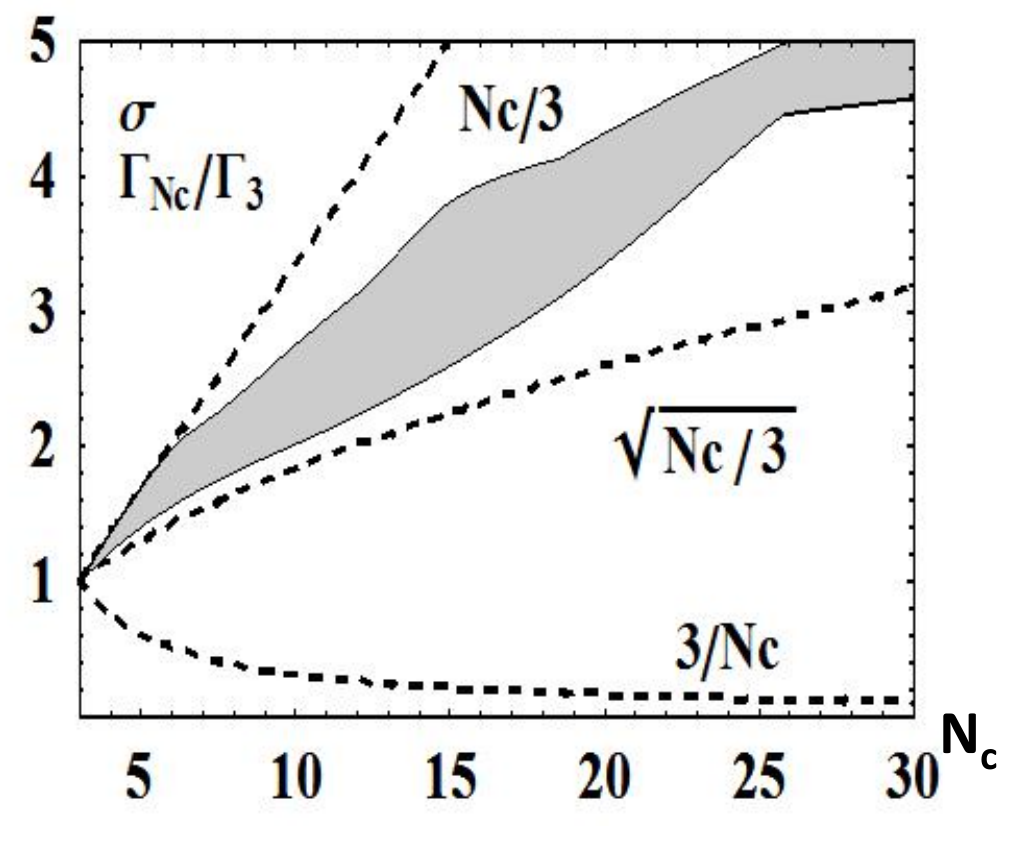}
\caption{ Mass and width $N_c$-behavior of the $\rho(770)$ and
$\sigma$ resonances form the NLO IAM \cite{Pelaez:2003dy}. 
All values are normalized at their $N_c=3$ physical values.
As explained in the main text, the bands cover the uncertainty due to the choice of renormalization scale where to apply the $N_c$ scaling.
Left: the $\rho(770)$ mass and width compared to their respective
expected $q\bar q$ behavior (dashed lines). Center: 
$\sigma$-meson mass versus $N_c$ compared to the expected constant behavior of
an ordinary meson (dashed line). 
Right: $\sigma$-meson width versus $N_c$ compared to 
several different behaviors represented as dashed lines.
Figures taken from \cite{Pelaez:2003dy}.
}
  \label{fig:NcPelaez}
\end{figure}

The $N_c$ scaling of IAM resonances was studied using the IAM
 to one-loop (NLO) in coupled channels in \cite{Pelaez:2003dy}
and to two-loops (NNLO) in the elastic case in \cite{Pelaez:2006nj}.
Thus, Fig.\ref{fig:NcPelaez} shows the
behavior of the $\rho(770)$ and $\sigma$ masses and widths
found to NLO in  \cite{Pelaez:2003dy}. 
On the left panel it can be seen that the 
$\rho(770)$ mass and width
follow nicely the expected behavior for 
a $\bar qq$ state: $M\sim 1$, $\Gamma\sim 1/N_c$ (dashed curves).
The bands just cover the uncertainty
in the renormalization scale 
 where the LECs are scaled with $N_c$.

In contrast, also in Fig.\ref{fig:NcPelaez} it can be seen that
the $\sigma$ shows a 
different behavior from that of a pure $\bar qq$
\emph{near $N_c$=3}, particularly because its width, seen in the right panel,
grows with $N_c$, i.e. its pole moves
away from the real axis. 
We can nevertheless see that the width $N_c$ dependence is relatively close
to the $\sqrt{N_c}$ behavior found with the simple unitarization of the tree level ChPT amplitude, which we commented at the beginning of this subsection.
The $N_c$ dependence  to NLO of the mass, shown in the central panel, is more uncertain.

A rather similar behavior to that of the $\sigma$ was also found for the $\kappa$ resonance in \cite{Pelaez:2003dy}, once again suggesting a very similar nature for these two scalar mesons.
Actually, within the same coupled-channel 
NLO IAM calculation \cite{Pelaez:2003dy} it was found that the 
whole lightest scalar octet made of the $\sigma$, $\kappa$, $f_0(980)$ 
and $a_0(980)$
have $N_c$ behaviors at odds with those of ordinary mesons, although in the case of the $a_0(980)$ there was some region of parameter space where it could still behave as such. The $K^*(892)$ vector resonance, not shown here, was also calculated in \cite{Pelaez:2003dy}
and has an ordinary behavior very similar to the $\rho(770)$.

\begin{figure}
  \centering
  \includegraphics[width=0.49\textwidth]{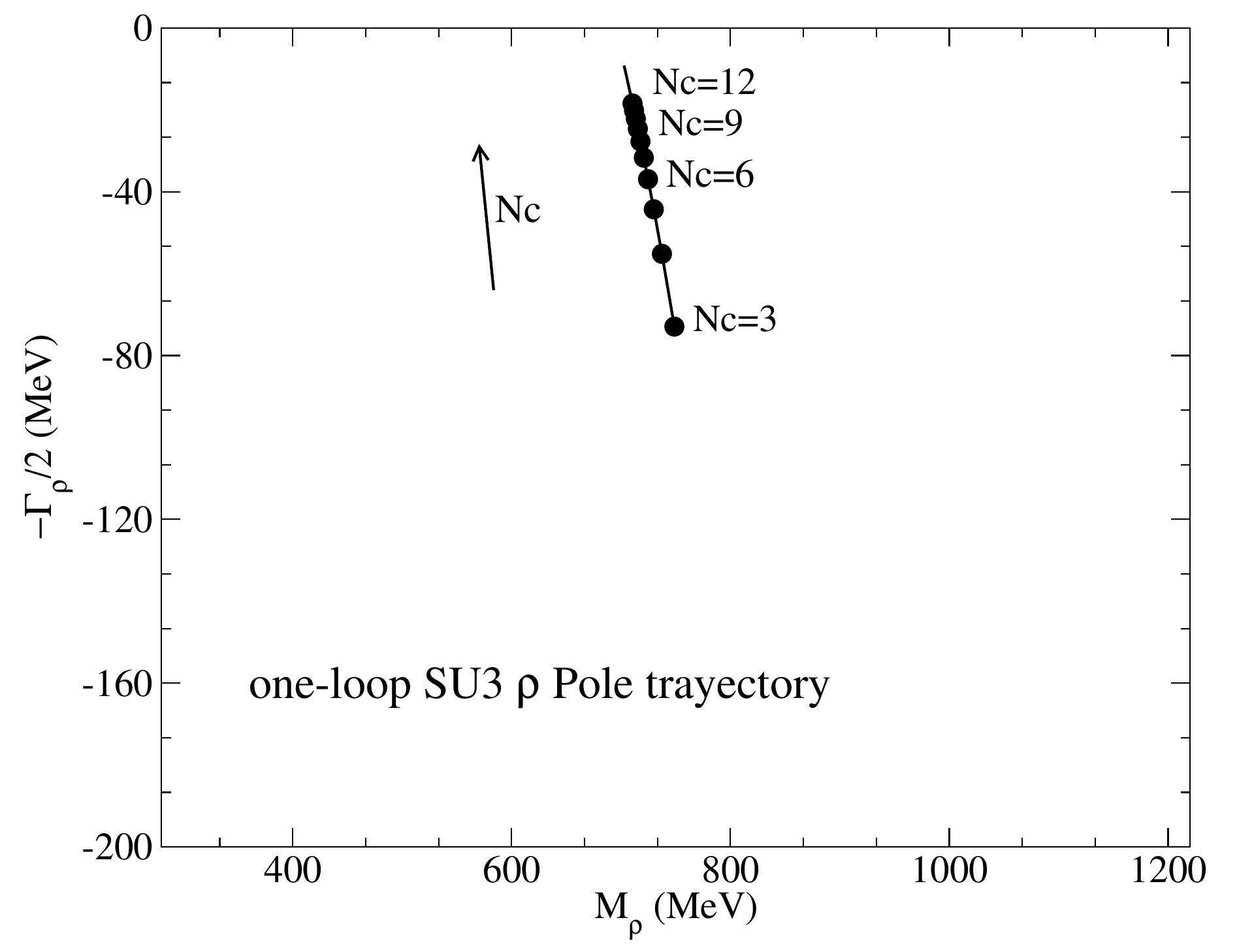}
  \includegraphics[width=0.49\textwidth]{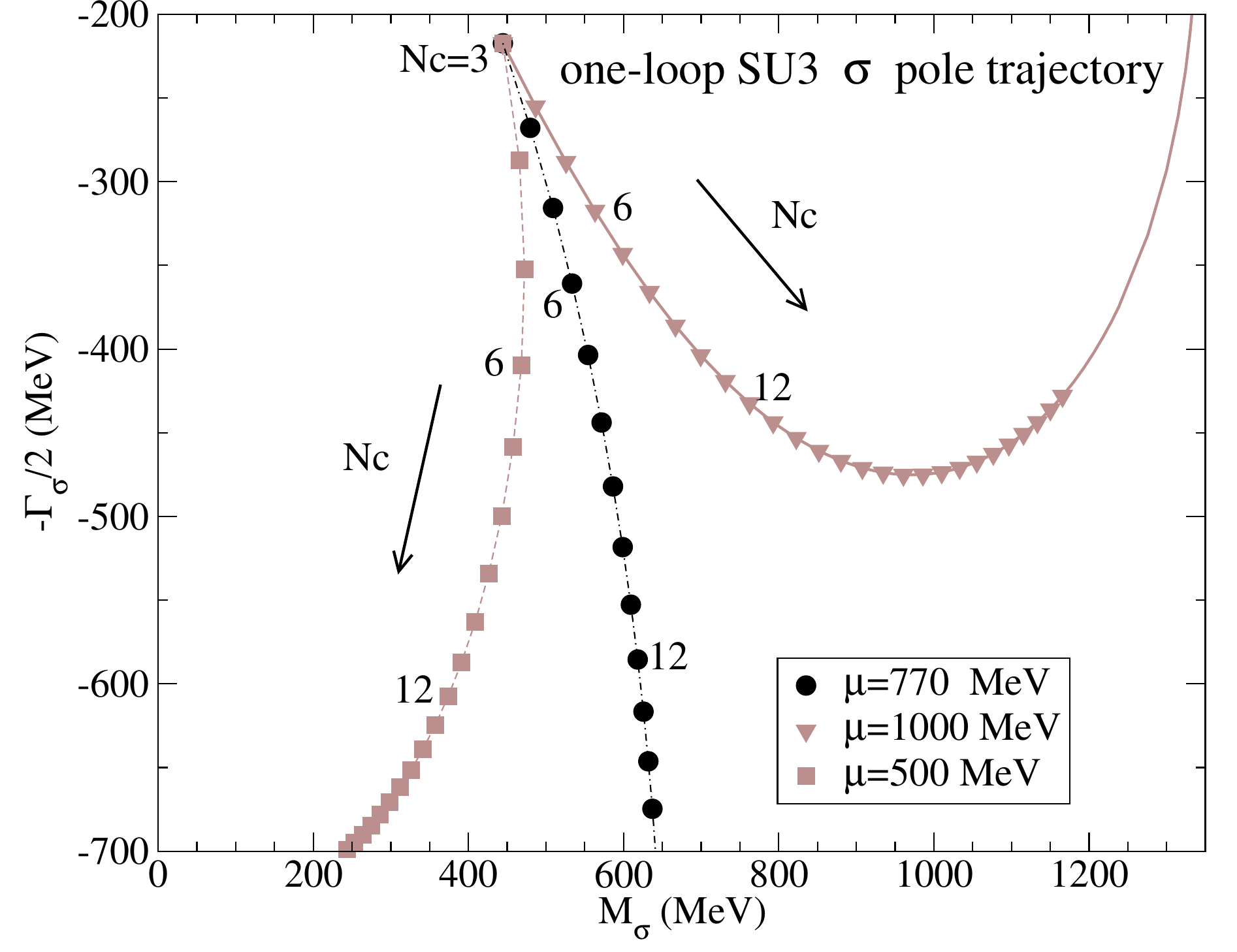}
\vspace*{-.5cm}
\caption{Trajectories of the $\rho(770)$ (left) and $\sigma$ (right) poles in the complex plane as $N_c$ is varied away form 3.
The lighter curves on the right plot
indicate the uncertainties when varying the regularization
scale $\mu$ in the usual range, as recalculated in \cite{RuizdeElvira:2010cs}.
In the case of the $\rho(770)$ the three lines almost overlap and are not plotted in the left figure. 
Figure taken from \cite{RuizdeElvira:2010cs}.
}
  \label{fig:Ncuncertainties}
\end{figure}

The trajectories of the $\rho(770)$ and $\sigma$ poles in the 
complex plane are shown in Fig.\ref{fig:Ncuncertainties},
obtained with updated NLO IAM fits in \cite{RuizdeElvira:2010cs}.
On the left panel the $\rho(770)$ pole is found to move towards the real axis as soon as $N_c$ is increased,
becoming a narrower resonance, but keeping an almost constant mass.
It should be noted that it was found in \cite{Pelaez:2010er}
that if $\mu$ is made $\sim 1.2\,\gev$, the $\rho(770)$ stops behaving as 
a $q \bar q$ meson, which reinforces the choice of 0.5 to 1 GeV as the range where the 
renormalization scale $\mu$ should be varied to scale the LECs with $N_c$. 
Back to the figure, on the right panel we observe the trajectory followed
by the $\sigma$ pole, which  
moves instead deep into the complex plane as $N_c$ is increased from 3 to $N_c\sim 10$ or 12. However, beyond that $N_c$ the uncertainty is so big that
the $\sigma$ trajectory could keep moving 
far from the real axis or it may even fall back \cite{Pelaez:2006nj,Nieves:2009ez,Sun:2004de}.
Let us therefore separate the analysis into 
the near-$N_c$=3 and large-$N_c$ regions.

In the near-$N_c$=3 region, the non-ordinary behavior of the pole
that we have just seen
was already found in the simplest unitarization model of the LO ChPT amplitude 
as discussed at the beginning of this subsection.  Basically, it
can be traced back to 
the dominant role of two-meson loops in the dynamics 
responsible for the $\sigma$ \cite{Oller:1998zr,Nieves:2009ez}.
The  IAM approach has been revisited several times with slightly different fits to data or some simplifications
and the non-ordinary behavior near $N_c$=3 seems very robust 
\cite{Nieves:2009ez,RuizdeElvira:2010cs,Uehara:2004es}.
Moreover, there is also a NNLO IAM calculation
 within $SU(2)$ ChPT \cite{Pelaez:2006nj}. 
In this case a $\chi^2$-like function was defined to measure how close 
 a resonance is from a $\bar qq$-like $N_c$  behavior.
But if this function is minimized one can try to impose 
an ordinary behavior into a resonance.
Thus, in \cite{Pelaez:2006nj} it was first shown that to NLO 
it is not possible to make
$\sigma$ behave predominantly as a $\bar q q$
while describing simultaneously the data
and the $\rho(770)$ $\bar qq$ behavior, thus
confirming the robustness of 
the NLO conclusions for $N_c$ close to 3.

\begin{figure}
  \centering
  \includegraphics[width=0.49\textwidth]{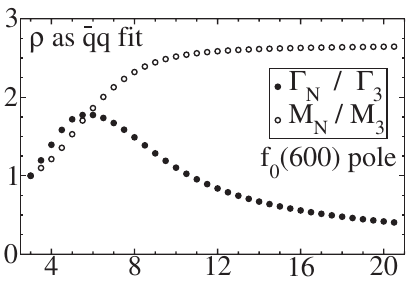}
\caption{
$N_c$ dependence of the $f_0(500)$ pole mass and width, normalized to their physical values, as obtained with the elastic NNLO IAM \cite{Pelaez:2006nj} imposing the $\rho(770)$ to behave as a $\bar qq$. Note that up to $N_c\simeq 12$ the $\sigma$ mass and width do not behave as for ordinary mesons. Beyond that value the sigma shows an ordinary behavior, but for a mass 
roughly 2.5 times larger than its physical value.
Figure taken from \cite{Pelaez:2006nj}.}
  \label{fig:Pelaez-Rios-PRL-largen-2}
\end{figure}

In order to check this robustness, a NNLO IAM fit was made to data 
but, since there are four more parameters 
at NNLO than NLO, it was also constrained to reproduce  the $N_c$ 
$\bar qq$ behavior for the $\rho(770)$ \cite{Pelaez:2006nj}. 
This is called a ``$\rho$ as a $\bar qq$ fit''.
The resulting $N_c$ behavior for the 
$\sigma$ mass and width is displayed 
in Fig.\ref{fig:Pelaez-Rios-PRL-largen-2}.
There we can see that, once again, in the near-$N_c$=3 region
both the mass and the width grow, i.e.
the $\sigma$ is not behaving as an ordinary meson.
Moreover, the non-ordinary behavior of the $\sigma$ in the near-$N_c=3$ region has also been found in other unitarization techniques,
as in \cite{Nieves:2009ez,Sun:2004de}.
In particular, in \cite{Guo:2011pa,Guo:2012yt}, using a chiral unitary approach but within the leading order $1/N_c$ $U(3)$ formalism, 
a similar non-ordinary behavior was found once more 
from the $N_c$ behavior close to three. 
As expected the $\eta'(980)$ does not seem to play 
a significant role in this rather robust statement.
Also, the inclusion of explicit heavy and ordinary
resonances  does not make the $\sigma$ to behave predominantly as an ordinary meson \cite{Nieves:2011gb}.

Therefore, the $\sigma$ pole behavior in the near-$N_c=3$ region 
can be considered a very strong hint of a 
predominantly non-ordinary nature 
of the $\sigma$ meson in particular, 
and of the whole light meson nonet in general.
As commented in the introduction, by 2007 this behavior was considered 
``... the only reliable identifications of observed effects that may be
examples of a different class of hadrons" \cite{Jaffe:2007id}. 

Of course, the above statement does not exclude the existence 
of some mixing with ordinary or glueball components, 
as long as they do not dominate the dynamics that generate the physical $f_0(500)$. There are actually hints about these 
possible subdominant components, but since they come from the
behavior of the pole for relatively larger values of $N_c$,
they are not so conclusive.

Let us then consider the large-$N_c$ region.
Turning back to the right panel of Fig.\ref{fig:Ncuncertainties},
we see examples of the two generic scenarios found in the literature,
exemplified by the borders of the uncertainty band. 
In the first scenario, after an initial non-ordinary behavior 
the curve labeled ``$\mu=1000\,\mev$'' 
falls back into the real axis, the mass stabilizes 
and the width decreases. As a matter of fact this is the behavior
expected for ordinary mesons and may be a hint 
of a subdominant ordinary component,
arising as loop diagrams become suppressed when $N_c$ grows.
Actually, as it can be seen in Fig.\ref{fig:Pelaez-Rios-PRL-largen-2}
this was the preferred behavior of the NNLO IAM
analysis \cite{Pelaez:2006nj} above  $N_c\simeq12$.  
As commented above, it was not possible to make this component dominant without spoiling the fit or the ordinary $\rho(770)$ behavior.
Hence, it is very important to emphasize that {\it this possible 
ordinary subdominant component appears at a mass about 2 to 2.5 times larger than the physical $f_0(500)$ mass}. Therefore this behavior 
would suggest that there is a
small mixing of the $\sigma$ with an ordinary meson component whose mass is around 1 or 1.5 GeV, precisely where it is widely accepted that 
an ordinary scalar nonet exists. This behavior bears a remarkable resemblance to unitarized quark model calculations \cite{vanBeveren:2006ua}
in which
a quark-antiquark state above 1 GeV 
can be deformed into the physical $\sigma$ meson pole by
making a $\pi\pi-q\bar{q}$ interaction sufficiently 
strong.

On the second scenario, exemplified by the curve labeled ``$\mu=500\,\mev$''
it behaves rather differently, and the $\sigma$ pole moves deep into the complex $\sqrt{s}$ plane. But note that this is actually the third quadrant
of the $s$ plane, so that there is no clear physical interpretation for such a pole \cite{Nieves:2009ez,Nieves:2011gb}.

Either one or both of these two general behaviors 
have been found in other unitarization approaches.
For instance, this has been revisited in \cite{Nieves:2011gb}, by 
unitarizing the $\pi\pi$ scattering amplitude within a Bethe-Salpeter like method, but including explicitly 
one heavy ordinary multiplet for each allowed $J^P$ quantum numbers.
In particular the lightest ordinary scalar multiplet is set around 1 GeV.
In this way one is also including the $1/N_c$ leading contribution of these resonances to all orders in the chiral expansion, instead of just
to NLO as in the IAM. 
The conclusions obtained from the region
near  $N_c=3$ are the same, the
predominant component of the $\sigma$ is not of an ordinary nature.
However, for {\it large} $N_c$  
two generic scenarios were found, similar to the scenarios described above. In this case, in the first scenario
the $\sigma$ subdominant ordinary component
coincides with the heavier ordinary scalar included explicitly in the Lagrangian.
In contrast, in \cite{Guo:2011pa} only the second large-$N_c$ scenario
is found, where the $\sigma$ moves away from the physical cut as $N_c$ increases far from 3. 
 Nevertheless, within this approach there is also a 
scalar state around 1 GeV that survives at large $N_c$, although 
in this case it is naturally identified  as a component of the $f_0(980)$. Note, however, that 
 there is some controversy \cite{Nieves:2011gb}
on the correct identification of the leading $1/N_c$ terms 
and the extrapolation of the amplitude to $N_c\neq3$ in \cite{Guo:2011pa}.

The $N_c$ behaviors of scalars has also received recent attention from lattice QCD \cite{Bali:2013kia,Bali:2013fya}. In this case the large $N_c$ limit is calculated, which is not exactly the same approach followed throughout this section ---recall we are rescaling all parameters with their leading $N_c$ behavior--- 
but might be compared with the largest $N_c$ values.
The scenario where the lightest  scalar mass 
for large $N_c$ is comparable or above the $\rho(770)$ mass, is qualitatively consistent
with the latest lattice calculations \cite{gunnarprivate}
finding that the ratio $M_{scalar}/M_\rho\simeq 1.13\pm0.10$
 in the $N_c\rightarrow \infty$ limit.

In conclusion, the predominantly non-ordinary nature of the sigma
is also very strongly supported by its leading-$1/N_c$ behavior
near $N_c$=3 from unitarized ChPT. Meson loops are responsible for this dynamics. For larger values of $N_c$ the uncertainties in the unitarization methods do not allow a conclusive statement. However, one possible 
scenario suggests the existence of mixing with a subdominant ordinary component with a mass around 1 GeV. For the other scenario, where the sigma pole moves deep in the complex plane, there is not any clear physical interpretation yet.

There is, however, a further semi-quantitative
argument that we will review next, which the first scenario fulfills naturally
but not the second unless some additional contributions from other resonances are taken into account.

\subsection{Semi-local duality}
\label{subsec:semilocal}

A well-known feature of the real world (at $N_c$ = 3) is that of ``local duality'' \cite{SLD} (see \cite{Donnachie:2002en,CollinsRegge} for textbook introductions). At low energies the scattering amplitude is fairly well represented by the exchange of resonances
(with a background), which as energy increases become wider and increasingly overlap. This overlap generates a smooth Regge 
behavior described by a small number of crossed channel Regge exchanges. 
Detailed studies of hadron-hadron scattering show that the sum of resonance contributions at all energies ``averages'' 
 the higher energy Regge behavior. Thus, s-channel resonances
are related to Regge exchanges in the t-channel and are ``dual'' to each other. Namely, one can use either formalism to describe data, at least ``on the average" (to be defined below). Regge exchanges are also built from $\bar{q}q$ and multi-quark
contributions. However, in the isospin 2 $\pi\pi$-scattering channel there are no  resonances, 
and so Regge exchanges with these quantum number necessarily involve multi-quark components. 
Experiment tells us that even at $N_c$=3 these components are suppressed 
compared to the dominant $\bar{q}q$ exchanges. Hence, semi-local
duality means that in $\pi^{+}\pi^{+} \rightarrow \pi^{+}\pi^{+}$, which is an $I=2$ process, 
the contribution from low energy resonances must cancel ``on the average'', or at least be much smaller than in processes with other isospins. Now, 
using the crossing relations in Eqs.\ref{eq:param:crossing-matrices} the $I=2$ t-channel amplitude can be recast as
a function of s-channel amplitudes: 
\begin{equation}
\mathrm{Im}\,T^{(I_t=2)}(s,t)=\frac{1}{3}\mathrm{Im}\,T^{(I_s=0)}(s,t)-\frac{1}{2}\mathrm{Im}\,T^{(I_s=1)}(s,t)+\frac{1}{6}\mathrm{Im}\,T^{(I_s=2)}(s,t).
\end{equation}
However $T^{(I_s=2)}$ is repulsive and small. This is for
instance seen in panel ``e"  of Fig.\ref{fig:IAMcoupled}, which shows the largest $I=2$ partial wave.
Therefore the strong cancellation occurs between 
$T^{(I_s=0)}$ and $T^{(I_s=1)}$, which are dominated at low energies by 
the  $f_0(500)$ and $\rho(770)$ resonances, respectively.
Hence, semi-local duality requires the contributions from these two resonances to cancel
``on the average''.  Since no other resonances appear in the $I=2$ wave 
at large $N_c$, it remains small and there is no reason why this suppression should disappear as $N_c$ increases.
 
The relevance of this feature for the nature of the $\sigma$ is that, as seen in the previous subsections,
 it seems to have a rather different $N_c$ behavior compared to ordinary mesons like the $\rho(770)$. 
Semi-local duality implies that the $\sigma$ behavior cannot
be such that the cancellation just described disappears at larger $N_c$. But this might happen if the $\sigma$
disappeared completely from the spectrum as $N_c$ increases, unless the
large-$N_c$ contributions of other resonances are fine tuned to cancel that of the $\rho(770)$. Such a fine tuning is unlikely since these other resonances are already subdominant at $N_c$=3.

This ``on the average cancellation'' is properly defined via Finite Energy Sum
Rules like:   
\begin{equation}\label{FESR}
F(t)^{21}_n\equiv \frac{\int_{\nu_{th}}^{\nu_{\mathrm{max}}}{d\nu\;\mathrm{Im}\, T^{(I_t=2)}(s,t)/\nu^n}}{\int_{\nu_{th}}^{\nu_{\mathrm{max}}}{d\nu\;\mathrm{Im}\,T^{(I_t=1)}(s,t)/\nu^n}},\;\;\;\nu=(s-u)/2.
\end{equation}

Then semi-local duality means that on the ``average'' and at least over one resonance tower, we have:
\begin{equation}\label{ReggeLocal}
  \int_{\nu_{\mathrm{th}}}^{\nu_{\mathrm{max}}}{d\nu\;\nu^{-n}\mathrm{Im}\,T^{(I_t)}(s,t)_{\mathrm{Data}}}\sim \int_{\nu_{\mathrm{th}}}^{\nu_{\mathrm{max}}}{d\nu\;\nu^{-n}\mathrm{Im}\,T^{(I_t)}(s,t)_{\mathrm{Regge}}},
\end{equation}
where the Regge amplitudes can be calculated with parameterizations available in the literature,
which are expected to work for $t\ll s$. For this reason semi-local duality integrals were evaluated at $t=0$ and $t=4m_\pi^2$ \cite{RuizdeElvira:2010cs}.
As we discussed in Subsec.\ref{subsec:highenergydata}, for very small $t$ and particularly for $t=0$, different Regge parameterizations of $\pi\pi$ scattering are fairly compatible, as seen in Fig.\ref{fig:reggedata}. This fair agreement is enough for the ``on the average" estimates needed for semi-local duality arguments. 
Therefore, the rigorous definition of  semi-local duality is 
 that $|F(t)^{21}_n| \ll 1$. Since the Regge trajectories do not depend on $N_c$, still one should find
$|F(t)^{21}_n| \ll 1$ when increasing $N_c$, due to a strong cancellation between the
$\rho(770)$ and the $f_0(500)$. In principle this would not occur if the $f_0(500)$ disappeared completely from the spectrum.

\begin{table}
\centering
    \begin{tabular}{c|c||c|c|c}
      $\nu_{max}$ & 400 GeV$^2$ &2.5 GeV$^2$&2 GeV$^2$&1 GeV$^2$ \\\hline 
      $F^{21}_1$ &0.021 $\pm$ 0.016 & 0.180 $\pm$ 0.066 &0.199 $\pm$ 0.089 & -0.320 $\pm$ 0.007\\\hline
      $F^{21}_2$ &0.057 $\pm$ 0.024 & 0.068 $\pm$ 0.024 &0.063 $\pm$ 0.025 & -0.115 $\pm$ 0.013\\\hline
      $F^{21}_3$ &0.249 $\pm$ 0.021 & 0.257 $\pm$ 0.022 &0.259 $\pm$ 0.022 & 0.221  $\pm$ 0.021\\\hline
    \end{tabular}
    \caption{Values of $F^{21}_n$  using the $\pi\pi$ scattering  parameterization in \cite{Kaminski:2006qe} and different cutoffs.
      All $F^{21}_n$ ratios for a 20 GeV cutoff turn out very small, but it can be seen that the suppression has already occurred when $s_{max}$ is still
      $\sim$1 or 2 GeV$^2$. Table taken from \cite{RuizdeElvira:2010cs}.}\label{tab:SLD1}
\end{table}

The NLO and NNLO IAM have been used \cite{RuizdeElvira:2010cs} to check the $N_c$ dependence of semi-local duality suppression in $\pi\pi$
scattering amplitudes. As expected, the $I=2$ s-channel amplitude remains
repulsive as $N_c$ increases, and still there is no resonance exchange. 
Since the IAM is only valid in the low energy region
the influence of the high energy part on this cancellation was first checked explicitly for $N_c=3$.
In particular, in Table \ref{tab:SLD1} we  show
the value of the FESR for different \textit{cutoffs} using the 
dispersive
$\pi\pi$ \cite{Kaminski:2006qe} data parameterization  
 as input, instead of the IAM. Note
that local duality is satisfied for $N_c$=3 since $|F(t)^{21}_n|
\ll 1$ and, at least for n=2,3, the main  
suppression already occurs below 1 GeV, where the IAM can be applied. 
Therefore, the IAM can be used to check the main part of the FESR
suppression with $N_c$, by simply calculating the unitarized partial waves $t_J(s)$
and reconstructing $T^{I_s=I}$ using the partial wave series in Eq.\ref{ec:pwdef}. 
However, we already commented that to NLO and NNLO, 
the IAM can only be unitarized in the S and P-waves, since higher waves have a non-vanishing imaginary part only at higher orders in ChPT. 
Thus, in \cite{RuizdeElvira:2010cs} it was checked that the effect of higher waves on Eq.\ref{FESR} is around
10$\%$, and it is dominated by the $f_2(1270)$ resonance, which is a nice Breit-Wigner like ordinary resonance, whose $N_c$ behavior was easy to implement.

\begin{figure}
\vspace{-4.5cm}
  \begin{center}
  \includegraphics[width=0.49\textwidth]{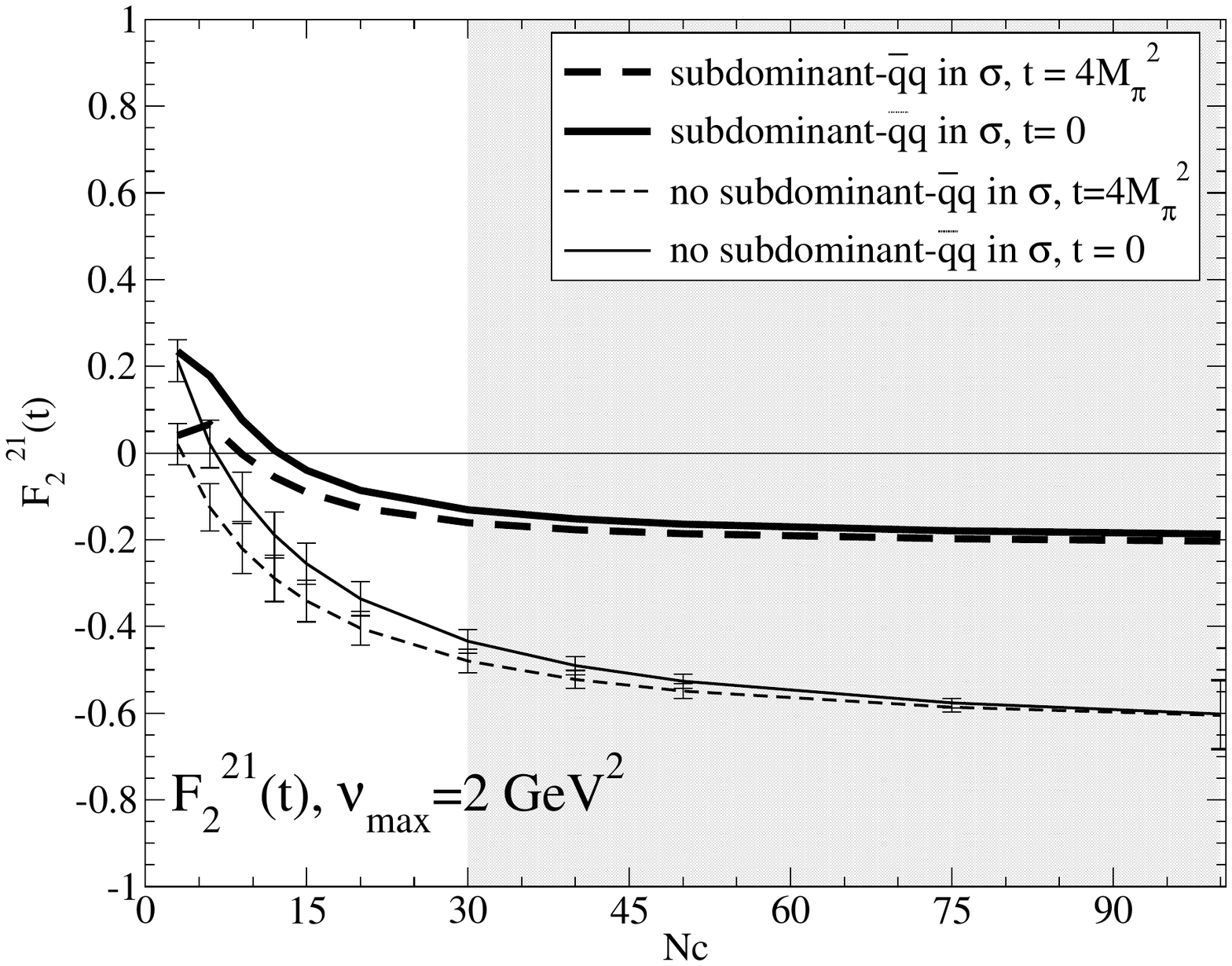}
  \includegraphics[width=0.49\textwidth]{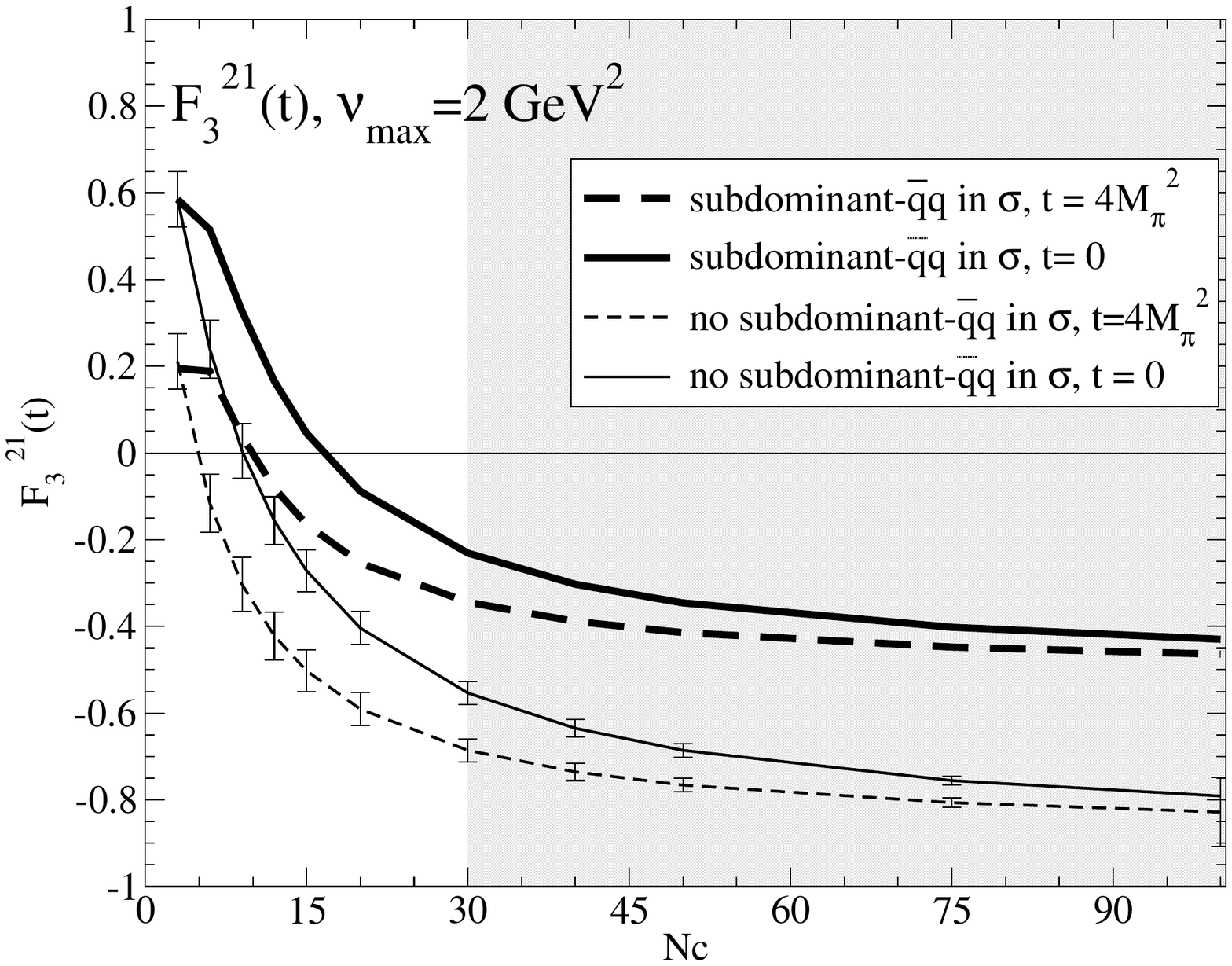}
  \caption{In the scenario where the $f_0(500)$ disappears from the spectrum there is no FESR suppression and local duality
    fails as $N_c$ grows. However, in the scenario where 
the $\sigma$ has a subleading ordinary component around 1 GeV, this component ensures local
    duality even when increasing $N_c$. Recall that the IAM is only expected to give good quantitative results up to $N_c=30$ at most.
    Beyond that, the curves are just extrapolations to show that the scalar contribution above 1 GeV would be enough to stabilize the ratios at the required small values. Figures taken from \cite{RuizdeElvira:2010cs}.}
\label{fig:Fevolution}
\end{center}
\end{figure}

Once the $N_c$=3 framework is set, if $N_c$ is increased 
the $\rho(770)$ mass remains roughly  constant and its width becomes narrower.
Also, for not too large $N_c$ we have already seen that the $\sigma$ moves deeper into the complex plane.
However, for larger $N_c$ there were basically two scenarios that were presented at the end of the previous subsection.
In the second scenario the  $f_0(500)$ disappears from the spectrum and its contribution to the total cross section below 1 GeV
becomes less and less important. Hence the ratios $|F(t)^{21}_n|$ grow and there is a 
conflict with semi-local duality. This is shown by the thin lines
of Fig.\ref{fig:Fevolution}. 
It is important to remark that this is a generic problem for any model where the $f_0(500)$ contribution vanishes at large $N_c$, not just for the IAM.
For instance, as pointed out in \cite{Nieves:2011gb},
 it could be an issue for the model described in \cite{Zhou:2010ra} built within the unitarized quark model
proposed in \cite{Tornqvist:1995kr}, where all scalars below 2 GeV except the glueball could
be described with $\bar qq$ seeds dressed by hadron loops, but the $\sigma$ had no correspondence whatsoever with these seeds. 
 Potentially this may lead to violations of semi-local duality at large $N_c$.

However, in the first scenario found in the previous subsection, there is a subdominant ordinary
component for the $f_0(500)$ with a mass somewhat above 1 GeV. This  
behavior corresponds to curves like that labeled $\mu=1000\,\mev$ in Fig.\ref{fig:Ncuncertainties} and 
occurred naturally within the two-loop IAM \cite{Pelaez:2006nj},
as seen in Fig.\ref{fig:Pelaez-Rios-PRL-largen-2}.
In \cite{RuizdeElvira:2010cs} it was found that this heavier subdominant component emerging at larger $N_c$ was enough
to ensure the  cancellation with the $\rho$(770) contribution required by semi-local duality.
The suppression effect is shown by the thick lines of Fig.\ref{fig:Fevolution}.
Note the gray area above $N_c=30$, where we have already discussed that the IAM is not reliable
and  therefore the results there are merely qualitative. It is just displayed in order to show that the presence of a scalar contribution
above 1 GeV stabilizes the ratios at small values, but that without such a component the suppression required by semi-local duality disappears..

Therefore, semi-local duality seems to favor naturally the first scenario,  in which the $\sigma$ has a
 subdominant ordinary component with a mass around 1 GeV. 
In such case it would be a component of the $f_0(980)$ the one that 
becomes relevant at $N_c>>3$.
Thus, the support from semi-local duality for the subdominant component is based on naturalness and the absence of fine tuning, apart from the fact that the $N_c$ behavior has a simple explanation in terms of the relatively well established 
ordinary scalar nonet above 1 GeV.
 However, as shown in  \cite{Guo:2011pa},
if one also takes into account part of the 
subleading $N_c$ scaling contributions (corresponding to the NLO in the so-called $\epsilon$ expansion), a particular choice of the vector coupling $G_V$, tunes the $\rho$ and $S_1$ 
masses to be equal in the large $N_c$ limit
and considers the variation with $N_c$ of the $f_0(980)$,  
it is still possible to satisfy semi-local duality 
at large $N_c$ even if the $\sigma$ does not have 
any subdominant ordinary component around 1 GeV \cite{Guo:2011pa}.
However, if one also includes the $J=2$ $f_2(1270)$ resonance the $F^{21}_3$ ratio is spoilt again,
so that one would have to assume that additional cancellations from even heavier resonances must occur. Note also that, as previously commented, the implementation of the $N_c$ scaling in \cite{Guo:2011pa} has been questioned in \cite{Nieves:2011gb},
where the first scenario is also found. 

\subsection{Regge theory and the $f_0(500)$}
\label{subsec:reggesigma}

Another well-known feature of $\bar qq$ hadrons is that they can be classified into linear $(J,M^2)$
trajectories relating the angular momentum $J$ and the mass squared with a universal slope of the order of $0.8-0.9\,\gev^{-2}$. In the case of mesons these trajectories are intuitively interpreted in terms of quark-antiquark states, since they are similar to those obtained from the relativistic rotation of a  rigid rod or 
of a flux tube connecting a quark and an antiquark. 
Strong deviations from this linear behavior would suggest a rather different nature and the scale of
the trajectory would also indicate the scale of the mechanism responsible for the presence of a resonance. 
In particular, the $\sigma$ is almost never listed as a member of linear  trajectories because it does not fit very well into this classification. 
For instance, in the very complete study of meson Regge trajectories in \cite{Anisovich:2000kxa} the ``enigmatic" $\sigma$ meson
was omitted from the  ``$\bar qq$ trajectory supposing it is alien to this classification". 

Recently, it has been shown \cite{Londergan:2013dza} how
to calculate, instead of fit,  the 
Regge trajectories of resonances appearing in elastic two-meson scattering, using a dispersive formalism \color{black} for the trajectory. The phenomenological input is the pole position and residue associated to the resonance. The only chiral constraint is that in the scalar scattering amplitude
the Adler zero is introduced explicitly.
Within this approach,  trajectories of resonances widely accepted as $\bar qq$ mesons like the $\rho(770)$, $K^*(892)$ and $K_1(1400)$ vectors, the $f_2(1270)$ and $f'_2(1525)$ tensors, and
even the $K_0^*(1430)$ scalar \color{black} come out almost real and linear with a $\sim 0.8-0.9 \,\gev^{-2}$ slope \cite{Londergan:2013dza,Carrasco:2015fva,Pelaez:2016ntz}, as expected. In contrast the $\sigma$
and $\kappa$  trajectories come out very different, confirming their non-ordinary nature.
Let us briefly review this approach and how the resulting $\sigma$  trajectory looks like, because it also provides a relevant hint on the $\sigma$ nature and structure.

A textbook introduction to Regge amplitudes and their analytic properties,
with almost the same notation we follow here, can be found in \cite{CollinsRegge}.
An elastic $\pi\pi$ partial wave with angular momentum $l$  near a Regge pole can be written as 
\begin{equation}
t_l(s)  = \beta(s)/(l-\alpha(s)) + f(l,s),
\label{Reggeliket}
\end{equation}
where $f(l,s)$ is a regular function of $l$, and the Regge trajectory $\alpha(s)$ and 
residue $\beta(s)$ are analytic functions, the former having a cut along the real axis for $s$ above threshold. 
Recall that in Regge Theory $l$ is promoted to a complex variable.
Nevertheless, if the pole dominates in Eq.\ref{Reggeliket}, the elastic unitarity condition in Eq.\ref{ec:unit} implies that, for real $l$, 
\begin{equation}
\mbox{Im}\,\alpha(s)   = \sigma(s) \beta(s).   \label{reggeunit} 
\end{equation}
 
In the residue it is convenient to make explicit the threshold behavior
as well as the cancellation of the poles of the Legendre function appearing in the full amplitude. Thus the $\beta(s)$ function is rewritten as~\cite{Chu:1969ga}
\begin{equation}
\beta(s) =  \gamma(s) \hat s^{\alpha(s)} /\Gamma(\alpha(s) + 3/2) , \label{reduced} 
\end{equation}
where $\hat s =( s-4M_\pi^2)/s_0$. The dimensional scale $s_0=1\,$ GeV$^2$ is introduced for 
convenience and the so-called reduced residue $\gamma(s)$ is an analytic function, whose phase is known because $\beta(s)$ is real in the real axis.

It is then possible to write dispersion relations for $\alpha(s)$ and $\beta(s)$, which are related via Eq.\ref{reggeunit},
leading to \cite{Londergan:2013dza,Chu:1969ga}:
\begin{eqnarray}
\mbox{Re}\, \alpha(s) & = &  \alpha_0 + \alpha' s +  \frac{s}{\pi} PV \int_{4m_\pi^2}^\infty ds' \frac{ \mbox{Im}\,\alpha(s')}{s' (s' -s)}, \label{iteration1}\\
\mbox{Im}\,\alpha(s)&= & \frac{ \sigma(s)  b_0 \hat s^{\alpha_0 + \alpha' s} }{|\Gamma(\alpha(s) + \frac{3}{2})|}
 \exp\Bigg( - \alpha' s[1-\log(\alpha' s_0)]\nonumber\\
&+& \frac{s}{\pi} PV\!\int_{4m_\pi^2}^\infty\!\!ds' \frac{ \mbox{Im}\,\alpha(s') \log\frac{\hat s}{\hat s'} + \mbox{arg }\Gamma\left(\alpha(s')+\frac{3}{2}\right)}{s' (s' - s)} \Bigg), 
\label{iteration2}
 \end{eqnarray}
where $PV$ denotes ``principal value'' and $\alpha_0, \alpha'$ and $b_0$ are free parameters to be determined phenomenologically.  
As already seen in Sec.\ref{subsec:chpt}, scalar $\pi\pi$ scattering
partial waves
have Adler zeros below threshold. This can also be made explicit in
 $\beta(s)$. 
In practice it is enough to multiply
the right hand side of Eq.\ref{iteration2} by $2s-M_\pi^2$,
which ensures the Adler zero position is correct to leading order in ChPT.
In such case the $3/2$ has to be replaced by $5/2$ inside the Euler $\Gamma$
functions in order not to spoil the large $s$-behavior.
Then $b_0$ has dimensions of GeV$^{-2}$.
For a given set of $\alpha_0, \alpha'$ and $b_0$ parameters 
the above equations, or those modified to have an Adler zero, 
can be solved iteratively. 

The Regge trajectory of a resonance is then obtained \cite{Londergan:2013dza} by 
imposing that the solution of Eqs.\ref{iteration2}
must yield  a pole in Eq.\ref{Reggeliket}
for the angular momentum of that resonance,
and varying $\alpha_0, \alpha'$ and $b_0$ to fit
the pole position and residue to the values
of the physical resonance pole. 
For the  $\sigma$ and $\rho(770)$ calculations \cite{Londergan:2013dza}, their pole parameters
are taken from the precise dispersive representations of $\pi\pi$ scattering data \cite{GarciaMartin:2011jx}
that we already discussed in Sec.\ref{subsec:precisepoles}. 
In particular, the $\sigma$ pole position and residue correspond to those in the last line of Table~\ref{tab:dispersivepoles}. 
In contrast, for the $f_2(1270)$, $f_2'(1525)$, \color{black} $K^*(892)$, $K_1(1400)$ and $K_0^*(1430)$ calculations \cite{Carrasco:2015fva,Pelaez:2016ntz}, their scattering
pole parameters are taken from phenomenological parameterizations: in the case of the $f_2(1270)$ from
the dispersively constrained fit of \cite{GarciaMartin:2011cn}, also commented
in Sec.\ref{subsec:precisepoles},
and in the rest of cases  from their RPP 2012 values \cite{PDG12}. \color{black}

\begin{figure}
  \centering
  \includegraphics[width=0.9\textwidth]{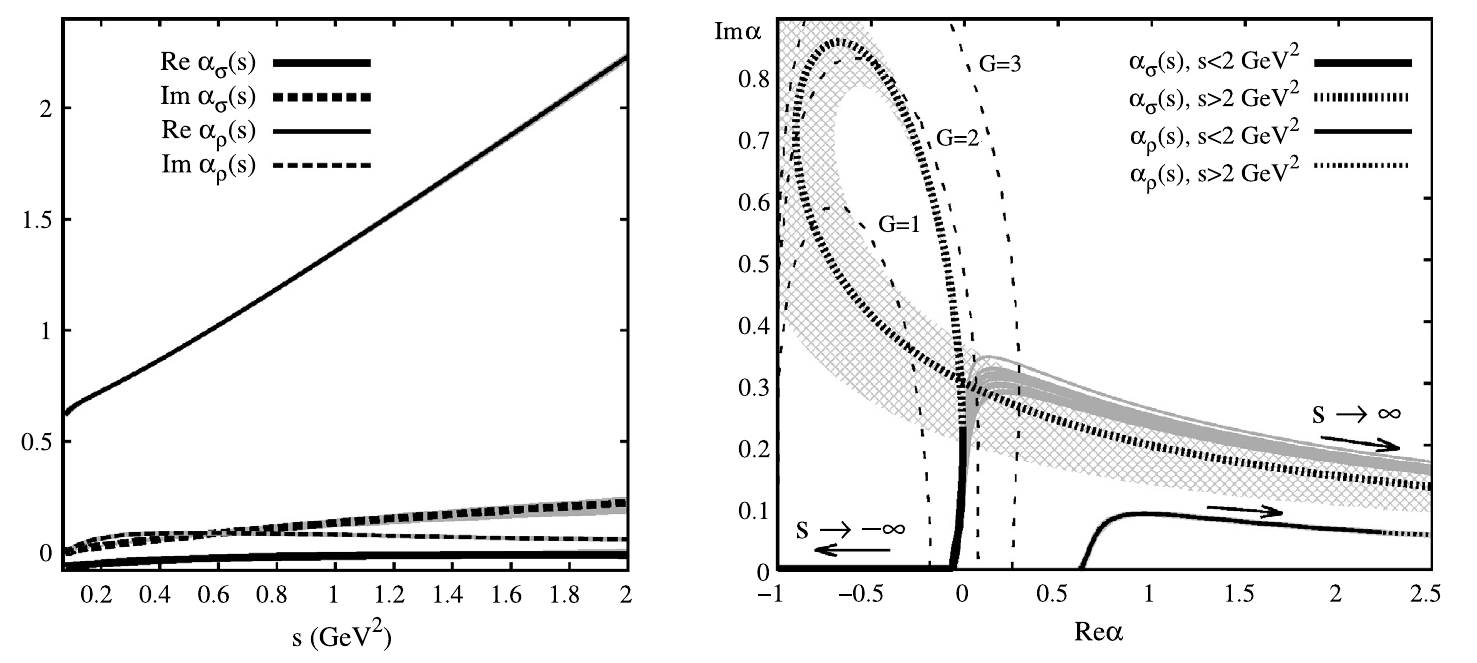}
\vspace*{-.5cm}
\caption{Left: $\alpha_\rho(s)$ and $\alpha_\sigma(s)$ Regge trajectories, 
$s$ dependence as calculated in \cite{Londergan:2013dza}.
 Right: The same $\alpha_\sigma(s)$ and $\alpha_\rho(s)$, plotted
in the complex plane. 
The dotted lines are the extrapolation of the \cite{Londergan:2013dza}
results beyond $s=2\,$GeV$^2$. 
 Within the input pole parameter error bands, in the case of the $\sigma$, 
two types of solutions are found. One set (pattern-filled band)
  has a loop in the $\mbox{Im}\alpha - \mbox{Re}\alpha$ plane. The other (gray lines), having slightly higher $\alpha'$ does not form a loop. 
At energies below $s=2\,$GeV$^2$, both are similar to the trajectories of the Yukawa potential $V(r)=-{\rm G} a \exp(-r/a) /r$, shown here for three different values of  G  \cite{Lovelace}. For the G=2 Yukawa curve 
\cite{Lovelace}
$a\simeq 0.5 \,$GeV$^{-1}\simeq 0.1 {\rm fm}$ can be estimated. This could be compared, for instance, to the S-wave $\pi\pi$ scattering length $\sim 1.6\, $GeV$^{-1}\simeq 0.3 {\rm fm}$. Figure taken from \cite{Pelaez:2010fj}
\label{fig:Reggesigma}
}
\end{figure}

Thus, the left panel of Fig.~\ref{fig:Reggesigma} shows
the resulting Regge trajectories for the $\sigma$ and $\rho(770)$
resonances, whose parameters are given in Table \ref{tab:reggeparam} \cite{Londergan:2013dza}. 
The imaginary part of $\alpha_\rho(s)$ comes out much smaller than the real part, and the latter grows linearly with $s$. 
This is the behavior expected for the $\rho(770)$ as an ordinary meson.
Taking into account the approximations, and that the  errors in the parameters only correspond to
the uncertainty in the input pole parameters, 
the agreement with previous $\rho(770)$ trajectory determinations is remarkable. 
Actually, fits of the $\rho(770)$ trajectory in the literature yield:
 $\alpha_\rho(0)=0.52\pm0.02$~\cite{Pelaez:2003ky}, $\alpha_\rho(0)=0.450\pm0.005$ 
\cite{PDG12}, $\alpha'_\rho\simeq 0.83\,$GeV$^{-2}$ \cite{Anisovich:2000kxa}, $\alpha'_\rho=0.9\,$GeV$^{-2}$ \cite{Pelaez:2003ky}, or $\alpha'_\rho\simeq 0.87\pm0.06$GeV$^{-2}$ \cite{Masjuan:2012gc}. The results for the 
 $K^*(892)$, $K_1(1400)$, $f_2(1270)$, $f_2'(1525)$ and $K_0^*(1430)$ are also very consistent with an ordinary behavior. \color{black} The trajectories also come out almost real and linear, with the slope $\alpha'$ consistent with the 0.8-0.9 GeV$^{-2}$ expected value. However, for the latter three resonances
the uncertainties are larger, mostly due to the fact that they have a small inelasticity, which has been considered as a source of uncertainty \cite{Carrasco:2015fva,Pelaez:2016ntz}. \color{black}

\begin{table}
\centering
\begin{tabular}{cccc}\hline
& $\alpha_0$ & $\alpha'$ (GeV$^{-2}$)  & \hspace{5mm}$b_0$\hspace{5mm} \\\hline
\rule[-1.5mm]{0mm}{6mm} $\rho(770)$ & $0.520\pm0.002$  &  $0.902\pm0.004$ & $0.52$ \\
\rule[-1.5mm]{0mm}{6mm} $f_2(1270)$ & $0.9^{+0.2}_{-0.3}$  & $0.7^{+0.3}_{-0.2}$ & $1.3^{+1.4}_{-0.8}$ \\
\rule[-1.5mm]{0mm}{6mm} $f_2'(1525)$ &  $0.53^{+0.09}_{-0.45}$ & $0.63^{+0.20}_{-0.04}$ & $1.33^{+0.64}_{-0.07}$\\
\rule[-0.15cm]{0cm}{.55cm} $K^*(892)$&0.32$\pm$0.01  & 0.83$\pm$0.01 & 0.48$\pm$0.03\\
\rule[-0.15cm]{0cm}{.55cm} $K_1(1400)$&$-0.72^{+0.13}_{-0.03}$  & $0.90^{+0.01}_{-0.07}$ & $6.02^{+0.39}_{-1.13}$\\
\rule[-0.15cm]{0cm}{.55cm} $K^*_0(1430)$&$-1.15^{+0.23}_{-0.15}$& $0.81^{+0.08}_{-0.1}$ & $4.04^{+1.26}_{-2.43}$ GeV$^{-2}$\\\hline
\rule[-1.5mm]{0mm}{6mm} $\sigma/f_0(500)$ &  $-0.090\,^{+\,0.004}_{-\,0.012}$ & $0.002^{+0.050}_{-0.001}$ & $0.12$ GeV$^{-2}$\\
\rule[-0.15cm]{0cm}{.55cm} $\kappa/K^*_0(800)$&0.28$\pm$0.02  & 0.15$\pm$0.01 & 0.44$\pm$0.04 GeV$^{-2}$\\
\hline
\end{tabular}
\caption{Parameters of Regge trajectories 
{\it calculated}, not fitted, for different resonances in \cite{Londergan:2013dza}, using the integral equations in Eq.\ref{iteration2}
fitted to their pole parameters. The  $\sigma$ and $\rho(770)$ trajectories were calculated in \cite{Londergan:2013dza}, 
whereas the $f_2(1270)$, $f_2'(1525)$, \color{black} $K^*(892)$, $K_1(1400)$, $K_0^*(1430)$ and $K_0^*(800)$ results come from \cite{Carrasco:2015fva,Pelaez:2016ntz}. The trajectories of the first six resonances 
came out linear and consistent with a universal slope $\alpha'\simeq 0.8-0.9\, \gev^{-2}$. In contrast the $\sigma$ and $\kappa$ scalar mesons do not follow a linear trajectory and their slope, calculated at their pole masses, is much smaller. \color{black}
 } \label{tab:reggeparam}
\end{table}

In contrast, as seen in the left panel of Fig.\ref{fig:Reggesigma}, the $f_0(500)$ trajectory is not evidently linear and its $\alpha'$ is about two orders of magnitude smaller than for ordinary linear trajectories. 
 This provides strong support for a non-ordinary nature of the $\sigma$ meson. 
 In addition, the resulting  scale of tens of MeV, or at most hundreds, for  $\alpha_\sigma'$, 
 is more typical of meson physics than of quark and gluon interactions.  Moreover, the very small  slope excludes the possibility 
 that any of the other known isoscalar resonances may lie on its trajectory.  
 As a further check, when in \cite{Londergan:2013dza} an ordinary linear trajectory 
 was imposed on the $\sigma$, the
fit to the $\sigma$ pole became very poor, particularly for the coupling, and the resulting
Regge-pole amplitude in Eq.\ref{Reggeliket} in the physical region was
qualitatively very different from the data.   \color{black} The situation is very similar for the $\kappa$, which once again suggests a relatively similar composition for both resonances \cite{Pelaez:2016ntz}. \color{black}

Furthermore, in the right panel of Fig.~\ref{fig:Reggesigma}  the striking similarities
 between trajectories of the $f_0(500)$ and those of Yukawa potentials
in non-relativistic scattering \cite{Lovelace} can be noticed, particularly below $s=2\,$GeV$^2$. 
From the Yukawa G=2 curve in that plot, which lies closest the low-$s$ part of the $f_0(500)$ trajectory, it is possible to estimate $a\simeq 0.5 \,$GeV$^{-1}\simeq 0.1 {\rm fm}$, following \protect \cite{Lovelace}.  Thus it seems that the range of a Yukawa potential that would mimic the $\sigma$ trajectory at low energies is comparable,
but smaller than the $\pi\pi$ scattering length in the scalar isoscalar channel $\simeq 1.6\, $GeV$^{-1}\simeq 0.3 \,{\rm fm}$ or the 
charge radius of the pion $\langle r^2\rangle ^{1/2}\simeq 0.66\pm0.02\,$fm. 
Thus, the trajectory is similar to a meson-meson Yukawa potential
but with a characteristic length smaller than expected. 
This is not necessarily related to the size of the $\sigma$ meson,
among other things because being a non-normalizable state it does not have a well-defined size. Actually, the
recent calculation \cite{Albaladejo:2012te} of its quadratic scalar radius leads to a complex quantity $\langle r^2\rangle_s^\sigma=(0.19\pm0.02)-i(0.06\pm0.02)\, {\rm fm}^2$,
which suggests a rather compact scale as well.

Of course, these results are most accurate at low energies (thick continuous line) and the extrapolation should be interpreted cautiously. Nevertheless, the Regge trajectory found for the
$f_0(500)$, apart from explaining why the $\sigma$ is alien to the ordinary liner Regge trajectory classification,
suggests that it looks more like a low-energy resonance of a short range potential,  {\it e.g.}\ between pions, than a bound state of a confining force between quarks and gluons.

\subsection{Quarks and gluons within the $f_0(500)$}
\label{subsec:models}

The main problem to interpret the $\sigma$ as well as 
other light scalar states is that,
due to its strong character, the QCD perturbative expansion 
in powers of its coupling does not work 
at energies below 1.5 or 2 GeV. Unless using other schemes 
like the $1/N_c$ expansion or lattice-QCD, one has to 
rely on relatively crude  models. These can be very informative 
and provide some intuitive understanding, but several caveats 
have to be kept in mind. For instance these models usually start 
by assuming either a confining potential or some boundary conditions 
or a constituent mass for quarks and gluons in order to generate a 
bound state spectrum. These confining potentials
are often motivated by some QCD considerations and symmetries.
One is then tempted to talk about $\bar qq$ mesons, glueballs, 
tetraquarks, etc, and one may even 
define mixing between these states. But these confined  or bare 
states cannot be directly compared to the physical states
because the latter decay and have widths. In particular, the 
$\sigma$ has a width which 
is even larger than its mass. As we will see, the most usual way 
to deal with this fact is either to bosonize
the quark model and end up with some form of low energy effective Lagrangian or 
to introduce purely phenomenological amplitudes at best remotely based on QCD, sometimes even reduced to just one parameter,
to describe the coupling
of bound states of quark and/or gluons with the observed mesons. At this point meson-meson states or  ``meson-molecules" should be
considered together with the previous list of possible states that make up a meson, although the definition of these states is very model dependent. 
Actually, dealing  simultaneously with quarks, gluons and meson states should raise some 
concerns about double-counting of states, not
always addressed in the literature.
In any case, the most elaborated models, those really trying to obtain a somewhat realistic description, 
include rescattering effects, i.e. some sort of unitarization with the scattering coupled channel, 
which is absolutely needed if scattering observables  are to be described. 
At this point the connection with the initial QCD quarks and gluons has blurred considerably.
For instance, it is very hard to disentangle tetraquarks from meson-molecules or 
rescattering effects and sometimes this just depends on how they are defined on each work.
Moreover, the mixing pattern of bare states could be drastically changed when the states acquire a large width
and mix with another state state through an additional mechanism.
This should not be understood as a criticism, 
since all these caveats are well known and frequently emphasized 
by the very authors of the models,
whose simplifications are deliberately made for the sake of simplicity and intuition.
With these caveats in mind, some of these models have actually 
provided an intuitive understanding, starting with QCD degrees of 
freedom, of the   dynamics that give raise to 
$\sigma$ and other scalars mesons.
For this reason the most popular
and relevant models, 
which have become benchmark references in the 
literature, will be reviewed briefly in the final part of this section.

But before that, let us briefly comment on lattice-QCD calculations.

\subsubsection{Lattice QCD}

\color{black}
Many lattice studies for the $\rho(770)$, which appears in the isovector $\pi\pi$ scattering channel
exist in the literature \cite{Aoki:2007rd,Dudek:2012xn,Fahy:2014jxa,rholattice}. Actually, in Sect.\ref{subsec:qm} we have already commented how the $\rho(770)$ quark mass dependence obtained from unitarized ChPT agrees reasonably well with lattice results, given the differences between both approaches. In particular the 
very small quark mass dependence of the $g_{\rho\pi\pi}$ observed in UChPT 
was also seen on the lattice \cite{Aoki:2007rd,Dudek:2012xn,Fahy:2014jxa}.
\color{black}
 
However, the study of the $\sigma$ has some well-known difficulties on the lattice and lattice studies are more scarce.
First of all, this state has a very large width and we have already seen that meson-meson
dynamics play a very important, if not the most important, role
in the generation of the $\sigma$. 
This is not easy to implement on the lattice. 
Quenched approximations should therefore be treated with care.

In addition, lattice calculations are often made with large quark masses and
we saw in Subsec.\ref{subsec:qm} that strong non-analyticities
might be expected if the quark mass dependence is to be extrapolated from higher masses. The $\sigma$ 
at high masses could appear as a bound state or a virtual state before it becomes
the wide object that we observe at the physical pion mass.
Moreover, there is the added complication of the so called ``disconnected diagrams''. i.e. closed quark loops, which present an additional challenge to the calculations. These are important because otherwise
the states being calculated are not genuine flavor-singlet states.

In a pioneering work, Alford and Jaffe \cite{Alford:2000mm}, ignoring quark-loops and quark annihilation
found that diquark-antidiquark
 correlations were sufficient to create bound $0^+$  states with 
no exotic quantum numbers and  for sufficiently large quark masses.

A few years later the SCALAR Collaboration, 
made a full lattice QCD calculation of scalar mesons \cite{Kunihiro:2003yj}.
In this work a clear $\sigma$ state appears for which
the  inclusion of the disconnected diagrams is necessary.
It should be noted, however, that this is for relatively 
large pion mass, since $M_\pi/M_\rho\simeq 0.7-0.77-0.83$ for 
different values of the lattice parameters. Since $M_\sigma/M_\rho=1.11-1.34-1.6$, 
respectively, the $\sigma$ cannot decay into two pions, 
so that there is no estimate of its width.
However, if the naive chiral limit is taken the $\sigma$ mass becomes much smaller than the $\rho$ mass. 
Thus, although the lattice calculation is still far
from describing the physical $f_0(500)$
the results are nevertheless very encouraging.
Finally, it is worth remarking that in a very recent work
by the same group \cite{Wakayama:2014gpa}, the significance of  four-quark components 
in isoscalar mesons has also been studied, by calculating 
the propagators of ``molecular'' and tetraquark states 
as well as singly connected diagrams. Once again it is shown that within this 
framework, disconnected diagrams are
essential  for four-quark states to exist.  Their conclusion is that {\it ``the light iso-singlet scalar meson $\sigma$ may be the molecular state''}, with a mass of approximately $2M_\pi$.
Of course, these results are not obtained for the physical pion mass, but for a much a larger one, but this seems to be in 
qualitative agreement 
 with a naive
 extrapolation of the unitarized ChPT results in Subsec.\ref{subsec:qm},
whose applicability region is limited to lower pion masses.

Lattice calculations at lower pion masses exist. 
As we already commented in Subsec.\ref{subsec:qm},
for $M_\pi\simeq 325\,$MeV a bound state in $\pi\pi$ is found
in \cite{Prelovsek:2010kg}, which compares well with the extrapolation of the unitarized ChPT results. However, the authors warn about the absence of 
disconnected diagrams in their calculations.
Moreover, previous lattice
studies \cite{Mathur:2006bs}, in the quenched approximation, had also suggested the existence of a
tetraquark/mesonium one-particle state for $M_\pi\sim 180-300\,$MeV.

We have also commented in Subsec.\ref{subsec:ncUChPT}
that there are recent lattice studies for QCD in the large-$N_c$ limit
\cite{Bali:2013kia,Bali:2013fya,gunnarprivate}.
This is not exactly the same LO $1/N_c$ behavior studied 
within unitarized ChPT, but it shows that the scalar mass
in this limit is not below the $\rho$ mass. 

In addition, lattice results for other light scalars have important implications for the  $\sigma$. For instance, the SCALAR Collaboration has also performed a study of the $\kappa$
meson within a quenched lattice approach \cite{Wada:2007cp} 
in order to check whether
such a state, without the contribution of the disconnected diagrams, could have mass as small as 800 MeV. They found that it is not possible and that such a low mass $\kappa$ ``may have another unconventional structure''.
Moreover, we also commented in Subsec.\ref{subsec:qm}
that a virtual $\kappa$ state has been found in \cite{Dudek:2014qha}
 at high pion masses,
consistently with the findings of unitarized ChPT.
As we have repeatedly emphasized, the existence of the $\kappa$ 
at low energies and with a similar structure to the $\sigma$
discards by itself the $\sigma$ glueball interpretation.
Furthermore, the $\sigma$ as a glueball interpretation is 
hard to maintain in view of the lattice results, which, as we have already commented,
all place the glueball around 1.5 to 1.8 GeV \cite{latticeglueball}.

 The methods to 
extract scalar resonance parameters from lattice are developing very fast. In this sense,
the study conducted in \cite{Doring:2011vk} using the L\"usher formalism \cite{Luscher:1986pf}
and synthetic lattice data generated with the Chiral Unitary Approach, in order  to obtain scattering results in the continuum, is very illustrating. Indeed, it is found that many lattice energy levels of different volumes and with high accuracy are needed to determine with only relative precision the $f_0(500)$ pole. That work also offers a strategy to find phase shifts in the continuum from lattice levels using an auxiliary potential based on the Chiral Unitary Approach. 
Using L\"usher's formalism, lattice data was also obtained
 for isospin-2 $\pi\pi$ scattering phase shifts \cite{Dudek:2010ew}. This 
channel has less complications than that with isospin 0, due to the
double charge and the absence of the so-called "disconnected diagrams''.

\color{black}
Moreover, in a recent lattice work \cite{Howarth:2015caa}, which studied the two-pion correlator within
the quenched approximation, an $f_0(500)$-like $0^{++}$ state 
was found in the two-pion channel with a mass of $609\pm80\,$MeV, 
when  a linear extrapolation to the physical values of only their lowest quark mass results
was performed. This work included
the most important annihilation diagram, i.e. the ``partially disconnected'' or ``singly connected'' one, but  omitted the ``doubly connected'' one.  The authors considered their ``exploratory'' findings consistent
with the $f_0(500)$ meson, given the approximations of the approach.

Finally, while finishing this report, it has  appeared  the first lattice calculation \cite{Briceno:2016mjc}
of $\pi\pi$ scattering in the scalar-isoscalar channel
including all such contributions and evaluated at two different  unphysical pion masses.
For $M_\pi=391\,$MeV a bound state at $758\pm4$ is found,
qualitatively consistent with the NNLO IAM predictions commented in Sec.\ref{subsec:qm}. 
It would be interesting to know if a mirror pole in the second sheet appears in a very symmetric position or not, to study the molecular nature of this state.
For $M_\pi=236\,$MeV, the lattice phase shift
is later described with different parameterizations that 
fulfill unitarity (but without a left cut), like a K-matrix. It is found that all 
them present a pole in the second Riemann sheet of the energy-squared
complex plane, with a very large imaginary part, thus representing a very wide resonance. 
\color{black}

In summary, there are very promising results on the lattice that
find a $\sigma/f_0(500)$ and those which have addressed its nature claim that its dominant component is of a non-ordinary nature.
\color{black} 
This technique is evolving impressively fast and the last achievements are very remarkable.
Still, it would be desirable to have calculations at more pion masses, particularly as close as possible to the physical mass. This might help understanding the evolution of the $\sigma$ from a bound state at large pion masses into a wide resonance for the physical mass, as well as confirm the present dispersive determinations of $\pi\pi$ scattering. 
\color{black}
Lattice QCD 
is making a continuous and impressive progress and this is one of the directions we hope can be explored thoroughly in the near future.

\subsubsection{Tetraquarks-molecules}
\label{subsec:tetraquarkmolecules}

The title of this subsection emphasizes the difficulty in disentangling 
different configurations involving two quarks and two antiquarks at the microscopic level.
As commented above models lack a description of the hadronization process, so that
the tetraquark label just corresponds to a convenient 
way of parameterizing symmetry transformations or a particular mixing or mass hierarchy
scheme. Very often the 
authors of ``classic'' tetraquark models admit that 
their states, after hadronization and final-state meson-meson rescattering,
may also be understood as some kind of meson-meson state or molecule.

The first model proposing a non-ordinary nature for the $\sigma$
together with the $\kappa$, $a_0(980)$ and $f_0(980)$ 
was given by Jaffe already in 1977 \cite{Jaffe:1976ig} 
and this has become a reference model.
Up to that moment tetraquark states were expected to have a mass around 1400 MeV, 
just by extrapolating from the masses of 
ordinary mesons like the $\rho(770)$, with two constituents, and of baryons with three. However following a semi-classical approach to the
MIT bag model, which had been relatively successful describing ordinary mesons and baryons so far, it was shown that 
S-wave $q^2\bar q^2$  bound states appeared as a light scalar nonet below 1 GeV.

In this first model  confinement is built {\it ab initio} by assuming a constant energy density $\sim 50\, {\rm MeV/fm}^3$ inside an spherical cavity (the ``bag") defined by some boundary conditions to guarantee that only color singlet hadrons exist. 
Inside the bag non-strange quarks are massless 
and couple to gluons as in QCD. The parameters of the model were fixed by the analysis of $\bar qq$ mesons and $q^3$ baryons.
The success of the approach relies on the dominance of the magnetic contribution of the gluon interaction; without it tetraquarks would appear around 1400 MeV. The model predicts two $0^+$ nonets, one below 1 GeV, including the $\sigma$, $\kappa$, $a_0(980)$ and $f_0(980)$ 
(see Fig.\ref{fig:scalarmultiplet} for the old notation)
plus an entire nonet of pure $\bar qq$ states bound in a P-wave, sitting above 1 GeV.

It is at this point that the widths and decays have to be estimated. Note that contrary to the decay of
a $\bar qq$ meson into a pair of $\bar qq$ mesons, which require the creation of a quark pair (OZI-allowed transition), for tetraquarks to decay into two-mesons no quark-line has to be created, they simply have to fall-apart (OZI-superallowed transition). Sometimes this mechanism is also called {\it dissociation}. 
Then the decays are introduced through a ``change of coupling transformation" which consists in 
rewriting the $q^2\bar q^2$ neutral states as a linear superposition of two mesons with the same total quantum numbers
multiplied by some coefficients. These coefficients are determined up to a universal multiplicative factor and provide 
the amplitude for a meson to decay into a particular two-meson channel. Since these decays proceed by dissociation,
 these resonances become very broad as soon as their mass is well above the threshold of the two-meson state to which they couple the strongest.
This naturally describes the large width of the $\sigma$ and the $\kappa$ versus the narrowness of the $f_0(980)$ and $a_0(980)$, which lie so close to the $\bar KK$ threshold. 

One of the main concerns about this model is the proliferation of states that have not been observed, which nevertheless could be due to the fact that they are very wide. Relatively soon this problem was addressed \cite{JaffeLow} by using the $R$-matrix method \cite{Rmatrix} to relate discrete quark-model spectra calculated with boundary conditions to scattering data. The conclusion was that  quark-model eigenstates  should not  automatically be interpreted as evidence for $S$-matrix poles, but just as estimates of forces between scattering mesons. Actually
 it was recently argued, also by Jaffe in \cite{Jaffe:2007id}, that 
in the case  when the mass of the quark model $q^2\bar q^2$ state is below the dominant $\bar KK$ decay threshold,
 the tetraquark state might be interpreted as a bound $\bar KK$ state coupled to the $\pi\pi$ continuum and that there is no clear distinction between the proposed $q^2\bar q^2$ state and a meson-meson molecule.

In the literature there are different interpretations of what a tetraquark is. Not only does the tetraquark/molecule dichotomy exist, but 
also the ``diquark-antidiquark''  tetraquark configuration \cite{Maiani:2004uc}
has been discussed in the literature. In this case  a fully antisymmetrized 
quark pair, denoted by $[q_1q_2]$,
is said to form  a diquark, 
which is then combined with the antisymmetrized antidiquark to form a color singlet.
By considering all flavor configurations one 
then obtains a flavor nonet which is identified with the usual light scalar nonet.
For instance, the neutral members of the octet are:
\begin{eqnarray}
f_0=\frac{([su][\bar s \bar u]+[sd][\bar s\bar d])}{\sqrt{2}},\quad \sigma_0= [ud][\bar u\bar d], \quad
a_0=\frac{([su][\bar s \bar u]-[sd][\bar s\bar d])}{\sqrt{2}},\quad \kappa=[ud][\bar s \bar d].
\label{eq:baretetraquarks}
\end{eqnarray}
The authors of \cite{Maiani:2004uc} then assume octet symmetry breaking and write 
the most general expression for the mass in terms now of meson fields, which depends on four parameters. Then they diagonalize the
mass matrix and identify the eigenvalues with four physical masses
of the $\sigma$, $f_0(980)$, $a_0(980)$ and $\kappa$. 
The physical $\sigma$ and $f_0(980)$ come as eigenstates of the mass matrix and 
 are actually a mixture:
\begin{equation}
\vert \sigma \rangle = -\sin \phi \vert f_0\rangle+\cos \phi \vert \sigma_0\rangle,\quad
\vert f_0(980)\rangle = \cos \phi \vert f_0\rangle+\sin \phi \vert \sigma_0\rangle,
\label{eq:rotation}
\end{equation}
with $\tan 2\phi\simeq -0.19$, i.e. $\phi\simeq-5.4^\degree$. 
But this mixing angle has to be interpreted with care because so far 
these states do not have a width. 
\color{black} In particular, when a width is
included, the spacial wave function becomes non-normalizable and 
strictly speaking the above rotation between normalized states is ill defined. 
In addition, the states do not even have a well defined mass.
Thus, Eq.\ref{eq:rotation} should be interpreted as the mixing that would exist if no meson-meson decay and rescattering were present, i.e. if the $\sigma$ and $f_0(980)$ were stable.
It is true that mixing equations like Eq.\ref{eq:rotation} are 
frequently used for other pairs of particles like $\omega-\phi$, 
$\eta-\eta'$ or even $\omega-\rho$,
but compared to their masses these are much narrower than the $\sigma$, and neglecting 
their width one does not expect a dramatic change in their formation mechanism, structure or dynamics, particularly in the first two cases.
 But this is obviously a bad approximation for the $\sigma$, 
whose width is comparable to its mass. Still, that equation
could be of some interest for the $\sigma$ if the kets 
only represent the flavor part of the state, as given for instance by the well defined two-meson asymptotic states  
(i.e., neglecting the pion and kaon electroweak decay), 
stripped of the spacial part describing the resonance. 
This is a common abuse of notation, but
outside that flavor context its use is not really justified \cite{Amato:2016xjv}. 

For the above reasons, the recent LHCb claims \cite{Aaij:2013zpt} about 
the non-tetraquark nature of the sigma based on 
ratios of $B$ meson decays into 
light scalars, only hold within a very particular and unrealistic 
model with strong assumptions,
as emphasized in the very abstract of the second reference in \cite{Aaij:2013zpt}. 
In particular, the model \cite{Fleischer:2011au,Stone:2013eaa} makes use of 
Eq.\ref{eq:rotation} above. For the tetraquark case 
the model identifies  the ``bare'' $\vert f_0\rangle$ and $\vert \sigma_0\rangle$
with the definitions in Eq.\ref{eq:baretetraquarks}, and in the 
$\bar qq$ case 
$\vert f_0\rangle=\vert{\bar ss}\rangle$ and $\vert\sigma_0\rangle=\vert{\bar nn}\rangle$, where 
$\vert{\bar nn}\rangle=( \vert{\bar uu}\rangle +\vert{\bar dd}\rangle)/ 2$.
No meson-meson state is considered and therefore the large width of the $\vert\sigma_0\rangle$
and the meson-meson state are neglected in the mixing.
After this assumption, the model arrives, for instance for the case of quarkonia, to:
\begin{equation}
\frac{{\cal B} (\bar B^0\rightarrow J/\Psi f_0(980))}{{\cal B} (\bar B^0\rightarrow J/\Psi f_0(500))}
\frac{\Phi(500)}{\Phi(980)}
=\tan^2\phi,
\label{eq:cagada}
\end{equation}
where it is implicitly assumed that the dynamics to create a $\sigma$ and an $f_0(980)$ 
are the same and cancel in the ratios. But 
looking at the $\sigma$ huge with of hundreds of MeV against 
the tiny tens of MeV width of the $f_0(980)$, as well as to the difference in masses, 
this assumption does not seem realistic. 
Actually, using this kind of ratios to extract mixing angles 
or other meson properties implies two radical assumptions:
I) that neglecting the huge $\sigma$ width does not change its nature and the dynamics of its formation,
the internal structure and distribution of its components, 
and II) that these dynamics, structure, distribution of components, etc, are exactly the same for the $f_0(980)$. In the model \cite{Stone:2013eaa} used by LHCb \cite{Aaij:2013zpt},
this is done in two steps. First, in the perturbative diagrams made with quarks, all the complications of hadronization and the formation of mesons are included
in 
some $F_B^f$ form factors and some $\cal Z_f$ constants representing "the coupling amplitude that depends on the quark configuration after the $\bar B$ meson decay and the quark content of the light meson in either the $\bar qq$ or tetraquark model". Second, it is claimed that all
these form factors and amplitudes cancel in ratios in such a way that only the kinematics and the
simple flavor factors remain in their ratios of amplitudes.
Thus, these $F_B^{f_0}{\cal Z}_{f_0}/F_B^\sigma {\cal Z}_\sigma$ ratios become 1 in the $\bar qq$ case and $1/2$ for a naive tetraquark.
But for instance, decays and formation processes 
might depend on the value of the quark wave functions at the origin 
(or their derivatives, or many other things) and there is no 
reason why this effect should be the same for the $\sigma$ and $f_0(980)$. Moreover
there are effects of pion/kaon loops, which can 
contribute differently to the $\sigma$ and $f_0(980)$ and, at least for the sigma, are huge.
Something similar might happen in order to form the $\sigma$ and the $f_0(980)$.
Therefore, there should be a ratio
$F_B^{f_0}{\cal Z}_{f_0}/F_B^\sigma {\cal Z}_\sigma$, 
which {\it for each particular  model} may contain flavor factors 
in the form of sines or cosines, but apart from that, there is not even a
 reliable calculation of its order of magnitude, and assuming a particular
 value of order one is definitely not justified \cite{Amato:2016xjv}.

We already pointed out in the introduction to this section that it is a relatively 
usual weakness of some quark-level analyses to treat hadronization with a single parameter, 
common to all resonances, with no energy dependence, etc, assuming 
it does not affect much the results obtained without hadronization.

Moreover, for these branching ratios measured at LHCb \cite{Aaij:2013zpt}, 
there is the additional caveat that
 the $\sigma$ is described by means of a Breit-Wigner formalism, which as we have seen
repeatedly in this report, is incorrect for the sigma. Therefore, the separation between decays into $\sigma$ and $f_0(980)$ is strongly model dependent.
Nevertheless, if one wants to take 
those $\sigma$ and $f_0(980)$ branching ratios at face value, 
 and  only
within this crude mixing model, which neglects widths and assumes identical hadronization amplitudes for the sigma and the $f_0(980)$, the tetraquark nature is strongly disfavored. Surprisingly, a recent reevaluation of this model including 
isospin breaking, which one would naively expect to be a minor correction,  reaches the opposite conclusion \cite{Close:2015rza}, favoring the tetraquark structure.
Hence, even ignoring the above caveats it is hard to reach a robust conclusion.
In any case, the main problem is that there is no systematic way to include 
meson-meson states and the $\sigma$ width into these naive mixing pictures, since the 
naive mixing equations at the very beginning would be ill defined and 
the simple relation between 
the measured branching fractions and the nature of light scalar states would be lost.
Actually, the ratio of hadronization amplitudes in Eq.\ref{eq:cagada} most likely has a complicated energy dependence given the different shapes of the two resonances and is in practice unknown. 
These are fundamental caveats to the use of $B_0$ decays in order to discern the nature of the $\sigma$ and $f_0(980)$ in a model independent way, by means of naive
mixing schemes. To conclude, the use of branching ratios as in Eq.\ref{eq:cagada} does not  provide any meaningful information about the physical $\sigma$ composition.

A model independent dispersive formalism has recently become
 available for the analysis of the $\bar B_{d/s}^0 \rightarrow J/\psi \pi \pi$ decays \cite{Daub:2015xja}. Being model independent, this analysis avoids Breit-Wigner parameterizations, does 
not discuss any naive mixing scheme of the form of Eq.\ref{eq:rotation} nor the
interpretation of the nature of scalar states. However, it is worth remarking that the hadronization form factor of the $\bar ss$ into the 
two pions (the ones resonating into a $\sigma$ or an $f_0(980)$ meson), is much more elaborated than just a constant. 
\color{black}

A strong diquark-antidiquark component mixed with conventional $\bar qq$ configurations were
also found for several scalar states within a non-relativistic quark model in  \cite{Vijande:2005jd}.
In this case the $\sigma$ is found to be predominantly a $\bar qq$, but this model only calculates masses of bound states within a Schr\"odinger equation which does not contain two-meson states and rescattering so that this $\sigma$ does not have a width,  which is an essential for the understanding of the physical $\sigma$. The model is well suited though for heavy quark mesons and when 
some results are rescaled for narrow states like the $f_0(980)$ and $a_0(980)$ it compares well with experimental data on two-photon decays.
No attempt of such a calculation for the $\sigma$ is done.

Actually, following the general pattern discussed in the introduction to 
this section, a phenomenological parameter 
is usually needed to connect bound or bare states with the physical ones. This is particularly important for the $\sigma$ given its very large width.
Such a parameter is included in the model \cite{Maiani:2004uc} in order to connect with hadron physics.
In this case the ``fall apart'' mechanism is not at work,
but here dissociation is understood as the process where 
a $\bar qq$ pair is switched between the diquark and the antidiquark to form two colorless $\bar qq$ mesons. In the exact
SU(3) limit there is only one such amplitude, which is then approximated by a single parameter. A convenient choice is $A=t_{a^+\rightarrow\bar K^0K^+}$. 
Therefore, we once again meet this usual feature of quark-level analysis, 
which is that hadronization is frequently described in a simple way by a single parameter.
In this case it is explicitly
assumed that this decay mechanism does not change the previously calculated masses. 
Then, 
``without attempting a systematic fit''
a very crude description of the scalar widths is tried.
However, the $\sigma$ width comes out as $325\pm50$, which lies outside
the RPP2012 very conservative estimate and  even farther from the conservative dispersive estimate in Eq.\ref{myestimate}.
In addition, the resulting $\kappa$ width is 138 MeV, to be compared with $557\pm24\,$MeV 
from the rigorous Roy-Steiner dispersive analysis in \cite{DescotesGenon:2006uk}. 
Of course, we are comparing here with pole parameters
which do not correspond precisely to the mass and widths quoted in \cite{Maiani:2004uc}.  
By adding other SU(3)-allowed phenomenological couplings some improvement can be found
for OZI allowed decays, but the $\kappa$ comes 
out even narrower, $\sim 60$ MeV,
the $f_0(980)\rightarrow\pi\pi$  
coupling is too small and the $a_0(980)\rightarrow\eta\pi$ too large.

Thus, in order to solve these drawbacks, 
in a subsequent work \cite{Hooft:2008we} this model was improved by considering 
an instanton-induced, six-fermion effective Lagrangian. 
The light  scalars are still predominantly described as diquark-antidiquark configurations with a phenomenological transition parameter allowing their decay into two mesons. 
The addition of the new interaction, apart from providing an
additional amplitude to bring the $f_0(980)$ and $a_0(980)$
decays into a better agreement with data,
also  produces a mixing 
between the 
``tetraquarks''  and a nonet of $\bar qq$ mesons identified around 1200-1700 MeV.

An introduction to instantons is far beyond the scope of this report and for that purpose \cite{Coleman} is highly recommended. 
Suffices it to say that instantons are solutions 
of the Euclidean gauge-field equations,  corresponding to
stationary points of the action\footnote{For fixed winding number. See \cite{Coleman} for details.}. Thus, they dominate the functional integral in
the path integral formalism and for the semi-classical approximation they are 
the most relevant kind of solutions. 
As it is argued in  \cite{Hooft:2008we}, QCD instantons generate an effective interaction of the form ${\cal L}_I\sim det(\bar q^i_L q^j_R)$, where $i,j$ are flavor indices and summation over color is understood. With three light flavors, this includes a 
diquark-antidiquark-$\bar qq$ mixing
term of the type Tr$([\bar q\bar q]_i[qq]_j \bar q_j q_i)$.
If the diquark-antidiquark mesons are collected into an $U(3)$ $S$ matrix
and the $\bar qq$ mesons into their corresponding $S'$ matrix,
this mixing can be recast into the form of Eq.\ref{ec:instantonmixing} already 
introduced when discussing 
the chiral Lagrangian approach of the Syracuse group \cite{Black:1999yz} in Subsec.\ref{subsec:Syracuse}.
This provides more flexibility in the fits through an additional parameter,
but it also requires a consistent identification of 
$\bar qq$ mesons above 1 GeV. 
Within this improved model \cite{Hooft:2008we}, 
the new numerical analysis of masses and decays leads again to a 
relatively small $\sigma-f$ mixing angle $\vert \phi\vert < 5^\degree$,
and the problems with the $a_0(980)$ and $f_0(980)$ decays are fixed. However, in their best fit
the $\sigma$ and $\kappa$ decay amplitudes  are still close to a factor of 2 too small. Thus, this model gives just a qualitative picture 
for the two lightest scalar resonances, 
but of course, one has to take into account the simplicity
of the approach. 
The model is nevertheless important to confirm 
the relevance of the mixing with the ordinary nonet above 1 GeV.

Although discussed in more detail in section in Subsec.\ref{subsec:other}
it is worth mentioning here that the models using chiral Lagrangian formalisms, including those of the Syracuse group, also 
indicate that in order to understand the masses and decays scalar mesons below 2 GeV,
a two-nonet scenario, plus possibly a glueball state, seems unavoidable. The preferred solutions
suggest that the main component of the states of the lowest nonet involve two-quarks and two-antiquarks, 
although their specific arrangement is unclear.
In addition, the unitarized ChPT approach in coupled channels, explained in detail in Subsec.\ref{subsec:coupleduchpt}, also supports strongly 
the existence of two nonets below 2 GeV, the lightest one being dynamically generated, which means that 
the two-meson loops play the most relevant role in its formation. In contrast, the nonet above 1 GeV would 
have a predominantly ordinary nature.

Consistently with the latter approach, the $f_0(980)$ and $a_0(980)$ 
can be very naturally understood as meson-meson molecules \cite{isgur1}, although they have a small decay width due to
the existence of another open state to which they can also decay. 
This is one of the possible pictures we have in mind when talking about states involving two quarks and two antiquarks.
This molecular scenario has been explored in the literature \cite{isgur1,Baru:2003qq,mol} 
and gives a very successful account of the two-meson and electromagnetic decays of these resonances. 
In addition, it has been observed in the chiral unitary approach \cite{Oller:1997ti} or the Krakow-Paris model \cite{Kaminski:1998ns}, which describe very well the existing 
$\pi\pi$ and $\bar KK$ scattering data, that if the $\pi\pi$ channel is decoupled from $\bar KK$, the $f_0(980)$ becomes a true
bound state below the $\bar KK$ threshold.
Of course, the $f_0(500)$ pole does not lie below $\pi\pi$ threshold nor even very close to it.
For these reasons, the  ``molecule" name is less adequate, but the meson-meson loops definitely play a big role in its existence and in 
the values taken by its mass and width, consistently with the $f_0(980)$ and $a_0(980)$ molecular interpretation. 
Moreover, we have seen in Subsec.\ref{subsec:qm} when discussing the quark mass dependence of the $f_0(500)$ within unitarized ChPT,
that for sufficiently high quark mass it becomes a virtual state, but that if that behavior is extrapolated to higher masses,
the $\sigma$ ends up becoming a molecular state.

By molecular state one often means ``composite'' or ``made of other mesons'', and therefore sometimes 
the non-molecular tetraquark is called ``elementary'', i.e. not made of other mesons. 
Obviously all them are composite objects made of quarks and gluons.
Frequently ``molecules'' are identified with extended objects, 
larger than the typical hadronic size, whereas 
the word ``tetraquark'' is kept for ``compact'' states 
of typical hadronic size, like the ordinary $q\bar{q}$ mesons. 
Unfortunately size is only well defined for stable states, not resonances.
In particular, we already commented that 
for the sigma the classical mean square scalar radius calculation yields
 a complex number $\langle r^2\rangle_s^\sigma=(0.19\pm0.02)-i(0.06\pm0.02)\, {\rm fm}^2$ and, 
 if it could be used as a measure of the scales involved
in the formation and decay  of the sigma, comes out rather small \cite{Albaladejo:2012te}. We also saw in the previous subsection that 
the pion-pion Yukawa potential that would mimic the Regge trajectory calculated from the $\sigma$ pole position and residue
also has a rather small scale of the order of 0.1 fm. Since this scales are smaller than the pion-pion scattering length
they suggest that there is some dynamical effect at work in a scale intermediate between the pure meson physics and the pure quark-level dynamics.

For a bound state strongly coupled to a single scattering channel,
as in the case of deuterium, Weinberg \cite{weinbergdeuterium} 
developed a model-independent criterion to determine whether
it was closer to a composite state of the scattering particles, or if it was
a state that would exist even in the absence of the scattering process.
The two pure-elementary or pure-molecular states
are just the extreme cases of a continuous of configurations
between them. 
 A considerable amount of work has been devoted to extend this 
model-independent criterion to resonances \cite{Baru:2003qq,YamagataSekihara:2010pj}, but sooner or later there is a requirement of vicinity to the threshold that the $f_0(500)$ does not fulfill. \color{black} A very recent generalization of this criterion, using a particular definition of compositeness, suggests that the $f_0(500)$ may have a
 40\% of "two-body compositeness" \cite{Guo:2015daa}, which in other models would be identified with a "molecular" component. The nature of the remaining elementary component cannot be addressed with this criterion and could still be any mixture of compact tetraquarks, quarkonia or even gluonia.\color{black}

Of course, it is  likely that the predominant composition of the $\sigma$ might be a superposition of different possible arrangements of two quarks and two antiquarks, plus some subdominant  $\bar qq$ or even glueball components. Let us remark that that in Subsec.\ref{subsec:twononets} we have already discussed proposals
where all these arrangements also respond to a different spacial distribution. However, given that
 the present experimental data on the $\sigma$
does not seem enough to resolve the composition or mixings, the possibility to disentangle an spacial distribution within the $\sigma$ looks very far fetched.

As a final comment, one might be worried that conventional tetraquarks 
have been recently shown to depend on $N_c$ in the same way as ordinary mesons,
which plays against them being the main component of the $f_0(500)$. However, as repeatedly stated, that counting
only applies in the large $N_c$ limit and only to the most straightforward and intuitive tetraquark definition in the large $N_c$ limit.
There are other arrangements, like the so-called polyquark or the molecule, that do not become narrow at large $N_c$.

\subsubsection{Unitarized quark-models}
\label{subsec:uqm}

Another interesting  model of light scalar mesons was proposed already in 1986 \cite{vanBeveren:1986ea} 
and has been refined or reformulated over the years \cite{EffRuppgeneral}. We will briefly sketch
it here following its application to discuss the nature of light scalars in \cite{vanBeveren:2006ua}.
It should be remarked that the very authors of \cite{vanBeveren:2006ua} state that 
``The model is undoubtedly an over-simplification; this is deliberate, so as to expose the essential features''. 
Nevertheless, despite not being aimed at precision, the model provides an acceptable phenomenological description of data
and captures the different dynamical origins of two multiplets of mesons: a standard one and a non-standard one. 
As already discussed, relatively similar conclusions were also obtained from the $1/N_c$ expansion of unitarized ChPT.

The model considers relativistic Schr\"odinger equations matched on both sides of a sphere of radius $r_0$. 
Inside the sphere, there is a confining potential for $\bar qq$ pairs, which is basically 
described by a flavor-blind harmonic oscillator. This choice is not crucial and any other confinement spectrum could be used \cite{EffRuppgeneral}.
Actually most of the attention is paid to the ground states and most potentials around the ground state
are well approximated to first order by a harmonic oscillator.
The solutions are joined at $r_0$ with plane waves for meson pairs. There is a universal coupling $\lambda$ between
$\bar qq$ pairs. The solution of a Schr\"odinger equation outside the sphere is recast in terms of  a multichannel $T$-matrix
which, due to the simplicity of the model can be written in closed form (see \cite{vanBeveren:2006ua} for details).
\color{black} The transition radius $r_0$ between the $\bar{q} q$   
and meson-meson sectors aims at modeling string breaking at a sharp
   distance, as approximately observed on the lattice \cite{Bali:2005fu}\color{black}.
Although there is no explicit inclusion of chiral symmetry in the potentials, the use of relativistic reduced masses for the
two-meson channels leads to zeros in the amplitudes near the position of the actual Adler zeros (for instance at $s=0$ for $\pi\pi$ scattering, which would be the Adler zero position in the chiral limit). The parameters $r_0$ and $\lambda$ are determined from fits to data, including
$\pi\pi$ and $K \pi$ scattering phase shifts, $K_{e4}$ and $J/\psi\rightarrow\omega \pi\pi$ decays, the scattering length from ChPT in \cite{CGL}, as well as $\phi\rightarrow\gamma a_0(980)$ and $\bar pp\rightarrow \omega a_0(980)$ line shapes.

Once the data has been fitted and the parameters have been determined it is possible to look for poles
in the $T$ matrix for different quantum numbers. It is found that the model is able 
to generate poles for the $\sigma$, $\kappa$, $f_0(980)$ and $a_0(980)$,
i.e. the whole light scalar nonet. In particular, the $\sigma$ pole is located 
in the range $(476-628)-i(226-346)\,$MeV, which has a considerably good overlap with the 
RPP 2012 revised estimate given in Eq.\ref{rpp2012sigmapole}, 
although it does not overlap by very little with the Conservative Dispersive Estimate suggested 
in this report in Eq.\ref{myestimate}. Therefore the $\sigma$ pole is quite acceptable, keeping in mind
the deliberate simplifications of the model.

The interesting features of this model are that, on the one hand, and as remarked by the authors,
{\it ``resonance formation arises from rescattering processes..., i.e, unitarization effects''},
thus bearing some resemblance to the results already reviewed on unitarized ChPT approaches in Sec.\ref{sec:chiralsigma}.
On the other hand, there is no one-to-one correspondence between the full-model spectrum
and the ``bare'' spectrum of their inner  confining potential. Actually, for each scalar ''bare'' $\bar qq$ 
ground-state, two resonances are produced by the $\lambda$ coupling to the continuum. Moreover
the lightest one turns out to be a ``non-regular'' meson, whereas the heavier one, appearing around 1.3-1.5 GeV
is a ``regular'' one. This was a remarkable piece of evidence for the two-nonet picture,
when this feature was first observed in \cite{vanBeveren:1986ea}.

Within this approach, a ``regular'' meson is defined as having a resonance width proportional to the square of its
coupling.
In this model, the $\sigma$ and the lightest scalar nonet do not show this behavior. 
This is illustrated in Table~\ref{tab:Effpoles} where the pole positions of all light scalars are shown
for different values of the $\bar qq$-meson-meson coupling $\lambda$. The preferred value of the coupling is
$\lambda\sim3\,{\rm GeV}^{-3/2}$. The interesting observation is that {\it if the coupling $\lambda$ is made weaker, the $\sigma$ mass
and width grow}. Therefore, it seems that the ``bare'' $\bar qq$ seed inside the sigma is heavier than 1 GeV.
However, as already commented, the physical $\sigma$-meson is dominated by rescattering effects, which make
the pole appear at a much lower mass.

\begin{table}[htb]
\begin {center}
\small
\begin{tabular}{r|cccc}
\hline\\[-4mm]
$\lambda$ & $\sigma$ & $\kappa$ & $f_0(980)$ & $a_0(980)$\\\hline\\[-4mm]
 1.5 &  942 - i794 &     ---     &     ---     &     ---      \\ 
 2.0 &  798 - i507 &     ---     &     ---     &     ---      \\ 
 2.2 &  738 - i429 &  791 - i545 &     ---     & 1081 - i8.0  \\ 
 2.4 &  682 - i368 &  778 - i472 &     ---     & 1051 - i25   \\ 
 2.6 &  633 - i319 &  766 - i409 & 1041 - i13  & 1024 - i45   \\ 
 2.8 &  589 - i278 &  754 - i355 & 1028 - i26  &  998 - i61   \\ 
 3.0 &  549 - i243 &  743 - i309 & 1015 - i35  &  978 - i60   \\ 
 3.5 &  468 - i174 &  717 - i219 &  976 - i37  &  896 - i142   \\ 
 4.0 &  404 - i123 &  693 - i155 &  948 - i38  &  802 - i103  \\ 
 5.0 &  308 - i50  &  651 - i69  &  889 - i34  &  711 - i40   \\ 
 7.5 &  216 + i0   &  610 + i0   &  752 - i25  &  632 + i0    \\
10.0 &  142 + i0   &  560 + i0   &  633 - i17  &  577 + i0    \\[-5mm]
\end{tabular}
\end {center}
\caption{Movement of the $\sigma$, $\kappa$, $f_0(980)$ and $a_0(980)$ 
poles as the coupling constant $\lambda$ is varied in the unitarized quark-model of \cite{vanBeveren:2006ua}. 
Bound states are
indicated by ``+i0''. Units are MeV for the poles and GeV$^{-3/2}$ for
$\lambda$. Reprinted from  E.~van Beveren, D.~V.~Bugg, F.~Kleefeld and G.~Rupp, ``The Nature of sigma, kappa, a(0)(980) and f(0)(980),''
Phys.\ Lett.\ B {\bf 641}, 265 (2006). Copyright 2006, with permission from Elsevier.}
\label{tab:Effpoles}
\end{table}

In previous sections we have seen that the $\sigma$ pole also describes a trajectory in the complex plane as $N_c$ is changed
within unitarized ChPT. As a matter of fact, increasing $N_c$ 
amounts to make QCD weaker, since its coupling scales as $g/\sqrt{N_c}$. In particular, increasing $N_c$
suppresses meson loops, i.e. rescattering, faster than other contributions to the meson-meson amplitudes.  
Now, when $N_c$ is increased not too far from its physical value of 3, the $\sigma$ is seen
to increase its pole mass and its pole width, which is a very similar effect to the one described here.
For larger $N_c$ the behavior is more uncertain, although the most favored behavior still has an increasing 
mass but a decreasing with beyond some given $N_c$. Of course, 
the $\lambda$ coupling is hard to relate to specific QCD parameters like $N_c$. But it seems that
at least around their physical values both approaches seem to suggest a similar predominantly non-$\bar qq$ behavior.

Further interest on this model comes from other applications to hadron physics, and particularly from its success describing
some charmed mesons, which are hard to fit in simple quark models, as ``cousins'' 
of light scalar mesons \cite{vanBeveren:2003kd}.

A similar ``unitarized'' quark-model was also proposed by Tornqvist in \cite{Tornqvist:1995ay},
in which fits to $\pi\pi$ and $K\pi$ s-wave scattering data as well as to the $a_0(980)$ resonance peak 
were performed. In this model a scalar nonet was included {\it a priori} to represent the quark-level ``bare'' states 
which are also mixed
with the meson-meson open channels. 
In addition, the model includes all
two-light-pseudoscalar thresholds, constraints from Adler zeros and flavor symmetric
couplings. Unitarity and some basic analyticity 
properties were implemented through a formalism closely related to the K-matrix,
or to a very deformed Breit-Wigner parameterization.
A similar ``doubling''  of the poles that 
took place in the previous model also   occurs here, 
namely, when the effective coupling with the continuum 
becomes large enough, twice as many poles can appear in the physical spectrum, compared to the number of ``bare'' poles originally put in.
At first \cite{Tornqvist:1995kr}, the $\sigma$ was missing because poles deep in the complex plane were not searched for.   The $\sigma$ 
pole within this model was later found in \cite{Tornqvist:1995ay} 
at $\sqrt{s_{\sigma}}\sim 470-i 250\,$MeV, which is consistent  with 
the present RPP2012 conservative estimate and even the conservative dispersive estimate 
proposed in Eq.\ref{myestimate}. In the words of the author \cite{Close:2002zu}, ``The new poles can then be interpreted as being mainly 
meson-meson bound states, but mixed with the states which are put in''.

However, it should be noticed that within this model a $\kappa$ meson was not found in $K\pi$ scattering. Actually, in a later work \cite{Tornqvist:1999tn} within the $SU(3)$ L$\sigma$M, that we briefly commented
 in Sec.\ref{subsec:ulsm}, the author identified the $K^*(1430)$ as the member of the lightest nonet.
 This is because the  $K\pi$ scattering pole in the s-wave was obtained around 1100 MeV. 
 However since all other light meson masses, except the $a_0(980)$ 
 came out about 200 MeV too high, it would have been even more natural to identify it with the $\kappa$. 
Unfortunately at that time the $\kappa$ was searched around 900 MeV, 
and the work in \cite{Cherry:2000ut} was understood \cite{Close:2002zu}
as a strong argument against a $\kappa$ resonance, when, as we commented in the introduction, 
it was only ruling out a $\kappa$ at 900 MeV, but not its existence below, which is not disputed in \cite{Cherry:2000ut}. 
In a 2002 review \cite{Close:2002zu} Tornqvist and Close actually 
claimed growing evidence for two nonets, where the lightest one would be made of the $\sigma$, $a_0(980)$, $f_0(980)$, and a $\kappa$, although its mass was left undetermined.
In that review they actually suggested that the tetraquark and molecule configuration may actually coexist
 within light scalars, the former predominantly in the inner part and the latter in outer layers. This has already been discussed in sufficient  detail in  Subsec.\ref{subsec:tetraquarkmolecules}.
 
 Thus, unitarized quark-models once again provide further support for the two-nonet picture of light scalars, with the lightest one having a predominantly non-ordinary nature.

\color{black}
\subsubsection{Sum rules}

Before commenting on the different results in the literature let us very briefly sketch  some concepts
of the sum rule formalism. For a general introduction we recommend the textbook \cite{bookNarison} or the seminal reference \cite{Shifman:1978bx}. The approach is very general and here we will only concentrate on works related to the $\sigma$ meson.
Generically, in this approach one extracts hadron masses and couplings from two-point correlator functions of 
hadronic currents $J_h(x)$, defined as:
\begin{equation}
  \label{eq:2point}
  \Pi_h(q^2)\equiv\int d^4x e^{iqx}\langle 0|{\cal T} J_h(x) J_h(0) | 0 \rangle,
\end{equation}
where the currents are chosen with the desired numbers to be studied, which in our case are those of the sigma.
Namely, the vacuum numbers. On the one hand, these correlators can be calculated within QCD by means
of the operator product expansion (OPE)
\begin{equation}
  \label{eq:ope}
  \Pi^{OPE}_h(q^2)\simeq \sum_{D=0,2,4...}\frac{1}{-q^2}\sum_{{\rm dim}\,{\cal O}=D} C_{\cal O}(q^2,\Lambda) \langle {\cal O}(\Lambda)\rangle,
\end{equation}
where $\Lambda$ is an arbitrary  scale (of the order of 1 or 2 GeV) that separates the low and high-energy regimes,  $C_{\cal O}$ are the so-called Wilson coefficients, calculable within perturbative QCD, 
and $\langle {\cal O}(\Lambda)\rangle$ are the condensates of operators ${\cal O}$ of dimension $D$, 
made of gluons and quarks, which encode the non-perturbative QCD contributions. Hence, the first
critical issue to be considered is the truncation order of the OPE. 
Moreover, although the values of the lowest condensates have been determined from phenomenology, those of higher order are customarily obtained from the lowest condensates
using the factorization ansatz, which, strictly speaking is only valid in certain limits, like the large-$N_c$ limit \cite{GomezNicola:2010tb}.

On the other hand, such a calculation can be
recast in terms of  a phenomenological representation, which is done by means of the
following dispersion relation:
\begin{equation}
  \Pi_h(q^2)= \frac{1}{\pi}\int_0^\infty ds \frac{\im \Pi_h(s)}{s-q^2-i\epsilon} \equiv 
\int_0^\infty ds \frac{\rho_h(s)}{s-q^2-i\epsilon},
\end{equation}
where we have introduced the ``spectral function'' $\rho_h(s)=\im \Pi_h/\pi$.
Note that the above dispersion relation is unsubtracted but subtractions can also be considered.
Thus, the second critical issue regarding
 sum rules is the phenomenological representation of the spectral function.
Customarily it is approximated by the ``duality'' ansatz, in which the lowest resonance
that couples to those currents is represented as a pole that dominates the low energy region below some energy $s_0$, above which the ``QCD-continuum'' obtained from the OPE is again used. Namely:
\begin{equation}
  \label{eq:dualityansatz}
  \rho_h(s)= f^2\delta(s-M^2)+\theta(s-s_0)\frac{1}{\pi}\Pi_h^{OPE}(s),
\end{equation}
where $M$ and $f$ are the mass and coupling constant (or residue) of the lightest resonance that couples to that specific current. Sometimes the residue is normalized differently by extracting an additional factor of $M^2$ in front of the delta function.
In principle heavier resonances could also be described explicitly by delta functions and setting the
onset of the continuum at higher energies. Let us remark that representing the coupling of the current to a resonance by a delta function is nothing but its narrow width limit and this is an important caveat
for the use of sum rules with a resonance as wide as the sigma. We will see below that this issue 
has been addressed in the literature.

Having the OPE correlator both in the theoretical and phenomenological sides is very inconvenient. 
This is why the final ingredient in the sum rule approach is to perform a Borel transformation
to suppress the continuum contribution, arriving to the following sum rule equation:
\begin{equation}
  \label{eq:SRBorel}
  f^2 e^{-M^2/M_B^2}=\int_0^{s_0}ds \, \rho_{OPE}e^{-s/M_B^2},
\end{equation}
where $M_B$ is an arbitrary parameter called ``Borel mass'', which is conveniently chosen in the region 
where the parameters of the analysis become reasonably stable, i.e., the ``Borel window''.

Within the SR approach the $\sigma$ meson was studied in 2001 in \cite{Narison:2000dh} by using the 
scalar-anomaly current. For our purposes it is enough
to remark that this current is made of glueball and $\bar qq$ operators, so that the analysis was based on
gluonia and quarkonia components. Both the narrow width and factorization hypothesis were used \color{black} either in a Borel transform (there called a Laplace transform 
because it is actually the inverse Laplace transform) or sometimes also considering
subtractions for the energy suppression. \color{black}
The results of this work showed that {\it ``unmixed scalar quarkonia ground states are not wide, which excludes the interpretation of the $\sigma$ as a $\bar qq$''}. 
 While preparing this review, an additional work \cite{Afonin:2016hia}
 on sum rules and the sigma has obtained
that the mass of the lightest $\bar qq$ scalar resonance has a lower and upper bound of
$0.78\,{\rm GeV}< M_\sigma<1.28\,{\rm GeV}$. These bounds are obtained from dimension 4 and 6 operators although the main contribution stems from the latter. The author states that the "analysis confirms a widespread idea that the $f_0(500)$-meson represents an exotic state". 

Concerning unmixed gluonia, in \cite{Narison:2000dh}
it was pointed out that 
for a sigma mass around 500 MeV the gluonia width should be relatively narrow, i.e., less that 100 MeV (very consistent with the expected $1/N_c^2$ suppression of glueball decay into two pions). 
This work then questioned the light sigma interpretation
 of the scattering data, since it could not be accommodated within the $\bar qq$ nor the glueball interpretation. Actually, since in 2011
the PDG still allowed for a 400 to 1200 MeV sigma, the glueball interpretation
with  a 1 GeV mass and 0.8 GeV width for the $\sigma$
was fairly reasonable. However, throughout this report we have reviewed
how we finally know that the $\sigma$ resonance
mass lies around 500 MeV and has a roughly 550 MeV width. 
Therefore, as of today, the conclusion to be drawn of the
sum rule study in \cite{Narison:2000dh} is that the sigma 
cannot be an unmixed gluonia nor $\bar qq$. Actually, 
given the gluonia very small width for a 500 MeV mass, 
not even a dominant glueball component seems 
likely.
To this, of course, we have to add the facts repeatedly discussed in previous sections, that the glueball appears well above 1 GeV both in quenched and unquenched lattice calculations and that the $\kappa$ resonance, very similar to the $\sigma$ but with strangeness,
has been firmly established with dispersion relations.
In short, when reading 
the sum rule analysis in \cite{Narison:2000dh} with our present knowledge about the sigma mass and width,  one must conclude that it strongly disfavors that the $f_0(500)$ may have a predominant quarkonia or gluonia nature.

Therefore, once more one is led to study tetraquark structures.
This was actually carried out by the end of 2004 in \cite{Brito:2004tv}, by following \cite{Maiani:2004uc} and
considering different diquark-antidiquark currents with the quantum numbers of the whole light scalar nonet, including of course, a $\sigma(500)$. The ``pole plus continuum'' approximation, Eq.\ref{eq:dualityansatz}, was used and the OPE was considered up to dimension six. The decay constants were then determined
with the sum rules from the values of the meson masses. In particular for a 500 MeV $\sigma$ mass it was found that $g_{\sigma\pi\pi}=(3.1\pm0.5)\,\gev$, in remarkably
 good agreement with the ``conservative dispersive estimate'' 
given in Eq.\ref{myfullestimate}.
 
However, following indications \cite{Lee:2005ny} that for multiquark states potentially important contributions to the sum rules could arise from operators with dimension $D>6$, the sum rules for
diquark-antidiquak of \cite{Brito:2004tv} were revisited in 2006 \cite{Lee:2005hs} with operators up to $D=8$, assuming factorization. The effect of those operators lead to the ``destruction'' of the sum rule and 
no evidence for the coupling of such tetraquark structure to the light scalar nonet, although the possibility that this could change with other interpolating diquark currents was also suggested. Actually, an alternative current structure, called ``instanton'' current, was studied up to $D=10$ in \cite{Lee:2006vk} and shown not to spoil the sum rule, giving support for a $\sigma$ around 780 MeV and a smaller residue than in \cite{Brito:2004tv}, so that the agreement with the present knowledge is somewhat worse. It was nevertheless pointed out that mixing with other states may change this result. As usual this result made use of the ``pole plus continuum'' ansatz, but the effect of two-pion intermediate states was included in a later work \cite{Lee:2007mva}, showing that the resulting $\sigma$ state had a very strong coupling to the two-pion state.

The study of all possible combinations of tetraquark currents was 
started by the Beijing-Osaka group in \cite{Chen:2006zh}, where it was shown that five independent local tetraquark currents exist.
Studying all possible combinations of these currents is very relevant because it allows to optimize the linear combination to have  a good Borel window and other relevant properties. 
The study was carried out first with linear combinations
of only two currents and to dimension 8 in the OPE. 
In addition, the general criticism about the use of delta functions to represent resonances was addressed. By including a finite Gaussian width in the resonance representation 
the results did not change much. Actually, 
it was found that there was a significant component of the scalar meson coupled to the tetraquark current without going through two-mesons, thus justifying the use of the narrow width approximation. In this work the masses of all members of the light scalar multiplet came about right (600 MeV for the $\sigma$) and following the mass hierarchy $M_\sigma<M_\kappa<M_{f_0},M_{a_0}$, giving support for the tetraquark structure of these states. Moreover, when following the same approach with a $\bar qq$ current, up to dimension six, the resulting scalar masses came out, once again, larger than 1 GeV, thus disfavoring the 
dominant $\bar qq$-component interpretation of light mesons.

In 2009  this SR approach \cite{Chen:2009gs} was completed by considering linear combinations of all the currents, thus exploring the whole space of local tetraquark currents. Interestingly, it was pointed out 
that, although in \cite{Chen:2006zh} the current basis was expressed
in terms of diquark-antidiquark currents, one could have chosen 
a basis of meson-meson currents, which is equivalent 
and expands the same space of local tetraquark currents. 
This implies that for generic
 linear combinations  one cannot distinguish whether the scalars are diquark-antidiquark or meson-meson molecules. However, the tetraquark currents with a single term do not lead to good results meaning that the $\sigma$ ``probably has a complicated structure''. This is the usual problem we have already found with other approaches, where by ``tetraquark'' different kinds of structures are meant, possibly mixed between them. In addition, the $\pi\pi$ continuum contribution was found necessary to obtain a good sum rule signal. The final mass of the
$\sigma$ within this very general scheme was $(530\pm40)\,\mev$, 
well within the present PDG range, but somewhat higher than the conservative dispersive estimate, although one has to take into account that this ``mass'' determination does not necessarily correspond to a pole mass.

Finally, in \cite{Kojo:2008hk} the singlet and octet scalar-isoscalar tetraquark currents were 
studied separately and up to dimension $D=12$ in the OPE, finding that they have masses around 700-850 MeV and  600-750 MeV respectively.
If the sigma is assumed to correspond to the ideally mixed state with just up and down quarks,
its mass comes around 600-800 MeV. Interestingly, in this work the effect of a large width in 
the determination of resonance parameters within the sum rule approach was studied, by going beyond the pole approximation for the spectral function. It was shown that there is a sizable effect in the size and location of the Borel stability window, but that the difference in the mass determination is of the order of 20\% between considering a zero width or a 400 MeV width for a Breit-Wigner like resonance. This shows the robustness of the sum rule prediction of a sigma mass below 800 MeV from tetraquark currents, suggesting, once more, that ``the $\sigma$ meson has largely four-quark components''.

In summary, as it happened with previous formalisms,  sum rule approaches show that the quarkonia interpretation of the $f_0(500)$ is
strongly disfavored. Once again, such ordinary $\bar qq$ states appear above 1 GeV. 
The glueball interpretation might have been acceptable if the $\sigma$ lied around 1 GeV, or above, but 
it is also strongly disfavored by our present knowledge that its mass is around 
500 MeV.  The sum-rule description in terms of tetraquark currents yields sigma mesons below 1 GeV
(as well as the other light scalars) 
with values closer to 500-600 MeV and reasonable couplings to two-pions. At this point there are no studies of tetraquark-$\bar qq$ mixings within the sigma, although some sum-rule
studies have been performed for non-singlet light scalars  \cite{Sugiyama:2007sg}.

\subsubsection{Schwinger-Dyson/Bethe-Salpeter approach}
\label{subsec:SDBS}

Schwinger-Dyson equations are just the full Quantum Field Theory equations for 
the Green or correlation functions of different operators. They are derived within the functional formalism from the vanishing of total derivatives of currents associated to 
symmetries or conserved quantities. They can also be interpreted as resummations of the perturbative series, if these were to converge. Their solution does not correspond to the systematic perturbative series expansion and usually involves some kind of truncation, although respecting the main symmetries. Typically the equations are formulated in a given gauge and with some phenomenologically sound ansatz for the non-perturbative vertices of the theory. 
The first approximation is called the ``rainbow-ladder'' (RL), in reference to the kind of diagrams it contains, which visually resemble a ladder.
While the Dyson-Schwinger equations generally couple Green functions with different number of particles, the Bethe-Salpeter equations instead organize each n-body Green function in terms of its irreducible part (i.e., interaction) and the free propagation of the n particles. They are widely used near a pole of the Green function to obtain the coupling of that resonance or bound state to its n constituents, the resulting equation being similar to a Schr\"odinger one in four dimensions. Since a description of this approach is rather technical and very far from our scope,
we just refer the reader to \cite{SDBS-intro} for a textbook introduction to these equations and to \cite{Roberts:1994dr}
for a review of the application to hadron physics.

In recent times, as the situation with light scalars and particularly the $f_0(500)$ 
became clearer, these techniques have also been applied to these resonances.   
Thus, $\pi\pi$ scattering was studied within this formalism in \cite{Bicudo:2001jq,Cotanch:2002vj}
in a chirally symmetric way using the RL approximation. In \cite{Cotanch:2002vj}, when comparing 
the results with simple one-meson exchange amplitudes,
both a $\rho$ and a scalar meson were found around 741 and 670 MeV, respectively,
although with a very small width of 172 MeV for the $\sigma$. However,
it was pointed out that significant corrections beyond RL as well as from pion loops were expected.
Actually, in 
\cite{Chang:2009zb} it was confirmed that
by identifying the light scalars as quark-antiquark $^3P_0$ states, 
the lightest scalar meson came around 600-700 MeV within the RL  approximation, 
but it was then shown that beyond 
this approximation the lightest scalar
comes out between 1 and 1.1  GeV.  Relatively similar results were found in 
\cite{Fischer:2009jm}, using a different ansatz: The lightest scalar-isoscalar $q\bar q$ meson appears around 600 MeV within the RL approach, but around 800 to 880 MeV beyond that approximation. It should be noted that the $\rho(770)$ mass appears around 880 MeV too, although once again it was argued that ``pion-cloud'' effects may bring its mass 
closer to its physical value. Therefore, once more it seems that the physical $\sigma$ meson is far too light to be accommodated  as an ordinary $q\bar q$-meson within this formalism. 

Thus, the attention has also turned very recently to tetraquark operators
\cite{Heupel:2012ua} treated within a Bethe-Salpeter formalism. In particular, in 
\cite{Heupel:2012ua} 
the full four-body equation for the tetraquark was approximated by a coupled set of two-body equations with meson
and diquark constituents. In this way,  the lightest scalar tetraquark
was found around 400 MeV. It is important to remark that
its wave function was dominated by the $\pi\pi$ constituents,
in agreement with the $f_0(500)$ molecule picture. 
A very recent calculation that has appeared when finishing this report \cite{Eichmann:2015cra},
indicates that the full four-body calculation 
does not change much these results, with the lightest
tetraquark showing up again around 400 MeV and
with a meson-meson molecule interpretation still preferred.  

\subsection{The $\sigma\rightarrow \gamma\gamma$ decay}
\label{subsec:ggpipi}

The relevance of radiative decays of scalar mesons is that, naively, 
they would be proportional to the squared charges of the meson constituents as well 
as to the distribution of those constituents inside the meson. In this sense, by studying the $\sigma\rightarrow\gamma\gamma$ decay one may expect to learn  about the $f_0(500)$ composition. 
However, life is not that simple.
On the one hand, a  significant improvement in the
determination of the ``decay width'' has been achieved over the last decade, mainly due to new data and to the use of model independent dispersive techniques. On the other hand, as of today, the interpretation of the result in terms of constituents is inconclusive 
and still a matter of debate. If anything, it may once again suggest that intermediate $\pi\pi$ loops
play a very significant role.

Let us then review first  the determination of the ``decay width'':
\begin{equation}
\Gamma(\sigma\rightarrow\gamma\gamma)=\frac{\alpha^2\vert \sigma(M_\sigma)g_{\sigma\gamma\gamma}\vert^2}{4 M_\sigma^2}.
\label{ec:Gggdef}
\end{equation}
Note that that we write ``decay width'' between inverted commas, because, as correctly pointed out in the RPP ``Note on scalar mesons'' \cite{PDG12} and emphasized in almost every work on this issue, this equation actually corresponds to a narrow width approximation. Nevertheless, once the coupling is known, the above 
equation provides an unambiguous definition that, 
following \cite{Morgan:1987gv,Pennington:2006dg}, has become standard in all references.

Given that this is a narrow width approximation, in the last years 
it has also become usual to identify $M_\sigma$ with the real part of the $f_0(500)$ pole position, $s_\sigma$, as we have been doing throughout this review.
Next, the coupling is obtained as the
residue of the $f_0(500)$ pole in the second Riemann sheet of
the S-wave isospin zero $\gamma\gamma\pi\pi$ scattering amplitude. Around the pole
this amplitude can be parameterized as 
$F_{II}^{I=0}(s)\sim g_{\sigma\gamma\gamma}g_{\sigma\pi\pi}/(s-s_\sigma)$. 
Normalizations vary in the literature. Of course, one needs to know
$g_{\sigma\pi\pi}$, which in turn is obtained from the sigma-pole residue in 
the S-wave isospin zero $\pi\pi$ scattering partial wave, $t_0^{(0)}(s)$, thoroughly discussed in Sec.\ref{subsec:cfd} of this review. 
In particular, some recent determinations of $\vert g_{\sigma\pi\pi}\vert$ have been listed in Tables \ref{tab:dispersivepoles} and \ref{tab:otherpoles}, which are all fairly consistent with our conservative dispersive estimate $\vert g_{\sigma\pi\pi} \vert=3.45^{+0.25}_{-0.22}$ in Eq.\ref{myfullestimate}.

\begin{figure}
\centering
\includegraphics[scale=0.78]{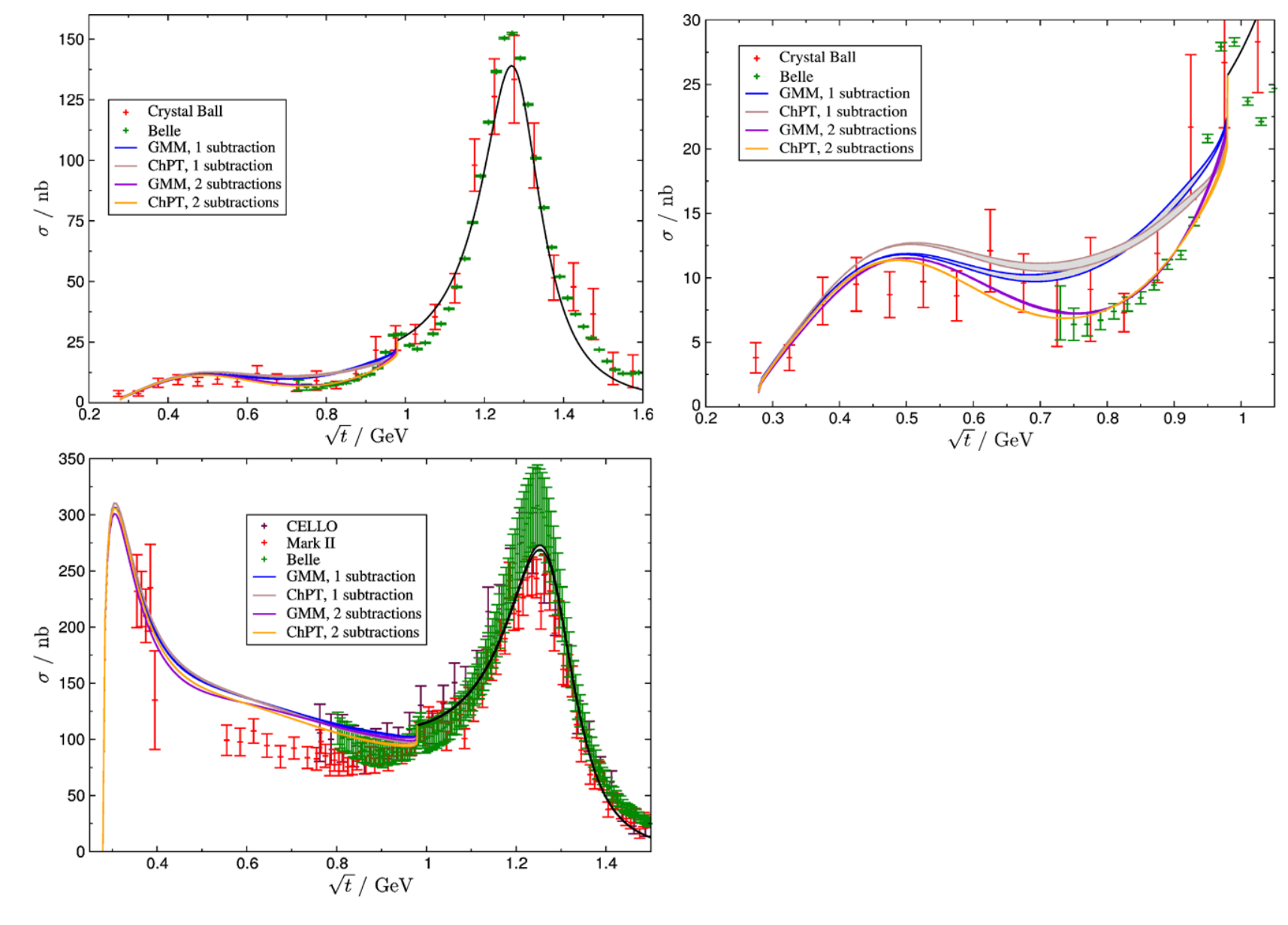}
 \caption{\rm \label{fig:ggpipi} 
Data on of $\gamma\gamma\rightarrow\pi\pi$ together with their description using a dispersive Roy-Steiner formalism.  
Left Panels:  Figure 5 from \cite{Hoferichter:2011wk}. "Total Cross sections for $\gamma\gamma\rightarrow\pi^0\pi^0$
\cite{Marsiske:1990hx,Uehara:2008ep}  and $\gamma\gamma\rightarrow\pi^+\pi^-$
 \cite{Boyer:1990vu,Behrend:1992hy,Mori:2006jj} for $\vert {\rm cos} \theta\vert\leq 0.8$ and 
 $\vert {\rm cos} \theta\vert\leq 0.6$, respectively". Right panel:  Figure 6 from \cite{Hoferichter:2011wk}. "Total cross section for $\gamma\gamma\rightarrow\pi^0\pi^0$ for 
 $\vert {\rm cos} \theta\vert\leq 0.6$ in the low energy region". With kind permission from Springer Science and Business media, Springer The European Physical Journal C, (2011) 71:1743, 
 "Roy-Steiner equations for $\gamma\gamma\rightarrow\pi\pi$", M. Hoferichter, D. Phillips and C. Schat, Figures 5 and 6.
}
\end{figure}

In 1990, data on $\gamma\gamma\rightarrow\pi^+\pi^-$ 
between 350 MeV  and 1.6 GeV was provided by the MarkII collaboration at SLAC \cite{Boyer:1990vu} and later on by the CELLO detector at PETRA above 750 MeV \cite{Behrend:1992hy}, whereas data on 
$\gamma\gamma\rightarrow\pi^0\pi^0$ 
from threshold to about 2 GeV
was published by the Crystal Ball Collaboration at DESY \cite{Marsiske:1990hx}. 
More recently, very high statistics analyses have been 
provided by the Belle Collaboration  for the charged final state 
 above 0.8 GeV \cite{Mori:2006jj}
as well as for the neutral one above 0.6 GeV \cite{Uehara:2008ep}.
These data can be found in Fig.\ref{fig:ggpipi}. 
Roughly speaking, data is dominated by  S-waves, which only couple to the two-photon 
helicity state with positive parity. In particular the prominent shape of the $f_2(1275)$
resonance can be easily identified.
The smaller D-waves also couple to the two-photon helicity state with
negative parity. 
The charged reaction, shown in the lower left panel,
 is dominated by the Born amplitude at low energies.
In contrast, the Born term is absent in $\gamma\gamma\rightarrow\pi^0\pi^0$,
 which roughly explains why its cross section, shown in the two upper panels,
is about an order of magnitude smaller at low energies. 
As a consequence, the strong rescattering effects become particularly relevant for the
$\pi^0\pi^0$ final state, making it 
remarkably sensitive to the $\sigma$ meson.

From the theory side the neutral case turns out to be particularly interesting because its lowest order within ChPT vanishes, making the NLO independent of the low-energy constants \cite{Bijnens:1987dc,Donoghue:1988eea}. However,
the NLO result does not yield a very good description of data, although it improves
considerably with unitarization techniques \cite{Donoghue:1988eea,Donoghue:1993kw,Oller:1997yg}, in which the $\sigma$ and/or the $f_0(980)$ are generated or, when going beyond 1 GeV, other resonances
are included explicitly. The interest on this process triggered the first
two-loop full calculation within ChPT \cite{Bellucci:1994eb}, which yields a fairly reasonable
agreement with data within the low-energy region. Of course, plain ChPT cannot reproduce 
the $\sigma$ pole and for this some dispersive or unitarization treatment is customarily used.

A relatively simple but rigorous formalism implementing
some basic dispersive constraints was already formulated 
in \cite{Morgan:1987gv} using the Mushkelishvili-Omn\`es method \cite{Omnes}.
This method has been applied with slight variations \cite{Donoghue:1993kw,Pennington:2006dg}.
For the sake of concreteness, we follow here \cite{Morgan:1987gv,Pennington:2006dg,Oller:2007sh,Pennington:2008xd,Mao:2009cc,GarciaMartin:2010cw,Moussallam:2011zg} and illustrate it with scalar waves, which are the most relevant for this report. As usual, the $\gamma\gamma\rightarrow \pi\pi$ 
partial wave $F^I(s)$, has a right cut above $s\geq 4M_\pi^2$ and 
a left one from $s=0$ to $-\infty$. Let us define a $L_I(s)$ 
function with the same left cut as $F^I(s)$, so that $F^I(s)-L^I(s)$ only has a right cut. As $s\rightarrow 0$ Low's theorem \cite{Low:1954kd} 
requires $F^I(s)$ to be equal to the
one-pion-exchange Born term, which therefore dominates $L^I(s)$ at low energies.
Now, Watson's final state theorem, requires
the phase of $F^I(s)$ to be the same as that of the corresponding $\pi\pi$ partial wave in the elastic regime (in principle up to multiples of $\pi$, but since both phases tend to zero at threshold this ambiguity is removed). This constraint can be implemented through 
the Mushkelishvili-Omn\`es function defined as:
\begin{equation}
\Omega^I(s)=\exp\Big[ \frac{s}{\pi} \int^\infty_{4M_\pi^2}\frac{\phi^I(s')}{s'(s'-s)}ds'\Big],
\end{equation}
where $\phi^I(s)$ is the phase of $F^I(s)$ along the right cut.
In particular, in the $\pi\pi$ elastic region the corresponding 
scattering phase shift is equal to $\phi^I(s)$. 
Now, a twice-subtracted dispersion relation for $(F^I(s)-L^I(s)) /\Omega^I(s)$, which only has right hand cut, can be written as:
\begin{equation}
F^I(s)=L^I(s)+s c_I\Omega^I(s)+\frac{s^2}{\pi}\Omega^I(s)\int^\infty_{4M_\pi^2}\frac{L^I(s')\sin \phi^I(s')}{s'^2(s'-s)\vert \Omega^I(s')\vert}.
\end{equation}
Note that Low's theorem \cite{Low:1954kd}, which implies $F^I(s)-L^I(s)\rightarrow0$
when $s\rightarrow 0$ has been used to get rid of a subtraction constant. 
The problem, of course, is to determine the subtraction constants as well as  $L^I(s)$ and $\phi^I(s)$ above the $K\bar K$ threshold.  

\begin{figure}
\centering
\includegraphics[scale=0.47]{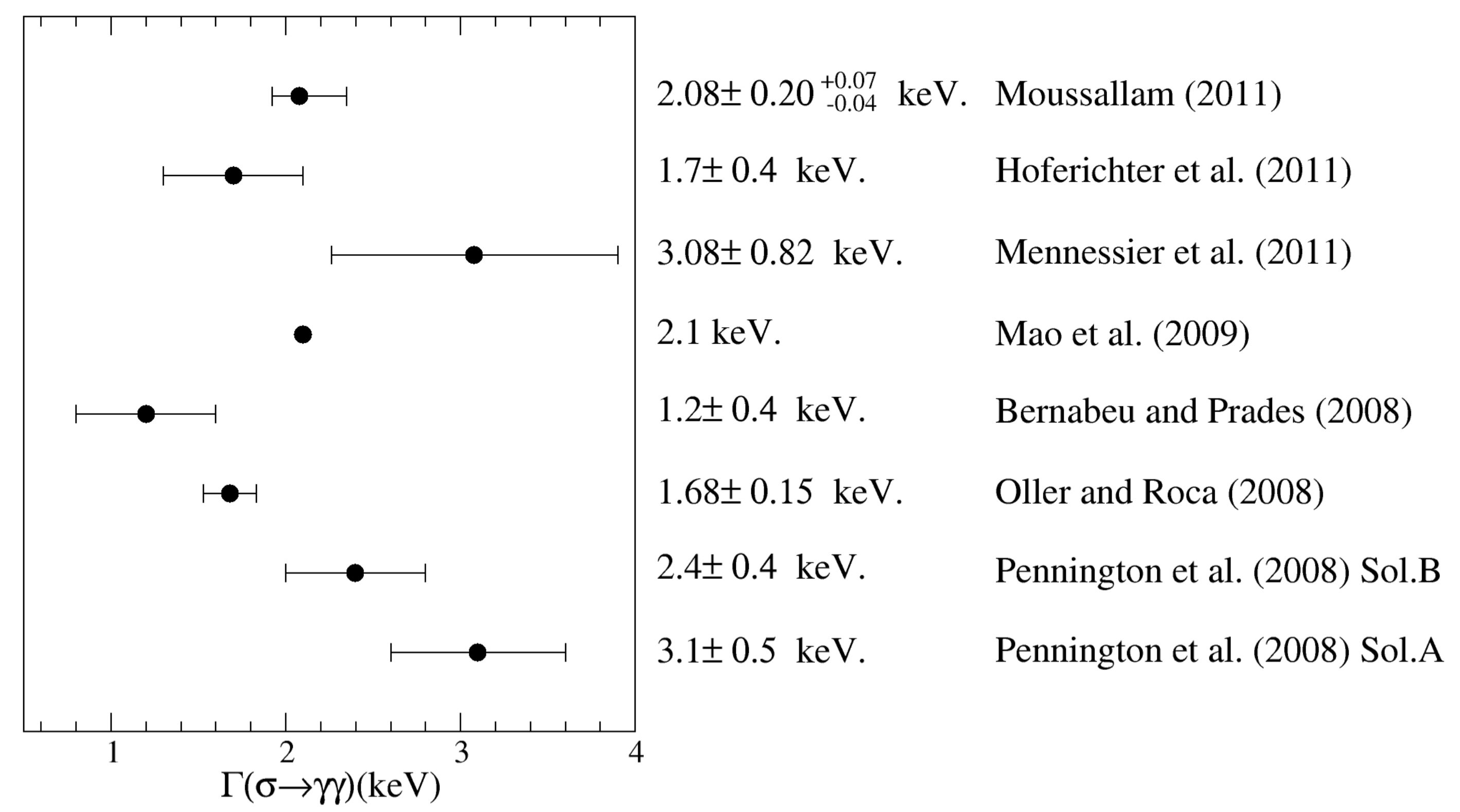}
 \caption{\rm \label{fig:Gsgg} 
Recent determinations of $\Gamma(\sigma\rightarrow\gamma\gamma)$. Only the latest updates of each group of authors are given: Pennington et al Sols. A and B \cite{Pennington:2008xd}, Oller and Roca \cite{Oller:2007sh}, Bernabeu and Prades \cite{Bernabeu:2008wt}, Mao et al. \cite{Mao:2009cc}, Mennessier et al. \cite{Mennessier:2010ij}, Hoferichter et al. \cite{Hoferichter:2011wk} and Moussallam \cite{Moussallam:2011zg}. All them use $\gamma\gamma\rightarrow\pi\pi$ data, except Bernabeu and Prades, which only use
the nucleon electromagnetic polarizabilities. 
}
\end{figure}

The precise $f_0(500)$ pole prediction in \cite{Caprini:2005zr} renewed the interest in this process and thus M. Pennington in 2006 \cite{Pennington:2006dg} fitted the Crystal Ball data using the 
formalism we have just described. In particular, he fixed  the Adler zero as in $\pi\pi$ scattering to LO in ChPT, used the $\pi\pi$ phase from a reevaluation of Roy equations and decay data,
and assumed that the left cut was only relevant up to $s\simeq -0.5\,\gev^2$, which in turn 
was dominated by the  Born term for pion exchange. 
With the dispersive representation it was possible to extract the pole and the residue and therefore, using Eq.\ref{ec:Gggdef},
a value of $\Gamma(\sigma\rightarrow\gamma\gamma)\simeq(4.09\pm0.29)\,{\rm keV}$ was found. A very similar approach was 
followed in 2008 \cite{Oller:2007sh}, which formally amounts to adding one more subtraction which results in a 10\% decrease of the ratio of residua.
In addition, a more elaborated estimation 
of the left cut contribution was also provided, showing the relevance of  axial-vector resonance exchange, which leads to an additional 10\% decrease in the radiative decay width. Moreover, 
when the value of $\vert g_{\sigma\pi\pi}\vert$ from the rigorous Roy Eq. analysis was used, an additional reduction of the order of 40\% occurred with
respect to \cite{Pennington:2006dg}, finally leading to $\Gamma\simeq 1.68\pm 0.15\,$keV. Actually, M. Pennington and collaborators, when 
analyzing the Belle results in that same year, but using updated values for the $\sigma$ pole parameters and including axial-vector exchanges, obtained two possible values: Solution A yields $\Gamma\simeq 3.1\pm 0.5\,$keV, whereas 
Solution B yields $\Gamma\simeq 2.4\pm 0.4\,$keV. Both significantly lower than the older estimate
in \cite{Pennington:2006dg} by M. Pennington alone.
In 2008 a different approach was presented in which, instead of fitting the $\gamma\gamma\rightarrow\pi\pi$ data to determine the subtraction constants and the left cut, these were constrained
from a sum rule involving nucleon electromagnetic polarizabilities, finding  
$\Gamma\simeq 1.2\pm 0.4\,$keV. Thus, this independent check also favored a relatively low 
value of the radiative width.
Later analyses of the Belle data using variations of the Mushkelishvili-Omn\`es 
method yielded values around
$\Gamma\simeq 2.1\,\kev$ \cite{Mao:2009cc,Moussallam:2011zg}.
This radiative width was also studied within 
a simple analytic K-matrix model \cite{Mennessier:2010ij}, yielding a somewhat larger value with a large uncertainty $\Gamma\simeq 3.08\pm 0.82\,$keV.
Nevertheless, it is worth remarking that the most advanced dispersive analysis to date 
 finds $\Gamma\simeq 1.7\pm 0.4\,\kev$ \cite{Hoferichter:2011wk}. It makes use of the
Roy-Steiner equations and therefore
implements model-independent dispersive constraints not only 
on the physical cut as in the Mushkelishvili-Omn\`es method, but
 also on the left cut, which makes it particularly reliable. 
This result should be taken as the reference value 
if we consistently follow the same criterion we applied to the determination of 
the $\sigma$ pole parameters and $g_{\sigma\pi\pi}$ coupling.
To summarize, in 
Fig.\ref{fig:Gsgg} we list the latest determinations of $\Gamma(\sigma\rightarrow\gamma\gamma)$
from different groups. Note that all values in that plot overlap within 1.5 standard deviations with that obtained using Roy-Steiner equations \cite{Hoferichter:2011wk}, which may be taken as a relatively conservative reference value. The only exception is Sol. A of Pennington et al. which overlaps at two, although this value is at odds with the nucleon polarizability sum rule result from \cite{Bernabeu:2008wt}.

However, despite having determined reliably $\Gamma(\sigma\rightarrow\gamma\gamma)$, 
a strong debate lingers on concerning the interpretation of these results in terms of 
the $\sigma$ composition. 

Let us start with the expected value of $\Gamma(\sigma\rightarrow\gamma\gamma)$ for 
ordinary quarkonia. Within vector meson 
dominance constrained with single quark selection rules this decay 
was studied many years ago in \cite{Babcock:1976hr} 
to find a ``nearly complete suppression of $\gamma\gamma$ couplings of  $0^{++}$ $q \bar q$ states'',
with their very crude estimate being $\simeq40$ eV.
 An actual calculation of a $0^+ \bar qq$ meson
decaying to $\gamma\gamma$ was described in \cite{Barnes:1985cy}. 
That work was not dedicated to
the $\sigma$, but to the $f_0(980)$, for which the $\bar qq$ composition 
yielded $\Gamma(f_0(980)\rightarrow\gamma\gamma\simeq 4.5\,$keV.
However,  in that work it was found that:
\begin{equation}
\frac{\Gamma(\bar qq (0^{++})\rightarrow\gamma\gamma)}{\Gamma(\bar qq (2^{++})\rightarrow\gamma\gamma)}=
\frac{15}{4}\left(\frac{M_{0^{++}}}{M_{2^{++}}}\right)^n,
\label{gamma02ratio}
\end{equation}
where the $15/4$ factor comes from the different polarization sums for each angular momentum resonance \cite{15/4}, 
with $n=3$ due to the short distance Coulombic component. This relation allows us to perform a modern
re-evaluation for the $f_0(500)$, by considering the present and very conservative value of
$M_\sigma\simeq 400-550\,$MeV in the RPP 2012 (see Eq.\ref{rpp2012sigmapole}),
so that the non-relativistic quark model of \cite{Barnes:1985cy} would yield today
 $\Gamma(\sigma\rightarrow\gamma\gamma)\simeq0.3-0.8\,\kev$. 
However, in \cite{Chanowitz:1988bp} and also within
the naive non-relativistic quark model, it was argued that the confining linear potential
would give $n=-1/3$. It was then suggested that the phenomenological exponent should lie somewhere in between. Thus, 
 based on the good agreement of the phenomenological choice $n=0$ 
for the $\rho$, $\omega$ and $\phi$, it was then estimated that
 $\Gamma(f_0(980)\rightarrow\gamma\gamma)\sim 10\, \kev$ for the $(u\bar u +d\bar d)/\sqrt{2}$ ( and 
$0.4 \,\kev$ for  $s\bar s$-like isoscalars) respectively. But if $n=0$ then we obtain the 
same prediction for the $\sigma$ meson: $\Gamma(\sigma\rightarrow\gamma\gamma)\sim 10\, \kev$.
However, in \cite{Bergstrom:1982qv} it was shown that the $15/4$ factor 
in Eq.\ref{gamma02ratio} could be reduced by a factor of two due to relativistic quark-model corrections.
Therefore the above numbers from the quark model should be roughly divided by 2. Still this 
leaves a prediction, within such a model, 
of  $\Gamma(\sigma\rightarrow\gamma\gamma)\simeq0.15-0.4\,\kev$ if $n=3$
or $\Gamma(\sigma\rightarrow\gamma\gamma)\simeq 5\,\kev$ if $n=0$. 
Moreover, the analysis of scalar-quarkonia decays into two photons 
was also carried out within the quark-level $L\sigma M$
 \cite{Kleefeld:2001ds}, which was applied to the $\sigma$ meson in 
\cite{vanBeveren:2002mc}. 
The NJL value $M_\sigma=2m_q$ yields a sigma mass of 675
   MeV (beyond the chiral limit), with a constituent quark mass of 337.5
   MeV given by the Goldberger-Treiman relation $m_q=g f_\pi$, where
   $g=2\pi/\sqrt{N_c} \approx 3.6$. 
Still one has to keep in mind that this is a manifestly real mass and 
does not correspond exactly to the real mass of the pole position that has been given in previous sections, since, as we have repeatedly emphasized, the $\sigma$ does not exhibit a Breit-Wigner shape.
Moreover,
the very authors have remarked the sensitivity of the decay to the mass of the $\sigma$ and
in a more recent evaluation \cite{vanBeveren:2008st} they find that 
$\Gamma(\sigma\rightarrow\gamma\gamma)\simeq 0.7\,\kev$ when using $M_\sigma=440\,\mev$. 
One of the relevant observations within this 
framework
is that two-pion loops play a very significant role increasing $\Gamma(\sigma\rightarrow\gamma\gamma)$. This is in line with the 
findings by the same group and collaborators (see Sec.\ref{subsec:uqm}),
that the $\sigma$ pole position in the real world is largely due to the strong rescattering or unitarization effects, with the ``bare'' pole being at a mass around 1 GeV. Another interesting observation 
about the two-photon quarkonium decay \cite{Giacosa:2007bs} is that 
a term in the triangle quark-loop diagram, which is often omitted in the literature, results in destructive interference and as a consequence the two photon decay of the $\sigma$ as pure quarkonium has to be $\Gamma(\sigma\rightarrow\gamma\gamma)<1 \,\kev$, for $M_\sigma=0.40-0.6\,\mev$ \footnote{Note that in the quark-level 
$L\sigma M$ of \cite{vanBeveren:2002mc,vanBeveren:2008st} this term vanishes due to the 
choice $M_\sigma=2 m_q$, so that their result is not affected by this correction.}.
In summary, all modern calculations within the pure quarkonium interpretation and the present
sigma mass,  yield a too low
value of $\Gamma(\sigma\rightarrow\gamma\gamma)<1 \,\kev$, although it could be made larger 
considering pion loops.

Let us then turn to the tetraquark and molecule configurations. 
Also in \cite{Barnes:1985cy}, the $f_0(980)$ case
was studied within 
a $\bar KK$-molecule model where $\Gamma(f_0(980)\rightarrow\gamma\gamma)\simeq 0.6\,$keV,
which is favored by the existing bound from Crystal Ball 
$\Gamma(f_0(980)\rightarrow\gamma\gamma)\simeq 0.8\,$keV. However, 
it is unclear how this molecular description could be translated to the sigma case if made of pions, since the $\sigma$ pole is above threshold, or to a $\bar KK$ molecule with such a big binding energy 
and such a large decay to the two-pion state.
The $f_0(980)$ 
radiative width could also be explained within the tetraquark model of \cite{Achasov:1982bt},
where it was estimated $\Gamma(f_0(980)\rightarrow\gamma\gamma)\simeq  0.27 \,\kev$. 
Concerning the $\sigma$, this latter group,  in a more recent analysis \cite{Achasov:2007fz} 
within the $SU(2)$ $L\sigma M$, which agrees ``qualitatively'' with the poles obtained in advanced dispersive analyses,  
showed that when separating within that model the dominant meson-meson loop contribution to $\Gamma(\sigma\rightarrow\gamma\gamma)$
from the ``direct coupling'' to the $\sigma$, the direct
coupling would yield $\Gamma(\sigma\rightarrow\gamma\gamma)\simeq 0.0034 \,\kev$ in agreement
with the prediction in \cite{Achasov:1982bt}.
Therefore, within that model,
a possible inner tetraquark structure could be hidden in a  
meson-meson component, consistently with the existing data.

Within the analytic K-matrix model of \cite{Mennessier:2008kk}
 and particularly in its recently improved version \cite{Mennessier:2010ij} 
an ``important contribution from meson loops'' is found in the determination of $\Gamma(\sigma\rightarrow\gamma\gamma)$. Within this model the authors separate a ``direct'' width 
$\Gamma(\sigma\rightarrow\gamma\gamma)^{\rm dir}\simeq0.16\pm0.03\,\kev$ and a 
``rescattering'' width $\Gamma(\sigma\rightarrow\gamma\gamma)^{\rm dir}\simeq1.89\pm0.81\,$keV.
\footnote{Both mechanisms interfere and the total width is not the sum of each partial width.}
From this the authors conclude that since such a large direct width is hard to obtain from quarkonia or tetraquark 
components, then the $\sigma$ may have a large glueball component. Although some glueball component
cannot be discarded, we have already discussed that a dominant glueball component in the $\sigma$ seems 
hard to reconcile with its large width, lattice results, or the existence of the kappa.
Moreover, one should recall that the ``direct versus rescattering'' separation is model dependent.
Furthermore, when extracting the total width from data with this model these authors obtain one of the largest values presently available in the literature: 
$\Gamma(\sigma\rightarrow\gamma\gamma)\simeq  3.08\pm0.82 \,\kev$, whereas  
model-independent dispersive approaches favor a much lower value (see Fig.\ref{fig:Gsgg}).

Finally, within the chiral unitary approach update in  \cite{Oller:2007sh} 
it is also possible to describe the $\gamma\gamma\rightarrow\pi\pi$ data and obtain a 
theoretical value of $\Gamma(\sigma\rightarrow\gamma\gamma)\simeq0.168\pm0.15\,\kev$.
Within this approach the $\sigma$ is basically a dynamically generated state from the meson-loops
and it explains the bulk of the width obtained by the same group from a Mushkelishvili-Omn\'es 
dispersive treatment of the existing data. No trace of further components beyond meson loops is seen.

To summarize this section, in principle the two-photon radiative width of the $f_0(500)$
could provide a test on its composition. After some initial determinations that got higher values,
the most recent dispersive determinations of $\Gamma(\sigma\rightarrow\gamma\gamma)$ seem to prefer lower values between 1 and 2.5 keV. Concerning the $\sigma$ composition, 
all analyses seem to indicate a very important, frequently dominant,  
contribution of meson-loops to this value. This confirms the important role of such mesons in the $\sigma$ properties.
Depending on what value of $\Gamma(\sigma\rightarrow\gamma\gamma)$ is chosen and how much
of it is left when the meson-loop
contribution is removed in each model, 
different groups advocate different additional or ``direct'' substructure, which always comes out smaller than the meson-loop contribution.
This ranges from no observed substructure to compatibility with tetraquarks or gluonia. In general 
quarkonia appears somewhat disfavored. Therefore, at this stage no conclusive 
statements can be made apart from the large contribution from meson loops.

\color{black}

\section{SUMMARY}

Before stating the conclusions in the next section, let us first summarize this review. The
$\sigma$ meson, nowadays called $f_0(500)$,  plays a relevant role in our understanding of nucleon-nucleon attraction and the spontaneous symmetry breaking of QCD. In addition, it is a firm candidate for a non-ordinary meson, in the sense that it is not intuitively made of a quark and an antiquark.
The progress on our understanding of this meson is a tale of cumulative theoretical and experimental efforts that started almost 60 years ago. In Sect.~\ref{sec:intro}  we have provided a historical perspective on the longstanding controversy about the existence, parameters, classification  and nature of the $\sigma$. This frequently confusing 
story as been partially illustrated by following the developments in the Review of Particle Physics (RPP). With this criterion, the existence of a relatively light and broad scalar meson was finally settled in the mid 90's, although a very large uncertainty in its parameters was kept, i.e. the RPP estimated range for its mass was 400 to 1200 MeV. Nevertheless, by that time the scalar meson physics community was already working with a light sigma in the 400-600 MeV range. It was only the experimental 
confirmation from heavy meson decays that triggered
the change of name to $f_0(600)$ in 2002. However,  the use of too simple and model-dependent parameterizations (Breit-Wigner shapes, isobar models, etc) in these analyses seemed not convincing enough to change the very large uncertainty attached to its parameters.
Finally, a major revision has taken place in the 2012 RPP edition. The present RPP values of the $\sigma$ mass and width, obtained from its pole position, are $\sqrt{s_\sigma}\simeq M_\sigma- i \Gamma_\sigma/2=(400-550)-i(200-350)\,$MeV. Accordingly,
the name of the resonance has been changed to $f_0(500)$. This is a very welcome improvement which has motivated the writing of this report.

Thus, we have devoted Sec.\ref{sec:parameters} to explain the experimental and theoretical developments that have triggered this dramatic change in the RPP. On the experimental side, the 2010 results on  $K\rightarrow\pi\pi e \nu$ from NA48/2 at CERN have provided very precise
low energy data on scalar-isoscalar $\pi\pi$ scattering, where the $\sigma$ resonance appears. Models incompatible with these data have been discarded from the RPP $\sigma$ parameter determination. On the theoretical side, in the RPP it is also suggested that a more ``radical point of view" can be 
taken by considering only the ``most advanced dispersive analysis". The reason is that dispersion theory provides the only model-independent
and consistent extension to the complex plane in order to determine precisely the $\sigma$ pole present in the amplitudes describing the existing data. Most of Sec.\ref{sec:parameters}
has been dedicated to introduce both the data and these rigorous dispersive techniques, including a detailed account of their uncertainties. 
As a result, we have proposed the following  ``Conservative Dispersive Estimate" of the $f_0(500)$ parameters
\begin{equation}
\sqrt{s_\sigma}\simeq M_\sigma- i \Gamma_\sigma/2=449^{+22}_{-16}-i(275\pm12)\, {\rm MeV},
\end{equation}
which is a more conservative estimate than the ``radical" one at the RPP 2012, since we have taken into account
the systematic origin of some of the uncertainties in different dispersive analysis.
As explained in the main text, {\it this estimate is based on precise data and dispersive techniques, which guarantee a model-independent 
extraction of the pole, whose parameters are also process-independent.} The ``most advanced'' dispersive analyses, according to the RPP, are included within these uncertainties.
Realistic models of the $\sigma$ meson should be consistent with this conservative estimate, or at least with the precise scattering data
from which it is extracted.

Sec.\ref{sec:chiralsigma} is dedicated to the relation between the $\sigma$ meson and chiral symmetry. 
The classic Linear Sigma Model is presented first as a pedagogical introduction to chiral symmetry, although explaining the reasons why we already know that {\it it does not correspond to the low energy effective theory of QCD}. Nevertheless such an effective theory exists and is known under the name of Chiral Perturbation Theory (ChPT), which is introduced next. ChPT provides the most general low energy expansion of amplitudes
in terms of pions, kaons and the eta, which is consistent with QCD symmetries. The structure and meaning of the ChPT low energy constants and scattering amplitudes is then explained. Since a series expansion cannot generate resonance poles, 
the largest fraction of Sec.\ref{sec:chiralsigma} has been dedicated to different unitarization 
techniques of the ChPT series. These methods are able to generate the poles associated to lightest resonances, while satisfying 
unitarity and matching the ChPT expansion at low energies. We have shown that they all basically yield the same results as far as the $\sigma$ and other light scalars are concerned, whose existence received a strong support with these techniques.
Unitarized ChPT methods are not as well suited for precision studies
as the rigorous dispersive approaches of Sec.\ref{sec:parameters}, but provide a connection with QCD through their matching with ChPT parameters. These techniques clearly show that the dynamics responsible for the formation of light scalars is dominated by
meson loops, i.e., rescattering, in contrast with other conventional resonances, like vectors or heavier scalars, 
which owe their existence to quark-level dynamics. Actually, unitarized ChPT in a coupled channel formalism, strongly suggests that 
the  light scalar nonet is formed by the $f_0(500)$, $f_0(980)$, $a_0(980)$ and $K_0^*(800)$ resonances,
in good agreement with very early claims about the non-$\bar qq$ structure of these mesons. 
Moreover, within this formalism,the octet members have been shown to become degenerate in the SU(3) limit. It also allows for a study of the mass dependence of light resonances, which shows some encouraging agreement with recent lattice results. This dependence seems to indicate that at very high pion masses the $f_0(500)$ would become a virtual bound state first and then  a $\pi\pi$ molecule.
 
 The rest of Sect.\ref{sec:chiralsigma} reviews  other popular models. These are based on 
 chiral Lagrangians with a priori choices of fields and couplings. Despite being comparatively simpler than previously described
  approaches, and lacking a power counting full generality, some of them are able to describe rather nicely  the existing data on masses and decays, and sometimes
  meson-meson scattering data in the scalar waves.  They are particularly interesting to illustrate some of the
  mechanisms, like mixing between different states, that could be at work in the scalar sector. The most successful models suggest the existence
  of two scalar nonets.  The lightest nonet would appear below 1 GeV, once again formed by the $f_0(500)$, $f_0(980)$, $a_0(980)$ and $K_0^*(800)$,
  whose predominant composition involves two quarks and two antiquarks  at the microscopic level, although their actual 
configurations as classic tetraquarks, diquark-antidiquarks or 
meson-meson molecules cannot be discerned by chiral 
symmetry transformation properties alone. The heavier nonet would have masses in the 1.3-1.7 GeV region and would be 
predominantly of a conventional  $\bar qq$ nature.
 Moreover, all these models involve some degree of 
mixing between these two nonets and possibly a glueball state, also with a mass well above 1 GeV.
 We also reviewed recent models
dealing with the relation of the $f_0(500)$ with conformal symmetry.
 
 In Sect.\ref{sec:nature}, we have first studied the information on the $\sigma$ nature that comes from 
the $1/N_c$ expansion of QCD, both around the physical value of $N_c=3$ and at larger values. This expansion strongly disfavors that the dominant component of the $f_0(500)$ and other scalars might be a glueball or the most straightforward generalization of a tetraquark to arbitrary $N_c$. This is done in a model-independent way. In addition, we discuss the very strong evidence, obtained from ChPT combined with dispersion theory, that the $f_0(500)$ predominant component is not of an ordinary nature. This is also confirmed by a recent calculation of the $f_0(500)$ Regge trajectory, which comes out non-linear 
with scales more typical of meson physics than quark-level dynamics.
Finally, when larger values of $N_c$ are considered, there is a hint 
of an ordinary-meson subdominant component inside the $f_0(500)$, but 
with a mass around 1 GeV or higher. This is consistent with most chiral 
and microscopic models and it also solves naturally a possible problem with semi-local duality.
 
 In the second part of  Sect.\ref{sec:nature} we have addressed the composition of the $\sigma$ in terms of quarks and gluons.
Unfortunately, lattice calculations are still of little help 
in scalar meson physics, although there are promising results 
indicating that this may change in the future. Nevertheless, they strongly disfavor a glueball interpretation for the $f_0(500)$. Moreover the recent  
dispersive calculation that confirms the existence of a $K_0^*(800)$ state,
very similar to the $f_0(500)$ also plays strongly against a 
dominant or even sizable  glueball component.
Thus, in the final part of Sect.4, we have very  briefly reviewed some of the most widely used 
models or those more representative of different approaches, that provide a description of light scalar mesons at the quark level. 
These models usually contain a QCD-inspired 
confining potential and some phenomenological coupling to take into account
hadronization and rescattering in decays. 
The most popular and successful models also agree with 
the chiral descriptions reviewed in the previous section in the need for two scalar nonets: a lighter one, predominantly  involving two quarks and two antiquarks, and a heavier conventional $\bar qq$ nonet, which are nevertheless mixed.
Many of these models describe mixing patterns before the hadronization process,
for bare states, and therefore their conclusions, although illustrative, should be interpreted cautiously. Those quark models that also describe rescattering via some unitarization technique, show that when the meson-loops are removed, 
the lightest scalar and the $\sigma$ poles move to much higher masses than the observed ones. This 
suggests again that at least in the $\sigma$ there is 
predominant molecular/tetraquark component mixed with an ordinary $\bar qq$ component
with a mass higher than 1 GeV. In addition, in this Section we have also reviewed the sum rule and Bethe-Salpeter or Schwinger-Dyson results, which, once again seem to disfavor the quarkonia interpretation in terms of a non-ordinary behavior, typically of a tetraquark, or meson-meson nature. Finally, we have reviewed the radiative decay of the sigma meson. Over the last decade a considerable improvement has been achieved in the determination of the decay into two photons, in part due to new data, but also to the appearance of more sophisticated and detailed approaches, particularly those based on dispersion theory, which are model independent and have basically settled its value. Unfortunately the interpretation of this value in terms of the inner composition of the sigma in terms of quarks and gluons is still under debate. Generically, pure quarkonia, gluonia or tetraquark estimates come somewhat low and, once again, the role of pion-pion loops or component seems very relevant.

\section{CONCLUSIONS}

The $\sigma$ has been a controversial state for almost six decades, due to the 
use of models and the scarce and sometimes conflicting sets of  data. There has been an intense debate over its
existence, its parameters and its nature, to which many experimental and theoretical groups have contributed. 
Here we have reviewed the present status of our knowledge, paying particular attention to the most recent developments, and
 the following conclusions can be reached:

\begin{itemize}
\item[1)] The existence of the $\sigma$ meson was already settled in the mid 90's. This was achieved from
cumulative evidence from better models with chiral symmetry and unitarity as well as new data on heavy meson decays.

\item[2)] Over the last decade, even better data from different sources, particularly the $\pi\pi$ scattering 
low energy data coming from $K_{e4}$ decays, as well as  model independent dispersive analyses 
have allowed for precise and rigorous determinations of the $\sigma$ parameters. This has led to a major revision 
of the $\sigma$ uncertainties in the Review of Particle Properties. Even its name has changed to $f_0(500)$. In view of the existing
model-independent dispersive analysis we have left the previously existing confusion behind and entered a new era of precision.
For this reason in this review we have strongly encouraged the use of pole parameters, for which we 
have provided
a conservative dispersive estimate.
This should be taken into consideration in future model building or experimental analysis. In particular,
too simple parameterizations do not describe well our present knowledge and should be avoided.

\item[3)] Concerning its spectroscopic classification, there is a well-established picture in which it belongs to a scalar
nonet together with the $a_0(980)$, $f_0(980)$, and $K_0^*(800)$. 
The latter resonance is strongly supported by recent experimental data,
appears naturally in unitarized Chiral Perturbation Theory
and in a rigorous dispersive formalism, 
which implies that models without a $K_0^*(800)$ are simply inconsistent with causality and unitarity. Together with lattice results showing that the lightest glueball lies above 1GeV, this no longer 
leaves room for a dominant glueball interpretation of the $f_0(500)$. 

\item[4)] Concerning its nature, all approaches respecting unitarity as well as some basic requirements of analyticity and chiral symmetry,
indicate that meson-meson dynamics is relevant for the generation of the physical $\sigma$. 
Its $N_c$ and Regge behaviors have been shown not to correspond predominantly to ordinary $\bar qq$ mesons.
The evidence in this respect is very robust and compelling. Most chiral-meson and quark-level  models indicate also
that the main component of the members of this lightest nonet involves some arrangement of two quarks and two antiquarks.

\item[5)] In addition, unitarized ChPT in coupled channels, other chiral Lagrangian approaches as well as most quark-level models
 suggest that the lightest scalar nonet may have some mixing with a heavier scalar nonet above 1 GeV and possibly with a glueball state. 
There are also hints of this scenario from the $f_0(500)$ large-$N_c$ behavior, but not so compelling.
The details of this two-nonet mixing mechanism, the assignment of observed resonances to the second nonet, as well as the additional mixing with a possible glueball candidate still depend strongly on the model.

\end{itemize}

Future developments are expected from lattice QCD, further confirmation of the $K_0^*(800)$ parameters, decays of 
light scalar mesons as well as the construction or update of models aiming at a more realistic and precise description of the $f_0(500)$ and 
their multiplet partners. 

\begin{flushright}
{\it ``It was easier to know it than to explain why I know it.''\\
Sherlock Holmes Quote.\\
A Study in Scarlet, Sir Arthur Conan Doyle 1886. }
\end{flushright}

\section*{Acknowledgements}
I want to especially thank J.Nebreda for her very careful reading of the whole manuscript,
her endless patience with corrections and help with many figures,
as well as J. Ruiz de Elvira and F. J. Llanes-Estrada for carefully reading and helping me 
to correct parts of it.
I am particularly indebted to C. Hanhart for his 
critical reading of most of the manuscript, particularly those sections
concerning the RPP and chiral symmetry as well as for his suggestions and constructive feedback.
Many thanks also to 
J.A. Oller and E. Oset for their careful reading of the whole manuscript, his numerous corrections and their suggestions and clarifications 
in several specific sections: the former on the
different mixing schemes, the chiral unitary approach and the radiative sigma decay,
and the latter  on the section of the sigma from decays of heavier mesons.
Thanks also to A. G\'omez Nicola for 
his advice with the chiral symmetry restoration subsection and the references therein.
In addition, and more or less in order of appearance at different stages of the elaboration of the manuscript, I also thank I. Caprini for her comments on rigorous analytic continuations and extractions of the $\sigma$ pole,  H. Leutwyler for pointing out the usefulness of Roy Eqs. in different aspects, 
and the two of them together with G. Colangelo for kindly providing their
files to display the Regge behavior. Thanks also to: J. Nieves for his explanations on his results on the $N_c$ behavior, F. Giacosa and D. Parganlija for their explanations of their extended L$\sigma$M, M. Nielsen for some clarifications on sum rules as well as G.Rupp for his comments on the Quark-level L$\sigma$M, the
Chiral Quark Model and on NJL models in general. I also want to thank W. Ochs for his
comments on the sigma as a glueball, on his critical view of the $\kappa$ resonance situation as well as on the difficulties of his Ph.D. thesis $\pi\pi$ scattering analysis with the CERN-Munich Collaboration, which is the closest one to the Madrid-Krakow final dispersive result. 
This work has been partially supported by the Spanish projects FPA2011-27853-C02-02, 
FPA2014-53375-C2-2 and FIS2014-57026-REDT.

\section*{References}

\begin{flushright} 
\begin{minipage}[8cm]{12.1cm}
{\it \small ``Now let us come to those references to authors which other books have, and you want for yours. The remedy for this is very simple: You have only to look out for some book that quotes them all, from A to Z as you say yourself, and then insert the very same alphabet in your book, and though the imposition may be plain to see, because you have so little need to borrow from them, that is no matter; there will probably be some simple enough to believe that you have made use of them all in this plain, artless story of yours. At any rate, if it answers no other purpose, this long catalogue of authors will serve to give a surprising look of authority to your book. Besides, no one will trouble himself to verify whether you have followed them or whether you have not,...'' 
\begin{flushright} El ingenioso hidalgo don Quijote de la Mancha.
Miguel de Cervantes 1605\\
Translated to English in 1885 by John Ormsby.
\end{flushright}
}
\end{minipage} 
\end{flushright}

\end{document}